\pdfoutput=1

\documentclass[12pt,twoside,singlespace]{mitthesis}
\usepackage{lgrind}
\usepackage{pdfpages}
\usepackage{cmap}
\usepackage[T1]{fontenc}
\def\all{all}
\ifx\files\all  \else 

\usepackage{amsmath}
\usepackage{amssymb}
\usepackage{braket}
\usepackage{tikz-feynman}
\usepackage{contour}
\usepackage{caption}
\usepackage{subcaption}
\usepackage{bm}
\usepackage{dsfont}
\usepackage{bbold}
\usepackage{rotating}
\usepackage{booktabs}
\usepackage{multirow}
\usepackage[colorlinks]{hyperref}
\hypersetup{
  pdfcreator = {},
  pdfproducer = {}
}
\usepackage{mathastext}
\usepackage[noabbrev]{cleveref}
\usepackage{cite}

\begin{document}

\title{Experimental and Phenomenological Investigations of the MiniBooNE Anomaly}

\author{Nicholas Kamp}
       \prevdegrees{B.S.E., University of Michigan (2019)}
\department{Department of Physics}

\degree{Doctor of Philosophy}

\degreemonth{June}
\degreeyear{2023}
\thesisdate{May 23, 2023}

\copyrightnoticetext{\copyright~2023 Nicholas Kamp.
All rights reserved.

The author hereby grants to MIT a nonexclusive, worldwide, irrevocable, royalty-free license to exercise any and all rights under copyright, including to reproduce, preserve, distribute and publicly display copies of the thesis, or release the thesis under an open-access license.}

\supervisor{Janet M. Conrad}{Professor}

\chairman{Lindley Winslow}{Associate Department Head of Physics}

\maketitle



\cleardoublepage
\setcounter{savepage}{\thepage}
\begin{abstractpage}
%
%
%

The $4.8\sigma$ excess of electron neutrino-like events reported by the MiniBooNE experiment at Fermilab's Booster Neutrino Beam (BNB) is one of the most significant and longest standing anomalies in particle physics.
This thesis covers a range of experimental and theoretical efforts to elucidate the origin of the MiniBooNE low energy excess (LEE).
We begin with the follow-up MicroBooNE experiment, which took data along the BNB from 2016 to 2021.
The detailed images produced by the MicroBooNE liquid argon time projection chamber enable a suite of measurements that each test a different potential source of the MiniBooNE anomaly.
This thesis specifically presents MicroBooNE's search for $\nu_e$ charged-current quasi-elastic (CCQE) interactions consistent with two-body scattering.
The two-body CCQE analysis uses a novel reconstruction process, including a number of deep-learning based algorithms, to isolate a sample of $\nu_e$ CCQE interaction candidates with $75\%$ purity.
The analysis rules out an entirely $\nu_e$-based explanation of the MiniBooNE excess at the $2.4\sigma$ confidence level.
We next perform a combined fit of MicroBooNE and MiniBooNE data to the popular $3+1$ model; even after the MicroBooNE results, allowed regions in $\Delta m^2$-$\sin^2 2_{\theta_{\mu e}}$ parameter space exist at the $3\sigma$ confidence level.
This thesis also demonstrates that, due to nuclear effects in the low-energy cross section behavior, the MicroBooNE data are consistent with a $\overline{\nu}_e$-based explanation of the MiniBooNE LEE at the $<2\sigma$ confidence level.
Next, we investigate a phenomenological explanation of the MiniBooNE excess involving both an eV-scale sterile neutrino and a dipole-coupled MeV-scale heavy neutral lepton (HNL).
It is shown that a 500~MeV HNL can accommodate the energy and angular distributions of the LEE at the $2\sigma$ confidence level while avoiding stringent constraints derived from MINER$\nu$A elastic scattering data.
Finally, we discuss the Coherent CAPTAIN-Mills (CCM) experiment--a 10-ton light-based liquid argon detector at Los Alamos National Laboratory.
The background rejection achieved from a novel Cherenkov-based reconstruction algorithm will give CCM world-leading sensitivity to a number of beyond-the-Standard Model physics scenarios, including dipole-coupled HNLs.

\end{abstractpage}


\cleardoublepage

\section*{Acknowledgments}

Development from a starry-eyed first year graduate student into a competent researcher is like the MSW effect: it doesn't happen in a vacuum.
There are many people who have helped me along the way in both physics and life, without whom I would never have gotten to the point of writing this thesis.

First and foremost, I owe an immense debt of gratitude to Janet Conrad.
Janet has been an incredible mentor to me during my time as a graduate student; her advice and wisdom have helped me become the scientist I wanted to be when I came to MIT four years ago.
I'd like to specifically thank Janet for taking my interests seriously and working with me to develop research projects that matched them.
Creativity, ingenuity, enthusiasm, and kindness run rampant in the Conrad group--I will always be grateful for being offered a spot in it.
I look forward to many years of fruitful collaboration to come.

To my partner, Wenzer: thank you for the love, support, and patience over the last two years.
Life is not so hard when we can act as a restoring force for one another--I am grateful for having been able to rely on it while writing this thesis.
I look forward with great excitement to our future adventures together.

Thank you to my past mentors: Christine Aidala for introducing me to particle physics through POLARIS, Robert Cooper for asking me about my research at APS DNP 2016, Bill Louis for answering my many questions about neutrino physics in my first summer at LANL, Richard Van de Water for teaching me to follow the data (which greatly influenced my choice of graduate research), and Josh Spitz for helping develop confidence as a neutrino physicist on JSNS$^2$.
I'd also like to thank Christopher Mauger and the rest of the Mini-CAPTAIN team for an introduction to what it takes to run a particle physics experiment at an accelerator.

Thank you to members of the Conrad group past and present: those I worked with, Lauren Yates, Adrian Hourlier, Jarrett Moon, Austin Schneider, Darcy Newmark, Alejandro Diaz, and John Hardin, for being fantastic collaborators, and those I did not, Loyd Waits, Joe Smolsky, Daniel Winklehner, Philip Weigel, and Josh Villareal, for making MIT a brighter place.
Thank you especially to Austin for your infinite patience in answering my many questions on statistics and software--each of our projects has been a great pleasure.

To my MicroBooNE collaborators not mentioned above, Taritree Wongjirad, Katie Mason, Joshua Mills, Polina Abratenko, Ran Itay, Mike Shaevitz, Georgia Karagiorgi, Davio Cianci, Rui An, and everyone else: thank you for your excellent teamwork in putting together the Deep Learning analysis.

To my CCM collaborators not mentioned above, Edward Dunton, Mayank Tripathi, Adrian Thompson, Will Thopmson, Marisol Ch\'avez Estrada, and everyone else: thank you for the invigorating research and discussion over the last couple of years, and best of luck with CCM200!

Thank you to Carlos Arg\"uelles, Mike Shaevitz, Matheus Hostert, Stefano Vergani, and Melissa Uchida for your excellent mentorship and collaboration in our phenomenological endeavors together.
Thank you specifically to Carlos for giving me a welcome introduction to neutrino phenomenology, Mike for ensuring the robustness of each analysis, and Matheus for patiently answering my many model-building questions.

To all of my friends not mentioned above; Jack, Ryan, Alexis, Melissa, Ben, Bhaamati, Charlie, Vincent, Patrick, Caolan, Ouail, Artur, Sam, Zhiquan, Rebecca, Felix, Lila, Rahul, Brandon, Field, Kelsey, Woody, Joey, Rory, Cooper, Daniel, Kaliro\"e, Elena, and everyone else: thank you for all of the great memories--climbing, hiking, skiing, playing music, eating, drinking, commiserating, and laughing--over the past four years.
Thank you especially to the last three for making preparation for the oral exam significantly more enjoyable.

Thank you to the MIT administrative staff, including (but not limited to) Lauren Saragosa, Karen Dow, Catherine Modica, Sydney Miller, Alisa Cabral, and Elsye Luc, for helping make graduate school a more manageable endeavor.
Thank you also to the rest of my thesis committee, Joseph Formaggio and Washington Taylor, for helping me get through the final part of graduate school.

Finally, thank you to my parents, Jim and Carla, my siblings, Serafina and Daniel, and the rest of my family.
From elementary school science fairs to Saturday Morning Physics to today, none of this would have been possible without your love and support.


\pagestyle{plain}
\tableofcontents
\newpage
\listoffigures
\newpage
\listoftables

\chapter{Introduction}
\label{ch:intro}

We begin with a brief primer on neutrinos, the surprises they have given physicists throughout recent history, and the mysteries that remain today.
Readers already familiar with the mathematical details of massive neutrinos and the Standard Model may wish to read only \cref{sec:history} and \cref{sec:anomalies} before continuing. 

\section{A Brief History of the Neutrino} \label{sec:history}

\begin{figure}
    \centering
    \includegraphics[width=0.6\textwidth]{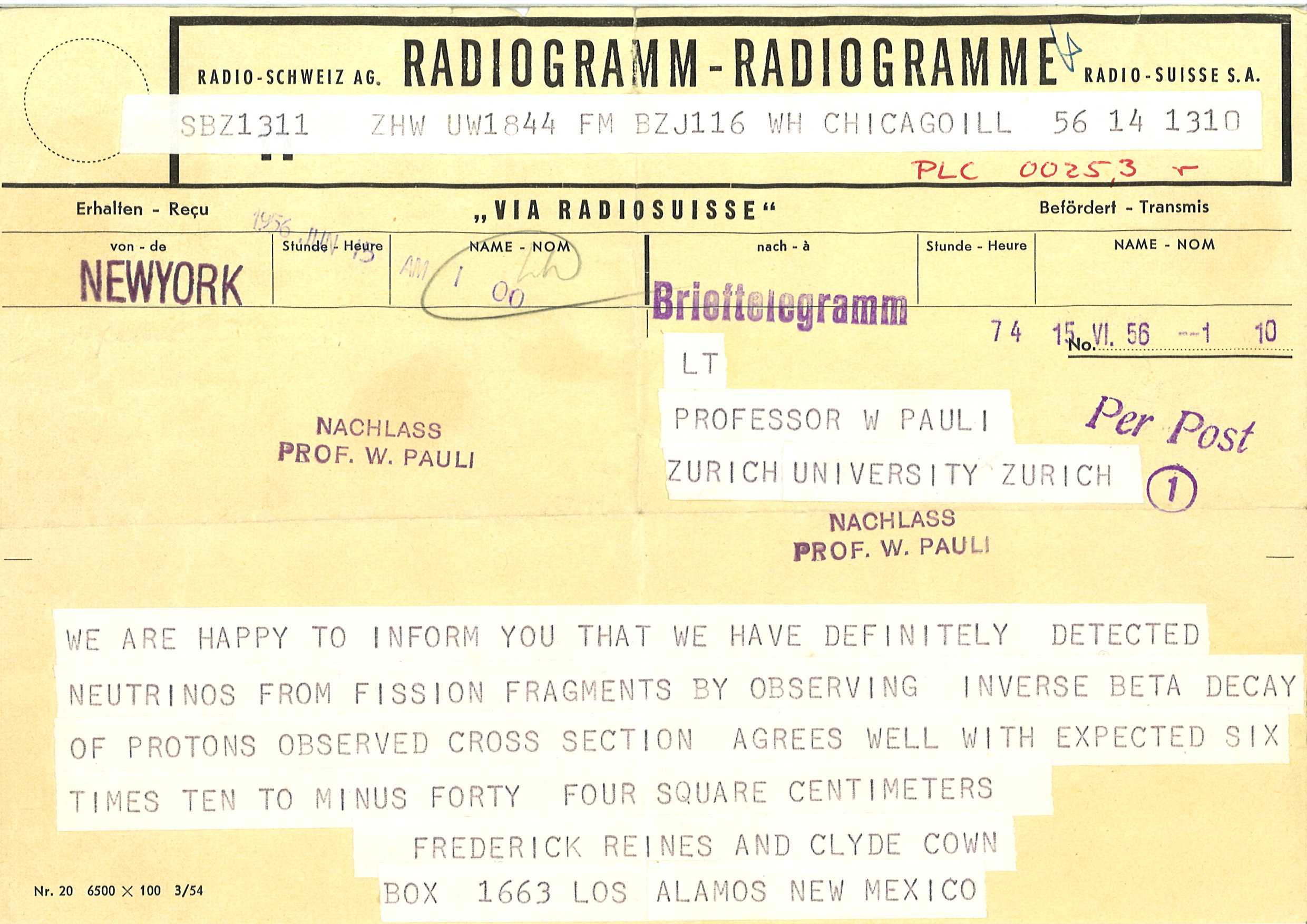}
    \caption{Telegram from Fred Reines and Clyde Cowan informing Wolfgang Pauli of their detection of neutrinos from a nuclear reactor.}
    \label{fig:pauli_telegram}
\end{figure}

The first indication of what would come to be known as the neutrino came from Wolfgang Pauli in 1930~\cite{Pauli:1930pc}.
Addressing the ``radioactive ladies and gentlemen'' of Tübingen, Germany, he appealed to the existence of new electrically neutral particles to save the law of energy conservation in nuclear beta decays.
This idea was developed further by Enrico Fermi in 1934, who calculated the transition probability for $\beta$-decay with a neutrino in the final state~\cite{Fermi:1934hr}.
Fermi's theory represents the first study of the weak interaction--the only Standard Model gauge group under which neutrinos are charged.

As the name ``weak interaction'' suggests, neutrinos interact very feebly with particles in the Standard Model.
Thus, it wasn't until 1956 that the neutrino was observed in an experimental setting for the first time.
A team of scientists from Los Alamos Scientific Laboratory, led by Frederick Reines and Clyde Cowan, detected a free neutrino from a nuclear reactor via the inverse beta decay interaction ($\bar{\nu}_e p \to e^+ n$)~\cite{Reines:1953pu,Cowan:1956rrn}.
Though it was not known at the time, they had detected electron antineutrinos ($\bar{\nu}_e$).
Electron (anti)neutrinos represent one of the three weak-flavor eigenstates neutrinos can occupy in the Standard Model--specifically, the eigenstate that couples to the $e^\pm$ charged leptons through the charged-current weak interaction.
Upon confirmation of their discovery, Reines and Cowan sent the telegram shown in \cref{fig:pauli_telegram} to Pauli, alerting him of the definitive existence of the neutral particles he proposed in Tübingen.

Shortly after this, the phenomenon of neutrino oscillations--periodic transitions between different types of neutrinos--started to appear in the literature.
In 1958, Bruno Pontecorvo discussed the possibility of mixing between right-handed antineutrinos $\bar{\nu}_R$ and ``sterile'' right-handed neutrinos $\nu_R$, in analogy with $K^0$--$\bar{K}^0$ mixing observed in the quark sector~\cite{Pontecorvo:1957qd}.
A second possible source of neutrino oscillations came following the 1962 experimental discovery of a second neutrino weak-flavor eigenstate--the muon neutrino ($\nu_\mu$)~\cite{Danby:1962nd}.
After this, the notion of mixing between neutrino flavor and mass eigenstates was introduced by Ziro Maki, Masami Nakagawa, and Shoichi Sakata~\cite{Maki:1962mu}.
In a 1967 paper~\cite{Pontecorvo:1967fh}, Pontecorvo introduced the possibility of vacuum $\nu_e$--$\nu_\mu$ oscillations, even predicting a factor of two suppression in the total solar neutrino flux before such a deficit would actually be observed~\cite{Bilenky:2016pep}.

\begin{figure}
    \centering
    \includegraphics[width=0.6\textwidth]{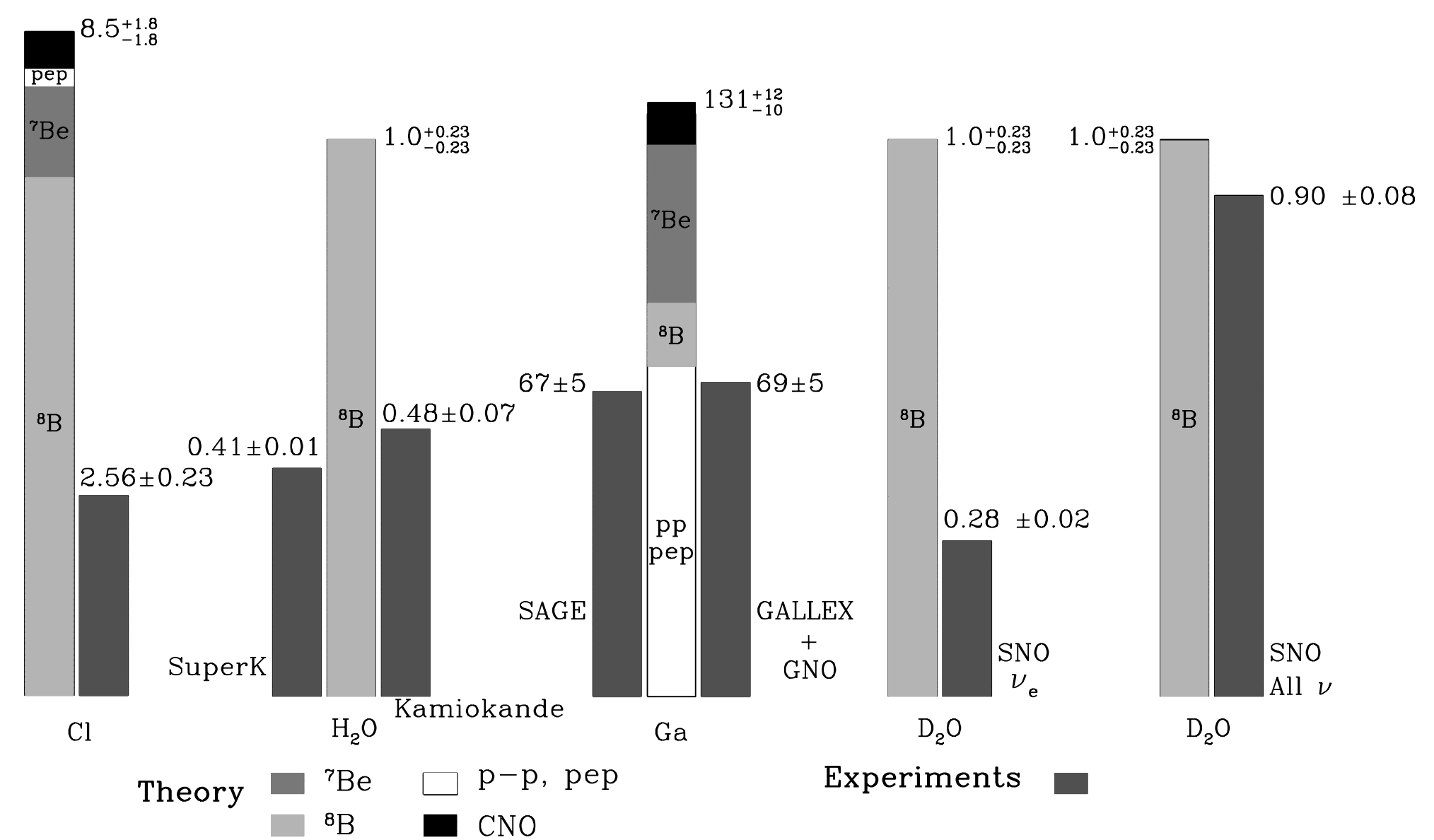}
    \caption{The deficit of the observed solar $\nu_e$ flux compared with the theoretical expectation. The Homestake experiment is shown on the far left; follow-up solar neutrino measurements confirming the deficit are also shown, including the 2002 SNO result which brought forth a solution to the solar neutrino problem. Figure from Ref.~\cite{Bahcall:2004qv}.}
    \label{fig:solar_neutrino_problem}
\end{figure}

The aforementioned deficit, known as the ``solar neutrino problem'', was established in 1968 through a now-famous experiment at the Homestake Mine in South Dakota led by Raymond Davis~\cite{Davis:1968cp}.
Davis and his colleagues detected the capture of electron neutrinos from the sun on ${}^{37}$Cl nuclei, allowing a measurement of the solar $\nu_e$ flux.
Their result was about a factor of $\sim 1/3$ lower than the leading prediction from John Bachall~\cite{Bahcall:1968hc}.
This is shown in \cref{fig:solar_neutrino_problem}, including confirmations of the deficit following the Homestake experiment.
The solution was not a mistake in the experimental measurement or theoretical prediction, as physicists expected at the time; rather, it was a deficiency in our understanding of neutrinos.
This was the first piece of the puzzle that would eventually lead to the discovery of neutrino oscillations and nonzero neutrino masses.

The next piece of the puzzle came from atmospheric neutrinos, i.e. neutrinos coming from the decay of mesons created from the interactions of primary cosmic rays in the atmosphere.
Around the mid-1980s, two water Cherenkov detectors, IMB-3~\cite{Casper:1990ac} and Kamiokande~\cite{Kamiokande-II:1992hns}, began to measure the interactions of atmospheric $\nu_\mu$ and $\nu_e$ events (initially just as a background for their main physics goal, the search for nucleon decay).
The ratio of $\nu_\mu:\nu_e$ interactions was found to be lower than the theoretical expectation by a factor of $\sim 2/3$~\cite{Learned:2000qq}.
This was known as the ``atmospheric neutrino anomaly''.
The source of this anomaly was not clear at the time; it could have been a deficit of muon neutrinos, an excess of electron neutrinos, or some of both.
Systematic issues in the flux prediction or muon identification were also suggested~\cite{Learned:2000qq}.
It was far from clear that neutrino oscillations could be responsible for the observed deficit.

\begin{figure}
    \centering
    \includegraphics[width=0.6\textwidth]{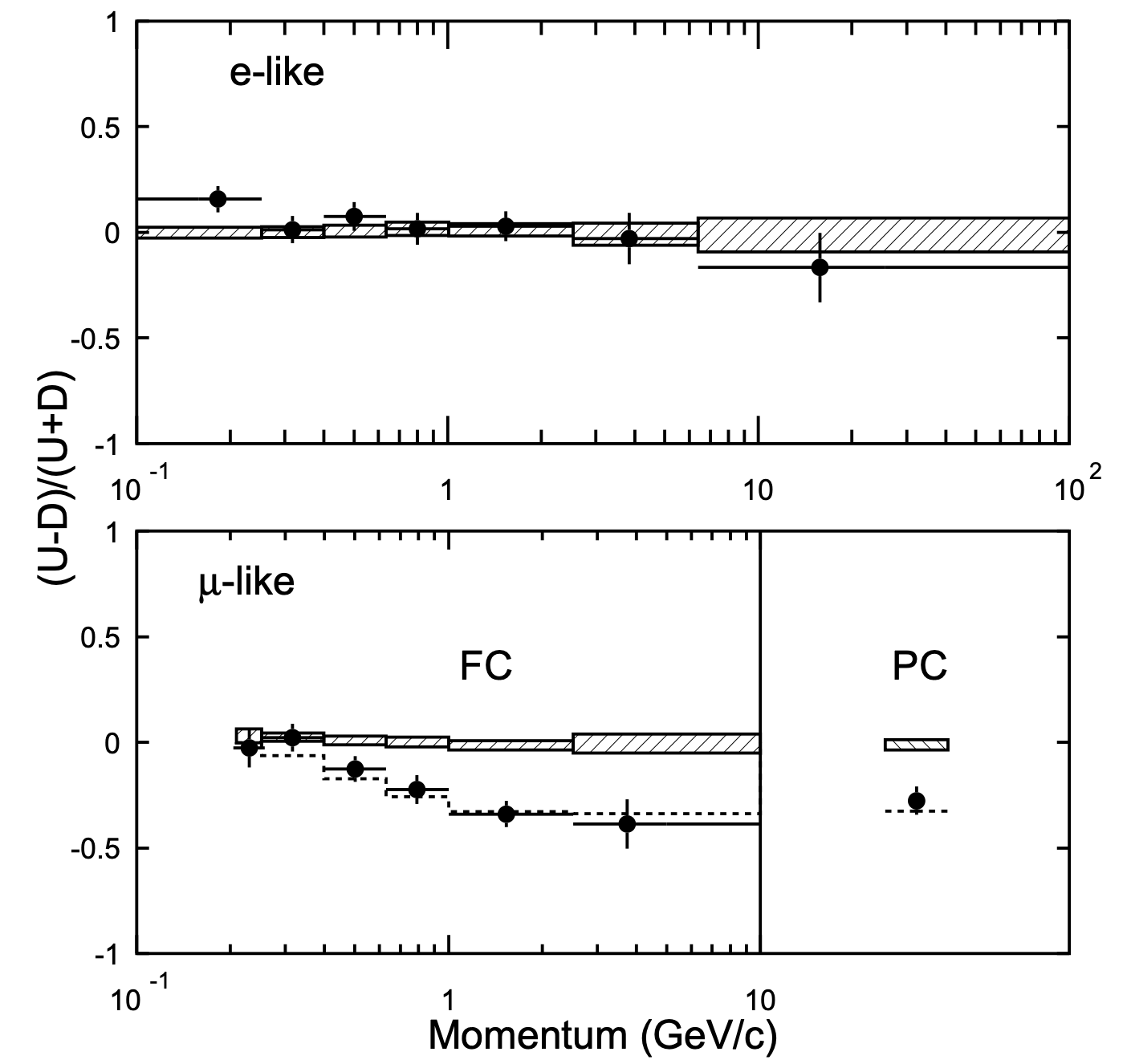}
    \caption{The up-down asymmetry measured in SuperK as a function of lepton momentum, separated into $e$-like and $\mu$-like events as well as fully-contained (FC) and partially-contained (PC) events. The dashed line indicates the best fit to $\nu_\mu \to \nu_\tau$ oscillations. Figure from Ref.~\cite{Super-Kamiokande:1998kpq}.}
    \label{fig:superK}
\end{figure}

The solution to the atmospheric neutrino anomaly came from the Super-Kamiokande (SuperK) experiment~\cite{Super-Kamiokande:2002weg}.
SuperK was a much larger version of the Kamiokande detector, allowing the detection of higher energy muons (up to $E_\mu \sim 5\;{\rm GeV}$).
SuperK also measured the up-down asymmetry of muon-like and electron-like events in their detector, $(N_{\rm up} - N_{\rm down})/(N_{\rm up} + N_{\rm down})$.
Upward-going events have traveled a much longer distance than downward-going events before reaching the SuperK detector--thus positive detection of an asymmetry would be smoking-gun evidence for a baseline-dependent effect like neutrino oscillations.
This is precisely what SuperK observed~\cite{Super-Kamiokande:1998kpq}.
As shown in \cref{fig:superK}, an up-down asymmetry is observed in the muon-like channel, the magnitude of which increases with the observed muon momentum.
Such behavior is consistent with muon neutrino oscillations to a third flavor eigenstate, $\nu_\tau$ (the mathematical details of neutrino oscillations will be described in \cref{sec:massive_nu}).
No such effect was observed in the electron-like channel.
Thus, the atmospheric neutrino anomaly is a result of muon neutrino disappearance, specifically coming from $\nu_\mu \to \nu_\tau$ oscillations.

\begin{figure}
    \centering
    \includegraphics[width=0.6\textwidth]{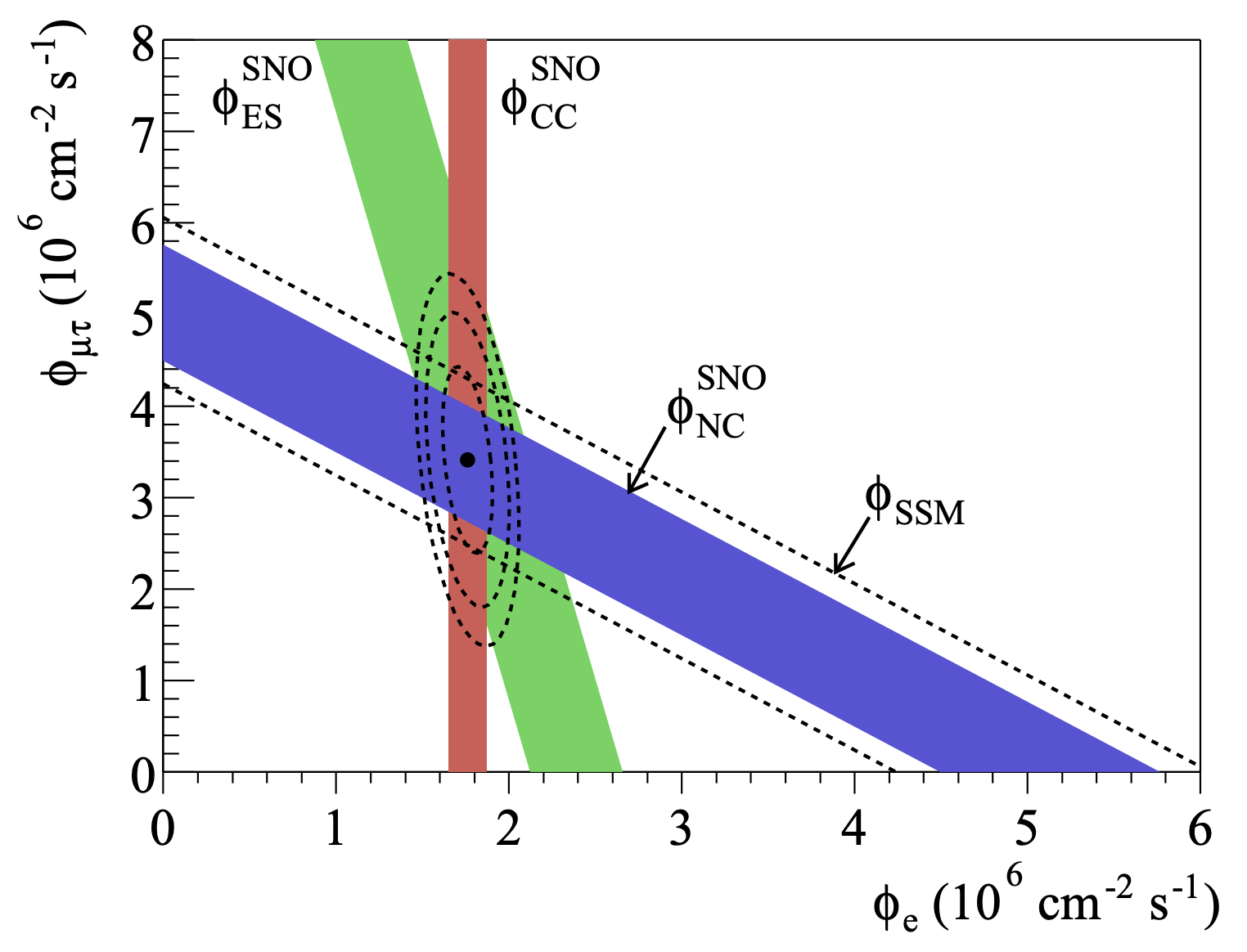}
    \caption{Measurement of the solar ${}^8$B flux from the SNO collaboration, broken down into the $\nu_e$ and $\nu_{\mu,\tau}$ sub-components. Measurement of the CC, NC, and ES channels show up as slices in the two-dimensional flux parameter space. Figure from Ref.~\cite{SNO:2002tuh}.}
    \label{fig:SNO}
\end{figure}

The solution to the solar neutrino problem came in 2002 from the Sudbury Neutrino Observatory (SNO)~\cite{SNO:1999crp}.
The SNO experiment used a heavy water Cherenkov detector, specifically relying on the use of deuterium target nuclei to be sensitive to three different neutrino interactions,
\begin{equation}
\begin{split}
&\nu_e + d \to p + p + e^- ~~~ (\rm{CC}), \\
&\nu_x + d \to p + n + \nu_x ~~~ (\rm{NC}), \\
&\nu_x + e^- \to \nu_x + e^- ~~~~~ (\rm{ES}).
\end{split}
\end{equation}
Charged-current (CC), neutral-current (NC), and elastic scattering (ES) interactions were separated based on the visible energy and scattering angle of the final state particles.
NC events were further separated by tagging the 6.25\;MeV photon released from neutron capture on deuterium.
By measuring all three channels, SNO was able to measure the ${}^8$B solar neutrino flux broken down into the $\nu_e$ and $\nu_{\mu,\tau}$ components.
SNO's 2002 result showed that the missing neutrinos from the Homestake experiment were in fact showing up in the $\nu_{\mu,\tau}$ component~\cite{SNO:2002tuh}.
\Cref{fig:SNO} shows the flux of each component as constrained by the measured CC, NC, and ES interaction rate.
The flavor transitions here come not from vacuum oscillations but rather from matter-enhanced resonant behavior as neutrinos travel through the dense solar medium--a phenomenon known as the Mikheyev–Smirnov–Wolfenstein (MSW) effect~\cite{Mikheyev:1985zog,Wolfenstein:1977ue}.
The MSW effect still, however, requires mixing between the neutrino flavor and mass eigenstate as well as non-zero squared differences between the mass eigenstates.
It is worth noting here that the KamLAND reactor neutrino experiment was essential in determining the oscillation parameters which led to the SNO observation~\cite{KamLAND:2004mhv}.
Thus, the SNO solution to the solar neutrino problem and the SuperK solution to the atmospheric neutrino anomaly were both evidence for the existence of neutrino oscillations and thus non-zero neutrino masses.
The collaborations shared the 2015 Nobel Prize in physics for this discovery~\cite{McDonald:2016ixn,Kajita:2016cak}.

Since SuperK and SNO, neutrino oscillations have been measured extensively by a global program of reactor, accelerator, atmospheric, and solar neutrino experiments.
The mixing angle and mass-squared splittings of the three Standard Model neutrinos have been measured to few-percent-level precision in most cases~\cite{deSalas:2020pgw,Esteban:2020cvm,Capozzi:2021fjo}.
There are a number of open questions in the standard three-neutrino mixing paradigm, including the ordering of the three mass eigenstates and the value of the charge-parity-violating complex phase $\delta_{CP}$.
Though preliminary results exist on both fronts~\cite{T2K:2019bcf,NOvA:2019cyt,Esteban:2020cvm,deSalas:2020pgw,Capozzi:2021fjo}, definitive answers to each will come from next-generation neutrino experiments, including Hyper-K~\cite{Hyper-Kamiokande:2016srs}, DUNE~\cite{DUNE:2015lol} and JUNO~\cite{JUNO:2015zny}.

\section{Neutrinos in the Standard Model}

The arguments and notation presented in this section follow closely from chapter 2 of Ref.~\cite{Mohapatra:1998rq}.

The interactions of the known fundamental particles of our Universe are described by a specific quantum field theory known as the Standard Model (SM).
Above the electroweak scale, there are three gauge groups contained within the SM:
\begin{itemize}
    \itemsep0em
    \item $SU(3)_c$, which governs the gluon-mediated ``strong interactions'' of color-charged fields.
    \item $SU(2)_L$, one part of the ``electro-weak interaction'', mediated by the $W^{\pm}_\mu$ and $W^0_\mu$ vector bosons.
    \item $U(1)_Y$, the other part of the ``electro-weak interaction'', mediated by the $B_\mu$ gauge boson.
\end{itemize}
After electro-weak symmetry breaking (EWSB) via the Higgs mechanism, the $SU(2)_L \times U(1)_Y$ subgroup breaks down to $U(1)_Q$, which describes the electromagnetic (EM) interactions of charged fields mediated by the $A_\mu$ gauge boson, also known as the photon.

Of the three fundamental interactions of the SM, neutrinos are only charged under the weak $SU(2)_L$ gauge group--they are singlets under the $SU(3)_c$ and $U(1)_Q$ gauge groups.
Thus, neutrinos only appear in the electro-weak part of the SM Lagrangian, which is given by
\begin{equation} \label{eq:SMlagrangian}
\LL = \frac{g}{\sqrt{2}} (J^\mu W^+_\mu + J^{\mu\dagger} W^-_\mu) + \frac{g}{\cos \theta_W} K^\mu Z_\mu,
\end{equation}
where $g = e/\cos \theta_W$ is the $SU(2)_L$ gauge coupling of $W_\mu$ and Higgs field, $\theta_W$ is the Weinberg angle describing the rotation that occurs during EWSB between the neutral parts of the $SU(2)_L$ and $U(1)_Y$ gauge boson fields, and $W^{\pm}_\mu$ ($Z_\mu$) is the charged (neutral) piece of $SU(2)_L$ after EWSB.
The currents coupled to $W^{\pm}_\mu$ and $Z_\mu$ bosons are given by
\begin{equation}
\begin{split}
J^\mu &= 
\begin{pmatrix}
\overline{u}^0 & \overline{c}^0 & \overline{t}^0
\end{pmatrix}
\gamma^\mu P_L
\begin{pmatrix}
d^0 \\
s^0 \\
b^0
\end{pmatrix}
+
\begin{pmatrix}
\overline{\nu_e} & \overline{\nu_\mu} & \overline{\nu_\tau}
\end{pmatrix}
\gamma^\mu P_L
\begin{pmatrix}
e \\
\mu \\
\tau
\end{pmatrix} 
\\
K^\mu &=  \sum_f \overline{f} \gamma^\mu [I_{3L} P_L - \sin^2 \theta_W Q_f] f \\
&= \sum_q [\epsilon_L(q) \overline{q} \gamma_\mu P_L q + \epsilon_R(q) \overline{q} \gamma_\mu P_R q] \\
&+ \frac{1}{2}\sum_{\alpha \in \{e,\mu,\tau\}} [\overline{\nu_\alpha} \gamma^\mu P_L \nu_\alpha + \overline{\ell}_\alpha \gamma_\mu (g_V^\alpha - \gamma_5 g_A^\alpha) \ell_\alpha],
\end{split}
\end{equation}
where $P_R (P_L) = (1 \pm \gamma^5)/2$ is the projection operator onto the right-handed (left-handed) chiral state, and the subscript $0$ on the quark fields indicates that these are the weak flavor eigenstates rather than the mass eigenstates.
The first generation coupling constants in $K^\mu$, which derive from the specified EM charge and $SU(2)_L$ representation of each field, are given by
\begin{equation}
\begin{split}
&\epsilon_L(u) = \frac{1}{2} - \frac{2}{3} \sin^2 \theta_W
~~~~~
\epsilon_R(u) =  - \frac{2}{3} \sin^2 \theta_W \\
&\epsilon_L(d) = - \frac{1}{2} + \frac{1}{3} \sin^2 \theta_W
~~~
\epsilon_R(d) =  \frac{1}{3} \sin^2 \theta_W \\
&g_V^e = -\frac{1}{2} + 2 \sin^2 \theta_W
~~~~~~~
g_A^e = -\frac{1}{2}.
\end{split}
\end{equation}

The Lagrangian in \cref{eq:SMlagrangian} can be used to calculate cross sections for the various SM interactions of the neutrino.
The first term describes the charged-current interactions of neutrinos such as nuclear beta decay, while the second term describes neutral current interactions such as $\nu_\mu e^-$ elastic scattering.
At energy scales below the electro-weak scale, one can integrate out the $W_\mu$ and $Z_\mu$ gauge bosons and describe interactions in terms of the dimensional Fermi constant
\begin{equation}
G_F = \frac{g^2}{4 \sqrt{2} M_W^2} = 1.166 \times 10^{-5} \;\rm{GeV}^{-2}.
\end{equation} 
The low-energy Lagrangian describing 4-fermion interactions can be derived from \cref{eq:SMlagrangian} as
\begin{equation} \label{eq:Fermi_Lagrangian}
\LL_{4f} = \frac{-4 G_F}{\sqrt{2}} [J_\mu J^{\mu \dagger} + K_\mu K^\mu].
\end{equation}

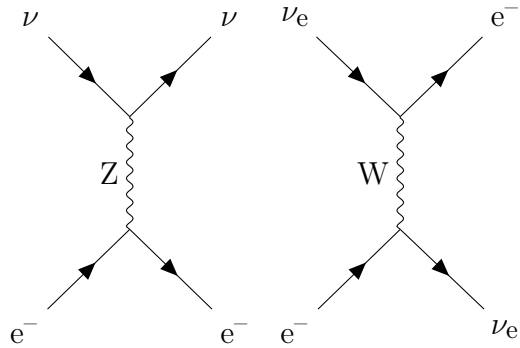
\begin{figure}
\begin{centering}
\begin{tikzpicture}
  \begin{feynman}
    \vertex  (a);
    \vertex [above left=of a] (f1) {\(\nu\)};
    \vertex [above right=of a] (f2) {\(\nu\)};
    \vertex [below=of a] (c);
    \vertex [below left=of c] (f3) {\(e^{-}\)};
    \vertex [below right=of c] (f4) {\(e^{-}\)};

    \diagram* {
      (f1) -- [fermion] (a) -- [fermion] (f2),
      (a) -- [boson, edge label'=\(Z\)] (c),
      (f3) -- [fermion] (c) -- [fermion] (f4),
    };
  \end{feynman}
\end{tikzpicture}
\begin{tikzpicture}
  \begin{feynman}
    \vertex  (a);
    \vertex [above left=of a] (f1) {\(\nu_e\)};
    \vertex [above right=of a] (f2) {\(e^{-}\)};
    \vertex [below=of a] (c);
    \vertex [below left=of c] (f3) {\(e^{-}\)};
    \vertex [below right=of c] (f4) {\(\nu_e\)};

    \diagram* {
      (f1) -- [fermion] (a) -- [fermion] (f2),
      (a) -- [boson, edge label'=\(W\)] (c),
      (f3) -- [fermion] (c) -- [fermion] (f4),
    };
  \end{feynman}
\end{tikzpicture}
\caption{\label{fig:nu_e_ES} Diagrams contributing to $\nu e^-$ elastic scattering}
\end{centering}
\end{figure}

As an example, we consider low-energy neutrino electron elastic scattering (ES) ($\nu e^- \to \nu e^-$).
This is a purely leptonic process and is therefore relatively clean; specifically, ES models do not need to account for the complex dynamics of the nuclear medium.
The Feynman diagrams for the contributing interactions are shown in \cref{fig:nu_e_ES}.
Both the charged-current (CC) and neutral-current (NC) diagrams contribute to $\nu_e e^-$ scattering, while only the NC diagram contributes to $\nu_{\mu,\tau} e^-$ scattering.
Using the Feynman rules associated with \cref{eq:Fermi_Lagrangian}, one can calculate the cross sections to be~\cite{Mohapatra:1998rq}
\begin{equation}
\begin{split}
\sigma_{\nu_e e^- \to \nu_e e^-}(E_\nu) &= \frac{G_F^2 m_e E_\nu}{2 \pi} 
\bigg[(2 \sin^2 \theta_W + 1)^2 + \frac{4}{3} \sin^4 \theta_W \bigg] \\
&\approx 0.9 \times 10^{-43} \bigg( \frac{E_\nu}{10\;{\rm MeV}} \bigg) {\rm cm}^2 \\
\sigma_{\nu_{\mu,\tau} e^- \to \nu_{\mu,\tau} e^-}(E_\nu) &= \frac{G_F^2 m_e E_\nu}{2 \pi} 
\bigg[(2 \sin^2 \theta_W - 1)^2 + \frac{4}{3} \sin^4 \theta_W \bigg] \\
&\approx 0.15 \times 10^{-43} \bigg( \frac{E_\nu}{10\;{\rm MeV}} \bigg) {\rm cm}^2,
\end{split}
\end{equation}
which is valid for $E_\nu >> m_e$. 
Similarly, one can calculate the cross section for antineutrino electron ES ($\overline{\nu} e^- \to \overline{\nu} e^-$).
The diagrams contributing for this process are shown in \cref{fig:antinu_e_ES}, and the cross section is given by~\cite{Mohapatra:1998rq}
\begin{equation}
\begin{split}
\sigma_{\overline{\nu}_e e^- \to \overline{\nu}_e e^-}(E_\nu) &= \frac{G_F^2 m_e E_\nu}{2 \pi} 
\bigg[\frac{1}{3}(2 \sin^2 \theta_W + 1)^2 + 4 \sin^4 \theta_W \bigg] \\
&\approx 0.378 \times 10^{-43} \bigg( \frac{E_\nu}{10\;{\rm MeV}} \bigg) {\rm cm}^2 \\
\sigma_{\overline{\nu}_{\mu,\tau} e^- \to \overline{\nu}_{\mu,\tau} e^-}(E_\nu) &= \frac{G_F^2 m_e E_\nu}{2 \pi} 
\bigg[\frac{1}{3}(2 \sin^2 \theta_W - 1)^2 + 4 \sin^4 \theta_W \bigg] \\
&\approx 0.14 \times 10^{-43} \bigg( \frac{E_\nu}{10\;{\rm MeV}} \bigg) {\rm cm}^2.
\end{split}
\end{equation}

We now turn to the interaction at the core of this thesis: neutrino-nucleon charged-current quasi-elastic (CCQE) scattering.
The relevant Feynman diagrams for this process are shown in \cref{fig:nu_CCQE_diagrams}.
Unlike ES, models of CCQE do need to account for the nuclear environment surrounding the target nucleon.
As the final state nucleon travels through the nuclear medium, it may scatter off of other nucleons and/or produce additional mesons through a process known as final state interactions (FSIs).
As shown in \cref{fig:nu_CCQE}, CCQE is dominant for  $E_\nu \lesssim 1\;{\rm GeV}$.
Above this energy, nucleon resonance processes start to take over, in which Delta resonances decay to final state mesons.
In the regime $E_\nu \gtrsim 10\;{\rm GeV}$, neutrinos start to undergo deep inelastic scattering (DIS) off of the constituent quarks within the nucleon.

In order to calculate the CCQE cross section, one considers a theory containing nucleon degrees of freedom.
The original calculation for free nucleons (i.e., not bound within a nucleus) was carried out by Llewellyn-Smith in 1972; the differential cross section as a function of the squared four-momentum transfer $Q^2$ is given by~\cite{Formaggio:2012cpf,LlewellynSmith:1971uhs}
\begin{equation}
\begin{split}
\frac{d\sigma}{dQ^2} = \frac{G_F^2 M^2 |V_{ud}|^2}{8\pi E_\nu^2}
\bigg[A \pm \frac{s-u}{M^2}B + \frac{(s-u)^2}{M^4}C\bigg],
\end{split}
\end{equation}
where +(-) refers to (anti)neutrino scattering, $M$ is the nucleon mass, $m$ is the lepton mass, $(s-u) = 4ME_\nu - Q^2 - m^2$, and $A$, $B$, and $C$ are functions of the vector, axial-vector, and pseudoscalar form factors of the nucleon (see equations 58, 59, and 60 of Ref.~\cite{Formaggio:2012cpf} for complete expressions).
These form factors describe the composite nature of nucleons under interactions with different Lorentz structures.

For $E_\nu << M$, the $\nu_e$ CCQE cross section is approximately~\cite{Mohapatra:1998rq}
\begin{equation}
\begin{split}
\sigma_{\nu_e n \to e^- p}(E_\nu) &\approx \frac{G_F^2 E_\nu^2}{\pi} (g_V^2 + 3 g_A^2) \\
&\approx 9.75 \times 10^{-42} \bigg[ \frac{E_\nu}{10\;{\rm MeV}} \bigg]^2\;{\rm cm}^2.
\end{split}
\end{equation}
In the regime $E_\nu \gtrsim 1\;{\rm GeV}$, the $\nu_e$ and $\nu_\mu$ CCQE cross sections are no longer suppressed by threshold effects and are thus the same, approximately $10^{-38}\;{\rm cm}^2$~\cite{Formaggio:2012cpf}.
This cross section is significantly larger than the elastic scattering and lower energy $\nu_e$ CCQE cross sections and is the dominant neutrino interaction for many accelerator-based neutrino experiments, including two at the heart of this thesis: MiniBooNE and MicroBooNE.
Finally, we note that the cross section for antineutrino CCQE tends to be smaller; this will be important in \cref{ch:neutrissimos}.

\begin{figure}
\begin{centering}
\begin{tikzpicture}
  \begin{feynman}
    \vertex  (a);
    \vertex [above left=of a] (f1) {\(\overline{\nu}\)};
    \vertex [above right=of a] (f2) {\(\overline{\nu}\)};
    \vertex [below=of a] (c);
    \vertex [below left=of c] (f3) {\(e^{-}\)};
    \vertex [below right=of c] (f4) {\(e^{-}\)};

    \diagram* {
      (f1) -- [anti fermion] (a) -- [anti fermion] (f2),
      (a) -- [boson, edge label'=\(Z\)] (c),
      (f3) -- [fermion] (c) -- [fermion] (f4),
    };
  \end{feynman}
\end{tikzpicture}
\begin{tikzpicture}
  \begin{feynman}
    \vertex  (a);
    \vertex [above left=of a] (f1) {\(\overline{\nu}_e\)};
    \vertex [below left=of a] (f2) {\(e^{-}\)};
    \vertex [right=of a] (c);
    \vertex [below right=of c] (f3) {\(e^{-}\)};
    \vertex [above right=of c] (f4) {\(\overline{\nu}_e\)};

    \diagram* {
      (f1) -- [anti fermion] (a) -- [anti fermion] (f2),
      (a) -- [boson, edge label'=\(W^-\)] (c),
      (f3) -- [anti fermion] (c) -- [anti fermion] (f4),
    };
  \end{feynman}
\end{tikzpicture}
\caption{\label{fig:antinu_e_ES} Diagrams contributing to $\overline{\nu} e^-$ elastic scattering}
\end{centering}
\end{figure}

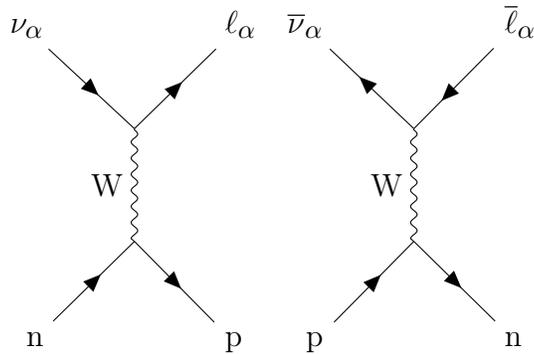
\begin{figure}
\begin{centering}
\begin{tikzpicture}
  \begin{feynman}
    \vertex  (a);
    \vertex [above left=of a] (f1) {\(\nu_\alpha\)};
    \vertex [above right=of a] (f2) {\(\ell_\alpha\)};
    \vertex [below=of a] (c);
    \vertex [below left=of c] (f3) {\(n\)};
    \vertex [below right=of c] (f4) {\(p\)};

    \diagram* {
      (f1) -- [fermion] (a) -- [fermion] (f2),
      (a) -- [boson, edge label'=\(W\)] (c),
      (f3) -- [fermion] (c) -- [fermion] (f4),
    };
  \end{feynman}
\end{tikzpicture}
\begin{tikzpicture}
  \begin{feynman}
    \vertex  (a);
    \vertex [above left=of a] (f1) {\(\overline{\nu}_\alpha\)};
    \vertex [above right=of a] (f2) {\(\overline{\ell}_\alpha\)};
    \vertex [below=of a] (c);
    \vertex [below left=of c] (f3) {\(p\)};
    \vertex [below right=of c] (f4) {\(n\)};

    \diagram* {
      (f1) -- [anti fermion] (a) -- [anti fermion] (f2),
      (a) -- [boson, edge label'=\(W\)] (c),
      (f3) -- [fermion] (c) -- [fermion] (f4),
    };
  \end{feynman}
\end{tikzpicture}
\caption{\label{fig:nu_CCQE_diagrams} Diagrams contributing to neutrino-nucleon charged-current quasielastic scattering}
\end{centering}
\end{figure}

\begin{figure}
    \centering
    \includegraphics[width=0.6\textwidth]{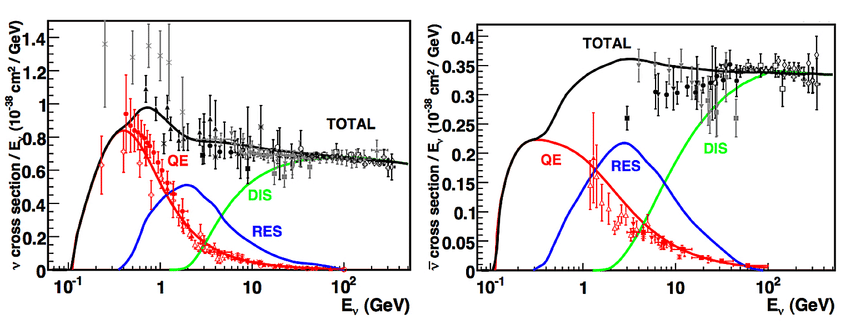}
    \caption{CC inclusive neutrino and antineutrino nucleon scattering cross sections as a function of neutrino energy. Figure from Ref.~\cite{Formaggio:2012cpf}.}
    \label{fig:nu_CCQE}
\end{figure}

\section{Massive Neutrinos} \label{sec:massive_nu}

The arguments and notation presented in this section follow closely from section 2.5, chapter 4, and chapter 5 of Ref.~\cite{Mohapatra:1998rq} as well as chapter 11 of Ref.~\cite{Schwartz:2014sze}.

Neutrinos are massless in the SM.
To see this, we will exhaust the two possible forms for a neutrino mass term in the SM Lagrangian: Dirac and Majorana.
These refer to the two possible fermionic spinor representations in which neutrinos can be found.
Dirac spinors in general have four complex components, or degrees of freedom, while Majorana spinors have only two.
The critical question is whether the right-handed chiral projection of the neutrino field, $\nu_R$, is the same as $\overline{\nu}_R$, the right-handed chiral projection of the antineutrino field (Majorana case), or if it is a new degree of freedom (Dirac case).

The definition of a free Dirac fermion field is
\begin{equation} \label{eq:dirac_field}
\psi(x) = \int \frac{d^3 p}{\sqrt{(2\pi)^3 2 E_p}} \sum_{s = \pm \frac{1}{2}} \Big(
f_s(\mathbf{p}) u_s(\mathbf{p}) e^{-i \mathbf{p} \cdot \mathbf{x}}
+ 
\overline{f_s}^\dagger (\mathbf{p}) v_s(\mathbf{p}) e^{i \mathbf{p} \cdot \mathbf{x}}
\Big),
\end{equation}
where $f_s(\mathbf{p})$ annihilates a single particle of momentum $\mathbf{p}$ while $\overline{f_s}^\dagger$ creates the corresponding antiparticle state, and $u_s(\mathbf{p})$ and $v_s(\mathbf{p})$ are spinors 
with positive and negative energy, respectively, satisfying the Dirac equations
\begin{equation} \label{eq:spinor_dirac_eq}
\begin{split}
&(\gamma^\mu p_\mu - m) u_s(\mathbf{p}) = 0 \\
&(\gamma^\mu p_\mu+ m) v_s(\mathbf{p}) = 0,
\end{split}
\end{equation}
where $\gamma^\mu$ are a set of Lorentz-indexed matrices satisfying $\{\gamma^\mu,\gamma^\nu\} = 2 g^{\mu \nu}$.
There are many possible representations for the $\gamma$-matrices.
We consider the Weyl basis, in which~\cite{Schwartz:2014sze}
\begin{equation}
    \gamma_\mu = 
    \begin{pmatrix}
    0 & \sigma_\mu \\
    \overline{\sigma}_\mu & 0 \\
    \end{pmatrix},
\end{equation}
where $\sigma_\mu = (\mathbb{1}, \vec{\sigma})$, $\overline{\sigma}_\mu = (\mathbb{1}, -\vec{\sigma})$, and $\vec{\sigma} = (\sigma_1, \sigma_2, \sigma_3)$ are the Pauli matrices.
This representation is convenient for understanding the different chiral components of the Dirac spinor $\psi$.
The Lorentz generators $S^{\mu \nu} \equiv \frac{i}{4}[\gamma^\mu,\gamma^\nu]$ become block diagonal, such that we can write the Dirac spinor of \cref{eq:dirac_field} as a doublet of two-component Weyl spinors with a different chiral nature,
\begin{equation}
    \psi =
    \begin{pmatrix}
    \psi_L \\
    \psi_R \\
    \end{pmatrix},
\end{equation}
which transform under different irreducible representations of the Lorentz group~\cite{Schwartz:2014sze}.
Here $\psi_L$ and $\psi_R$ refer to the left-handed and right-handed Weyl spinor, respectively.
We can isolate the different chiral components of the Dirac spinor using the matrix $\gamma^5 \equiv i \gamma^0 \gamma^1 \gamma^2 \gamma^3$, which takes the form
\begin{equation}
\gamma^5 = 
\begin{pmatrix}
    -\mathbb{1} & 0 \\
    0 & \mathbb{1} \\
\end{pmatrix}
\end{equation}
in the Weyl basis.
We can define projection operators $P_L = \frac{1}{2}(1 - \gamma^5)$ and $P_R = \frac{1}{2}(1 + \gamma^5)$ such that $P_L \psi = \psi_L$ and $P_R \psi_R$.
It is worth noting that while the behavior of these projection operators is especially clear in the Weyl representation, they will isolate the chiral components of $\psi$ in any representation of $\gamma^\mu$.


Dirac mass terms couple left-handed and right-handed chiral states.
To see this, consider the Dirac equation in the Weyl basis, which takes the form~\cite{Schwartz:2014sze}
\begin{equation}
(\gamma^\mu p_\mu -  m) \psi =
\begin{pmatrix}
-m & \sigma^\mu p_\mu \\
\overline{\sigma}^\mu p_\mu & -m \\
\end{pmatrix}
\begin{pmatrix}
\psi_L \\
\psi_R \\
\end{pmatrix}
= 0.
\end{equation}
It is evident that this matrix equation mixes the left-handed and right-handed components of $\psi$.
Dirac mass terms take the form $m\psi_L^\dagger \psi_R$ and $m\psi_R^\dagger \psi_L$, thus requiring both chiral components.
After EWSB, the non-neutrino fermions in the SM acquire a Dirac mass from their interactions with the Higgs field.
Neutrinos, however, do not have a right-handed chiral state in the SM; therefore, the SM cannot include a Dirac mass term for neutrinos.

Now we turn to the Majorana mass term.
The expression for a Majorana field is the same as \cref{eq:dirac_field}, subject to a condition relating particles and antiparticles.
We see that the expression for $\psi^*(x)$ would involve $f_s^\dagger(\mathbf{p})$, which creates a particle state, and $\overline{f_s}(\mathbf{p})$, which annihilates an antiparticle state.
It turns out the relationship $\psi(x) = \psi^*(x)$ is not Lorentz invariant~\cite{Mohapatra:1998rq}.
To remedy this, we must define the conjugate Dirac field
\begin{equation}
\psi^C(x) \equiv \gamma_0 C \psi^*(x),
\end{equation}
where the representation-dependent conjugation matrix $C$ is defined by the equation
\begin{equation}
\begin{split}
&\gamma_0 C \sigma^*_{\mu \nu} = -\sigma_{\mu \nu} \gamma_0 C, \\
& \sigma_{\mu \nu} \equiv \frac{i}{2}[\gamma_\mu,\gamma_\nu].
\end{split}
\end{equation}
In the Weyl representation, for example, $C = i \gamma_2 \gamma_0$.
This requirement for $C$ ensures that $\psi^C(x)$ transforms in the same way as $\psi(x)$ under the Lorentz group~\cite{Mohapatra:1998rq}.
The Lorentz-invariant Majorana condition specifically requires
\begin{equation} \label{eq:majorana_condition}
\psi(x) = e^{i\theta} \psi^C(x),
\end{equation}
where $\theta$ is an arbitrary phase, which we can take to be $\theta = 0$.
It is important to note that this condition can only be satisfied for fields that carry no additive quantum numbers~\cite{Schwartz:2014sze}.

In the Weyl basis, \cref{eq:majorana_condition} relates the left-handed and right-handed components of $\psi(x)$ such that~\cite{Schwartz:2014sze}
\begin{equation}
\psi(x) =
\begin{pmatrix}
\psi_L \\
i \sigma_2 \psi^*_L \\
\end{pmatrix},
\end{equation}
where the number of degrees of freedom has been reduced from four to two.
Since $i \sigma_2 \psi^*_L$ transforms like a right-handed spinor, we can now write mass terms of the form $i m \psi_L^\dagger \sigma_2 \psi_L^*$ and $i m \psi_L^\dagger \sigma_2 \psi_L^*$.
These are Majorana mass terms.
Note that they couple the same chiral component of the fermion.

The impossibility of a neutrino Majorana mass term is a bit more nuanced.
Majorana mass terms for neutrinos in the SM contain the bi-linear expression $\nu_L^T \sigma_2 \nu_L$.
However, $\nu_L$ belongs to an $SU(2)_L$ doublet in the SM, thus this Majorana mass term transforms as a triplet under $SU(2)_L$.
It also breaks lepton number by two units, hence it also violates baryon minus lepton number ($B-L$), which is conserved to all orders of the SM gauge couplings~\cite{Mohapatra:1998rq}.
Therefore, neutrinos also cannot have a Majorana mass term in the SM.

Despite these arguments, neutrino oscillations have given physicists definitive evidence that at least two of the three SM neutrino masses are nonzero (as discussed in \cref{sec:history}).
This requires the presence of physics beyond the Standard Model (BSM).
The minimal extension of the SM which can accommodate nonzero neutrino masses introduces additional right-handed neutrino states $N_R$~\cite{Mohapatra:1998rq,Schwartz:2014sze}.
These fields, which are singlets under the SM gauge group, can generate both Dirac and Majorana mass terms for neutrinos.
The most general expression for the neutrino mass Lagrangian is then
\begin{equation} \label{eq:general_nu_mass}
-\LL_{\rm mass} = \frac{1}{2} 
\begin{pmatrix}
\overline{\nu}_L & \overline{N^C_L}
\end{pmatrix}
\begin{pmatrix}
0 & M \\
M^T & B \\
\end{pmatrix}
\begin{pmatrix}
\nu_R^C \\
N_R
\end{pmatrix}
+ {\rm h.c.},
\end{equation}
where $M$ and $B$ are the Dirac and Majorana mass matrices of the neutrino sector, respectively, and $\nu_L$ and $N_R$ are column vectors containing the left-handed and right-handed projections of each neutrino generation.

In order to obtain the mass eigenstates of this theory, one must diagonalize the mass matrix in \cref{eq:general_nu_mass}.
If we assume one generation of neutrinos, the eigenvalues of this mass matrix are
\begin{equation}
m_{1,2} = \frac{1}{2} (\sqrt{B^2 + 4M^2} \mp B).
\end{equation}
In the limit $B>>M$, the eigenvalues are approximately given by
\begin{equation} \label{eq:seesaw}
m_1 \approx \frac{M^2}{B},
~~~
m_2 \approx B.
\end{equation}
This is the famous ``seesaw mechanism'' for neutrino mass generation~\cite{Yanagida:1979as}.
One can see that if $B$ is at roughly the GUT scale ($10^{16}$~GeV) and $M$ is at roughly the electro-weak scale (100~GeV), we see that $m_1 < 1$~eV.
This is the right order-of-magnitude regime predicted by neutrino oscillation data and is consistent with existing upper bounds on the neutrino mass from KATRIN~\cite{KATRIN:2021uub}.
Thus, this model is an elegant explanation of the observed neutrino oscillation phenomenon, though experimental confirmation of right-handed neutrino fields at the GUT scale is probably not feasible for quite a long time.

While we do not know the mechanism through which neutrinos acquire mass, it is relevant to ask whether the resulting mass terms are Dirac or Majorana in nature.
An extensive worldwide experimental program is currently underway to answer this question by searching for neutrino-less double beta decay, a rare decay process in which a nucleus undergoes two simultaneous beta decays without emitting any neutrinos in the final state~\cite{KamLAND-Zen:2022tow,CUORE:2017tlq,GERDA:2018pmc}.
A positive observation would imply that neutrinos are Majorana.

As discussed in \cref{sec:history}, perhaps the most famous consequence of massive neutrinos is the phenomenon of neutrino oscillations~\cite{Pontecorvo:1957qd,Pontecorvo:1967fh,Maki:1962mu}.
This arises because the three weak flavor eigenstates $\nu_\alpha$ are not aligned with the three mass eigenstates $\nu_i$.
The two bases are related by the unitary Pontecorvo--Maki--Nakagawa--Sakata (PMNS) mixing matrix $U_{\alpha i}$,
\begin{equation} \label{eq:PMNS_matrix}
\begin{pmatrix}
\nu_e \\
\nu_\mu \\
\nu_\tau
\end{pmatrix}
=
\begin{pmatrix}
U_{e1} & U_{e2} & U_{e3} \\
U_{\mu 1} & U_{\mu 2} & U_{\mu 3} \\
U_{\tau 1} & U_{\tau 2} & U_{\tau 3} \\
\end{pmatrix}
\begin{pmatrix}
\nu_1 \\
\nu_2 \\
\nu_3
\end{pmatrix}.
\end{equation}

As seen in \cref{eq:SMlagrangian}, neutrinos interact in the weak flavor eigenstates $\nu_\alpha$.
Thus, a neutrino a produced alongside a charged anti-lepton $\overline{\ell}$ is in the state
\begin{equation}
\ket{\nu(t=0)} = \ket{\nu_\ell} = \sum_{i \in \{1,2,3\}} U_{\ell i} \ket{\nu_i}.
\end{equation}
Neutrinos propagate, however, in their mass eigenstates.
Each mass eigenstate $\nu_i$ is associated with a mass $m_i$ and four-momentum $(p_i)_\mu = (E_i, \Vec{p_i})$ satisfying the on-shell requirement $(p_i)^2 = m_i^2$.
Thus, after a time $t$, the neutrino will be in the state
\begin{equation}
\ket{\nu(t)} = \sum_i e^{- i p_i \cdot x} U_{\ell i} \ket{\nu_i}.
\end{equation}

The overlap with a different weak flavor eigenstate $\nu_{\ell'} \neq \nu_\ell$ is non-trivial, given by the expression
\begin{equation}
\begin{split}
\braket{\nu_{\ell'}|\nu(t)} &= \sum_{i,j} \bra{\nu_j} U^\dagger_{j \ell'} e^{- i p_i \cdot x} U_{\ell i} \ket{\nu_i} \\
&= \sum_i e^{- i p_i \cdot x} U_{\ell i} U_{\ell' i}^*,
\end{split}
\end{equation}
where we have invoked the orthonormality of the mass basis in the last line.
The probability of finding a neutrino in flavor eigenstate $\nu_{\ell'}$ given an initial $\nu_\ell$ state is then
\begin{equation} \label{eq:osc_prob}
\begin{split}
P_{\nu_\ell \to \nu_{\ell'}}(t) &= |\braket{\nu_\ell'|\nu(t)}|^2 \\
&= \sum_{i,j} |U_{\ell i} U_{\ell' i}^* U_{\ell j}^* U_{\ell' j}| e^{-i(p_i-p_j)\cdot x + i \phi_{\ell \ell' i j}}.
\end{split}
\end{equation}
where $\phi_{\ell \ell' i j} \equiv \rm{arg}(U_{\ell i} U_{\ell' i}^* U_{\ell j}^* U_{\ell' j})$.

We now make a simplifying assumption, in which all neutrino mass eigenstates propagate with the same momentum, i.e. $\Vec{p}_i = \Vec{p}_j \equiv \Vec{p} \forall i,j$.
This treatment is not necessarily physical.
However, for the parameters relevant to most laboratory neutrino experiments, it leads to the same result as the correct but complicated full treatment of the quantum mechanical neutrino wave packet~\cite{Kayser:1981ye}.
Given this assumption along with the approximation that $m_i << p_i$ (which should hold for all existing and near-future experiments), we can show
\begin{equation}
\begin{split}
(p_i- p_j) \cdot x &= (E_i - E_j)t \\
&= \Big(\sqrt{\Vec{p}^2 +m_i^2} - \sqrt{\Vec{p}^2 +m_j^2}\Big)t \\
& \approx \frac{\Delta m_{ij}^2 t}{2|\Vec{p}|},
\end{split}
\end{equation}
where $\Delta m_{ij}^2 = m_i^2 - m_j^2$. Working in natural units ($c = \hbar = 1$), we note that ultra-relativistic neutrinos satisfy $|\Vec{p}| \approx E$ and $t \approx L$, where $L$  is the distance traveled by the neutrino.
Taking only the real part of the exponential in \cref{eq:osc_prob}, we have
\begin{equation}\label{eq:osc_prob_simple}
P_{\nu_\ell \to \nu_{\ell'}}(t) = \sum_{i,j} |U_{\ell i} U_{\ell' i}^* U_{\ell j}^* U_{\ell' j}| \cos\Big( \frac{\Delta m_{ij}^2 L}{2E} - \phi_{\ell \ell' i j}\Big).
\end{equation}

If we consider a two-neutrino paradigm, the unitary mixing matrix is real and can be parameterized by a single ``mixing angle'' $\theta$,
\begin{equation}
U \equiv
\begin{pmatrix}
U_{\ell 1}  & U_{\ell 2} \\
U_{\ell' 1}  & U_{\ell' 2}
\end{pmatrix}
=
\begin{pmatrix}
\cos \theta  & \sin \theta \\
-\sin \theta & \cos \theta
\end{pmatrix}.
\end{equation}
In this scenario, summing over the two mass eigenstates as in \cref{eq:osc_prob_simple} gives
\begin{equation} \label{eq:two_nu_osc}
P_{\nu_\ell \to \nu_{\ell'}}(t) = \sin^2 2\theta \sin^2 \Big( \frac{\Delta m^2 L}{4E} \Big).
\end{equation}

The extension to the standard three neutrino paradigm can be found in any text on neutrino oscillations.
We quote the result here.
Three mass eigenstates lead to two independent mass-squared splittings, $\Delta m_{12}^2$ and $\Delta m_{23}^2$.
The mixing matrix in \cref{eq:PMNS_matrix} can be parameterized by three real mixing angles $\theta_{ij}$ and one complex CP-violating phase $\delta$,
\begin{equation}
U = 
\begin{pmatrix}
1 & 0 & 0 \\
0 & c_{23} & s_{23} \\
0 & -s_{23} & c_{23} \\
\end{pmatrix}
\begin{pmatrix}
c_{13} & 0 & s_{13} e^{-i \delta} \\
0 & 1 & 0 \\
-s_{13} e^{i \delta} & 0 & c_{13}  \\
\end{pmatrix}
\begin{pmatrix}
c_{12} & s_{12} & 0 \\
-s_{12} & c_{12} & 0 \\
0 & 0 & 1 \\
\end{pmatrix}
\end{equation}
where $c_{ij} \equiv \cos \theta_{ij}$ and $s_{ij} \equiv \sin \theta_{ij}$.
The three mixing angles ($\theta_{12}$, $\theta_{13}$, $\theta_{23}$) and two relevant mass squared splittings $\Delta m_{12}^2$ and $|\Delta m_{23}^2|$ have been measured to a precision of $\mathcal{O}(1\%-10\%)$ over the past two decades~\cite{deSalas:2020pgw,Esteban:2020cvm,Capozzi:2021fjo}.
An extensive experimental program is planned to measure $\delta$ to similar precision, as well as the neutrino hierarchy (i.e., the sign of $\Delta m_{23}^2$) and the octant of $\theta_{23}$~\cite{ParticleDataGroup:2020ssz}.  

\section{Anomalies in the Neutrino Sector} \label{sec:anomalies}

Despite the success of the three-neutrino mixing paradigm, several anomalous results have appeared.
Perhaps the most famous of these is the excess of $\bar{\nu}_e$ candidate events observed by the Liquid Scintillator Neutrino Detector (LSND) experiment~\cite{LSND:2001aii}.
LSND took data at Los Alamos Meson Physics Facility (LAMPF) from 1993-1998, observing neutrino interactions from a high-intensity decay-at-rest (DAR) source.
The LSND detector was a 167-ton cylindrical tank of mineral oil that collected scintillation and Cherenkov light produced in neutrino interactions.
The LAMPF accelerator provided a $\sim 1\;{\rm mA}$ beam of 798 MeV protons, which were then focused on a water or high-Z target.
This process created a large number of pions, which then decayed to produce neutrinos.
Most $\pi^-$ came to rest and were captured by nuclei in and around the target, and the $\pi^+ \to  \nu_e e^+$ decay chain is helicity-suppressed due to the interplay between angular momentum conservation and the left-chiral nature of the weak interaction.
Thus the dominant neutrino production process was $\pi^+ \to \nu_\mu (\mu^+ \to \bar{\nu}_\mu \nu_e e^+)$.

\begin{figure}
    \centering
    \includegraphics[width=0.6\textwidth]{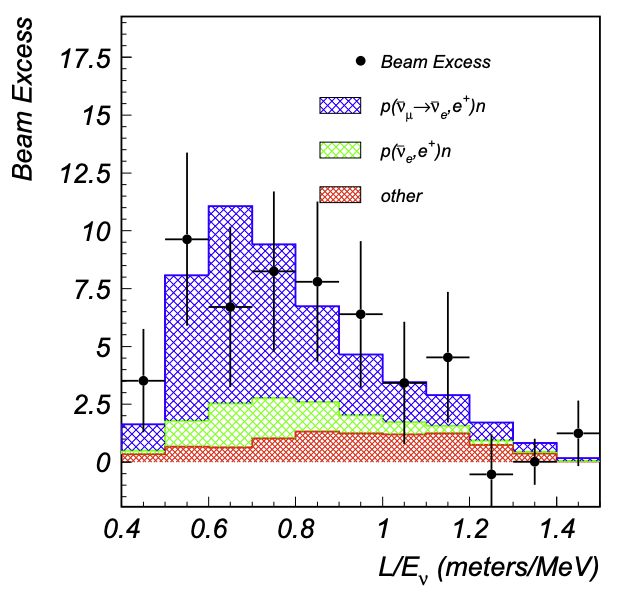}
    \caption{The LSND excess of $\overline{\nu}_e$ events on top of the predicted SM background (green and red regions). The blue region indicates the best fit to $\overline{\nu}_\mu \to \overline{\nu}_e$ oscillations via a sterile neutrino state. Figure from Ref.~\cite{LSND:2001aii}}
    \label{fig:LSND_excess}
\end{figure}

LSND looked specifically for $\bar{\nu}_\mu \to \bar{\nu}_e$ conversion using $\bar{\nu}_\mu$ from $\mu^+$ DAR.
The $\bar{\nu}_e$ events were observed via the inverse beta decay (IBD) process.
This is a very clean channel, as one can require a coincidence between the initial positron emission and the subsequent neutron capture on hydrogen, which releases a characteristic $2.2\;{\rm MeV}$ photon.
The intrinsic $\bar{\nu}_e$ flux, coming predominately from $\pi^-$ decay-in-flight (DIF), was suppressed compared to intrinsic $\bar{\nu}_\mu$ by a factor of $\sim 8 \times 10^{-4}$.
Any significant excess of $\bar{\nu}_e$ would be evidence for $\bar{\nu}_\mu \to \bar{\nu}_e$ oscillations.
This is exactly what LSND observed, as shown in \cref{fig:LSND_excess}.
However, the neutrino energies $\mathcal{O}(30\;{\rm MeV})$ and baselines $(\mathcal{O}(30\;{\rm m})$ required a mass-squared-splitting of $\Delta m^2 \sim 1\;{\rm eV^2}$.
This is larger than the measured values of $\Delta m^2_{12}$ and $|\Delta m^2_{23}|$ by at least three orders of magnitude--therefore, the LSND result cannot be explained by the standard three neutrino oscillation paradigm.
One must introduce a fourth neutrino to the SM neutrinos in order to facilitate such oscillations.
Measurements of the invisible width of the $Z$~boson forbid this neutrino from coupling to the weak force in the same way as the three SM neutrinos~\cite{Abrams:1989yk}.
Thus, this fourth neutrino is typically referred to as a ``sterile neutrino'' ($\nu_s$).
The sterile neutrino paradigm will be introduced in more detail in \cref{sec:anomalies} and discussed thoroughly throughout this thesis.
The LSND anomaly is currently under direct investigation by the follow-up JSNS$^2$ experiment~\cite{JSNS2:2013jdh,Maruyama:2022juu}, which will use a gadolinium-loaded liquid scintillator detector~\cite{JSNS2:2021hyk} to measure IBD interactions at the J-PARC Materials and Life Science Experimental Facility.

The Mini Booster Neutrino Experiment (MiniBooNE) was designed to follow up on the LSND anomaly~\cite{MiniBooNE:2007uho}.
MiniBooNE took data at Fermilab's Booster Neutrino Beam (BNB) from 2002-2019, observing the interactions of neutrinos with energy $E \sim 500\;{\rm MeV}$ in an 800-metric-ton mineral oil (CH$_2$) detector~\cite{MiniBooNE:2008paa}.
The Fermilab Booster accelerates protons to a kinetic energy of 8\;GeV, at which point they collide with the beryllium target of the BNB.
This produces a cascade of mesons, predominately pions.
The charged mesons are focused using a magnetic horn and decay in a 50\;m decay pipe; in the nominal ``neutrino mode'', the magnetic field is generated to create a flux of mostly muon neutrinos from $\pi^+$ decay-in-flight~\cite{MiniBooNE:2008hfu}.
The current in the magnetic horns can be reversed to instead focus $\pi^-$ along the beamline, thus creating a beam of mostly muon antineutrinos--this is referred to as ``antineutrino mode''.
MiniBooNE was situated at a baseline of $L \sim 500\;{\rm m}$ from the BNB target, resulting in a similar characteristic $L/E$ as that of LSND, $\approx 1\;{\rm m/MeV}$.
By \cref{eq:osc_prob_simple}, this means MiniBooNE would also be sensitive to oscillations at $\Delta m^2 \sim 1\;{\rm eV}^2$.

In 2007, MiniBooNE began to report results from their flagship analysis: the search for an excess of $\nu_e$ events in the BNB~\cite{MiniBooNE:2007uho}.
MiniBooNE relied primarily on the reconstruction of Cherenkov light from charged final state particles to identify neutrino interactions.
Thus, $\nu_e$ CC interactions would show up as a ``fuzzy'' Cherenkov ring due to multiple scattering of the electron as well as the induced EM shower~\cite{Patterson:2009ki}.
These fuzzy Cherenkov ring events are hereafter referred to as ``electron-like'' events.
Starting with the initial results~\cite{MiniBooNE:2007uho,MiniBooNE:2008yuf}, MiniBooNE has consistently observed an excess of electron-like events above their expected SM background, the significance of which has grown over the 17-year data-taking campaign of the experiment~\cite{MiniBooNE:2020pnu}.
\Cref{fig:miniboone_enuqe} shows the $4.8\sigma$ MiniBooNE electron-like excess considering the total neutrino mode dataset, corresponding to $18.75 \times 10^{20}$ protons-on-target (POT)~\cite{MiniBooNE:2020pnu}.
A similar excess was observed in the antineutrino mode dataset~\cite{MiniBooNE:2018esg}.
The as-yet-unexplained MiniBooNE excess represents one of the most significant disagreements with the SM to date.

Though the origin of the MiniBooNE excess remains unknown, neutrino physicists have converged on a number of potential explanations.
The most famous explanation involves sterile neutrino-driven $\nu_\mu \to \nu_e$ oscillations consistent with the LSND result ($\Delta m^2 \sim 1\;{\rm eV^2}$).
While this model can explain at least some of the MiniBooNE excess, the excess in the lowest energy region ($E_\nu \lesssim 400\;{\rm MeV}$) sits above even the best-fit sterile neutrino solution.
Due to the Cherenkov nature of the detector, electrons and photons are essentially indistinguishable--both seed EM showers which appear as fuzzy Cherenkov rings.
Thus, the MiniBooNE excess could also come from a mismodeled photon background.
Though not the subject of this thesis, there have been extensive experimental and theoretical efforts, both within and outside of the MiniBooNE collaboration, to validate the MiniBooNE SM photon background prediction~\cite{MiniBooNE:2020pnu,MicroBooNE:2021zai,Brdar:2021ysi,Kelly:2022uaa}.
One can also consider BSM sources of electron-like events in MiniBooNE.
Typical models introduce additional sources of photons and/or $e^+ e^-$ events in MiniBooNE through couplings to new dark sector particles.
Resolution of the LSND and MiniBooNE anomalies, often referred to as the short baseline (SBL) anomalies, is a major goal within the particle physics community~\cite{Karagiorgi:2022fgf}.
This thesis specifically investigates the MiniBooNE anomaly in further detail, covering both experimental and phenomenological studies into the origin of the excess.

\begin{figure}
    \centering
    \includegraphics[width=0.6\textwidth]{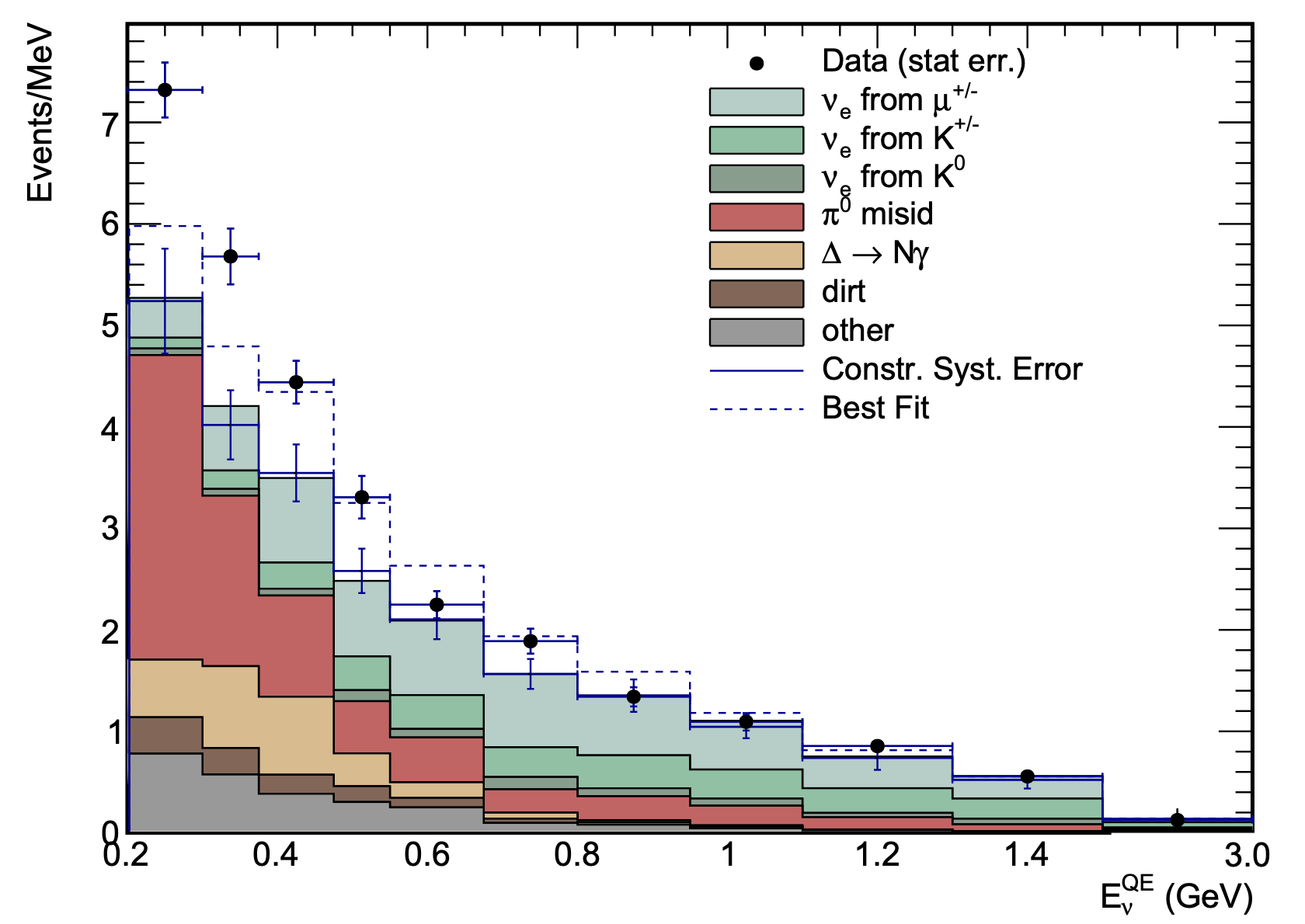}
    \caption{The MiniBooNE electron-like channel data and SM background prediction for the entire neutrino mode dataset, as a function of the reconstructed neutrino energy. }
    \label{fig:miniboone_enuqe}
\end{figure}

\begin{figure}
    \centering
    \includegraphics[width=0.6\textwidth]{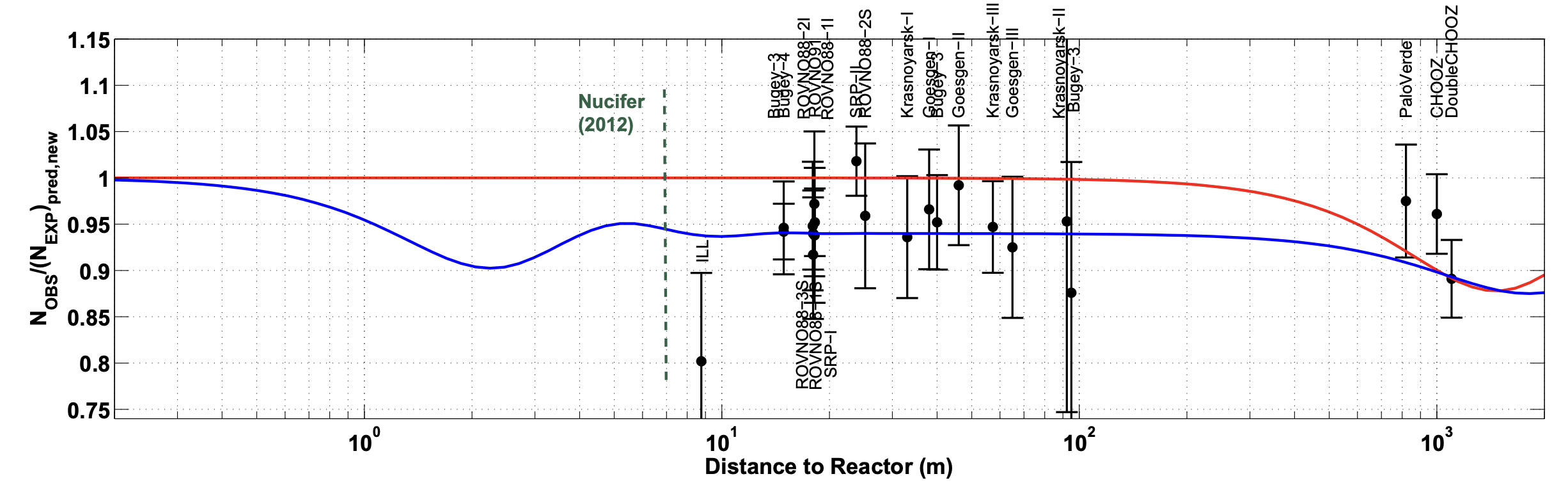}
    \caption{Data contributing to the reactor antineutrino anomaly, indicating the $\sim 5\%$ flux deficit observed by short-baseline reactor neutrino experiments. The red line indicates the prediction incorporating SM neutrino oscillations only, while the blue line shows an example prediction including a sterile neutrino. Figure from Ref.~\cite{Abazajian:2012ys}.}
    \label{fig:RAA}
\end{figure}

We now briefly touch on two additional classes of anomalies that have surfaced over the years: the reactor antineutrino anomaly (RAA) and the gallium anomaly.
The RAA~\cite{Mention:2011rk} is a $\sim 5\%$ deficit in the total $\overline{\nu}_e$ rate observed from nuclear reactors compared to the theoretical expectation from the Huber-Mueller (HM) model~\cite{Huber:2011wv,Mueller:2011nm}.
The HM model combines results using the summation method (summing the contributions of all beta-decay branches in the reactor) and the conversion method (relying on older measurements of the $\overline{\nu}_e$ flux from the different fissionable isotopes in the reactor).
The data contributing to the RAA mostly come from reactor neutrino experiments operating at baselines short enough that the effects of SM neutrino oscillations are negligible.
One can interpret the RAA as due to $\overline{\nu}_e$ disappearance via oscillations involving a sterile neutrino.
Coincidentally, due to the relevant neutrino energies and baselines, such a solution requires $\Delta m^2 \gtrsim 1\;{\rm eV}^2$, similar to the LSND and MiniBooNE solution~\cite{Abazajian:2012ys}.
\Cref{fig:RAA} shows an overview of the RAA circa 2012, including the suite of short baseline reactor experiments which observe a deficit with respect to the HM model with SM neutrino oscillations (red line), as well as an example sterile neutrino solution to the RAA (blue line).
Recently, the reactor $\overline{\nu}_e$ flux calculation has been revisited by various groups, each of which improves upon some aspect of the summation or conversion method used in the HM flux model~\cite{Estienne:2019ujo,Hayen:2019eop,Kopeikin:2021ugh,Giunti:2021kab}.
The significance of the RAA either diminishes or disappears in some of these models; however, these improved models have difficulty removing the RAA while also explaining the ``5-MeV bump'' observed by most short baseline reactor experiments with respect to the HM model~\cite{Giunti:2021kab}.
Thus, while the RAA story is quickly evolving, our understanding of reactor neutrino fluxes is far from clear.

\begin{figure}
    \centering
    \includegraphics[width=0.6\textwidth]{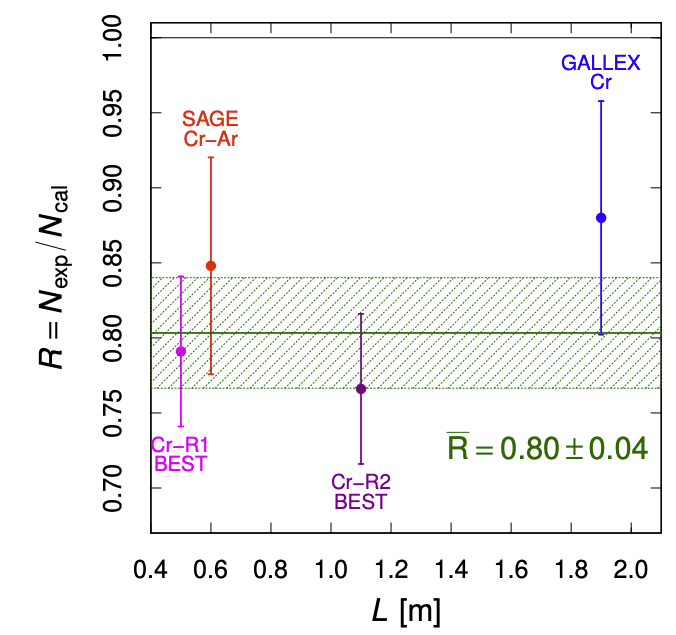}
    \caption{Data contributing to the gallium anomaly, indicating the $\sim 20\%$ deficit in the ${}^{71}$Ge production rate observed by SAGE, GALLEX, and BEST. Figure from Ref.~\cite{Giunti:2022btk}.}
    \label{fig:gallium_anomaly}
\end{figure}

The gallium anomaly refers to a series of gallium-based detectors that have observed a deficit of $\nu_e$ capture events on ${}^{71}$Ga with respect to the theoretical expectation.
The original harbingers of the anomaly, SAGE~\cite{SAGE:2009eeu} and GALLEX~\cite{Kaether:2010ag}, were designed to measure solar neutrinos using the ${}^{71}{\rm Ga} \nu_e  \to {}^{71}{\rm Ge} e^-$ capture process.
Each detector was calibrated using electron capture $\nu_e$ sources, including ${}^{51}$Cr and ${}^{37}$Ar.
Combining all available calibration data across both experiments, the observed ${}^{71}$Ge production rate was lower than the expectation by a factor of $0.87 \pm 0.05$~\cite{SAGE:2009eeu}.
Though the statistical significance of the anomaly was only modest ($2-3\sigma$), the community was already beginning to interpret the anomaly as $\nu_e \to \nu_s$ transitions via an eV-scale sterile neutrino~\cite{Giunti:2006bj}.
A follow-up experiment to the SAGE and GALLEX anomaly, BEST~\cite{Barinov:2021asz}, released their first results in 2021.
BEST placed a 3.414~MCi ${}^{51}$Cr $\nu_e$ source at the center of two nested ${}^{71}$Ga volumes, each with a different average distance from the source.
The ratio of observed to the predicted ${}^{71}$Ge production rate was $R_{in} = 0.79 \pm 0.05$ ($R_{out} = 0.77 \pm 0.05$) for the inner (outer) volume, thus reaffirming the gallium anomaly~\cite{Barinov:2021asz}.
No evidence for a difference in the deficit between the inner and outer volumes was observed, which would have been a smoking gun signature of a baseline-dependent effect like $\nu_e \to \nu_s$ oscillations.
However, the statistical significance of the gallium anomaly is now much stronger; the combined SAGE, GALLEX, and BEST results give evidence for a deficit at the $5.0\sigma$ level~\cite{Giunti:2022btk}.
The datasets contributing to this anomaly are summarized in \cref{fig:gallium_anomaly}.

As alluded to above, the most common BSM interpretation of the SBL, reactor antineutrino, and gallium anomalies is the ``3+1 model'', which involves the addition of a new neutrino state--the sterile neutrino--at the eV scale.
The sterile neutrino introduces a fourth weak interaction eigenstate $\nu_s$ and mass eigenstate $\nu_4$ to the standard three-neutrino mixing paradigm.
Thus, \cref{eq:PMNS_matrix} becomes
\begin{equation} \label{eq:extended_PMNS_matrix}
\begin{pmatrix}
\nu_e \\
\nu_\mu \\
\nu_\tau \\
\nu_s
\end{pmatrix}
=
\begin{pmatrix}
U_{e1} & U_{e2} & U_{e3} & U_{e4}  \\
U_{\mu 1} & U_{\mu 2} & U_{\mu 3} & U_{\mu 4} \\
U_{\tau 1} & U_{\tau 2} & U_{\tau 3} & U_{\tau 4} \\
\end{pmatrix}
\begin{pmatrix}
\nu_1 \\
\nu_2 \\
\nu_3 \\
\nu_4
\end{pmatrix}.
\end{equation}

As we are interested in an eV-scale sterile neutrino, the mass-squared splittings between the three active neutrinos are smaller by at least 2-3 orders of magnitude compared to their mass-squared splittings with the fourth mass eigenstate.
This means that the active neutrino mass splittings are negligible for short-baseline experiments, i.e. those in which the argument of the second $\sin^2$ term in \cref{eq:two_nu_osc} is small.
Experiments contributing to the aforementioned anomalies all satisfy this condition.
Thus, when considering sterile neutrino explanations for these anomalies, we can make the approximation
\begin{equation}
\Delta m^2_{41} \approx \Delta m^2_{42} \approx \Delta m^2_{43} \equiv \Delta m^2,
\end{equation}
where we hereafter use $\Delta m^2$ to refer to the mass-squared splitting of the fourth mass eigenstate.
This approximation holds regardless of the hierarchy of SM neutrino mass eigenstates.

The experiments discussed in this thesis are sensitive only to $\barparen{\nu}_e$ and $\barparen{\nu}_\mu$ interactions.
The sterile neutrino can facilitate short-baseline oscillations between these flavor states; the oscillation probability expressions, which can be derived using \cref{eq:osc_prob_simple} within the 3+1 framework, are given by~\cite{Diaz:2019fwt}
\begin{equation} \label{eq:sterile_osc_prob}
\begin{split}
&P_{\nu_e \to \nu_e} = 1 - 4 \sin^2 2\theta_{ee} \sin^2 (1.27 \Delta m^2 L / E) \\
&P_{\nu_\mu \to \nu_\mu} = 1 - 4 \sin^2 2\theta_{\mu \mu} \sin^2 (1.27 \Delta m^2 L / E) \\
&P_{\nu_\mu \to \nu_e} = 4 \sin^2 2\theta_{\mu e} \sin^2 (1.27 \Delta m^2 L / E),
\end{split}
\end{equation}
where $\Delta m^2$, $L$, and $E$ are in units of eV$^2$, km, and GeV, respectively, and
\begin{equation}
\begin{split}
&\sin^2 2 \theta_{ee} = 4 (1 - |U_{e4}|^2) |U_{e4}|^2 \\
&\sin^2 2 \theta_{\mu \mu} = 4 (1 - |U_{\mu 4}|^2) |U_{\mu 4}|^2 \\
&\sin^2 2 \theta_{\mu e} = 4 |U_{\mu 4}|^2 |U_{e4}|^2.
\end{split}
\end{equation}

\begin{figure}
    \centering
    \includegraphics[width=0.6\textwidth]{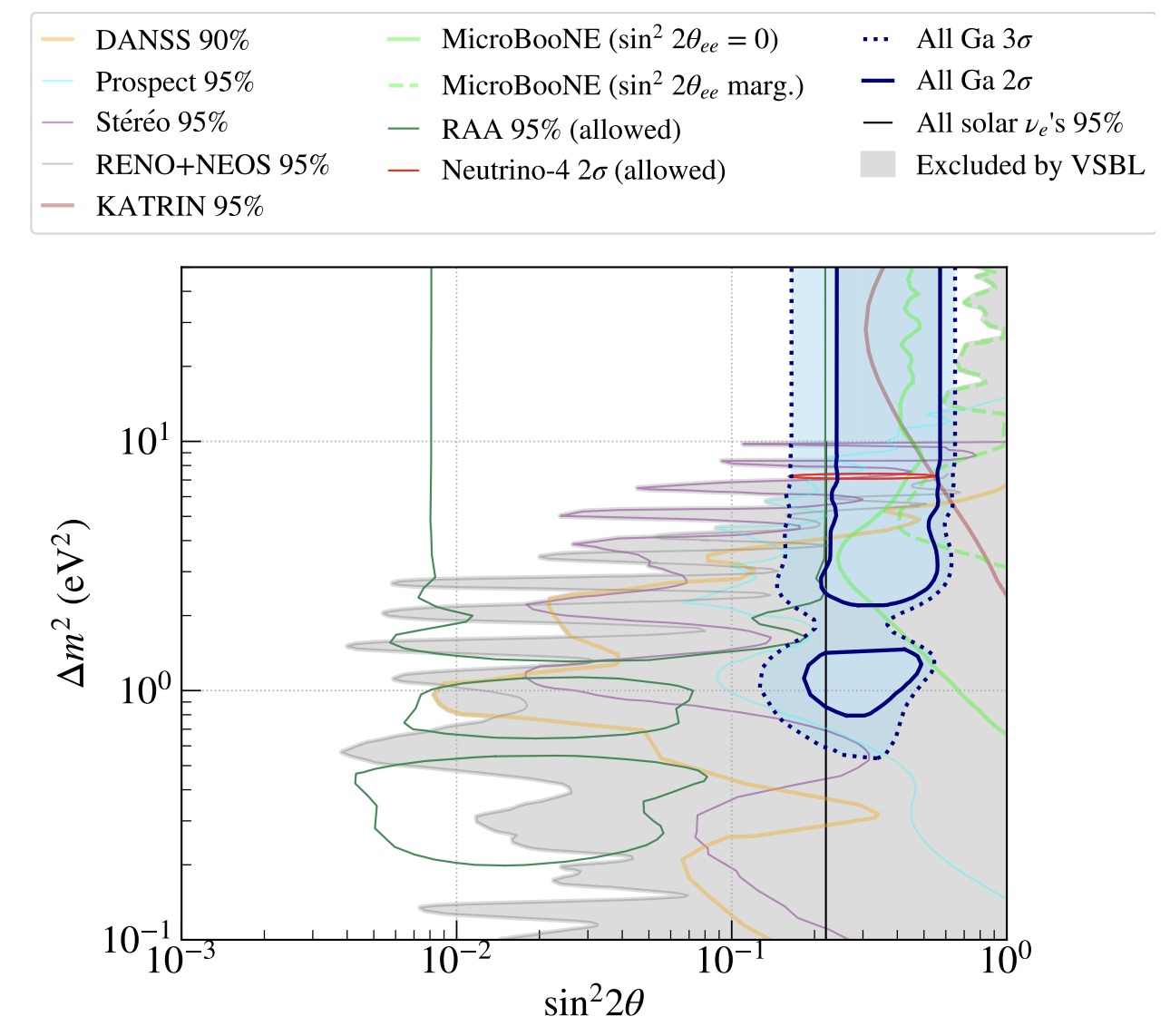}
    \caption{Preferred regions in $\sin^2 2\theta_{ee}$--$\Delta m^2$ parameter space to explain the RAA~\cite{Mention:2011rk} (green contour) and gallium anomaly~\cite{Barinov:2021asz} (blue regions). The total excluded region from other experiments (grey region) is also shown. Figure from Ref.~\cite{Barinov:2021asz}.}
    \label{fig:RAA_GA_sterile}
\end{figure}

The first expression in \cref{eq:sterile_osc_prob} can potentially explain the deficit of $\overline{\nu}_e$ and $\nu_e$ events observed in the RAA and gallium anomaly, respectively.
Though both anomalies stem qualitatively from the same phenomenon--$\barparen{\nu}_e$ disappearance at short baseline--the gallium anomaly in general prefers a larger value of $\sin^2 2\theta_{ee}$ than the RAA.
This is evident in \cref{fig:RAA_GA_sterile}, which shows the regions in $\sin^2 2\theta_{ee}$--$\Delta m^2$ parameter space preferred by the RAA and gallium anomalies, as well as global constraints from other experiments.
These constraints come from short-to-medium-baseline reactor experiments, including NEOS~\cite{NEOS:2016wee}, RENO~\cite{RENO:2020uip}, and Daya Bay~\cite{DayaBay:2016qvc}, as well as very-short-baseline reactor experiments, including STEREO~\cite{STEREO:2022nzk}, DANSS~\cite{DANSS:2021raa}, and PROSPECT~\cite{PROSPECT:2020sxr}.
Each of these experiments searches for $\overline{\nu}_e$ disappearance in a reactor-flux-agnostic way: the former though comparisons of the reactor $\overline{\nu}_e$ spectra measured by different detectors~\cite{RENO:2020hva}, and the latter through the use of modular or movable detectors capable of comparing $\overline{\nu}_e$ interaction rates across different baselines.
The KATRIN experiment, which is sensitive to the neutrino mass via an extremely precise measurement of the tritium beta spectrum endpoint, also places strong constraints on $\sin^2 2\theta_{ee}$ in the $\Delta m^2 \gtrsim 10~{\rm eV}^2$ region~\cite{KATRIN:2022ith}.

\begin{figure}
    \centering
    \includegraphics[width=0.5\textwidth]{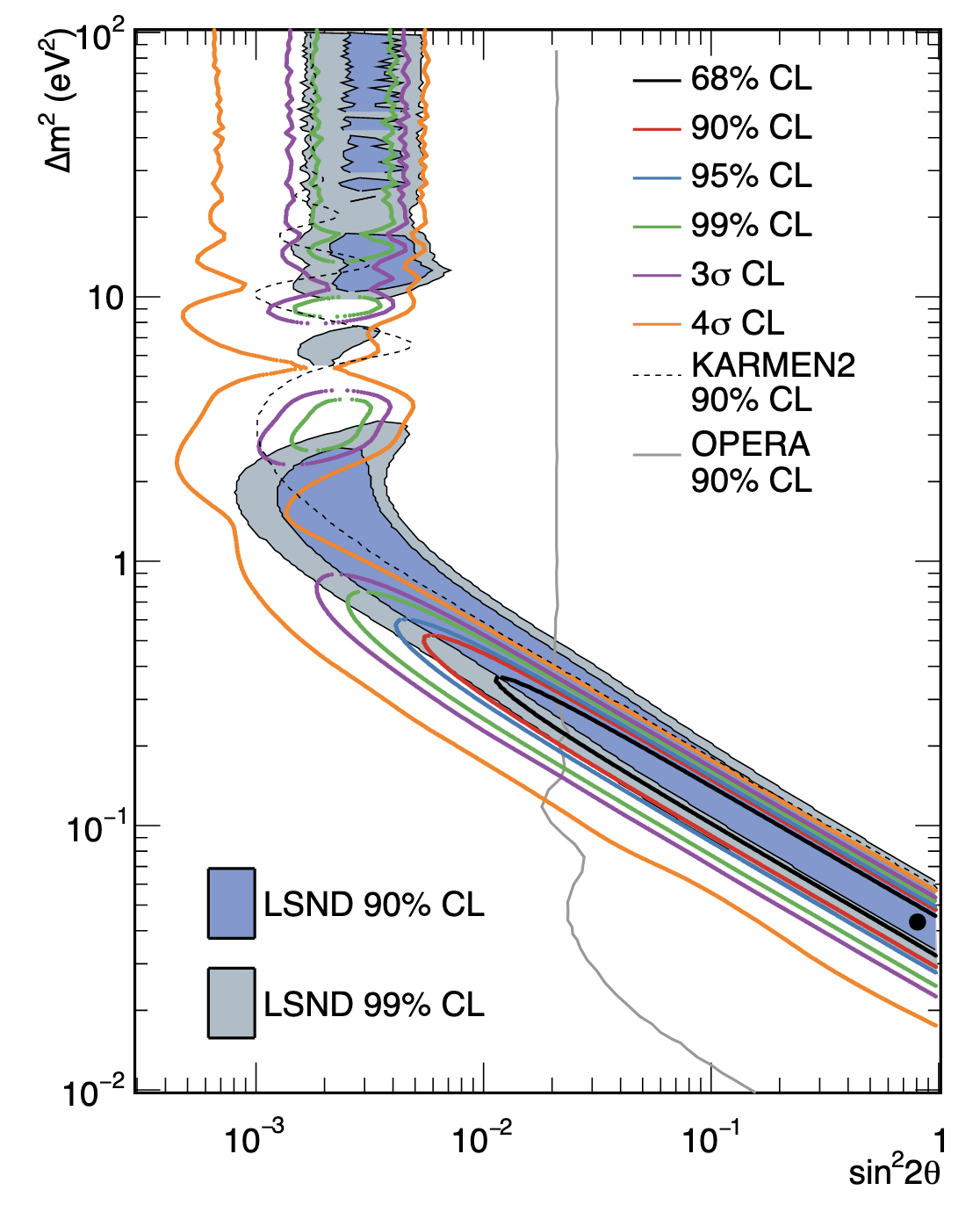}
    \caption{Preferred regions in $\sin^2 2\theta_{\mu e}$--$\Delta m^2$ parameter space to explain the LSND anomaly~\cite{LSND:2001aii} (filled contours) and MiniBooNE anomaly~\cite{MiniBooNE:2020pnu} (open contours). Figure from Ref.~\cite{MiniBooNE:2020pnu}.}
    \label{fig:MB_LSND_sterile}
\end{figure}

The second expression in \cref{eq:sterile_osc_prob} can potentially explain the SBL anomalies.
This is because both LSND and MiniBooNE operated at accelerator neutrino sources for which the neutrino flux was generated mainly by charged pion decay~\cite{LSND:2001aii,MiniBooNE:2008hfu}; thus, due to helicity suppression, the flavor composition was dominated muon-flavored (anti)neutrinos.
This means that even a small value of $\sin^2 2 \theta_{\mu e}$ could generate an observable level of $\barparen{\nu}_e$ appearance on top of the SM $\barparen{\nu}_e$ flux prediction.
\Cref{fig:MB_LSND_sterile} shows the allowed regions in $\sin^2 2\theta_{\mu e}$--$\Delta m^2$ parameter space from LSND and MiniBooNE~\cite{MiniBooNE:2020pnu}.
Strikingly, both anomalies generally prefer the same region of parameter space.
However, as the MiniBooNE excess tends to peak more sharply at lower energies, the 3+1 fit prefers lower values of $\Delta m^2$ compared to the LSND result.

It is important to note that the fits performed in \cref{fig:MB_LSND_sterile} account only for $\barparen{\nu}_\mu \to \barparen{\nu}_e$ oscillations, ignoring any potential $\barparen{\nu}_e$ or $\barparen{\nu}_\mu$ disappearance in the SM background prediction.
This is a reasonable approximation, however, the inclusion of the latter effects does indeed impact the MiniBooNE allowed regions.
This effect was only accounted for recently in Ref.~\cite{MiniBooNE:2022emn}, which is presented in \cref{sec:MBuB_sterile_paper} of this thesis.

While there are indications of short baseline $\barparen{\nu}_\mu \to \barparen{\nu}_e$ appearance and $\barparen{\nu}_e$ disappearance in the global anomaly picture, direct observation of $\barparen{\nu}_\mu$ disappearance via the third expression in \cref{eq:sterile_osc_prob} remains elusive.
Long baseline experiments such as MINOS/MINOS+~\cite{MINOS:2017cae,MINOS:2020iqj} and CCFR84~\cite{Stockdale:1984cg} have searched for muon neutrino disappearance from an accelerator neutrino source.
Additionally, the IceCube experiment has searched for a sterile-induced matter resonance impacting muon neutrinos as they transit through the earth~\cite{IceCube:2020phf}.
So far, no definitive evidence for $\barparen{\nu}_\mu$ disappearance has been found (up to a $\sim 2\sigma$ preference in the IceCube results~\cite{IceCube:2020phf}).

\begin{figure}
     \centering
     \begin{subfigure}[b]{0.45\textwidth}
         \centering
         \includegraphics[width=\textwidth]{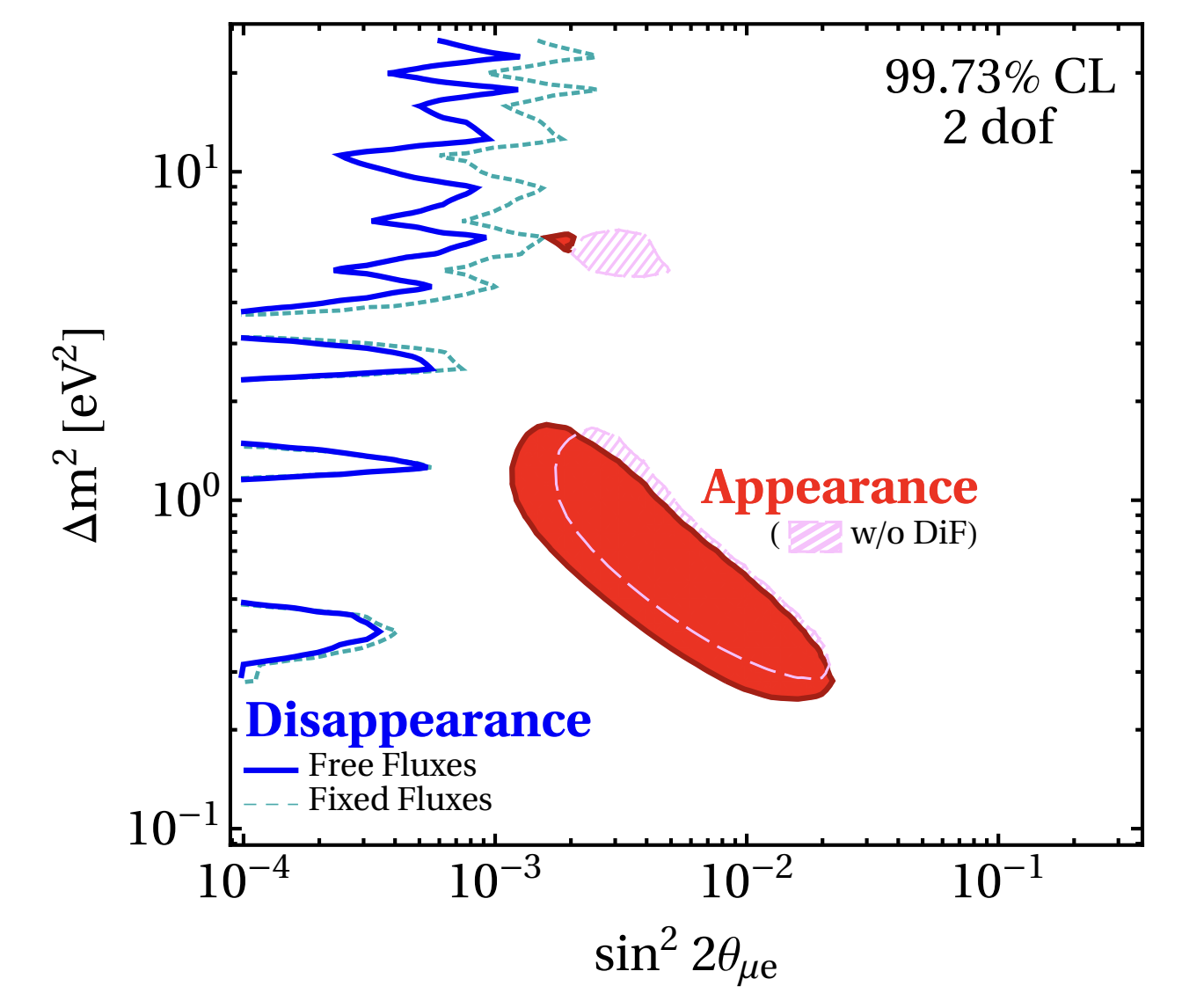}
         \caption{From Ref.~\cite{Dentler:2018sju}}
         \label{fig:global_tension_dentler}
     \end{subfigure}
     \hfill
     \begin{subfigure}[b]{0.45\textwidth}
         \centering
         \includegraphics[width=\textwidth]{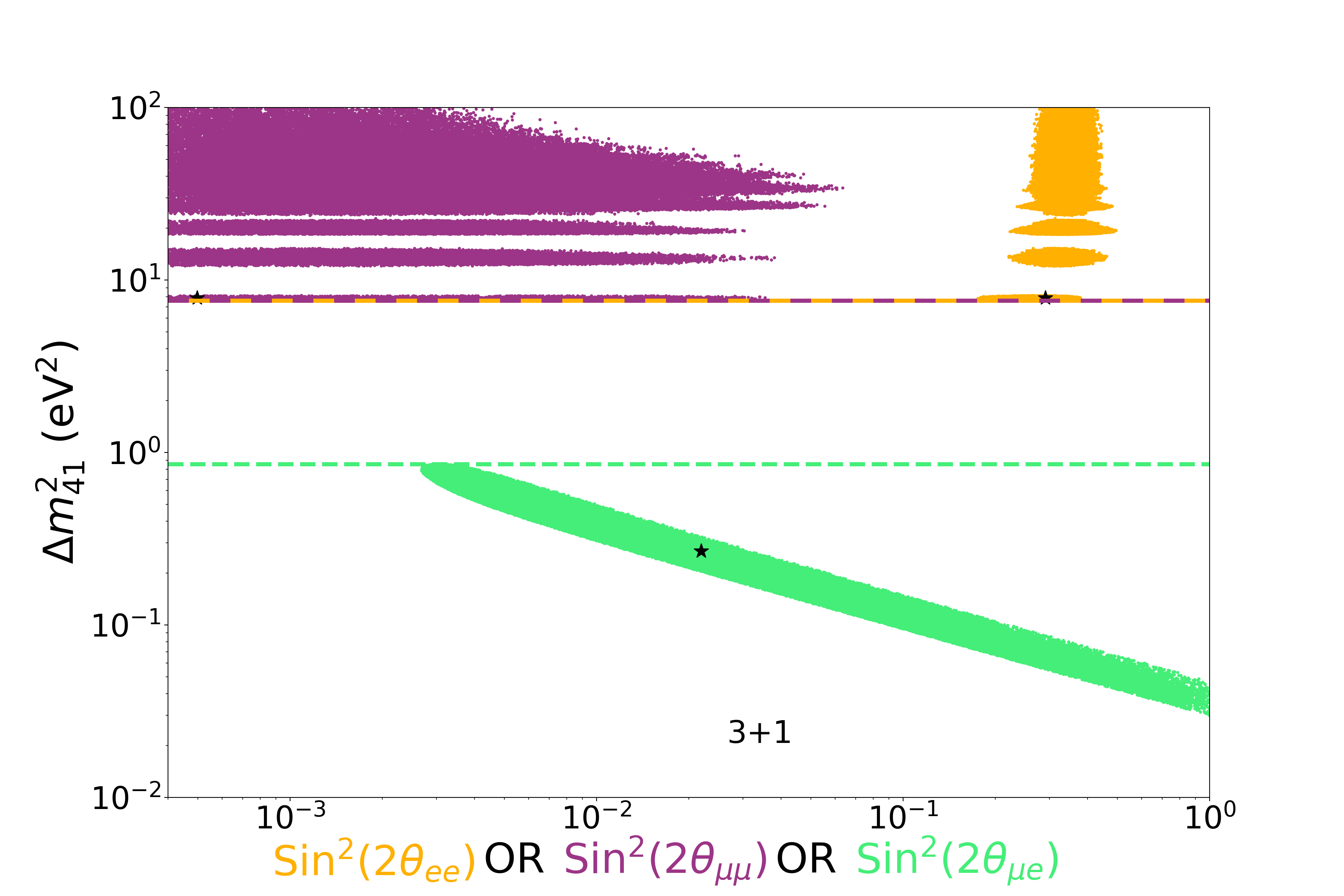}
         \caption{From Ref.~\cite{Hardin:2022muu}}
         \label{fig:global_tension_hardin}
     \end{subfigure}
        \caption{Graphical representation of the tension observed in 3+1 global fits between different subsets of the experimental landscape. \Cref{fig:global_tension_dentler} shows the tension between $\nu_e$ appearance experiments and $\nu_e$/$\nu_\mu$ disappearance experiments observed in Ref.~\cite{Dentler:2018sju}.
        \Cref{fig:global_tension_hardin} shows the tension between allowed regions from $\nu_e$ appearance (lower right), $\nu_e$ disappearance (upper right), and $\nu_\mu$ disappearance (upper left) experiments observed in Ref.~\cite{Hardin:2022muu}, which includes the latest results from the BEST experiment.}
        \label{fig:global_tension}
\end{figure}

The lack of $\barparen{\nu}_\mu$ disappearance introduces significant tension when one tries to fit global neutrino data within a consistent 3+1 model.
This conclusion has been reached by multiple 3+1 global fitting efforts~\cite{Diaz:2019fwt,Dentler:2018sju,Hardin:2022muu}; \cref{fig:global_tension} shows a representation of the tension between appearance and disappearance experiments observed in global fits.
This tension persists even with the inclusion of the recent BEST result, which prefers larger values of $|U_{e4}|^2$ (thus allowing lower values of $|U_{\mu 4}|^2$ to fit the $\barparen{\nu}_e$ appearance anomalies)~\cite{Hardin:2022muu}.
Thus, the 3+1 model, while still an important benchmark BSM scenario, has become disfavored as a solution to all observed anomalies in the neutrino sector.
The state of the sterile neutrino explanation of the SBL anomalies is discussed in more detail throughout this thesis.

In recent years, neutrino physicists have begun to turn toward alternative explanations of the anomalies, often involving dark sector particles with additional interactions. 
\Cref{ch:neutrissimos} of this thesis covers one such explanation of the MiniBooNE anomaly, involving heavy right-handed neutrinos with a transition magnetic moment coupling to the active neutrinos.

\chapter{The MiniBooNE Experiment}
 \label{ch:miniboone}

This chapter is intended to give the reader an overview of the Mini Booster Neutrino Experiment (MiniBooNE), specifically concerning the excess of electron-like events observed by MiniBooNE in data taken between 2002--2019 at Fermilab's Booster Neutrino Beam (BNB).
The MiniBooNE excess is at the center of this thesis; the research presented here covers both experimental follow-up and theoretical interpretations of this anomaly.
Thus, the remaining chapters require a thorough discussion of the MiniBooNE experiment and its most famous result.

\section{Overview of MiniBooNE}

MiniBooNE was originally designed as an experimental follow-up to the LSND excess of $\overline{\nu}_e$ events observed at the Los Alamos Meson Physics Facility (LAMPF)~\cite{LSND:2001aii}.
As described in \cref{sec:anomalies}, the LAMPF flux comprised mostly of $\nubar{\mu}$, which dominated over the $\nubar{e}$ flux by three orders of magnitude~\cite{Conrad:2013mka}.
Because of this, LSND was able to perform a low-background search for the IBD interaction $\nubar{e} p \to e^+ n$.
An excess of IBD events was observed above the intrinsic $\nubar{e}$ flux prediction from the beam dump source--this is known as the ``LSND amonaly''~\cite{LSND:2001aii}.

The LSND anomaly has traditionally been interpreted as evidence for $\nubar{\mu} \to \nubar{e}$ oscillations at $\Delta m^2 \approx 1\;{\rm eV}^2$.
The LSND detector sat relatively close to the LAMPF nuclear target; the characteristic length-to-energy ratio in the experiment was $L/E \sim 30\;{\rm m}/30\;{\rm MeV}$.
According to \cref{eq:two_nu_osc}, in order to be sensitive to the oscillation-based interpretation of the LSND anomaly, one must maintain the same ratio $L/E$.
This was the design strategy of the MiniBooNE experiment, which observed the interactions of neutrinos from the BNB with characteristic energy $E_\nu \sim 500\;{\rm MeV}$, at a baseline of $L \sim 500\;{\rm m}$ from the BNB beryllium target.
The BNB produced mostly $\nu_\mu$ from pion decay-in-flight; thus, MiniBooNE searched for $\nu_\mu \to \nu_e$ oscillations in the BNB at $\Delta m^2 \approx 1\;{\rm eV}^2$.

\subsection{The Booster Neutrino Beam} \label{sec:bnb}

\begin{figure}[h!]
    \centering
    \includegraphics[width=0.6\textwidth]{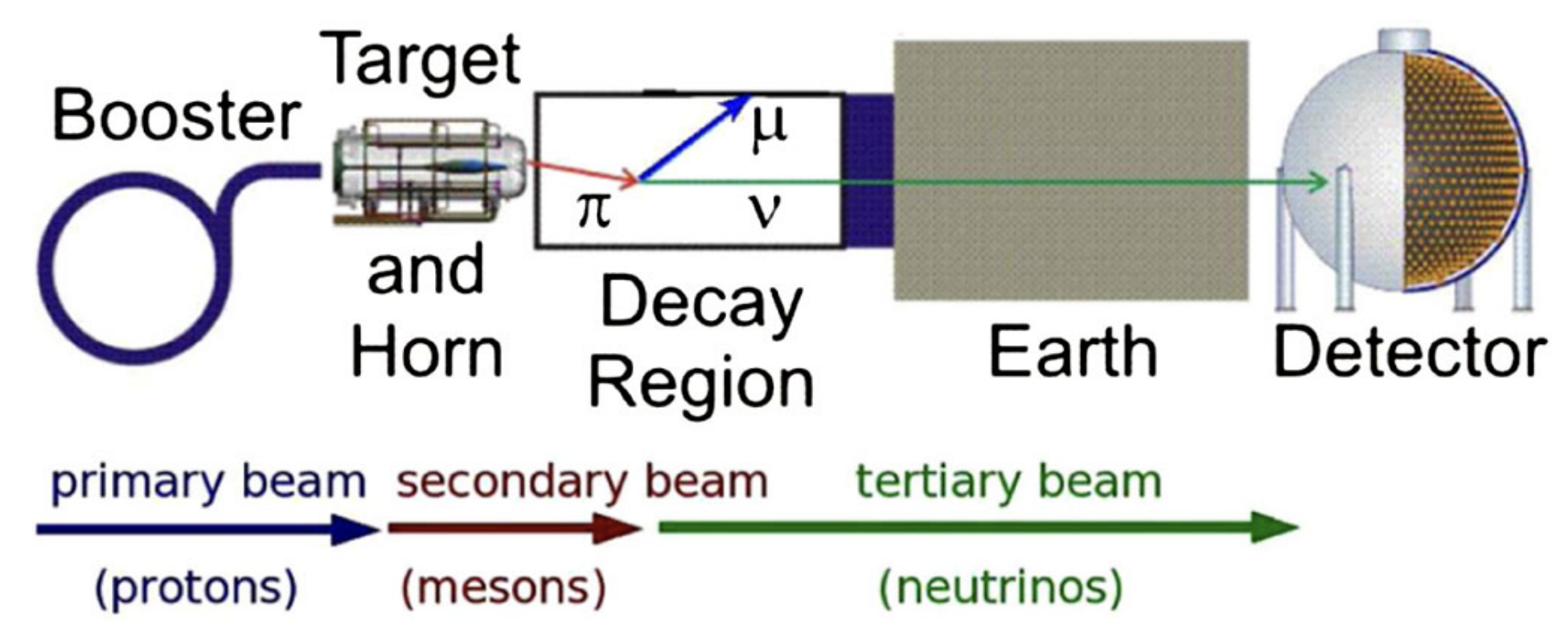}
    \caption{A schematic depiction of the BNB at Fermilab, including the downstream MiniBooNE detector. Figure from Ref.~\cite{MiniBooNEDM:2018cxm}.}
    \label{fig:bnb}
\end{figure}

The BNB, which is still operational, follows the typical design of a neutrino beamline~\cite{Conrad:2013mka}.
Protons are accelerated in a synchrotron up to a momentum of 8.89\;GeV, at which point they are kicked out of the synchrotron and interact within the Be target of the BNB, producing a cascade of secondary particles~\cite{MiniBooNE:2008hfu}.
The charged mesons in this cascade are then focused using a toroidal magnetic field from an aluminum horn.
By switching the direction of the current in the horn (and thus the direction of the magnetic field), one can choose whether to focus positively-charged mesons and de-focus negatively-charged mesons (``neutrino mode''), or vice-versa (``antineutrino mode'').
After focusing, charged mesons pass through a concrete collimator and enter a 50-meter-long air-filled region where they decay to neutrinos.
The neutrinos travel through another 474 meters of bedrock before reaching the MiniBooNE detector.
A schematic depiction of this process is shown in \cref{fig:bnb}.

\begin{figure}[h!]
    \centering
    \includegraphics[width=0.4\textwidth]{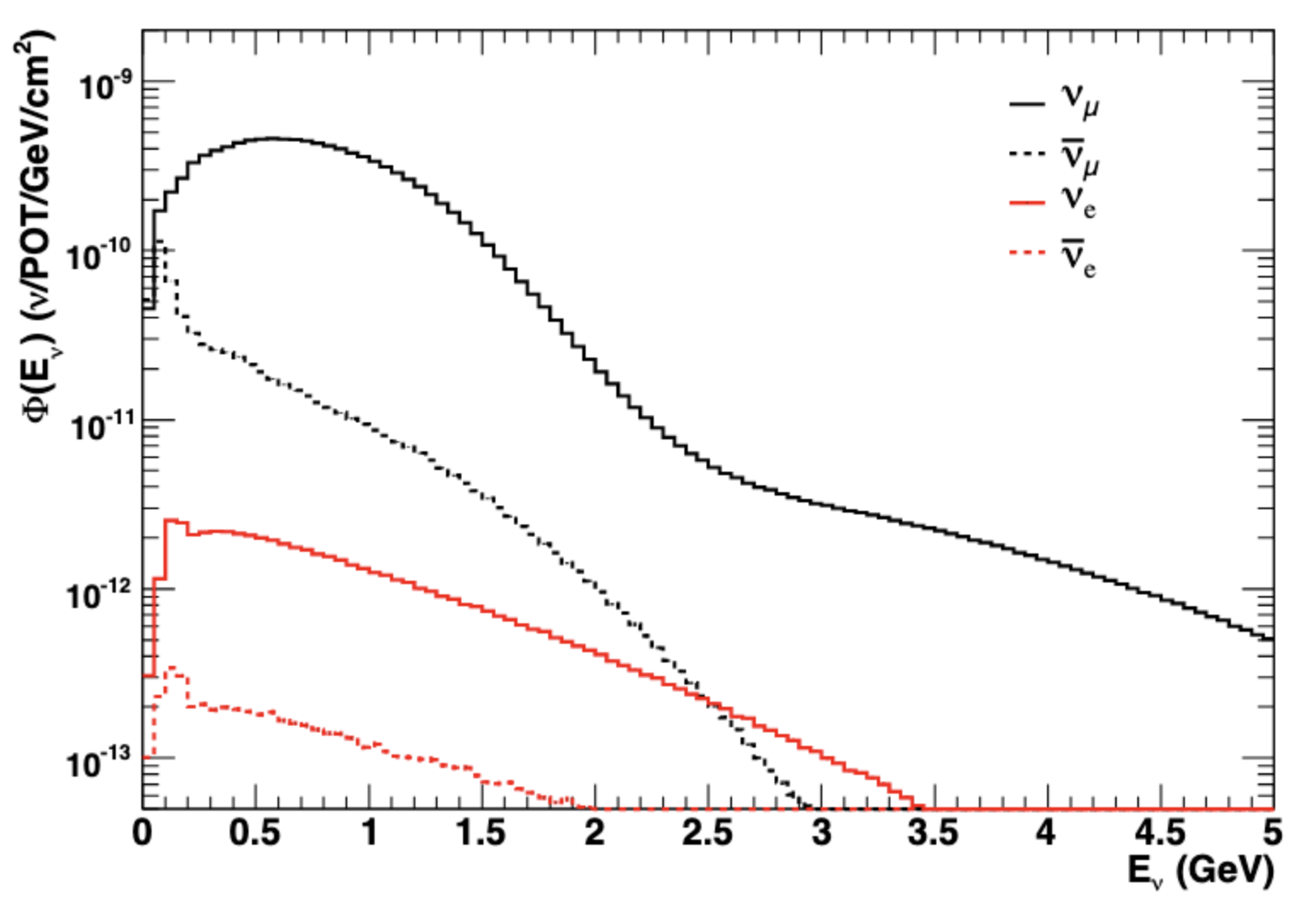}
    \includegraphics[width=0.4\textwidth]{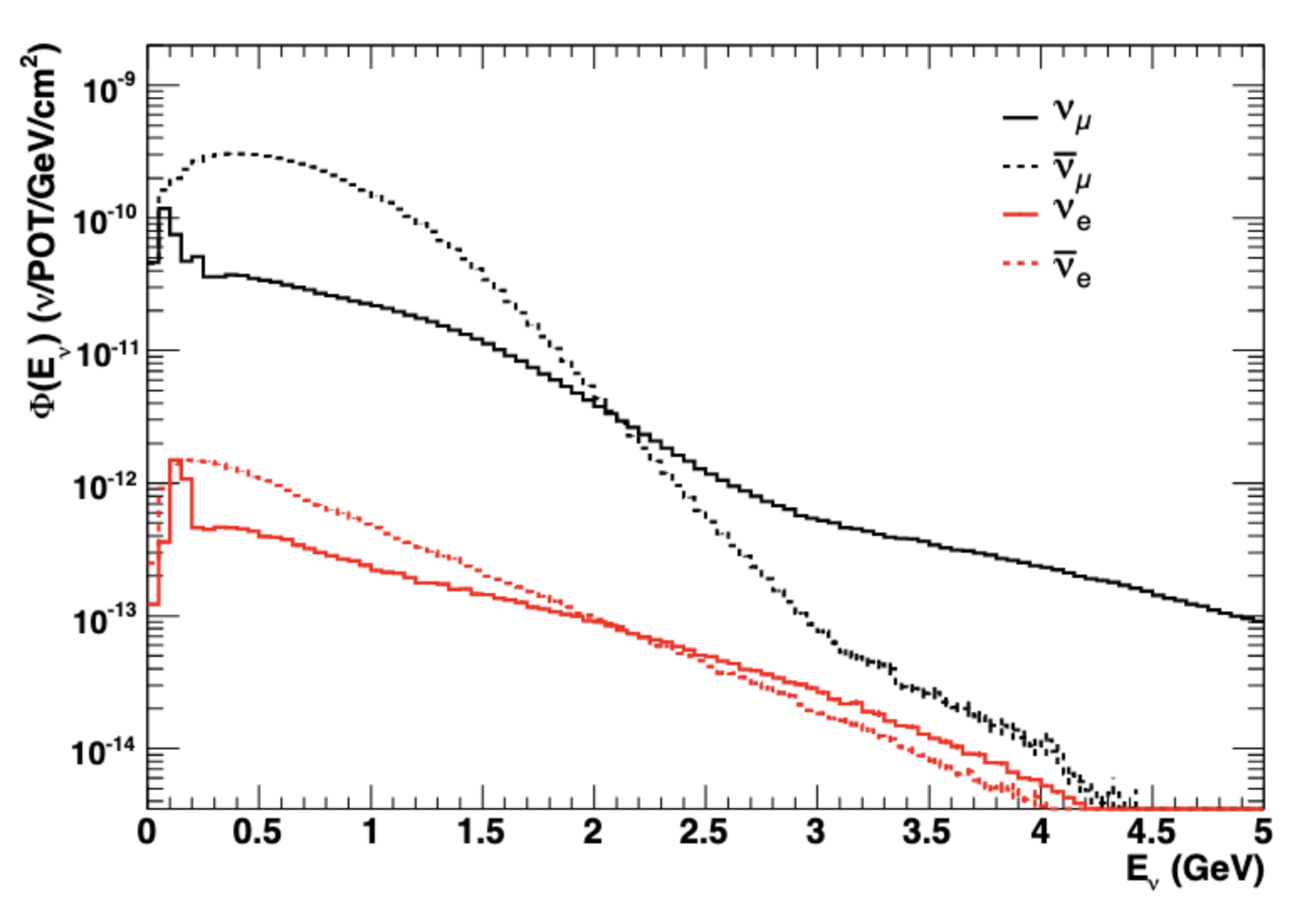}
    \caption{Breakdown of the neutrino flux at the BNB in neutrino (left) and antineutrino (right) mode. Figure from Ref.~\cite{MiniBooNE:2008hfu}}
    \label{fig:bnb_flux}
\end{figure}

The MiniBooNE flux is described in detail in Ref.~\cite{MiniBooNE:2008hfu}; we summarize the most important details here.
In neutrino (antineutrino) mode, the flux is dominated by $\nu_\mu$ ($\nubar{\mu}$) from $\pi^+$ ($\pi^-$) decay.
Wrong-sign $\nubar{\mu}$ ($\nu_\mu$), coming mostly from the decay of oppositely-charged pions, contribute at the 5\% (15\%) level.
Two and three-body kaon decays also contribute to the $\nu_\mu$ and $\nubar{\mu}$ flux at the few-percent level.
The BNB also produces electron (anti)neutrinos, which represent $<1\%$ of the total flux in both neutrino and antineutrino mode.
These come from two main sources: the decay of the secondary muon produced in the original charged pion decay, which is dominant for $E_\nu \lesssim 1\;{\rm GeV}$, and two-body kaon decay, which is dominant for $E_\nu \gtrsim 1\;{\rm GeV}$.
The BNB flux breakdown in neutrino and antineutrino mode are shown in \cref{fig:bnb_flux}.

The $\pi^\pm$ production rate from p-Be interactions has been measured by the HARP~\cite{HARP:2007dqt} and BNL E910~\cite{E910:2007puw} experiments.
HARP took data at the BNB proton incident momentum (8.89\;GeV/c) with a replica of the BNB beryllium target, while E910 took data at varying incident proton momenta above and below the nominal BNB value.
These data were used to constrain a Sanford-Wang parameterization of the $\pi^\pm$ differential production cross section in the BNB~\cite{Wang:1970bn}.
The charged and neutral kaon production rates in p-Be were constrained by measurements from other experiments at proton momenta around 8.89\;GeV/c; the Feynman scaling hypothesis was used to relate these measurements to the BNB proton momentum~\cite{MiniBooNE:2008hfu}.

\subsection{The MiniBooNE Detector} \label{sec:miniboone_detector}

\begin{figure}[h!]
    \centering
    \includegraphics[width=0.6\textwidth]{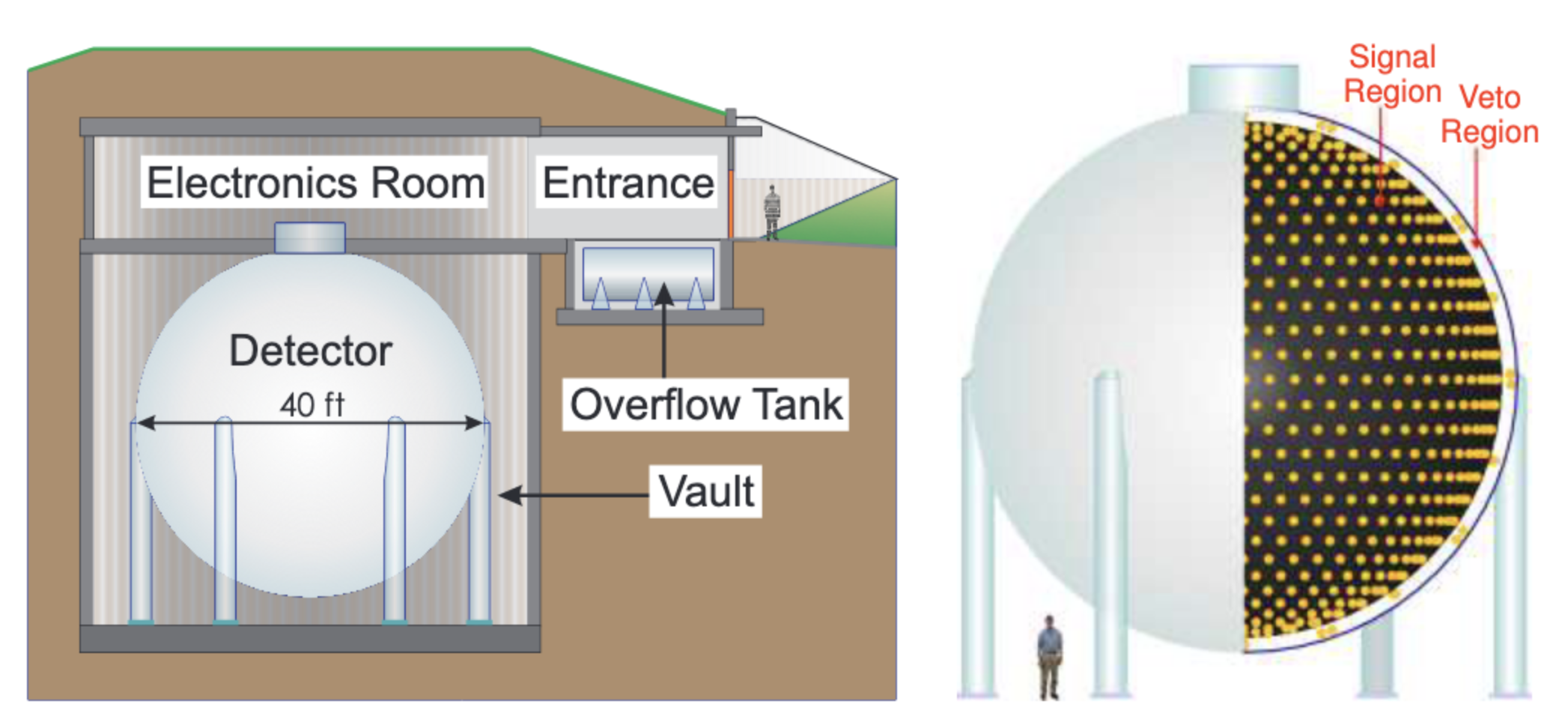}
    \caption{The MiniBooNE detector situated in the cylindrical detector hall (left) and an image of the interior of the MiniBooNE detector (right), showing the PMTs in both the signal and veto regions. Figure from Ref.~\cite{MiniBooNE:2008paa}.}
    \label{fig:miniboone_detector}
\end{figure}

The MiniBooNE detector is an 818-ton, 6.1-meter-radius spherical volume filled with mineral oil (approximately CH$_2$)~\cite{MiniBooNE:2008paa}.
It was designed to measure the Cherenkov light produced from charged particles created in the charged-current interactions of BNB neutrinos within the detector volume.
To do this, the inner surface of the sphere was instrumented with 1280 photo-multiplier tubes (PMTs), corresponding to a photocathode coverage of 11.3\%.
An additional 240 PMTs were used to instrument the surrounding veto region, which rejected cosmic muons and neutrino interactions outside the detector volume with an efficiency of $\sim 99.99\%$~\cite{MiniBooNE:2008paa}.
Mineral oil was chosen as the detector medium due to its high index of refraction (n=1.47), leading to more Cherenkov light production by electrons traversing the detector volume.
The exact mineral oil mixture, Marcol 7, was selected by optimizing the behavior of photons with wavelengths between 320\;nm and 600\;nm (e.g., requiring an extinction length greater than 20\;m)~\cite{MiniBooNE:2008paa}.
The detector was situated in a cylindrical vault just below ground level, under $\sim 3\;{\rm m}$ of dirt.
A schematic depiction of the MiniBooNE detector is shown in \cref{fig:miniboone_detector}.

The reconstruction of the final state from a neutrino interaction in MiniBooNE relied on the detection of Cherenkov light.
Specifically, the collaboration developed reconstruction algorithms that turned the spatiotemporal distribution of photon hits on the PMTs into kinematic information on each observable final state particle~\cite{Patterson:2009ki}.
These algorithms used maximum likelihood estimators to estimate the starting location, direction, and energy of final state particles using the observed photon hits in each PMT, relying on the known transport properties of Cherenkov photons within the detector medium.
Cherenkov photons are emitted when a charged particle travels faster than the speed of light in a medium, at an angle of $\cos \theta_C = 1/n\beta$ with respect to the charged particle track.
This results in a characteristic ring-like pattern on the detector wall.
Such Cherenkov rings formed the basis of the MiniBooNE reconstruction algorithm.

\begin{figure}[h!]
     \centering
     \begin{subfigure}[b]{0.3\textwidth}
         \centering
         \includegraphics[width=\textwidth]{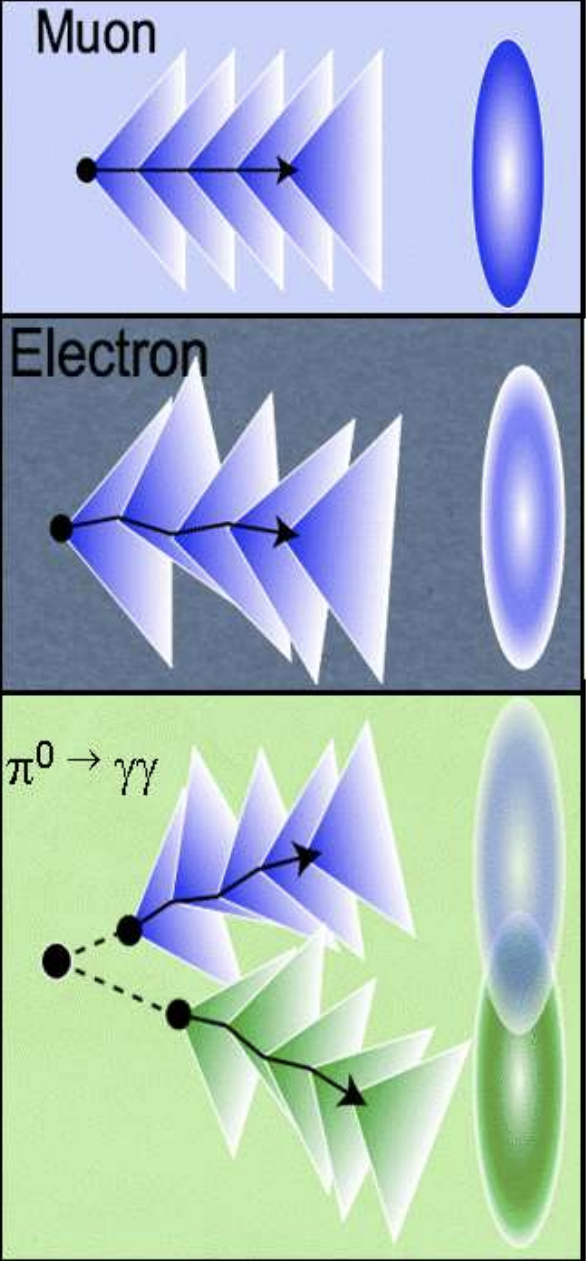}
         \caption{From Ref.~\cite{MiniBooNE:2007uho}}
         \label{fig:miniboone_PID_diagram}
     \end{subfigure}
     \hfill
     \begin{subfigure}[b]{0.45\textwidth}
         \centering
         \includegraphics[width=\textwidth]{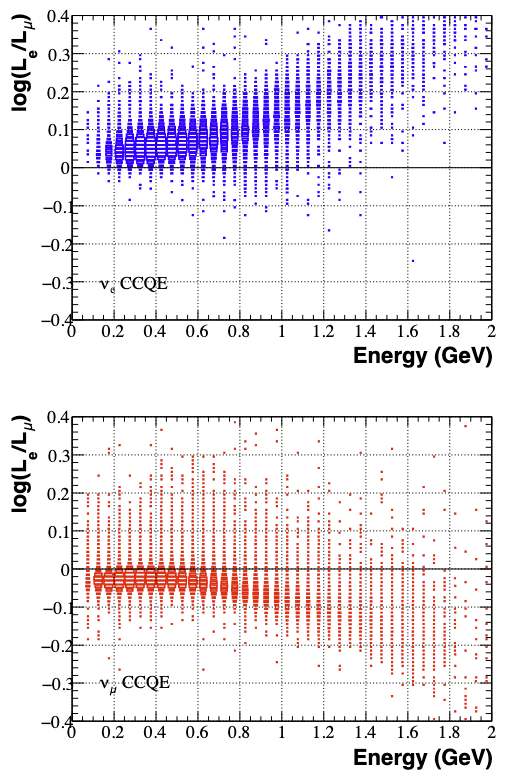}
         \caption{From Ref.~\cite{Patterson:2009ki}}
         \label{fig:miniboone_likelihood_ratio}
     \end{subfigure}
        \caption{Visual representations of particle identification in MiniBooNE. \Cref{fig:miniboone_PID_diagram} shows a schematic representation of the detector signature from the three main particle classes in MiniBooNE: muons, electrons, and neutral pions. \Cref{fig:miniboone_likelihood_ratio} shows the MiniBooNE log-likelihood-ratio between the $e$-like and $\mu$-like hypothesis as a function of reconstructed neutrino energy, considering both simulated $\nu_e$~CCQE (top) and $\nu_\mu$~CCQE (bottom) interactions.}
        \label{fig:miniboone_PID}
\end{figure}

There were two main classes for observable final state particles in MiniBooNE: muon-like ($\mu$-like) and electron-like ($e$-like).
Each elicits a different Cherenkov ring pattern~\cite{Patterson:2009ki}.
At MiniBooNE energies, muons are typically minimum-ionizing particles and thus would appear as a uniform ring in the PMT array.
The ring would be filled in if the muon exits the detector volume before going below the Cherenkov threshold, and would be open otherwise.
Electrons, on the other hand, undergo radiative processes as they travel, emitting photons via the Bremsstrahlung process, which then undergo pair-production to $e^+ e^-$, which then emit more photons, and so on.
This process is typically called an ``electromagnetic (EM) shower''.
The multiple constituent electrons and positrons in this EM shower would result in a distorted Cherenkov ring in the PMT array.
Importantly, high energy photons also produced these distorted rings after undergoing an initial pair-production interaction; thus, electrons and photons were essentially indistinguishable in MiniBooNE and were both classified as $e$-like.
Another relevant final-state particle in MiniBooNE was the neutral pion, which could be identified via two separate distorted Cherenkov rings via the $\pi^0 \to \gamma \gamma$ decay.
It is also important to note that $\pi^0$ events could be misclassified as $e$-like if one of the photons was not reconstructed, which could happen if one of the photons escaped the detector without pair producing or had energy below the detection threshold.
A schematic diagram of the detector signature of muons, electrons, and neutral pions in MiniBooNE is shown in \cref{fig:miniboone_PID_diagram}. 

A separate likelihood was calculated for three different final state particle hypotheses: electron, muon, and neutron pion~\cite{Patterson:2009ki}.
Ratios of these likelihoods were used to distinguish one particle from another in MiniBooNE.
As an example, we show the separation of electron and muon events as characterized by the log-likelihood-ratio as a function of reconstructed neutrino energy in \cref{fig:miniboone_likelihood_ratio}.
This ratio was the main selection tool in selecting $e$-like events for MiniBooNE's flagship search for $\nu_\mu \to \nu_e$ oscillations in the BNB.

\section{The MiniBooNE Low Energy Electron-Like Excess} \label{sec:miniboone_explanations}

As stated above, MiniBooNE was designed to test the LSND excess of $\nubar{e}$ events discussed in \cref{sec:anomalies}.
To do this, MiniBooNE isolated a sample of $e$-like events using the likelihood ratios described in the previous section~\cite{MiniBooNE:2007uho}.
This sample was optimized to select $\nu_e$ CCQE interactions within the detector while rejecting $\nu_\mu$ interaction backgrounds, thus maximizing sensitivity to potential $\nu_\mu \to \nu_e$ oscillations within the BNB.
MiniBooNE's flagship $e$-like analysis, which has remained stable over the lifetime of the experiment, achieved a $\nu_e$ CCQE efficiency of $\sim 20\%$ while rejecting $\sim 99.9\%$ of $\nu_\mu$ backgrounds~\cite{MiniBooNE:2020pnu}.
The full MiniBooNE dataset consists of $18.75 \times 10^{20}$ ($11.27 \times 10^{20}$) protons-on-target (POT) in neutrino (antineutrino) mode collected over 17 years of operation.
In this dataset, the $e$-like analysis observes 2870 (478) data events in neutrino (antineutrino) mode, compared to an SM prediction of 2309.4 (400.6) events~\cite{MiniBooNE:2020pnu}.
Combining neutrino and antineutrino mode data, MiniBooNE observes an excess of $638.0  \pm 52.1\;({\rm stat.})\pm 122.2\;({\rm sys.})$ $e$-like events, corresponding to a Gaussian significance of $4.8\sigma$~\cite{MiniBooNE:2020pnu}.

\Cref{fig:miniboone_enuqe} shows the reconstructed neutrino energy distribution of the MiniBooNE $e$-like excess in both neutrino and antineutrino mode.
The stacked histogram corresponds to the SM prediction from the NUANCE event generator~\cite{Casper:1990ac}, while the data points correspond to the observed number of $e$-like events in each bin.
The error bars on the stacked histogram correspond to the systematic uncertainty on the SM prediction, calculated within a covariance matrix formalism.
The dominant sources of systematic uncertainty include neutrino cross section modeling (derived largely using MiniBooNE's own cross section measurements~\cite{MiniBooNE:2010bsu,MiniBooNE:2007iti,MiniBooNE:2009dxl,MiniBooNE:2008mmr}), nuclear effects, detector response and optical modeling, and BNB flux estimation~\cite{MiniBooNE:2018esg,MiniBooNE:2020pnu}.
The presented error in each bin of the $\nu_e$ and $\nubar{e}$ sample incorporates a constraint from MiniBooNE's dedicated $\nu_\mu$ and $\nubar{\mu}$ CCQE samples.
The dashed line corresponds to the best fit of the MiniBooNE excess to the $3+1$ sterile neutrino model described in \cref{sec:anomalies}.
As one can see, the excess in data events is strongest in the lowest energy bins; for this reason, this anomaly is often referred to as the MiniBooNE low-energy excess (LEE). 

As MiniBooNE used a Cherenkov detector, it was not sensitive to the final state hadronic activity in a neutrino interaction.
Thus, kinematic reconstruction of the original neutrino relied entirely on the final state lepton.
Under the assumption that the neutrino underwent a CCQE interaction off of a nucleon at rest within the nucleus, the original neutrino energy is given by~\cite{MiniBooNE:2010bsu}
\begin{equation}
E_\nu^{\rm QE} = \frac{2(M_n') E_\ell - ((M_n')^2 + m_\ell^2 - M_p)}{2[(M_n') - E_\ell + \sqrt{E_\ell^2 - m_\ell^2} \cos \theta_\ell]},
\end{equation}
where $E_\ell$ is the total lepton energy, $\cos \theta_\ell$ is the lepton scattering angle, and $M_n$, $M_p$, and $m_\ell$ are the neutron, proton, and lepton mass, respectively.
The adjusted neutron energy is defined as $M_n' \equiv M_n - E_B$, where $E_B$ is the nuclear binding energy of the initial state neutron.
An analogous relation exists for antineutrino energy reconstruction in a CCQE interaction~\cite{MiniBooNE:2013qnd}.
This is the reconstructed energy definition used to generate the histograms in \cref{fig:miniboone_enuqe}.

\Cref{fig:miniboone_evis_costheta} shows the visible energy and $\cos \theta$ distributions of the final state lepton in MiniBooNE's $e$-like neutrino mode sample.
The visible energy distribution shows the strongest discrepancy for softer lepton kinetic energies, as expected for a low-energy excess.
For the angular distribution, it is worth noting that while there is an excess across the entire range, the largest deviation above the SM prediction comes from the $\cos \theta \in [0.9,1.0]$ bin.
The angular distribution of the MiniBooNE LEE is an important piece of information for potential solutions to the anomaly--as we will discuss throughout this thesis, BSM physics models often cannot explain the energy and angular distributions of the MiniBooNE LEE simultaneously.

The green contributions to the stacked histograms of \cref{fig:miniboone_enuqe} represent the interactions of intrinsic $\nu_e$ or $\nubar{e}$ in the BNB.
At low energies, these events come mostly from the decay of the secondary muon in the  $\pi^+ \to \mu^+$ or $\pi^- \to \mu^-$ decay chain, while $\nu_e$ and $\nubar{e}$ from kaon decays start to contribute more at higher energies~\cite{MiniBooNE:2008hfu}.

The red and brown contributions to the stacked histograms represent misidentified photon backgrounds that are reconstructed as a single distorted Cherenkov ring.
The largest photon background comes from misidentified $\pi^0$ created via $\nu_\mu$ and $\nubar{\mu}$ neutral-current (NC) resonant scattering, in which the initial state nucleon is excited to a $\Delta$ resonance before decaying to nucleon-pion pair.
The $\pi^0$ decay promptly to a pair of photons, which should nominally appear as a pair of distorted Cherenkov rings as in \cref{fig:miniboone_PID_diagram}.
However, if one of the photons exits the detector before converting to an $e^+ e^-$ pair, or if the original pion energy is distributed asymmetrically such that the visible energy of one of the photons sits below the reconstruction threshold of 140\;MeV~\cite{MiniBooNE:2007uho}, the $\pi^0$ decay will be misidentified as an $e$-like event.
An enhancement of the NC $\pi^0$ background in \cref{fig:miniboone_enuqe} could in principle explain the observed excess.
However, MiniBooNE constrained the rate of this NC $\pi^0$ background \textit{in situ} via a measurement of the two-gamma invariant mass peak in well-reconstructed NC $\pi^0$ events~\cite{MiniBooNE:2020pnu}.
Additionally, the radial distribution of the excess peaks toward the center of the detector, while misidentified NC $\pi^0$ backgrounds happen more often toward the edge of the detector, where it is more likely for a photon to escape before pair-producing.

The next-largest photon background comes from rare $\Delta \to N \gamma$ decays in $\nu_\mu$ and $\nubar{\mu}$ NC resonant scattering interactions.
As this process has never been observed directly, it was not possible for MiniBooNE to constrain the $\Delta \to N \gamma$ event rate \textit{in situ}.
It was instead constrained indirectly by the NC $\pi^0$ two-gamma invariant mass distribution~\cite{MiniBooNE:2020pnu}.
A factor 3.18 enhancement in $\Delta \to N \gamma$ events could explain the MiniBooNE LEE; however, this hypothesis has since been disfavored by recent results from the MicroBooNE experiment~\cite{MicroBooNE:2021zai}.
The MicroBooNE experiment will be covered in more detail in \cref{ch:microboone_detector,ch:microboone_selection,ch:microboone_results}.

MiniBooNE has also studied neutrino interactions outside the detector volume which result in a single photon entering the detector (the ``dirt'' backgrounds in \cref{fig:miniboone_enuqe}).
The timing distribution of the MiniBooNE $e$-like dataset suggests that the excess comes primarily in time with the beam, while dirt background events are often delayed by $\sim 10\;{\rm ns}$~\cite{MiniBooNE:2020pnu}.
This result disfavors an enhancement of external neutrino interactions as an explanation of the MiniBooNE excess. 

Therefore, the $4.8\sigma$ MiniBooNE excess remains unexplained.
Resolution of the MiniBooNE LEE is one of the major goals of the neutrino community~\cite{Karagiorgi:2022fgf}.

\begin{figure}[h!]
     \centering
     \begin{subfigure}[b]{0.45\textwidth}
         \centering
         \includegraphics[width=\textwidth]{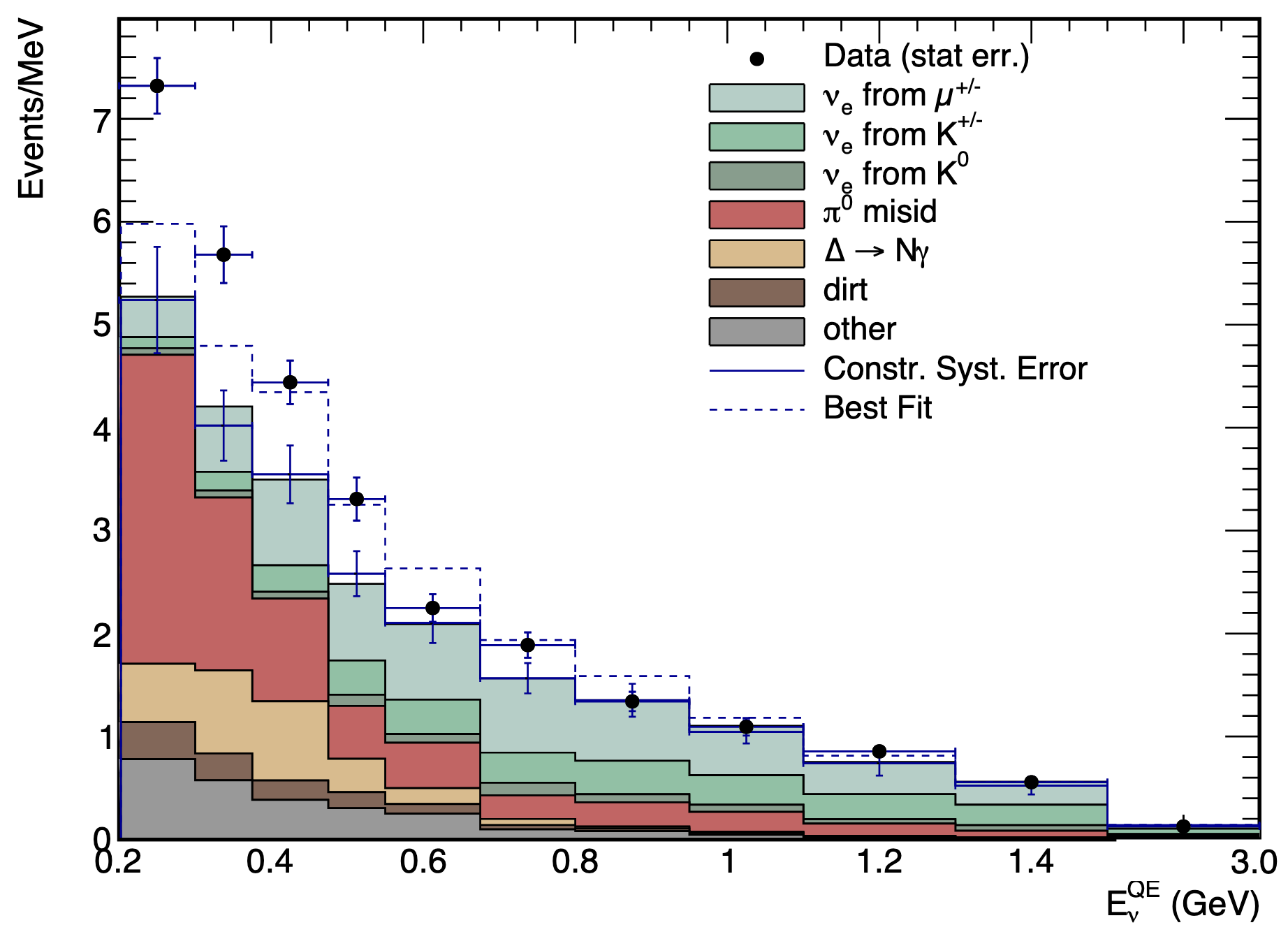}
         \caption{$\nu_e$ sample, from Ref.~\cite{MiniBooNE:2020pnu}}
         \label{fig:miniboone_nu_enuqe}
     \end{subfigure}
     \hfill
     \begin{subfigure}[b]{0.45\textwidth}
         \centering
         \includegraphics[width=\textwidth]{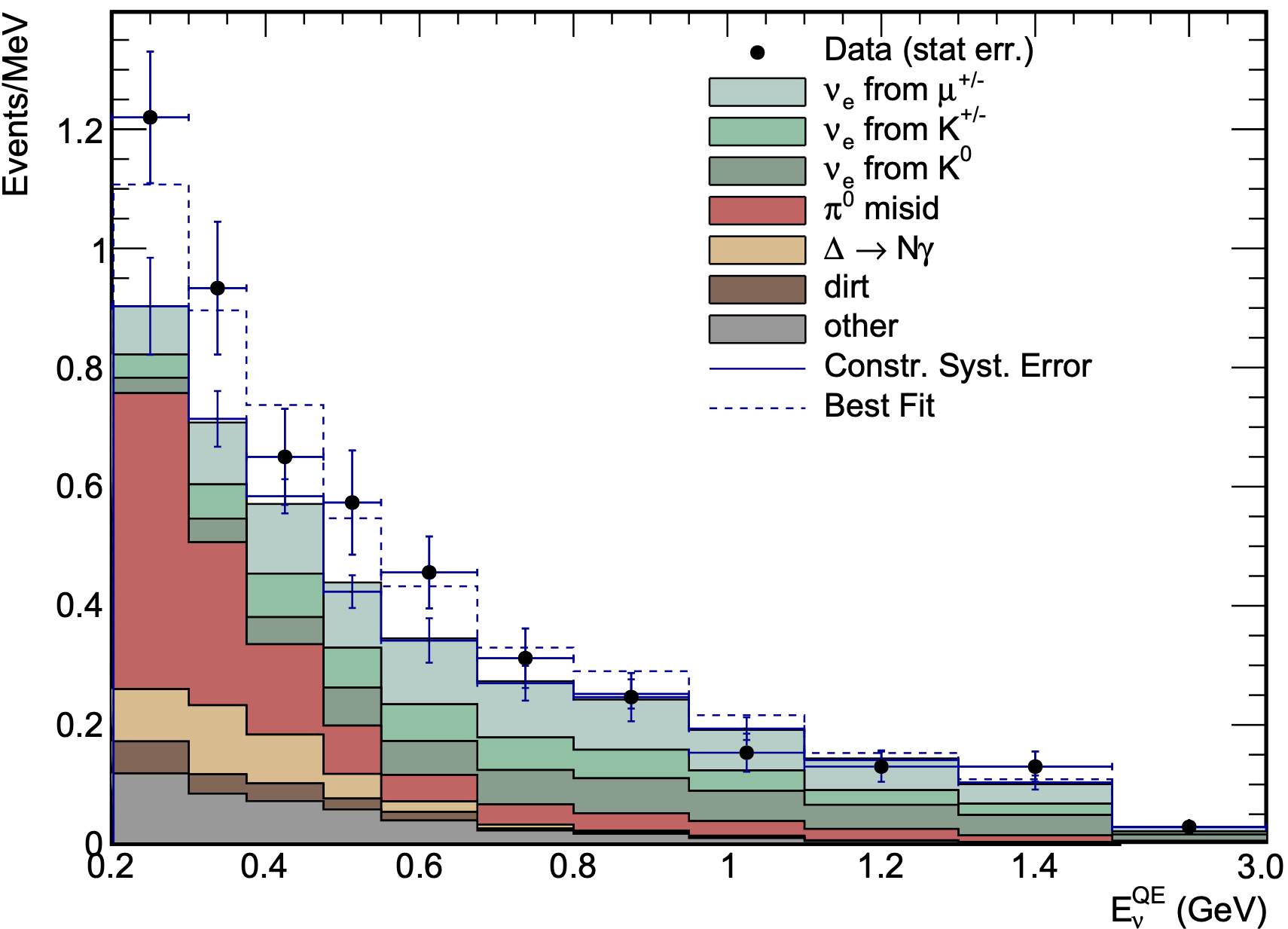}
         \caption{$\nubar{e}$ sample, from Ref.~\cite{MiniBooNE:2018esg}}
         \label{fig:miniboone_antinu_enuqe}
     \end{subfigure}
        \caption{The $E_\nu^{\rm QE}$ distribution of the MiniBooNE $e$-like excess in the total neutrino mode (\cref{fig:miniboone_nu_enuqe}) and antineutrino mode (\cref{fig:miniboone_antinu_enuqe}) datasets. The observation and SM prediction in each bin are shown by the data points and colored histograms, respectively.}
        \label{fig:miniboone_enuqe}
\end{figure}

\begin{figure}[h!]
     \centering
     \begin{subfigure}[b]{0.45\textwidth}
         \centering
         \includegraphics[width=\textwidth]{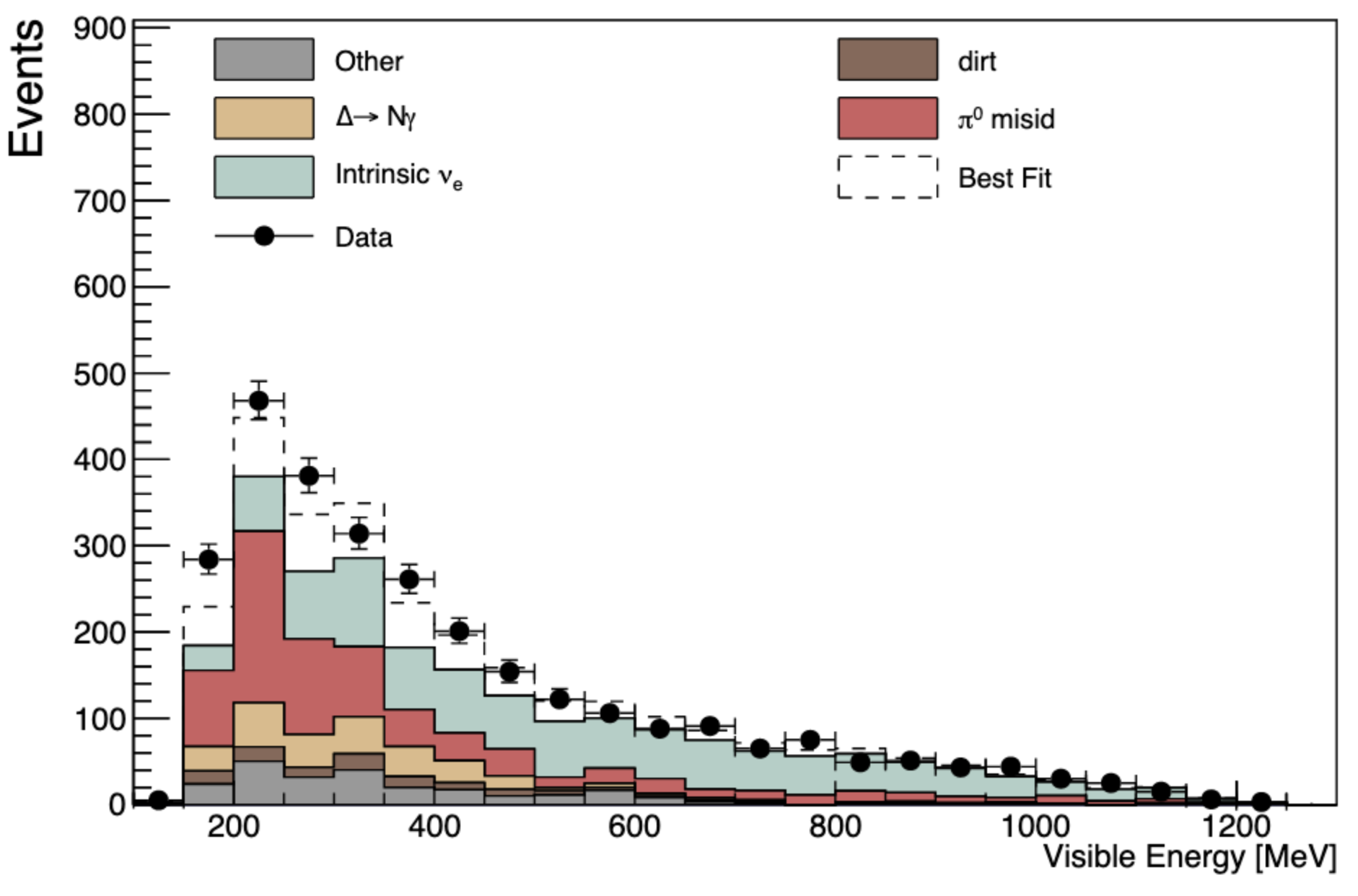}
         \caption{Lepton $E_{\rm vis}$ distribution}
         \label{fig:miniboone_evis}
     \end{subfigure}
     \hfill
     \begin{subfigure}[b]{0.45\textwidth}
         \centering
         \includegraphics[width=\textwidth]{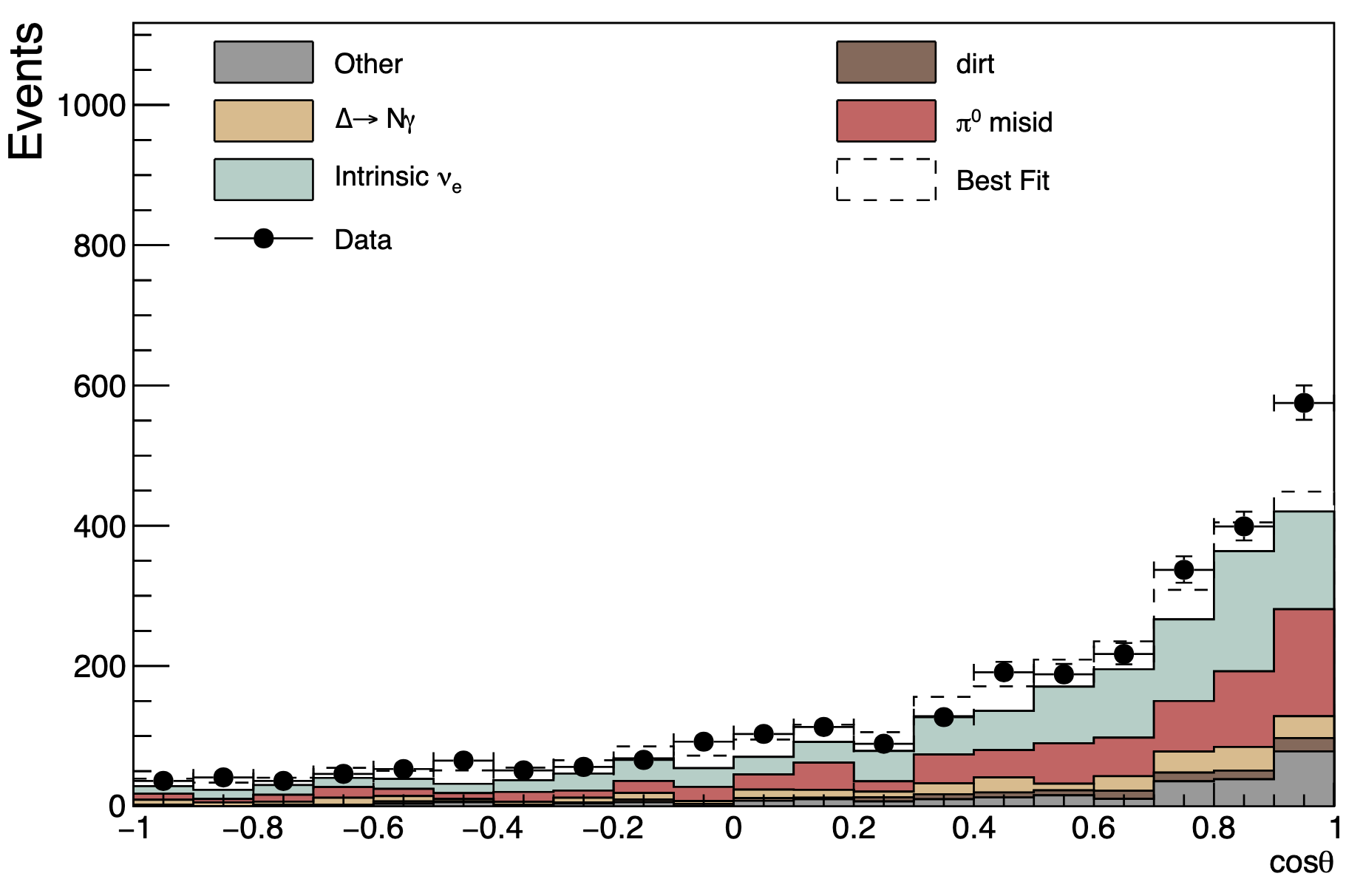}
         \caption{Lepton $\cos \theta$ distribution}
         \label{fig:miniboone_costheta}
     \end{subfigure}
        \caption{The lepton visible energy (\cref{fig:miniboone_evis}) and $\cos \theta$ (\cref{fig:miniboone_costheta}) distributions of the MiniBooNE $e$-like excess in the total neutrino mode dataset. The observation and SM prediction in each bin are shown by the data points and colored histograms, respectively. Figures from Ref.~\cite{MiniBooNE:2020pnu}.}
        \label{fig:miniboone_evis_costheta}
\end{figure}

\begin{figure}
    \centering
    \includegraphics[width=0.6\textwidth]{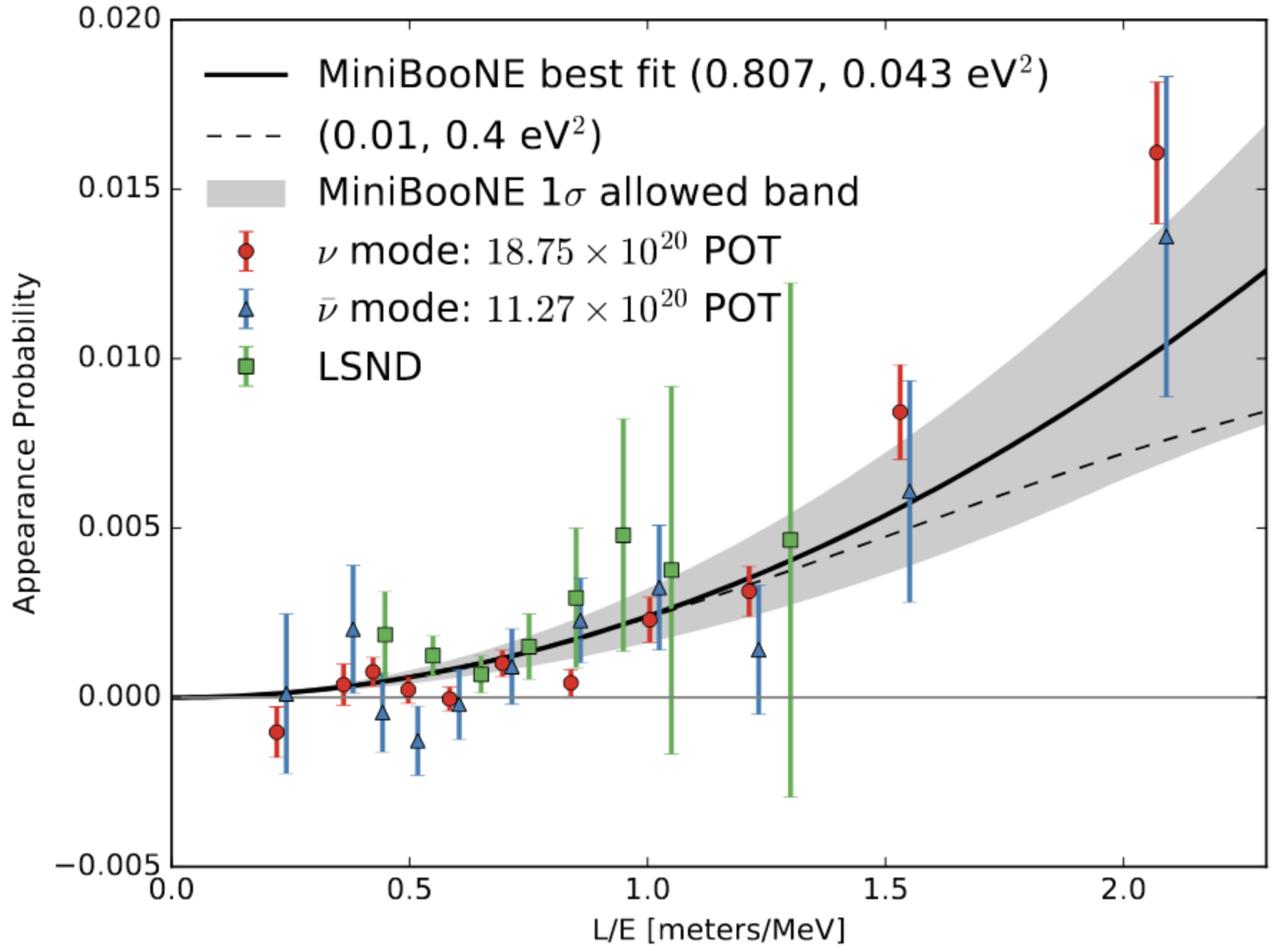}
    \caption{The MiniBooNE and LSND excesses as a function of the ratio $L/E$. The MiniBooNE data is separated into neutrino and antineutrino mode. Figure from Ref.~\cite{MiniBooNE:2020pnu}.}
    \label{fig:miniboone_LSND_LoverE}
\end{figure}

The MiniBooNE LEE is most commonly interpreted within the context of the 3+1 model introduced in \cref{sec:anomalies}.
This is primarily because the MiniBooNE excess has historically been considered alongside the LSND excess, as both results can be explained by short-baseline $\nu_\mu \to \nu_e$ and $\nubar{\mu} \to \nubar{e}$ appearance.
Strikingly, the MiniBooNE and LSND anomalies both prefer similar regions in sterile neutrino parameter space, as shown in \cref{fig:MB_LSND_sterile}.
This is further supported by \cref{fig:miniboone_LSND_LoverE}, which shows the rising nature of the MiniBooNE and LSND excesses as a function of the ratio $L/E$, behavior that is consistent with a sterile neutrino explanation. 

There are, however, complications regarding a sterile neutrino explanation of the MiniBooNE excess.
The 3+1 model has difficulty reproducing the lowest energy and lowest scattering angle region of the excess.
\Cref{fig:miniboone_evis_costheta} shows that the best fit 3+1 prediction, indicated by the dotted line, still falls below the observed data in the lowest lepton $E_{\rm vis}$ bin and highest lepton $\cos \theta$ bin.
Additionally, as discussed in \cref{sec:anomalies}, there is significant tension between the MiniBooNE and LSND observation of $\nu_e$/$\nubar{e}$ appearance and experiments searching for $\nu_e$/$\nubar{e}$ and $\nu_\mu$/$\nubar{\mu}$ disappearance.
Finally, the follow-up MicroBooNE experiment has not observed an excess of $\nu_e$ events consistent with the expectation from the MiniBooNE LEE~\cite{MicroBooNE:2021tya}.
The MicroBooNE $\nu_e$ analysis is one of the main results of this thesis and will be explored in more detail in \cref{ch:microboone_selection,ch:microboone_results}.
While the non-observation of a MiniBooNE-like excess of $\nu_e$ events in the BNB does set constraints on $3+1$ parameter space, it does not fully exclude the MiniBooNE allowed regions~\cite{MiniBooNE:2022emn}.
This point will be discussed further in \cref{ch:microboone_results}.

These complications with the eV-scale sterile neutrino interpretation of the MiniBooNE LEE have prompted the community to explore alternative explanations.
Many of these are relatively simple extensions beyond $3+1$, such as $3+N$ models involving $N$ additional sterile neutrino states~\cite{Diaz:2019fwt}, decaying sterile neutrino models~\cite{Bai:2015ztj,deGouvea:2019qre,Dentler:2019dhz}, and sterile neutrinos with altered dispersion relations from large extra dimensions~\cite{Pas:2005rb,Carena:2017qhd,Doring:2018ncz}.
Other explanations for the MiniBooNE LEE introduce a number of new particle species prescribed with new interactions that create an additional source of photons or $e^+e^-$ pairs in MiniBooNE.
Such models include heavy neutral leptons (HNLs) which decay to photons via a transition magnetic moment~\cite{Gninenko:2009ks,Gninenko:2010pr,Dib:2011jh,Gninenko:2012rw,Masip:2012ke,Radionov:2013mca,Ballett:2016opr,Magill:2018jla,Balantekin:2018ukw,Balaji:2019fxd,Balaji:2020oig,Fischer:2019fbw,Vergani:2021tgc,Alvarez-Ruso:2021dna,Kamp:2022bpt} and models with heavy neutrinos coupled to a ``dark sector'' involving, for example, new vector or scalar mediators~\cite{Bertuzzo:2018ftf,Ballett:2018ynz,Bertuzzo:2018itn,Ballett:2019pyw,Abdullahi:2020nyr,Datta:2020auq,Dutta:2020scq,Abdallah:2020biq,Abdallah:2020vgg,Dutta:2021cip,Arguelles:2018mtc}.
\Cref{ch:neutrissimos} of this thesis explores one such explanation of MiniBooNE involving an HNL with a transition magnetic moment coupling to active neutrinos, which we hereby refer to as a ``neutrissimo''.
Neutrissimo decays in MiniBooNE provide an additional source of single photons which could explain the $e$-like excess~\cite{Vergani:2021tgc,Kamp:2022bpt}.

Thus, there are many potential explanations for the MiniBooNE anomaly.
Distinguishing between these explanations requires careful consideration of the kinematic distributions of the MiniBooNE excess~\cite{Jordan:2018qiy,Kamp:2022bpt}.
Further, these models are often subject to constraints from existing accelerator neutrino experiments, such as MINERvA and NA62~\cite{Arguelles:2018mtc,Ballett:2019pyw}.
A complete evaluation of constraints from existing data is essential in determining the most viable models among the many proposed MiniBooNE explanations.
The work presented in \cref{ch:neutrissimos} takes a step in this direction by calculating constraints from MINERvA on the neutrissimo model.

\chapter{The MicroBooNE Detector} \label{ch:microboone_detector}

In order to ascertain the nature of the MiniBooNE excess, one needs a detector capable of providing more detailed event-by-event information than MiniBooNE's Cherenkov detector.
This is the concept behind the MicroBooNE experiment.
The MicroBooNE detector is a large-scale liquid argon time projection chamber (LArTPC) with the ability to record high-resolution images of neutrino interactions.
MicroBooNE recently released its first results investigating the nature of the MiniBooNE excess~\cite{MicroBooNE:2021tya,MicroBooNE:2021zai}, which will be presented in \cref{ch:microboone_selection,ch:microboone_results}.
This chapter introduces the detector that made this measurement possible.

\section{Liquid Argon Time Projection Chamber}

MicroBooNE used an 85-metric-ton fiducial volume LArTPC detector to observe the interactions of neutrinos in the BNB~\cite{MicroBooNE:2016pwy,MicroBooNE:2021pvo}.
This makes MicroBooNE the first $\mathcal{O}(100~{\rm t})$ LArTPC operated in the United States.
The idea for a LAr-based total absorption detector originated in the 1970s~\cite{Willis:1974gi}.
The introduction of the LArTPC detector concept came from Carlo Rubbia in 1977~\cite{Rubbia:1977zz}, extending earlier work from David Nygren~\cite{Nygren:1974nfi} and Georges Charpak~\cite{Charpak:1970az}.
The first operational large-scale LArTPC was the 500-metric-ton active volume ICARUS T600 detector~\cite{ICARUS:2004wqc}, which came online in 2010.
ICARUS observed cosmic ray and neutrino interactions at the Gran Sasso underground National Laboratory~\cite{Rubbia:2011ft} and even set constraints on $\nu_\mu \to \nu_e$ interpretations of the LSND and MiniBooNE anomalies using the CERN to Gran Sasso neutrino beam~\cite{ICARUS:2013cwr}.
On a smaller scale, the ArgoNeuT experiment operated a 0.25-metric-ton LArTPC at Fermilab's Neutrino Main Injector beamline from 2009-2010, where it performed the first measurements of neutrino-argon cross sections~\cite{ArgoNeuT:2014rlj}.

The MicroBooNE detector is situated $70$~m downstream of the MiniBooNE detector along the BNB and operated from 2015 to 2021, observing a total of approximately $1.5 \times 10^{21}$~POT~\cite{MicroBooNE:2021pvo}.
MicroBooNE LArTPC data come in the form of high-resolution three-dimensional images of the ionization energy deposited by final state charged particles in neutrino interactions.
The information contained in these images allows for the event-by-event separation of photons and electrons--an essential capability for determining the source of the MiniBooNE excess.
MicroBooNE can also reconstruct hadronic activity in the final state of the neutrino interaction, which helps further distinguish between the possible sources of the MiniBooNE excess.

\begin{figure}[h!]
     \centering
     \begin{subfigure}[b]{0.6\textwidth}
         \centering
         \includegraphics[width=\textwidth]{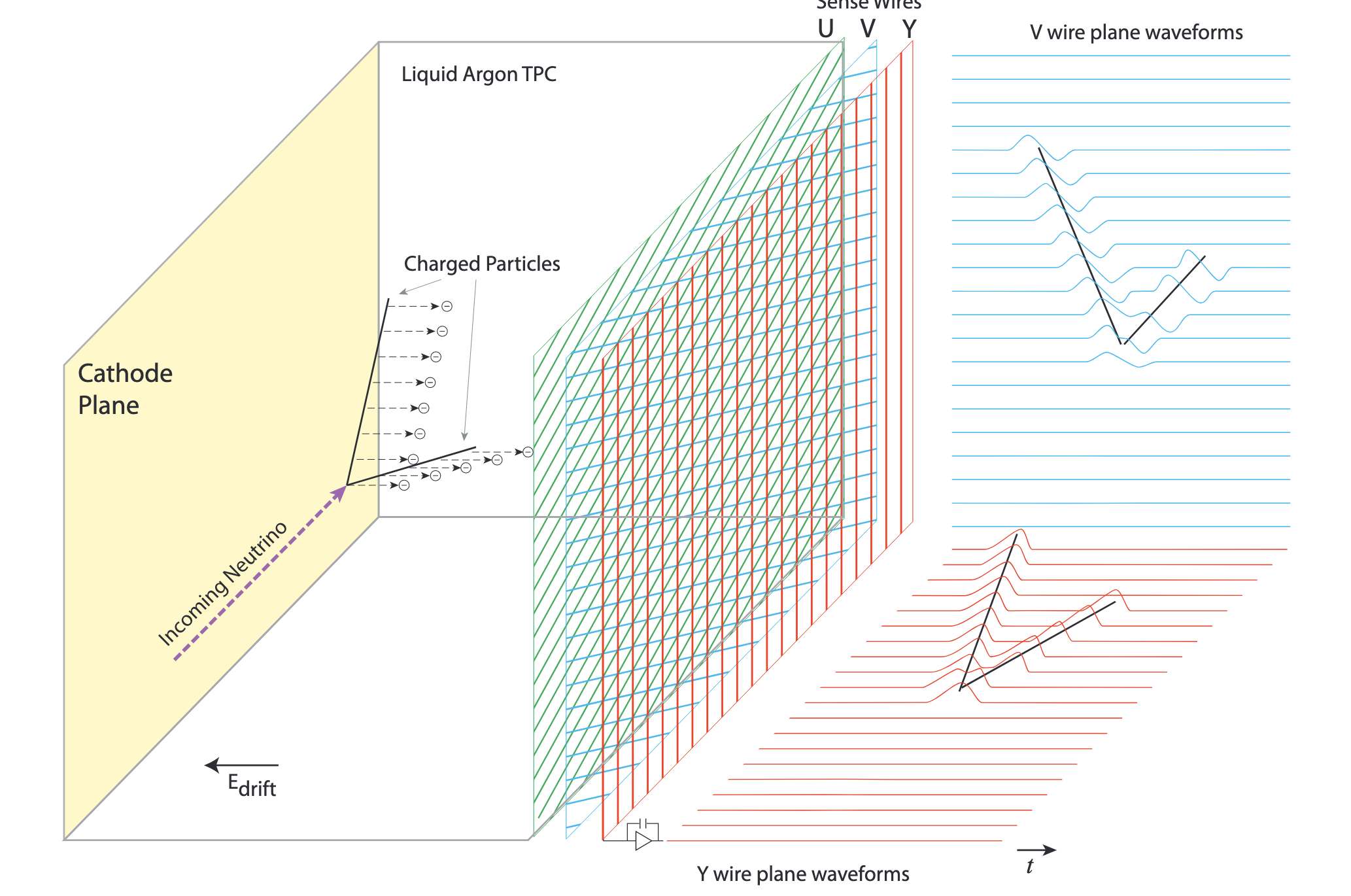}
         \caption{}
         \label{fig:lartpc_detection}
     \end{subfigure}
     \hfill
     \begin{subfigure}[b]{0.35\textwidth}
         \centering
         \includegraphics[width=\textwidth]{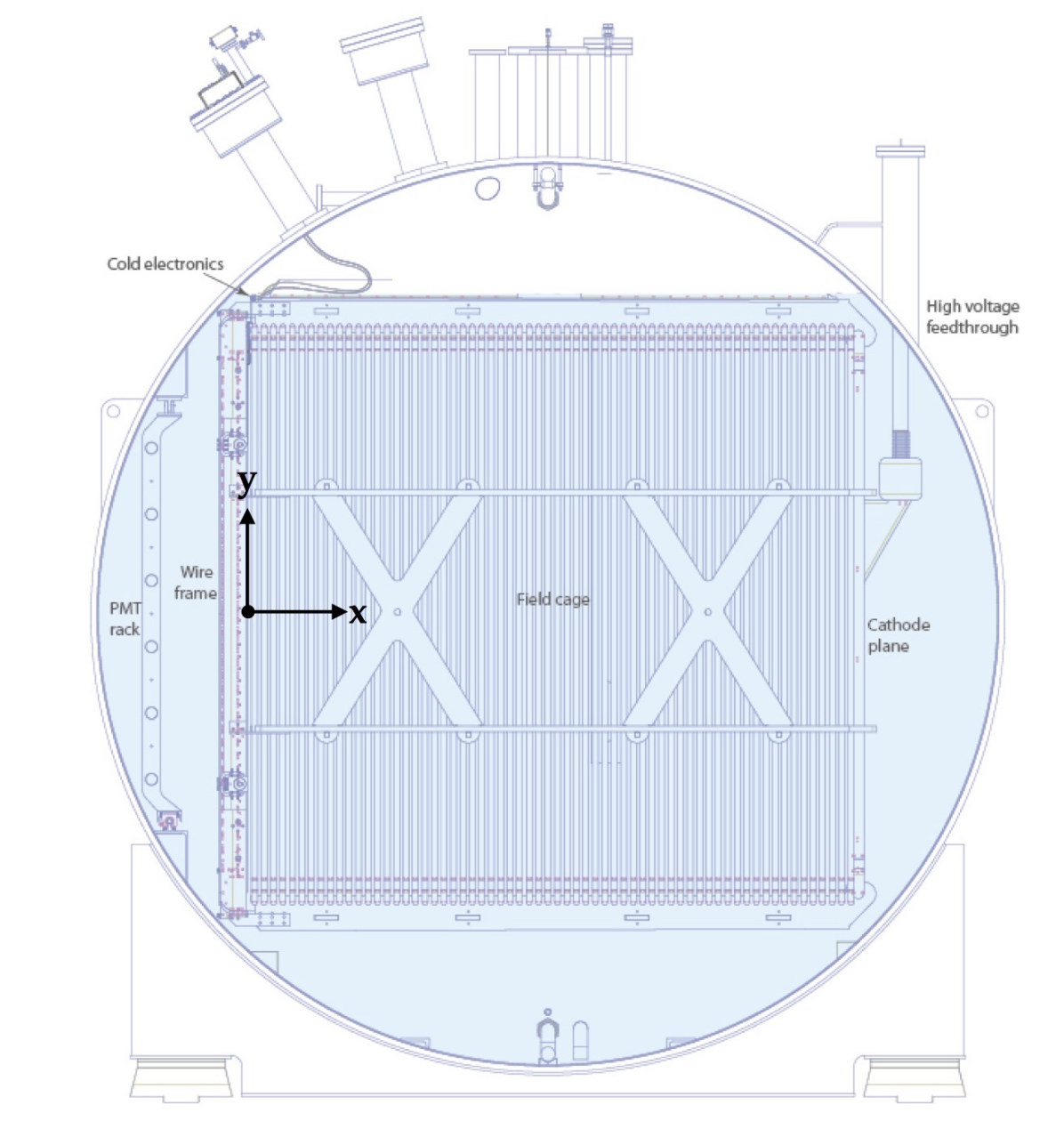}
         \caption{}
         \label{fig:lartpc_beamview}
     \end{subfigure}
        \caption{Schematic depictions of the MicroBooNE LArTPC. \Cref{fig:lartpc_detection} shows the detection process for charged particles from a neutrino interaction in a MicroBooNE-like LArTPC. \Cref{fig:lartpc_beamview} shows a cross-sectional view of the MicroBooNE detector along the $-\hat{z}$ direction. Figures from Ref.~\cite{MicroBooNE:2016pwy}.}
        \label{fig:lartpc}
\end{figure}

We begin with a brief overview of the MicroBooNE reconstruction procedure.
Charged-current neutrino interactions in the LAr volume produce charged particles in the final state, which ionize argon atoms as they traverse the detector.
Thus, each charged particle leaves behind a trail of ionized electrons which, in theory, can drift freely through the noble element detector medium without being captured.
This drift is controlled via an external electric field with strength $|E| \sim 273$~V/cm, accelerating the ionized electrons to a final velocity $v \sim 0.11$~cm/$\mu$s toward three anode wire planes~\cite{MicroBooNE:2021pvo}.
MicroBooNE employs a right-handed coordinate system, in which BNB neutrinos travel along the $\hat{z}$ direction, ionization electrons drift along the $-\hat{x}$ direction, and $\hat{y}$ represents the vertical direction~\cite{MicroBooNE:2016pwy}.
The anode planes consist of two induction planes and one collection plane, each containing a series of wires spaced 3~mm apart and oriented at $\pm 60^{\circ}$ and $0^{\circ}$ with respect to the $\hat{y}$ direction for the induction and collection planes, respectively.
Each plane is biased such that ionization electrons drift past the induction plane wires, generating a signal via induction, and terminate on the collection plane wires, generating a signal via direct charge collection.
The signals on the anode wire planes allow for two-dimensional reconstruction of the charged particle trajectory in the $\hat{y}-\hat{z}$ plane, transverse to the drift direction.
The $\hat{x}$ dimension of the charged particle trajectory can be reconstructed using the arrival time of signals on the anode wires in conjunction with the known drift time of the ionization electrons.
In order for this technique to work, one must know the initial time at which the charged particle entered the detector.
This can be established using either an external beam trigger or an internal trigger from the light collection system, which operates on much shorter time scales, $\mathcal{O}({\rm ns})$, compared to characteristic electron drift times of $\mathcal{O}({\rm ms})$.
A schematic of this process is shown in \cref{fig:lartpc}.

\subsection{Cryogenics}

The MicroBooNE detector is relatively large--the LArTPC volume spans 2.6~m, 2.3~m, and 10.4~m in the $\hat{x}$, $\hat{y}$, and $\hat{z}$ direction, respectively~\cite{MicroBooNE:2016pwy}.
Thus, ionization electrons must drift through $\mathcal{O}(m)$ of LAr before reaching the anode wire planes.
Reconstruction of these ionization electrons requires careful control of the drift process.
This is the main objective of the MicroBooNE cryogenic system.

The LArTPC is housed within a larger cylindrical cryostat, which itself is supported by an argon purification system and nitrogen refrigeration system~\cite{MicroBooNE:2016pwy}.
The purification system consists of two recirculation pumps and two filter skids that remove electronegative impurities from the LAr, mainly oxygen (O$_2$) and water (H$_2$O).
These impurities must be kept below the 100 parts-per-trillion O$_2$-equivalent level in order to maintain electron drift lengths of at least 2.5~m~\cite{Baibussinov:2009gs,MicroBooNE:2016pwy}.
Additionally, the nitrogen contamination must be kept below 2 parts-per-million in order to maintain an argon scintillation light attenuation length greater than the size of the detector~\cite{Jones:2013bca}.
Nitrogen cannot be appreciably removed from the argon via the purification system; rather, the initial nitrogen contamination is fixed by the quality of the delivered argon, and additional contamination must be controlled by minimizing the atmosphere leakage rate into the cryostat.

The nitrogen refrigeration system is designed to combat the heat load on the LAr from the environment and electrical power systems, maintaining thermal homogeneity throughout the active volume.
It consists of two condensers, each designed to handle a heat load of approximately 9.5~kW~\cite{MicroBooNE:2016pwy}.
The temperature of the LAr volume must be stable to $\pm 0.1$~K in order to keep the $\hat{x}$ direction resolution of charged particle tracks below 0.1\%~\cite{MicroBooNE:2016pwy}.

\subsection{LArTPC Drift System}

The drift system inside the LArTPC volume consists of three major subsystems: the cathode plane, the field cage, and the three anode wire planes.
The purpose of the drift system is to maintain a uniform electric field throughout the active volume such that ionization electrons are transported to the anode plane at a stable drift velocity.

\begin{figure}[h!]
     \centering
     \begin{subfigure}[b]{0.45\textwidth}
         \centering
         \includegraphics[width=\textwidth]{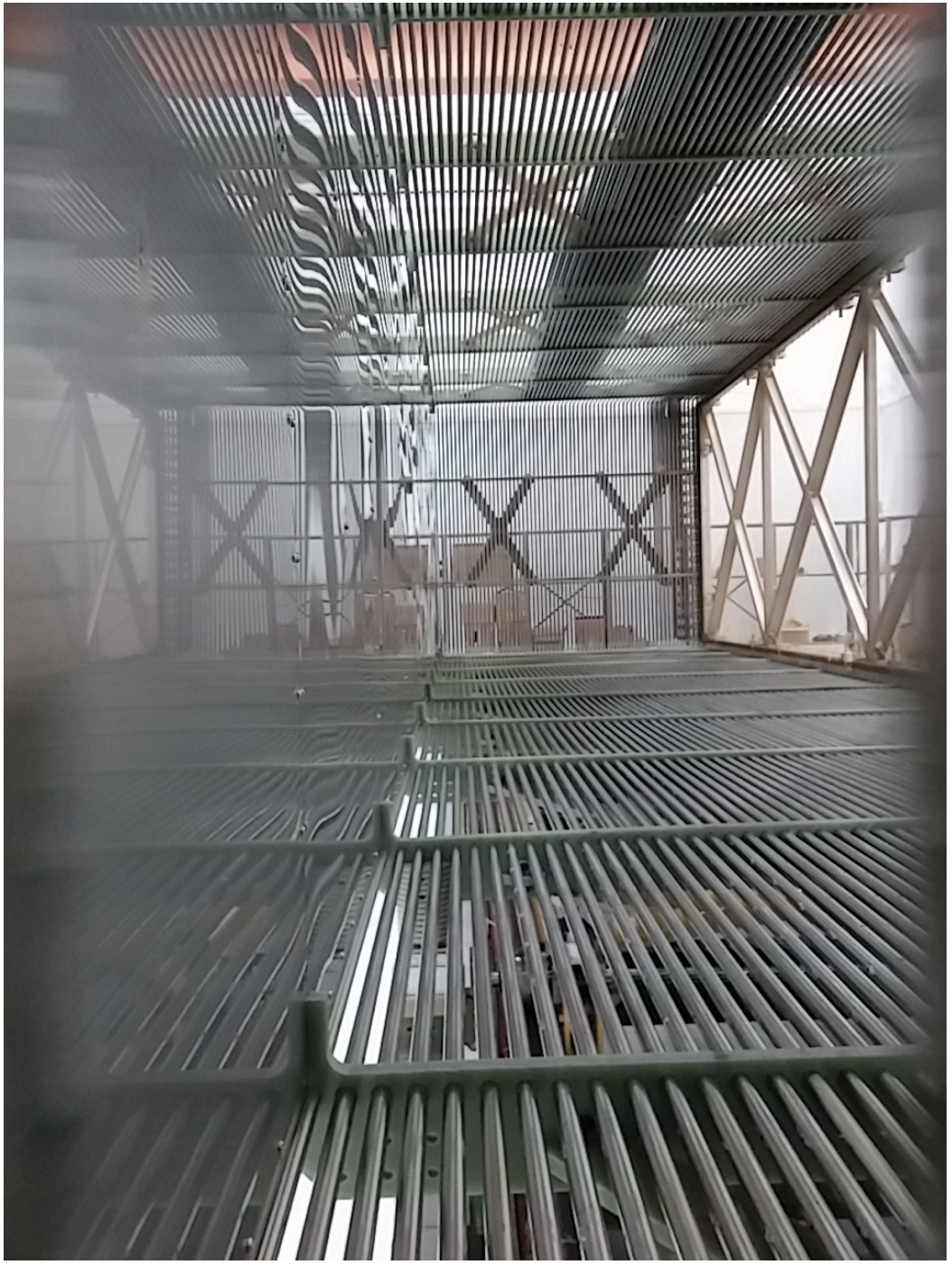}
         \caption{}
         \label{fig:lartpc_interior}
     \end{subfigure}
     \hfill
     \begin{subfigure}[b]{0.45\textwidth}
         \centering
         \includegraphics[width=\textwidth]{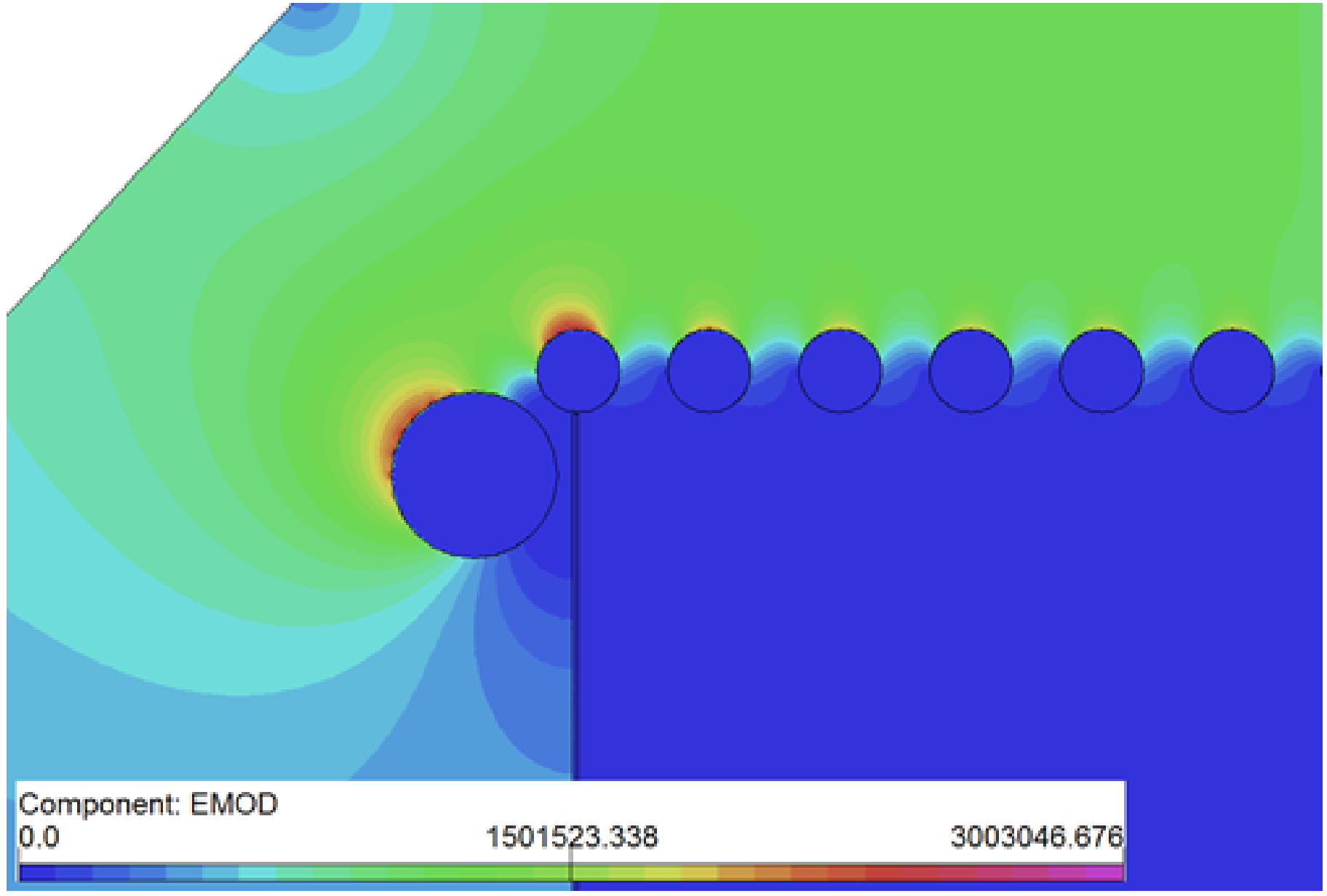}
         \caption{}
         \label{fig:lartpc_Efield}
     \end{subfigure}
        \caption{\Cref{fig:lartpc_interior} shows a close-up image of the cathode plane of the MicroBooNE LArTPC. The stainless steel field cage tubes can also be seen surrounding the active volume. \Cref{fig:lartpc_Efield} shows a cross-sectional map of the electric field at the edge of the active volume, considering a cathode plane voltage of -128~kV. The legend shows the field strength in units of V/m. Figures from Ref.~\cite{MicroBooNE:2016pwy}.}
        \label{fig:lartpc_cathode_fieldcage}
\end{figure}

The cathode consists of nine stainless steel sheets connected to a supporting frame to form a single plane.
Laser tracker measurements indicate that a majority of the cathode plan is flat to within $\pm 3$~mm~\cite{MicroBooNE:2016pwy}.
The cathode plane is kept at a negative potential of approximately $-70$~kV via a high voltage feedthrough on the cryostat.
The field cage maintains a uniform electric field between the cathode plane and anode planes.
It consists of 64 stainless steel tubes wrapped around the LArTPC active volume.
A resistor divider chain connects each tube to its neighbor, sequentially stepping the voltage from $-70$~kV to ground in $1.1$~kV increments.
The chain provides a resistance of 250~M$\Omega$ between adjacent tubes such that the current flow is approximately 4.4 $\mu$A, much larger than the $\mathcal{O}({\rm nA})$ current from signals on anode plane wires~\cite{MicroBooNE:2016pwy}.
\Cref{fig:lartpc_cathode_fieldcage} shows the MicroBooNE cathode and field cage, as well as a simulated map of the electric field within the LArTPC active volume.

\begin{figure}[h!]
     \centering
     \begin{subfigure}[b]{0.45\textwidth}
         \centering
         \includegraphics[width=\textwidth]{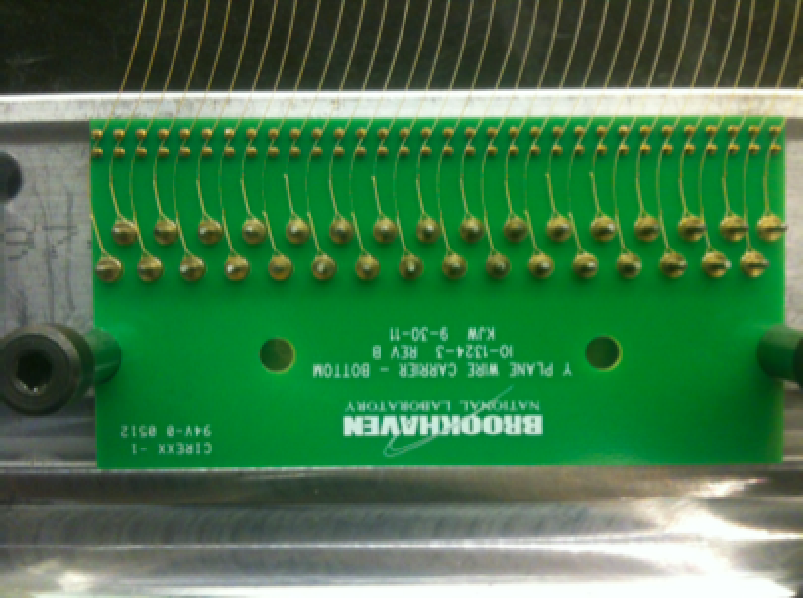}
         \caption{}
         \label{fig:lartpc_wire_board}
     \end{subfigure}
     \hfill
     \begin{subfigure}[b]{0.45\textwidth}
         \centering
         \includegraphics[width=\textwidth]{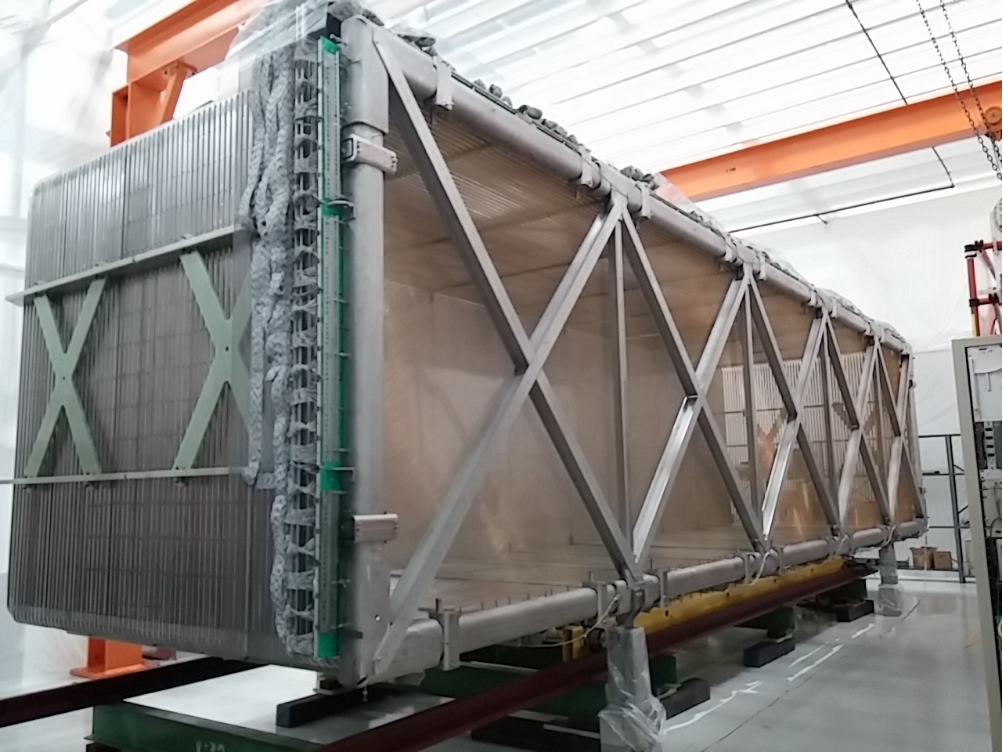}
         \caption{}
         \label{fig:lartpc_complete}
     \end{subfigure}
        \caption{\Cref{fig:lartpc_wire_board} shows a photograph of a single wire carrier board with 32 mounted wires. \Cref{fig:lartpc_complete} shows the fully-assembled MicroBooNE LarTPC, highlighting the anode plane mounted on the stainless steel frame. Figures from Ref.~\cite{MicroBooNE:2016pwy}.}
        \label{fig:lartpc_anode}
\end{figure}

Perhaps the most critical components of the MicroBooNE detector are the three anode wire planes.
The U and V induction planes contain 2400 wires each, while the Y collection plane contains 3456 wires.
As mentioned above, the U and V plane wires are oriented at $\pm 60^{\circ}$ with respect to the vertical, while the Y plane is oriented vertically.
The U, V, and Y planes are biased at -200~V, 0~V, and +440~V, respectively, to ensure termination of ionization electrons on the Y collection plane.
Each wire is 150~$\mu$m in diameter and is spaced 3~mm from its neighbors.
The planes themselves are spaced 3~mm from one another.
The wires are held in place by wire carrier boards, which house 16 wires each in the U and V planes and 32 wires in the Y plane.
Each wire is terminated using a semi-automated wrapping procedure around a 3~mm diameter brass ferrule.
On the wire carrier boards, each wire makes contact with a gold pin that connects to the electronic read-out system.
The anode planes are held in place by a single stainless steel frame, which houses each wire carrier board via an array of precision alignment pins.
Wires are tested to withstand three times the nominal load of 0.7~kg without breakage, both before and after placement onto the wire carrier board.
\Cref{fig:lartpc_wire_board} shows an image of a single Y plane wire carrier board with 32 mounted wires.
An image of the fully-assembled MicroBooNE LArTPC is shown in \cref{fig:lartpc_complete}, specifically highlighting the anode planes mounted on the stainless steel frame.

\subsection{Light Collection System} \label{sec:ub_light_collection}

Liquid argon is a prolific scintillation medium due to its low cost, high scintillation yield ($\mathcal{O}(10^4)$ photons per MeV of deposited energy), and transparency to its own scintillation light~\cite{Jones:2013bca}.
This last feature comes from the scintillation mechanism in LAr: when argon atoms are ionized, they combine with one another to form singlet and triplet excimer states.
When these excimer states decay, they emit 128~nm photons which pass unattenuated through the surrounding atomic argon~\cite{Suzuki:1979km}.
The decay of the singlet (triplet) state happens on timescales of $\mathcal{O}$(ns) ($\mathcal{O}$($\mu$s))~\cite{DEAP:2020hms,Kubota_1978}.
Thus, scintillation light emission happens on much shorter timescales than the $\mathcal{O}$(ms) drift time of the ionization electrons.

\begin{figure}
    \centering
    \includegraphics[width=0.6\textwidth]{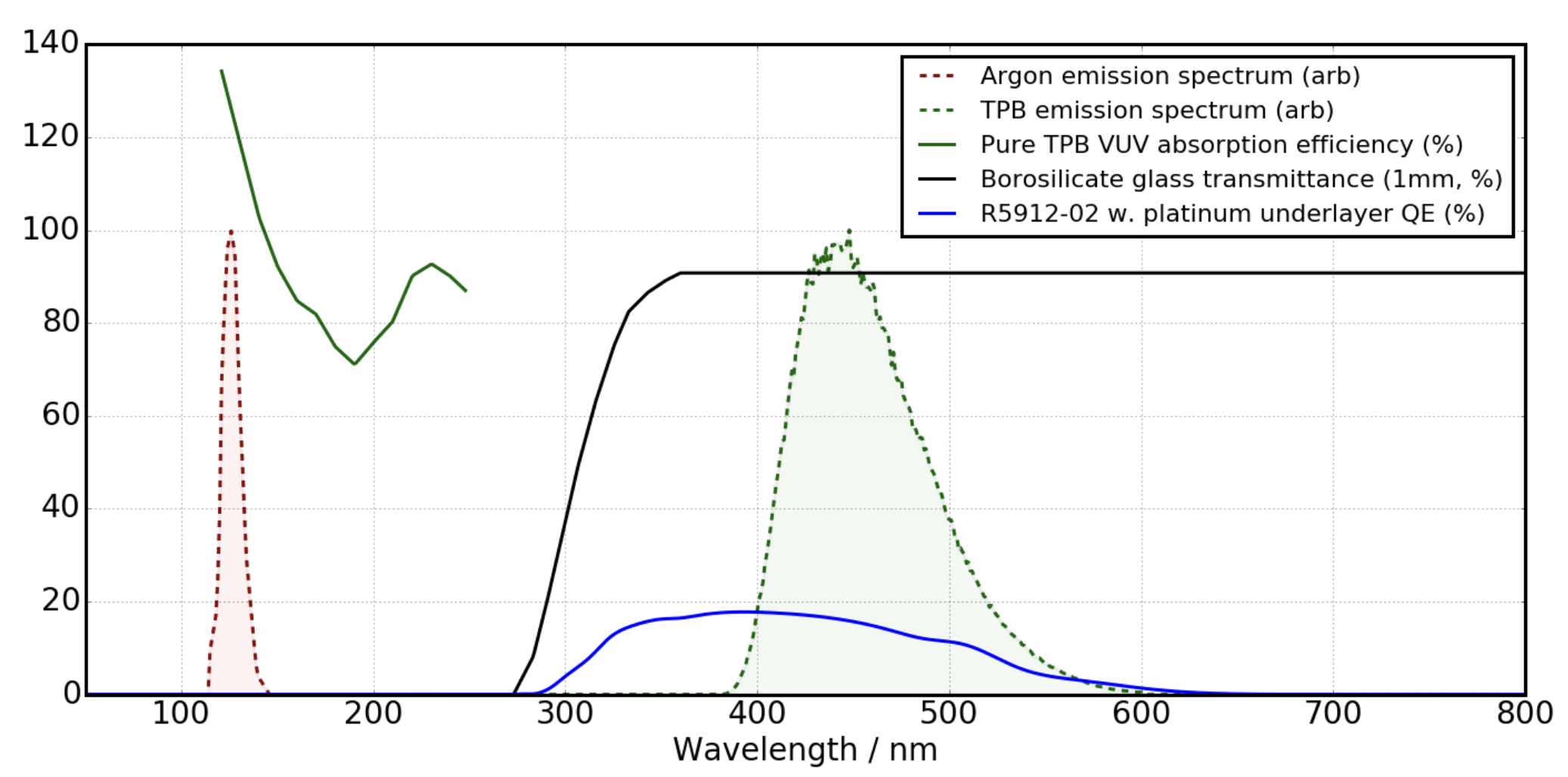}
    \caption{The LAr scintillation light spectrum, TPB ultra-violet absorption spectrum, TPB emission spectrum, PMT quantum efficiency, and PMT surface transmission efficiency as a function of photon wavelength. Figure from Ref.~\cite{Jones:2015bya}.}
    \label{fig:ar_wavelengths}
\end{figure}

\begin{figure}[h!]
     \centering
     \begin{subfigure}[b]{0.3\textwidth}
         \centering
         \includegraphics[width=\textwidth]{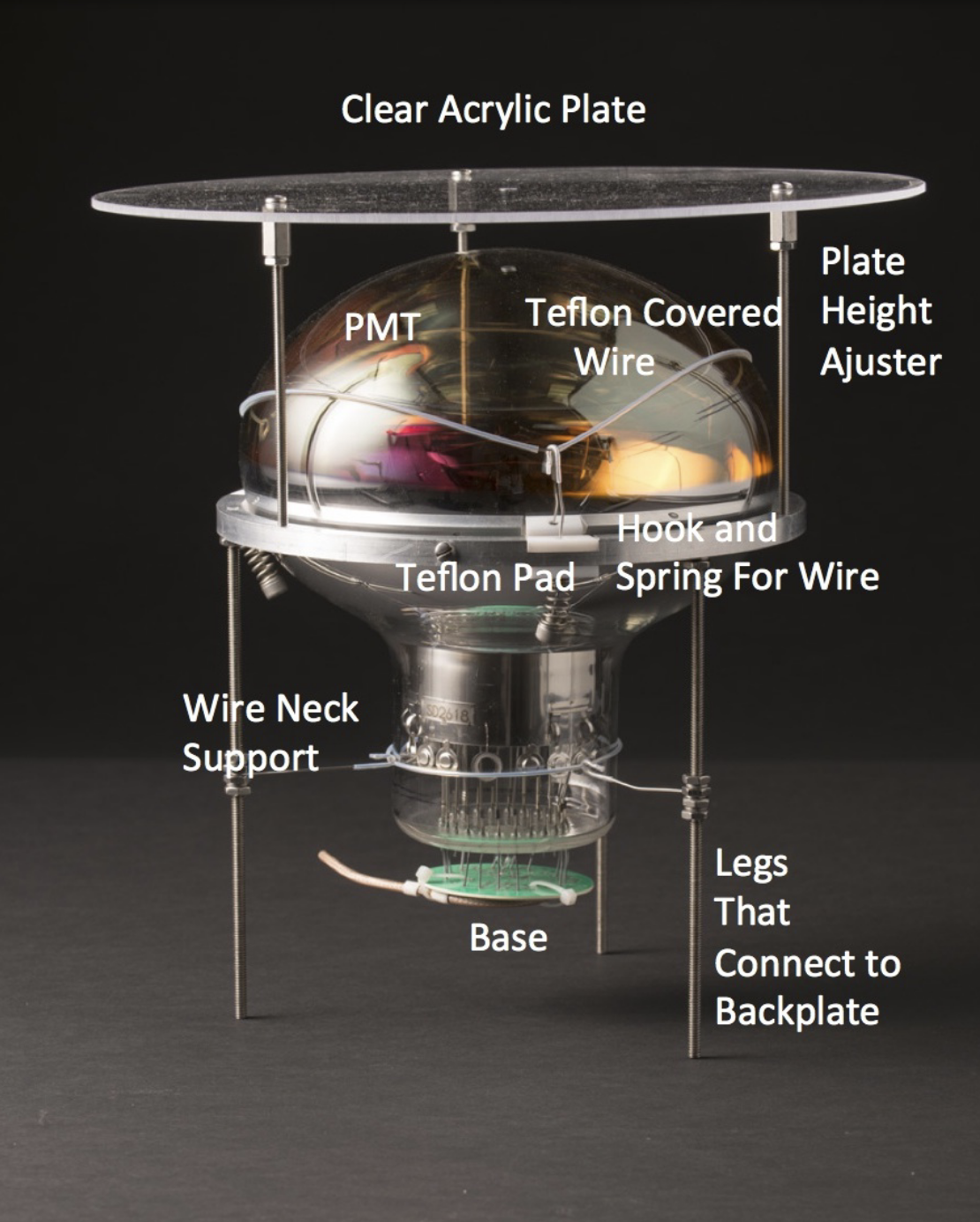}
         \caption{}
         \label{fig:pmt}
     \end{subfigure}
     \hfill
     \begin{subfigure}[b]{0.55\textwidth}
         \centering
         \includegraphics[width=\textwidth]{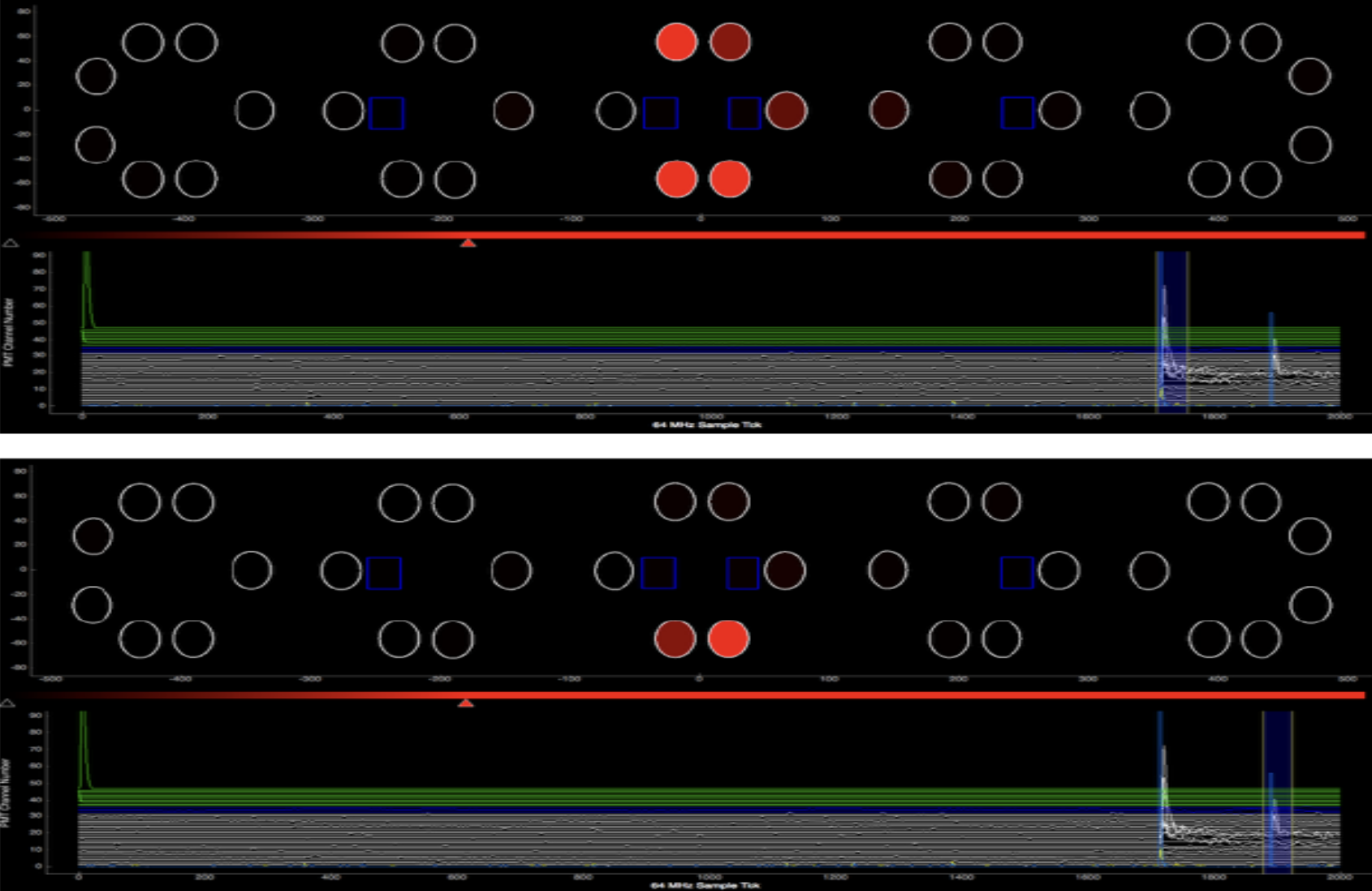}
         \caption{}
         \label{fig:pmt_muon}
     \end{subfigure}
        \caption{\Cref{fig:pmt} shows a photograph of a single PMT system used in the MicroBooNE detector. The acrylic window here has not yet been coated with TPB. \Cref{fig:pmt_muon} shows the PMT signal from a stopped cosmic muon (top) that decays to a Michel (bottom). Figures from Ref.~\cite{MicroBooNE:2016pwy}.}
        \label{fig:pmt_system}
\end{figure}

The light collection system in MicroBooNE is designed to detect the scintillation photons produced in a neutrino interaction.
It consists of 32 8-inch Hammamatsu R5912-02mod cryogenic PMTs situated behind an acrylic plate coated with tetraphenyl butadiene (TPB)~\cite{MicroBooNE:2016pwy}.
An image of one such PMT assembly is shown in \cref{fig:pmt}.
TPB is a wavelength shifter that absorbs the 128~nm argon scintillation light and re-emits a photon in the visible range.
The necessity of this procedure is shown in \cref{fig:ar_wavelengths}, which demonstrates that, unlike direct LAr scintillation light, TPB emission is well within the wavelength acceptance range of the PMTs.

When a photon hits the bi-alkali photocathode surface of an R5912-02mod PMT, an electron is released via the photoelectric effect~\cite{Einstein:1905cc}--this electron is often referred to as a ``photoelectron'' (p.e.).
Each p.e. is focused toward a dynode chain--a series of electrodes designed to produce a number $n>1$ electrons for each incident electron, resulting in an avalanche of electrons by the final anode which can be read out in the form of a current.
The wavelength-dependent probability with which a given p.e. enters the dynode chain is known as the ``quantum efficiency`` of the PMT.
At the nominal operating temperature of 87~K, the MicroBooNE PMTs have an average quantum efficiency of 15.3\%~\cite{MicroBooNE:2016pwy}.
Each PMT was tested in a liquid nitrogen (77~K) cryogenic environment, in which the gain and rate of thermal emission (``dark current'') of the PMT were measured as a function of the supplied high voltage (HV) across the dynode chain~\cite{Briese:2013wua}.
The HV for each PMT was set to produce a gain of $3 \times 10^7$ at 77~K~\cite{MicroBooNE:2016pwy}, corresponding to a dark current of $\mathcal{O}$(kHz)~\cite{Briese:2013wua}.
The current output from each PMT passed through preamp/shaper boards before being digitized via an analog-to-digital converter (ADC) with a sampling rate of 64~MHz~\cite{MicroBooNE:2016pwy}.
Thus, light is collected in time ticks with a length of 15.625~ns~\cite{MicroBooNE:2021pvo}.

The light collection system was critical in detecting activity in the detector coincident with a beam spill, indicating the presence of a neutrino interaction.
In order to record a given event, at least 5 photoelectrons must have been detected across all PMTs~\cite{MicroBooNE:2021pvo}.
Additionally, a ``common optical filter'' was applied to reduce the non-neutrino trigger rate--this filter required at least one string of six time ticks with greater than 20 photoelectrons detected during the 1.6~$\mu$s beam spill window, and no such strings of six time ticks in the 2~$\mu$s prior to the beam spill.
The spatiotemporal distribution of light observed by the PMTs can also be used to augment the signal from the wire planes.
For example, the PMT signal from a cosmic muon which stops in the detector and decays to a Michel electron is shown in \cref{fig:pmt_muon}.

The scintillation light yield and detection efficiency in liquid argon are impacted by a number of phenomena.
Impurities can dissociate the argon excimers before they have a chance to emit scintillation light.
For example, O$_2$ molecules in the LAr volume can undergo a two-body collision with an argon excimer~\cite{WArP:2008dyo},
\begin{equation}
{\rm Ar}_2^* + {\rm O}_2 \to 2{\rm Ar} + {\rm O}_2.
\end{equation}
This interaction mainly decreases the decay probability of the longer-lived triplet state.
Because of this, measurements of the delayed scintillation lifetime in LAr are sensitive to the concentration of impurities in the detector~\cite{DEAP:2020hms,WArP:2008dyo,MicroBooNE:2016pwy}.

Scintillation light can also be absorbed by impurities within the detector.
At concentrations of around 2~ppm, dissolved nitrogen will decrease the attenuation length of 128~nm scintillation light to around 30~m~\cite{WArP:2008rgv,Jones:2013bca}.
TPB emanation from the painted acrylic plates can also lead to a bulk fluorescence effect within the liquid argon~\cite{Asaadi:2018ixs}.
Additionally, Rayleigh scattering can deflect scintillation photons, diminishing the detector's capability to translate photon detection into spatial information regarding the path of the original charged particle.
The Rayleigh scattering length for 128~nm photons in liquid argon has been measured to be approximately 55~cm~\cite{Grace:2015yta}.
Given the size of the MicroBooNE detector, scintillation photons will undergo around five Rayleigh scattering interactions on average before reaching a PMT.

\section{TPC Signal Processing} \label{sec:ub_signal_processing}

MicroBooNE has developed a custom cryogenic electronics system to amplify and digitize the analog signal on each wire of the LArTPC~\cite{MicroBooNE:2016pwy}.
The wire signals first pass through front-end ASICs located on cryogenic motherboards attached directly to the wire carrier boards on the LArTPC.
Each ASIC amplifies and shapes the signal from 16 wires--thus, 516 ASICs are needed to fully instrument the 8,256 readout channels of the LArTPC.
Cryogenic-tested cables carry signals from the ASICs to intermediate amplifiers directly outside the cryostat, after which the signals travel through 20~m cables to the readout system.
This system consists of 130 readout modules, each containing 8 custom-designed 12-bit ADCs and a Front End Module (FEM).
Each ADC digitizes the signal from 8 wires at a rate of 16~MHz.
The dynamic range of the ADC is set to enable the detection both of a minimum-ionizing particle far away from the anode planes and of a highly-ionizing-particle close to the anode planes without saturation.
The FEM uses a field-programmable gate array (FPGA) to store and compress the digitized signals.
Once a trigger is received, the FPGA sends 4.8~ms of TPC data to the data acquisition (DAQ) system.
The 4.8~ms window is digitized into 9600 time ticks on each wire, with each time tick corresponding to 500~ns of integrated charge~\cite{MicroBooNE:2021pvo}.

\subsection{Noise Filtering}

The MicroBooNE LArTPC underwent an initial engineering run from October 2015 to July 2016.
Various sources of noise in the TPC were identified using data from this run, all of which are described in detail in Ref.~\cite{MicroBooNE:2017qiu}.
The frequency dependence of the noise was studied to inform the creation of an offline noise filter.

Some of this noise is inherent to the electronics system, coming from the cold, front-end ASIC or the two RC circuits in the intermediate amplifier and ADC board.
The gain and peaking time of the front-end ASIC are configurable from among four pre-defined settings each, which determine the dynamic range and timing granularity of the ASIC.
This choice also fixes the level of irreducible noise in the ASIC, comprised of multiple sources with different frequency dependence~\cite{MicroBooNE:2017qiu}.
The timing response of the RC circuits can distort large amplitude, long duration signals such as a cosmic muon traveling parallel to a single wire.
This effect can be corrected via a deconvolution process~\cite{MicroBooNE:2017qiu}.

\begin{figure}
    \centering
    \includegraphics[width=0.6\textwidth]{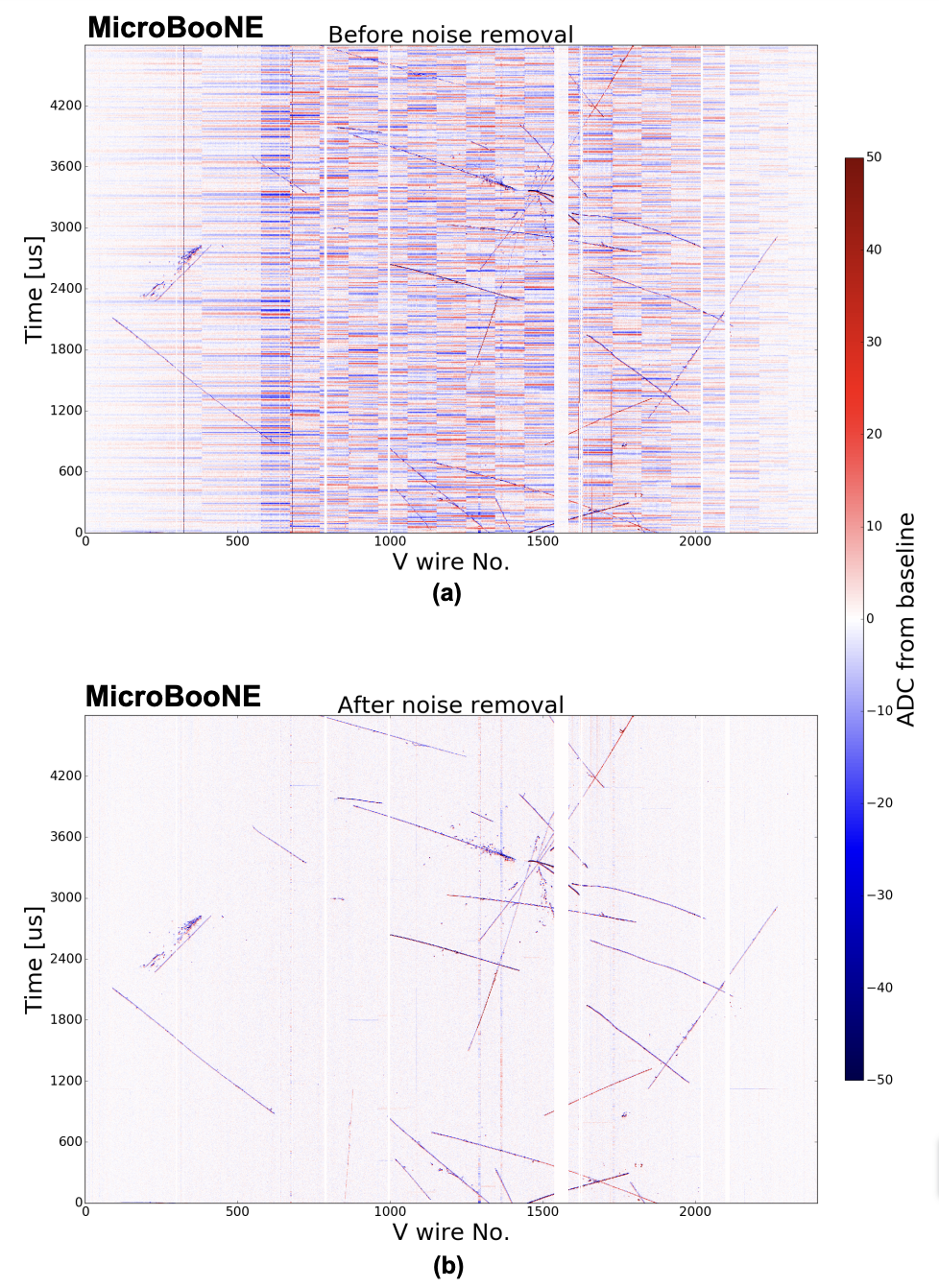}
    \caption{2D displays of the signal from wires in one of the induction planes in a single data event, before and after the application of the offline noise filters. Figure from Ref.~\cite{MicroBooNE:2017qiu}.}
    \label{fig:tpc_noise}
\end{figure}

The engineering run also revealed a number of excess noise sources above the inherent electronics noise.
The most prominent of these was low frequency ($\lesssim 30$~kHz) noise from the low voltage regulators on the front-end ASICs.
Harmonic noise was also observed corresponding to the ripple frequency of the HV power supply for the field cage.
Offline filters were developed to subtract the noise induced from both of these sources.
As shown in \cref{fig:tpc_noise}, these filters are able to remove most of the noise in the wire planes while having minimal impact on the true signal from ionization electrons.
After the engineering run, hardware upgrades were made to the low voltage regulators and the HV power supply which greatly reduced the excess noise in the TPC readout signals.
Even after these upgrades, the offline noise filters were still applied to optimize the TPC noise level.

\subsection{Deconvolution}

Even after noise filtering, the signal from the TPC is a complex combination of a number of factors: the ionization electron distribution, the induced current in each wire due to passing ionization electrons, and the response of the electronics system.
The signal must undergo deconvolution in order to obtain the quantity of physical interest--the number of ionization electrons passing each wire as a function of time.
The signal processing algorithm designed to extract this information is described in detail in Ref.~\cite{MicroBooNE:2018swd} and validated using MicroBooNE data in Ref.~\cite{MicroBooNE:2018vro}.
We give a brief overview of the algorithm here.

The measured signal $M(t)$ on a given wire is a convolution of the original signal $S(t')$ and the detector response function $R(t,t')$, which is intended to capture all the factors impacting signal formation discussed above.
This can be expressed by the integral
\begin{equation}
    M(t) = \int_{-\infty}^\infty R(t,t') S(t') dt'.
\end{equation}
For a time-invariant detector response function $R(t,t') = R(t-t')$, one can express this relationship in the frequency domain via a Fourier transform.
In a realistic setup, however, this treatment tends to amplify the effect of high-frequency components of the inherent electronics noise.
Typically one introduces a filter function $F(\omega)$ to suppress high frequency response, such that the original signal can be determined from
\begin{equation}
S(\omega) = \frac{M(\omega)}{R(\omega)} F(\omega).
\end{equation}
In this setup, one can consider $F(\omega)$ as a replacement for the detector response function.

The filter function used in MicroBooNE is inspired by the Weiner filter~\cite{10.7551/mitpress/2946.001.0001}, which is designed to optimize the signal-to-noise ratio while minimizing variance and bias and the resulting $S(\omega)$.
The exact Weiner filter is not applicable in the realistic MicroBooNE setup, as it does not conserve the total number of ionization electrons and can lead to non-local charge smearing due to suppression of low frequencies~\cite{MicroBooNE:2018swd}.
MicroBooNE instead constructs a Weiner filter $F(\omega)$ from simulation, and then fits the result to a functional form designed to preserve the normalization of the measured signal and remove the unwanted low-frequency suppression behavior.

The procedure discussed so far only accounts for the signal generated in a given wire due to ionization electrons passing by that wire.
In reality, electrons passing by neighboring wires can also induce a current in a given wire.
This is addressed by using a two-dimensional (2D) deconvolution in both the time and wire dimensions rather than the one-dimensional deconvolution in the time dimension introduced above.
An additional Weiner-inspired filter is constructed along the wire dimension when performing 2D deconvolution.
After the 2D deconvolution, the algorithm identifies a region of interest (ROI) around each potential signal.
These ROIs act as high-pass filters which suppress low-frequency noise in the deconvolved signal while preventing any non-local charge smearing.

\begin{figure}
    \centering
    \includegraphics[width=0.6\textwidth]{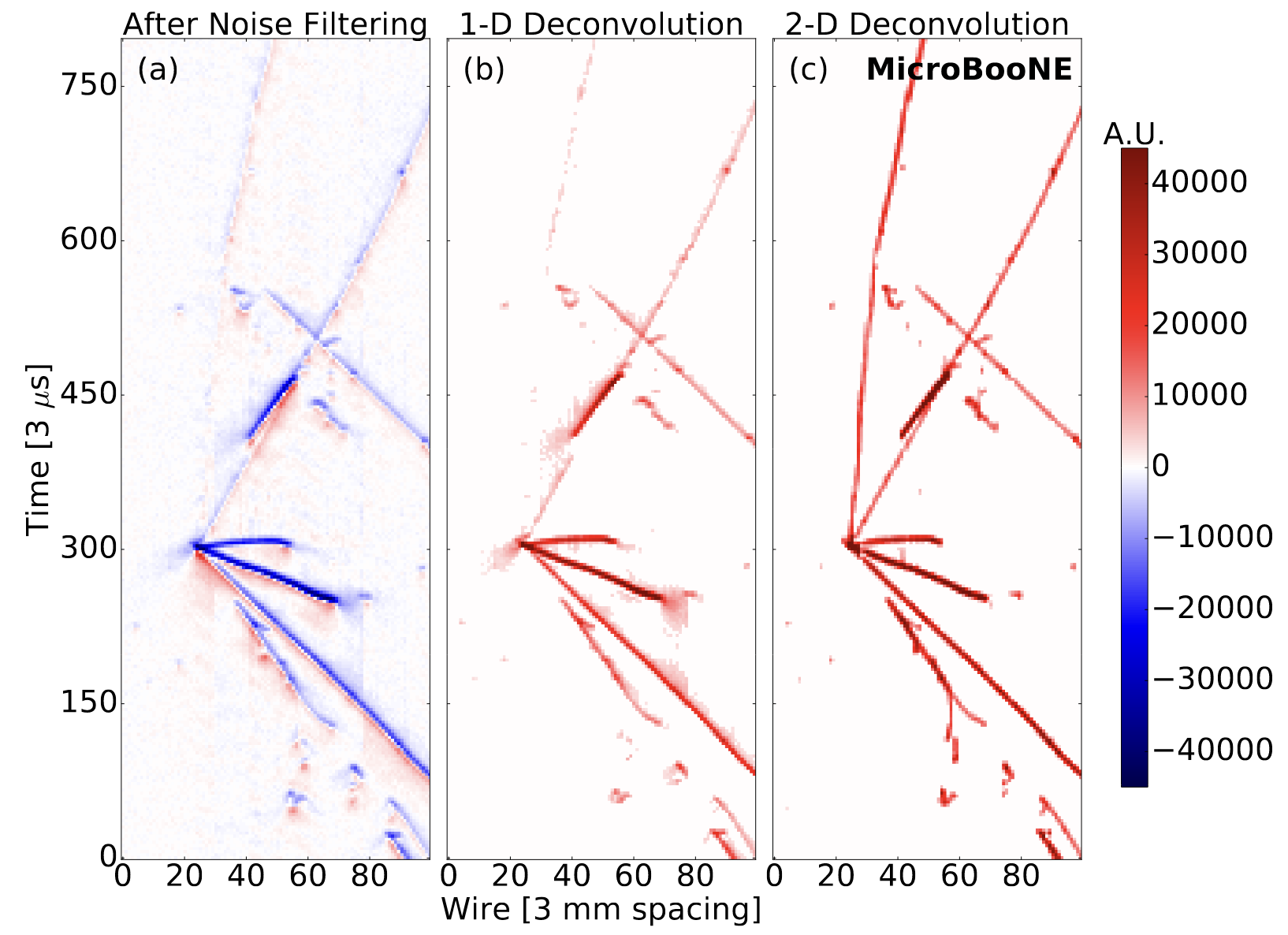}
    \caption{A U plane event display of a candidate neutrino interaction in the MicroBooNE data. The impact of the 1D and 2D deconvolution algorithms on the post-noise-filtering signal is shown. Figure from Ref.~\cite{MicroBooNE:2018swd}.}
    \label{fig:deconvolution}
\end{figure}

The full deconvolution algorithm results in a robust reconstruction of the amount of charge passing by or deposited on each wire of the LArTPC.
\Cref{fig:deconvolution} shows the U plane event display for a neutrino interaction candidate in MicroBooNE data after the application of the deconvolution algorithm.
After the 2D deconvolution, one can clearly identify all of the charged particle tracks coming from the neutrino interaction vertex.
The shape of the deconvolved signal for wires in each plane matches closely between data and simulation, as discussed in Ref.~\cite{MicroBooNE:2018vro}.
The signals produced by the deconvolution algorithm are ready to be used by MicroBooNE physics analyses, including the search for the MiniBooNE electron-like excess presented in \cref{ch:microboone_selection,ch:microboone_results}.

\chapter{The MicroBooNE Electron Neutrino Analysis: Overview and Selection} \label{ch:microboone_selection}

The MicroBooNE experiment was conceived to investigate the nature of the excess of low-energy electron-like events observed by the MiniBooNE experiment, described in detail in \cref{ch:miniboone}.
The MicroBooNE LArTPC, described in \cref{ch:microboone_detector}, produces high-resolution event displays of neutrino interactions which enable the separation of events with electrons in the final state from those with photons.
The collaboration released its first results searching for a MiniBooNE-like excess of electromagnetic events in October 2021.
Four different analyses were performed, one focusing on photons and three focusing on electrons.
The photon analysis searched for an excess of $\Delta \to N \gamma$ radiative decays~\cite{MicroBooNE:2021zai}--the only SM photon background not constrained \textit{in situ} by MiniBooNE.
The three electron analyses searched for an excess of $\nu_e$ charged-current (CC) interactions, each with a different final state signal definition~\cite{MicroBooNE:2021tya}.
One of the electron analyses, hereafter referred to as the ``inclusive analysis'', isolated all $\nu_e$ CC interactions using the WireCell reconstruction framework~\cite{MicroBooNE:2021nxr}.
Another electron analysis, hereafter referred to as the ``MiniBooNE-like analysis'', used the Pandora reconstruction framework to isolate all $\nu_e$ CC interactions without pions in the final state~\cite{TheMicroBooNECollaboration:2021cjf}.
This chapter describes the ``two-body CCQE analysis'', which used deep learning techniques to isolate charged-current quasi-elastic (CCQE) $\nu_e$ interactions consistent with two-body scattering~\cite{MicroBooNE:2021pvo}.
The results from this analysis are presented in \cref{ch:microboone_results}.
The two-body CCQE analysis represents much of my early work as a graduate student.

\textit{Publications covered in this chapter for which I either held a leading role or made major contributions: \cite{MicroBooNE:2021pvo,MicroBooNE:2021tya,MicroBooNE:2021nss}}

\section{Dataset and Simulation} \label{sec:data_and_sim}

The results covered in this section come from the first $6.88 \times 10^{20}$ POT of BNB data taken by the MicroBooNE experiment.
These data were separated into three run periods which spanned from 2016 to 2018: run 1 ($1.75 \times 10^{20}$~POT),  run 2 ($2.70 \times 10^{20}$~POT), and run 3 ($2.43 \times 10^{20}$~POT)~\cite{MicroBooNE:2021pvo}.
Data from two more MicroBooNE run periods exist on tape and are currently being processed by the collaboration, corresponding to another roughly $5 \times 10^{20}$~POT.
The entire dataset was taken during nominal neutrino mode running conditions for the BNB, in which 8~GeV kinetic energy protons from the Fermilab Booster strike the beryllium target to produce a cascade of mesons, and the positively-charged mesons are focused and decay to produce a beam of neutrinos.
The BNB was described in detail in \cref{sec:bnb}.
The intrinsic $\nu_e$ component of the beam, which makes up $\sim 0.5$\% of the total neutrino flux, can be seen in \cref{fig:bnb_flux}.

The MicroBooNE simulation relies on the \texttt{GENIE v3.00.06} \cite{Andreopoulos:2009rq,Andreopoulos:2015wxa,GENIE:2021npt,GENIE:2021zuu} neutrino event generator.
MicroBooNE specifically employs the G18\_10a\_02\_11a model, which uses the Valencia CCQE and meson-exchange-current (MEC) models~\cite{Nieves:2011pp} as well as a local Fermi gas model to describe the nuclear environment.
GENIE models the final state interactions (FSI) of particles produced in neutrino interactions with the nuclear environment using the typical $A^{2/3}$ scaling, including corrections from nuclear binding energy~\cite{Bodek:1980ar} as well as Pauli blocking and the velocity dependence of the nuclear mean field~\cite{Pandharipande:1992zz}.
To improve the nuclear modeling of CCQE and MEC interactions, MicroBooNE has developed a custom tune of the G18\_10a\_02\_11a model using a $\nu_\mu$ CC$0\pi$ cross section measurement from T2K~\cite{MicroBooNE:2021ccs,T2K:2016jor}.
This so-called ``MicroBooNE tune'' was determined by performing a two-dimensional fit of the T2K double-differential cross section data in muon momentum $p_\mu$ and angle $\cos \theta_\mu$.
The fit adjusts four parameters related to the CCQE and MEC cross section models in \texttt{GENIE v3.00.06}, resulting in a better fit to both T2K and MiniBooNE double-differential cross section data (though discrepancies exist at lower muon momenta in the latter case).
The MicroBooNE tune is used for all three MicroBooNE $\nu_e$ analyses; it gives better agreement with MicroBooNE $\nu_\mu$ data, which are essential for constraining the $\nu_e$ prediction in each analysis, and updates the predicted $\nu_e$ CCQE and MEC cross sections on argon~\cite{MicroBooNE:2021ccs}.

Electron scattering data can be used to evaluate the vector part of the vector-axial neutrino-nucleus cross section models in event generators such as \texttt{GENIE v3.00.06}~\cite{Ankowski:2020qbe,electronsforneutrinos:2020tbf}.
Earlier versions of \texttt{GENIE} were found to be in disagreement with electron scattering data in the sub-GeV energy transfer regime~\cite{Ankowski:2020qbe}.
Updates to the QE, MEC, and $\Delta$ excitation models in \texttt{GENIE} improved agreement with the QE peak and MEC-dominated dip region, though disagreements remain at higher momentum transfers~\cite{electronsforneutrinos:2020tbf}.
Dedicated electron scattering measurements using the CLAS detector at Jefferson Lab have also revealed biases in the \texttt{GENIE}-based reconstruction of the incident electron energy, especially in the large final state transverse momentum regime~\cite{CLAS:2021neh}.

Particle propagation in the liquid argon detector is handled by the \texttt{Geant4} toolkit V10.3.03c~\cite{GEANT4:2002zbu}, and the MicroBooNE TPC response is modeled using the general \texttt{LArSoft}~\cite{Snider_2017} package and the internal \texttt{uboonecode} package.
The effect of cosmic particles crossing the TPC during the readout window is modeled by overlaying off-beam data on top of the simulated events.
This is important, as around 20-30 cosmic rays can cross the detector in a single event due to the long readout window and near-surface location of the MicroBooNE detector~\cite{MicroBooNE:2021pvo}.

\section{The Electron Low Energy Excess Template} \label{sec:lee_template}

The hypothesis studied in each of the three electron analyses is that the MiniBooNE anomaly comes from an excess of $\nu_e$ interactions from the BNB.
While such an excess could come from $\nu_\mu \to \nu_e$ oscillations in a $3+1$ model, this is not the only possible source.
For example, mismodeling of $\mu^+$ decay-in-flight and $K^+$ decay-at-rest in the BNB~\cite{MiniBooNE:2008hfu} could lead to an enhancement of low energy $\nu_e$ compared with the expectation.

\begin{figure}[h!]
     \centering
     \begin{subfigure}[b]{0.4\textwidth}
         \centering
         \includegraphics[width=\textwidth]{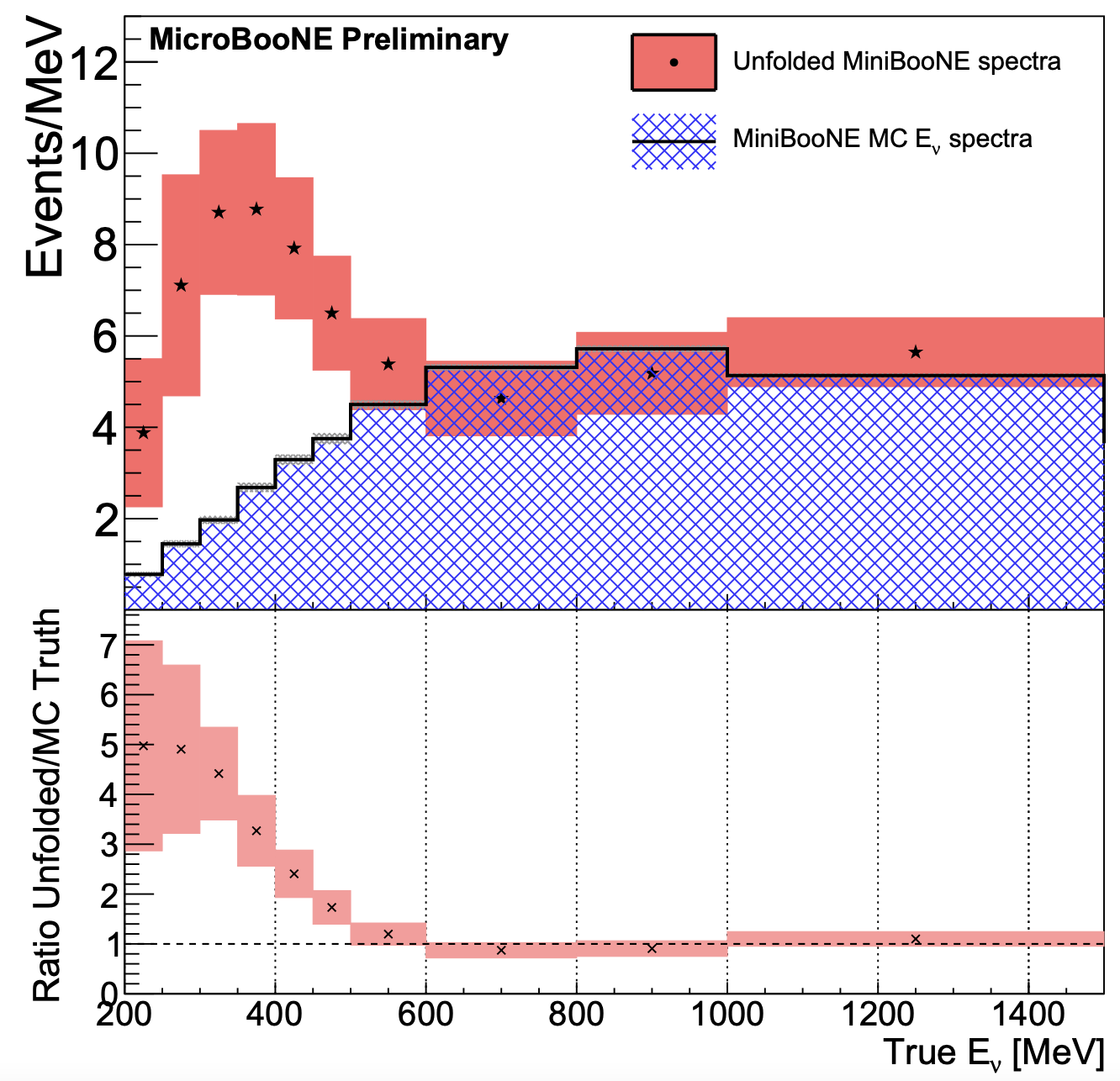}
         \caption{From Ref.~\cite{osti_1573217}}
         \label{fig:unfolding_technote}
     \end{subfigure}
     \hfill
     \begin{subfigure}[b]{0.55\textwidth}
         \centering
         \includegraphics[width=\textwidth]{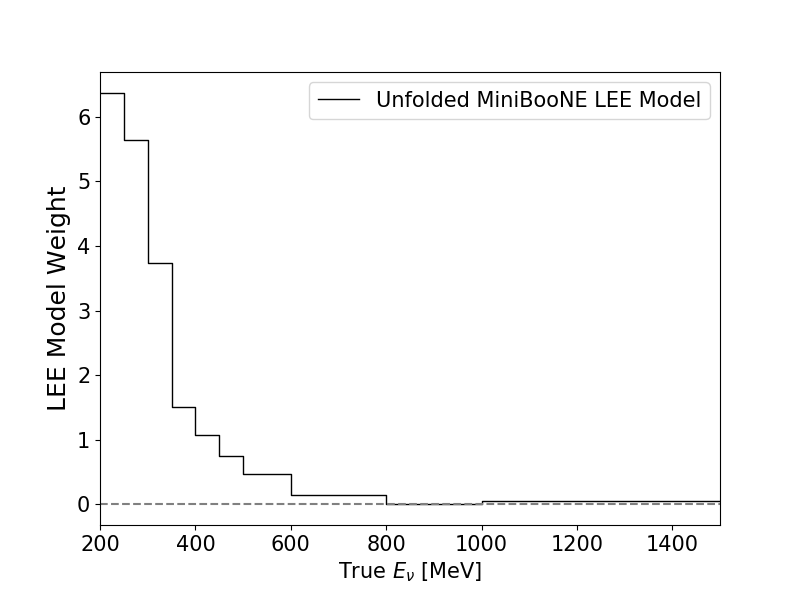}
         \caption{}
         \label{fig:unfolding_final}
     \end{subfigure}
        \caption{\Cref{fig:unfolding_technote} shows the unfolded $\nu_e$ prediction in MiniBooNE calculated using the first $6.46 \times 10^{20}$~POT of neutrino mode data~\cite{osti_1573217}. \Cref{fig:unfolding_final} shows the unfolded eLEE model weights derived from the first $12.84 \times 10^{20}$~POT of MiniBooNE neutrino mode data, which constitute the MicroBooNE eLEE model.}
        \label{fig:unfolding}
\end{figure}

MicroBooNE has elected to remain agnostic about possible sources of $\nu_e$ events for its first results.
To do this, the collaboration has developed an ``electron low energy excess (eLEE) model'' by unfolding the MiniBooNE excess under a $\nu_e$ hypothesis~\cite{MicroBooNE:2021pvo,osti_1573217}.
This is accomplished using the D'Agostini iterative unfolding procedure~\cite{DAgostini:1994fjx}, which updates a predicted distribution in true space using an observed distribution in reconstructed space and a known detector 
response matrix connecting the two.
The algorithm relies on Bayes theorem to update the predicted distribution over a specified number of iterations.
A small number of iterations will give an unfolded distribution biased toward the initial guess, while a larger number of iterations will give an unfolded distribution that is more sensitive to statistical fluctuations in the observed data~\cite{osti_1573217}.
For the eLEE model, the MiniBooNE observation as a function of reconstructed neutrino energy, as shown in \cref{fig:miniboone_enuqe}, is used to unfold the predicted $\nu_e$ interaction rate in MiniBooNE as a function of the true neutrino energy.
This procedure accounts for the detection efficiency; thus, the unfolded distribution can be interpreted as the number of $\nu_e$ events that interact in MiniBooNE.
One can take the ratio of the unfolded MiniBooNE $\nu_e$ prediction to the nominal $\nu_e$ prediction from the MiniBooNE Monte Carlo (MC), as shown in \cref{fig:unfolding_technote}.
This gives a set of binned weights as a function of the true $\nu_e$ energy, which can be applied to any BNB MC sample of $\nu_e$ interactions to represent a MiniBooNE-like excess.
These weights constitute the MicroBooNE eLEE model of the MiniBooNE anomaly.
\Cref{fig:unfolding_final} shows the actual eLEE model weights used in all three MicroBooNE $\nu_e$ analyses, calculated using the first $12.84 \times 10^{20}$~POT of the MiniBooNE neutrino mode dataset.

A few comments are necessary regarding the unfolding procedure.
First, the number of iterations is a regularization parameter that must be selected by the user.
MicroBooNE used three iterations of the D'Agostini algorithm to produce the eLEE model, a choice designed to minimize the variance and bias in the unfolded spectrum while remaining in good agreement with the MiniBooNE data, i.e. retaining a $\chi^2$ per degree-of-freedom less than 1 in the reconstructed distribution.
While it is true that these criteria select a unique choice for the number of iterations, the eLEE model shown in \cref{fig:unfolding_final} is not the unique unfolded distribution consistent with the MiniBooNE excess.
This point is discussed further in Ref.~\cite{Arguelles:2021meu}.
Additionally, the eLEE model was developed before MiniBooNE had looked at any $\nu_e$ data in the $E_\nu^{\rm QE} < 200\;{\rm MeV}$ region~\cite{MiniBooNE:2018esg}.
Thus, the unfolded $\nu_e$ prediction will lose events that reconstruct below 200 MeV.
It is also worth noting that while the MiniBooNE $e$-like dataset is only sensitive in principle to $\nu_e$ CC interactions without pions in the final state (as both charged and neutral pions are visible in a Cherenkov detector), the eLEE model weights are applied to all $\nu_e$ CC interactions in MicroBooNE.
The impact of this choice is mitigated by the development of three different $\nu_e$ analyses with an emphasis on different final states.
If the MiniBooNE excess only exists in one class of $\nu_e$ interactions, this should be evident when comparing results from all three analyses.
Finally, the eLEE model does not account for systematic uncertainties on the MiniBooNE prediction.
This is important, as MiniBooNE is a systematics-limited experiment~\cite{MiniBooNE:2020pnu}.
The eLEE model weights shown in \cref{fig:unfolding_final} can be thought to represent a ``median'' model of the MiniBooNE excess.
One can scale the weights by an overall factor, hereafter referred to as the ``LEE signal strength'' $x_{\rm LEE}$, to capture the 21\% statistical $\mathop{\oplus}$ systematic normalization uncertainty on the MiniBooNE prediction.
In \cref{ch:microboone_results}, we will cover results from a one-dimensional signal strength scaling test.

\section{Philosophy Behind the Two-Body CCQE Analysis}

The two-body CCQE analysis seeks to do what MiniBooNE could not--isolate a high-purity sample of $\nu_e$ interactions with a low SM photon background, especially at low energies.
To do this, we have selected a very specific final state definition: $\nu_e$ CCQE interactions with reconstructed kinematic variables that are consistent with two-body scattering.

\begin{figure}[h!]
    \centering
    \includegraphics[width=0.6\textwidth]{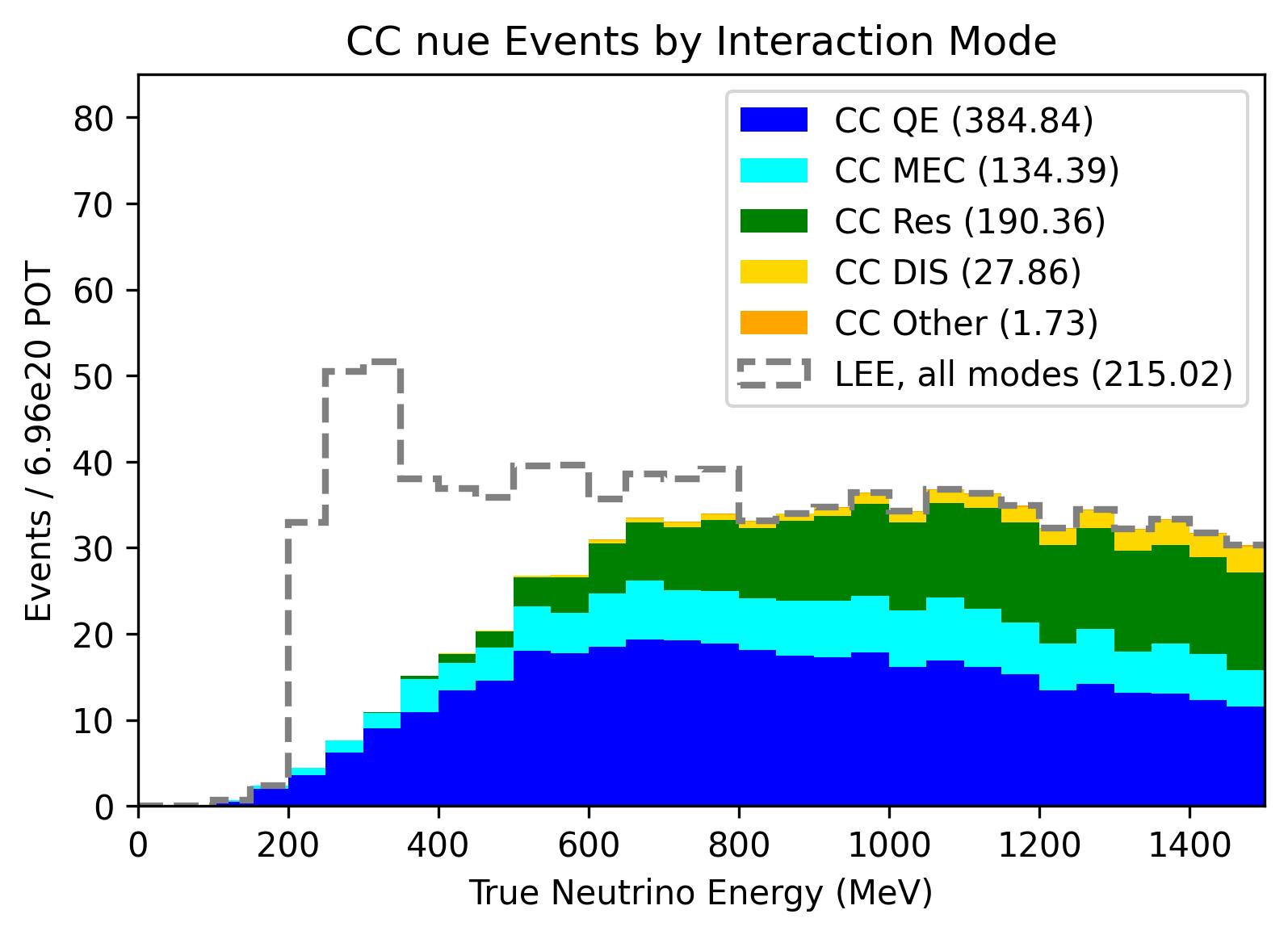}
    \caption{The expected distribution of $\nu_e$ interactions in MicroBooNE as a function of the true $\nu_e$ energy. The dotted line shows the expectation from the MicroBooNE eLEE model of the MiniBooNE excess discussed in \cref{sec:lee_template}.}
    \label{fig:ub_interaction_expectation}
\end{figure}

The $\nu_\ell$ CCQE interaction channel is a simple two-body scattering process,
\begin{equation}
\nu_\ell n \to \ell p.
\end{equation}
It was chosen because CCQE is the dominant neutrino interaction at lower neutrino energies, where the eLEE model weights in \cref{fig:unfolding_final} are the highest.
As shown in \cref{fig:ub_interaction_expectation}, CCQE interactions comprise 77\% of the $\nu_e$ CC interactions in MicroBooNE below 500 MeV~\cite{MicroBooNE:2021pvo}.
Other interaction channels, such as meson exchange current (MEC) and resonant (RES) scattering, become more significant at higher neutrino energies.
These channels are much more complicated than CCQE and thus come with larger modeling uncertainties.
MEC events involve pion exchange between two nucleons and are thus sometimes referred to as ``two-particle-two-hole'' interactions~\cite{Benhar:2015ula}.
RES events involve the excitation of a nucleon to a $\Delta$ resonance which will subsequently decay, most often to a pion and nucleon.
CCQE events are also easier to reconstruct; the original neutrino energy is simply the sum of the observed electron and proton kinetic energies, up to smearing from nuclear effects.
In MEC and RES events, one must account for the kinetic energy of additional final state protons and pions, and the potential existence of invisible final state neutrons makes it more difficult to connect the total kinetic energy of the observed final state to the initial neutrino energy.
Thus, the relative simplicity of CCQE interactions is another advantage of choosing such a signal definition.

\begin{figure}[h!]
    \centering
    \includegraphics[width=0.6\textwidth]{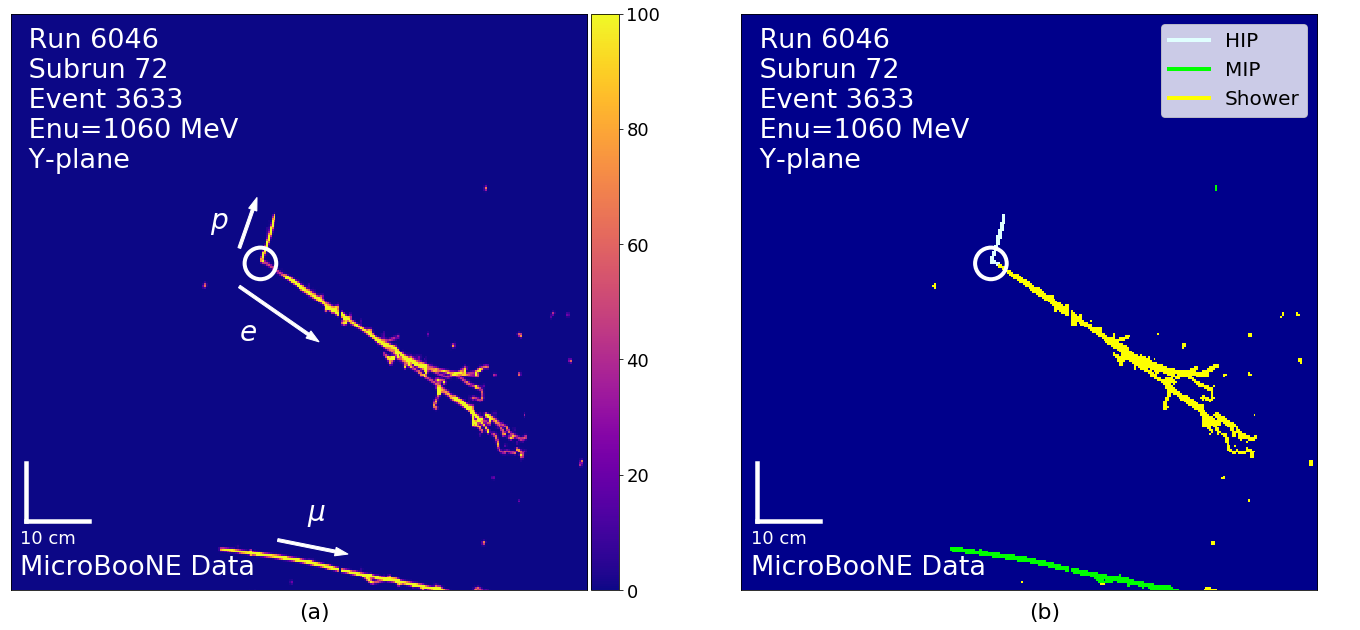}
    \caption{An example candidate $1e1p$ data event in MicroBooNE, including the raw LArTPC collection plane image (left) and the pixel labels assigned from \texttt{SparseSSNet} (right).}
    \label{fig:1e1p_example_EVD}
\end{figure}

The two-body CCQE analysis has developed a number of techniques to isolate $\nu_e$ CCQE events from other $\nu_e$ and $\nu_\mu$ interactions in MicroBooNE.
The most basic of these involves leveraging the clean $1e1p$ final state topology of the signal events.
The left panel of \cref{fig:1e1p_example_EVD} shows an example candidate $1e1p$ event in MicroBooNE data.
One can clearly see the proton track and electron EM shower emanating from the same vertex, presumably the location of the neutrino interaction in the LArTPC.
Our reconstruction algorithm takes advantage of this distinct ``vee'' shape, imposing a strict requirement that exactly two ``prongs'' of charge (i.e., tracks or showers) come from the same 3D-consistent interaction vertex.
This process is augmented by a powerful convolutional neural network to specifically find events with one track and one shower, indicative of a $1e1p$ final state.
The reconstruction algorithm for this analysis will be discussed in more detail in the next section.

\begin{figure}[h!]
    \centering
    \includegraphics[width=0.6\textwidth]{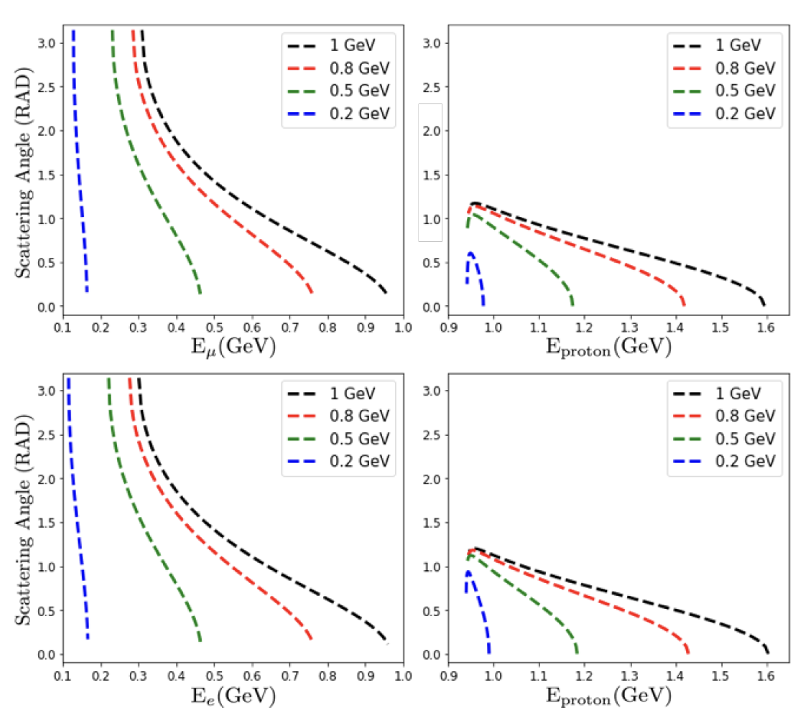}
    \caption{The relationship between the final state lepton (left) and proton (right) energy and scattering angle, for different neutrino energies, in $\nu_\mu$ (top) and $\nu_e$ (bottom) CCQE scattering.}
    \label{fig:twobody_angle}
\end{figure}

We also rely on two-body kinematics to select $\nu_e$ CCQE interactions.
In theory, the observed $1e1p$ final state from a $\nu_e$ CCQE interaction should obey kinematic constraints imposed in a two-body scattering process.
In a realistic MicroBooNE scenario, the struck neutron lives within the argon nuclear medium and has non-zero initial momentum.
The final state proton must also travel through the argon nuclear medium, in which it can undergo final state interactions that alter its momentum and/or create additional hadronic particles.
In order to optimize our reconstruction and minimize uncertainties, we are interested in events with minimal final state interactions.
We also want to remove backgrounds that will be especially inconsistent with two-body kinematics, such as $\nu_e$ MEC events and $\nu_\mu$ events with a $\pi^0$ in the final state in which one of the photons is reconstructed alongside a proton track.
This is achieved by introducing variables to the analysis that are especially sensitive to deviations from two-body kinematics.
For example, the total transverse momentum of the neutrino interaction ($p_T$) will be further from zero for non-CCQE events.
Additionally, as shown in \cref{fig:twobody_angle}, the proton in a two-body CCQE scattering interaction will always be forward-going.
While nuclear effects can in principle break this requirement in MicroBooNE, it is much more likely for a non-CCQE background event to result in a backward-going proton than a true CCQE event.
Therefore, we explicitly require forward-going protons in this analysis.
We can also consider the Bjorken scaling variable
\begin{equation}
x_{Bj} = \frac{Q^2}{2 p_p \cdot q}, 
\end{equation}
where $q^2 = -Q^2$ is the four-momentum transfer and $p_p$ is the final state proton four-momentum.
In the case of quasi-elastic scattering, $x_{Bj}$ should be close to unity up to nuclear effects~\cite{Thomson:2013zua}, while it can significantly differ from unity for non-CCQE backgrounds~\cite{MicroBooNE:2021pvo}.

Finally, we can harness the full power of the LArTPC technology to reconstruct the $\nu_e$ energy in multiple ways.
For $\nu_e$ traveling along the $z$ axis of the detector, the neutrino energy can be calculated from any of the following,
\begin{align}
E_\nu^{\rm range}&= \textnormal{K}_{p} + \textnormal{K}_{\ell} + M_{\ell} + M_{p} - (M_{n} - B),  \label{eq:erange} \\
E_\nu^{QE-p}&= \left(\dfrac{1}{2}\right) \/\dfrac{2\cdot(M_{n}-B)\cdot E_{p}-((M_n-B)^{2}+M_{p}^{2}-M_{\ell}^{2})}{(M_{n}-B)-E_{p}+\sqrt{(E_{p}^{2}-M_{p}^{2})}\cdot \cos\theta_{p}}, \label{enuqep} \\
E_\nu^{QE-\ell}&=\left(\dfrac{1}{2} \right)\/ \dfrac{2\cdot(M_{n}-B)\cdot E_{\ell}-((M_n-B)^{2}+M_{\ell}^{2}-M_{p}^{2})}{(M_{n}-B)-E_{\ell}+\sqrt{(E_{\ell}^{2}-M_{\ell}^{2})}\cdot \cos\theta_{\ell}}, \label{eq:enuqelepton}
\end{align}
where $K_{\ell/p}$ is the kinetic energy of the lepton or proton, $\theta_{\ell/p}$ is the angle of the shower or track with respect to the $z$ axis, $M_{\ell/p/n}$ is the particle mass, and $B$ is the average binding energy, which is around 40\;MeV in argon~\cite{Ankowski:2005wi}.
For a true $\nu_e$ CCQE interaction, all three of these formulae should yield a similar reconstructed neutrino energy.
To leverage this effect we define the ``QE consistency'' variable $\Delta^{QE}$ as
\begin{equation}
(\Delta^{QE})^2 \equiv (E_\nu^{\rm range} - E_\nu^{QE-p})^2 + (E_\nu^{\rm range} - E_\nu^{QE-\ell})^2 + (E_\nu^{QE-\ell} - E_\nu^{QE-p})^2,
\end{equation}
which should be close to zero for CCQE events.
To mitigate smearing from nuclear effects, $E_\nu^{QE-p}$ and $E_\nu^{QE-\ell}$ are calculated in the rest frame of the struck nucleon.
We note here that \cref{eq:erange} is the official reconstructed neutrino energy used for this analysis; $E_\nu$ and $E_\nu^{\rm range}$ are used interchangeably throughout this thesis. 

Finally, it is worth mentioning that the two-body CCQE analysis also isolates a control sample of $\nu_\mu$ CCQE interactions with a $1\mu1p$ final state.
The $1\mu1p$ sample is constructed analogously to the signal $1e1p$ sample, though with around two orders of magnitude more events (as $\nu_\mu$ comprise most of the BNB flux).
This sample is used to constrain the intrinsic $\nu_e$ prediction and uncertainties in the $1e1p$ sample, as the $\nu_e$ and $\nu_\mu$ flux and cross section are highly correlated.
We also make use of dedicated $\pi^0$ and Michel electron samples, as will be discussed in \cref{sec:shower_publication}.

In summary, the goal of the analysis presented here is the isolation of clean two-body $\nu_e$ CCQE interactions in MicroBooNE.
The following sections explain how we leverage the topological and kinematic patterns described above to identify such events.

\section{Reconstruction} \label{sec:ub_reco}

We begin with a broad overview of the reconstruction chain for the two-body CCQE analysis.
An initial set of neutrino event candidates are identified by requiring sufficient light detection in time with the beam trigger, as discussed in \cref{sec:ub_light_collection}.
These events are passed through the noise filter and deconvolution algorithm described in \cref{sec:ub_signal_processing} and shown in \cref{fig:deconvolution}.
Additionally, a ``good runs cut'' is employed to remove events during which the MicroBooNE detector response was especially noisy~\cite{MicroBooNE:2021pvo}.
These steps are common to all MicroBooNE analyses.
They result in one event display per wire plane for every event, similar to the rightmost panel of \cref{fig:deconvolution} and the left panel of \cref{fig:1e1p_example_EVD}.
The color axis in the left panel of \cref{fig:1e1p_example_EVD} is given in ``pixel intensity units'' (PIU), which are defined as the integrated charge in one wire over six 0.5~$\mu$s time ticks~\cite{MicroBooNE:2021pvo}.
Given the electron drift speed of 0.11~cm/$\mu$s, this corresponds to a spatial extent of 0.33~cm along the drift direction.
The wire spacing is such that each pixel in the left panel of \cref{fig:1e1p_example_EVD} represents a $0.3$~cm $\times$ $0.33$~cm 2D square in the detector.

In the two-body CCQE analysis, these event displays are first passed through the sparse semantic segmentation network (\texttt{SparseSSNet})~\cite{MicroBooNE:2020yze}.
\texttt{SparseSSNet} is a custom convolutional neural network (CNN) developed by the analysis team that labels the active pixels in each event according to five different charge deposition hypotheses: highly-ionizing particle (HIP), minimum-ionizing particle (MIP), EM activity, delta electron, or Michel electron.
For the purposes of this analysis, we group HIP and MIP labels into the parent ``track'' label and EM activity, delta electrons, and Michel electrons into the parent ``shower'' label.
An example \texttt{SparseSSNet} label is shown in the right panel of \cref{fig:1e1p_example_EVD}.

After going through \texttt{SparseSSNet}, the images are filtered for crossing cosmic interactions using the WireCell charge-to-light matching algorithm~\cite{MicroBooNE:2020vry,MicroBooNE:2021zul}.
This algorithm matches charge observed on the TPC to light observed by the PMTs to identify regions of charge which are not coincident in time with the beam spill, presumably coming from cosmic rays.
We can then mask these pixels to remove the bulk of cosmic-related charge deposition.
The cosmic-filtered \texttt{SparseSSNet} output is then used as input to a vertex and track identification algorithm, which looks for a shower-like and track-like prong of charge coming from a given 3D vertex~\cite{MicroBooNE:2020sar}.
Track-like prongs pass through a dedicated step of the algorithm that performs a step-by-step 3D reconstruction of the extent and direction of the track, which is then translated into a range-based energy and angle estimation.
The events are then passed through a dedicated EM shower identification algorithm that identifies cones of shower-like pixels pointing back to the neutrino interaction vertex and estimates the total visible energy and angle of the shower.~\cite{MicroBooNE:2021nss}.
This algorithm has the capability of finding up to two showers separated from the vertex, which is necessary for identifying events with a $\pi^0 \to \gamma \gamma$ decay.

After finding candidate neutrino vertices, we pass the event displays through the multiple particle identification (\texttt{MPID}) CNN~\cite{MicroBooNE:2020hho}.
This network assigns five scores to each image indicating the network's confidence that the event contains a proton, electron, muon, photon, or charged pion.
The MPID scores are not used until the final signal selection described in \cref{sec:1e1p_selection}.

At this point, the reconstruction chain has identified two-prong candidate events and reconstructed the relevant lowest-level kinematic variables of each final state particle, i.e. the energy and direction.
These kinematic variables are then combined to calculate more complicated kinematic variables such as $Q^2$, $p_T$, $x_{Bj}$, and the reconstructed $\pi^0$ mass (in the case of two-shower events), as well as the most important variable in this analysis: the reconstructed neutrino energy $E_\nu^{\rm range}$, as defined in \cref{eq:erange}.
The full suite of reconstructed kinematic variables used to define the $1e1p$ and $1\mu1p$ samples is given in \cref{tab:KinVarTable}.
These variables are used as input to the signal selection described in \cref{sec:1e1p_selection}.
The following subsections explain in more detail the most important steps of the reconstruction procedure.

\begin{sidewaystable}[h!]
\begin{center}
\begin{tabular}{c|c}
\hline
\textbf{Variable Name} & \textbf{Definition} \\
\hline
\multicolumn{2}{c}{Base Variables}\\ \hline
$\textnormal{K}_{p}$ & Kinetic energy of proton determined from range \cite{MicroBooNE:2020sar}\\
$\textnormal{K}_{\ell}$ & Kinetic energy determined from range for muons \cite{MicroBooNE:2020sar}
and from calorimetry for electrons ~\cite{MicroBooNE:2021nss}\\
$M_{\ell}$, $M_{n}$, $M_p$ & Masses of the lepton, neutron and proton\\ 
$\cos{\theta_p}$, $\cos{\theta_\ell}$ & $p_p^z / p_p$,~$p_\ell^z / p_\ell$ \\
$\phi_p$, $\phi_\ell$ & $atan2(p_p^y,p_p^x)$,~ $atan2(p_\ell^y,p_\ell^x)$ \\
$P_p=(E_{p},p_{p})$, $P_\ell=(E_{\ell},p_{\ell})$ & Reconstructed 4-vector of the proton, lepton\\
$B$ & Binding Energy for argon; the analysis assumes $B=40$~MeV \\
\addlinespace[0.15cm]
\hline
\multicolumn{2}{c}{Definitions Related to Neutrino Energy}\\ \hline
$E_\nu^{range}$ (Default value of $E_\nu$) & $ \textnormal{K}_{p} + \textnormal{K}_{\ell} + M_{\ell} + M_{p} - (M_{n} - B)$   \\
\addlinespace[0.15cm] 
$E_\nu^{QE-p}$ & $\big(\dfrac{1}{2}\big) \/\dfrac{2\cdot(M_{n}-B)\cdot E_{p}-((Mn-B)^{2}+M_{p}^{2}-M_{\ell}^{2})}{(M_{n}-B)-E_{p}+\sqrt{(E_{p}^{2}-M_{p}^{2})}\cdot \cos\theta_{p}}$ \\
\addlinespace[0.15cm]
$E_\nu^{QE-\ell}$ & $\big(\dfrac{1}{2} \big)\/ \dfrac{2\cdot(M_{n}-B)\cdot E_{\ell}-((Mn-B)^{2}+M_{\ell}^{2}-M_{p}^{2})}{(M_{n}-B)-E_{\ell}+\sqrt{(E_{\ell}^{2}-M_{\ell}^{2})}\cdot \cos\theta_{\ell}}$ \\
\addlinespace[0.15cm]
 $\Delta^{QE}$ (2-Body Consistency) & $\sqrt{ (E_\nu^{range} - E_\nu^{QE-p})^2 + (E_\nu^{range} - E_\nu^{QE-\ell})^2 + (E_\nu^{QE-\ell} - E_\nu^{QE-p})^2}$\\
 \addlinespace[0.15cm]
\hline
\multicolumn{2}{c}{Event Kinematics} \\ \hline
$Q^2$ & $2E_\nu^{range}(E_\ell - P_\ell^z) -M_\ell^2$ \\
\addlinespace[0.15cm]
Hadronic Energy (E$_{had}$) & $E_\nu^{range} - E_\ell$ \\
\addlinespace[0.15cm]
Bj{\"o}rken's Scaling x (x$_{Bj}$) & $Q^2 / 2M_n E_{had}$ \\
\addlinespace[0.15cm]
Bj{\"o}rken's Scaling y (y$_{Bj}$) & $E_{had} / E_\nu^{range}$ \\
\addlinespace[0.15cm]
Opening angle & $\cos^{-1} (\hat{p}_\ell \cdot \hat{p}_p)$\\
\addlinespace[0.15cm]
$p_T$ & $\sqrt{ (p_\ell^x+p_p^x)^2 + (p_\ell^y+p_p^y)^2 }$\\
\addlinespace[0.15cm]
$p_L$ & $p_p^z + p_\ell^z$\\
\addlinespace[0.15cm]
$\alpha_T$ & $\cos^{-1} \left(- \dfrac{\vec{P}_{T}^{l} \cdot \vec{P}_{T}}{ |\vec{P}_{T}^{l}| |\vec{P}_{T}| }\right)$ \\
\addlinespace[0.15cm]
$\phi_T$ & $\cos^{-1} \left(- \dfrac{\vec{P}_{T}^{l} \cdot \vec{P}_{T}^{p}}{|\vec{P}_{T}^{l}||\vec{P}_{T}^{p}|}\right)$\\
\addlinespace[0.15cm]
\hline
\multicolumn{2}{c}{Boosting Parameters}\\ \hline
$p_{fermi}^T$ & $p_p^T + p_\ell^T $ \\
\addlinespace[0.15cm]
$p_{fermi}^z$ & $p_p^z + p_\ell^z - E_\nu^{range}$ \\ 
\hline
\end{tabular}
\end{center}
\caption{The definition of kinematic variables used throughout the two-body CCQE analysis.
\label{tab:KinVarTable}}
\end{sidewaystable}

\subsection{Convolutional Neural Networks in LArTPCs}

As discussed above, The two-body CCQE analysis relies on two CNNs to assist in reconstructing the MicroBooNE LArTPC images: \texttt{SparseSSNet} and \texttt{MPID}.

\begin{figure}[h!]
     \centering
     \begin{subfigure}[b]{0.55\textwidth}
         \centering
         \includegraphics[width=\textwidth]{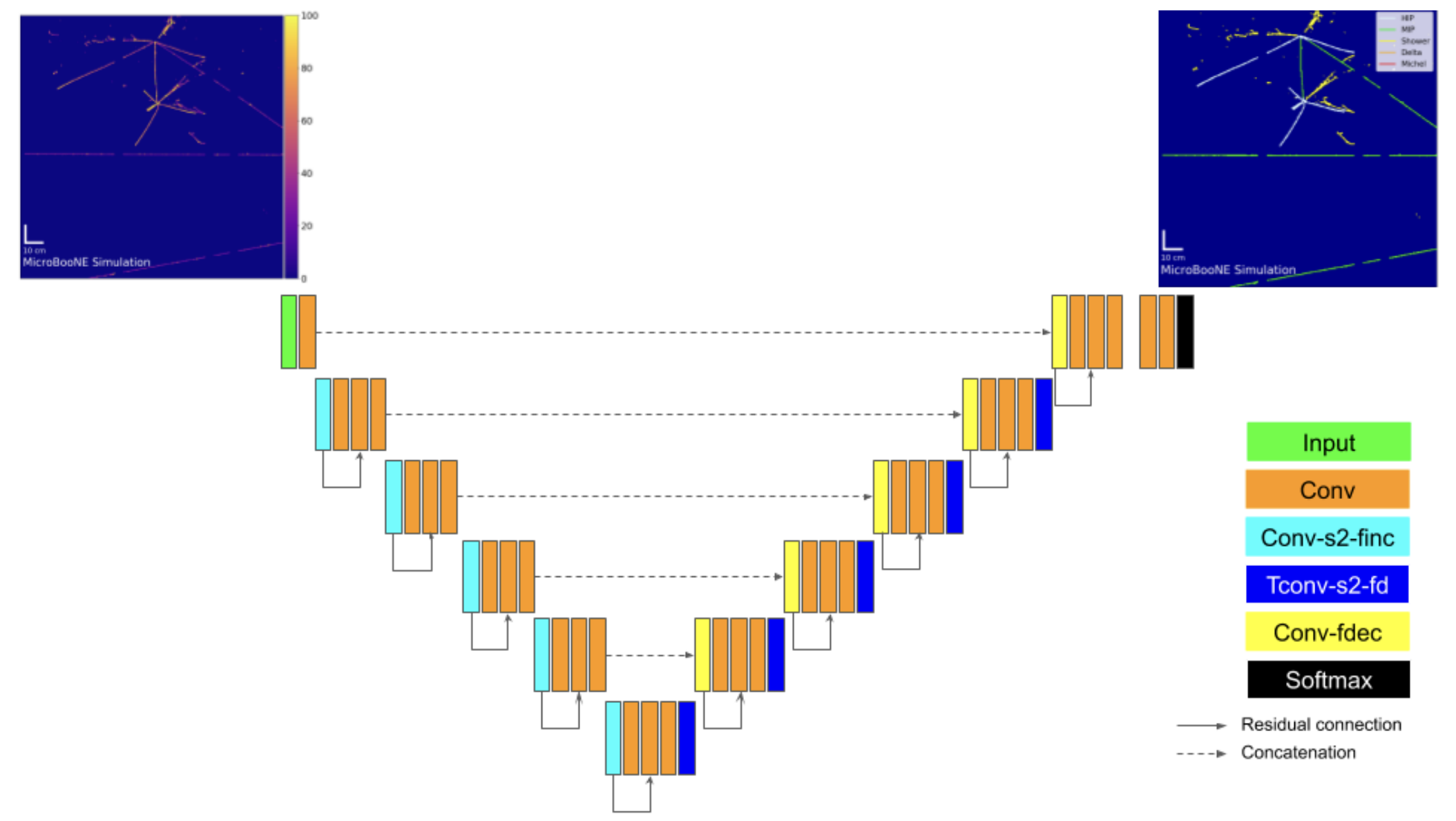}
         \caption{}
         \label{fig:sparsessnet_arc}
     \end{subfigure}
     \hfill
     \begin{subfigure}[b]{0.4\textwidth}
         \centering
         \includegraphics[width=\textwidth]{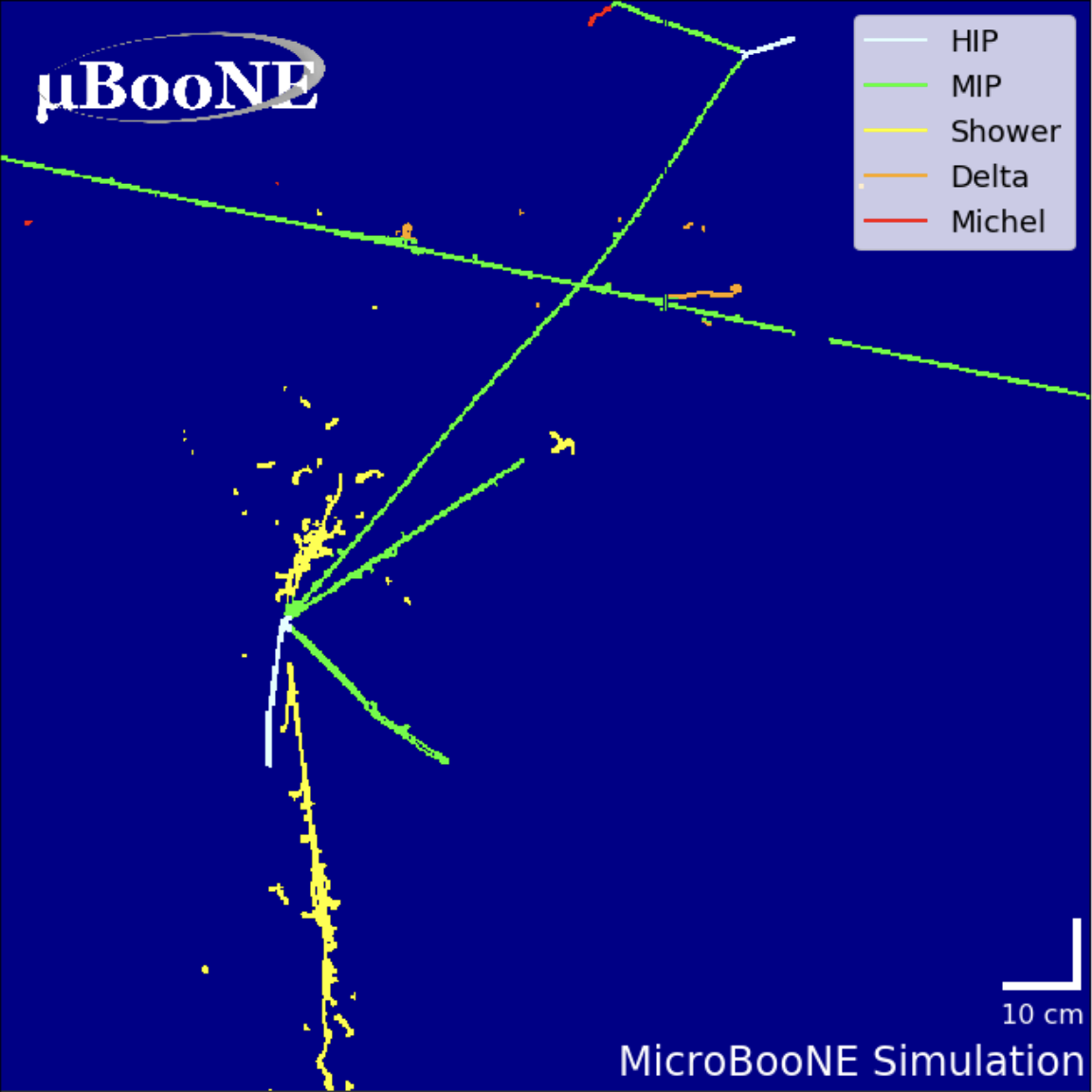}
         \caption{}
         \label{fig:sparsessnet_img}
     \end{subfigure}
        \caption{\Cref{fig:sparsessnet_arc} shows a diagram of the \texttt{SparseSSNet} U-ResNet architecture. \Cref{fig:sparsessnet_img} shows the \texttt{SparseSSNet} pixel labels on a simulated image in the training dataset. Figures from Ref.~\cite{MicroBooNE:2020yze}.}
        \label{fig:sparsessnet}
\end{figure}

\texttt{SparseSSNet} is described in detail in Ref.~\cite{MicroBooNE:2020yze}.
The network uses a combination of the U-Net~\cite{ronneberger2015u} and ResNet~\cite{he2016deep} architectures.
The U-Net uses an encoding step, in which the input image is downsampled to a smaller size, followed by a decoding step, in which the encoded image is upsampled to the same dimensionality and size as the original image.
This second step is required to perform pixel-level labeling.
The ResNet architecture employs residual connections that improve the training performance of deeper networks compared to regular CNNs~\cite{he2016deep}.
\texttt{SparseSSNet} also makes use of sparse submanifold convolutions~\cite{graham20183d}, which perform better on semantic segmentation tasks for sparse input data like LArTPC images (where very few pixels in a given image have nonzero charge).
The training sample is generated using the simulation described in \cref{sec:data_and_sim}; specifically, a random set of visible final state particles are generated at randomly distributed locations within the detector.
The network performs considerably well on the nominal MicroBooNE BNB simulation sample, as well as the dedicated intrinsic $\nu_e$ simulation sample, neither of which appeared in the \texttt{SparseSSNet} training sample~\cite{MicroBooNE:2020yze}.
While \texttt{SparseSSNet} has not been statistically evaluated against MicroBooNE data, hand scans of the network output on data event displays do not reveal any bias in the network performance~\cite{MicroBooNE:2020yze}.
The performance of an earlier iteration of \texttt{SparseSSNet} was evaluated using two data samples: Michel electrons and $\nu_\mu$ CC $\pi^0$ events~\cite{MicroBooNE:2018kka}.
Good agreement was observed between pixel labels in data and simulation for both samples, and no significant differences were found between network-labeled images and physicist-labeled images.
\Cref{fig:sparsessnet} shows the \texttt{SparseSSNet} architecture as well as an example labeled image in the training dataset.

\begin{figure}[h!]
     \centering
     \begin{subfigure}[b]{0.45\textwidth}
         \centering
         \includegraphics[width=\textwidth]{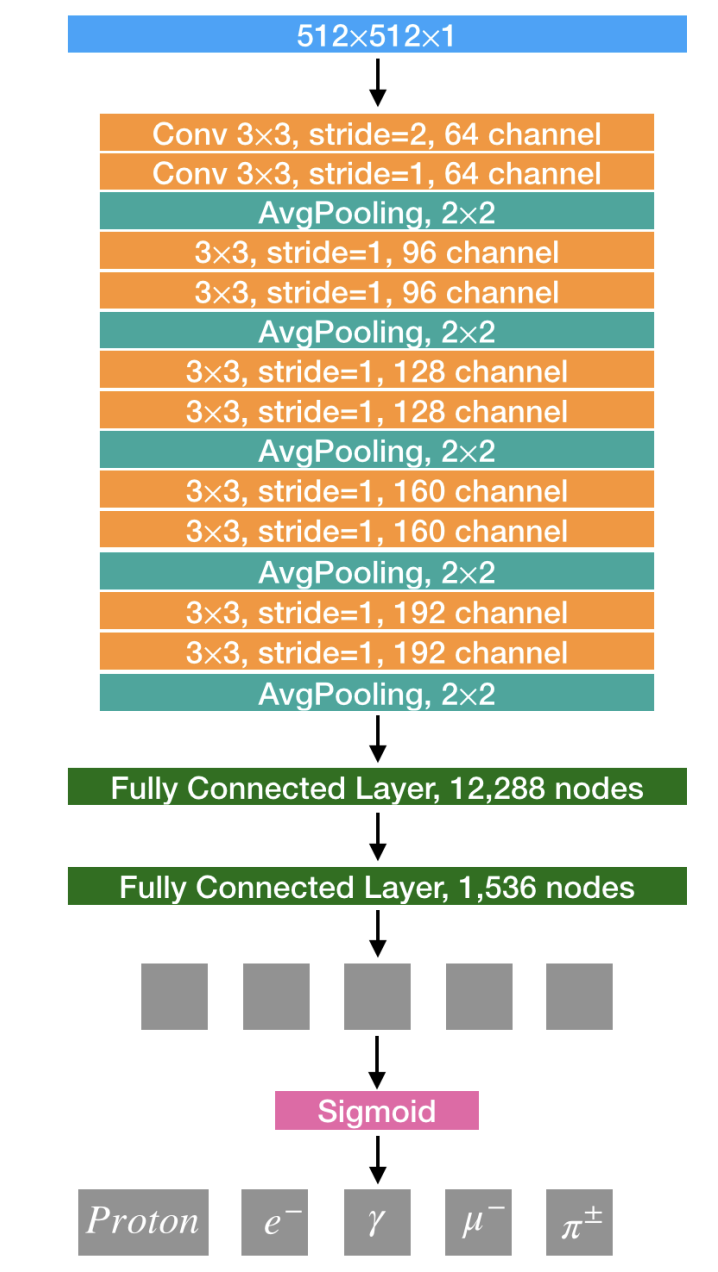}
         \caption{}
         \label{fig:mpid_arc}
     \end{subfigure}
     \hfill
     \begin{subfigure}[b]{0.5\textwidth}
         \centering
         \includegraphics[width=\textwidth]{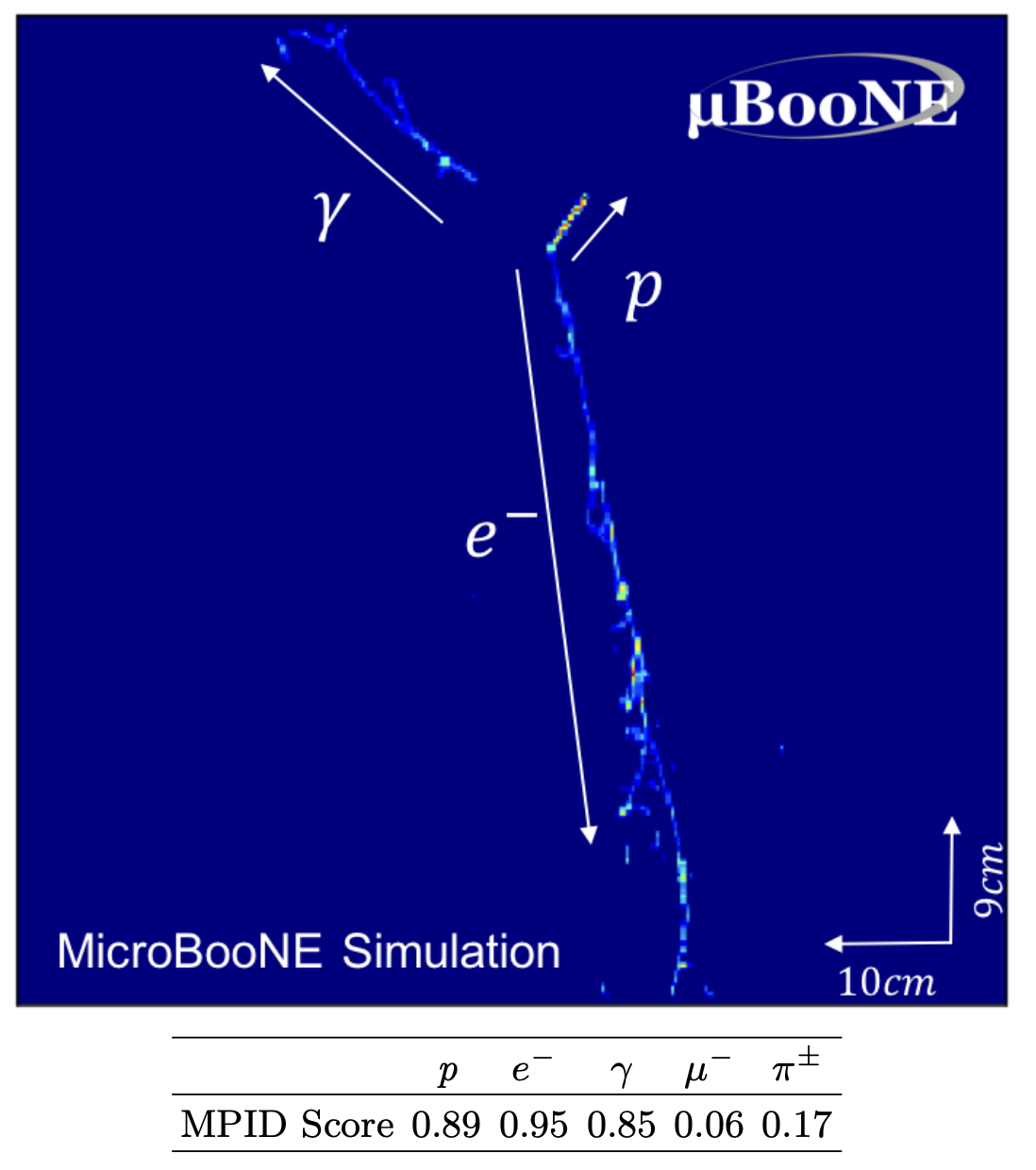}
         \caption{}
         \label{fig:mpid_img}
     \end{subfigure}
        \caption{\Cref{fig:mpid_arc} shows a diagram of the \texttt{MPID} architecture. \Cref{fig:mpid_img} shows the \texttt{MPID} image scores on an example simulated event. Figures from Ref.~\cite{MicroBooNE:2020hho}.}
        \label{fig:mpid}
\end{figure}

The \texttt{MPID} network is described in detail in Ref.~\cite{MicroBooNE:2020hho}.
It uses a standard CNN architecture which takes as input a 2D LArTPC image cropped around the neutrino interaction candidate vertex on a single wire plane and outputs five numbers representing different particle scores for protons, electrons, muons, photons, and charged pions.
The score is meant to represent the likelihood that a given particle exists in the provided image.
Two types of scores are generated: an ``image'' score, describing the probability that the particle exists anywhere in the image, and the ``interaction'' score, describing the probability that the particle is connected to the neutrino interaction vertex.
The training sample is generated similarly to that of \texttt{SparseSSNet}; random sets of particles are generated throughout the detector volume with energies in a range relevant to MicroBooNE.
The MPID score distributions exhibit excellent agreement between data and simulation in two different samples: $\nu_\mu$ CC $1\mu1p$ events and $\nu_\mu$ CC $\pi^0$ events~\cite{MicroBooNE:2020hho}.
\Cref{fig:mpid} shows the \texttt{MPID} architecture as well as an example simulated image with MPID image scores.

It is also worth briefly mentioning energy reconstruction in LArTPCs using CNNs.
My earliest work as a graduate student investigated the ability of the ResNet~\cite{he2016deep} and Inception network~\cite{szegedy2015going} CNN models to reconstruct EM shower energies in MicroBooNE.
Compared to the standard clustering-based shower reconstruction algorithm described in \cref{sec:shower_publication}, the CNN method appeared to better handle EM showers that passed through unresponsive wires in the MicroBooNE detector.
Despite the promise of the CNN method, it was not developed in time to be implemented in the two-body CCQE analysis.
Nevertheless, the MicroBooNE-specific network was extended by an MIT undergraduate, Kiara Carloni, to enable a systematic study of CNN-based EM shower energy reconstruction in a general LArTPC~\cite{Carloni:2021zbc}.
The corresponding publication is included in Appendix~A.
\Cref{fig:cnn_urw} shows the main takeaway of this work: a CNN-based reconstruction method is significantly more robust to unresponsive wires in the detector compared with a traditional EM shower reconstruction algorithm, which employs a linear calibration between the total shower charge and energy.

\begin{figure}[h!]
    \centering
    \includegraphics[width=0.6\textwidth]{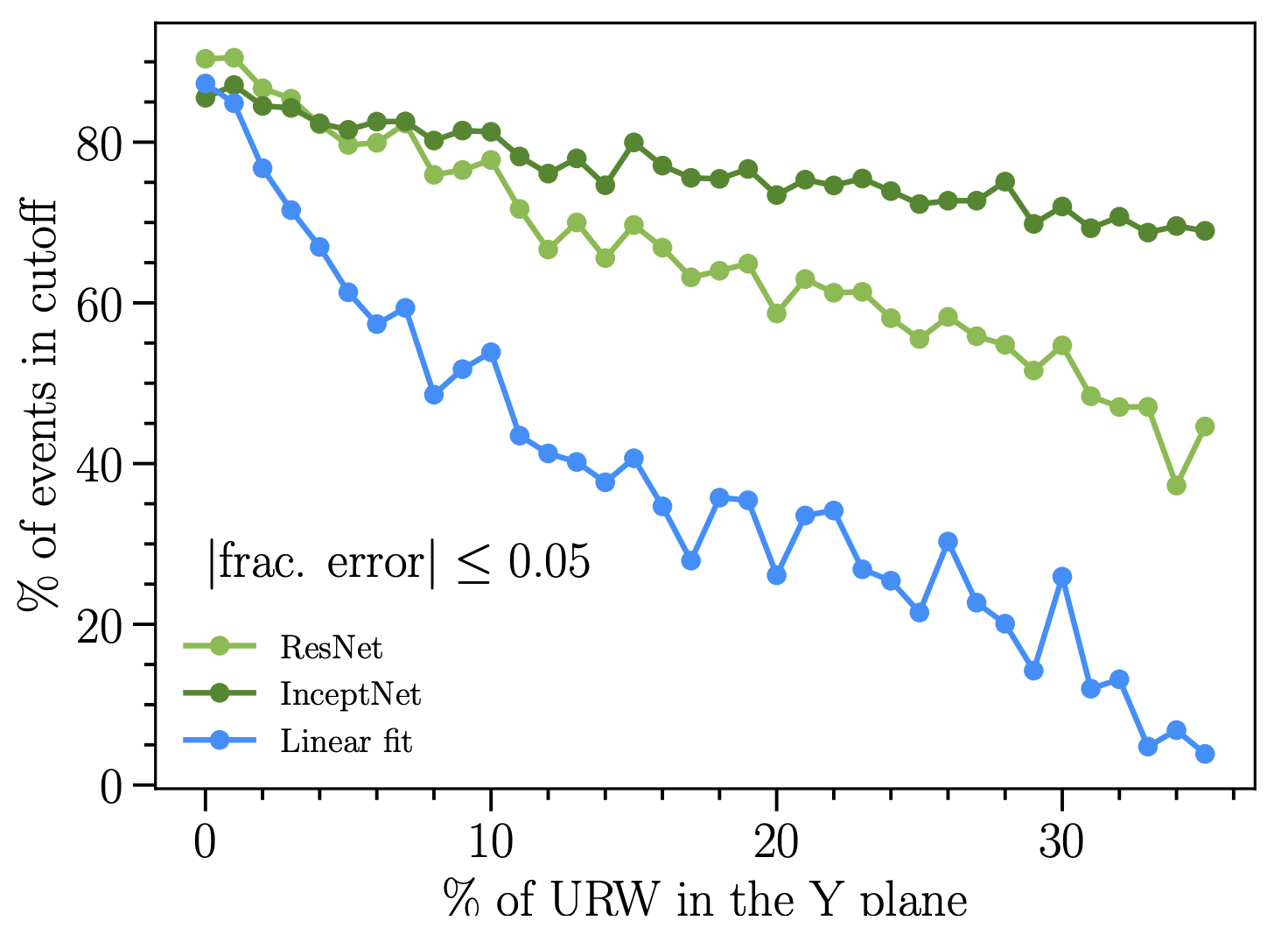}
    \caption{The fraction of EM showers reconstructed to within 5\% of the true deposited energy as a function of the fraction of unresponsive wires in the LArTPC collection plane, considering three different methods. The ResNet and Inception network significantly outperform the traditional linear calibration between charge and energy. Figure from Ref.~\cite{Carloni:2021zbc}.}
    \label{fig:cnn_urw}
\end{figure}

\subsection{Vertex and Track Reconstruction} \label{sec:vtx_track}

\begin{figure}[h!]
     \centering
     \begin{subfigure}[b]{0.45\textwidth}
         \centering
         \includegraphics[width=\textwidth]{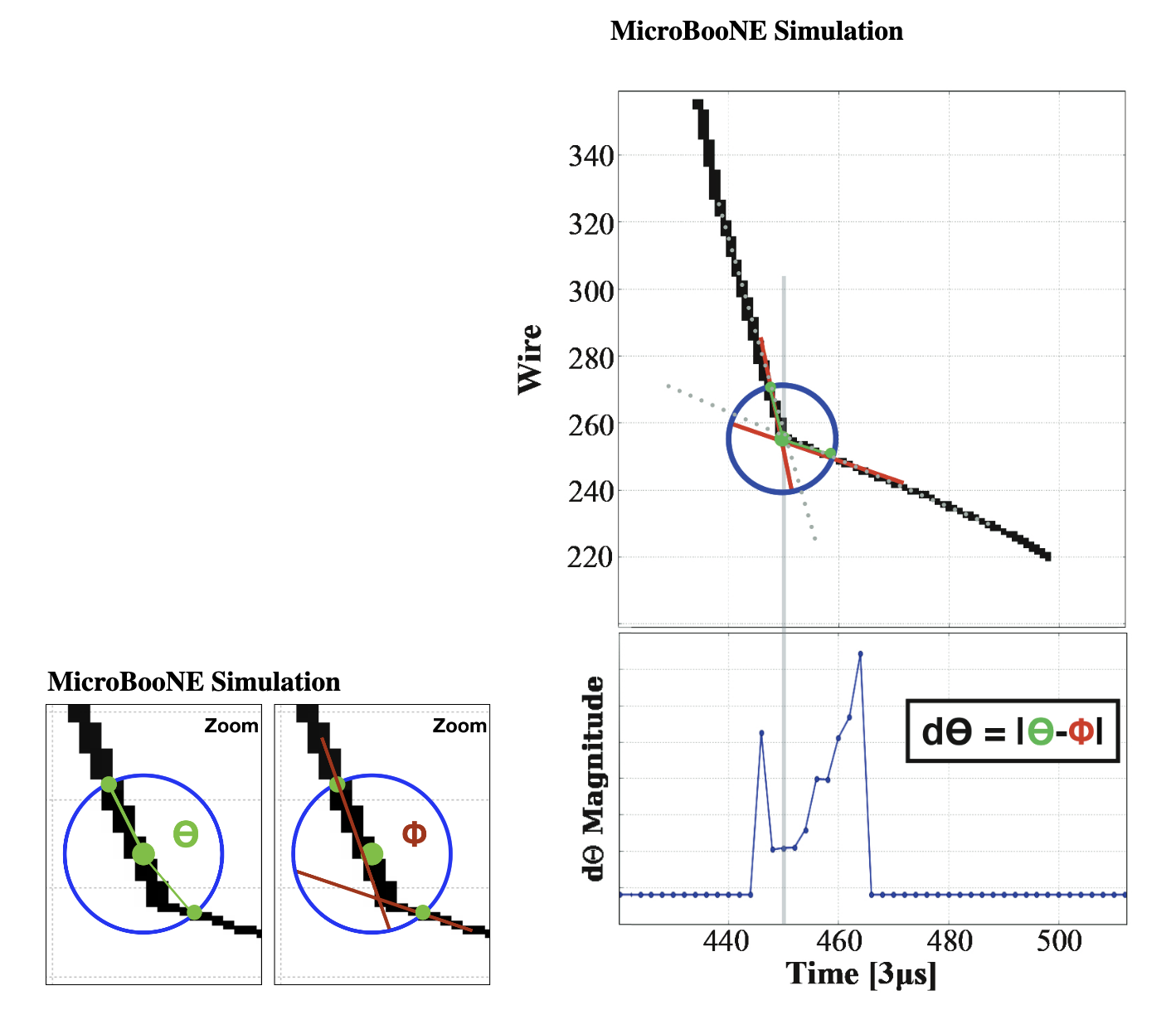}
         \caption{}
         \label{fig:vtx}
     \end{subfigure}
     \hfill
     \begin{subfigure}[b]{0.45\textwidth}
         \centering
         \includegraphics[width=\textwidth]{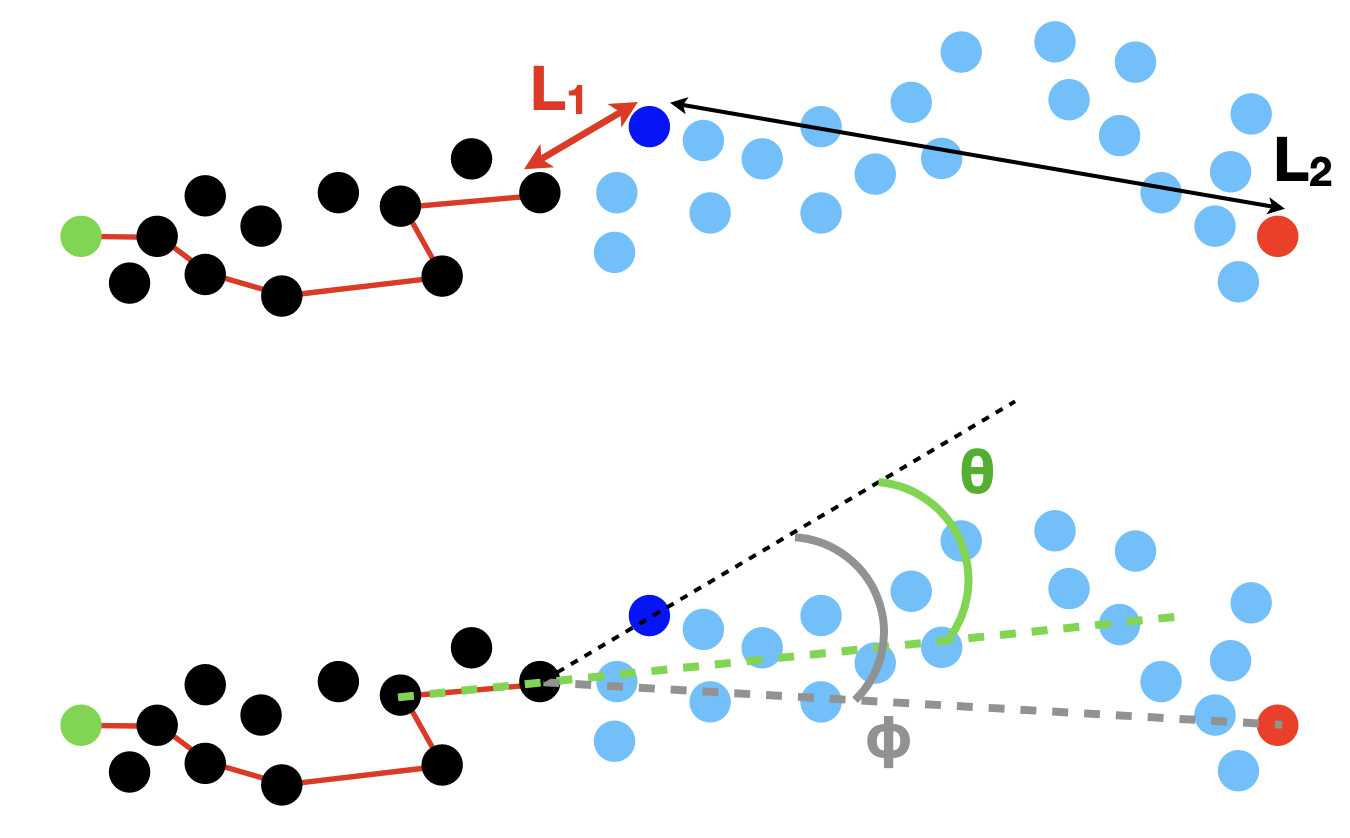}
         \caption{}
         \label{fig:track}
     \end{subfigure}
        \caption{\Cref{fig:vtx} shows the angular metric that is minimized to find a 3D neutrino vertex candidate. \Cref{fig:track} shows the iterative track reconstruction algorithm, which relies on calculating distances (L$_1$ and L$_2$) and angles ($\theta$ and $\phi$) with respect to the previous point and the end of the track. Figures from Ref.~\cite{MicroBooNE:2020sar}.}
        \label{fig:vtx_track}
\end{figure}

After running \texttt{SparseSSNet} and filtering cosmic-associated charge, the images are passed through the vertex reconstruction algorithm described in Ref.~\cite{MicroBooNE:2020sar}.
In brief, the algorithm looks for the characteristic ``vee'' shape of both $1e1p$ signal events, as shown in the left panel of \cref{fig:1e1p_example_EVD}, and $1\mu1p$ control sample events.
The algorithm begins by isolating a sample of 2D vertex seeds (i.e., candidate neutrino interaction locations) on each plane.
This is done by clustering regions of charge and then applying two techniques--contour defect identification using a convex hull and intersection identification using principal component analysis.
The set of vertex seeds is reduced to those that are consistent in the time dimension across all three planes.
From each triplet of time-consistent vertex seeds on the three wire planes, a final 3D vertex is found by minimizing an angular metric that represents the likelihood that two clusters of charge emanate radially from that vertex location.
\Cref{fig:vtx} shows an example minimization of this angular metric, which involves calculating two different estimates of the opening angle between the two prongs for a given vertex location.
Simulation studies indicate that this algorithm is able to reconstruct neutrino interaction vertices to within less than 1~cm with an average efficiency of 56\%~\cite{MicroBooNE:2020sar}.

After finding a neutrino vertex candidate, we perform a 3D reconstruction of the prongs of charge emitted from the vertex~\cite{MicroBooNE:2020sar}.
An iterative stochastic algorithm identifies a set of 3D spatial points associated with each prong by building upon previous points; these points comprise the track.
This process is shown diagrammatically in \cref{fig:track}.
It is worth noting that the ``tracks'' at this stage can correspond to either true tracks (HIPs and MIPs) or EM showers.

The 3D track can be projected onto each 2D plane.
At this point, the \texttt{SparseSSNet} labels on each plane are used to determine if the event is a ``track-track'' or ``track-shower'' event.
The latter case happens when greater than 20\% of the pixels in a given prong are labeled as shower pixels.
In the ``track-shower'' case, the shower-like prong is considered to be the electron while the track-like prong is the proton.
Thus, these are candidate $1e1p$ events.
The ``track-track'' case corresponds to candidate $1\mu1p$ event, where the shorter track is ascribed to the highly-ionizing proton.
Electron EM showers are not sufficiently reconstructed in this step and are left for a dedicated shower reconstruction algorithm described in \cref{sec:shower_publication}.
Muons and protons can, however, be reconstructed at this stage.
The polar and azimuth angles of each particle can be determined from the direction of the associated track. 
The length of the tracks is used in conjunction with the known stopping power of each particle in liquid argon to determine the kinetic energy.
Simulation studies of $1\mu1p$ events suggest an angular resolution of a few degrees and a kinetic energy resolution of a few percent~\cite{MicroBooNE:2020sar}.

It is also worth mentioning that two-prong events with an opening angle of less than 10$^\circ$ in any plane are rejected at this point.
Thus, signal events in this analysis consisted of well-separated leptons and protons.

\subsection{Publication: \textit{Electromagnetic shower reconstruction and energy validation with Michel electrons and $\pi^0$ samples for the deep-learning-based analyses in MicroBooNE}} \label{sec:shower_publication}

As mentioned in the previous section, a dedicated algorithm is used to reconstruct the EM showers coming from the neutrino interaction vertex.
The algorithm is described in detail in Ref.~\cite{MicroBooNE:2021nss}, which also covers a data-driven validation of the shower reconstruction using dedicated $\pi^0$ and Michel electron samples.
A brief overview of the shower reconstruction is as follows.
We first mask out pixels in each plane for which the \texttt{SparseSSNet} shower score is below a value of 0.5.
Next, we attach a triangle to the interaction vertex in each plane and optimize the direction, gap from the vertex, opening angle, and length of the triangle to encapsulate as many shower pixels as possible without becoming too large.
We calculate the energy of the shower by performing a linear calibration between the summed PIU in the collection plane shower triangle and the simulated electron/photon energy.
The collection plane is used because it produces the most robust signals, as discussed in \cref{sec:ub_signal_processing}.
We reconstruct 3D showers by looking for overlap in the time dimension between the shower triangles on all three planes.
This algorithm can be run a second time after masking out the optimized triangle from the first pass, which is essential for isolating a $\pi^0$ sample.

This MicroBooNE publication was led by myself and Katie Mason, a graduate student from Tufts University.
Katie focused on the triangle-fitting procedure of the shower reconstruction and the $\pi^0$ mass peak fit, while I focused on the charge-to-energy scaling of the shower reconstruction and the Michel energy spectrum fit.
The full \textit{JINST} publication is included below.
The most important result from this study is shown in Figure~16 of the paper, which demonstrates good agreement between the simulation-derived charge-to-energy scaling applied to EM showers in this analysis and two standard candles: the $\pi^0$ invariant mass peak and the cutoff of the Michel electron energy distribution.

\includepdf[pages=-]{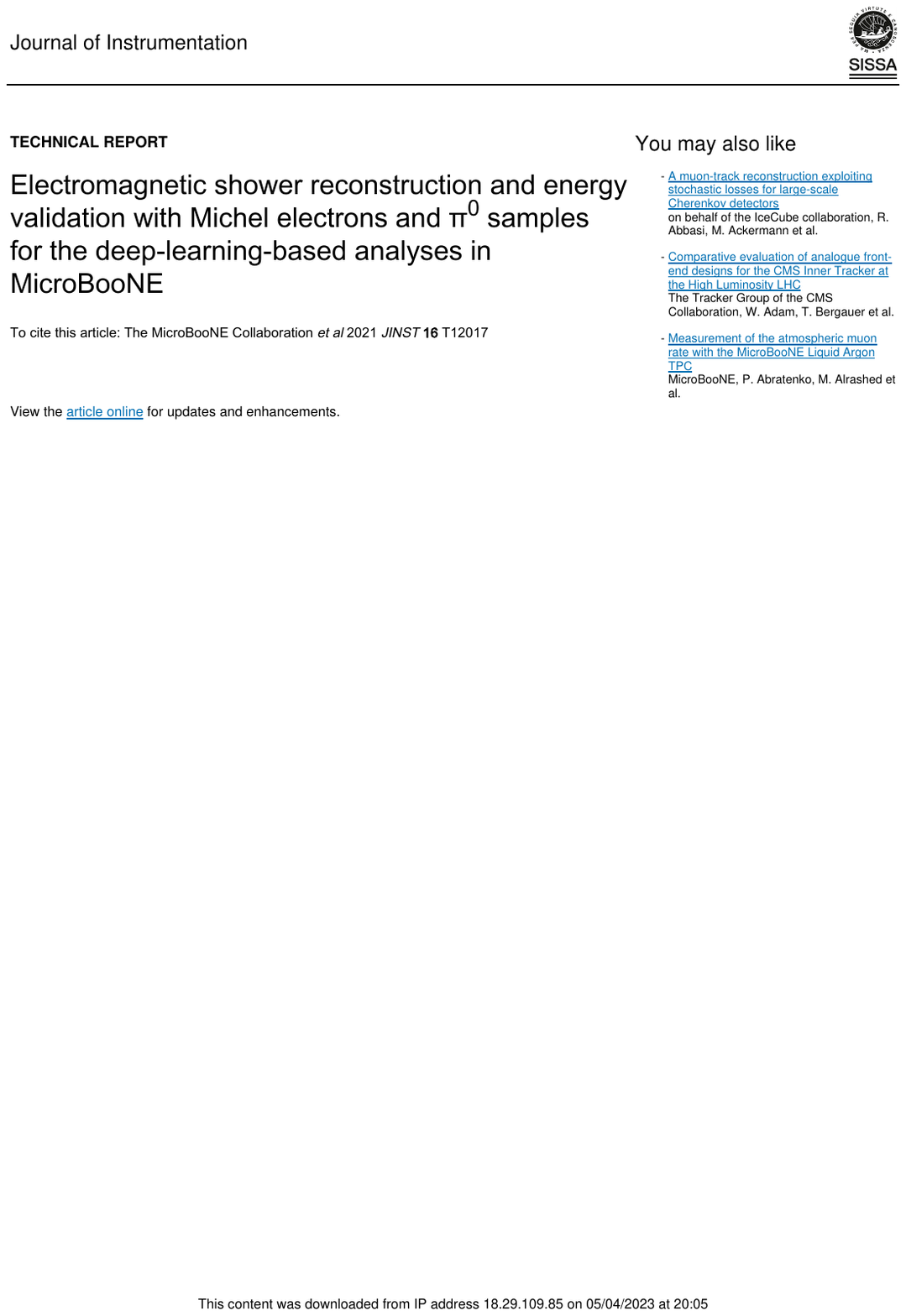}

\section{$1e1p$ Event Selection} \label{sec:1e1p_selection}

The signal events for this analysis, $1e1p$ events consistent with two-body CCQE scattering, are isolated using a series of cuts on kinematic, topological, and MPID-based variables as well as an ensemble of boosted decision trees (BDTs).
The $1e1p$ signal selection represents one of my main projects within the MicroBooNE experiment.
This section covers in detail the requirements which define the $1e1p$ sample.
The $1\mu1p$ control sample is isolated analogously to the $1e1p$ sample and will not be covered in this thesis, though more information can be found in Ref.~\cite{MicroBooNE:2021pvo}.

\subsection{Basic Data Selection Criteria}

We first apply a series of basic ``precuts'' meant to remove poorly reconstructed events and obvious backgrounds.
We begin with the collection of ``track-shower'' vertices selected by the algorithm described in \cref{sec:vtx_track}, which represent candidate $1e1p$ events.
We remove any vertices that are within 10~cm of the TPC edges or within a region of unresponsive wires between $700 < z~[{\rm cm}] < 740.$
This cut defines the fiducial volume.
We also require that the minimal distance of either the track or the shower to the edge of the TPC is greater than 15~cm.
In the case that both prongs are reconstructed close to an edge, we instead require that the distance of each is greater than 5~cm, which helps retain neutrino interactions that begin near the edge of the fiducial volume.
Events in the $1e1p$ are further required to be outside of an inefficient region of the U-plane.

\begin{figure}
    \centering
    \includegraphics[width=0.8\textwidth]{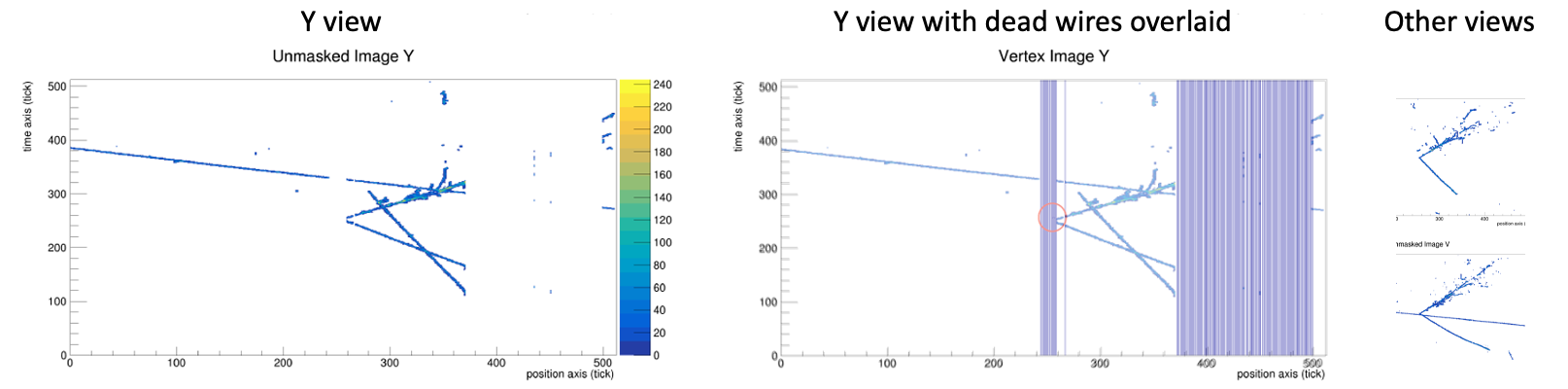}
    \caption{An example of an event that fails the shower energy consistency cut, because the EM shower passes through an unresponsive region of the collection plane.}
    \label{fig:econsist_img}
\end{figure}

\begin{figure}[h!]
    \centering
     \begin{subfigure}[b]{0.45\textwidth}
         \centering
         \includegraphics[width=\textwidth]{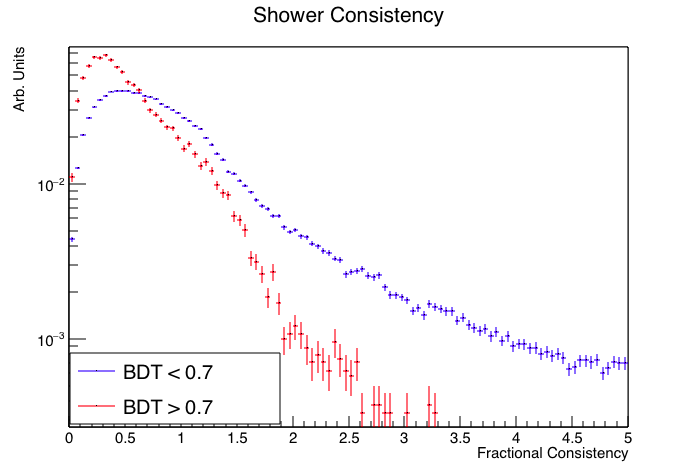}
         \caption{}
         \label{fig:econsist_dist}
     \end{subfigure}
     \hfill
     \begin{subfigure}[b]{0.45\textwidth}
         \centering
         \includegraphics[width=\textwidth]{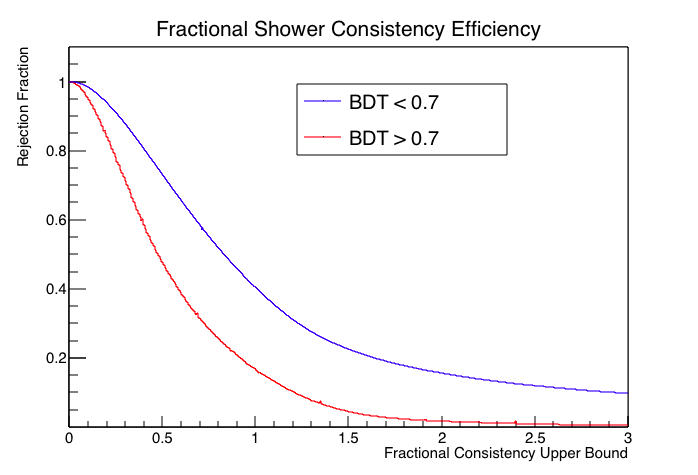}
         \caption{}
         \label{fig:econsist_eff}
     \end{subfigure}
        \caption{\Cref{fig:econsist_dist} shows the distribution of the fractional shower energy consistency variable for events with an old $1e1p$ BDT score above/below 0.7. \Cref{fig:econsist_eff} shows the efficiency with which events above/below the old $1e1p$ BDT cutoff pass the fractional consistency cut as a function of the chosen upper bound.}
        \label{fig:econsist}
\end{figure}

We also employ a cut requiring the reconstructed shower energy of the candidate electron to be consistent between the three planes.
Even though we rely on the collection plane for the official shower energy estimation, this cut is helpful in removing events that would not lead to a consistent 3D picture of the shower.
An example of such an event is shown in \cref{fig:econsist_img}.
The cut specifically employed in this analysis is
\begin{equation} \label{eq:consistency}
\frac{\sqrt{(E_e^U-E_e^V)^2 + (E_e^U-E_e^Y)^2 + (E_e^V-E_e^Y)^2}}{E_e^Y} < 2,
\end{equation}
where $E_e^{[U,V,Y]}$ denote the reconstructed electron candidate energy on the $U$, $V$, and $Y$ plane, respectively.
\Cref{fig:econsist_dist} shows the distribution of the fractional consistency variable (left-hand side of \cref{eq:consistency}) for events above and below a $1e1p$ BDT score cutoff of 0.7, where the BDT here is an older iteration of the final BDT ensemble used in this analysis.
As can be seen, signal-like events (with a higher BDT score) tend to have lower fractional consistency values compared to background-like events (with a lower BDT score).
The upper bound of 2 in \cref{eq:consistency} was chosen to retain $\sim95\%$ of signal-like events while rejecting $\sim20\%$ of background-like events, as shown in \cref{fig:econsist_eff}.

Next, we make cuts on the reconstructed neutrino energy ($200<E_\nu~[{\rm MeV}]<1200$), electron kinetic energy ($K_e>35~{\rm MeV}$), and proton kinetic energy ($K_p>50~{\rm MeV}$).
We also require a forward-going proton and an opening angle between the electron and proton greater than 0.5~rad.
Finally, we require that the event can be boosted to the rest frame of the nucleon using the $p_{fermi}^T$ and $p_{fermi}^z$ variables defined in \cref{tab:KinVarTable}.

\subsection{Boosted Decision Tree Ensemble}

The next stage of the $1e1p$ selection uses an ensemble of BDTs to identify signal-like events in MicroBooNE.
This is the most powerful cut used in the $1e1p$ analysis.
The ensemble consists of 20 BDTs each trained using the XGBoost gradient-boosting algorithm~\cite{chen2016xgboost}.
Each BDT takes as input a set of 19 kinematic variables, such as the reconstructed energies and angles of the electron and proton, and 4 topological variables related to the observed charge, such as the fraction of \texttt{SparseSSNet}-labeled shower pixels in the event.
The full suite of variables used to train the $1e1p$ and $1\mu1p$ BDTs is shown in \cref{tab:BDTCompare}.
Each BDT itself consists of a collection of regression trees which each assign a continuous score to an event based on a set of decision rules developed using the input variables; the assigned BDT score of the event is given by the sum over the collection~\cite{chen2016xgboost}.
BDTs are trained via gradient-based optimization of an objective function that includes a regularization term to penalize overfitting.

\begin{figure}[h!]
    \centering
     \begin{subfigure}[b]{0.3\textwidth}
         \centering
         \includegraphics[width=\textwidth]{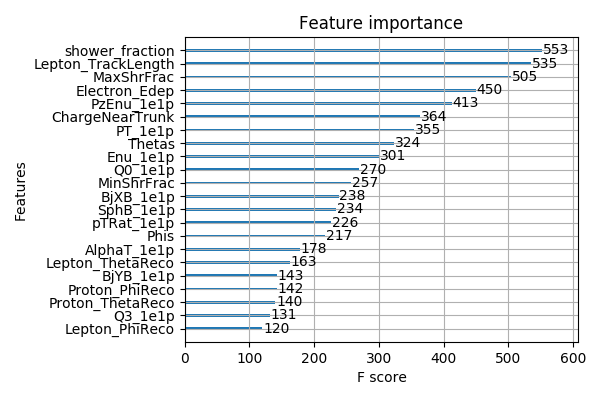}
         \caption{Run 1}
         \label{fig:fscore_r1}
     \end{subfigure}
     \hfill
     \begin{subfigure}[b]{0.3\textwidth}
         \centering
         \includegraphics[width=\textwidth]{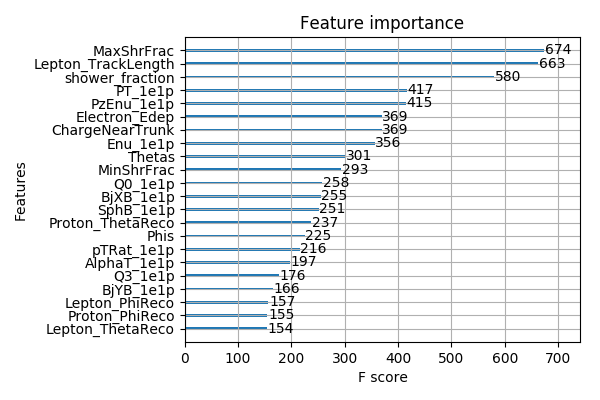}
         \caption{Run 2}
         \label{fig:fscore_r2}
     \end{subfigure}
     \hfill
     \begin{subfigure}[b]{0.3\textwidth}
         \centering
         \includegraphics[width=\textwidth]{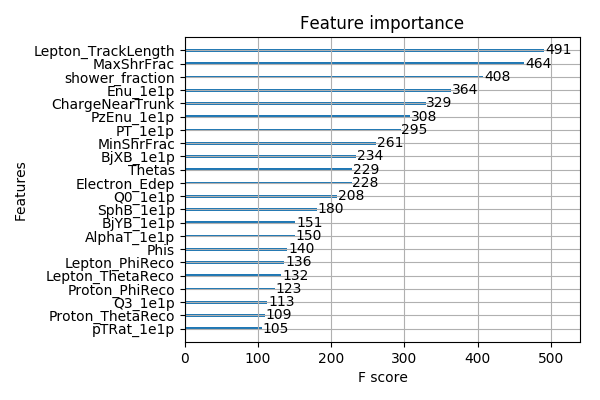}
         \caption{Run 3}
         \label{fig:fscore_r3}
     \end{subfigure}
     \hfill
        \caption{The F score of each variable for one of the $1e1p$ BDTs in the ensemble from run 1, run2, and run 3.}
        \label{fig:fscore}
\end{figure}

A different ensemble is generated for each of the three run periods, as the detector response changed slightly between them.
Thus, there are 60 total BDTs trained for the $1e1p$ analysis.
The BDTs are trained using the MicroBooNE MC samples, including the nominal ($\nu_\mu$-dominated) BNB sample as well as dedicated intrinsic $\nu_e$, $\pi^0$, and cosmic samples.
The signal for the training sample is defined as true $\nu_e$ CCQE interactions with one electron and one proton in the final state.
We further restrict to well-reconstructed events for which the vertex is reconstructed with 5~cm of the true vertex and the neutrino energy within 20\% of the true value.
Only non-$\nu_e$ CCQE $1e1p$ events are explicitly classified as background.
Each BDT in an ensemble is trained using a randomly-selected half of the MC sample for the given run period.
\Cref{fig:fscore} shows the F score distribution for each variable in one of the $1e1p$ BDTs from the ensemble for each run period.
The F score represents the number of times each variable is used across all trees in the BDT; thus, it can be thought of as a measure of the importance of each variable.

\begin{figure}[h!]
    \centering
     \begin{subfigure}[b]{0.49\textwidth}
         \centering
         \includegraphics[width=\textwidth]{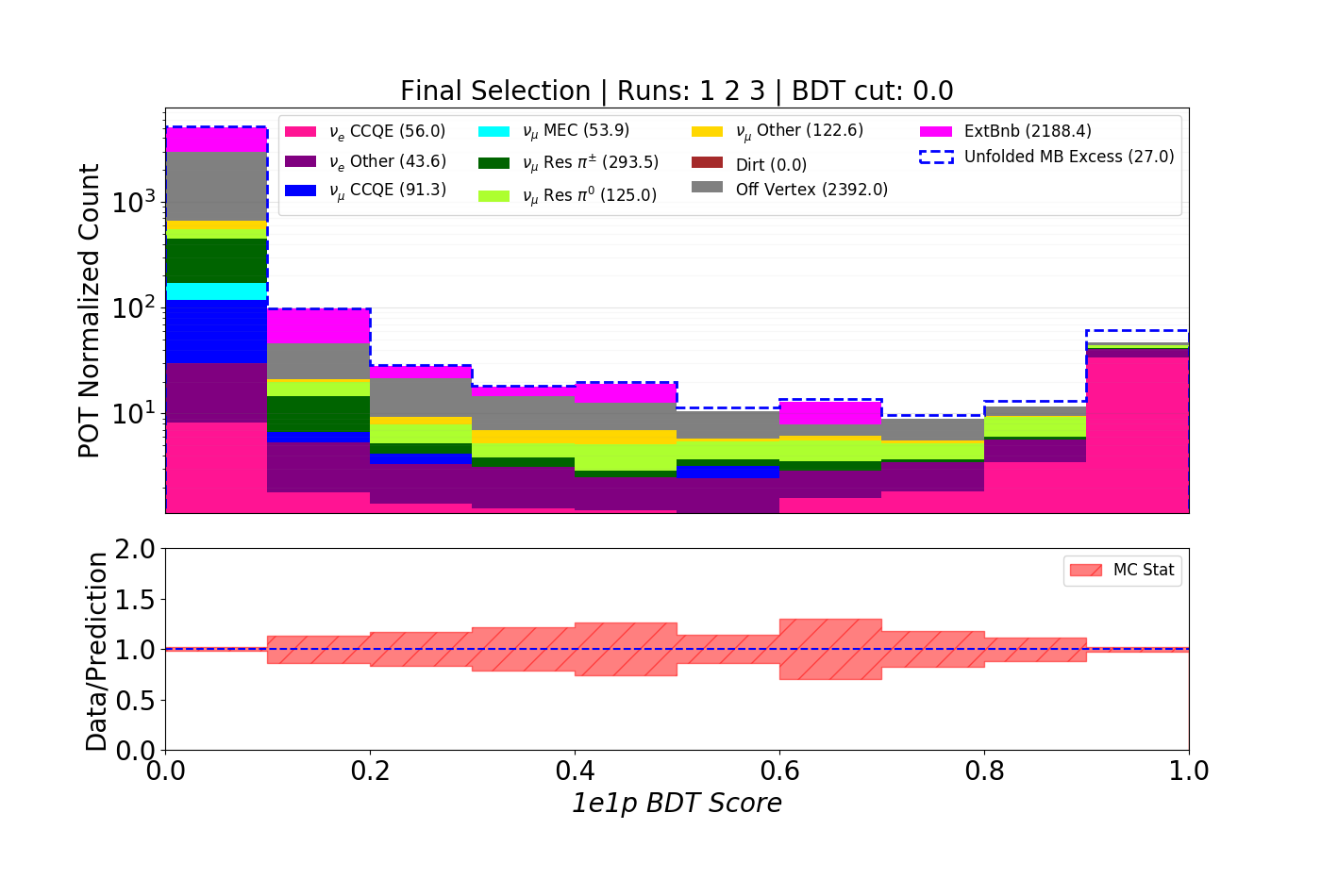}
         \caption{}
         \label{fig:BDTdist_full}
     \end{subfigure}
     \hfill
     \begin{subfigure}[b]{0.49\textwidth}
         \centering
         \includegraphics[width=\textwidth]{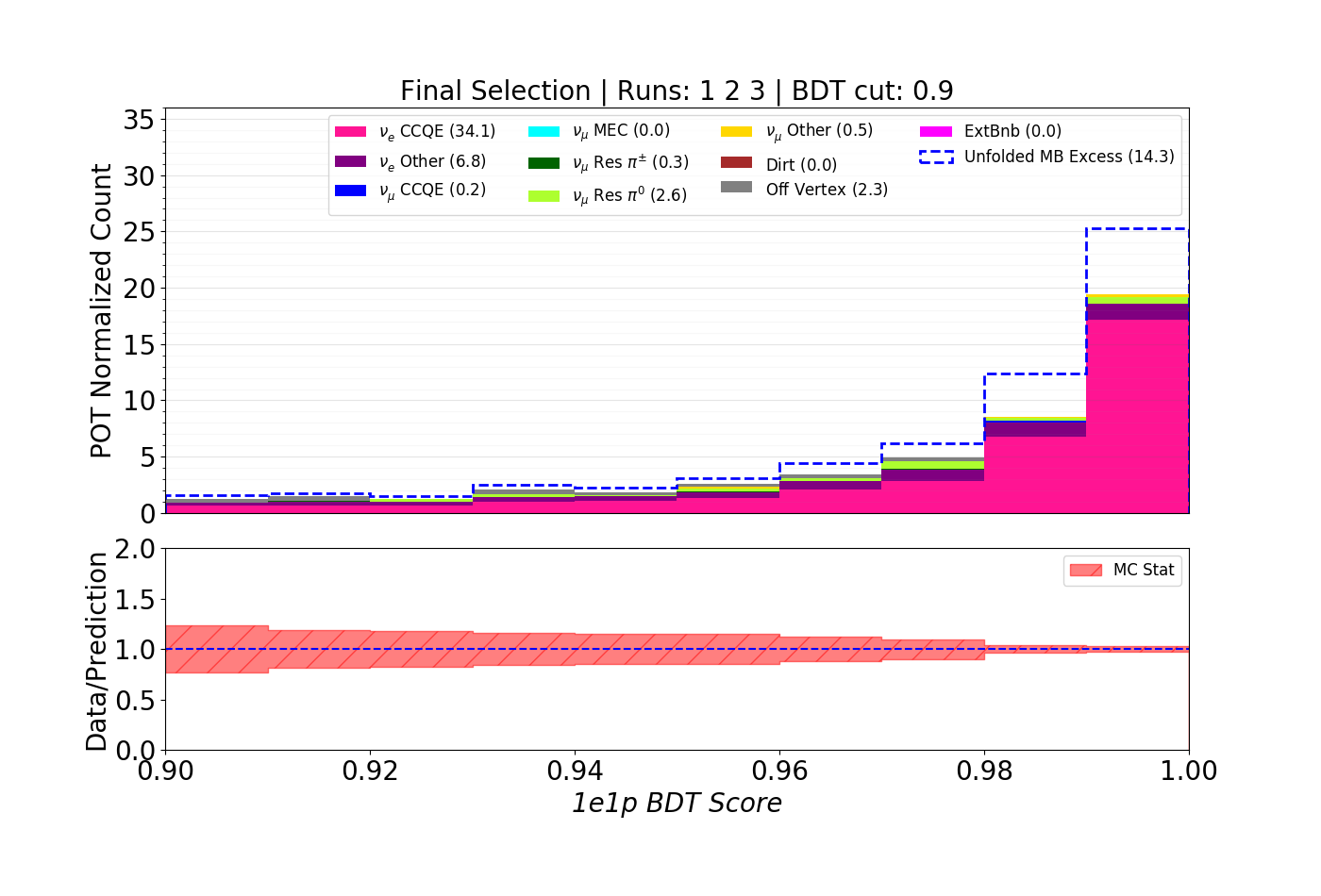}
         \caption{}
         \label{fig:BDTdist_signal}
     \end{subfigure}
     \hfill
        \caption{\Cref{fig:BDTdist_full} and \cref{fig:BDTdist_signal} show the MC distribution of the $1e1p$ BDT ensemble average score over all three run periods for the full [0,1] range and zoomed in to the [0.95,1] range, respectively.}
        \label{fig:BDTdist}
\end{figure}

The $1e1p$ score for an event is evaluated by calculating the average BDT score across the ensemble.
This reduces the dependence of the ensemble score on the specific training sample of any single BDT.
Ensemble-based methods have been shown to reduce variance on the output likelihood variable compared to a single classifier~\cite{10.1145/2009916.2009932}.
When computing the average score for events in the MicroBooNE MC sample, we omit BDTs for which the event appeared in that BDT's training sample.
It is extremely unlikely ($p = 2^{-20}$) for an event to appear in the training sample of all BDTs in an ensemble; thus, we can use the BDT ensemble to isolate a $1e1p$ signal sample prediction from the MicroBooNE MC without throwing away any events.
The score is normalized to the range [0,1], where higher scores indicate a higher signal likelihood.
We show the distribution of the ensemble-averaged $1e1p$ BDT score in \cref{fig:BDTdist}, which shows that true $\nu_e$ CCQE events tend to peak toward 1 while all other events peak toward 0.
Signal events are required to have an average score greater than 0.95, which was chosen to optimize sensitivity to the eLEE model described in \cref{sec:lee_template}.

We have performed a series of tests of the robustness of the BDT ensemble method.
We first examined the signal selection power of a BDT ensemble trained on simulated events from a given run period when used to infer the signal likelihood of simulated events from a different run period. 
This is useful for two reasons. 
For one, it provides a method for testing the ensemble on events that were in neither the training nor the validation sample of the constituent BDTs. 
Testing on a sample independent of the validation sample is important because the BDT training was halted when the classification error on the validation set didn't improve after 50 training iterations. 
It also tests the impact of removing training events from the MC prediction.
When using the BDT ensemble of a given run period to evaluate the signal likelihood of events from the same run period, one must remove events that appeared in the training sample of each constituent BDT.
In this study we use the BDT ensemble trained on run period 2 to evaluate simulated $\nu_e$ events from run period 3 and vice versa, so one does not need to remove training events.

\begin{figure}[h!]
    \centering
     \begin{subfigure}[b]{0.45\textwidth}
         \centering
         \includegraphics[width=\textwidth]{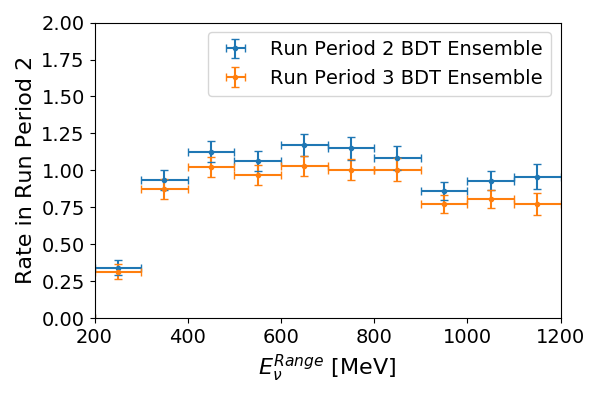}
         \caption{}
         \label{fig:BDTswap_r2}
     \end{subfigure}
     \hfill
     \begin{subfigure}[b]{0.45\textwidth}
         \centering
         \includegraphics[width=\textwidth]{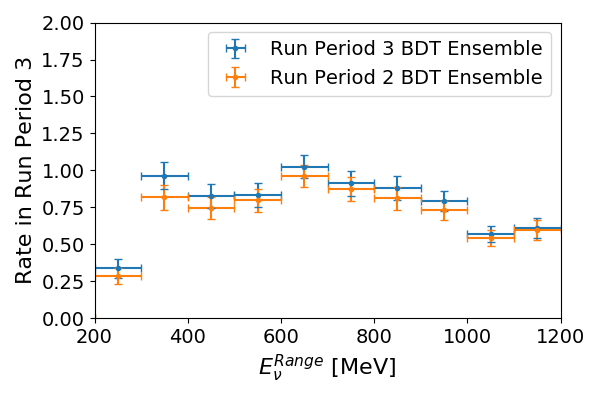}
         \caption{}
         \label{fig:BDTswap_r3}
     \end{subfigure}
     \hfill
        \caption{\Cref{fig:BDTswap_r2} and \cref{fig:BDTswap_r3} show the predicted $\nu_e$ event rate in run period 2 and run period 3, respectively, using both the run period 2 ensemble and the run period 3 ensemble.}
        \label{fig:BDTswap}
\end{figure}

\Cref{fig:BDTswap} shows the results of this run period swap study.
As one can see, the differences between the selection using the correct and incorrect run period ensemble are small. 
The correct run period performs slightly better, as expected due to small differences in the detector status specific to each run period. 
However, the relatively small change in $\nu_e$ selection efficiency suggests that the BDT ensemble is able to perform well on simulated events that do not appear in either the training or validation sample of constituent BDTs. 
One can interpret the loss in performance shown in \cref{fig:BDTswap} as a bound on the potential impact of applying the BDT ensemble to events that are not in the training or validation samples of the entire ensemble.
The drop in performance is $< 10\%$ over most of the energy range--an effect which is smaller than our systematic error before the $\nu_\mu$ constraint described in \cref{ch:microboone_results} ($\sim 15\%$) and much smaller than the statistical error in each reconstructed neutrino energy bin after the final selection ($\sim$ 50\%).

Next, we investigated the impact of using only a subset of the BDTs in the ensemble to calculate the average BDT score of an event.
This is relevant because we omit BDTs that contained a given MC event in the training sample when calculating the average BDT score for that event.
However, this analysis uses all 20 BDTs in the ensemble to calculate the average BDT score of events in the data.
Therefore, to verify that comparisons between data and simulation are robust, one needs to ensure that removing BDTs from the ensemble does not significantly bias the average score calculation. 

\begin{figure}[h!]
    \centering
     \begin{subfigure}[b]{0.45\textwidth}
         \centering
         \includegraphics[width=\textwidth]{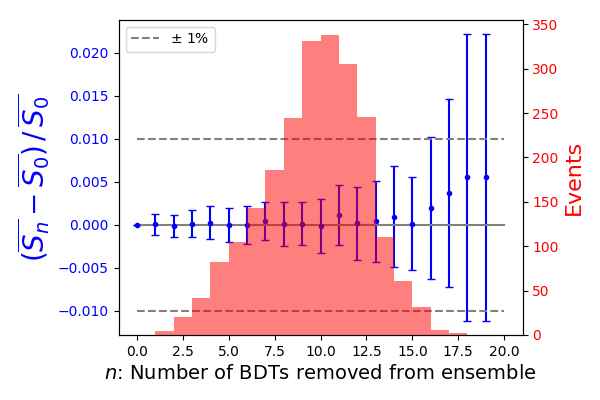}
         \caption{}
         \label{fig:BDTdrop_r2}
     \end{subfigure}
     \hfill
     \begin{subfigure}[b]{0.45\textwidth}
         \centering
         \includegraphics[width=\textwidth]{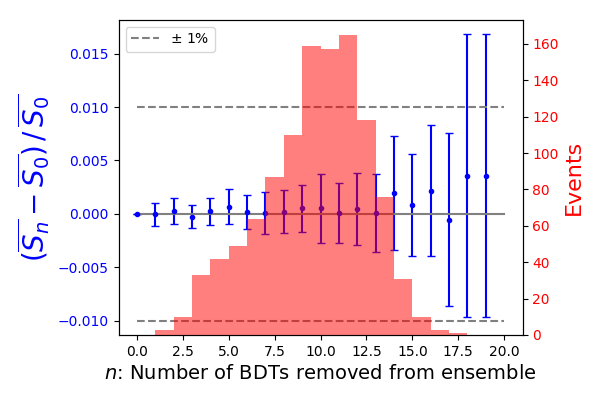}
         \caption{}
         \label{fig:BDTdrop_r3}
     \end{subfigure}
     \hfill
        \caption{\Cref{fig:BDTdrop_r2} and \cref{fig:BDTdrop_r3} show the fractional difference in average BDT score $(S_n - S_0)/S_0$ as a function of the number of omitted BDTs $n$ over the simulation from run period 2 and run period 3, respectively. The red histogram shows the actual distribution of the number of omitted BDTs over the run period 2 and run period 3 simulation samples, respectively. Scores are calculated using the run period 3 and run period 2 BDT ensemble, respectively.}
        \label{fig:BDTdrop}
\end{figure}

To this end, consider $S_n$ to be the $1e1p$ BDT ensemble average score after removing $n$ BDTs from the ensemble.
For events in data, the BDT score is $S_0$ while for simulated events, the BDT score is $S_n$ for some $n \in \{1,...,20\}$.
\Cref{fig:BDTdrop} shows the fractional difference $(S_n - S_0)/S_0$ as a function of the number of omitted BDTs $n$ over signal-like (BDT score $>$ 0.95) events in the simulation from run period 2 and run period 3, respectively.
As in the previous study, in order to avoid bias from BDT training we use the BDT ensemble trained on run period 2 to evaluate the signal likelihood of simulation events from run period 3 and vice versa.
The data points and error bars in \cref{fig:BDTdrop} indicate the average and standard deviation of the fractional difference over the simulation sample, respectively.
The red histograms show the actual distribution of omitted BDTs over the simulation from each run period.
One can see that the average BDT score does not exhibit significant bias upon removing BDTs from the calculation.
Also, the standard deviation of the fractional difference only becomes larger than $1$\% when removing $\gtrsim 17$ BDTs from the calculation.
This only happens for $<< 1$\% of simulated events, as expected.

\begin{table}[ht!]
{
\begin{center}
\begin{tabular}{|l|c|c|} \hline \hline \hline 

  Variable  &  Used in $1\mu 1p$ BDT &  Used in $1e 1p$ BDT\\
 \hline 
 \multicolumn{3}{c}{Variables Used in BDTs, Based on Ionization} \\ \hline
 Charge within 5\,cm of vertex & Yes & Yes \\
 Shower charge in event image / & & \\
 shower charge clustered as electron & No & Yes \\
 Proton shower fraction & No & Yes  \\
 Electron shower fraction & No & Yes  \\ \hline
  \multicolumn{3}{c}{Variables Used in BDTs, Related to Energy Measurements} \\ \hline
 Neutrino Energy  & Yes & Yes \\ 
Energy of electromagnetic shower & No & Yes \\
Lepton length & Yes & Yes  \\ 
Proton length & No & Yes \\
 $p_z - E_\nu$ & No & Yes \\ \hline
  \multicolumn{3}{c}{Variables Used in BDTs, Related to 2-Body Scattering Consistency} \\ \hline
Bjorken's $x$ & Yes * & Yes * \\
Bjorken's $y$ & Yes * & Yes * \\
QE Consistency & Yes * & Yes *  \\
 $Q_0$ & Yes & Yes \\
 $Q_3$ & Yes & Yes \\  \hline
  \multicolumn{3}{c}{Variables Used in BDTs, Related to Transverse Momentum} \\ \hline
$\alpha_T$  & Yes & Yes  \\
Event $p_T$  & Yes & Yes  \\
Event  $p_T/p$ (``PTrat") & Yes & Yes  \\
$\phi_T$  & Yes & No \\ \hline
  \multicolumn{3}{c}{Variables Used in BDTs, Related to Angles} \\ \hline
Proton $\phi$  & Yes & Yes \\
Proton $\theta$ & Yes & Yes \\
Lepton $\phi$ & Yes & Yes \\
Lepton $\theta$ & Yes & Yes \\
 $\phi_p - \phi_\ell$  & Yes & Yes\\
 $\theta_p + \theta_e$ & No & Yes \\ \hline 
   \multicolumn{3}{c}{Variables Useful for Comparison, Not Used in Either BDT} \\
   \hline
$\eta$ (Normalized average ionization difference)& No & No \\
Opening Angle & No & No \\
$x$ Vertex & No & No \\
$y$ Vertex & No & No \\
$z$ Vertex & No & No \\  
  \hline
\end{tabular}
\end{center}}
\caption{The suite of variables used to isolate and analyze the $1e1p$ and $1\mu1p$ samples. Variables used in the BDT ensemble for each sample are specified. The ``$*$'' character indicates that the variable is calculated in the rest frame of the struck nucleon.  The mathematical definitions of many of these variables appear in Table~\ref{tab:KinVarTable}.
\label{tab:BDTCompare}}
\end{table}

\subsection{Particle Identification Cuts}

\begin{figure}[h!]
    \centering
     \begin{subfigure}[b]{0.45\textwidth}
         \centering
         \includegraphics[width=\textwidth]{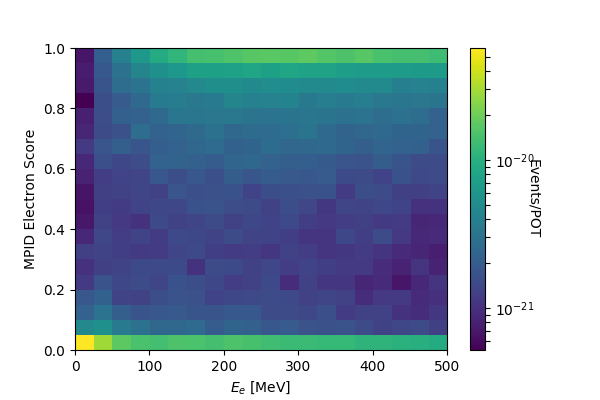}
         \caption{}
         \label{fig:MPIDe_vs_E}
     \end{subfigure}
     \hfill
     \begin{subfigure}[b]{0.45\textwidth}
         \centering
         \includegraphics[width=\textwidth]{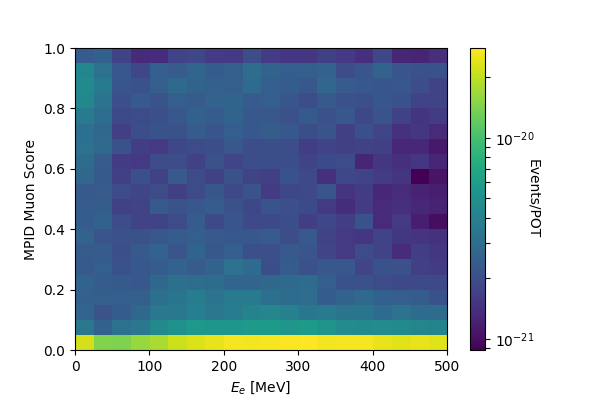}
         \caption{}
         \label{fig:MPIDm_vs_E}
     \end{subfigure}
     \hfill
        \caption{\Cref{fig:MPIDe_vs_E} and \cref{fig:MPIDm_vs_E} show the MPID electron and muon score, respectively, as a function of the reconstructed electron energy in intrinsic $\nu_e$ MC events.}
        \label{fig:MPID_vs_E}
\end{figure}

Finally, we employ a series of particle-identification requirements to clean up the remaining backgrounds that survive the $1e1p$ BDT ensemble cut.
The first of these is a cut on the invariant $\pi^0$ mass.
If the shower reconstruction algorithm described in \cref{sec:shower_publication} is able to identify a second EM shower, the reconstructed $M_{\pi^0}$ must be less than 50~MeV.
Another cut requires the ratio of the MPID $\gamma$ and $e^-$ image scores to be less than 2.
This helps remove remaining $\pi^0$ events that survive the $M_{\pi^0}$ cut.
We also require the MPID muon interaction score to be less than 0.2, which helps remove $\nu_\mu$ CC$\pi^0$ events where the muon gets mistaken for a proton.
This last cut is only applied for events in which $E_{e^-} > 100$~MeV, as MPID tends to assign low electron scores and high muon scores to true electrons below 100~MeV, as shown in \cref{fig:MPID_vs_E}.
This is because electrons stop radiating below this energy, becoming more track-like and thus difficult to distinguish from muons.
The specific values used for the cuts described here were chosen by optimizing the signal-to-background ratio in the $1e1p$ sample.
Finally, if an event has more than one reconstructed neutrino vertex candidate passing all selection criteria, we keep only the vertex with the highest $1e1p$ ensemble BDT score. 

\subsection{The Final $1e1p$ Sample}

\begin{figure}[h!]
    \centering
    \includegraphics[width=0.6\textwidth]{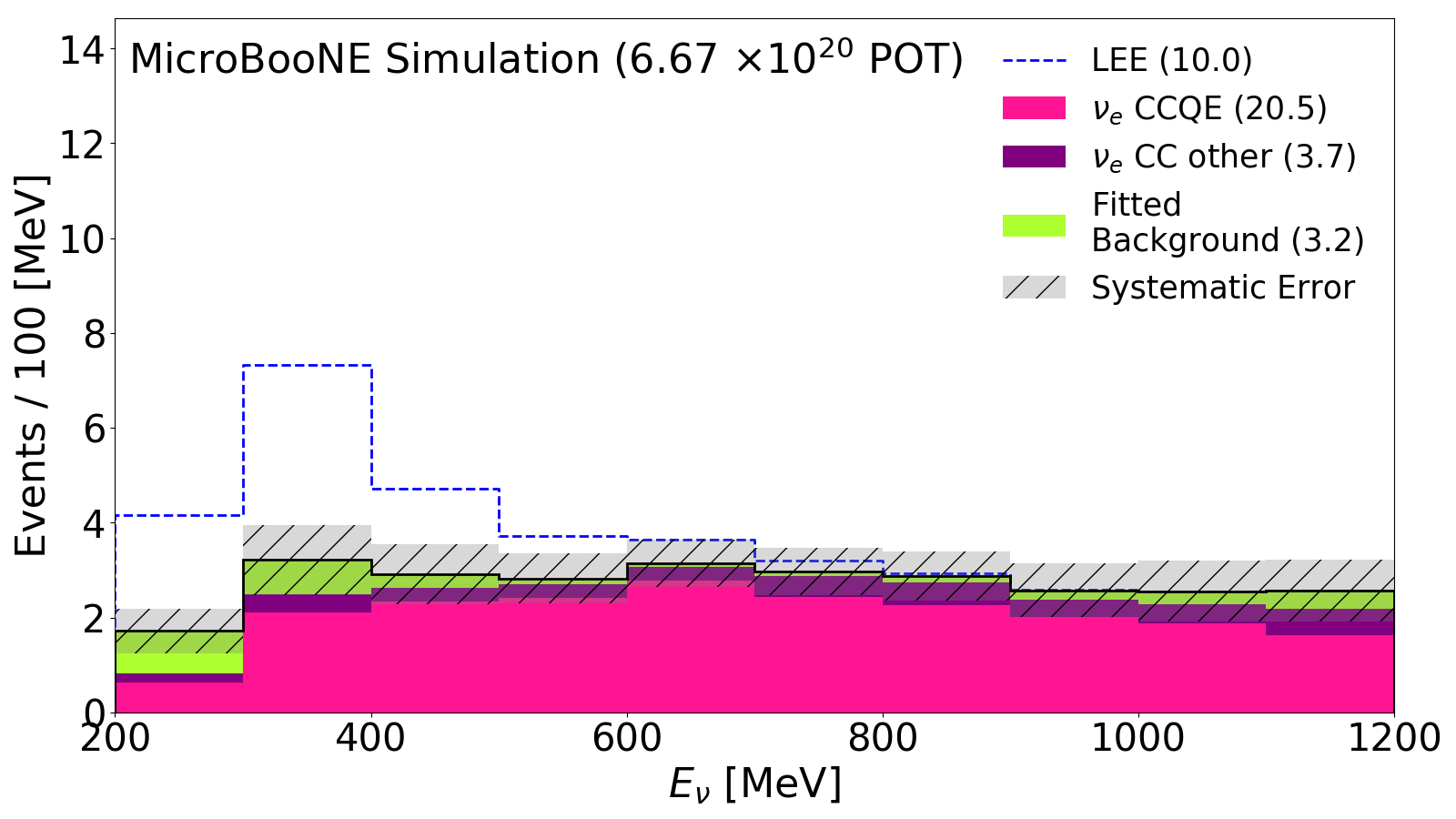}
    \caption{The $E_\nu^{\rm range}$ distribution for the $1e1p$ signal sample, showing only the predicted event rate from the MC. The prediction from the eLEE model is shown in the dashed blue line.}
    \label{fig:1e1p_enu_MC}
\end{figure}

The requirements described throughout this section define the $1e1p$ sample; they are summarized and compared with those in the $1\mu1p$ sample in \cref{tab:allcuts}.
This results in a highly-pure $1e1p$ signal sample--75\% of events are true $\nu_e$ CCQE interactions.
\Cref{fig:1e1p_enu_MC} shows the predicted $E_\nu^{\rm range}$ distribution for this sample, including the prediction from the eLEE model.
This figure also shows the systematic uncertainty and non-$\nu_e$ background prediction, which will be discussed further in \cref{ch:microboone_results}.
The neutrino energy distribution in \cref{fig:1e1p_enu_MC} forms the basis of the statistical results discussed in \cref{ch:microboone_results}.

\begin{figure}[h!]
    \centering
     \begin{subfigure}[b]{0.45\textwidth}
         \centering
         \includegraphics[width=\textwidth]{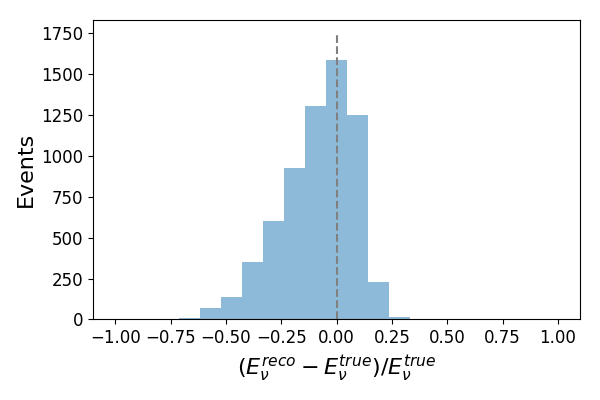}
         \caption{}
         \label{fig:NuEnergyFracError_Analysis}
     \end{subfigure}
     \hfill
     \begin{subfigure}[b]{0.45\textwidth}
         \centering
         \includegraphics[width=\textwidth]{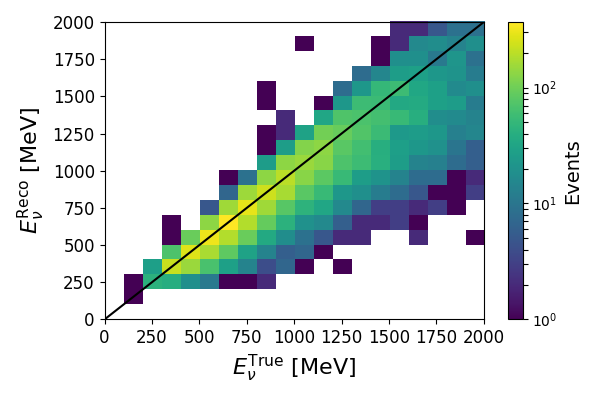}
         \caption{}
         \label{fig:NuEnergy2D_Full}
     \end{subfigure}
     \hfill
        \caption{\Cref{fig:NuEnergyFracError_Analysis} shows the distribution of the fractional error on the neutrino energy for MC events in the $1e1p$ signal sample, restricted to $200 < E_\nu^{\rm Range}\;[{\rm MeV}] < 1200$. \Cref{fig:NuEnergy2D_Full} shows the 2D distribution of fractional error as a function of the true neutrino energy.}
        \label{fig:NuEnergy}
\end{figure}

\begin{figure}[h!]
    \centering
     \begin{subfigure}[b]{0.45\textwidth}
         \centering
         \includegraphics[width=\textwidth]{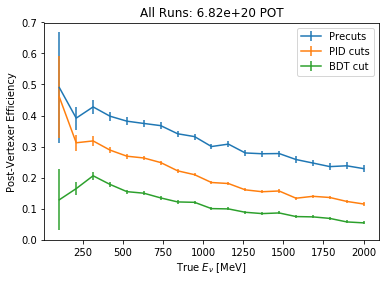}
         \caption{}
         \label{fig:EnsembleEfficiency}
     \end{subfigure}
     \hfill
     \begin{subfigure}[b]{0.45\textwidth}
         \centering
         \includegraphics[width=\textwidth]{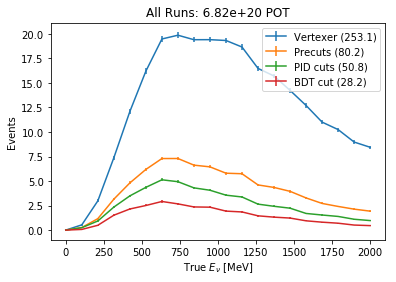}
         \caption{}
         \label{fig:EnsembleRetention}
     \end{subfigure}
     \hfill
        \caption{\Cref{fig:EnsembleEfficiency} shows the post-vertex-identification efficiency of true $\nu_e$ CCQE selection for subsequent stages of the $1e1p$ cuts. \Cref{fig:EnsembleRetention} shows the true $\nu_e$ CCQE event rates over the full run 1-3 dataset after subsequent stages of the $1e1p$ cuts.}
        \label{fig:ensemble_eff}
\end{figure}

The neutrino energy resolution is 16.5\% for selected $1e1p$ events, as shown in \cref{fig:NuEnergy}.
This figure also indicates a slightly higher rate of events with under-predicted neutrino energy, owing mainly to showers and tracks that pass through unresponsive regions of the detector. 
\Cref{fig:ensemble_eff} reports the signal efficiency and total number of $\nu_e$ CCQE events in this sample after applying each of the three sets of cuts.
The efficiency peaks at lower neutrino energies $\sim 300$~MeV, which is optimal for testing the eLEE model.
Note that these efficiencies in \cref{fig:EnsembleEfficiency} are reported with respect to all identified vertices.
If we include the effect of the vertex algorithm, the average efficiency for retaining signal events in the $1e1p$ sample is 6.6\%~\cite{MicroBooNE:2021pvo}.
This is by design--in choosing a very exclusive signal definition, we optimize for a high-purity sample at the price of lower efficiency.

\begin{sidewaystable}[p]
{\footnotesize{
\begin{center}
\begin{tabular}{|l|c|c|} \hline \hline \hline 

 Name/Description  of  Code  &  Requirement for $1\mu 1p$ &  Requirement for $1e 1p$\\
 \hline 
\multicolumn{3}{|c|}{ \bf {Cuts/Selections Applied in External Code}}\\  \hline 
Common Optical Filter & $>20$ PMT hits w/i 6 ticks in beam window &  Same \\ 
& \& $\le 20$ hits w/i 6 ticks in 2 $\mu s$ prior to spill& \\ \hline
WC tagger & Mask cosmic-associated charge & Same \\ \hline \hline 

\multicolumn{3}{|c|}{\bf {Preselection Cuts for Event Selection}}\\  \hline
Run Quality Requirements: & & \\
Good Run Flag &  True & Same   \\ \hline
Vertex Requirements: & &    \\
$1 \ell 1p$ consistent & Two prongs of at least 5cm length, no additional prongs & Same  \\
Edge fiducial &  $>10$\,cm from active volume edge & Same  \\
No dead region & outside $z=700$ to 740\,cm & Same   \\  \hline
Containment Requirements: & &  \\
Edge containment & closest approach $>15$\,cm from edge & Same  \\
Efficient region &  N/A & outside $y = (1/\sqrt{3})*z - 117$ \\
& & to $y = 1/\sqrt{3}*  z - 80$~cm. \\ \hline
Particle Energy Requirements: & &  \\
Neutrino energy (range-based) & $E_\nu<1200$ MeV & Same \\ 
Lepton kinetic energy & $K_\ell>35$ MeV & Same \\
Proton kinetic energy & $K_p>50$ MeV & Same \\ \hline
Analysis Orthogonality Cut & Max Shower Frac $<0.2$ & Max Shower Frac $> 0.2$ \\ \hline
Other Basic Quality Requirements: & &  \\
Opening Angle & $> 0.5$ radians & Same  \\
Proton $\theta$ & $\cos(\theta_{p}) > 0$ & Same \\
Boostable & $\gamma > 1$, $0<|\beta|<1$ & Same   \\  
Shower 3-view energy consistency & N/A  & Inconsistency $<$ 200\%  \\
Vertex with BDT score that is & lowest ({\it i.e.} best) &  highest ({\it i.e.} best)  \\ \hline \hline

\multicolumn{3}{|c|}{\bf {BDT Requirements for Event Selection}}\\  \hline
BDT requirement & $\nu_\mu$ BDT $< 0.4$ & $\nu_e$ BDT $> 0.95$   \\ \hline \hline

\multicolumn{3}{|c|}{\bf {PID Requirements for Event Selection}}\\  \hline
$\pi^0$ mass & N/A & $\pi^0$ mass $<50$ MeV   \\
muon MPID interaction score & N/A &  $<$ 0.2 if $E_e> 100$ MeV  \\
$\gamma/e$ image score ratio & N/A & $<2$  \\
proton MPID interaction (max of 3 planes) & $> 0.9$ if $E_\nu< 400$ MeV & N/A  \\ \hline

\end{tabular}
\end{center}}}
\caption{The specific used to define the $1\mu1p$ and $1e1p$ samples. For definitions of kinematic variables, see Table ~\ref{tab:KinVarTable}.
\label{tab:allcuts}}
\end{sidewaystable}

\chapter{The MicroBooNE Electron Neutrino Analysis: Results and Discussion} \label{ch:microboone_results}

This chapter presents the first results from the two-body CCQE analysis described in \cref{ch:microboone_selection}.
They come from data taken during the first three MicroBooNE run periods, corresponding to $6.67 \times 10^{20}$~POT after data quality cuts.
The signal data were examined for the first time in Summer 2021 and presented to the community in October 2021~\cite{MicroBooNE:2021pvo}.
The results from the two-body CCQE analysis were accompanied by results from the inclusive~\cite{MicroBooNE:2021nxr} and MiniBooNE-like~\cite{TheMicroBooNECollaboration:2021cjf} $\nu_e$ analyses.
All three MicroBooNE $\nu_e$ analyses are summarized in Ref.~\cite{MicroBooNE:2021tya}.

This chapter focuses on the statistical results from the $\nu_e$ two-body CCQE analysis.
We cover the background prediction, systematic error evaluation, $\nu_\mu$ $1\mu1p$ constraint procedure, and blinded analysis approach in detail.
We then discuss the series of statistical tests performed using our $1e1p$ signal sample.
We close with a discussion of the MicroBooNE results, including a presentation of two non-MicroBooNE publications examining the implications of the MicroBooNE data under two different explanations of the MiniBooNE LEE: $3+1$ oscillations and excess $\overline{\nu}_e$ interactions. 

\textit{Publications covered in this chapter for which I either held a leading role or made major contributions: \cite{MicroBooNE:2021pvo,MicroBooNE:2021tya,MiniBooNE:2022emn,Kamp:2023mjn}}

\section{First Results from the Two-Body CCQE Analysis}

\begin{figure}[h!]
    \centering
    \includegraphics[width=0.92\textwidth]{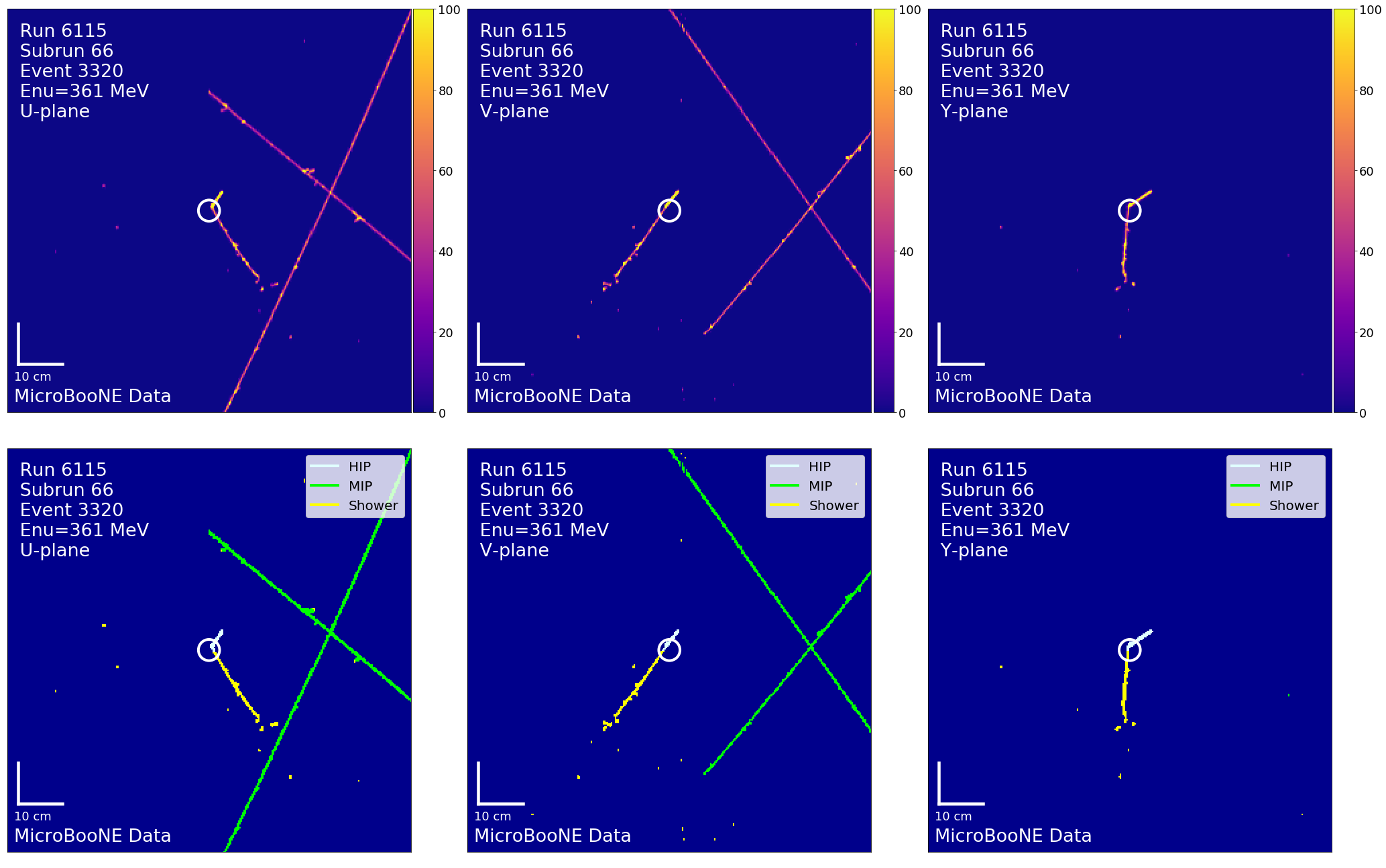}
    \caption{Top: pixel intensity (color scale is in PIU as defined in \cref{sec:ub_reco}); Bottom: \texttt{SparseSSNet} labels; Left to Right: U, V, Y, planes. The white circle indicates the reconstructed vertex. The horizontal axis corresponds to the wire plane direction and the vertical axis corresponds to the electron drift direction, which is measured using the arrival time of charge on the wires.}
    \label{fig:evd1}
\end{figure}

\begin{figure}[h!]
     \centering
     \begin{subfigure}[b]{0.45\textwidth}
         \centering
         \includegraphics[width=\textwidth]{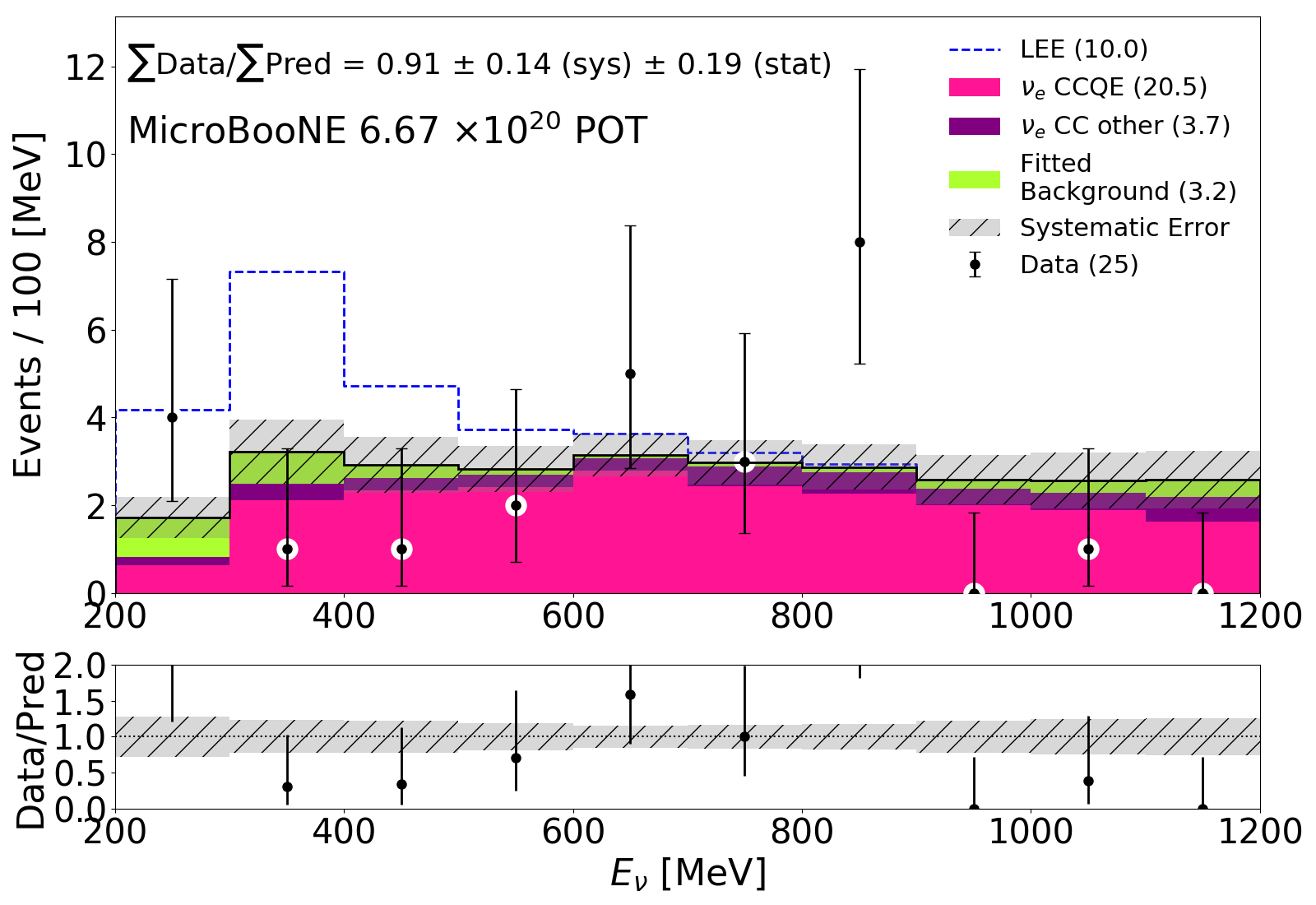}
         \caption{}
         \label{fig:enu_1e1p_prd}
     \end{subfigure}
     \hfill
     \begin{subfigure}[b]{0.45\textwidth}
         \centering
         \includegraphics[width=\textwidth]{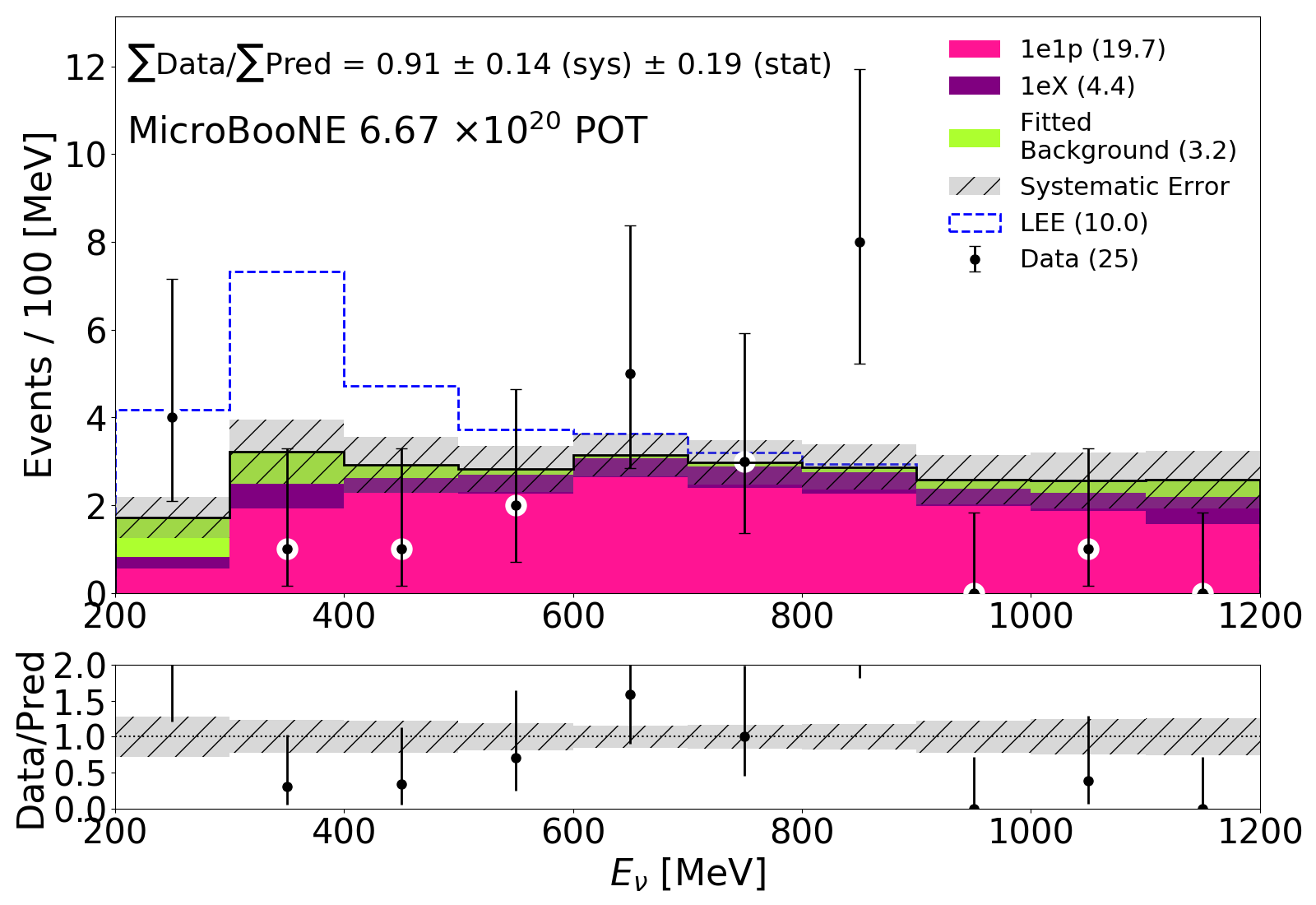}
         \caption{}
         \label{fig:enu_1e1p_top}
     \end{subfigure}
        \caption{The $1e1p$ sample $E_\nu$ distribution, comparing data (black points) to the unconstrained prediction (stacked histogram) in the $200<E_\nu<1200$\,MeV region. The eLEE model prediction is represented by the dashed blue line. The prediction is presented in terms of both interaction type (\cref{fig:enu_1e1p_prd}) and final state topology (\cref{fig:enu_1e1p_top}).
        }
        \label{fig:Enu_stacked}
\end{figure}

\begin{figure}[h!]
    \centering
    \includegraphics[width=0.6\textwidth]{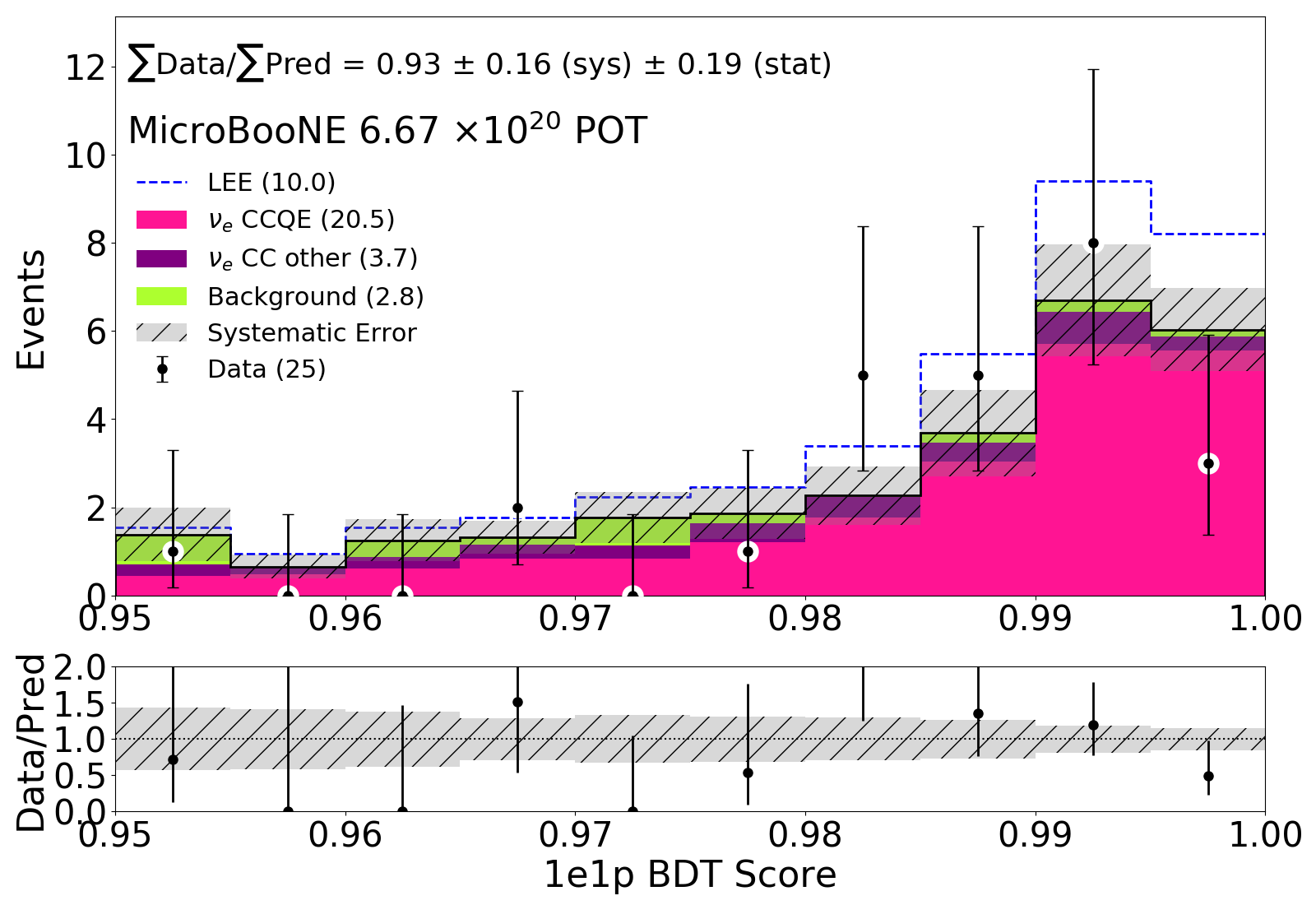}
    \caption{Average $1e1p$ BDT ensemble score distribution comparing data to the unconstrained prediction.}
    \label{fig:bdt_1e1p}
\end{figure}

\begin{figure}[h!]
     \centering
     \begin{subfigure}[b]{0.45\textwidth}
         \centering
         \includegraphics[width=\textwidth]{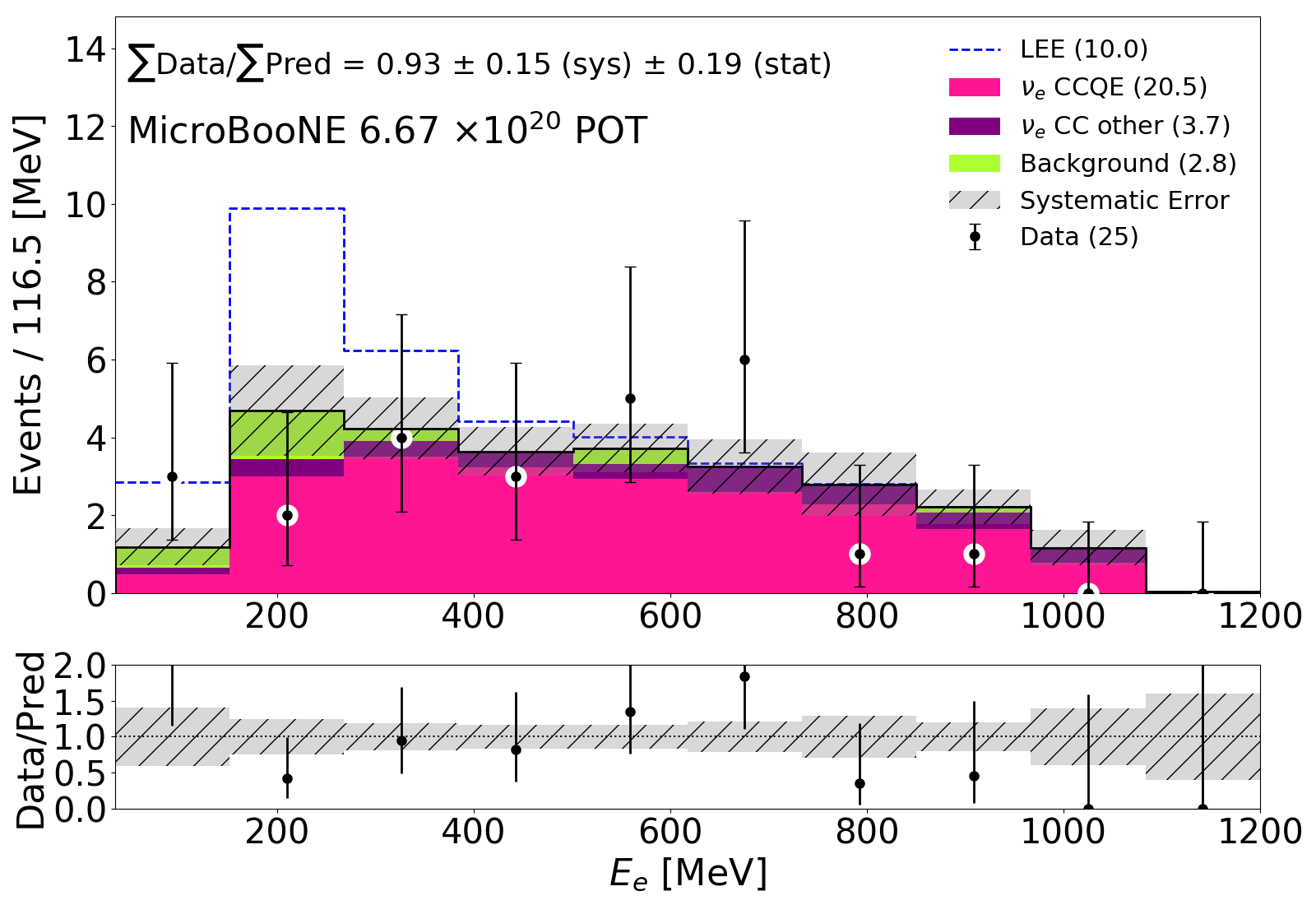}
         \caption{}
         \label{fig:electron_edep_1e1p}
     \end{subfigure}
     \hfill
     \begin{subfigure}[b]{0.45\textwidth}
         \centering
         \includegraphics[width=\textwidth]{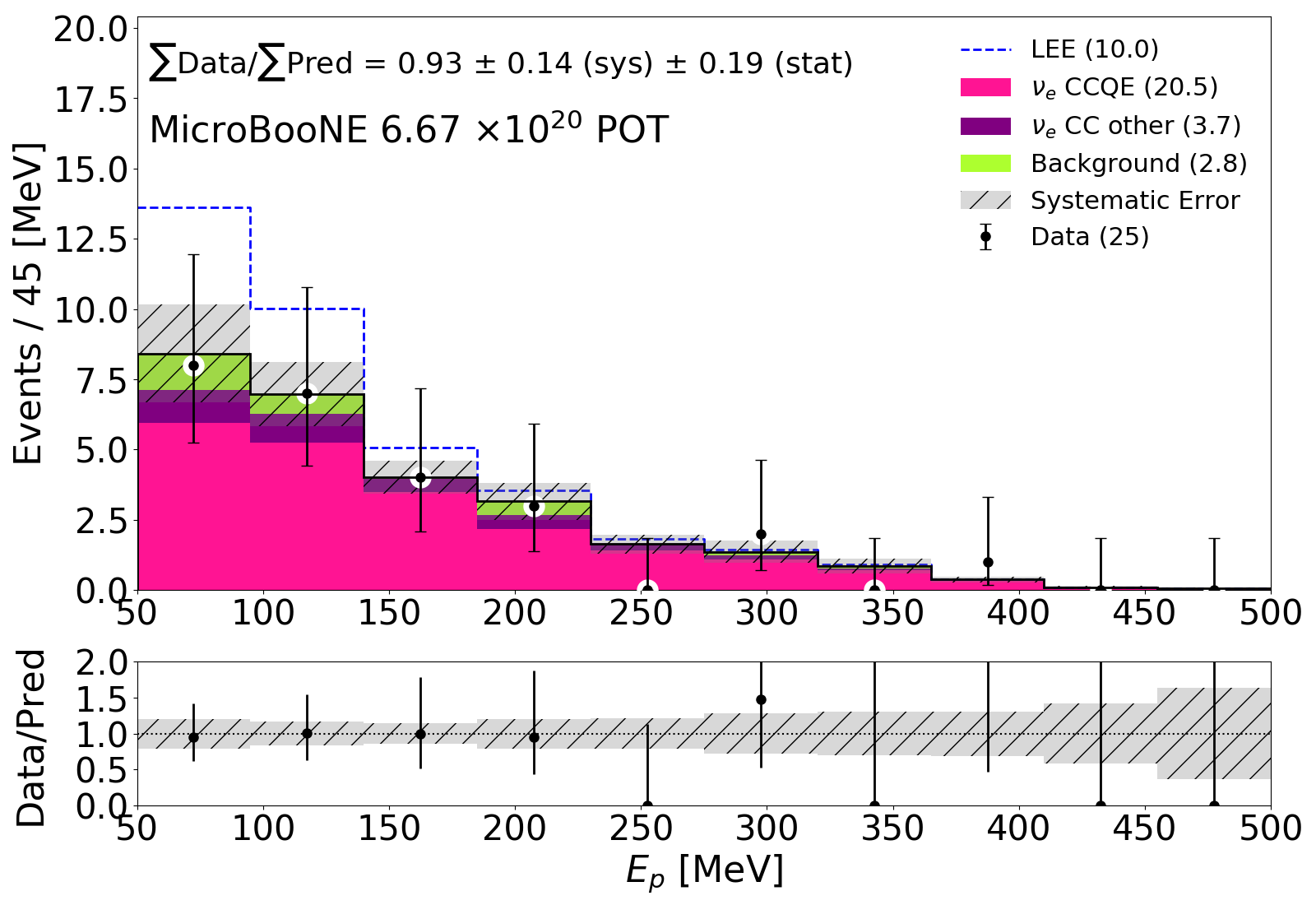}
         \caption{}
         \label{fig:proton_edep_1e1p}
     \end{subfigure}
     \hfill
     \begin{subfigure}[b]{0.45\textwidth}
         \centering
         \includegraphics[width=\textwidth]{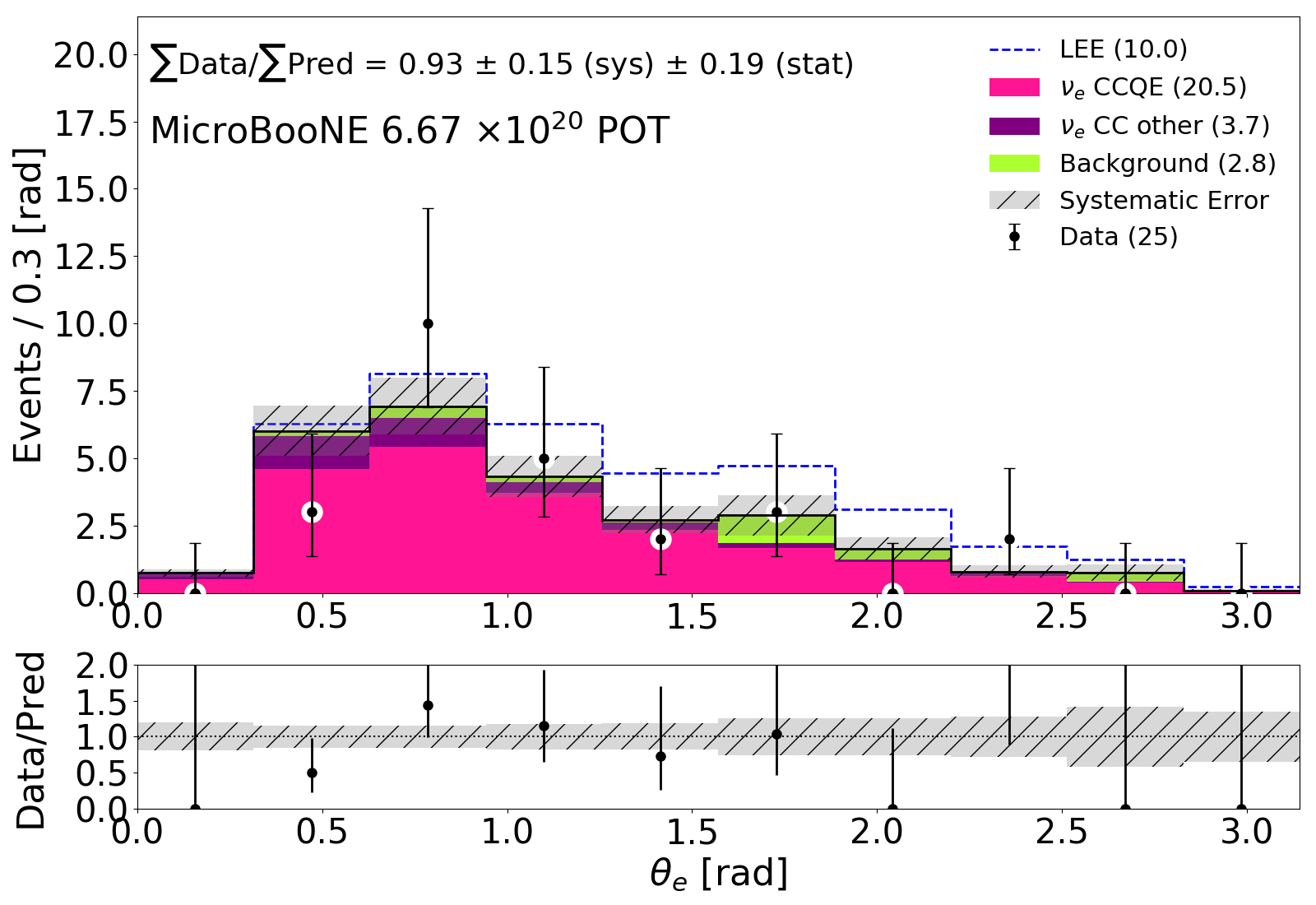}
         \caption{}
         \label{fig:lepton_theta_1e1p}
     \end{subfigure}
     \hfill
     \begin{subfigure}[b]{0.45\textwidth}
         \centering
         \includegraphics[width=\textwidth]{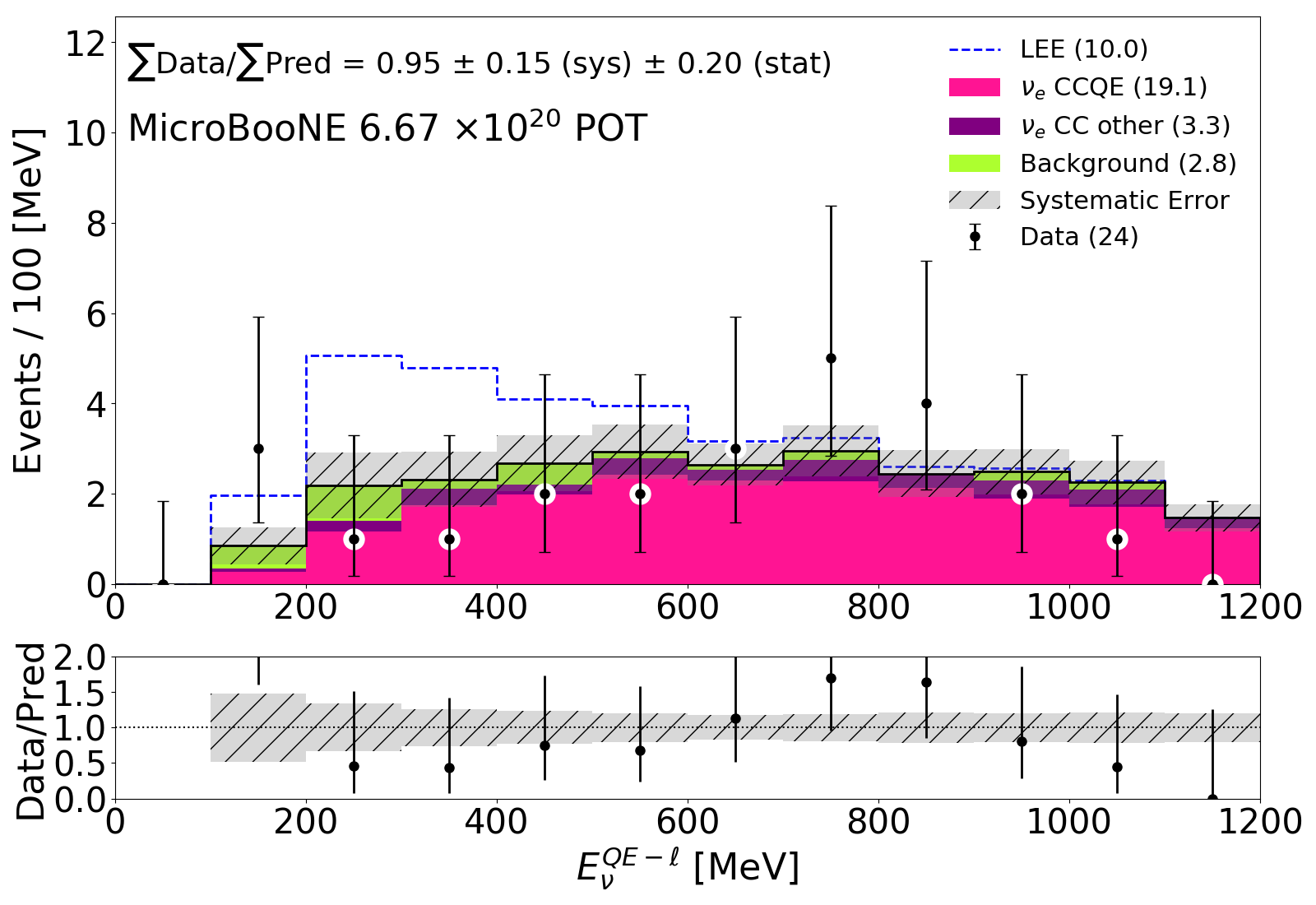}
         \caption{}
         \label{fig:enuqe_lep_1e1p}
     \end{subfigure}
     \hfill
        \caption{Comparison between data and unconstrained prediction in the $E_e$ (\cref{fig:electron_edep_1e1p}), $E_p$ (\cref{fig:proton_edep_1e1p}), $\theta_e$ (\cref{fig:lepton_theta_1e1p}), and $E_\nu^{QE-\ell}$ (\cref{fig:enuqe_lep_1e1p}) distributions of the $1e1p$ sample.
        }
        \label{fig:more_dists_1e1p}
\end{figure}

The two-body CCQE analysis observed 25 events across data from the first three run periods.
This is in good agreement with the total prediction in the $200-1200$~MeV range, which is $27.4 \pm 3.8_{\rm sys} \pm 5.2_{\rm stat}$ before the $1\mu1p$ constraint discussed in \cref{sec:constraint}~\cite{MicroBooNE:2021pvo}.
One of these events is shown in \cref{fig:evd1}, including both pixel intensity images from the TPC readout and \texttt{SparseSSNet} pixel-labeled images on all three planes.
Event displays for the remaining 24 selected events are included in Appendix~B.
The $E_\nu^{\rm range}$ distribution of these events is shown in \cref{fig:Enu_stacked}, where the prediction is given both in terms of the underlying interaction type and the final state event topology.
The $1e1p$ BDT ensemble average score distribution is shown in \cref{fig:bdt_1e1p}.
\Cref{fig:more_dists_1e1p} shows the distributions of four more interesting variables: $E_e$, $E_p$, $\theta_e$, and $E_\nu^{QE-\ell}$.
A full suite of data-prediction comparisons in 36 variables is provided in the Supplemental Material for Ref.~\cite{MicroBooNE:2021pvo}, including those described in \cref{tab:KinVarTable} and those used to train the BDT ensembles.

One can see clearly in \cref{fig:Enu_stacked} that no excess of $\nu_e$ $1e1p$ candidate events is observed at low energy in the two-body CCQE analysis.
Instead, we observe a slight deficit compared to even the nominal prediction at low energies, in conflict with the eLEE model represented by the blue dashed line.
This is a trend that is observed across all three $\nu_e$ analyses in MicroBooNE~\cite{MicroBooNE:2021tya}.
The extent to which we agree or disagree with the eLEE model will be discussed further in \cref{sec:stat_interp}.

One might also notice an apparent excess of data events in the 800-900~MeV bin in \cref{fig:Enu_stacked}.
This appears to be a statistical fluctuation; other kinematic variables do not exhibit such an excess of events, including the closely related $E_\nu^{QE-\ell}$ distribution shown in \cref{fig:enuqe_lep_1e1p}.
If this were a systematic effect rather than a fluctuation, one would expect similar behavior in the distributions of other variables.
While these statements will be quantified further in \cref{sec:stat_interp}, it is useful to include such a qualitative discussion at this point.

The following subsections discuss the details behind our prediction in the $E_\nu$ distribution.
It is important to note that while we have already shown the observed data in our signal channel, the procedures outlined in the following subsections were frozen before examining this dataset.
Specifically, we followed the blinded analysis approach outlined in \cref{sec:blind} to avoid biasing our selection.

\subsection{Background Estimation} \label{sec:bkgfit}

In \cref{tab:bkgbreakdown} we give a breakdown of the MC events in the ``background'' category of \cref{fig:bdt_1e1p,fig:more_dists_1e1p} by their interaction channel and event topology.
The ``off-vertex'' label in this table refers to events for which the reconstructed neutrino vertex is further than 5~cm away from the true neutrino vertex.
These events can happen when a cosmic muon track happens to cross a photon shower from a $\pi^0$ decay, or a muon decays to a Michel electron; in both cases, the muon is mistaken for a proton.
\Cref{tab:bkgbreakdown} indicates that $\nu_\mu$ resonant $\pi^0$ events are the most common background interaction channel, followed by off-vertex events.
One can also see that events with a $\pi^0$ in the final state are the dominant background topology.

There are not many MC events to inform the numbers reported in \cref{tab:bkgbreakdown}.
The tools developed for this analysis produce a highly pure $\nu_e$ CCQE sample, resulting in a low-statistics simulation sample for assessing the non-$\nu_e$ background to the $1e1p$ signal.
This is apparent in the fluctuations of the background prediction in \cref{fig:bdt_1e1p,fig:more_dists_1e1p}.
In order to obtain a more robust prediction of this background, we have elected to leverage information on the energy distribution of non-$\nu_e$ background events at a loose BDT score cutoff of 0.7 and extrapolate this to our signal cutoff of 0.95.
This is accomplished by fitting a parameterized Landau+linear probability density function (PDF) to the background energy distribution at the loose cutoff and scaling this prediction to the signal cutoff.
As most of these backgrounds come from $\nu_\mu$ interactions, this is referred to as the $\nu_\mu$ background fit.

\begin{figure}[h!]
    \centering
    \begin{subfigure}[b]{0.45\textwidth}
         \centering
         \includegraphics[width=\textwidth]{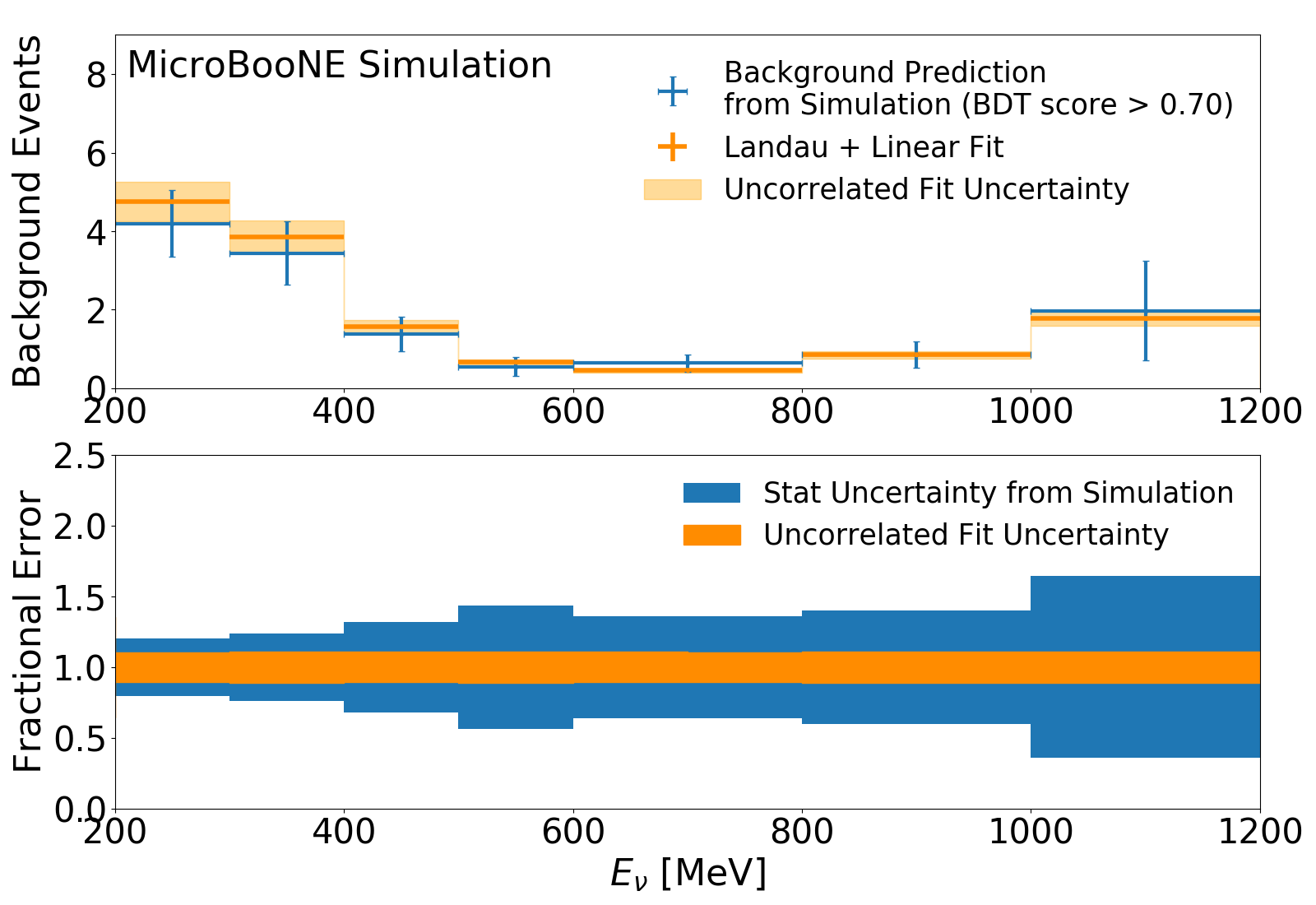}
         \caption{}
         \label{fig:bkg_fit0.7}
     \end{subfigure}
     \hfill
    \begin{subfigure}[b]{0.45\textwidth}
         \centering
         \includegraphics[width=\textwidth]{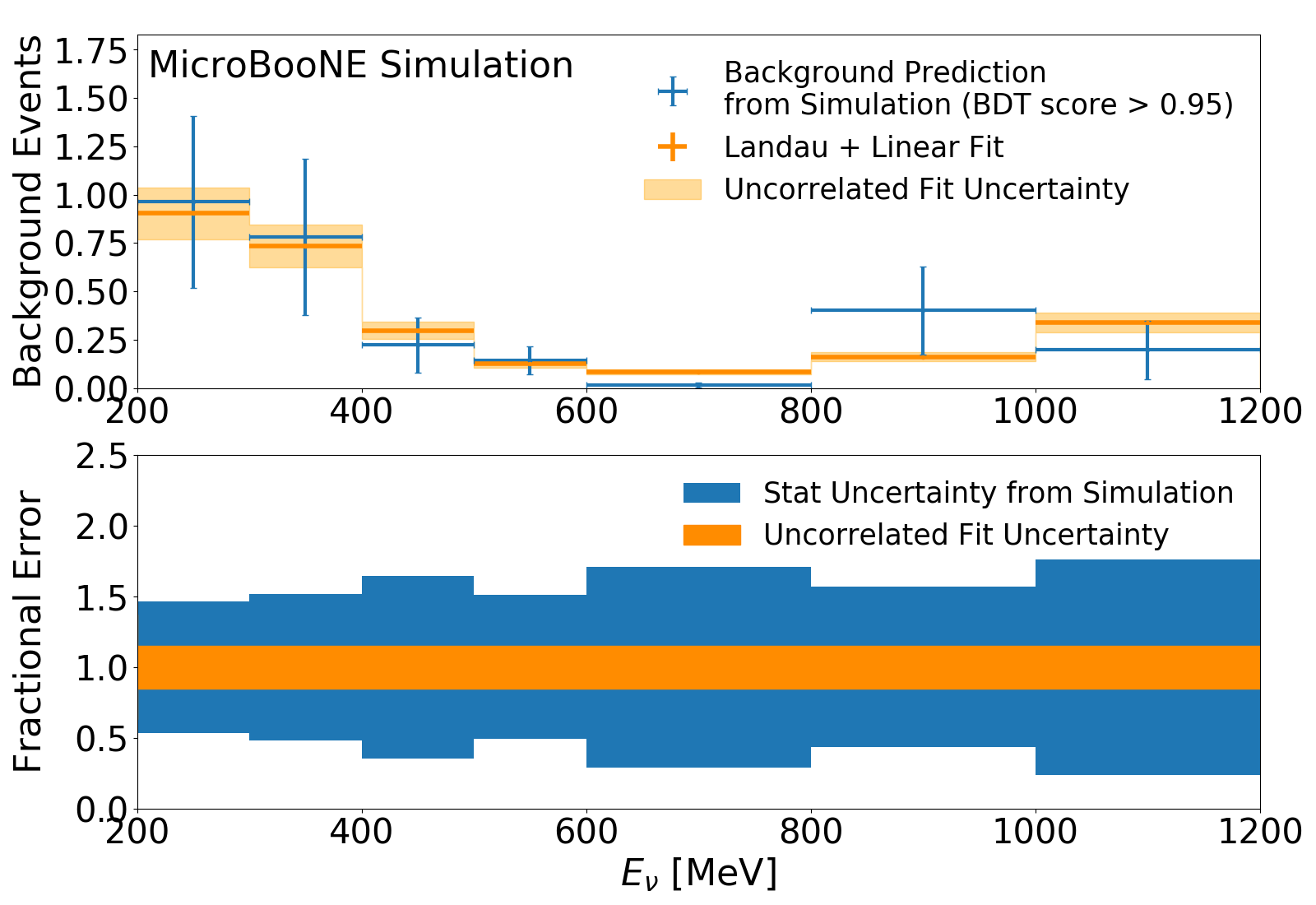}
         \caption{}
         \label{fig:bkg_fit0.95}
     \end{subfigure}
\caption{The fit to the $\nu_\mu$ background distribution to the $1e1p$ analysis. The shape fit is performed at a loose BDT score cutoff of 0.7 (\cref{fig:bkg_fit0.7}) and scaled to the signal cutoff of 0.95 (\cref{fig:bkg_fit0.95}). Blue points represent the prediction from the simulation, with error bars representing the Gaussian approximation of the statistical error (quadrature sum of event weights). The orange line and corresponding shaded region represent prediction and uncertainty, respectively, coming from the Landau+linear fit.} 
\label{fig:bkgfitresult}
\end{figure}

The Landau+linear shape is motivated empirically by the observation of a rise in the background rate toward the lowest energies ($\lesssim 500$\,MeV) and a smaller rise toward the highest energies ($\gtrsim 800$\,MeV), both for the loose BDT score cutoff and the signal cutoff, as shown by the blue points in \cref{fig:bkgfitresult}.
The Landau portion of the fit is additionally motivated by the observation that a majority of $\nu_\mu$ background events contain a $\pi^0$ in the final state, for which one of the decay photons is misinterpreted as an electron shower.
The reconstructed neutrino energy of these events is governed predominately by the energy of this photon, which has a tail extending to higher energies caused by pions with high momentum in the lab frame.
This tail is a characteristic feature of the Landau function.
The output of the fit and resulting error are used in place of the raw prediction and statistical error for the $\nu_\mu$ backgrounds in this analysis.
We note here that the simulated $\nu_\mu$ events with a $\pi^0$ in the final state are weighted according to the observation in our dedicated $\pi^0$ sample.
The method for calculating these weights is described in Ref.~\cite{MicroBooNE:2021pvo}.

The predicted background spectrum in each reconstructed neutrino energy bin $f(E_i)$ is generated by integrating the Landau+linear PDF $p(E)$ within that bin. Specifically,
\begin{equation} \label{eq:bkgshape}
\begin{split}
&p(E) = \exp\bigg[\frac{-(E' + e^{-E'})}{2}\bigg] + aE,\\
&E' = (E - \mu)/\sigma, \\
&f(E_i) = \int_{E_i - \delta E_i}^{E_i + \delta E_i} p(E) dE,
\end{split}
\end{equation}
where  $\mu$ and $\sigma$ are the center and width of the Moyal approximation of the Landau function~\cite{moyalapprox}, $a$ is the linear slope parameter, $E_i$ is the center of the i'th energy bin, and $\delta E_i$ is half of the bin width.
The Landau+linear fit is carried out using only shape information at a loose BDT score cutoff of 0.7.
In order to get the overall normalization, we fit the BDT score distribution of the backgrounds to a linear PDF $\Tilde{p}(x)$, which we can integrate to get the total expected number of background events $\Tilde{f}(x)$ for a given BDT score cutoff $x$,
\begin{equation} \label{eq:bkgnorm}
\Tilde{p}(x) = mx + b ~~~~ \Tilde{f}(x) = \int_x^1 \Tilde{p}(x) dx,
\end{equation}
where $m$ and $b$ are the slope and bias parameters for the BDT score distribution, respectively.
The resulting shape+normalization fit for the background distribution at the loose BDT cutoff of 0.7 and signal cutoff of 0.95 are shown in Fig.~\ref{fig:bkgfitresult}.
One can see that the fit agrees with the raw MC prediction within statistical error in both cases and that the error on the fit is generally reduced compared to the simulation statistical error. 

The errors on the fits in Fig.~\ref{fig:bkgfitresult} are obtained by simulating pseudo-experiments according to the covariance matrix of the fit parameters.
This is accomplished efficiently using Cholesky decomposition~\cite{cholesky}.
The overall normalization error on the background rate at a given BDT score cutoff is calculated using the error on the linear fit to the BDT score distribution of background events.
For the Landau+linear fit, we generate a shape-only covariance matrix for the $\nu_\mu$ backgrounds in the nominal reconstructed neutrino energy bins.
The overall normalization error has been added as a fully-correlated contribution to each bin of the shape-only $E_\nu$ covariance matrix.
The uncorrelated uncertainty from the fit, given by the square root of the diagonal entries of the full (shape $\oplus$ normalization) covariance matrix, is indicated by the orange band in Fig.~\ref{fig:bkgfitresult}.
This covariance matrix replaces the nominal statistical uncertainty on the simulated background, given by the quadrature sum of weights in each $E_\nu$ bin, which is indicated by the shaded blue band in Fig.~\ref{fig:bkgfitresult}.

\begin{figure}[h!]
    \centering
    \includegraphics[width=0.6\textwidth]{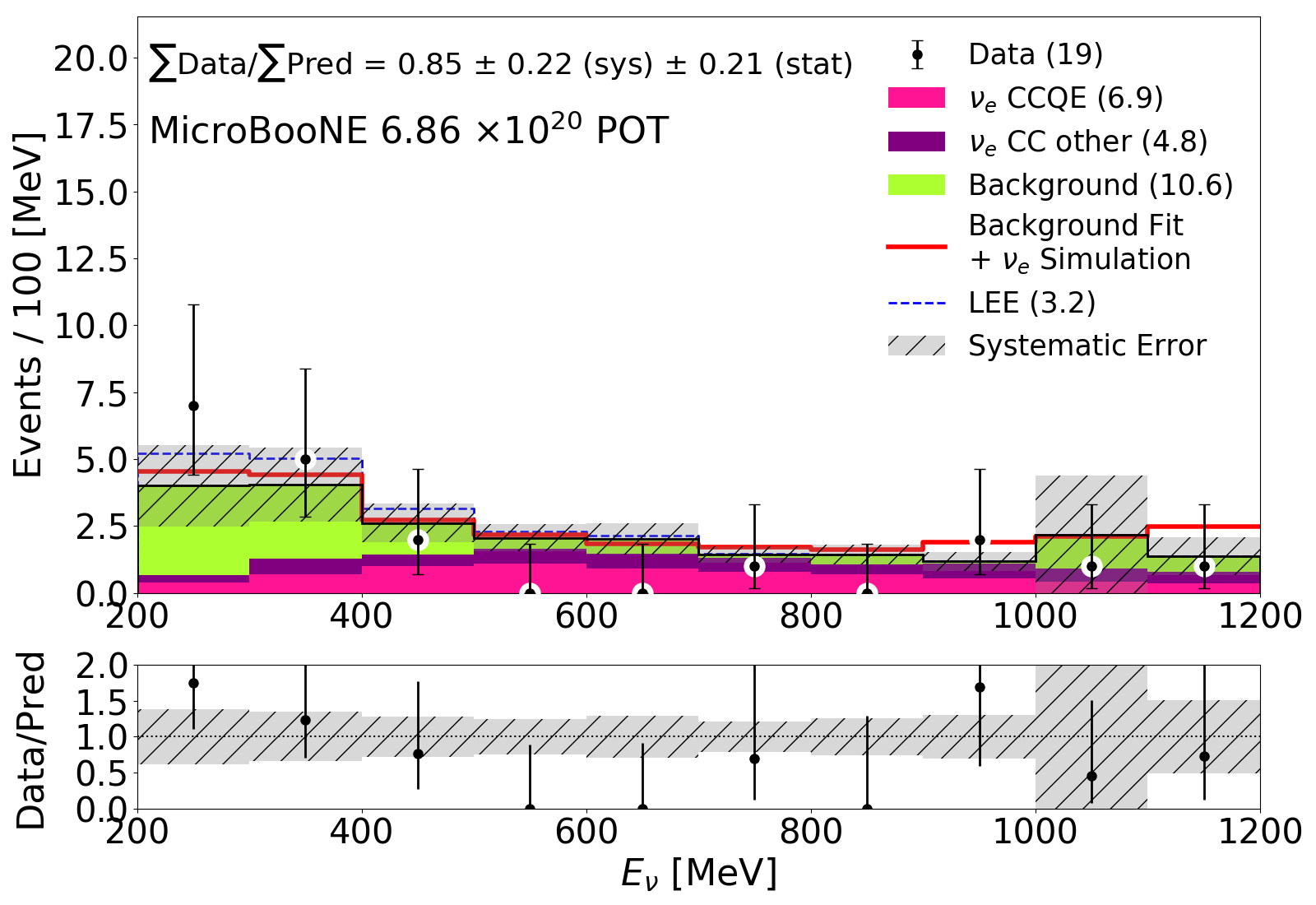}
    \caption{The data and MC prediction for events with a $1e1p$ BDT score inside $[0.7,0.95]$. Good agreement is observed between data and prediction. The prediction incorporating the Landau+linear background fit is shown by the red line.} 
    \label{fig:midBDTplot}
\end{figure}

The performance of the fit is evaluated on data by examining events slightly below the signal region, with a BDT score in the range $[0.7,0.95]$.
These events are shown in Fig.~\ref{fig:midBDTplot}.
The raw MC prediction is given by the stacked histogram, while the prediction incorporating the $\nu_\mu$ background fit on top of the simulated $\nu_e$ prediction is given by the red line.
The data agree well with both predictions, indicating the consistency of the fit method presented here with both the observed and MC-predicted background rate.

\begin{table}[h]
    \centering
    \begin{tabular}{|c|c|}
    \hline
    {\bf Interaction Channel} & {\bf Predicted Rate}  \\
    \hline
    $\nu_\mu$ resonant $\pi^0$     &  1.26 \\
    $\nu_\mu$ resonant $\pi^\pm$     &  0.21 \\
    $\nu_\mu$ CCQE     &  0.14 \\
    $\nu_\mu$ other     &  0.19 \\
    Off-vertex     &  0.93 \\
    \hline
    {\bf Event Topology} & {\bf Predicted Rate}  \\
    \hline
    $1\mu N \pi^0$     &  0.57 \\
    $0 \mu N \pi^0$     &  1.09 \\
    $1\mu 1p$     &  0.14 \\
    Off-vertex     &  0.93 \\
    \hline
    \end{tabular}
    \caption{Breakdown of MC events in the ``background'' category of \cref{fig:bdt_1e1p,fig:more_dists_1e1p} over the range $200<E_\nu<1200$\,MeV. The events are partitioned both by the interaction channel and the event topology.}
    \label{tab:bkgbreakdown}
\end{table}

\subsection{Evaluation of Systematic Uncertainties} \label{sec:systematics}

There are five main sources of systematic uncertainty in this analysis, related to the beam flux, neutrino-nucleus cross section modeling, hadron re-interaction modeling, detector simulation, and finite statistics in the MC-based prediction.
These are handled within a covariance matrix formalism.
For each binned distribution, the diagonal entries of the covariance matrix contain the variance in each bin, while the off-diagonal entries contain the covariance between different bins.
Thus, the covariance matrix is also symmetric.
The use of a covariance matrix is essential for encoding correlations between binned distributions of selected events in the $1e1p$ and $1\mu1p$ samples, which is relevant for the constraint procedure described in \cref{sec:constraint}.
The total covariance matrix for a given distribution is simply the element-wise sum of the covariance matrices for each source of systematic uncertainty.
A few definitions will be helpful for our discussion here; given a covariance matrix $M_{ij}$ and binned prediction $N_i$, the fractional covariance matrix is given by $F_{ij} \equiv M_{ij} / (N_i N_j)$ and the correlation matrix is given by $\rho_{ij} \equiv M_{ij} / \sqrt{M_{ii} M_{ij}}$.
The correlation matrix is meant to capture the degree of correlation between two bins normalized within the range $[-1,1]$, where diagonal entries are 1 by construction.

The flux, cross section, and re-interaction uncertainties are evaluated using a reweighting method, in which event weights are modified according to tunable parameters.
Variations in the distributions of reconstructed variables are used to determine the associated covariance matrix.
The detector systematic covariance matrix is evaluated using a dedicated set of simulation samples that each alter the detector response in a different way.
These alterations to the detector response are derived using cosmic muon data; the full procedure is described in detail in Ref.~\cite{MicroBooNE:2021roa}.

Flux uncertainties are related to hadron production in the target, secondary hadron interactions, and focusing efficiency of the BNB magnetic horn.
Neutrino events are reweighted according to the species and momentum of the parent meson~\cite{MiniBooNE:2008hfu,MicroBooNE:2019nio}.
The largest flux uncertainty comes from $\pi^+$ production in the target.
The flux uncertainties are also highly correlated between the $1e1p$ and $1\mu1p$ samples, as they both come predominately from the same meson decay chain, $\pi^+ \to \nu_\mu (\mu^+ \to e^+ \nu_e \overline{\nu}_\mu)$.

The cross section covariance matrix is calculated using both interaction-specific reweightable parameters and FSI-related reweightable parameters in \texttt{GENIE v3.00.06}.
Thus, each MC event is reweighted according to its interaction channel and any final state particles which underwent FSI.
The full suite of reweightable parameters is described in Ref.~\cite{MicroBooNE:2021ccs}.
We also evaluate uncertainties related to the difference between the $\nu_e$ CCQE and $\nu_\mu$ CCQE cross sections~\cite{Day:2012gb}.
The largest cross section uncertainties come from CCQE-related parameters, which is expected given our signal definition, and FSI-related parameters.
FSI uncertainties are especially important in this analysis, as FSI can cause $\nu_e$ CCQE events to deviate from two-body scattering kinematics and also cause non-$\nu_e$ CCQE events to look more consistent with two-body scattering.
Both CCQE-related and FSI-related effects impact the $1e1p$ and $1\mu1p$ samples similarly and thus are at least partially addressed by the constraint procedure in \cref{sec:constraint}.

Hadron re-interaction uncertainties describe the scattering of final state protons and $\pi^\pm$ off of argon nuclei.
The covariance matrix is calculated by reweighting events according to \texttt{Geant4} truth-level information about the hadron trajectory~\cite{Calcutt:2021zck}.
These uncertainties are a small effect in this analysis.

The detector systematic covariance matrix is calculated using a set of dedicated simulation samples that vary certain aspects of the detector response.
These include the amplitude and width of TPC signals as a function of the $x$ and $(y,z)$ positions as well the directions $\theta_{XZ}$ and $\theta_{YZ}$ of the charged particle; the electron-ion recombination rate; the electric field map in the TPC; and the light yield, attenuation, and Rayleigh scattering length.
As these samples are computationally expensive to generate, we are not left with many MC events with which to construct a covariance matrix.
Thus, we rely on kernel density estimation~\cite{10.1214/aoms/1177704472,10.1214/aoms/1177728190} to smooth the predictions in each detector variation sample.
This procedure does not work for the $\nu_\mu$ backgrounds to the $1e1p$ analysis since we are especially starved for MC statistics here, as discussed in \cref{sec:bkgfit}.
We instead assign a conservative uncorrelated 20\% fractional detector systematic uncertainty on these events, based on the extent of differences in the total rate of selected background events in the detector variation samples.

The variance due to finite MC statistics is simply given by the quadrature sum of event weights in each bin.
These uncertainties are uncorrelated and give a diagonal covariance matrix.
The exception is finite MC statistics uncertainties on the $\nu_\mu$ backgrounds in the $1e1p$ analysis, which are calculated using the procedure described in \cref{sec:bkgfit}.

Finally, we note that to evaluate the reweightable systematic uncertainties on the $\nu_\mu$ backgrounds to the $1e1p$ analysis, we use the selected MC events to calculate a fractional covariance matrix, which is then scaled by the parameterized background prediction derived in \cref{sec:bkgfit}.
When performing the constraint in \cref{sec:constraint}, we do not consider off-diagonal systematic uncertainties between $\nu_\mu$ backgrounds in the $1e1p$ sample and bins in the $1\mu1p$ sample, as it is not straightforward to compute this correlation.

\begin{figure}[h!]
    \centering
    \begin{subfigure}[b]{0.45\textwidth}
         \centering
         \includegraphics[width=\textwidth]{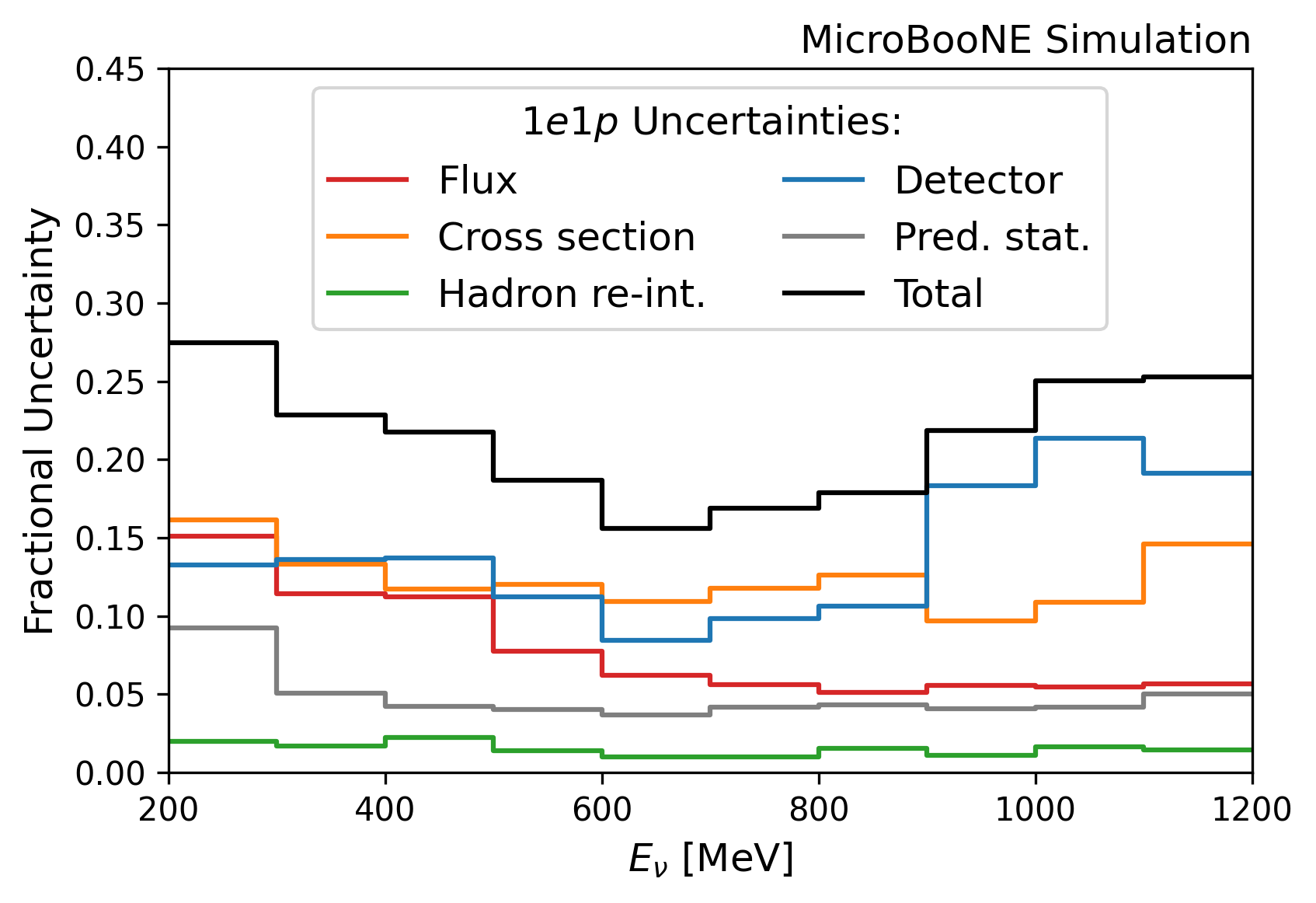}
         \caption{}
         \label{fig:errors_1e1p}
     \end{subfigure}
     \hfill
    \begin{subfigure}[b]{0.45\textwidth}
         \centering
         \includegraphics[width=\textwidth]{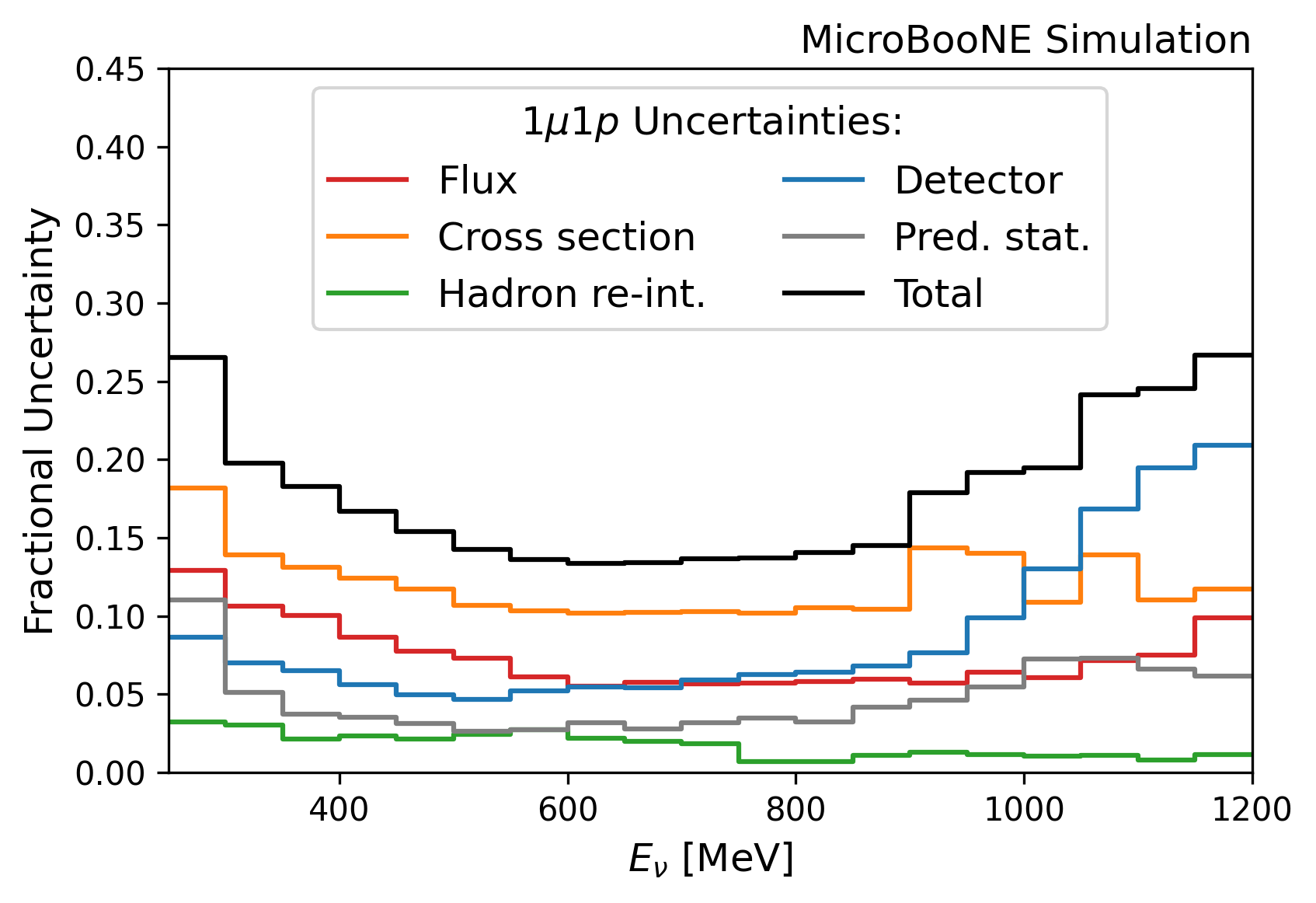}
         \caption{}
         \label{fig:errors_1mu1p}
     \end{subfigure}
\caption{The uncertainty in each bin of the $E_\nu^{\rm range}$ distribution of the $1e1p$ (\cref{fig:errors_1e1p}) and $1\mu1p$ (\cref{fig:errors_1mu1p}) samples.} 
\label{fig:diag_errors}
\end{figure}

\begin{figure}[h!]
    \centering
    \begin{subfigure}[b]{0.45\textwidth}
         \centering
         \includegraphics[width=\textwidth]{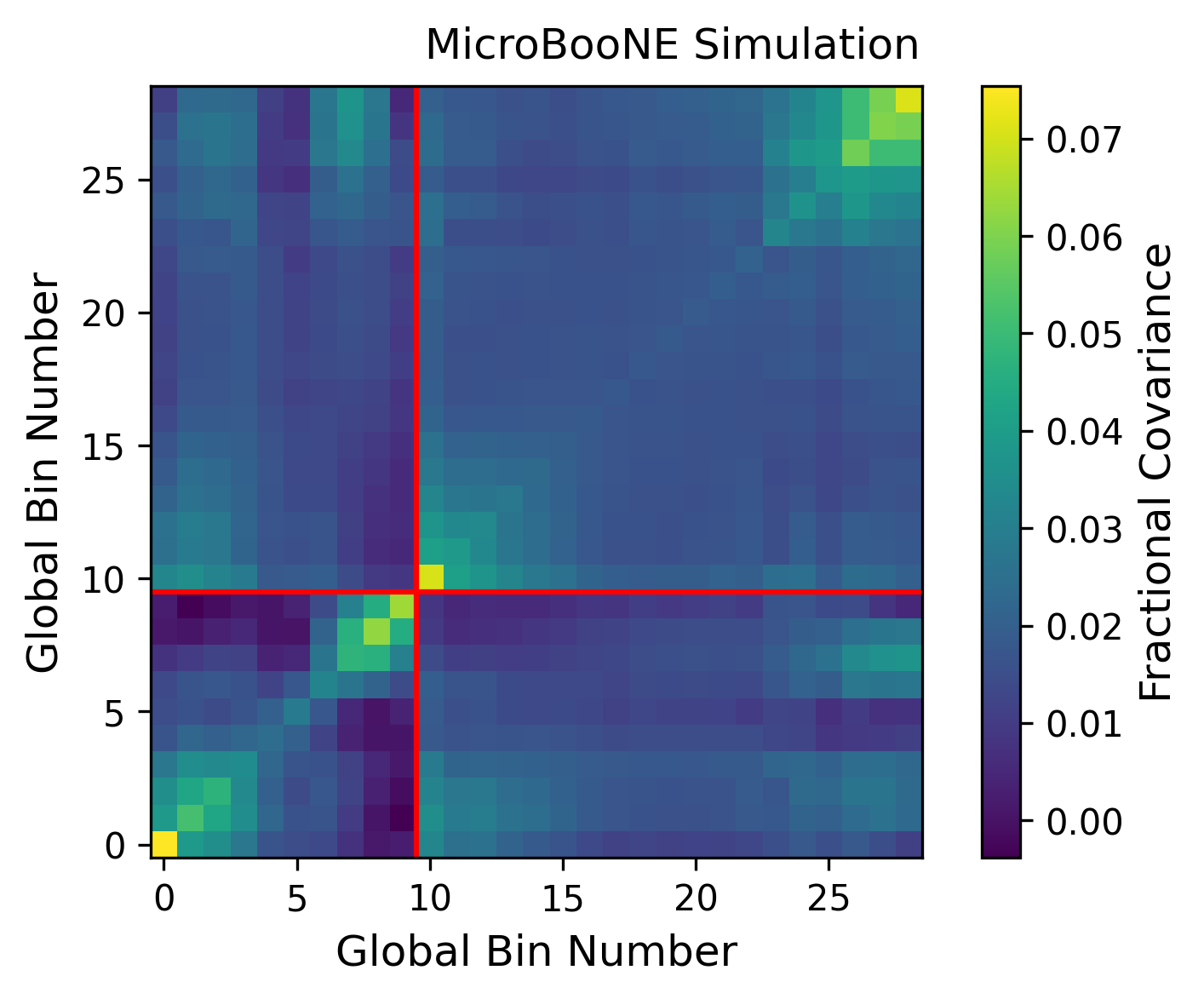}
         \caption{}
         \label{fig:covar}
     \end{subfigure}
     \hfill
    \begin{subfigure}[b]{0.45\textwidth}
         \centering
         \includegraphics[width=\textwidth]{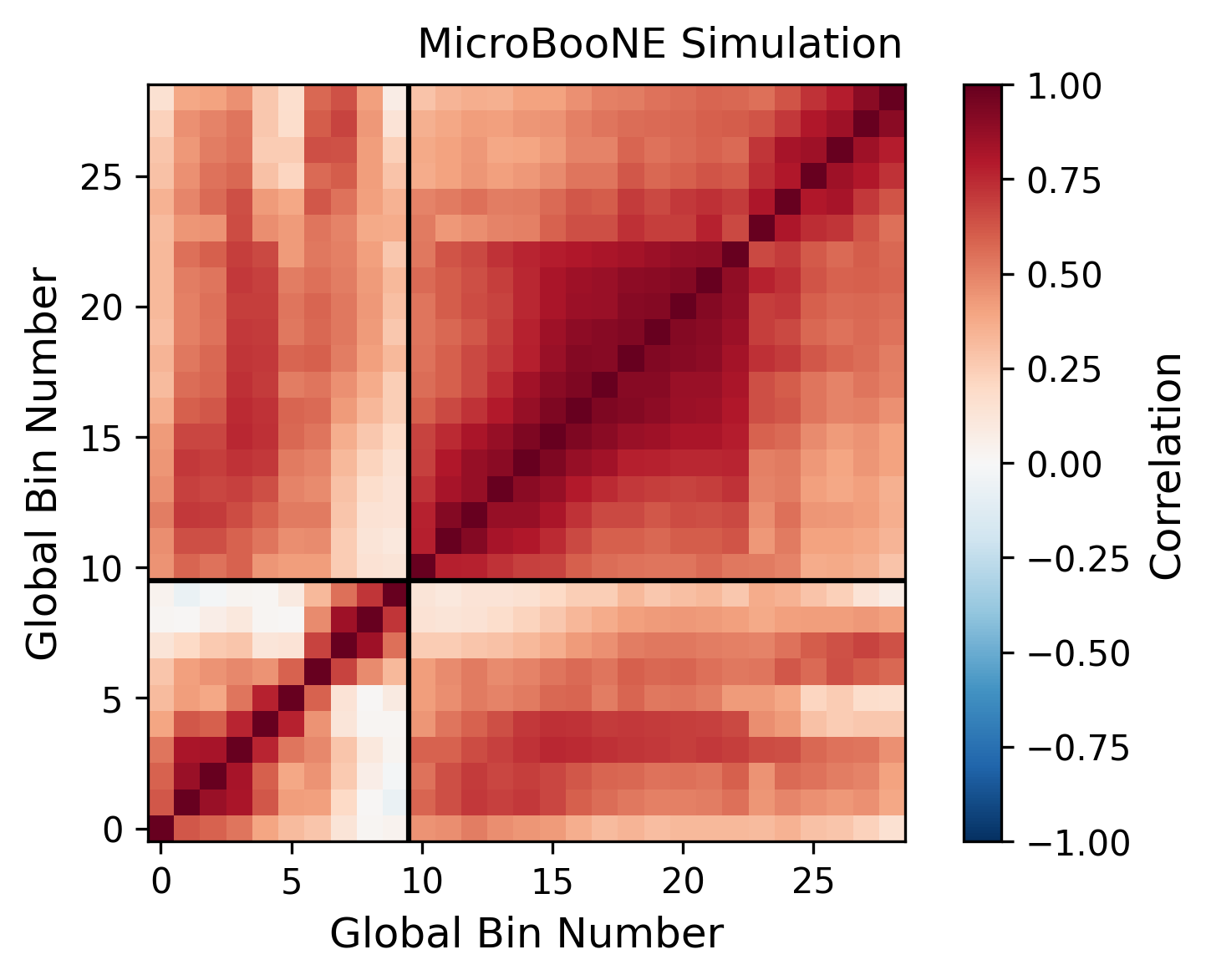}
         \caption{}
         \label{fig:correl}
     \end{subfigure}
\caption{The joint covariance (\cref{fig:covar}) and correlation (\cref{fig:correl})  matrices for the $E_\nu^{\rm range}$ distribution of the $1e1p$ and $1\mu1p$ samples.} 
\label{fig:error_matrices}
\end{figure}

The fractional uncertainty in each bin of the $E_\nu^{\rm range}$ distribution in the $1e1p$ and $1\mu1p$ samples is shown in \cref{fig:diag_errors}.
These correspond to the square root of the diagonal entries of the total systematic covariance matrix calculated from all sources described above, divided by the predicted event rate in each bin.
In the $1e1p$ sample, flux, cross section, and detector uncertainties dominate at low energies while detector uncertainties dominate at high energies.
The joint $1e1p$ and $1\mu1p$ covariance and correlation matrices are shown in \cref{fig:error_matrices}.
The large off-block-diagonal correlations between the $1e1p$ and $1\mu1p$ samples will be exploited in \cref{sec:constraint}.

\subsection{Constraint from the $1\mu1p$ Sample} \label{sec:constraint}

As alluded to thus far, we use the observation in the $1\mu1p$ control sample to constrain the prediction and uncertainties in the $1e1p$ signal sample.
We rely on the conditional covariance method, which has also been used by the MiniBooNE collaboration~\cite{MiniBooNE:2007uho}.
We perform the constraint in the $E_\nu^{\rm range}$ distribution of each sample.
Let us represent the joint covariance matrix between both samples shown in \cref{fig:covar} as
\begin{equation}
M = 
\begin{pmatrix}
M^{ee} & M^{e\mu} \\
M^{\mu e} & M^{\mu\mu} \\
\end{pmatrix}.
\end{equation}
Given a measurement $d_\mu$ and prediction $\mu_\mu$ in the $1\mu1p$ channel, the constrained prediction $\mu_{e,{\rm const}}$ and covariance matrix $M^{ee,{\rm const}}$ are given by~\cite{MicroBooNE:2021nxr}
\begin{equation} \label{eq:constraint}
\begin{split}
&\mu_{e,{\rm const}} = \mu_e + M^{e\mu} \cdot (M^{\mu \mu})^{-1} \cdot (d_\mu - \mu_\mu), \\
&M^{ee,{\rm const}} = M^{ee} - M^{e\mu} \cdot (M^{\mu \mu})^{-1} \cdot M^{\mu e}, 
\end{split}
\end{equation}
where $\mu_e$ is the nominal prediction in the $1e1p$ channel before the constraint.

\begin{figure}[h!]
    \centering
    \includegraphics[width=0.6\textwidth]{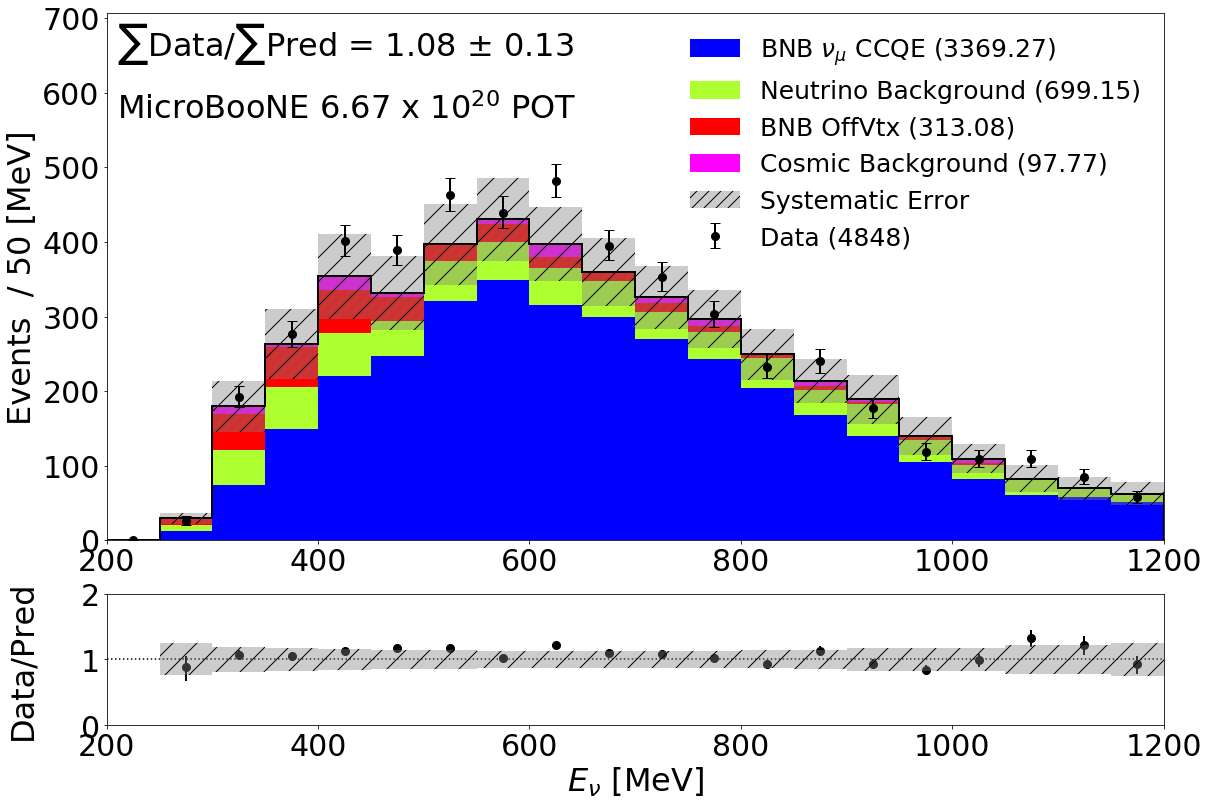}
    \caption{The $E_\nu^{\rm range}$ distribution in the $1\mu1p$ channel, comparing data to the MC prediction.}
    \label{fig:1m1p_enu}
\end{figure}

\begin{figure}[h!]
    \centering
    \includegraphics[width=0.6\textwidth]{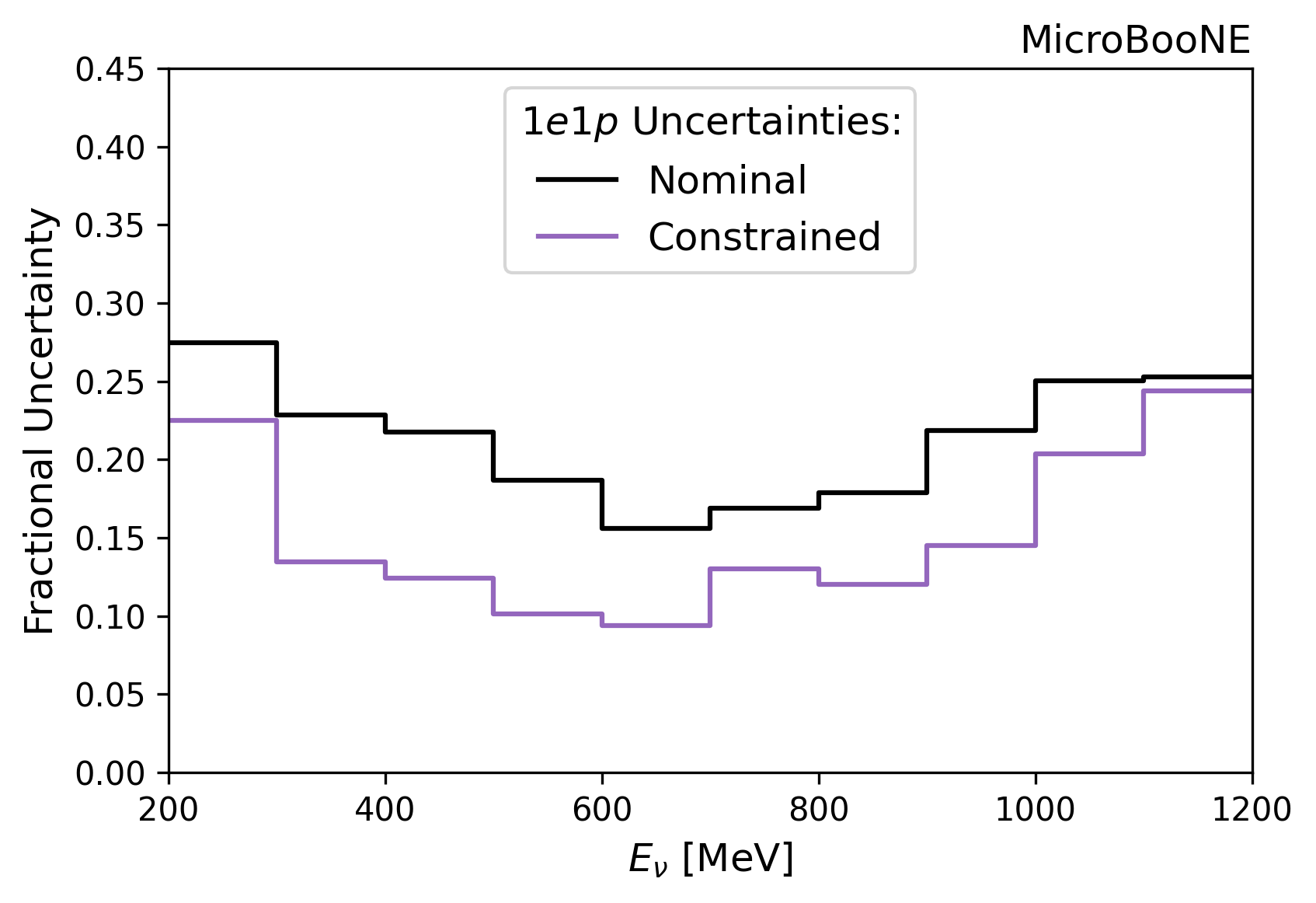}
    \caption{Fractional systematic uncertainty in the $1e1p$ $E_\nu^{\rm range}$ distribution before and after the $1\mu1p$ constraint.}
    \label{fig:constr_err}
\end{figure}

\begin{figure}[h!]
    \centering
    \includegraphics[width=0.6\textwidth]{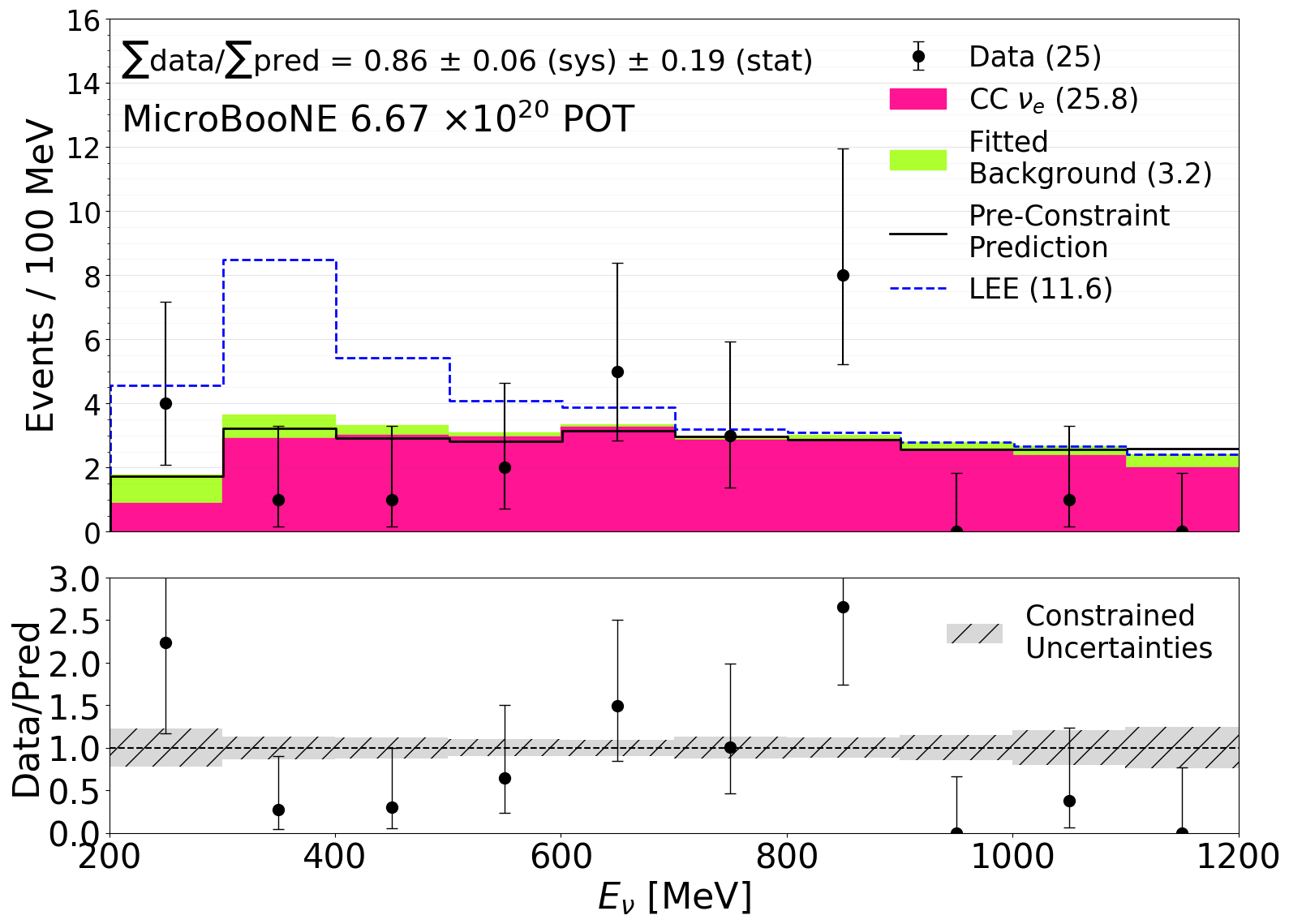}
    \caption{Comparison between data and prediction in the $1e1p$ $E_\nu^{\rm range}$ distribution after applying the $1\mu1p$ constraint procedure.}
    \label{fig:constr_1e1p}
\end{figure}

\Cref{fig:1m1p_enu} shows the comparison between data and prediction in the $1\mu1p$ $E_\nu^{\rm range}$ distribution.
This observation is used as input to the constraint procedure in \cref{eq:constraint}.
The diagonal uncertainties in the $1e1p$ $E_\nu^{\rm range}$ distribution are reduced by up to a factor of $\sim 2$ across most bins, as shown in \cref{fig:constr_err}.
In \cref{fig:constr_1e1p} we compare observed data to the final constrained prediction and uncertainties in the $1e1p$ $E_\nu^{\rm range}$ distribution.
As can be seen, the constraint procedure slightly increases the $1e1p$ prediction at lower $E_\nu^{\rm range}$ values, though this is not a large effect.
The result shown in \cref{fig:constr_1e1p} forms the basis of the statistical tests discussed in \cref{sec:stat_interp}.

\subsection{Blinded Analysis Approach} \label{sec:blind}

\begin{figure}[h!]
    \centering
    \includegraphics[width=0.6\textwidth]{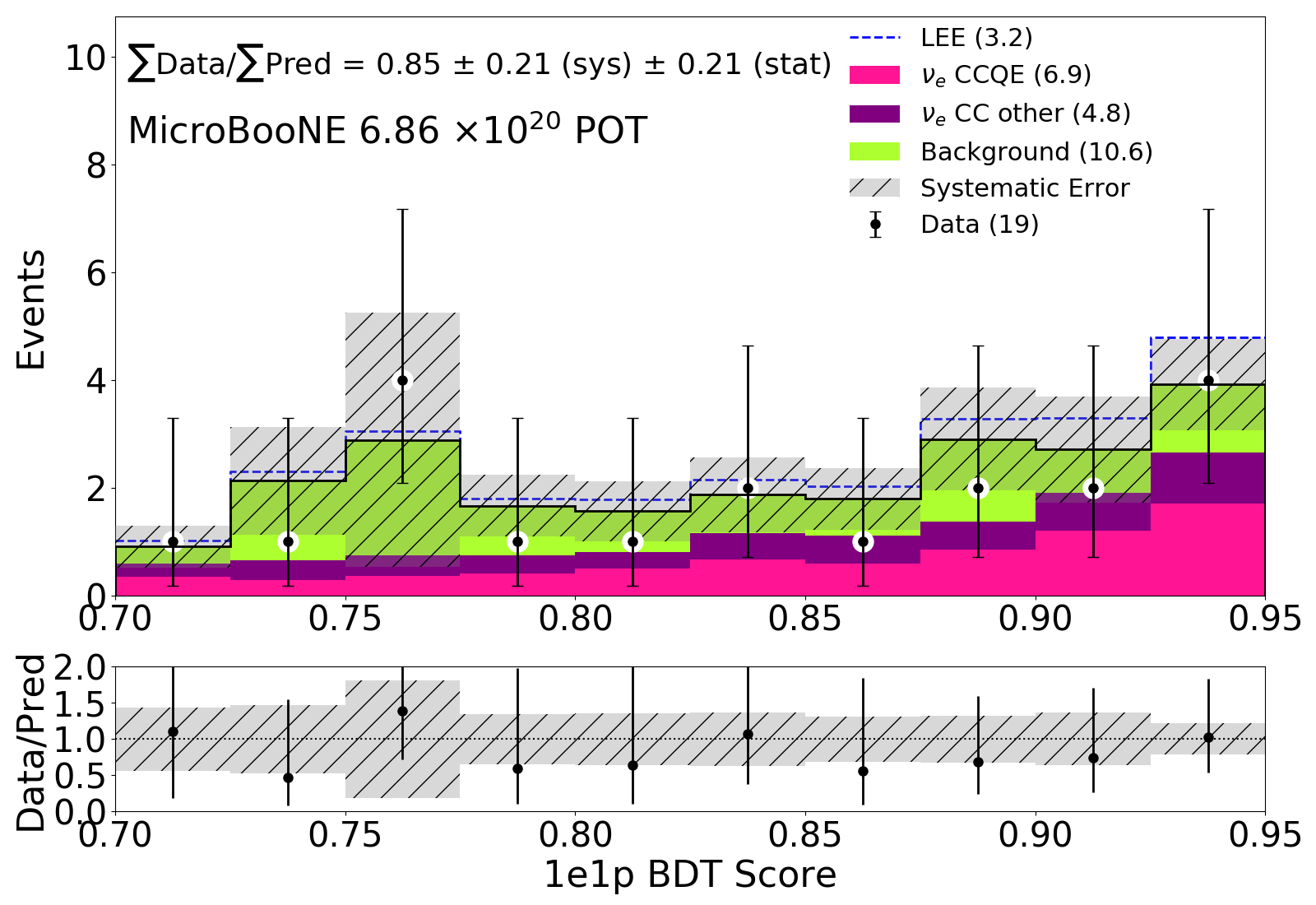}
    \caption{Comparison between data and prediction in the $1e1p$ BDT ensemble average score distribution within the range $[0.7,0.95]$.}
    \label{fig:bdtscore_midBDT}
\end{figure}

Here we briefly cover the blind analysis procedure and related cross-checks performed before examining data in the $1e1p$ signal sample.
The main strategy here was to use a sequential unblinding approach.
Data were first examined for events with a low BDT score ($<0.7$) or a high reconstructed neutrino energy ($> 700$~MeV).
Next, we examined data with a medium BDT score (within $[0.7,0.95]$) or medium energy ($500-700$~MeV).
Only after establishing good agreement between data and prediction in all of these samples were potential signal events first examined, where signal events are defined to have a $1e1p$ BDT score greater than 0.95 and reconstructed neutrino energy below 500~MeV.
This sequence was chosen because sensitivity to the eLEE model is minimized at lower BDT scores and higher neutrino energies.
Moving on to each subsequent stage required signoff from the entire MicroBooNE collaboration.
\Cref{fig:bdtscore_midBDT} shows the comparison between data and prediction in the $1e1p$ BDT ensemble average score distribution for this ``medium BDT score'' sample.
\Cref{fig:midBDTplot} is another example of one of the ``medium BDT score'' plots.

We also explored a sample that made a number of cuts on kinematic variables in order to mimic the $1e1p$ BDT ensemble cut.
As humans are poorer optimizes than BDTs, this sample contained a much larger rate of $\nu_\mu$ background events compared to the official signal selection.
Good agreement between data and prediction was observed across all variables explored within this sample, thus giving confidence in the MC description of backgrounds to the $1e1p$ signal sample.

Finally, before any $\nu_e$ analysis was able to look at data in their signal sample, we each had to pass a series of fake data tests.
In these tests, five different data-like MC samples were generated with various unknown alterations to the nominal MC.
These included, for example, the presence of a large $\nu_e$ eLEE signal and alterations to underlying cross section models.
All three $\nu_e$ analyses ran through the statistical results presented in \cref{sec:stat_interp} for each fake data sample.
In all cases, the statistical conclusions determined from each fake dataset were consistent with the truth-level injected events.
At this point, the $\nu_e$ analyses (including the two-body CCQE analysis) were ready to move on to the real signal dataset.

\section{Statistical Interpretation} \label{sec:stat_interp}

We perform three different statistical tests of the eLEE model: a $\chi^2$ goodness-of-fit test, a $\Delta \chi^2$ two-hypothesis test, and a signal strength scaling test.
We use the combined Neyman-Pearson (CNP) chi-square test statistic~\cite{Ji:2019yca},
\begin{equation}
\chi^2_{\rm CNP} = (\vec{d} - \vec{\mu})^T (M^{\rm stat}_{\rm CNP} + M^{\rm sys})^{-1} (\vec{d} - \vec{\mu}),
\end{equation}
where $\vec{d}$ and $\vec{\mu}$ are the observation and prediction for a particular distribution, $M^{\rm sys}$ is the systematic covariance matrix described in \cref{sec:systematics}, and 
\begin{equation}
(M^{\rm stat}_{\rm CNP})_{ij} \equiv \frac{3 \delta_{ij}}{\frac{1}{d_i} + \frac{2}{\mu_i}}.
\end{equation}
This formulation of the statistical error is meant to more closely approximate the Poisson likelihood, which is important given the low-statistics nature of the $1e1p$ analysis.
Unlike the Poisson likelihood, the CNP chi-square test statistic is compatible with the covariance matrix formalism developed for our systematic uncertainty treatment.

All statistical tests use the constrained $1e1p$ prediction and uncertainties shown in \cref{fig:constr_1e1p} when comparing to the 25 observed data events.
We define two distinct hypotheses which we will refer to in the following discussion: $H_0$, the prediction without the eLEE model, considering only intrinsic $\nu_e$ events and non-$\nu_e$ backgrounds (i.e. the stacked histogram in \cref{fig:constr_1e1p}); and $H_1$, the prediction including the eLEE model (i.e. the dashed blue line in \cref{fig:constr_1e1p}).
Throughout this section, we report probabilities, or $p$-values, calculated using the following frequentist method.
To test the distribution of a specific variable under a given hypothesis, $10^5$ pseudo-experiments are generated by sampling the multivariate distribution defined by the prediction and covariance matrix of the hypothesis.
This gives a predicted rate in each bin of the distribution; we use this to sample from a Poisson distribution to get an integer number of ``observed'' events in each bin.
We calculate the desired test statistic for each pseudo-experiment to build up the probability distribution of that test statistic, which incorporates both systematic and statistical uncertainties.
Then, we can compare the observed value of the test statistic in the real dataset to the calculated probability distribution to determine the probability of observing a more extreme value--this defines the $p$-value.

\subsection{Goodness of Fit} \label{sec:gof}

For the first test, we evaluate the goodness of fit in the $1e1p$ sample for $H_0$ and $H_1$ using the $\chi^2_{\rm CNP}$ test statistic.
This is done both before and after the $1\mu1p$ constraint.
We also consider the goodness-of-fit in both the full analysis range, $200 < E_\nu^{\rm range}\;[{\rm MeV}] < 1200$, and the low energy region  $200 < E_\nu^{\rm range}\;[{\rm MeV}] < 500$, where the eLEE model prediction is highest.
\Cref{tab:gof} shows the $\chi^2_{\rm CNP}$ per degree-of-freedom (DOF) for each of these scenarios, as well as the corresponding $p$-values calculated via the frequentist method described above.

The first thing to note there is significant tension with $H_1$, especially after the $1\mu1p$ constraint, as indicated by the low $p$-values.
This is true both over the full analysis energy range and in the low energy range.
After the constraint, there is greater than $3\sigma$ tension between $H_1$ and the observed data across the full energy range.

There is also disagreement between $H_0$ and the observed data, though not to the same level as the $H_1$ case.
Over the full analysis energy range, the $p$-values in \cref{tab:gof} indicate $2.1\sigma$ ($2.4\sigma$) tension between $H_0$ and the data before (after) the $1\mu1p$ constraint.
This seems to come from two sources: the deficit in data compared to prediction below $500$~MeV in \cref{fig:constr_1e1p}, and the excess in the 800-900~MeV bin in \cref{fig:constr_1e1p}.
The first of these is possibly a real effect--similar low-energy deficits are observed by all three $\nu_e$ analyses~\cite{MicroBooNE:2021tya}.
It is also observed in the $E_\nu^{QE-\ell}$ distribution in \cref{fig:enuqe_lep_1e1p} and is therefore not likely to be related to proton misreconstruction.
Our Michel electron sample covered in \cref{sec:shower_publication} indicates robust modeling of low-energy electron showers, so this is also unlikely to be the culprit.
The excess in the 800-900~MeV bin, however, appears to be a statistical fluctuation.
The other variables examined in this analysis all show good agreement between data and MC and do not have similar narrow excesses.
For instance, the $E_e$, $E_p$, $\theta_e$, and $E_\nu^{QE-\ell}$ distributions in \cref{fig:more_dists_1e1p} have a $\chi^2_{\rm CNP}$/DOF ($p$-value) of 10.68/10 (0.42), 5.77/10 (0.80), 10.19/10 (0.48), and 11.46/10 (0.52), respectively.
The $\chi^2_{\rm CNP}$/DOF and $p$-value of all 36 variables examined in this analysis are reported in the Supplemental Material of Ref.~\cite{MicroBooNE:2021pvo}.

\begin{table}[h!]
    \centering
    \begin{tabular}{c|cc|cc}
    \hline \hline
        \multicolumn{5}{c}{Nominal Predictions} \\
    \hline
        \multirow{2}{*}{Range} & \multicolumn{2}{c|}{$H_0$} & \multicolumn{2}{c}{$H_1$} \\
         & $\chi^2_\text{CNP}/\text{DOF}$ & $p$-value & $\chi^2_\text{CNP}/\text{DOF}$ & $p$-value \\
    \hline
        200--500\,MeV & 6.06/3 & 0.138 & 8.30/3 & 0.053 \\
        200--1200\,MeV & 23.02/10 & 0.024 & 25.37/10 & 0.014 \\
    \hline \hline
        \multicolumn{5}{c}{Constrained Predictions} \\
    \hline
        \multirow{2}{*}{Range} & \multicolumn{2}{c|}{$H_0$} & \multicolumn{2}{c}{$H_1$} \\
         & $\chi^2_\text{CNP}/\text{DOF}$ & $p$-value & $\chi^2_\text{CNP}/\text{DOF}$ & $p$-value \\
    \hline
        200--500\,MeV & 7.91/3 & 0.075 & 17.3/3 & 0.002 \\
        200--1200\,MeV & 25.28/10 & 0.014 & 36.35/10 & $5.0 \times 10^{-4}$ \\
    \hline \hline
    \end{tabular}
    \caption{Results from goodness-of-fit tests comparing observed $1e1p$ data to the $H_0$ and $H_1$ predictions, reported via the $\chi^2_\text{CNP}$ test statistic and the frequentist $p$-value. The top half of the table considers the nominal prediction and uncertainties before the $1\mu 1p$ constraint described in \cref{sec:constraint}, while the bottom half considers the post-constraint prediction and uncertainties.}
    \label{tab:gof}
\end{table}

\subsection{Two Hypothesis Test}

\begin{figure}[h!]
    \centering
    \includegraphics[width=0.6\textwidth]{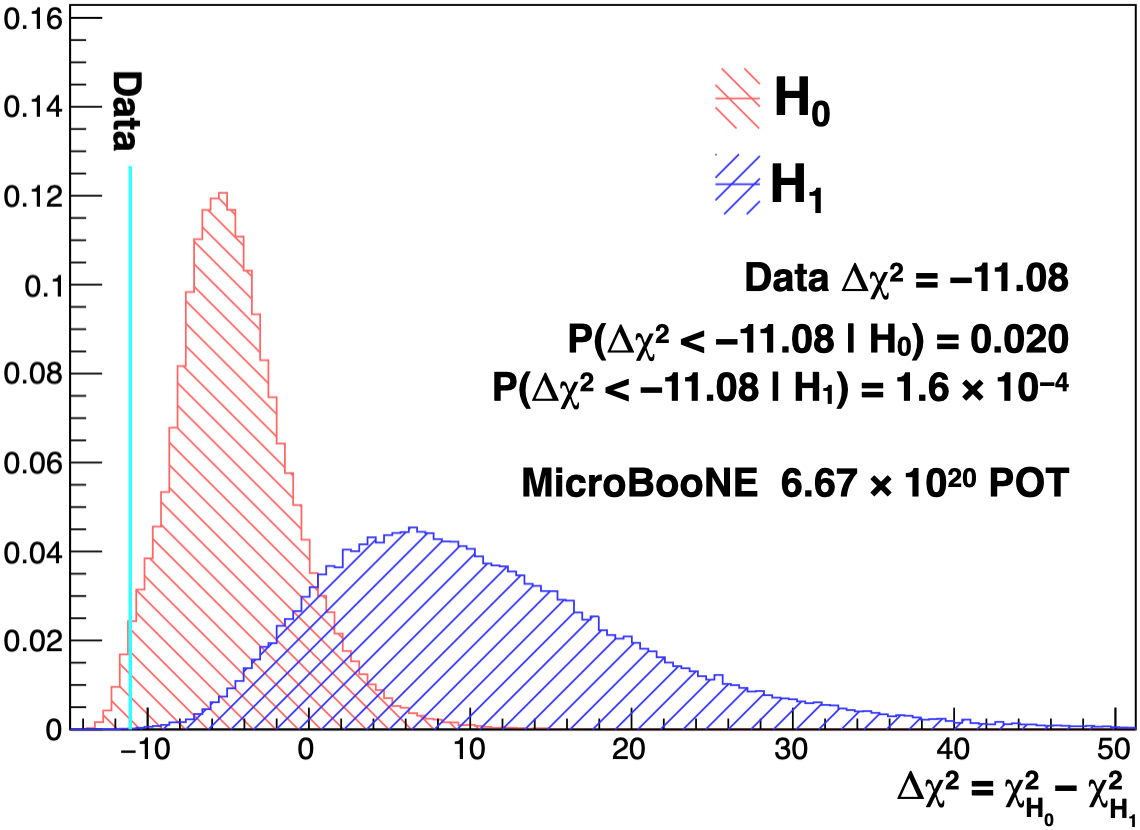}
    \caption{Distributions of the $\Delta \chi^2$ test statistic defined in \cref{eq:two_hypothesis} for $H_0$ (red) and $H_1$ (blue), calculated by generating $10^5$ pseudo-experiments under each hypothesis. The $\Delta \chi^2$ value of the data is also shown.}
    \label{fig:two_hypothesis}
\end{figure}

The next test we perform is a two-hypothesis $\Delta \chi^2$ test between $H_0$ and $H_1$.
Specifically, we use the test statistic
\begin{equation} \label{eq:two_hypothesis}
\Delta \chi^2 = \chi^2_{{\rm CNP},H_0} - \chi^2_{{\rm CNP},H_1}.
\end{equation}
Thus, a positive (negative) value indicates better agreement with $H_1$ ($H_0$).
For this test, we consider the post-$1\mu1p$-constraint prediction and errors for the $1e1p$ sample.
\Cref{fig:two_hypothesis} shows the distribution of this test statistic under both hypotheses, generated using the frequentist method described above.
As expected, the $H_0$ distribution peaks below zero while the $H_1$ distribution peaks above zero.
These distributions can be used to determine the one-sided exclusion sensitivity of each hypothesis in the two-body CCQE analysis.
If we denote the median $\Delta \chi^2$ value under $H_1$ as $(\Delta \chi^2)_{H_1}^{\rm med}$, the probability of $H_0$ generating a larger $\Delta \chi^2$ than $(\Delta \chi^2)_{H_1}^{\rm med}$ is 0.003.
This corresponds to an $H_0$ exclusion sensitivity of $2.7\sigma$.
Conversely, if we define $(\Delta \chi^2)_{H_0}^{\rm med}$ in an analogous way, the probability of $H_1$ generating a \textit{smaller} $\Delta \chi^2$ than $(\Delta \chi^2)_{H_0}^{\rm med}$ is 0.017, corresponding to an $H_1$ exclusion sensitivity of $2.1\sigma$.

As shown in \cref{fig:two_hypothesis}, the observed data result in a test statistic of $\Delta \chi^2 = -11.08$.
The probability of observing a smaller $\Delta \chi^2$ under $H_0$ and $H_1$ is 0.020 and $1.6 \times 10^{-4}$, respectively, corresponding to a Gaussian significance of $2.1\sigma$ and $3.6\sigma$.
Thus, the data exhibits mild tension with $H_0$ and significant tension with $H_1$.
A similar conclusion was reached for the goodness-of-fit test; the tension with $H_0$ likely comes from the same sources discussed in \cref{sec:gof}.
Due to the fact that non-negligible tension is observed with respect to both hypotheses, we report our exclusion significance for $H_1$ using the CL$_s$ method~\cite{Junk:1999kv,Read:451614}.
The CL$_s$ test statistic is defined as
\begin{equation}
{\rm CL}_s = \frac{p_{H_1}}{p_{H_0}},
\end{equation}
where ${\rm CL}_s = 0.008$ for our observed data.
Using this method, we reject $H_1$ in favor of $H_0$ with a Gaussian significance of $2.4\sigma$.

\subsection{Signal Strength Scaling Test}

The final statistical test we perform on our $1e1p$ dataset is a one-parameter signal strength scaling test.
Specifically, we multiply by the eLEE weights in \cref{fig:unfolding_final} by an overall normalization factor $x_{\rm LEE}$ which can vary within $[0,\infty]$.
In this construction, $x_{\rm LEE} = 0$ corresponds to $H_0$ and $x_{\rm LEE} = 1$ corresponds to $H_1$.
We can use the Feldman-Cousins procedure~\cite{Feldman:1997qc} along with the test statistic
\begin{equation}
\Delta \chi^2 (x_{\rm LEE}) = \chi^2_{\rm CNP} (x_{\rm LEE}) - \min_{x_{\rm LEE} \in [0,\infty]} \{ \chi^2_{\rm CNP} (x_{\rm LEE})\}
\end{equation}
to determine confidence intervals on $x_{\rm LEE}$.
We can use the Asimov dataset, in which the observation is exactly the predicted value in each bin, to determine the sensitivity to each hypothesis in this test.
Under $H_0$, the expected upper bound on $x_{\rm LEE}$ at the 90\% ($2\sigma$) confidence level is 0.75 (0.98).
Under $H_1$, the expected confidence interval at the $1\sigma$ ($2\sigma$) confidence level is $[0.53, 1.66]$ ($[0.28, 2.67]$), and the expected sensitivity to rule out $x_{\rm LEE} = 0$ is $2.8\sigma$.

\begin{figure}
    \centering
    \includegraphics[width=0.6\textwidth]{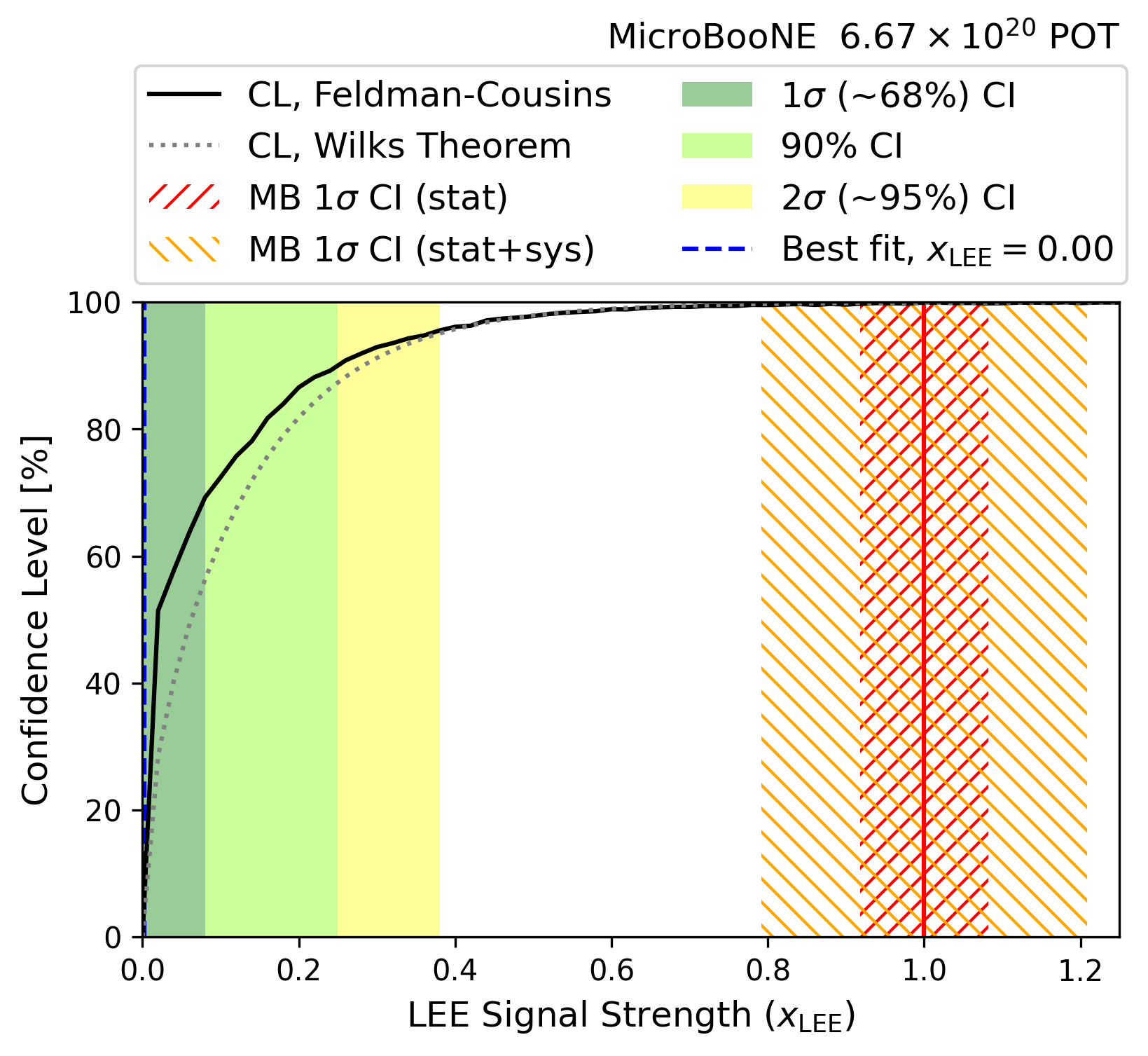}
    \caption{Confidence intervals on $x_{\rm LEE}$ calculating using the Feldman-Cousins procedure. The solid and dotted lines indicate the confidence level with which a given $x_{\rm LEE}$ is disfavored, calculated using the Feldman-Cousins method~\cite{Feldman:1997qc} and Wilks theorem~\cite{Wilks:1938dza}, respectively. The MiniBooNE statistical and systematic errors are shown as a band around $x_{\rm LEE} = 1$.}
    \label{fig:sig_scaling}
\end{figure}

\Cref{fig:sig_scaling} shows the confidence intervals on $x_{\rm LEE}$ derived from the observed data.
The best-fit value is $x_{\rm LEE} = 0$, which is expected given the deficit in data at the lowest neutrino energies where the eLEE model weights are the largest.
The 90\% ($2\sigma$) upper bound on $x_{\rm LEE}$ is 0.25 (0.38).
These upper bounds are stronger than the expected confidence intervals under $H_0$, which is mainly due to the aforementioned deficit at low neutrino energies.

\section{Discussion and Outlook}

The observation in the $1e1p$ signal channel of the two-body CCQE analysis is inconsistent with a MiniBooNE-like excess of $\nu_e$ CCQE interactions.
Under a two-hypothesis test, we rule out the nominal eLEE model ($H_1$) at the $2.4\sigma$ using the CL$_s$ method.
We can set an upper bound on the signal strength scaling parameter $x_{\rm LEE}$ of 0.38 at the $2\sigma$ level.

\begin{figure}[h!]
     \centering
     \begin{subfigure}[b]{0.45\textwidth}
         \centering
         \includegraphics[width=\textwidth]{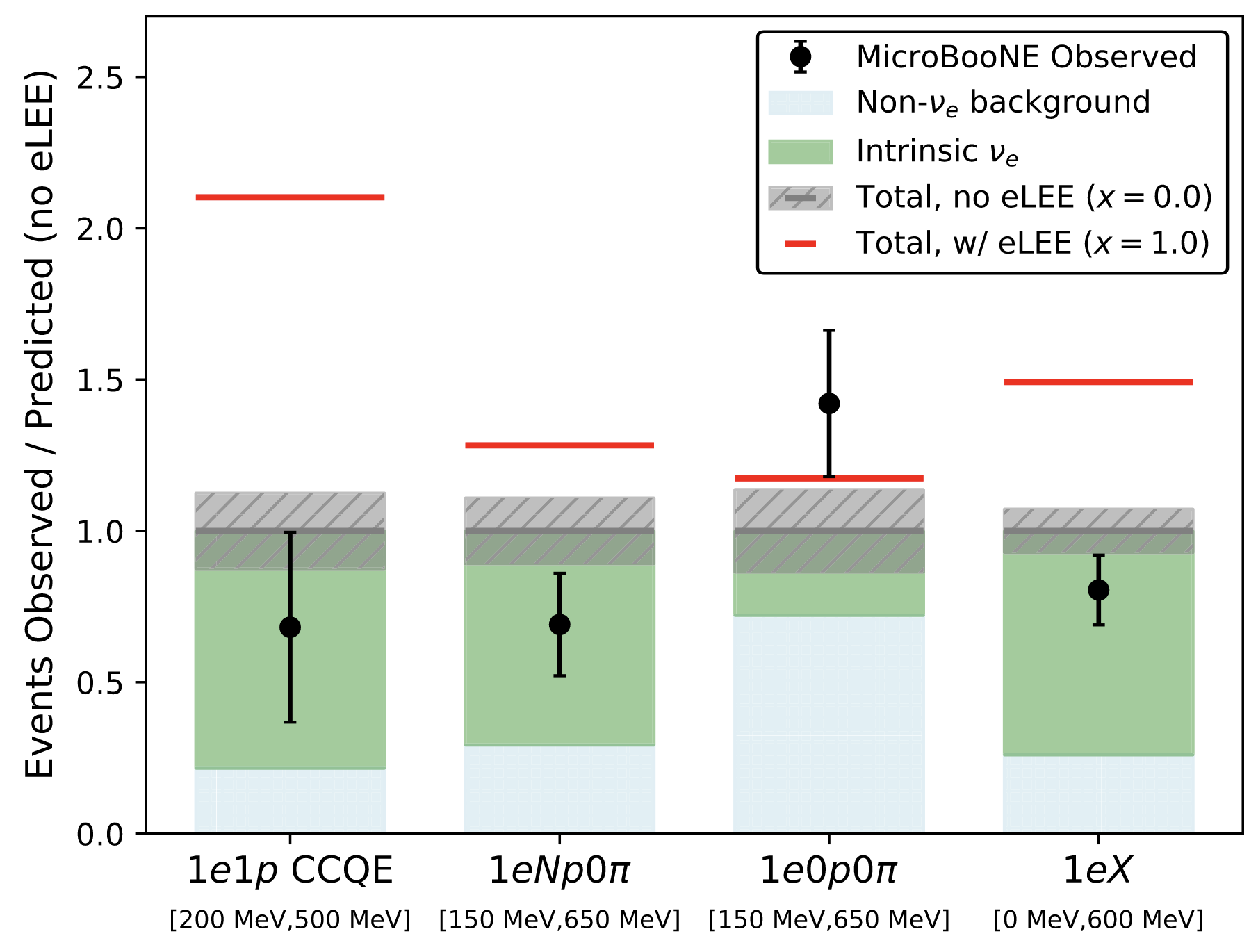}
         \caption{}
         \label{fig:ub_nue_1}
     \end{subfigure}
     \hfill
     \begin{subfigure}[b]{0.45\textwidth}
         \centering
         \includegraphics[width=\textwidth]{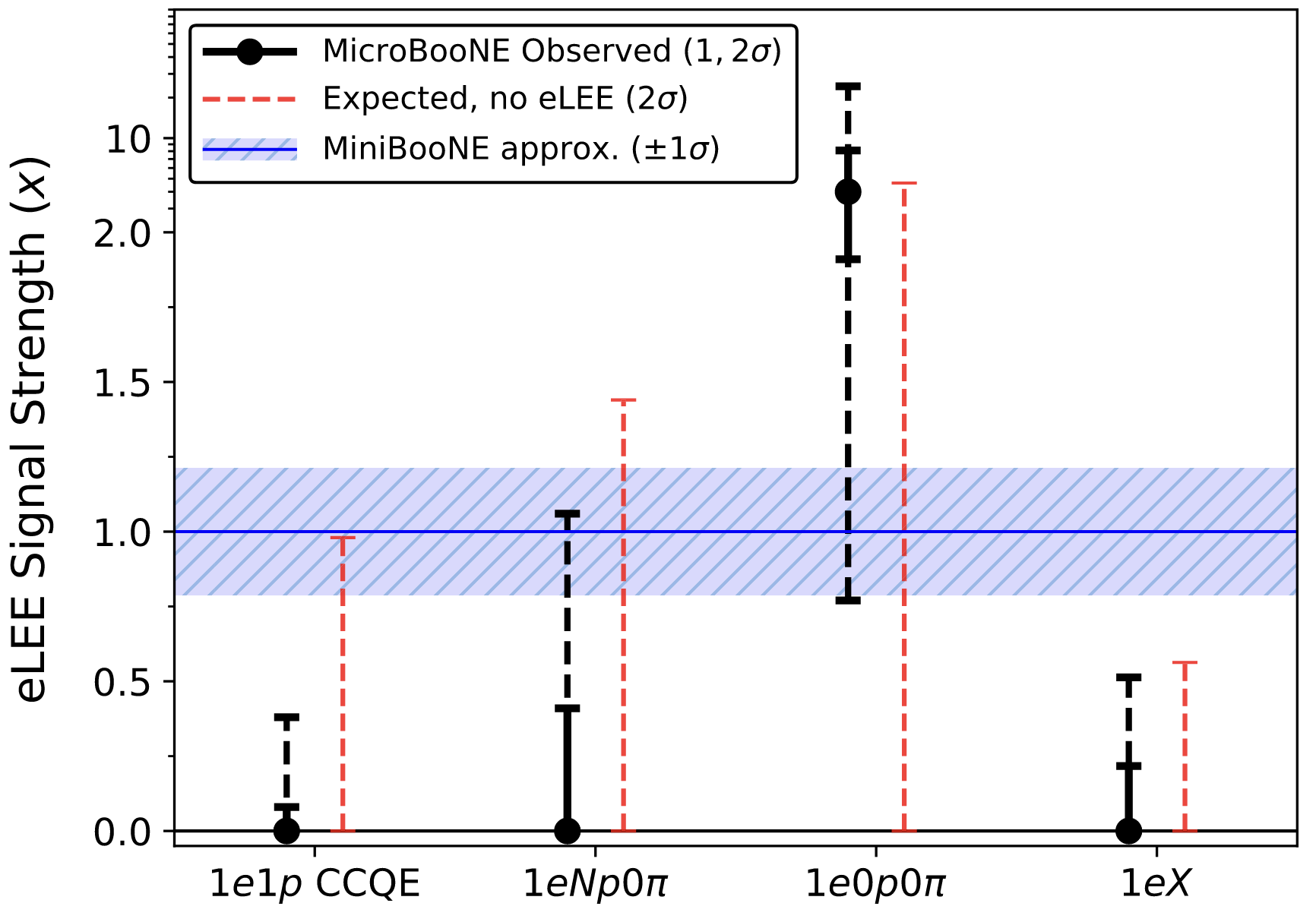}
         \caption{}
         \label{fig:ub_nue_2}
     \end{subfigure}
        \caption{\Cref{fig:ub_nue_1} shows the observation compared to the nominal ($H_0$) prediction in all four signal channels from the three MicroBooNE $\nu_e$ analyses, including statistical errors on the data points and systematic errors on the prediction. The eLEE prediction ($H_1$) is also indicated by the red line. \Cref{fig:ub_nue_2} shows the observed $1\sigma$ and $2\sigma$ confidence intervals on $x_{\rm LEE}$ from all four signal channels. The $2\sigma$ expected sensitivity of each channel is shown in red. 
        }
        \label{fig:ub_nue}
\end{figure}

This is consistent with the results from all three MicroBooNE $\nu_e$ analyses~\cite{MicroBooNE:2021tya}.
As shown in \cref{fig:ub_nue_1}, three of the four signal channels from the three $\nu_e$ analyses observe a deficit of selected events at the lowest neutrino energies, where the eLEE model is supposed to peak.
The exception is the $1e0p0\pi$ channel from the MiniBooNE-like sample; however, this is the least sensitive of the four channels, as it is dominated by non-$\nu_e$ backgrounds.
\Cref{fig:ub_nue_2} shows the results from the signal strength scaling test in all four channels.
With the exception of the $1e0p0\pi$ channel, all analyses rule out $x_{\rm LEE} = 1$ at greater than the $2\sigma$ confidence level.
The most sensitive analyses observe an upper bound $x_{\rm LEE} \lesssim 0.5$ at the $2\sigma$ confidence level.

The first MicroBooNE results also included a search for $\Delta \to N \gamma$ decays as an explanation for the MiniBooNE excess.
The analysis did not observe an excess of $\Delta \to N \gamma$ candidate events and ruled out the required enhancement to explain the MiniBooNE anomaly at the 95\% confidence level.
The full details of this analysis can be found in Ref.~\cite{MicroBooNE:2021zai}.

Thus, the MicroBooNE results disfavor both an enhancement of $\nu_e$ CC interactions and $\Delta \to N \gamma$ decays as the sole explanation of the MiniBooNE excess.
Potential explanations of the MiniBooNE anomaly must now remain consistent with these MicroBooNE results.
One promising avenue involves dark sector models which produce a photon or $e^+ e^-$ pair in the MiniBooNE detector.
Such event signatures have not yet been explored by MicroBooNE, though this work is in progress.
Cref{ch:neutrissimos} explores a photon-based explanation of MiniBooNE, involving the decay of a heavy neutral lepton to a single photon via a transition magnetic moment.
The theoretical and experimental neutrino physics communities are actively working to develop innovative models to explain the MiniBooNE LEE and equally innovative measurements to test those models.

The following subsections close with two projects I co-led with various collaborators to follow up on the state of the MiniBooNE excess after the MicroBooNE results.
The first examines the $3+1$ model simultaneously in MiniBooNE and MicroBooNE~\cite{MiniBooNE:2022emn}.
The second explores the implications of a $\overline{\nu}_e$-based MiniBooNE LEE for the MicroBooNE results~\cite{Kamp:2023mjn}.

\subsection{Publication: \textit{MiniBooNE and MicroBooNE Combined Fit to a $3 + 1$ Sterile Neutrino Scenario}} \label{sec:MBuB_sterile_paper}

The initial results from the MicroBooNE $\nu_e$ analyses did not explicitly test short-baseline $\nu_\mu \to \nu_e$ oscillations from the $3+1$ model.
Shortly after the MicroBooNE results came out, two studies appeared assessing the sensitivity of the MicroBooNE $\nu_e$ analyses to the $3+1$ model~\cite{Arguelles:2021meu,Denton:2021czb}.
Following up on this, the MiniBooNE collaboration released the first combined fit to a $3+1$ model using MicroBooNE and MiniBooNE data~\cite{MiniBooNE:2022emn}.
This study was led by myself and Austin Schneider.
The full \textit{Physical Review Letters} publication is included below.
We considered the two-body CCQE and inclusive MicroBooNE $\nu_e$ analyses, as these placed the most stringent constraints on the eLEE model and provided data releases with the requisite detail to perform a $3+1$ fit~\cite{MicroBooNE:2021tya}.
The most important result from this study is Figure~2, which indicates that there are still allowed regions in $3+1$ parameter space at the $3\sigma$ confidence level in the MiniBooNE-MicroBooNE combined fit.
This was also the first MiniBooNE analysis to account for all three oscillation probabilities in \cref{eq:sterile_osc_prob}: $\nu_e$ appearance, $\nu_\mu$ disappearance, and $\nu_e$ disappearance.
Previous $3+1$ fits to MiniBooNE data have only considered $\nu_e$ appearance, as this is the dominant effect in the electron-like channel.
However, $\nu_e$ and $\nu_\mu$ disappearance can have a non-negligible impact on the intrinsic $\nu_e$ and misidentified $\nu_\mu$ background rate in the MiniBooNE electron-like prediction.
The $\nu_e$ disappearance probability is also important when considering MicroBooNE data, as the prediction in each MicroBooNE $\nu_e$ analysis is dominated by intrinsic $\nu_e$ interactions~\cite{MicroBooNE:2021pvo,MicroBooNE:2021nxr}.
MicroBooNE has since released official constraints on the $3+1$ model using data from the inclusive $\nu_e$ analysis~\cite{MicroBooNE:2022sdp}--these constraints are consistent with our implementation.

\includepdf[pages=-]{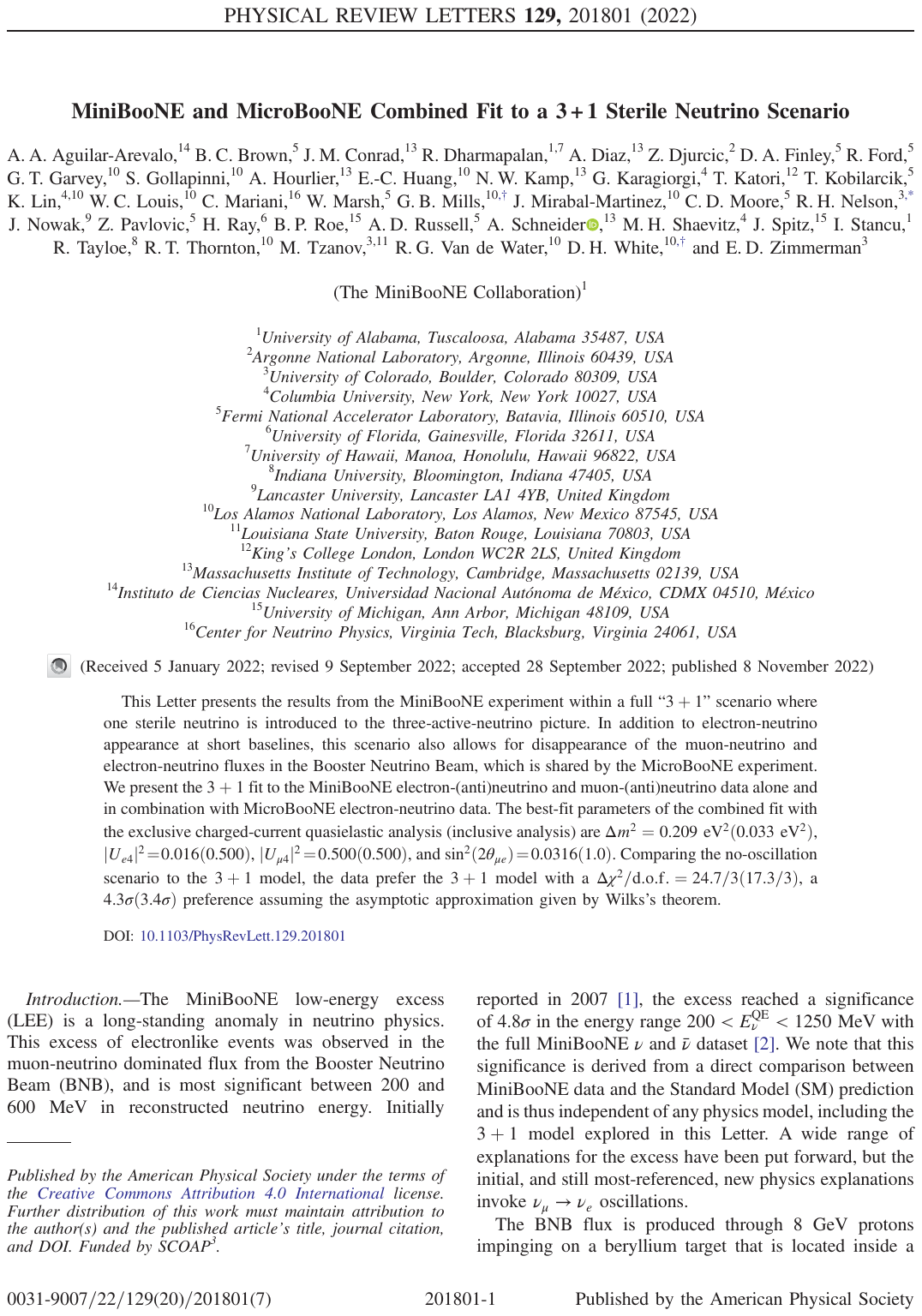}
\includepdf[pages=-]{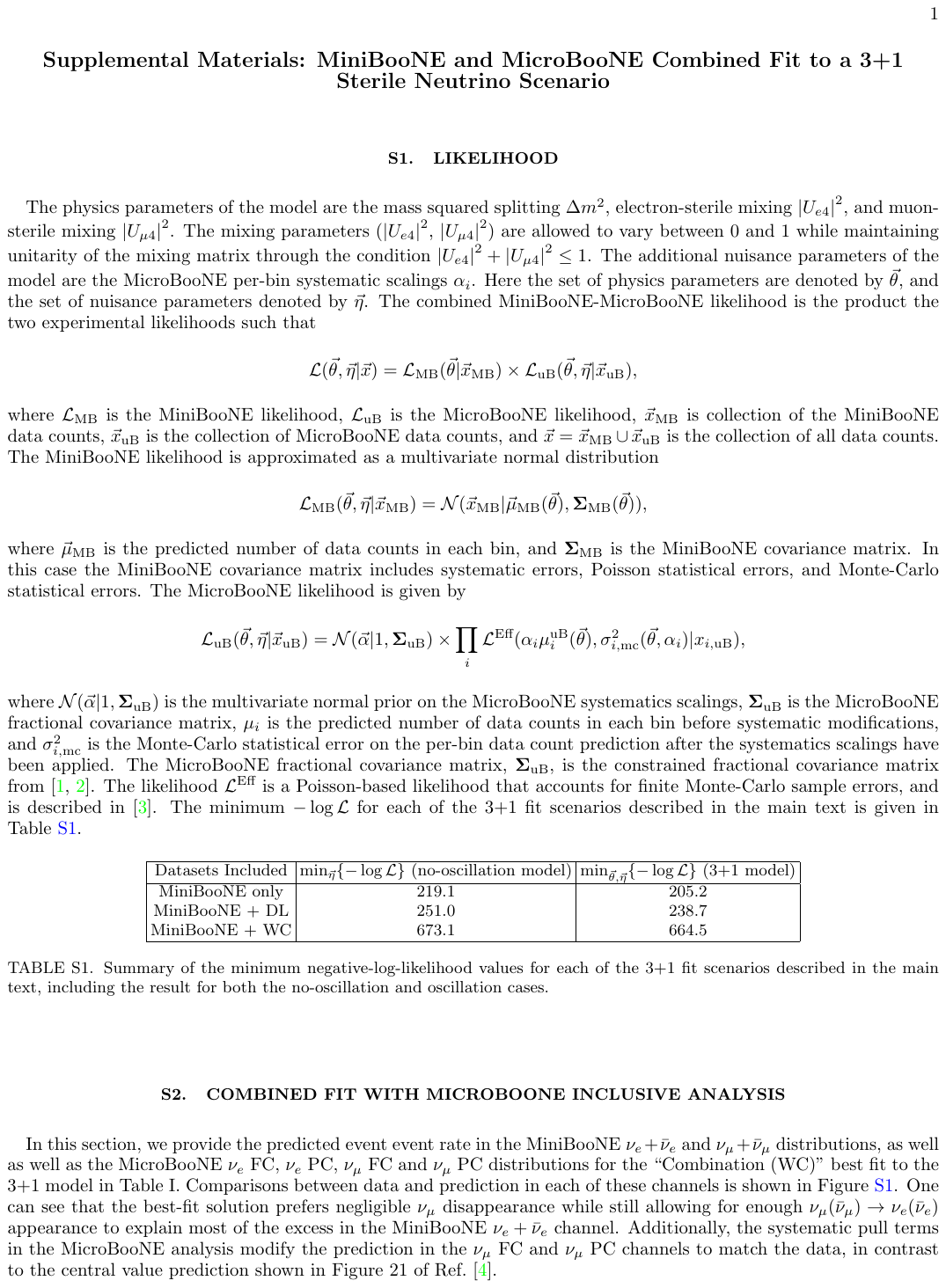}

\subsection{Publication: \textit{Implications of MicroBooNE’s low sensitivity to electron antineutrino interactions in the search for the MiniBooNE excess}} \label{sec:miniboone_antinu}

The first MicroBooNE results indeed set strong limits on $\nu_e$ CC interactions as an explanation for the entire MiniBooNE LEE.
However, as pointed out in Ref.~\cite{Kamp:2023mjn}, this conclusion changes if the MiniBooNE LEE comes instead from $\overline{\nu}_e$ CC interactions.
This is driven mainly by differences in the low-energy suppression of $\overline{\nu}_e$-nucleus and $\nu_e$-nucleus cross sections, coming from interference between the vector and axial form factors as well as nucleon binding energy effects.
Binding energy effects are especially important in MicroBooNE--within a non-isoscalar nucleus like ${}^{40}$Ar, it takes more energy for a $\nubar{e}$ to turn a proton into a neutron than it does for a $\nu_e$ to turn a neutron into a proton. 
This is not necessarily the case in MiniBooNE, as ${}^{12}$C is an isoscalar nucleus.
This study was led by myself and Matheus Hostert.
The full \textit{Physical Review D} publication can be found below.
The most important results are shown in Figure~5 and Figure~6, which indicate that the MicroBooNE sensitivity to the MiniBooNE LEE decreases as more of the excess is attributed to $\overline{\nu}_e$ rather than $\nu_e$ CC interactions.
Specifically, the MicroBooNE data are consistent at the $2\sigma$ confidence level with a scenario in which the entire MiniBooNE LEE comes from $\overline{\nu}_e$ interactions.

\includepdf[pages=-]{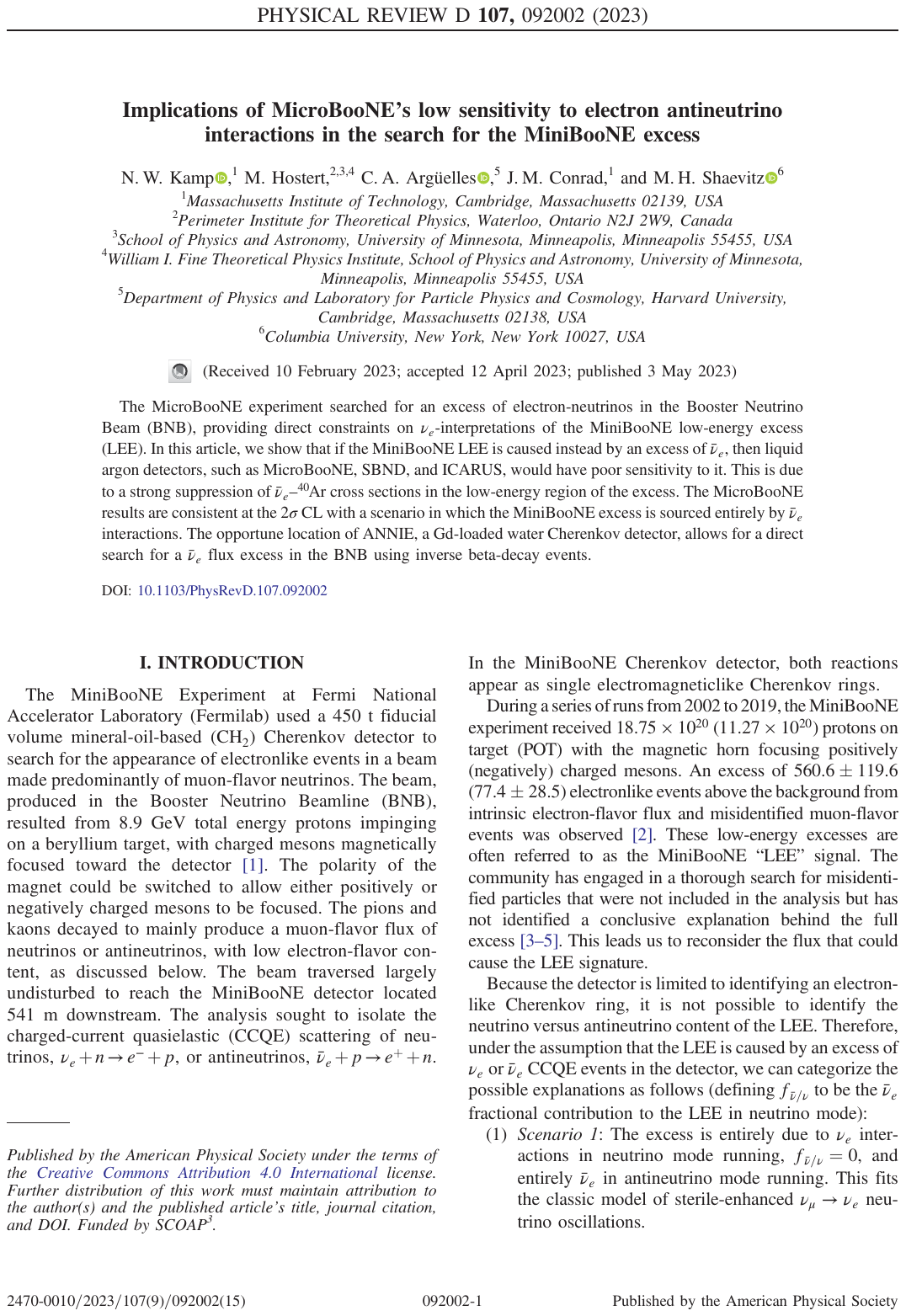}

\chapter{Neutrissimos: Heavy Neutral Leptons with a Dipole Moment} \label{ch:neutrissimos}

In this chapter, we discuss a phenomenological explanation of the MiniBooNE excess involving both an eV-scale sterile neutrino and an MeV-scale heavy neutral lepton (HNL) with an effective transition magnetic moment coupling to active neutrinos.
This ``mixed model'' was introduced in Ref.~\cite{Vergani:2021tgc} and refined in Ref.~\cite{Kamp:2022bpt}. 
The HNL in this model is hereafter referred to as a ``neutrissimo''~\cite{Kamp:2022bpt}, a term which has previously been used to describe interacting HNLs in the literature~\cite{Loinaz:2004qc,NuSOnG:2008weg}.

We begin with a brief overview of the theoretical framework behind neutrissimos.
We then motivate the mixed model and discuss its implications for global neutrino experiments.
A robust evaluation of the neutrissimo signal in MiniBooNE is performed.
For a neutrissimo mass $m_{\mathcal{N}} \sim 500~{\rm MeV}$, single photons from visible neutrissimo decay are shown to provide a reasonable explanation of the energy and angular distributions of the MiniBooNE excess.
World-leading constraints on neutrissimos are derived using elastic scattering data from the MINER$\nu$A experiment~\cite{MINERvA:2015nqi,Valencia:2019mkf,MINERvA:2022vmb}.
While the MINER$\nu$A data exclude a large region of parameter space for $m_{\mathcal{N}} = \mathcal{O}(100~{\rm MeV})$, they do not rule out the MiniBooNE solution at the 95\% confidence level.
The details behind this analysis are given in Ref.~\cite{Kamp:2022bpt}; the full published manuscript is included in \cref{sec:neutrissimo_paper}.

\textit{Publications covered in this chapter for which I held a leading role: \cite{Vergani:2021tgc,Kamp:2022bpt}}

\section{Dipole-Portal Neutrissimos} \label{sec:neutrissimo_model}

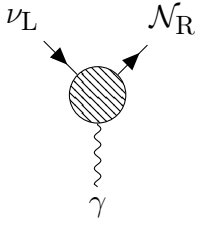
\begin{figure}
\begin{centering}
\centering
    \begin{tikzpicture}
    \begin{feynman}
      \vertex[blob] (m) at ( 0, 0) {\contour{white}{}};
      \vertex (a) at (-1,1) {$\nu_L$};
      \vertex (b) at ( 1,1) {$\mathcal{N}_R$};
      \vertex (c) at (0, -1.5) {$\gamma$};
      \diagram* {
        (a) -- [fermion] (m) -- [fermion] (b),
        (m) -- [boson] (c),
      };
    \end{feynman}
    \end{tikzpicture}
\caption{\label{fig:dipole_effective} Feynman diagram depicting the effective dipole operator of \cref{eq:dipole_operator}.}
\end{centering}
\end{figure}

\begin{figure}[h!]
    \centering
     \begin{subfigure}[b]{0.3\textwidth}
         \centering
         \includegraphics[width=\textwidth]{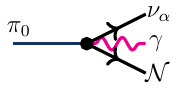}
         \caption{}
         \label{fig:neutrissimo_dalitz}
     \end{subfigure}
     \hfill
     \begin{subfigure}[b]{0.3\textwidth}
         \centering
         \includegraphics[width=\textwidth]{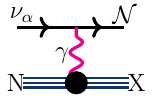}
         \caption{}
         \label{fig:neutrissimo_primakoff}
     \end{subfigure}
     \hfill
     \begin{subfigure}[b]{0.3\textwidth}
         \centering
         \includegraphics[width=\textwidth]{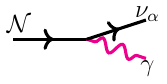}
         \caption{}
         \label{fig:neutrissimo_decay}
     \end{subfigure}
     \hfill
        \caption{New interactions involving the neutrissimo that are enabled by the dipole operator in \cref{eq:dipole_operator}, including three-body $\pi^0$ decay (\cref{fig:neutrissimo_dalitz}), Primakoff-like upscattering (\cref{fig:neutrissimo_primakoff}), and neutrissimo decay (\cref{fig:neutrissimo_decay}).}
        \label{fig:neutrissimo_feyn_diagrams}
\end{figure}

The neutrissimo $\mathcal{N}$ is a right-handed neutrino that couples to the active neutrinos through a transition magnetic moment.
This coupling is described by the dimension-five dipole operator
\begin{equation} \label{eq:dipole_operator}
\mathcal{L}_D  = d_{\alpha \mathcal{N}}\, \overline{\nu_\alpha} \sigma_{\mu\nu}  F^{\mu \nu} \mathcal{N}_R  + \text{ h.c.},
\end{equation}
where $d_{\alpha \mathcal{N}}$ is the dipole coupling between the right-handed neutrissimo $\mathcal{N}_R$ and the left-handed weak flavor eigenstate $\nu_\alpha$, and $F^{\mu \nu} = \partial^\mu A^\nu - \partial^\nu A^\mu$ is the EM field strength tensor.
The dipole coupling has units of inverse energy, as \cref{eq:dipole_operator} is an effective operator.
This means it is agnostic to the ultraviolet (UV) completion: the short-distance physics generating the dipole coupling at a higher energy scale.
Possible UV completions for the dipole operator include (but are not limited to) $SU(2)_L \times SU(2)_R$ theories~\cite{Shrock:1982sc}, leptoquark models~\cite{Brdar:2021ysi}, and models introducing additional Higgs multiplets or other charged scalars~\cite{Pal:1981rm,Babu:2020ivd,Georgi:1990za,Lindner:2017uvt}.
We will treat neutrissimos within the effective field theory framework throughout this chapter.
\Cref{fig:dipole_effective} shows a Feynman diagram of the effective operator in \cref{eq:dipole_operator}.
It facilitates a number of new interactions involving the neutrissimo, the most relevant of which are shown in \cref{fig:neutrissimo_feyn_diagrams}.

\Cref{eq:dipole_operator} is only valid below the electroweak scale, as it does not respect $SU(2)_L \times U(1)_Y$ gauge invariance.
Before electroweak symmetry breaking (EWSB), the dipole coupling is described by the gauge invariant dimension six operator
\begin{equation} \label{eq:dipole_electroweak}
\mathcal{L} \supset \frac{1}{\Lambda^2} \overline{L_\alpha}\widetilde{H} \sigma^{\mu \nu}  \mathcal{N}_R\left( C_B^\alpha \, B_{\mu \nu} + C_W^\alpha\,W^a_{\mu \nu}\sigma_a\right) + \text{ h.c.},
\end{equation}
where $\Lambda$ is the UV completion scale, $L_\alpha$ is the left-handed lepton $SU(2)_L$ doublet, $\widetilde{H} = i \sigma_2 H^*$ is the conjugate Higgs field, and $C_B^\alpha$ and $C_W^\alpha$ are the Wilson coefficients for couplings to the $B_\mu$ and $W_\mu$ vector gauge bosons, respectively.
The dipole coupling in \cref{eq:dipole_operator} after EWSB is then given by
\begin{equation}
d_{\alpha \mathcal{N}} = \frac{v_h}{\sqrt{2} \Lambda^2} (C_B^\alpha \cos \theta_W + C_W^\alpha \sin \theta_W ), 
\end{equation}
where $v_h$ is the Higgs vacuum expectation value and $\theta_W$ is the Weinberg angle.
Note that couplings are also generated with the $W$ and $Z$ bosons after EWSB, but interactions involving these couplings are suppressed by $G_F$ and are thus not as important as the photon coupling.

Transition magnetic moments between different neutrino species have been discussed extensively in the literature~\cite{Giunti:2008ve,deGouvea:2006hfo,Balantekin:2013sda,Vogel:1989iv,Kayser:1982br,Brdar:2021ysi,Magill:2018jla,Georgi:1990za,Babu:2021jnu}.
The imprint of heavy neutral leptons with transition magnetic moments in experimental data has received particular attention~\cite{Magill:2018jla,Brdar:2021ysi,Gninenko:1998nn,Gninenko:2009ks,Gninenko:2010pr,Gninenko:2012rw,Masip:2012ke,Coloma:2017ppo,Plestid:2020vqf,Schwetz:2020xra,Atkinson:2021rnp,Bolton:2021pey,Alvarez-Ruso:2021dna,Gustafson:2022rsz,Ovchynnikov:2022rqj,Zhang:2023nxy,Brdar:2023tmi}.
One potential complication of this model is that large transition magnetic moments often lead to large Dirac masses $m_{\nu \mathcal{N}}$ for neutrinos~\cite{Brdar:2021ysi,Magill:2018jla,Kamp:2022bpt}.
This is because the photon line can typically be removed from \cref{fig:dipole_effective}, resulting in a contribution to the Dirac mass term $\mathcal{L} \supset m_{\nu \mathcal{N}} \overline{\nu}_L \mathcal{N}_R$.
Dipole coupling strengths of interest to MiniBooNE ($d \sim 10^{-6}$~GeV) would generate Dirac mass contributions $m_{\nu \mathcal{N}} = \mathcal{O({\rm MeV}})$~\cite{Brdar:2021ysi}.
Considering a neutrissimo mass $m_{\mathcal{N}} \sim 500$~MeV, the typical seesaw relation of \cref{eq:seesaw} would predict active neutrino masses $m_\nu >> 1~{\rm eV}$, in conflict with existing limits.
This can be remedied by the inverse seesaw mechanism~\cite{Mohapatra:1986aw,Mohapatra:1986bd}, in which the smallness of active neutrino masses comes from approximate lepton number conservation without relying on large right-handed neutrino masses.~\cite{Brdar:2021ysi,Kamp:2022bpt}.

Another complication is the potential existence of mass mixing between the neutrissimo and the active neutrinos.
This mixing is heuristically given by $U_{\alpha \mathcal{N}} \sim m_{\nu \mathcal{N}}/m_{\mathcal{N}}$ and is large even in the inverse seesaw mechanism~\cite{Brdar:2021ysi}.
If this operator is not suppressed, the interactions of the neutrissimo are dominated by those involving the $W$ and $Z$ bosons.
Any potential mass mixing would be subject to constraints from accelerator neutrino experiments, which can be strong for $m_{\mathcal{N}} = \mathcal{O}(100~{\rm MeV})$~\cite{Arguelles:2021dqn}.
Specifically, searches for HNLs produced in kaon decay at T2K (using the ND280 near detector)~\cite{T2K:2019jwa}, E949~\cite{E949:2014gsn}, and PS191~\cite{Bernardi:1987ek} set upper bounds on the muon-flavor mixing of $|U_{\mu \mathcal{N}}|^2 \leq 10^{-8}-10^{-9}$ up to the kaon mass threshold of $m_{\mathcal{N}} \sim 400~{\rm MeV}$.
Within the range $400 \leq m_{\mathcal{N}}~[{\rm MeV}] \leq 2000$, NuTeV is able to set upper bounds of $|U_{\mu \mathcal{N}}|^2 \leq 10^{-6}-10^{-7}$ by searching for HNLs produced in $D$ meson decays.

In order to suppress mass mixing, the Dirac mass itself must be small compared to the neutrissimo mass.
One possible method to accomplish this is the approximate horizontal $SU(2)_H$ symmetry proposed by Voloshin~\cite{Voloshin:1987qy}.
The idea of an approximate $SU(2)_H$ symmetry initially gained popularity as a means to enable the $\nu_e$ diagonal/transition magnetic moment solution to the solar neutrino problem~\cite{Babu:1989wn,Leurer:1989hx,Babu:1989px}.
For the heavy neutral lepton case, we consider a scenario in which $(\nu_L^C,\mathcal{N}_R)$ transforms as a doublet under $SU(2)_H$~\cite{Barbieri:1988fh,Brdar:2021ysi,Lindner:2017uvt}.
The antisymmetric nature of the dipole operator $\overline{\nu}_\alpha \sigma_{\mu \nu} F^{\mu \nu} \mathcal{N}_R$ in flavor space means that it transforms like a singlet under $SU(2)_H$, while the symmetric nature of the Dirac mass term makes it a triplet under $SU(2)_H$.
Thus, approximate $SU(2)_H$ conservation will suppress the Dirac mass contribution from the photon-less version of \cref{fig:dipole_effective} while allowing for large contributions to the transition magnetic moment.
Though it is not clear the level of fine-tuning required within this model to accommodate the MiniBooNE solution~\cite{Lindner:2017uvt}, we consider this sufficient motivation to consider a scenario in which the mass mixing operator of the heavy neutrino is negligible compared to the dipole operator.
It is also worth pointing out the leptoquark UV completion of the dipole operator proposed in Ref.~\cite{Brdar:2021ysi}, which can naturally suppress the mass mixing of neutrissimos via an $SU(2)_H$ symmetry without the need for significant fine-tuning.

\section{Overview of the Mixed Model} \label{sec:mixed_model}

The mixed model consists of the eV-scale sterile neutrino described in \cref{sec:anomalies} and the MeV-scale neutrissimo described above in \cref{sec:neutrissimo_model}.
This model is motivated by the significant tension observed between appearance and disappearance experiments in $3+1$ global fits~\cite{Diaz:2019fwt,Hardin:2022muu,Dentler:2018sju,Gariazzo:2017fdh}, which comes from strong signals indicating $\nu_e$ appearance and $\nu_e$ disappearance coupled with the lack of such a signal in the $\nu_\mu$ disappearance channel.
This tension is measured using the parameter goodness-of-fit (PG) test~\cite{Maltoni:2003cu}, which uses the test statistic
\begin{equation}
\begin{split}
&\chi^2_{PG} = \chi^2_{\rm global} - (\chi^2_{\rm app} + \chi^2_{\rm dis}), \\
&N_{PG} = (N_{\rm app} + N_{\rm dis}) - N_{\rm global},
\end{split}
\end{equation}
where $\chi^2_{\rm global}$, $\chi^2_{\rm app}$, and $\chi^2_{\rm dis}$ are the total $\chi^2$ at the $3+1$ best-fit parameters for global, appearance-only, and disappearance-only experiments, respectively.
The number of degrees-of-freedom $N$ is defined similarly.
The probability that the appearance and disappearance experiments come from the same underlying $3+1$ model can be calculated using the $\chi^2_{PG}$ distribution for $N_{PG}$ degrees-of-freedom.

\begin{figure}[tb]
\begin{center}
\includegraphics[width=0.3\columnwidth]{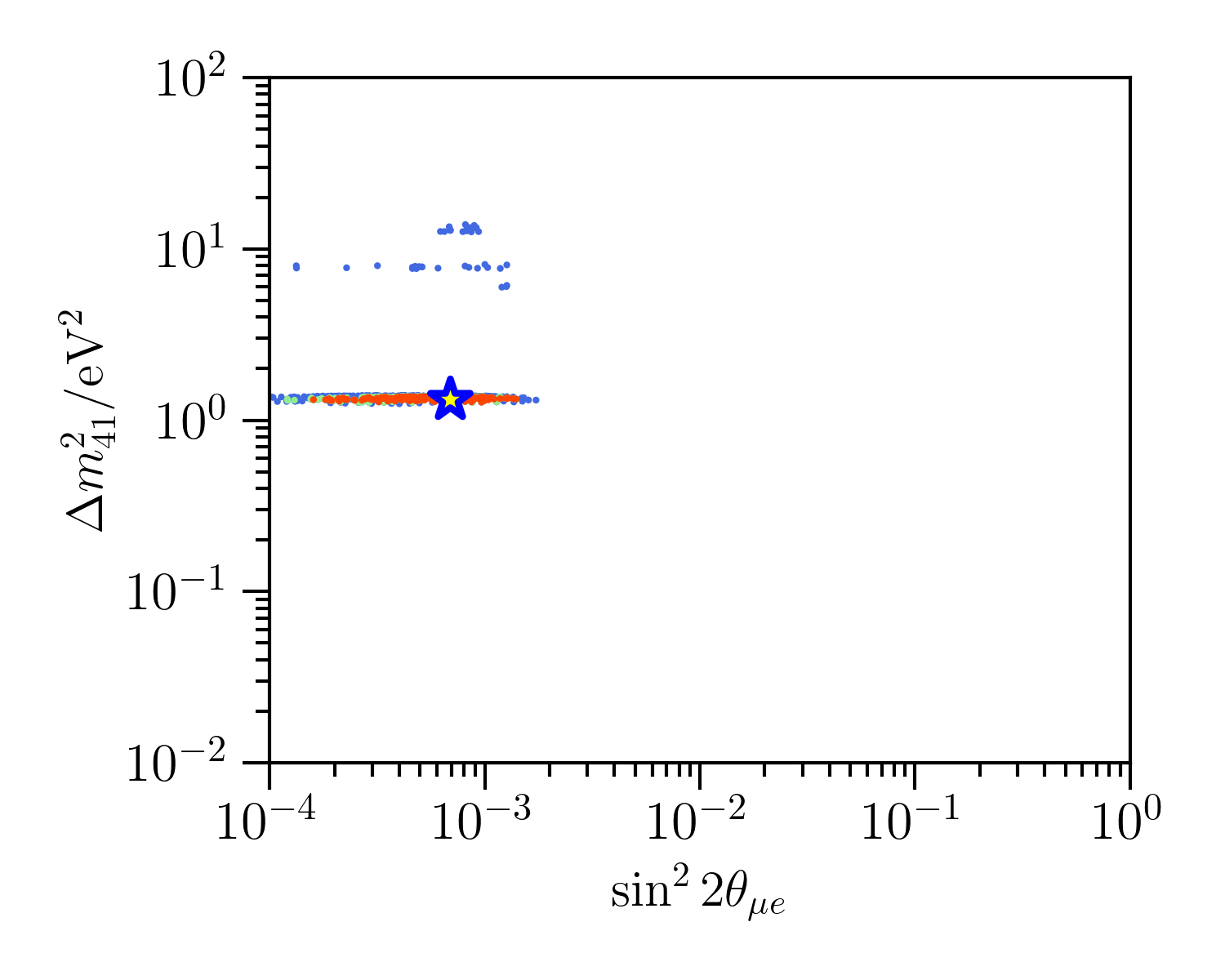}
\includegraphics[width=0.3\columnwidth]{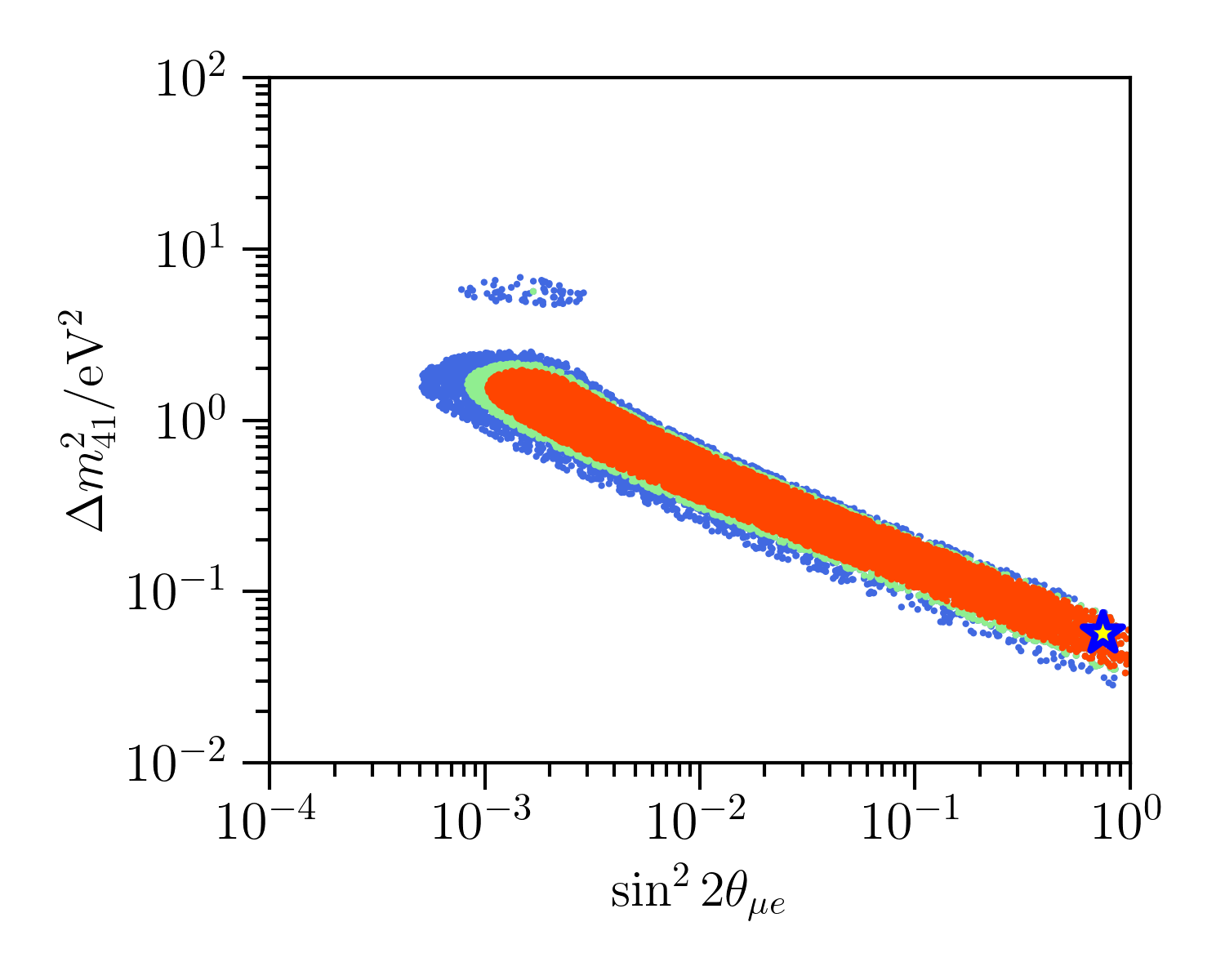}
\includegraphics[width=0.3\columnwidth]{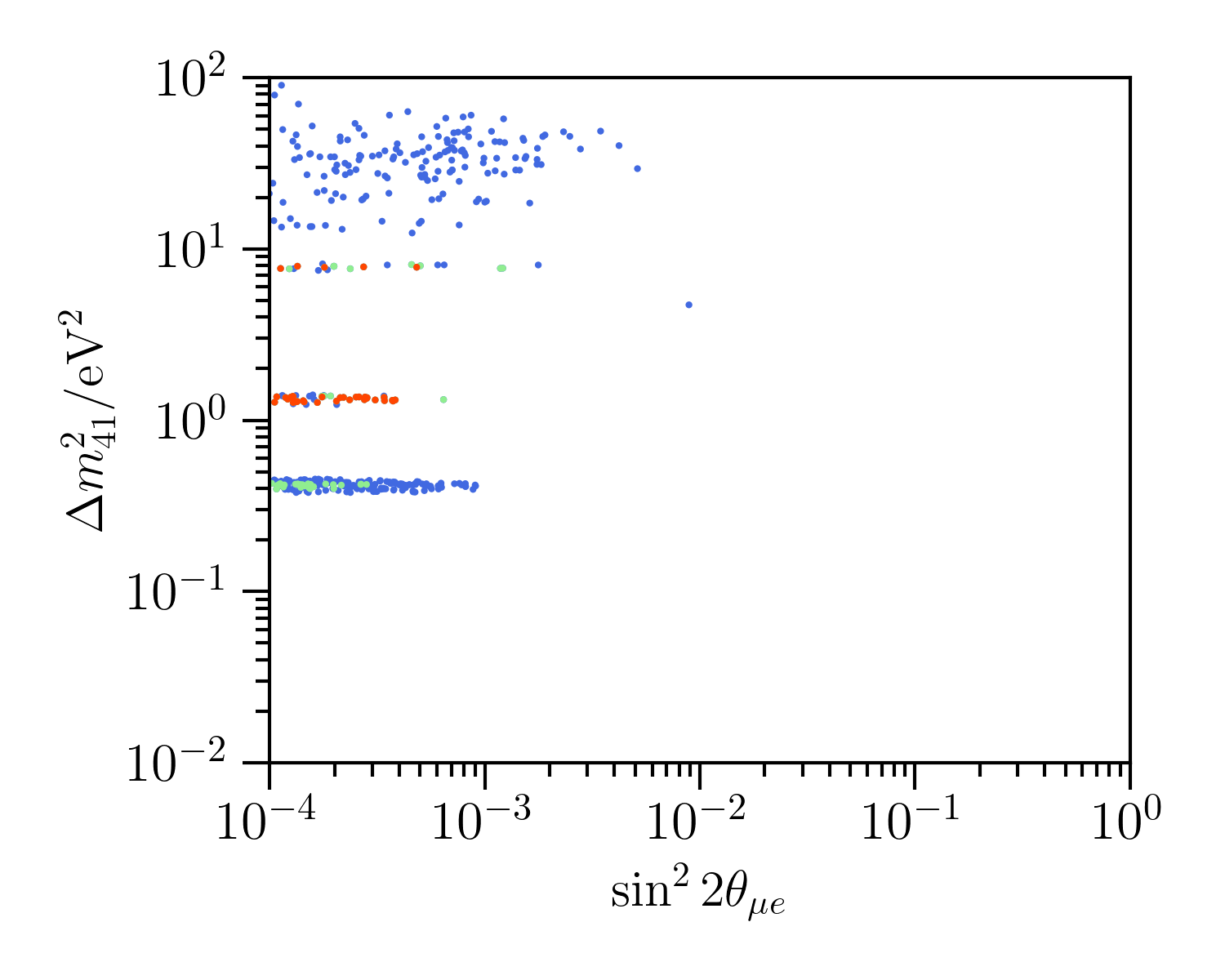}
\end{center}
\caption{$3+1$ global fits including MiniBooNE, considering global (left), appearance-only (middle), and disappearance-only (right) experiments. The allowed regions in $3+1$ parameter space at the 90\%, 95\%, and 99\% confidence levels are shown by the red, green, and blue points, respectively. The best-fit point is indicated by the star.
\label{fig:3plus1}}
\end{figure}

\begin{figure}[tb]
\begin{center}
\includegraphics[width=0.3\columnwidth]{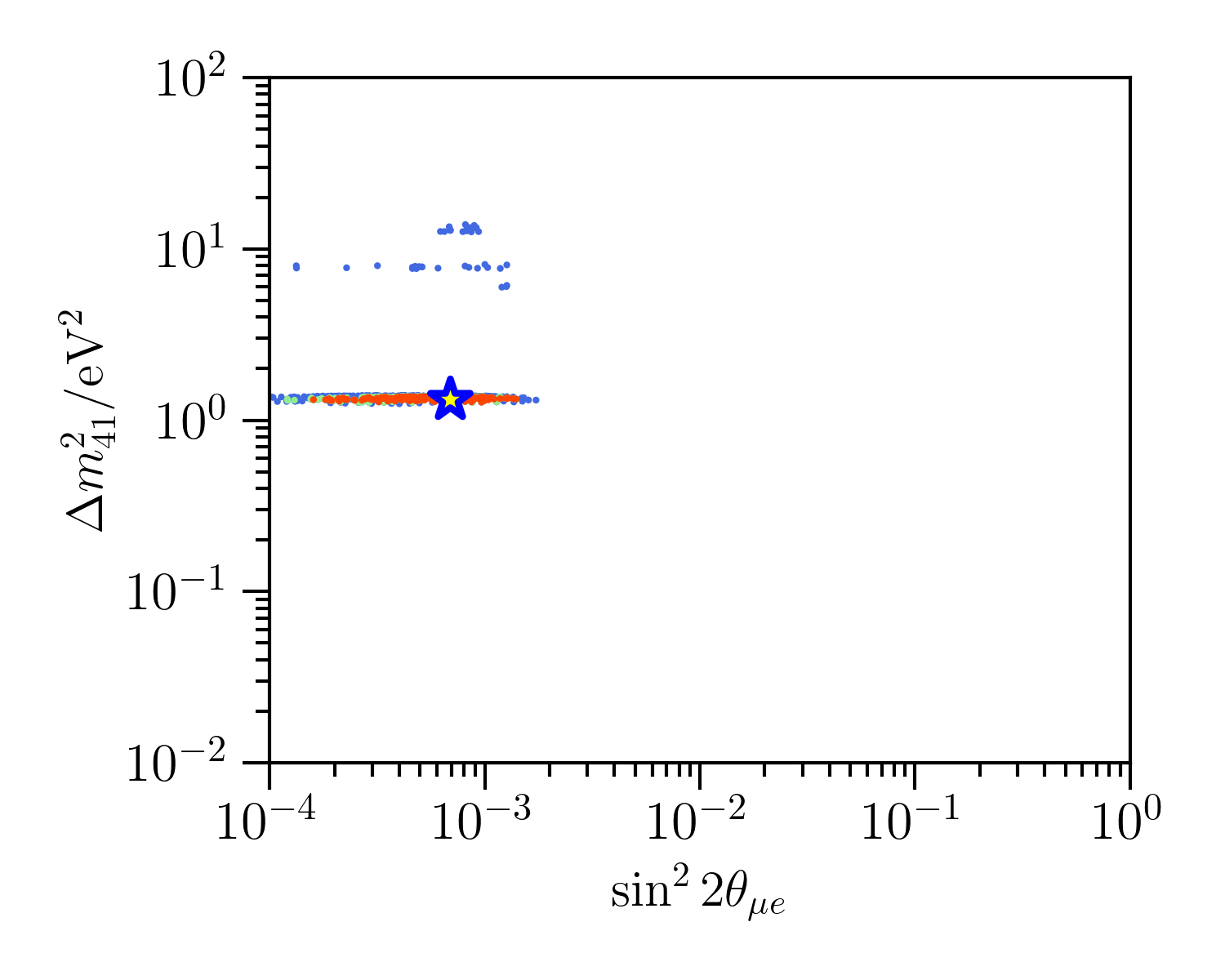}
\includegraphics[width=0.3\columnwidth]{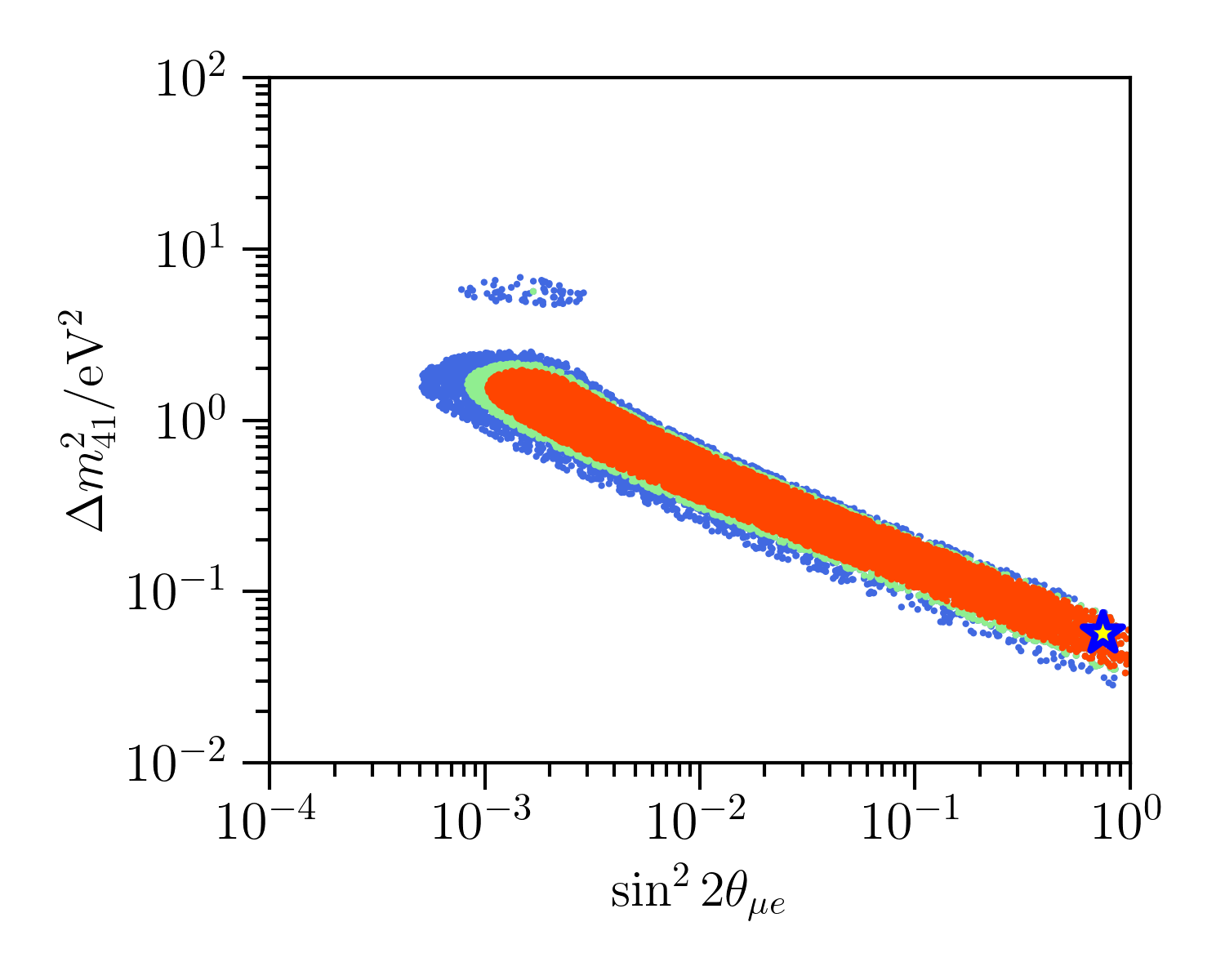}
\includegraphics[width=0.3\columnwidth]{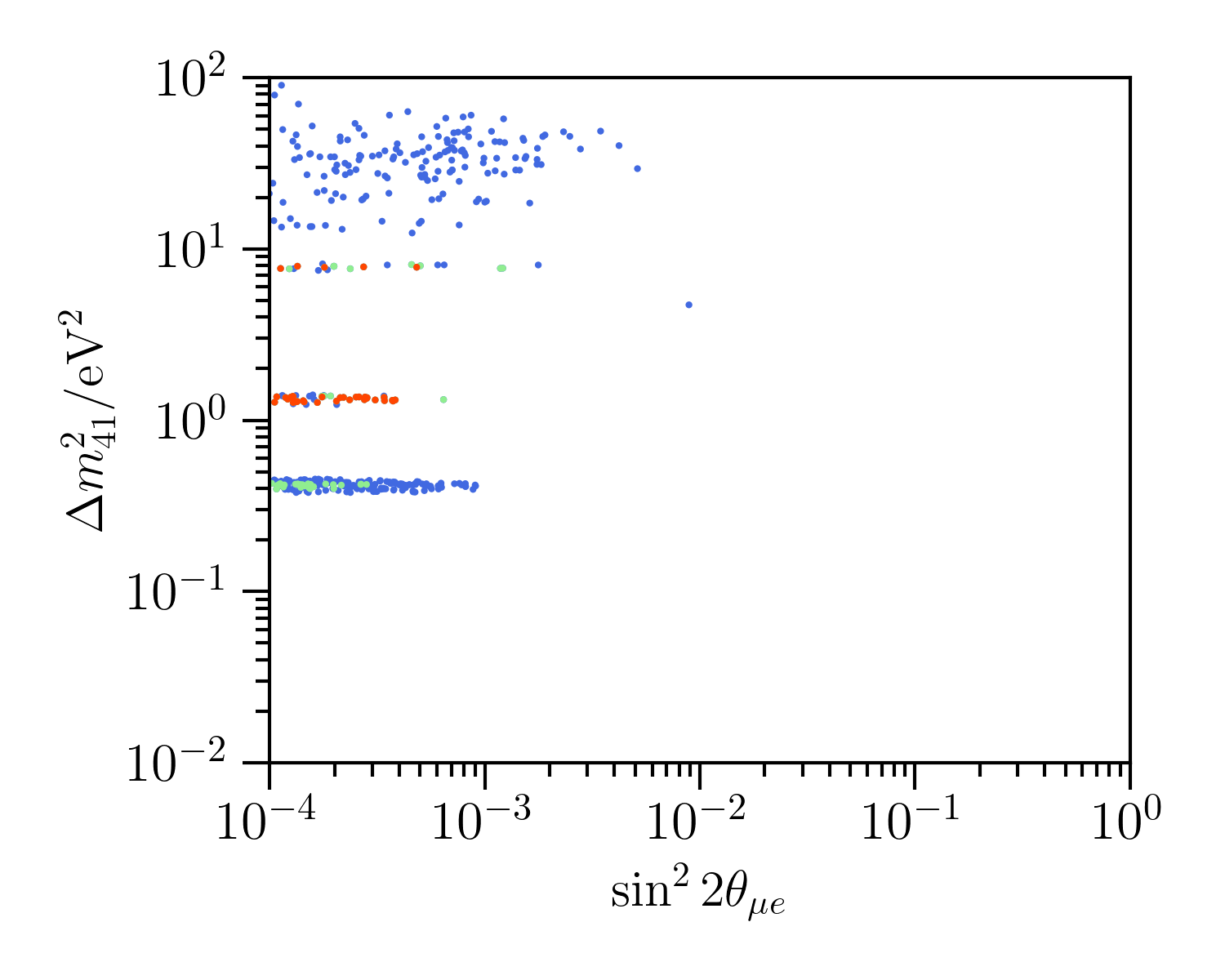}
\end{center}
\caption{$3+1$ global fits without MiniBooNE, considering global (left), appearance-only (middle), and disappearance-only (right) experiments. The allowed regions in $3+1$ parameter space at the 90\%, 95\%, and 99\% confidence levels are shown by the red, green, and blue points, respectively. The best-fit point is indicated by the star.
\label{fig:3plus1withoutMBNUMI}}
\end{figure}

This tension appears to be driven significantly by MiniBooNE data~\cite{Vergani:2021tgc,Hardin:2022muu}.
Within the Harvard-Columbia-MIT global fit that existed at the time of Ref.~\cite{Vergani:2021tgc}, the PG test tension decreased from $4.8\sigma$ to $2.5\sigma$ upon removing the MiniBooNE dataset.
The allowed regions in $\Delta m_{41}^2$-$\sin^2 2\theta_{\mu e}$ parameter space for the global, appearance-only, and disappearance-only experiments both before and after removing MiniBooNE are shown in \cref{fig:3plus1} and \cref{fig:3plus1withoutMBNUMI}, respectively.
One can see visually that there is more overlap in the $\Delta m^2$ dimension between the three different allowed regions after removing MiniBooNE.
This is likely because the MiniBooNE excess peaks strongly at low energies, an effect which is difficult to produce in the $3+1$ model, as discussed in \cref{sec:miniboone_explanations}.
The latest MiniBooNE $3+1$ fits discussed in \cref{sec:MBuB_sterile_paper} specifically show that the oscillation hypothesis under-predicts the excess in the lowest bins of the MiniBooNE electron-like dataset.
This behavior drives the global fit to the $\Delta m_{41}^2 < 1~{\rm eV}^2$ region, in contrast to the higher values ($\Delta m_{41}^2 \gtrsim 1~{\rm eV}^2$) preferred by the reactor and gallium anomalies.

The global fit without MiniBooNE, shown in the leftmost panel of \cref{fig:3plus1withoutMBNUMI}, strongly prefers a solution at $\Delta m_{41}^2 \sim 1~{\rm eV}^2$ and $10^{-4} \lesssim \sin^2 2\theta_{\mu e} \lesssim 10^{-3}$.
The best fit point is specifically at $\Delta m_{41}^2 = 1.3~{\rm eV}^2$ and $\sin^2 2\theta_{\mu e} = 6.9 \times 10^{-4}$.
We take these as our benchmark $3+1$ parameters, which fix the $\nu_\mu \to \nu_e$ oscillation contribution to the MiniBooNE electron-like channel.
The oscillation prediction alone is not sufficient to explain the MiniBooNE LEE; it can only accommodate $\sim 10\%$ of the total excess events and peaks at $E_\nu \sim 600~{\rm MeV}$, inconsistent with the low-energy nature of the MiniBooNE excess.
However, the benefit of these specific oscillation parameters is that they can explain LSND, reactor, and gallium data while remaining consistent with null results from $\nu_\mu$ disappearance searches.
The impact of different mixing angles within the allowed region of the global fit without MiniBooNE is discussed in Ref.~\cite{Vergani:2021tgc}, and the impact of the (larger) oscillation contribution predicted by the MiniBooNE-MicroBooNE combined best fit~\cite{MiniBooNE:2022emn} is discussed in Ref.~\cite{Kamp:2022bpt}.

The remaining majority of the MiniBooNE excess must then come from another source.
In the mixed model, we ascribe the bulk of the LEE to the visible decays of neutrissimos.
This model works well for MiniBooNE, as neutrissimos can be produced abundantly in the BNB through the first two diagrams in \cref{fig:neutrissimo_feyn_diagrams}: Primakoff upscattering $\nu A \to \mathcal{N} A$ and the three-body decay $\pi^0 \to \gamma \nu \mathcal{N}$.
The third diagram, neutrissimo decay $\mathcal{N} \to \nu \gamma$, could then provide a source of extra photons in MiniBooNE.
As discussed in \cref{ch:miniboone}, these photons would show up in the electron-like sample due to MiniBooNE's reliance on Cherenkov rings for particle identification.
Models of heavy neutrino decay to single photons have been previously considered as a solution to the MiniBooNE anomaly~\cite{Gninenko:2009ks,McKeen:2010rx,Gninenko:2010pr,Dib:2011jh,Gninenko:2012rw,Masip:2012ke,Radionov:2013mca,Ballett:2016opr,Magill:2018jla,Balantekin:2018ukw,Balaji:2019fxd,Balaji:2020oig,Fischer:2019fbw,Alvarez-Ruso:2021dna}.
We extend upon these studies by performing a robust analysis of the neutrissimo model in MiniBooNE within the context of the mixed model.
This analysis is described in detail in the next section.
It turns out that compared to the $3+1$ model, neutrissimo decays are more consistent with the energy and angular distributions of the MiniBooNE excess.
Thus there are two distinct advantages of the mixed model: (1) it can provide an explanation for other anomalies in the neutrino sector while relieving tension in global $3+1$ fits, and (2) it can provide a better explanation of the MiniBooNE excess itself.

\section{Neutrissimos in MiniBooNE} \label{sec:neutrissimos_in_MB}

The analysis presented in Ref.~\cite{Vergani:2021tgc} used a custom simulation to estimate the single photon event rate from neutrissimo decays in the MiniBooNE detector.
The updated MiniBooNE analysis in Ref.~\cite{Kamp:2022bpt} relied instead on the public \texttt{LeptonInjector}~\footnote{\href{https://github.com/Harvard-Neutrino/LeptonInjector}{github.com/Harvard-Neutrino/LeptonInjector}} software package, which will be described in more detail in \cref{sec:neutrissimo_lepinj}.
In Ref.~\cite{Vergani:2021tgc} we considered neutrissimo production via both $\pi^0 \to \gamma \nu \mathcal{N}$ decays and $\nu A \to \mathcal{N} A$ Primakoff upscattering.
The former process was estimated using the Sanford-Wang parameterization of the $\pi^0$ production rate at the BNB~]~\cite{Wang:1970bn,MiniBooNE:2008hfu} and was found to be negligible for the dipole parameters of interest to MiniBooNE.
We therefore explicitly ignored this process, considering Primakoff upscattering as the sole neutrissimo production mechanism.

The $\nu_\alpha A \to \mathcal{N} A$ upscattering rate was calculated using the differential cross section with respect to the Mandelstam variable $t = Q^2 = (p_\mathcal{N} - p_\nu)^2$~\cite{Vergani:2021tgc,Magill:2018jla,Shoemaker:2018vii,Coloma:2017ppo},
\begin{equation} \label{eq:neutrissimo_xsec}
\begin{split}
\frac{d \sigma_{\nu_\alpha A \to \mathcal{N} A}}{dt} = \frac{2 \alpha d_\alpha^2}{M} 
\bigg[ & F_1^2(t) \Big( \frac{1}{E_r} - \frac{1}{E_\nu} + m_{\mathcal{N}}^2 \frac{E_r - 2 E_\nu - M}{4 E_\nu^2 E_r M} + m_{\mathcal{N}}^4 \frac{E_r - M}{8 E_\nu^2 E_r^2 M^2} \Big) \\
+ &\frac{F_2^2(t)}{4 M^2} \Big( \frac{2 M }{E_\nu^2} ( (2 E_\nu - E_r)^2 - 2 E_r M ) + m_{\mathcal{N}}^2 \frac{E_r - 4 E_\nu}{E_\nu^2} + \frac{m_{\mathcal{N}}^4}{E_\nu^2 E_r}\Big)\bigg],
\end{split}
\end{equation}
where $\alpha$ is the fine structure constant, $d$ is the dipole coupling, $E_\nu$ is the SM neutrino energy, $m_{\mathcal{N}}$ is the mass of the heavy neutrino, $M$ is the target mass, $t = -(p_\mathcal{N} - p_\nu)^2$ is the momentum transfer, $E_r = -t/2M$ is the target recoil energy, and $F_{1/2}(t)$ are the electric/magnetic target form factors, respectively.
Note that the term proportional to $E_r m_N^4$ in the $F_1$ line only exists for spinless nuclei, and must be replaced for nonzero spin nuclei~\cite{Masip:2012ke}.
In the case of coherent scattering off of a nucleus, $F_1$ receives a $Z^2$ enhancement and is therefore dominant over $F_2$.
Thus, we explicitly neglect the $F_2$ term for this analysis.
In Ref.~\cite{Vergani:2021tgc}, $F_1$ was calculated using a dipole parameterization.
However, since the dipole approximation tends to underestimate the cross section at high squared momentum transfer $Q^2$, Ref.~\cite{Kamp:2022bpt} updated to the Fourier-Bessel parameterization,
\begin{equation}
F^{\rm FB}(Q) = Z^2 N \frac{\sin(QR)}{QR} \sum_n \frac{(-1)^n a_n}{n^2 \pi^2 - Q^2},
\end{equation}
where $Z$ is the proton number of the struck nucleus, $N$ is a normalization factor ensuring $F^{\rm FB}(0) = 1$, and $a_n$ are coefficients obtained from nuclear scattering data~\cite{Fricke:1995zz,DeVries:1987atn,DeJager:1974liz,VT_NDT}.
For the case of inelastic upscattering off of individual nuclei, we use the nucleon electric and magnetic form factors discussed in the appendix of Ref.~\cite{Vergani:2021tgc}.
Inelastic upscattering exhibits only a linear enhancement in the number of protons and neutrons, which suppresses its effect compared to the $Z^2$ enhancement of the coherent upscattering cross section.
Ref.~\cite{Kamp:2022bpt} also includes contributions from the sub-dominant helicity-conserving channel of both the elastic and inelastic upscattering cross sections.

The total upscattering cross section was obtained by integrating \cref{eq:neutrissimo_xsec} over the physical momentum transfer range~\cite{Magill:2018jla}.
The overall neutrissimo production rate via upscattering in the BNB was then calculated using this cross section in combination with the density of each nuclear and nucleon target available along the BNB.
These include nuclei in the bedrock and air between the BNB target and the detector, as well as carbon and hydrogen nuclei in the detector itself.
The bedrock composition is defined in \cref{tab:crustnuclei} and is considered to have a density of 2.9~g/cm$^3$.
Upscattering in the bedrock (detector) is more important for long-lived (short-lived) neutrissimos.
The kinematics of the outgoing neutrissimo were fixed by sampling a momentum transfer $t$ according to \cref{eq:neutrissimo_xsec}, which consequently determines the nuclear recoil energy $E_r = -t/2M$ as well as the neutrissimo energy and scattering angle~\cite{Masip:2012ke},
\begin{equation}
\begin{split}
&E_\mathcal{N} = E_\nu - E_r, \\
&\cos(\theta) = \frac{E_\nu - E_r - M E_r / E_\nu - m_{\mathcal{N}}^2/2E_\nu}{\sqrt{E_\nu^2 + E_r^2 - 2E_\nu E_r - m_{\mathcal{N}}^2}}.
\end{split}
\end{equation}
One can rewrite \cref{eq:neutrissimo_xsec} to isolate an overall $1/t^2$ prefactor, indicating a preference for $t \approx 0$ and thus $\cos \theta \approx 1$ for this process.

After neutrissimos are produced, they will decay via the process $\mathcal{N} \to \nu \gamma$.
The decay width $\Gamma$ and lab-frame decay length $L_{\rm decay}$ of this process are given by ~\cite{Magill:2018jla,Vergani:2021tgc,Kamp:2022bpt}
\begin{equation} \label{eq:neutrissimo_decay}
\begin{split}
&\Gamma = \frac{d^2 m_{\mathcal{N}}^3}{4\pi}, \\
&L_{\rm decay} = 4 \pi \frac{\beta E_\mathcal{N}}{d^2 m_{\mathcal{N}}^4},
\end{split}
\end{equation}
where
\begin{equation}
d \equiv \Big( \sum_\alpha |d_{\alpha N}|^2 \Big).
\end{equation}
Considering the dipole parameters preferred by the MiniBooNE excess, the typical neutrissimo decay length at BNB energies is $L \lesssim 1~{\rm m}$.
The probability of decaying within the MiniBooNE detector is then
\begin{equation}
P_{\rm decay} = \exp \bigg( \frac{-L_{\rm enter}}{L_{\rm decay}} \bigg) - \exp \bigg( \frac{-L_{\rm exit}}{L_{\rm decay}} \bigg),
\end{equation}
where $L_{\rm enter}$ and $L_{\rm exit}$ are the distance along the neutrissimo path from the upscattering location to the entrance and exit of the MiniBooNE detector, respectively.
Assume the neutrissimos are Dirac particles, their differential decay which is given by~\cite{Balantekin:2018ukw,Alvarez-Ruso:2021dna}
\begin{equation}
\frac{d \Gamma}{d \cos \theta} = \frac{1}{2}(1 + \alpha \cos \theta),
\end{equation}
where $\alpha = 1$ ($\alpha$ = -1) for the decay of a right-handed (left-handed) heavy neutrino, and vice versa for the antineutrino case.

In Ref.~\cite{Vergani:2021tgc} and Ref.~\cite{Kamp:2022bpt}, we used the combination of the production rate from \cref{eq:neutrissimo_xsec} and the decay rate from \cref{eq:neutrissimo_decay} to compute the single photon rate from neutrissimo decay in MiniBooNE.

\begin{table}[ht!]
    \centering
    \begin{tabular}{|c|c|c|c|c|c|}
        \hline
        Nucleus & Z & A & Crust Mass Fraction & Crust Atomic Fraction & Nuclear Radius [MeV$^{-1}$] \\  
        \hline
        O & 8 & 16 & 0.466 & 0.627 & 0.00218 \\
        Si & 14 & 28 & 0.277 & 0.213 & 0.00252 \\
        Al & 13 & 27 & 0.081 & 0.065 & 0.00247 \\
        Fe & 26 & 56 & 0.05 & 0.019 & 0.00301 \\
        Ca & 20 & 40 & 0.037 & 0.02 & 0.00281 \\
        K & 19 & 39 & 0.027 & 0.015 & 0.00277 \\
        Na & 11 & 23 & 0.026 & 0.024 & 0.00241 \\
        Mg & 12 & 24 & 0.015 & 0.013 & 0.00247 \\
        Ti & 22 & 48 & 0.004 & 0.002 & 0.0029 \\
        P & 15 & 31 & 0.001 & 0.001 & 0.00257 \\
        \hline
    \end{tabular}
    \caption{Relevant parameters of the ten most abundant nuclei in the Earth's upper crust according to \cite{EarthCrust}}
    \label{tab:crustnuclei}
\end{table}

\subsection{Simulation in \texttt{LeptonInjector}} \label{sec:neutrissimo_lepinj}

As stated above, Ref.~\cite{Vergani:2021tgc} built a custom simulation to assess the neutrissimo model in MiniBooNE.
Ref.~\cite{Kamp:2022bpt} instead performed the MiniBooNE simulation using the \texttt{LeptonInjector} software package.
\texttt{LeptonInjector} was originally developed within the IceCube collaboration~\cite{IceCube:2020tcq} to simulate the interactions of high-energy neutrinos within their detector; however, this version was not suitable to handle the neutrissimo model out-of-the-box.
A series of updates were made to \texttt{LeptonInjector}, led by Austin Schneider and myself, in order to carry out the study described here.
These include (but are not limited to):
\begin{itemize}
    \item Support for the input of total and differential cross section tables.
    \item Support for a wide variety of detector subsystem geometries, including extruded polygons.
    \item Ability to simulate a tree of secondary interactions following the initial interaction.
    \item Ability to calculate event weights for general interaction trees.
    \item Support for a combination of cross sections and decays as possible interactions for a given particle.
    \item Support for a fiducialization requirement to improve the computational efficiency. 
\end{itemize}
The current version of \texttt{LeptonInjector} is suitable for general purpose studies of particle interactions in a wide variety of detectors.
It was presented at the March 2023 Coherent-CAPTAIN-Mills Workshop~\cite{CCMWorkshop} and is publicly available on \texttt{GitHub}~\cite{LeptonInjector}.
While \texttt{LeptonInjector} is written in \texttt{C++}, the classes and methods are importable as a library in \texttt{Python}~\cite{10.5555/1593511}.

We now briefly review the \texttt{LeptonInjector}-based MiniBooNE analysis presented in Ref.~\cite{Kamp:2022bpt}.
Neutrinos are first injected along the beam axis with energies sampled from the BNB flux~\cite{MiniBooNE:2008hfu}.
The upscattering location is then sampled according to the material-dependent interaction length along the neutrino path.
The interaction length in each material is calculated via upscattering cross section tables computed using \texttt{DarkNews}~\cite{Abdullahi:2022cdw}.
In order to optimize the efficiency of the simulation, neutrinos are required to undergo a Priamkoff upscattering interaction within three neutrissimo decay lengths of the MiniBooNE detector.
The kinematics of the neutrissimo are fixed by sampling the differential cross section in \cref{eq:neutrissimo_xsec}, calculated in tabular format using \texttt{DarkNews}.
The decay location of the neutrissimo is then sampled according to the lab frame decay length in \cref{eq:neutrissimo_decay}.
If the neutrissimo path crosses the detector volume, it is required to decay within or just before the detector, further optimizing the efficiency of the simulation.
Once a photon is generated, we sample the pair-production location according to the radiation length of each traversed material, which we use to perform a robust fiducialization cut.
The weight of each event is calculated by comparing the generation probability of each sampled quantity (e.g. the energy/direction of the initial neutrino, the kinematics of the produced neutrissimo/photon, and the upscattering/decay locations) to the specified physical probability distribution.
The flux units provided in Ref.~\cite{MiniBooNE:2008hfu} allow us to compute event weights in units of POT$^{-1}$.

\begin{figure}[h!]
    \centering
     \begin{subfigure}[b]{0.45\textwidth}
         \centering
         \includegraphics[width=\textwidth]{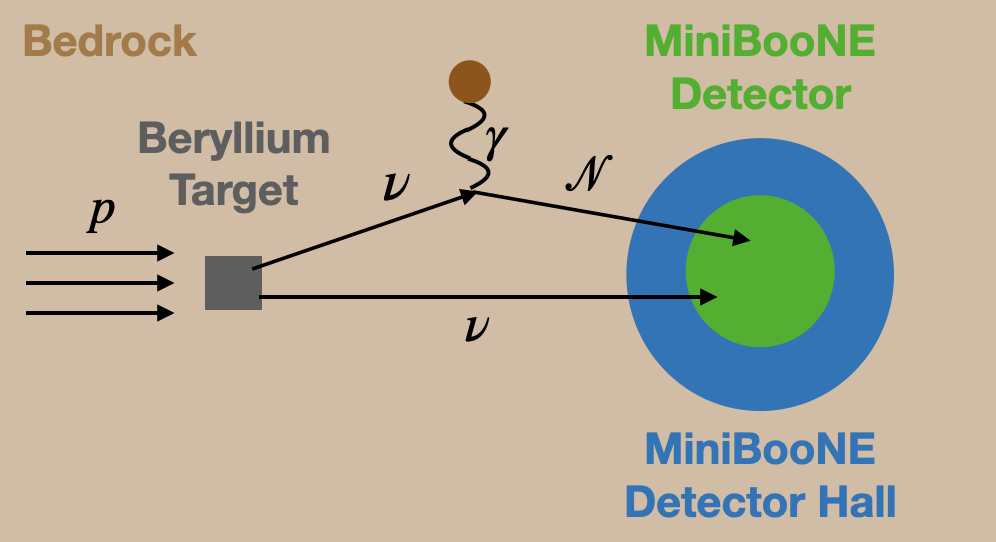}
         \caption{}
         \label{fig:mb_beamline}
     \end{subfigure}
     \hfill
     \begin{subfigure}[b]{0.45\textwidth}
         \centering
         \includegraphics[width=\textwidth]{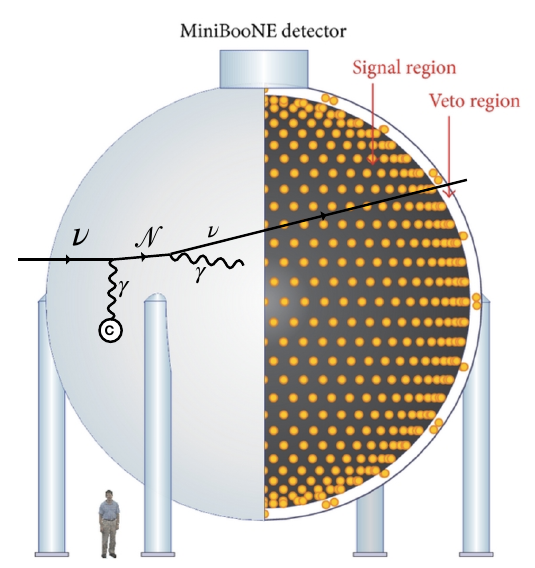}
         \caption{}
         \label{fig:mb_dipole}
     \end{subfigure}
     \hfill
        \caption{Schematic depiction of the neutrissimo model in MiniBooNE as simulated using \texttt{LeptonInjector}. \Cref{fig:mb_beamline} shows the simulation of Primkaoff upscattering along the beamline, and \cref{fig:mb_dipole} shows an example of upscattering, neutrissimo decay and pair-production within the MiniBooNE detector.}
        \label{fig:mb_neutrissimo_diagram}
\end{figure}

\Cref{fig:mb_neutrissimo_diagram} shows a schematic depiction of the \texttt{LeptonInjector} neutrissimo simulation in MiniBooNE, which incorporates 541~m of bedrock, a 9~m-radius spherical air-filled detector hall, and the 6.1~m-radius spherical mineral oil detector.
The result of this procedure is a robust Monte Carlo sample of single photons in MiniBooNE from neutrissimo decay.
The kinematic variables of each observable photon have been carefully computed, and the detailed treatment of the geometry of the MiniBooNE detector hall allows for a realistic estimation of the fiducialization requirement.
While the latter point is not necessarily vital for the relatively simple MiniBooNE detector, it is much more important when considering the complex MINER$\nu$A detector for the analysis discussed in \cref{sec:neutrissimo_paper}.
The careful treatment of new physics models enabled by \texttt{LeptonInjector} is essential for differentiating between the many proposed explanations of MiniBooNE, both when determining the preferred region in model parameter space from MiniBooNE data and when determining constraints from other experiments.
This is one of the intended use cases motivating our recent public release of \texttt{LeptonInjector}~\cite{LeptonInjector}.

\subsection{Fits to the MiniBooNE Excess}

After performing the procedure outlined in \cref{sec:neutrissimo_lepinj}, we down-weighted each event according to a linear parameterization of the MiniBooNE photon detection efficiency~\cite{Vergani:2021tgc,MiniBooNE_eff}.
We also imposed a reconstruction threshold on the true photon kinetic energy requiring $E_\gamma > 140~{\rm MeV}$.
The energy (angle) of the photon was smeared according to a power-law (quadratic) fit to a MiniBooNE single electromagnetic shower simulation sample~\cite{MBres}.
The typical resolution in the energy (angular) distribution was 10\% ($3^\circ$) for the photons generated in this model, consistent with values reported by MiniBooNE~\cite{Shaevitz:2008zza}.

After these steps, the simulated events could then be used to estimate the single photon rate from neutrissimo decays in the $E_\nu^{\rm QE}$ and $\cos \theta$ distributions of MiniBooNE's electron-like sample.
We performed a separate fit to the MiniBooNE excess in each distribution.
After subtracting the SM prediction and the oscillation prediction from the global fit without MiniBooNE, we calculated a $\chi^2$ test statistic between the data and the neutrissimo prediction throughout a grid of points in $d_{\mu \mathcal{N}}$-$m_{\mathcal{N}}$ parameter space.
For the $E_\nu^{\rm QE}$ distribution, we used the MiniBooNE electron-like covariance matrix provided in the data release~\cite{hepdata.114365} associated with Ref.~\cite{MiniBooNE:2020pnu}.
The MiniBooNE collaboration has not released any information regarding systematic uncertainty in the $\cos \theta$ distribution; therefore, we assign an uncorrelated fractional systematic uncertainty of 13\% in this distribution, consistent with the overall level in the $E_\nu^{\rm QE}$ distribution.
We compute confidence regions in neutrissimo parameter space using the $\Delta \chi^2$ test statistic, assuming Wilks' theorem~\cite{Wilks:1938dza} for two degrees of freedom.

\begin{figure}[h!]
    \centering
    \includegraphics[width=0.6\textwidth]{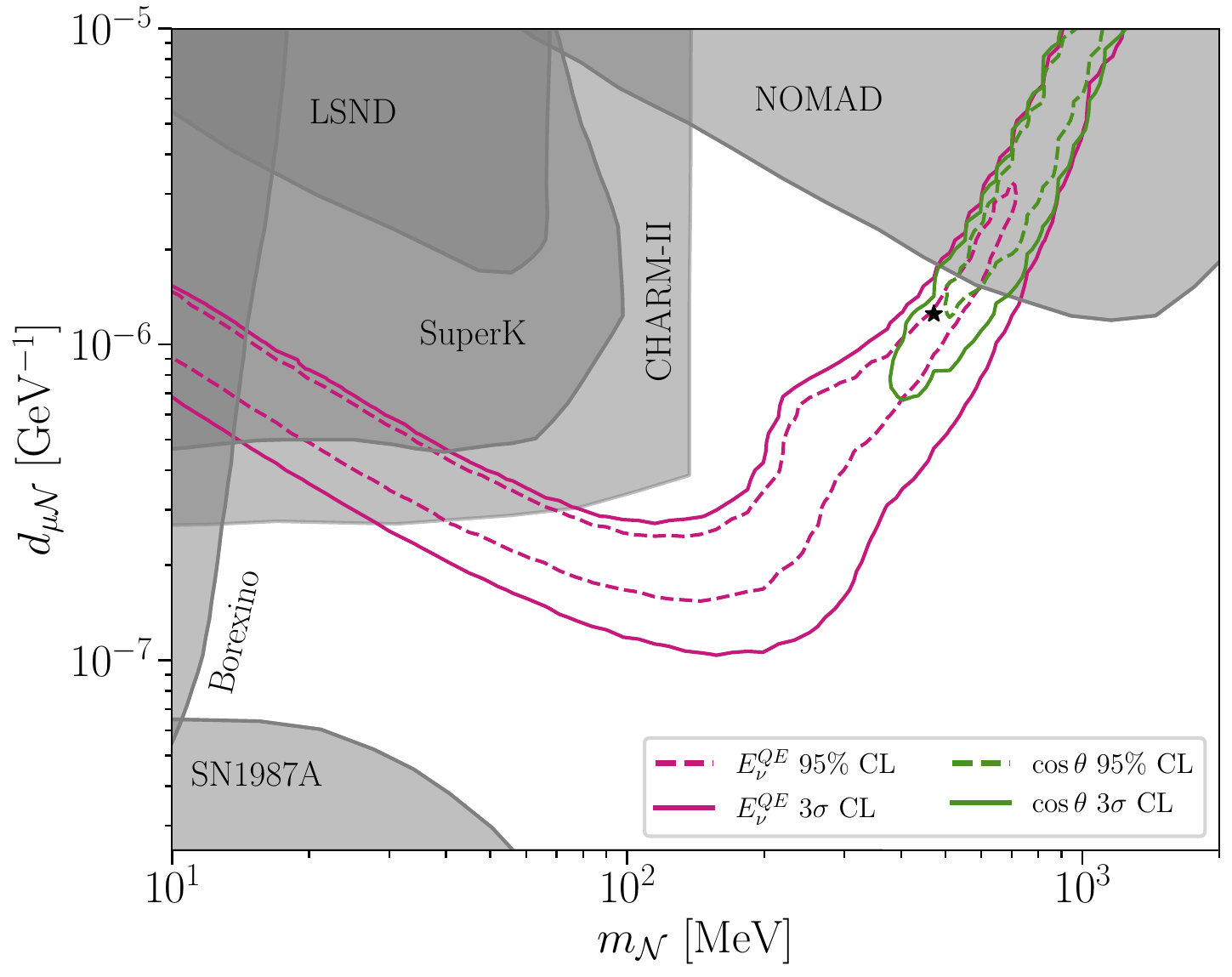}
    \caption{Allowed regions at the 95\% and $3\sigma$ confidence level in $d_{\mu \mathcal{N}}$-$m_{\mathcal{N}}$ obtained through fits to the MiniBooNE excess in the $E_\nu^{\rm QE}$ and $\cos \theta$ distributions. Existing $2\sigma$ constraints on this model are indicated by the grey regions.}
    \label{fig:neutrissimo_miniboone_confidence_regions}
\end{figure}

\begin{figure}[h!]
    \centering
     \begin{subfigure}[b]{0.45\textwidth}
         \centering
         \includegraphics[width=\textwidth]{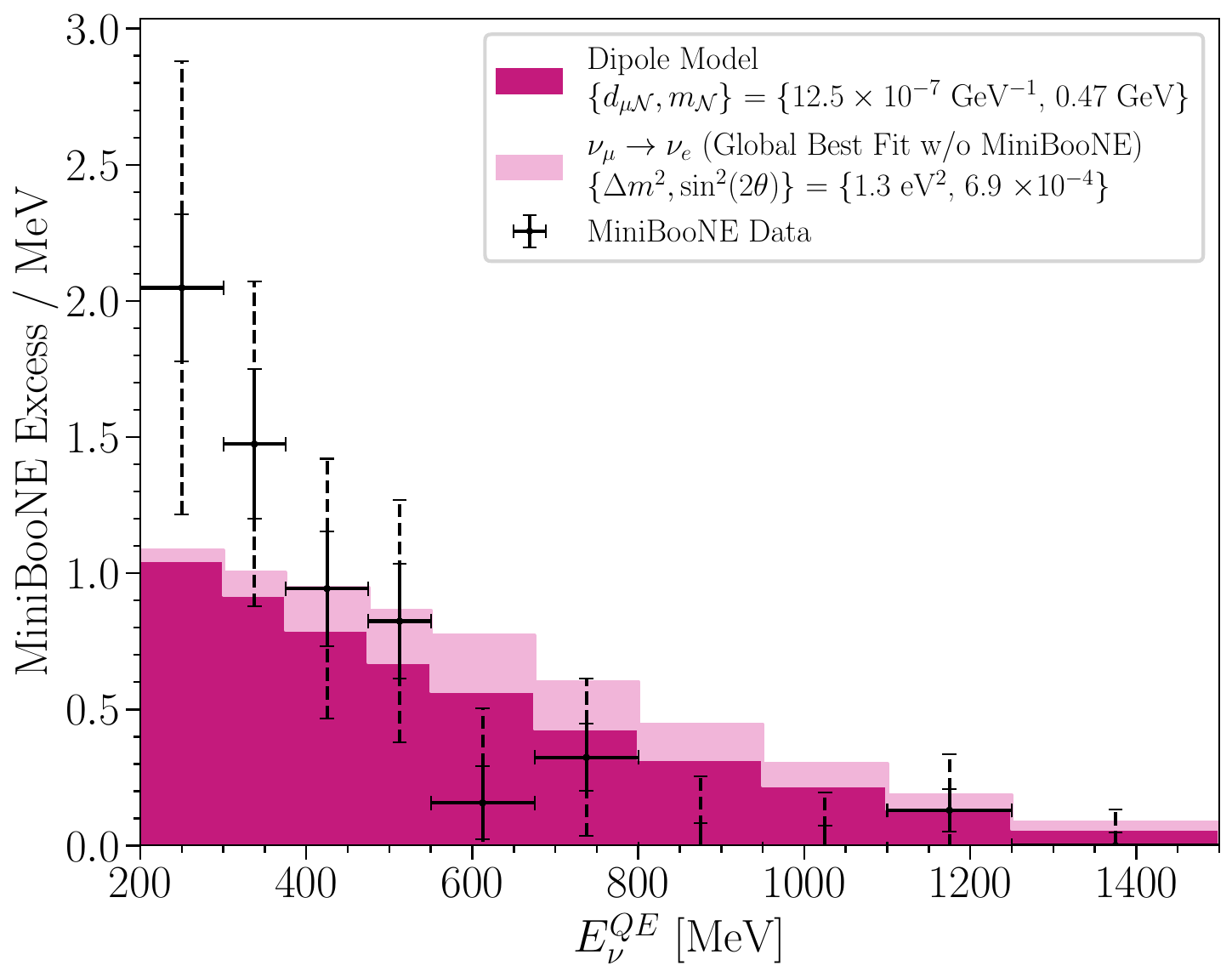}
         \caption{}
         \label{fig:mb_neutrissimo_enuqe}
     \end{subfigure}
     \hfill
     \begin{subfigure}[b]{0.45\textwidth}
         \centering
         \includegraphics[width=\textwidth]{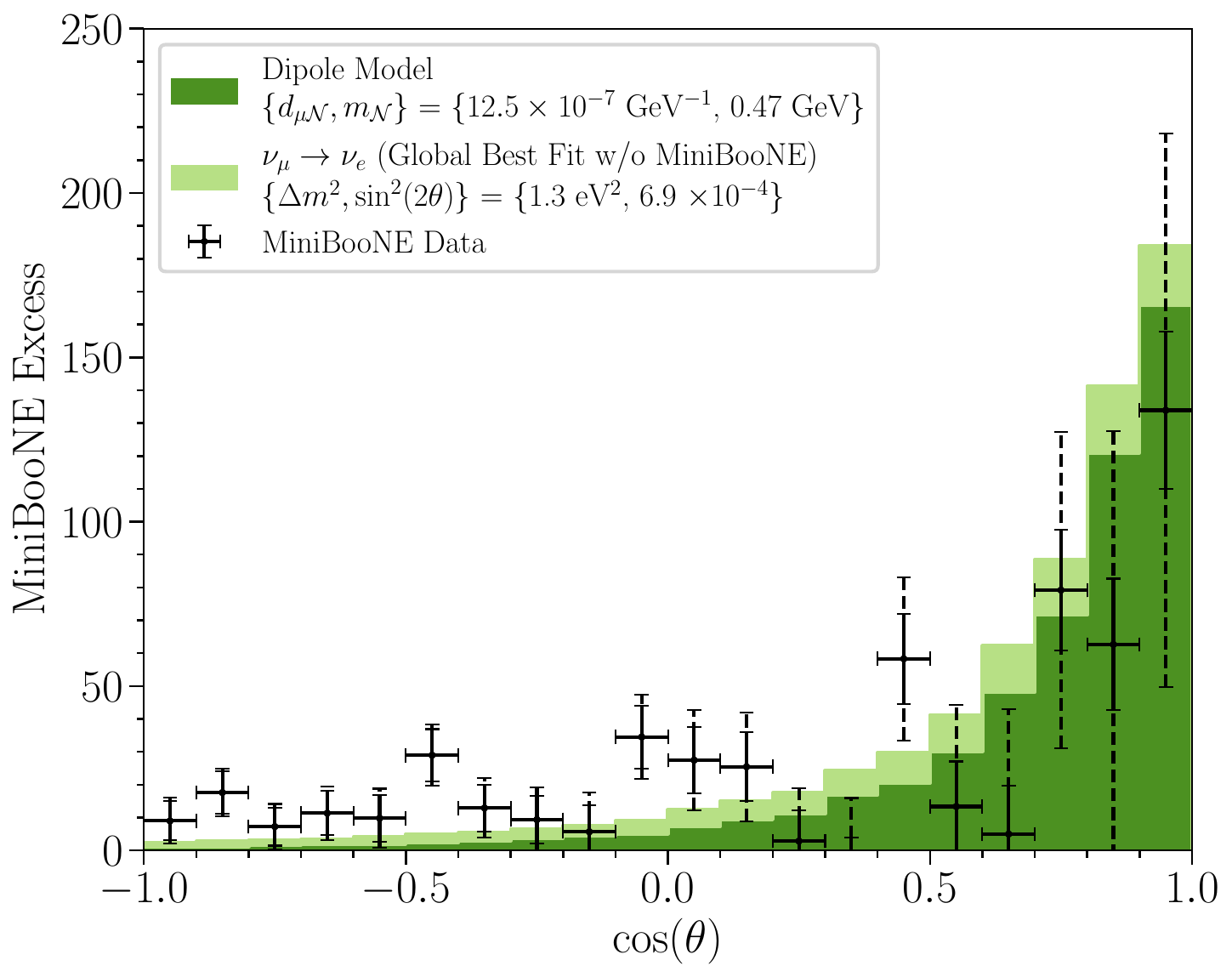}
         \caption{}
         \label{fig:mb_neutrissimo_costheta}
     \end{subfigure}
     \hfill
        \caption{\Cref{fig:mb_neutrissimo_enuqe} and \cref{fig:mb_neutrissimo_costheta} show the $E_\nu^{\rm QE}$ and $\cos \theta$ distributions of the MiniBooNE excess, respectively, compared with the prediction from the neutrissimo model indicated by the black star in \cref{fig:neutrissimo_miniboone_confidence_regions}. The oscillation contribution from the $3+1$ global fit without MiniBooNE is also shown.}
        \label{fig:mb_neutrissimo_dists}
\end{figure}

\Cref{fig:neutrissimo_miniboone_confidence_regions} shows the 95\% and $3\sigma$ confidence level allowed regions in neutrissimo parameter space from the MiniBooNE $E_\nu^{\rm QE}$ and $\cos \theta$ distributions.
These allowed regions come from Ref.~\cite{Kamp:2022bpt}, which is intended to supersede the results in Ref.~\cite{Vergani:2021tgc}.
This is because Ref.~\cite{Kamp:2022bpt} makes a number of improvements upon the analysis in Ref.~\cite{Vergani:2021tgc}, including the use of a more robust nuclear electromagnetic form factor and a more detailed simulation within \texttt{LeptonInjector}.
One can see from \cref{fig:neutrissimo_miniboone_confidence_regions} that the energy and angular distributions of the MiniBooNE excess prefer different regions of parameter space.
This is because the angular distribution has a not-insignificant component at large angles, $-1 \leq \cos \theta \leq 0$, as shown in \cref{fig:miniboone_costheta}.
Larger neutrissimo masses are required to explain this region, as such neutrissimos will have a lower Lorentz boost and thus result in a photon with broader lab-frame angles with respect to the beamline.
This is in contrast with the energy distribution, which prefers lower neutrissimo masses such that the lower energy part of the BNB neutrino flux is not kinematically forbidding from upscattering into a neutrissimo.
The difference between preferred regions of the energy and angular distributions in such visible heavy neutrino decay models has been discussed previously in the literature~\cite{Alvarez-Ruso:2021dna,Radionov:2013mca}; however, our study is the first comprehensive analysis of the compatibility of both distributions.
Our results in \cref{fig:neutrissimo_miniboone_confidence_regions} indicate an overlap between the allowed regions of the $E_\nu^{\rm QE}$ and $\cos \theta$ allowed regions at the 95\% C.L.
\Cref{fig:mb_neutrissimo_dists} shows comparisons between the MiniBooNE excess in each distribution and the mixed model prediction, considering the neutrissimo model indicated by the black star in \cref{fig:neutrissimo_miniboone_confidence_regions}, which is near this 95\% C.L. overlap.
One can see that the oscillation contribution to the MiniBooNE LEE in this model is indeed small compared with the neutrissimo contribution.
A larger oscillation contribution might help explain the $-1 \leq \cos \theta \leq 0$ region of the excess--this point is discussed further in \cref{sec:neutrissimo_paper}.
That being said, the mixed model does a better job of predicting the excess in the lowest-energy region.
This can be seen by comparing the compatibility between data and prediction in the first two bins of \cref{fig:mb_neutrissimo_enuqe} to that in \cref{fig:miniboone_nu_enuqe}.

There are a number of existing constraints on the neutrissimo model in the $10 \leq m_{\mathcal{N}}~[{\rm MeV}] \leq 1000$ regime, as indicated by the grey regions in \cref{fig:neutrissimo_miniboone_confidence_regions}.
These include Super-Kamiokande~\cite{Gustafson:2022rsz,Plestid:2020vqf}, Borexino~\cite{Plestid:2020vqf}, Supernova 1987A~\cite{Magill:2018jla,Brdar:2021ysi}, LSND~\cite{Magill:2018jla}, CHARM-II~\cite{Coloma:2017ppo}, and NOMAD~\cite{Gninenko:1998nn}.
See \cref{sec:neutrissimo_paper} for a more detailed discussion on each constraint.
We briefly note that the overlap in the 95\% allowed regions in MiniBooNE falls close to the NOMAD constraint derived in Ref.~\cite{Gninenko:1998nn}, which may motivate a re-analysis of the NOMAD data.

\begin{figure}[h!]
    \centering
    \includegraphics[width=0.6\textwidth]{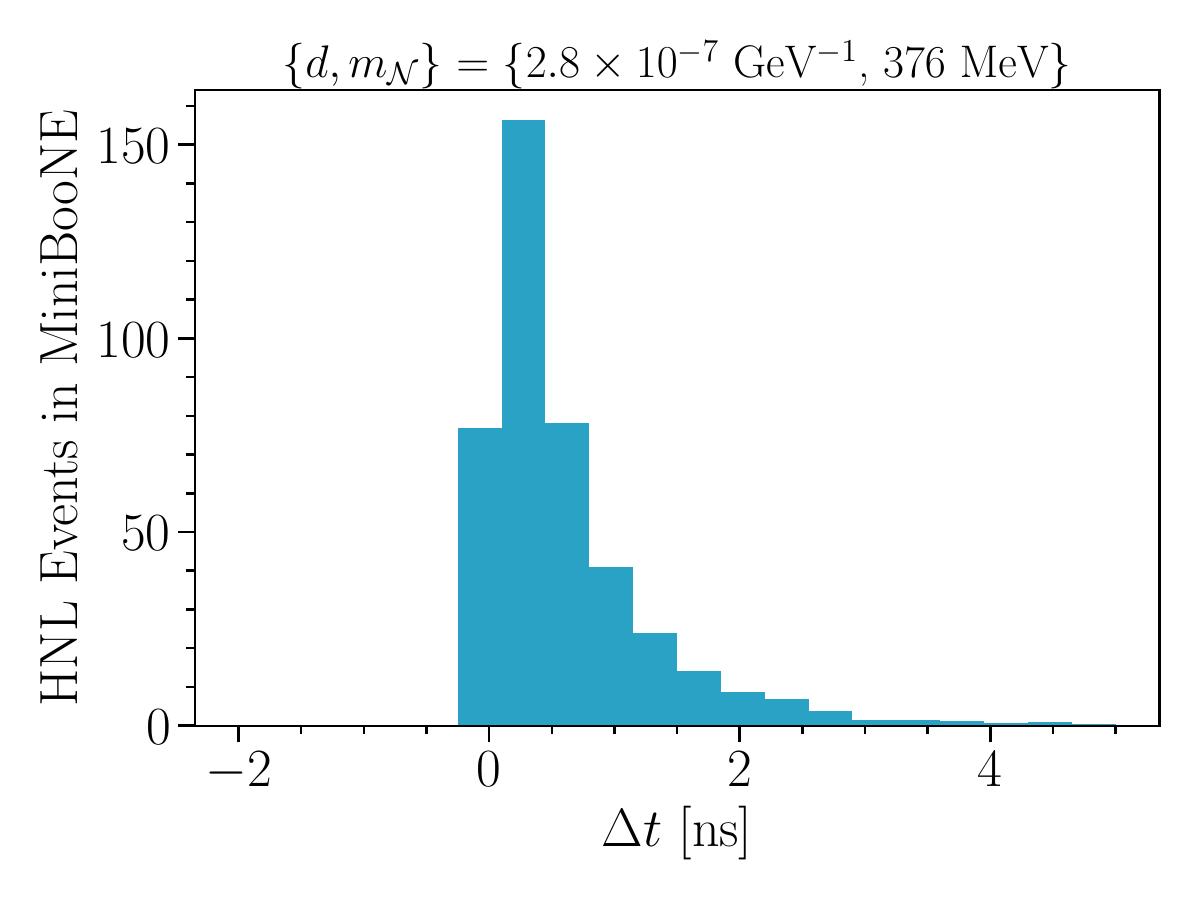}
    \caption{The added time delay in MiniBooNE for a neutrissimo with the indicated parameters, as calculated in Ref.~\cite{Vergani:2021tgc}.}
    \label{fig:neutrissimo_time_delay}
\end{figure}

Finally, we remind the reader that any explanation of MiniBooNE must be compatible not only with the energy and angular distributions of the excess but also with the timing distribution~\cite{MiniBooNE:2020pnu}.
Specifically, the new physics events cannot experience time delays $\Delta t \gtrsim 10~{\rm ns}$.
This was studied within the context of the neutrissimo model in Ref.~\cite{Vergani:2021tgc}.
As shown in \cref{fig:neutrissimo_time_delay}, neutrissimos with $m_{\mathcal{N}} \sim 500~{\rm MeV}$ have time delays $\Delta t \sim 1~{\rm ns}$, certainly within the constraint imposed by the MiniBooNE timing distribution.
This is because in the region of parameter space preferred by MiniBooNE, neutrissimos tend to be very short-lived and must therefore be produced via upscattering off of nuclei in the detector.
As such, the total travel time is dominated by the original speed-of-light neutrino.

\section{Publication: \textit{Dipole-coupled neutrissimo explanations of the MiniBooNE excess
including constraints from MINERvA data}} \label{sec:neutrissimo_paper}

We close this chapter with a presentation of Ref.~\cite{Kamp:2022bpt}.
The work was led by myself, Matheus Hostert, and Austin Schneider.
The full \textit{Physical Review D} publication is included below.
As described in \cref{sec:neutrissimos_in_MB}, this study performed a more detailed analysis of the mixed model described in \cref{sec:mixed_model} as a solution to the MiniBooNE excess.
We also derived world-leading constraints on the neutrissimo model using neutrino electron elastic scattering data from the MINER$\nu$A experiment~\cite{MINERvA:2015nqi,Valencia:2019mkf,MINERvA:2022vmb}.
While these constraints are strong, they do not rule out the overlap in the 95\% C.L. allowed regions from the MiniBooNE energy and angular distributions shown in \cref{fig:neutrissimo_miniboone_confidence_regions}.
As shown in Figure~8 of the paper below, this is a result of the strict kinematic cuts employed in the MINER$\nu$A elastic scattering analysis.
If these cuts are relaxed, MINER$\nu$A should be sensitive to the MiniBooNE neutrissimo solution in the mixed model.
The most important result from this study is given in Figure~10, which shows the surviving parts of the MiniBooNE allowed regions in neutrissimo parameter space after the MINER$\nu$A constraints are incorporated.
The dashed lines in the right panel indicate contours of constant number of neutrissimo decays in the MINER$\nu$A fiducial volume before any cuts are applied.
One can see that MINER$\nu$A should have $10^4-10^5$ events before cuts within the MiniBooNE-preferred region of parameter space.
We also test the implications of weak-flavor-conserving UV-completions of the dipole operator of \cref{eq:dipole_operator}, which predict $d_{\alpha \mathcal{N}} \propto m_{\ell_\alpha}$ and thus a large $d_\tau$.
This scenario mostly alters the allowed regions and constraints in the lower mass region of parameter space without impacting the MiniBooNE solution.

\textit{Note: while the published version of this article uses the phrase ``heavy-neutral-lepton'' in the title, the authors prefer the arXiv title given above.}

\includepdf[pages=-]{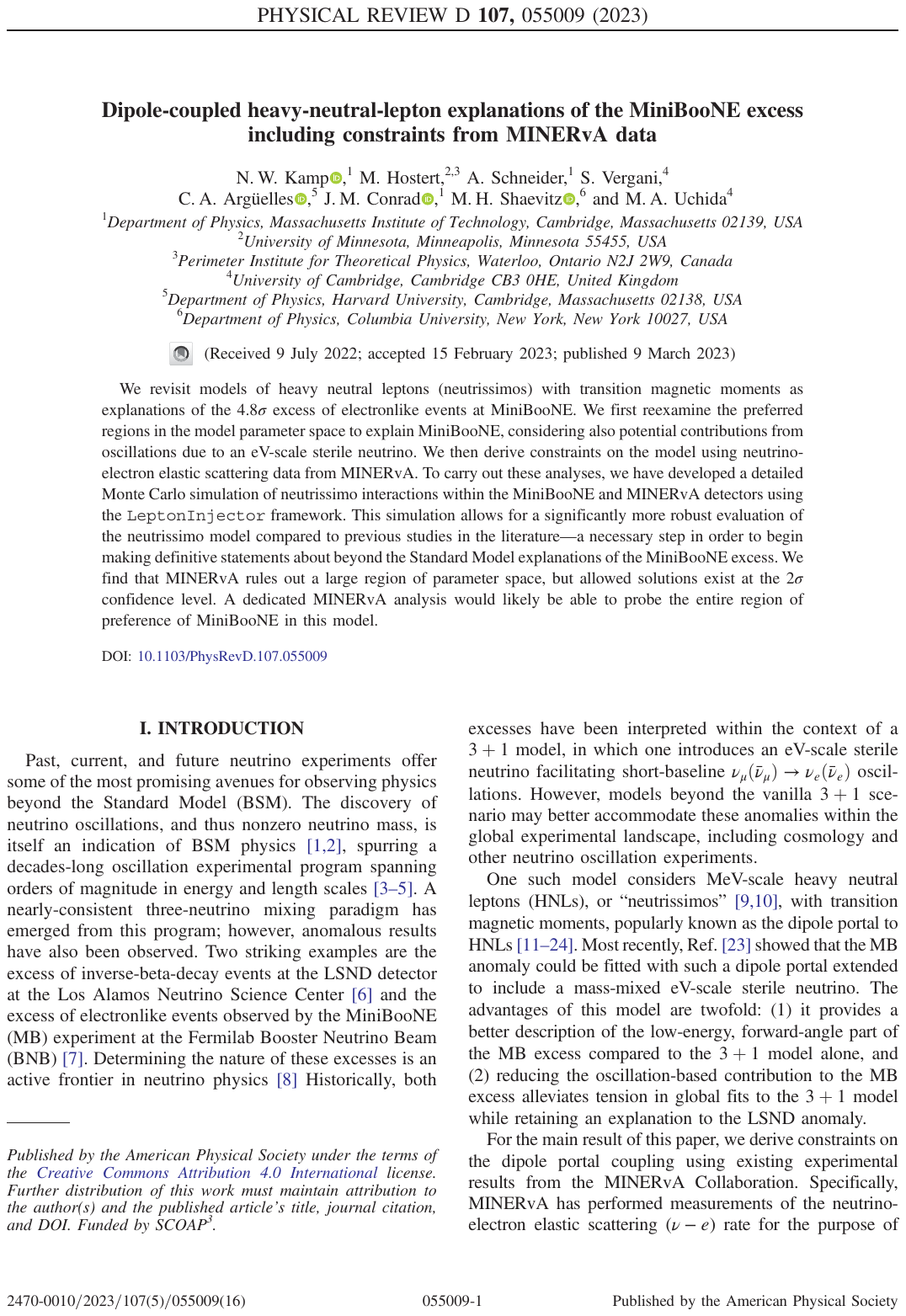}

\chapter{The Coherent CAPTAIN-Mills Experiment}
\label{ch:ccm}

We now turn to the Coherent CAPTAIN-Mills (CCM) experiment at Los Alamos National Laboratory (LANL).
CCM uses a 10-ton cylindrical liquid argon detector to observe the scintillation and Cherenkov light generated by the interactions of particles produced at the Lujan beam dump facility~\cite{CCM:2021leg}.
CCM was originally designed to detect $\nu_\mu$ produced by $\pi^+$ decay-at-rest via coherent elastic neutrino-nucleus scattering (CE$\nu$NS)~\cite{COHERENT:2017ipa}.
Given the baseline and neutrino energy accessible to the CCM detector, any deficit in this rate would be smoking-gun evidence for $\nu_\mu$ disappearance via a sterile neutrino~\cite{CCM:2021leg}.
In order to realize this physics goal, CCM was built to maximize photon yield and thus minimize the detection energy threshold.
This setup makes CCM an ideal detector to search for potential light dark matter (DM) particles produced in the Lujan beam dump, including $U(1)'$ vector-portal DM~\cite{CCM:2021leg}, leptophobic vector-portal DM~\cite{CCM:2021yzc}, and axion-like particles (ALPs)~\cite{CCM:2021lhc}.
CCM120, a 120-PMT prototype detector, completed a six-week engineering run in Fall 2019, which was used to set constraints on these models.
The full 200-PMT CCM200 detector is funded for a three-year physics run at the Lujan facility, which began in Summer 2022.
Results from this dataset will set strong constraints on the DM models listed above as well as a variety of additional beyond-the-Standard-Model (BSM) scenarios.

In this chapter, we discuss the projects within CCM in which I was involved during my time as a graduate student.
These include the fabrication of veto PMT assemblies for CCM200, the investigation of Cherenkov light detection as a powerful background rejection technique in CCM, and the evaluation of CCM's sensitivity to the neutrissimo model from \cref{ch:neutrissimos}.

\section{The CCM Beamline and Detector}

\begin{figure}[h!]
    \centering
    \includegraphics[width=0.6\textwidth]{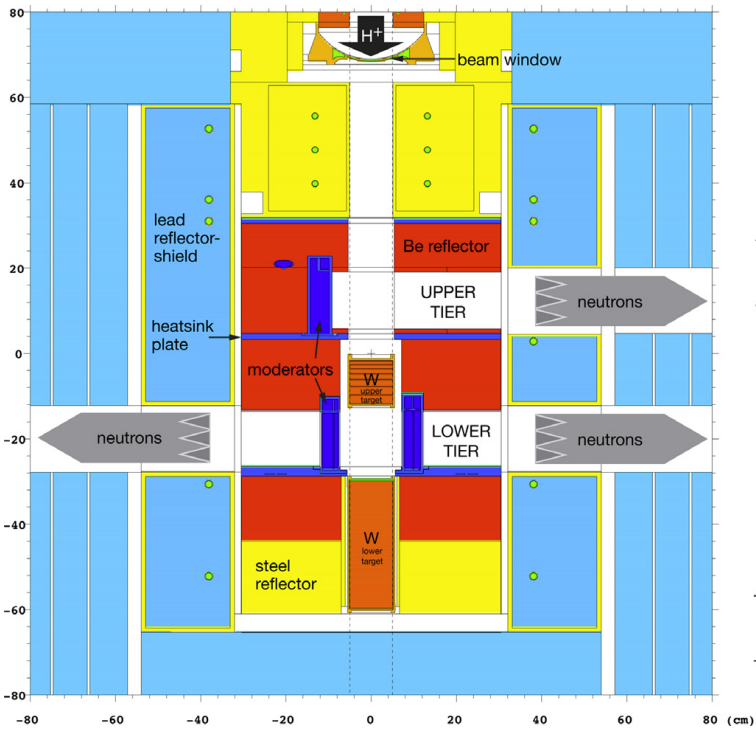}
    \caption{Schematic depiction of the Lujan TMRS Figure from Ref.~\cite{ZAVORKA2018189}.}
    \label{fig:lujan_target}
\end{figure}

\begin{figure}[h!]
    \centering
     \begin{subfigure}[b]{0.45\textwidth}
         \centering
         \includegraphics[width=\textwidth]{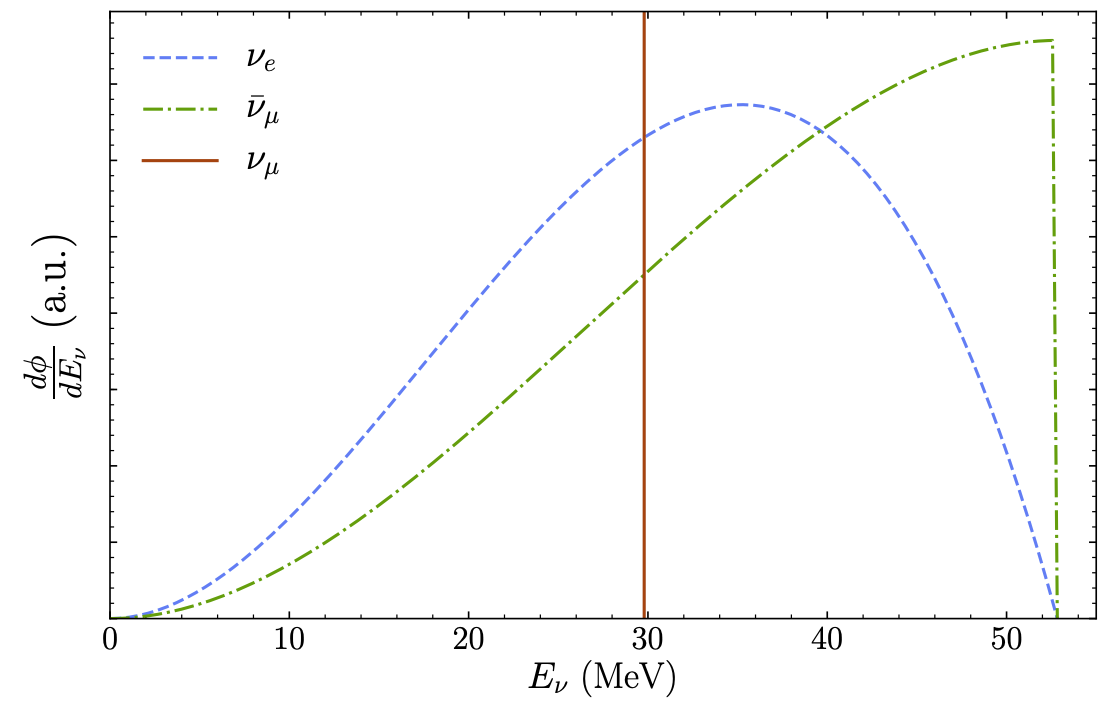}
         \caption{}
         \label{fig:lujan_energy}
     \end{subfigure}
     \hfill
     \begin{subfigure}[b]{0.45\textwidth}
         \centering
         \includegraphics[width=\textwidth]{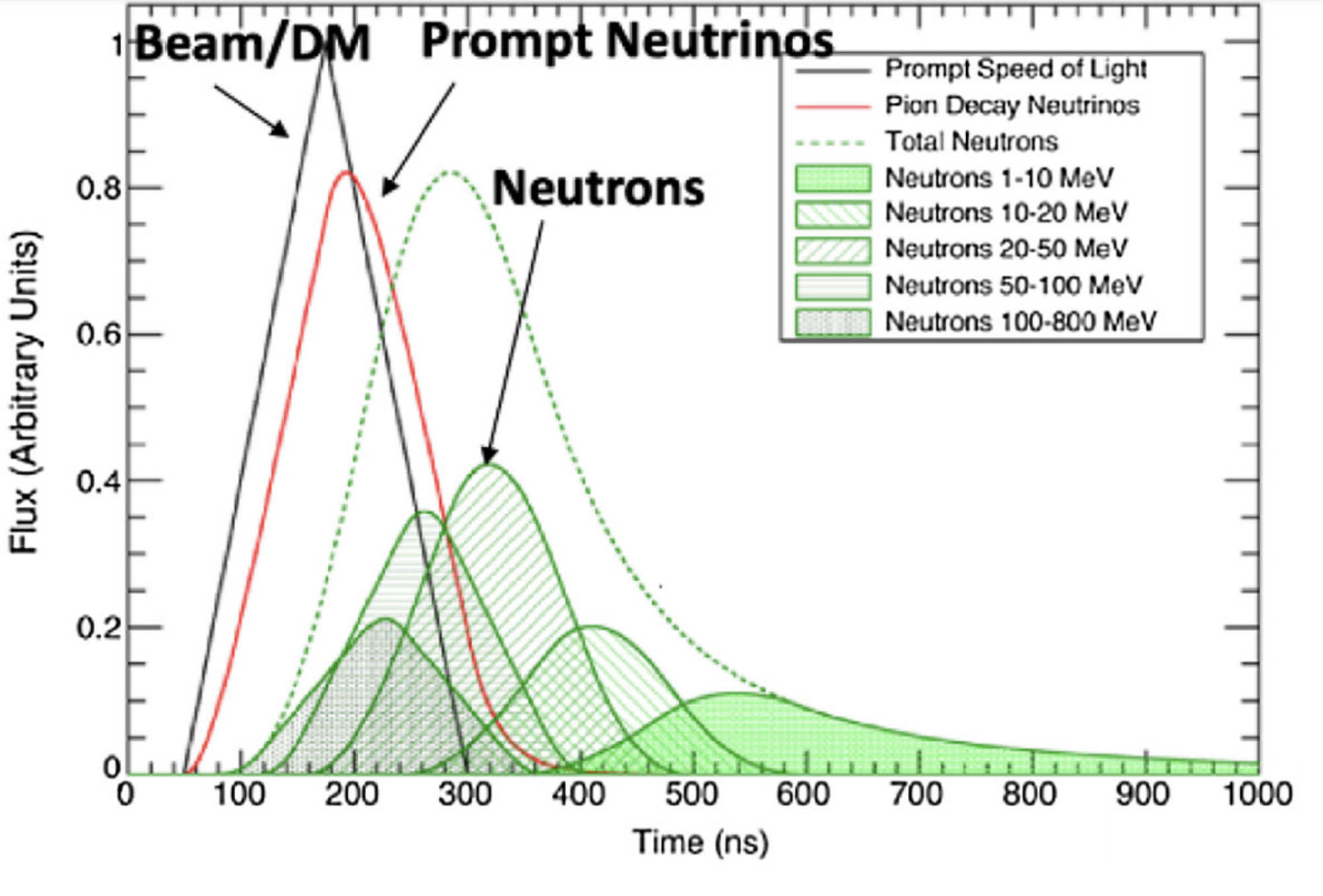}
         \caption{}
         \label{fig:lujan_timing}
     \end{subfigure}
     \hfill
        \caption{\Cref{fig:CCM200_render}, from Ref.~\cite{Baxter:2019mcx}, shows the energy distribution of $\pi^+$ decay-at-rest neutrinos from the Lujan beam dump source. \Cref{fig:lujan_timing}, from Ref.~\cite{CCM:2021leg}, shows the timing distribution of particles produced in the Lujan beam dump source after traveling through the TMRS.}
        \label{fig:lujan_dists}
\end{figure}

The CCM detector operates at the Lujan beam dump facility of the Los Alamos Neutron Science Center (LANSCE)~\cite{NELSON2012172,LISOWSKI2006910}.
At LANSCE, protons are accelerated to 800 MeV in the proton storage ring and bunched before impinging vertically downward on the Lujan tungsten target at a rate of 20~Hz.
The bunches contain $\sim 3.1 \times 10^{13}$ protons each and are delivered in a $280~{\rm ns}$ triangular pulse.
The beam spills occur at a rate of 20~Hz, corresponding to a duty factor of $\sim 5 \times 10^{-6}$.
This low duty factor is essential for isolating prompt signal events from background, setting the Lujan facility apart from similar beam dump sources.
A cascade of particles is produced when protons hit the tungsten target.
The target is housed within a cylindrical target-moderator-reflector-shield (TMRS) which provides immediate shielding around the tungsten disks as well as moderators to control neutron output~\cite{ZAVORKA2018189}, as shown in \cref{fig:lujan_target}.
The TMRS itself is surrounded by an additional 4~m of steel shielding.
Though the primary objective of the Lujan beam dump is neutron production, proton collisions also create pions, electrons, photons, and potential DM particles.
Neutrinos are created via the $\pi^+ \to \nu_\mu (\mu^+ \to \overline{\nu}_\mu \nu_e e^+)$ decay chain, as $\pi^-$ dominantly capture on nuclei instead of decaying.
The total neutrino flux produced by the Lujan source for each neutrino species in this decay chain is $4.74 \times 10^5~\nu/{\rm cm}^2/s$ at the CCM120 location (20~m from the tungsten target).
The $\pi^+$ come to rest in the target before decaying, leading to the neutrino energy distributions shown in \cref{fig:lujan_energy}.
\Cref{fig:lujan_timing} shows the timing profile of different particles produced in the beam dump at the CCM120 location.
This timing distribution is essential for separating $\nu_\mu$ from $\pi^+$ decay-at-rest and speed-of-light DM particles in the prompt region from delayed non-relativistic neutron backgrounds.
This is the main strategy behind the CCM physics analyses.

\begin{figure}[h!]
    \centering
     \begin{subfigure}[b]{0.35\textwidth}
         \centering
         \includegraphics[width=\textwidth]{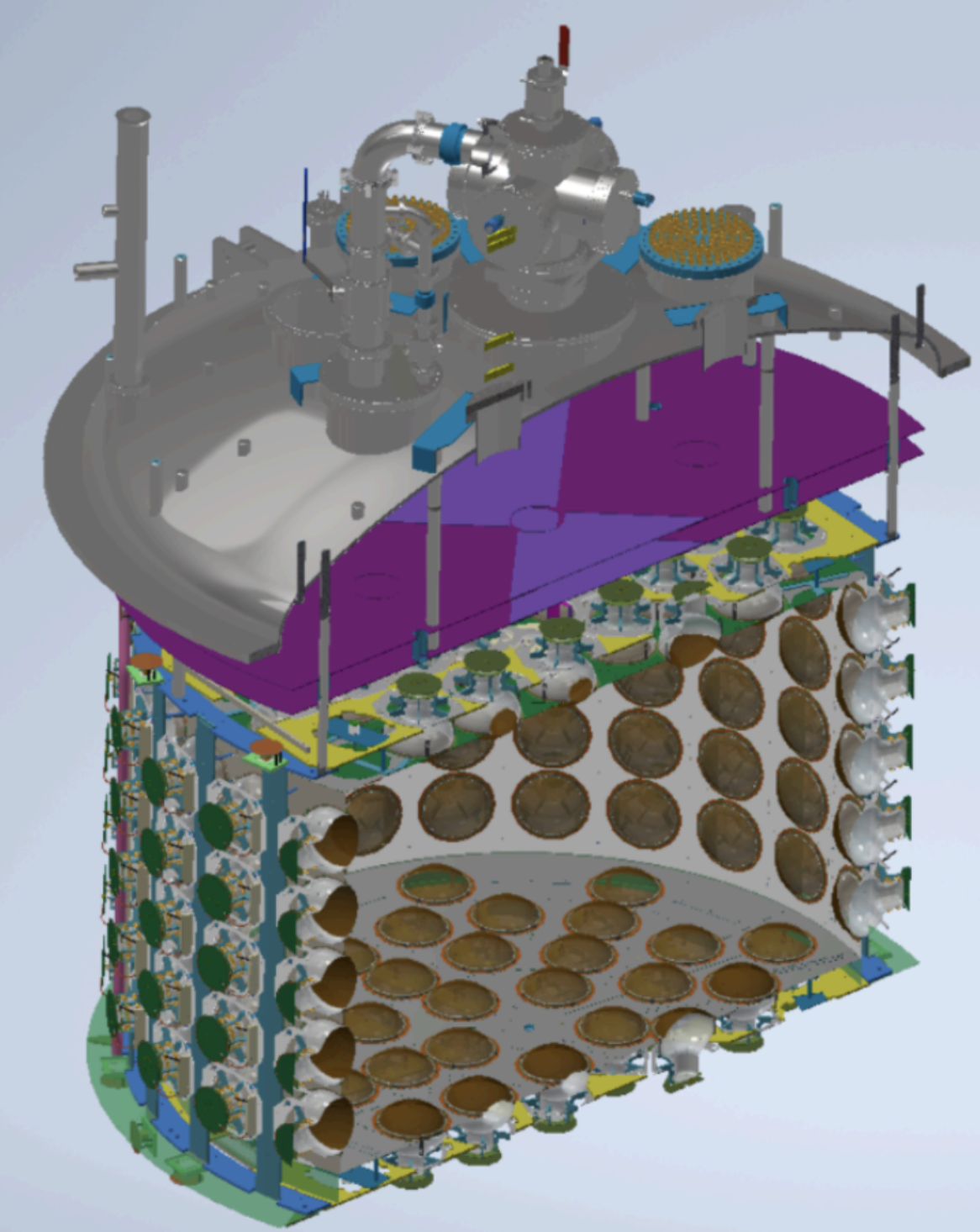}
         \caption{}
         \label{fig:CCM200_render}
     \end{subfigure}
     \hfill
     \begin{subfigure}[b]{0.55\textwidth}
         \centering
         \includegraphics[width=\textwidth]{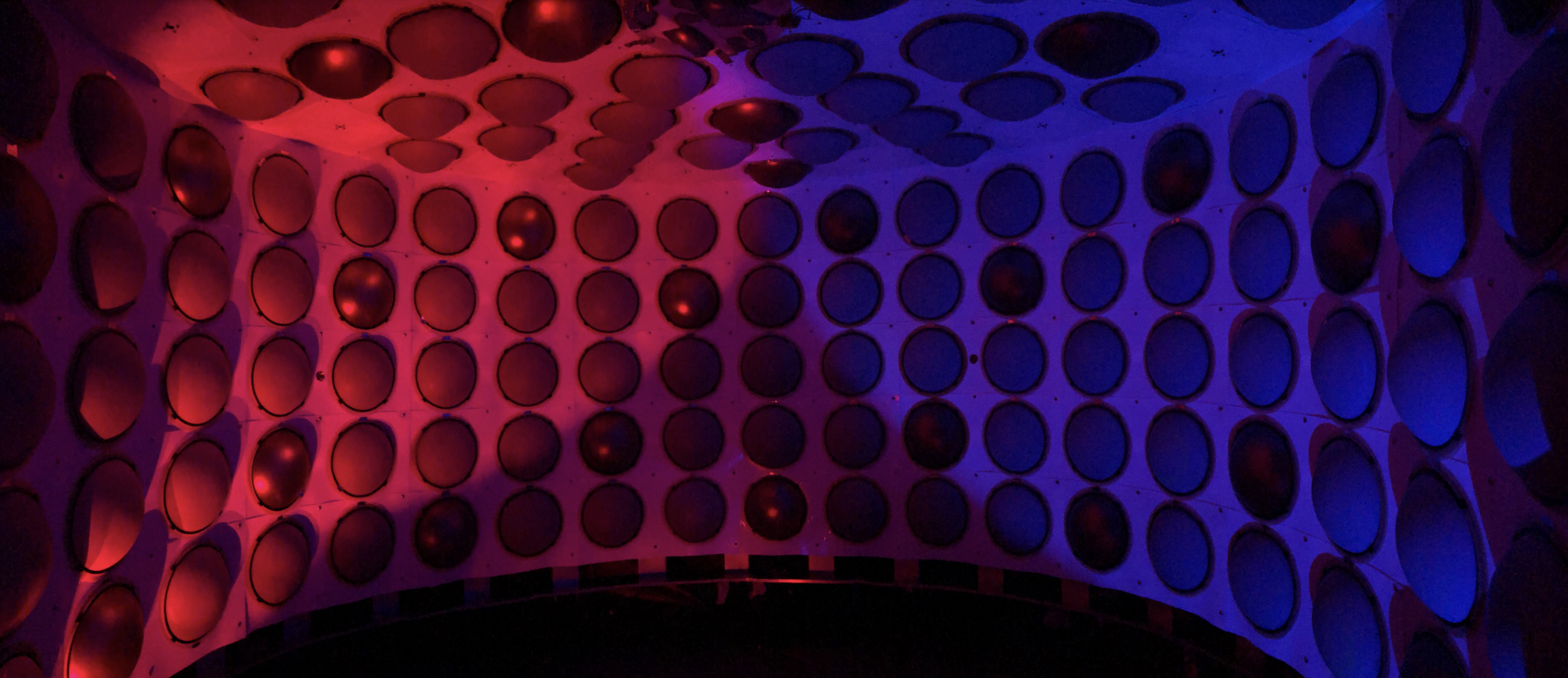}
         \caption{}
         \label{fig:CCM200_interior}
     \end{subfigure}
     \hfill
        \caption{\Cref{fig:CCM200_render} shows a schematic 3D rendering of the CCM200 detector. \Cref{fig:CCM200_interior} shows an image of the interior of the CCM200 detector.}
        \label{fig:CCM200_detector}
\end{figure}

The CCM detector is a 10-ton cylindrical volume of liquid argon 2.58~m diameter and 2.25~m in height.
CCM repurposes the cryostat used by the Cryogenic Apparatus for Precision Tests of
Argon Interactions with Neutrinos (CAPTAIN) experiment~\cite{CAPTAIN:2013irr}.
In order to take advantage of the timing information discussed above, CCM relies on the fast scintillation light produced when particles interact in the detector, which is emitted on a characteristic time scale of $\mathcal{O}(ns)$.
Scintillation light in liquid argon was discussed in \cref{sec:ub_light_collection} within the context of the MicroBooNE experiment.
In order to collect this scintillation light, the CCM200 detector is instrumented with 200 8-inch Hamamatsu R5912-mod2 PMTs, the same used in the MicroBooNE light collection system described in \cref{sec:ub_light_collection}.
A schematic drawing of this detector is shown in \cref{fig:CCM200_render}.
80 PMTs are arranged uniformly on the top and bottom of the detector, while 120 PMTs are arranged in five rows around the barrel.
The CCM120 engineering run only included the PMTs around the barrel.
80\% of the PMTs are coated with TPB, which shifts the 128~nm argon wavelength to the visible regime where it can be detected by the PMTs.
Additionally, the walls of the detector are covered in reflective foils that are evaporatively coated with TPB.
The remaining 20\% of PMTs are not coated with TPB and are thus not sensitive to the direct scintillation light produced in the detector.
They are only able to observe photons after they are re-emitted by a TPB-coated surface somewhere else in the detector, either on another PMT or on the reflective foils.
The presence of uncoated PMTs breaks the degeneracy of the detector response to 128~nm light, helping determine the TPB properties during the calibration process.
This point will be important later in \cref{sec:ccm_cherenkov}.
\Cref{fig:CCM200_interior} shows an image of the interior PMTs of the CCM200 detector, in which the dull (reflective) hemispheres correspond to coated (uncoated) PMTs.
The voltage output from each PMTs is carried through warm cables to a series of CAEN VX1730 boards, which digitize the PMT signals at a rate of 500~MHz.
While the beam is running, the digitizers are triggered at a rate of 22.2~Hz, which includes the beam trigger (20~Hz), random strobe trigger (1.1~Hz), and light-emitting-diode (LED) trigger (1.1~Hz).
For each trigger, 16~$\mu$s of waveform data on each PMT are saved.

\begin{figure}[h!]
    \centering
     \begin{subfigure}[b]{0.35\textwidth}
         \centering
         \includegraphics[width=\textwidth]{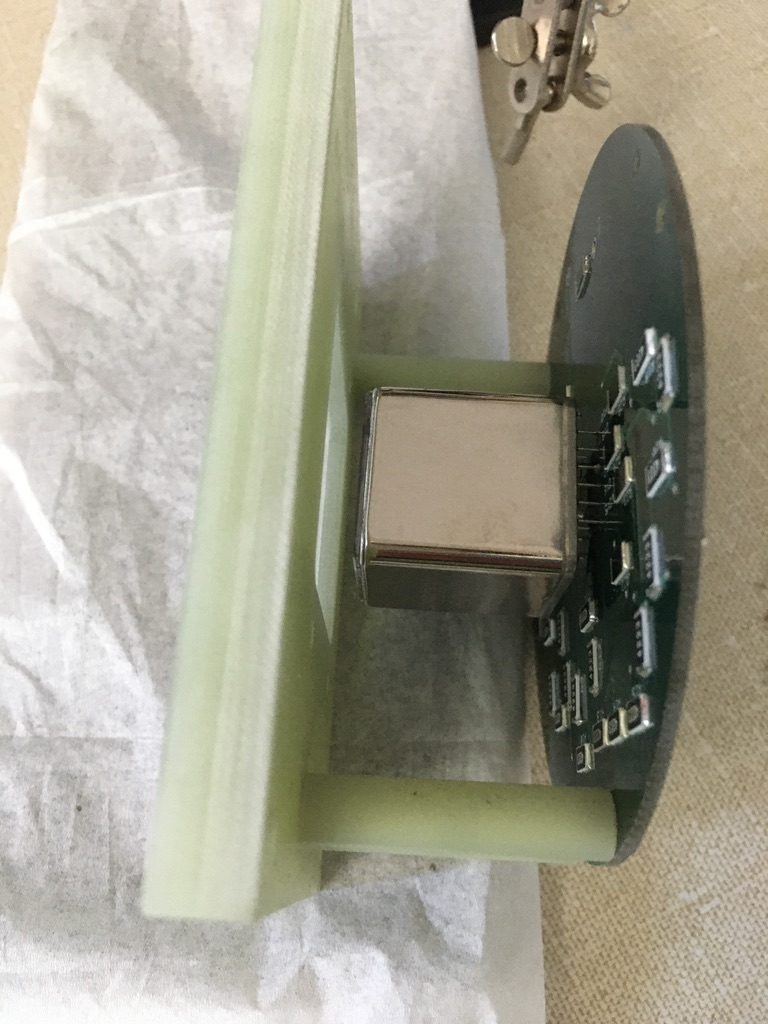}
     \end{subfigure}
     \hfill
     \begin{subfigure}[b]{0.45\textwidth}
         \centering
         \includegraphics[width=\textwidth]{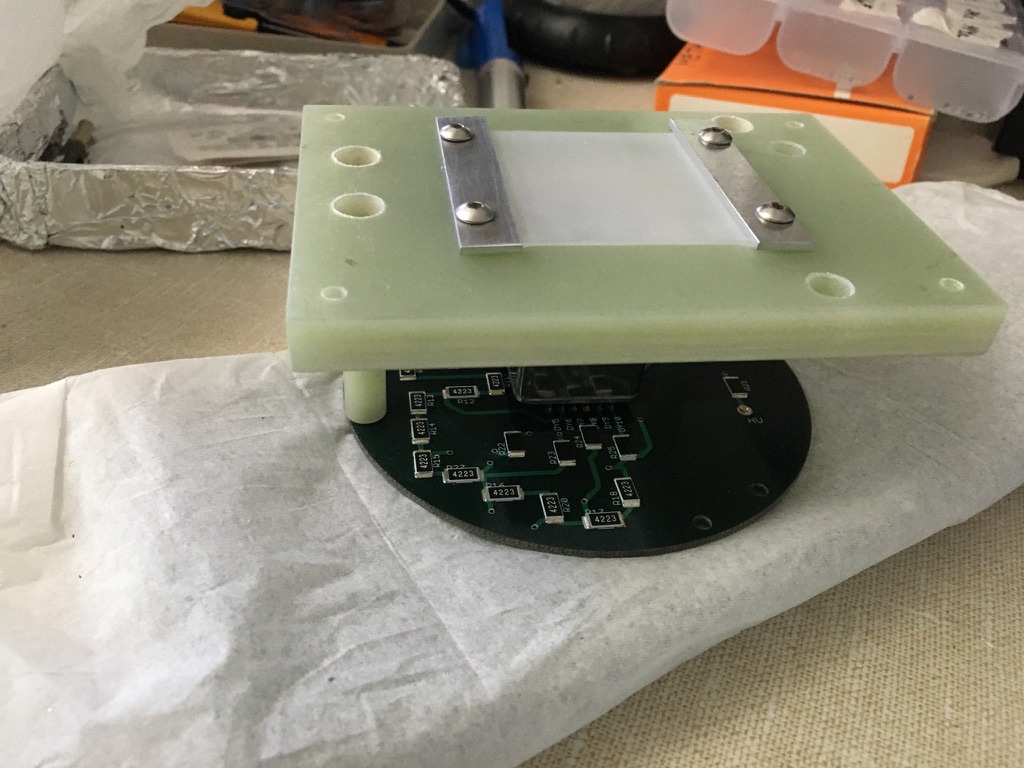}
     \end{subfigure}
     \hfill
        \caption{Two of the veto PMT assemblies constructed at MIT, including the 1-inch PMT, base circuit board, and TPB-coated acrylic window.}
        \label{fig:veto_pictures}
\end{figure}

\begin{figure}[h!]
    \centering
     \begin{subfigure}[b]{0.35\textwidth}
         \centering
         \includegraphics[width=\textwidth]{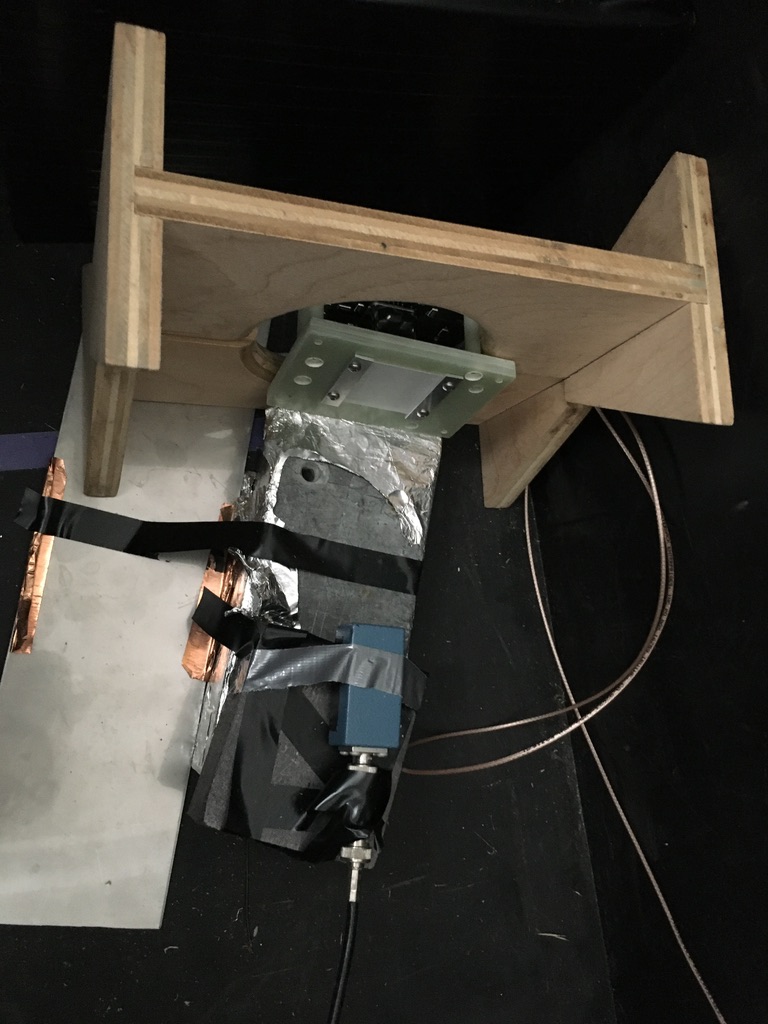}
         \label{fig:veto_box_image}
     \end{subfigure}
     \hfill
     \begin{subfigure}[b]{0.55\textwidth}
         \centering
         \includegraphics[width=\textwidth]{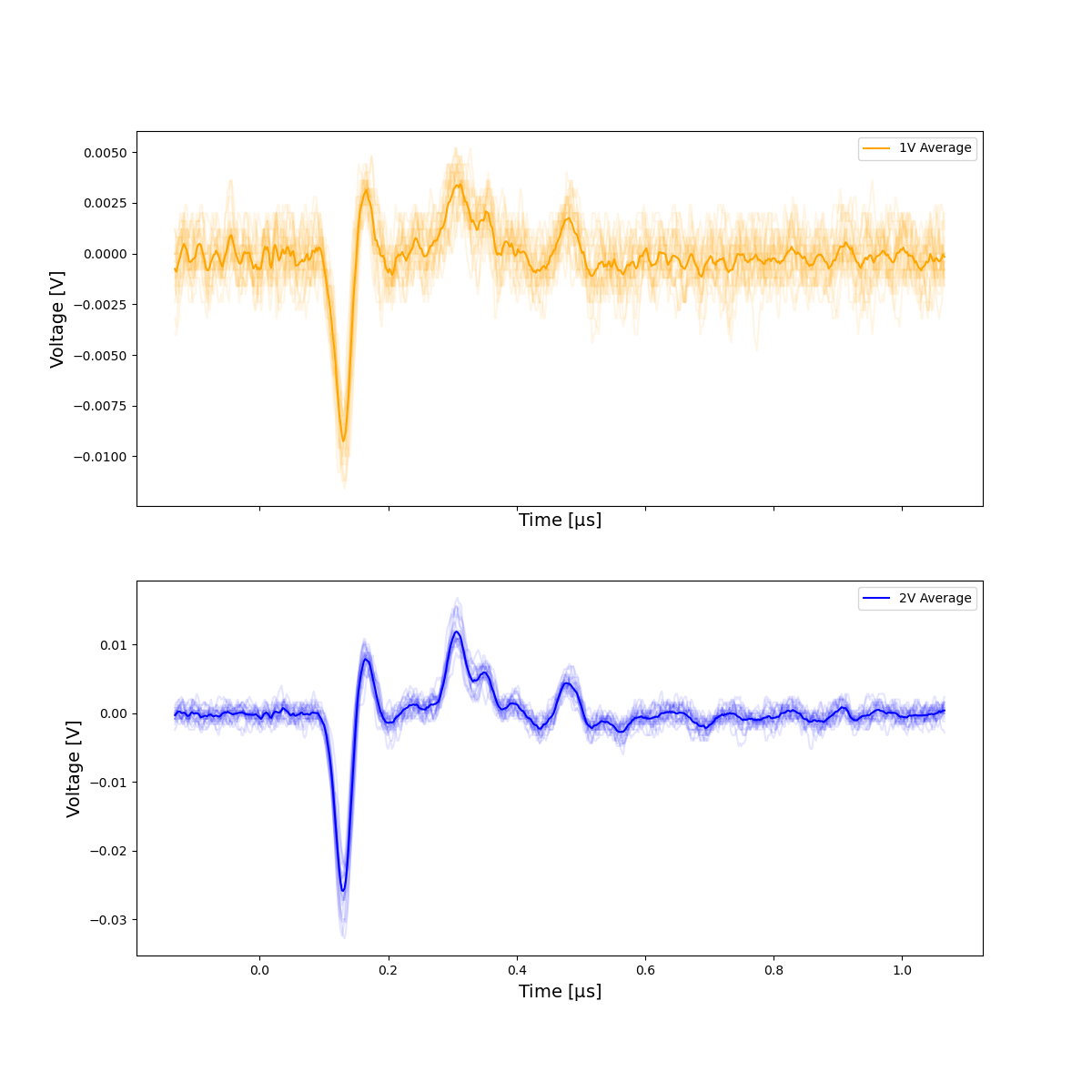}
         \label{fig:veto_avg_response}
     \end{subfigure}
     \hfill
        \caption{\Cref{fig:veto_box_image} shows one of the veto PMTs across from the LED in the light-tight box. \Cref{fig:veto_avg_response} shows the average response of the 20 veto PMTs to 1~V (top) and 2~V (bottom) LED pulses.}
        \label{fig:veto_test}
\end{figure}

CCM200 also houses an additional 40 veto PMTs outside of the fiducial volume to identify and reject events that originate outside of the fiducial volume.
These veto PMTs observe light 
The CCM120 detector used only 28 veto PMTs, of which 5 were the same 8-inch Hamamatsu R5912-mod2 PMTs used in the inner volume and 23 were 1-inch Hamamatsu PMTs.
An additional 20 1-inch veto PMTs were assembled and tested at MIT in Summer 2021.
Each PMT was assembled into the mounts shown in \cref{fig:veto_pictures}, which included an acrylic window in front of the photocathode and a custom base circuit for reading out the PMT signal.
The acrylic windows were coated with a mixture comprising 1~g of TPB, 1~g of polystyrene, and 50~mL of (very pungent) toluene.
The PMTs were tested in a light-tight box with an LED, as shown in \cref{fig:veto_test}.
The LED was pulsed via a function generator at 1~V and 2~V, leading to the average veto PMT response shown in \cref{fig:veto_avg_response}.
The overshoot observed in these waveforms came largely from the custom breadboard ``splitter'' circuit assembled for this study, which separated transient PMT signals from high voltage.
As the official splitter circuit used by CCM is more robust to ringing, this response was deemed sufficient for CCM200, and the 1-inch veto PMTs were shipped back to LANL.

\begin{figure}[h!]
    \centering
    \includegraphics[width=0.6\textwidth]{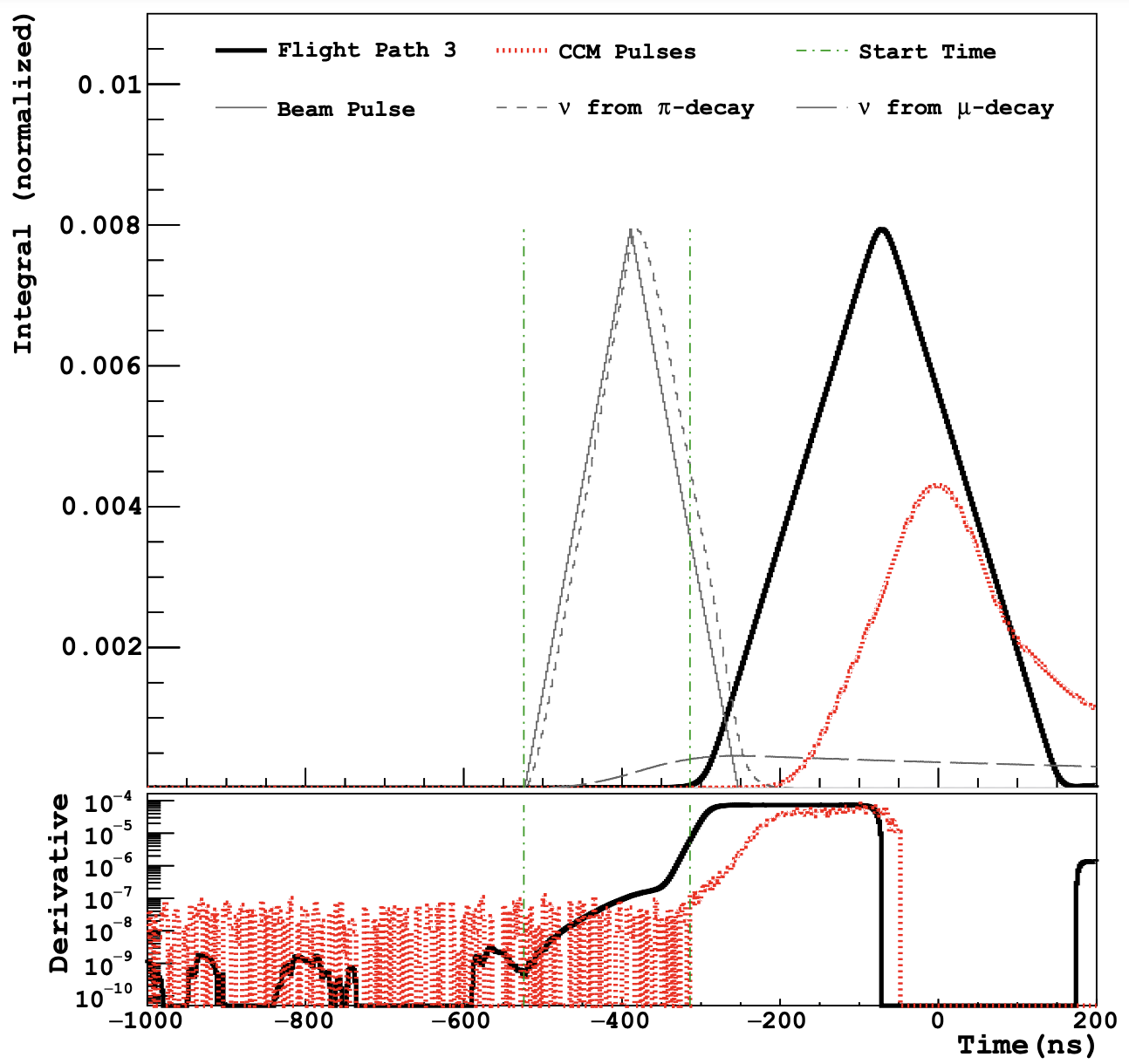}
    \caption{The timing distribution of photons from the Lujan source (solid black line) compared with that of neutrons measured in CCM120 (dashed red line). Figure from Ref.~\cite{CCM:2021leg}.}
    \label{fig:ccm_timing_final}
\end{figure}

CCM uses a variety of shielding between the target and detector to slow down and absorb beam-related neutrons, the most prominent beam-related background at the Lujan facility.
This reduces the overall background rate and also separates beam-related neutrons in time from prompt BSM signals.
\Cref{fig:lujan_timing} shows the timing distributions of neutrons, $\pi^+$ decay-at-rest neutrinos, and prompt DM from the Lujan source after the 4~m of steel shielding surrounding the TMRS.
The neutrons are pushed back further in time using additional shielding concrete, steel, and borated polyethylene shielding.
For the CCM120 run, 6~ft of concrete and 16~in. of steel were placed immediately outside of the steel surrounding the TMRS.
A 2~m-thick wall of steel was constructed in ER2, the room housing the CCM detector. 
Three additional walls of concrete bricks covered with borated polyethylene sheets were constructed around the front and sides of the CCM detector.
The borated polyethylene sheets were included to absorb neutrons that had been sufficiently slowed down by the other shielding.
An Eljen EJ-301 scintillation detector was deployed in a flight path neighboring CCM to observe the time of the prompt gamma flash from the Lujan source.
\Cref{fig:ccm_timing_final} shows the timing distribution of these photons compared to that of neutrons measured in the CCM120 detector.
The discrepancy between these distributions was used to determine a 190~ns background-free region of interest (ROI) in which to look for prompt signals from neutrinos or BSM particles.
The steady-state background expectation in the ROI is taken from a beam-out-of-time measurement before the ROI, which is about 22 times larger than the ROI itself~\cite{CCM:2021leg}.

A number of updates were made to the shielding for CCM200.
A new larger wall of concrete, steel, and borated polyethylene was constructed in ER1, and more steel and concrete shielding was placed in ER2.
Additionally, a steel and polyethylene roof was constructed over CCM to suppress backgrounds entering the detector from above.
Preliminary estimates suggest the shielding upgrades have reduced backgrounds in CCM200 by a factor of 7 compared with CCM120~\cite{EdThesis}.

CCM uses three different methods to calibrate the detector.
The first of these consists of two blue LEDs, which are used to determine the single photo-electron (p.e.) response of each PMT, as measured in units of analog-to-digital-conversion (ADC) counts.
The PMTs are gain-matched in order to make the p.e.-to-ADC values as similar as possible between them.
These LEDs are operated continuously while taking beam data in order to monitor changes in the single p.e. rate over time.
CCM also uses ${}^{57}$Co and ${}^{22}$Na radioactive sources, which emit 126~keV and 2.2~MeV photons, respectively, to determine the energy scale of the detector.
This is measured in units of p.e./MeVee (electromagnetic equivalent)--a quenching factor must be considered to determine the energy scale of the detector response to nuclear recoils~\cite{Simon:2002cw,COHERENT:2020iec}.
The sodium peak, which provided the more robust energy scale measurement, set the energy scale of CCM120 to $15.1 \pm 4.0$ p.e./MeVee.
Finally, a custom diode-pumped laser is used to determine the PMT response to 213~nm and 532~nm light.
The ability to operate at two different wavelengths is important for disentangling the TPB response in CCM.
Data from the radioactive sources and the laser were compared with output from a \texttt{Geant4} simulation of CCM120 in order to tune the optical model of the detector, as described in Ref.~\cite{EdThesis}.

During the operation of CCM120, the effective PMT quantum efficiency was about a factor of two lower than the expectation from Hamamatsu.
This was likely related to the oxygen contamination, which was measured at the 2~ppm level in the CCM120 liquid argon.
Such contamination can lead to a reduction of the scintillation light produced in liquid argon by about a factor of two~\cite{WArP:2008dyo}, similar to the apparent reduction in quantum efficiency.
In response to this, CCM200 will use a dedicated oxygen filtration system adapted from the Mini-CAPTAIN experiment~\cite{CAPTAIN:2020pup} to keep the oxygen level below 100~ppb.
Filtration is expected to increase the scintillation light output by a factor of around three~\cite{EdThesis}.

\section{Cherenkov Light Reconstruction} \label{sec:ccm_cherenkov}

CCM relies predominantly on scintillation light to reconstruct particle interactions in the detector.
This is because liquid argon is a powerful scintillation medium, as described in \cref{sec:ub_light_collection}.
That being said, the identification of Cherenkov radiation in CCM is a promising potential upgrade.
Cherenkov radiation refers to the light emitted by a charged particle as it travels faster than the speed of light in a given medium~\cite{Grupen:2008zz}.
The particle polarizes surrounding atoms as it moves through the material, creating a cloud of electric dipoles.
If $\beta > c/n$ (where $n$ is the wavelength-dependent index of refraction), the dipoles are not arranged symmetrically around the particle, resulting in a non-vanishing overall dipole moment.
The time variation of this dipole moment results in Cherenkov radiation.
The light is emitted in a cone with an opening angle of $\cos \theta_C = 1/n \beta$.
The number of photons emitted per unit length is given by~\cite{Grupen:2008zz}
\begin{equation} \label{eq:Cherenkov_light_yield}
\frac{dN}{dx} = 2\pi \alpha z^2 \int_{\lambda_1}^{\lambda_2} \bigg(1 - \frac{1}{(n(\lambda)\beta)^2} \bigg) \frac{d\lambda}{\lambda^2},
\end{equation}
where $z$ is the electric charge of the particle in units of the electron charge $e$, and $\lambda_1$ and $\lambda_2$ represent the photon wavelength range under consideration.
From \cref{eq:Cherenkov_light_yield}, one can see that the intensity of Cherenkov radiation peaks at lower wavelengths.

For liquid argon, $n(\lambda = 128~{\rm nm}) \sim 1.5$ and decreases for larger wavelengths~\cite{Grace:2015yta}.
This means that only electrons satisfy the $\beta > c/n$ requirement at typical CCM energies of $\mathcal{O}(10~{\rm MeV})$.
Though they are neutral, photons in this energy regime will produce Cherenkov light after undergoing pair production to an $e^+ e^-$ final state.
Crossing cosmic muons will also produce Cherenkov light, though such events are rare in the 1.6~$\mu$s beam window and can be tagged by the veto region.
Importantly, the most prominent backgrounds in CCM--beam-related neutrons and low-energy photons from radioactive impurities--will not produce Cherenkov light.
In contrast, signals from ALPs~\cite{CCM:2021lhc} and other BSM scenarios of interest to CCM~\cite{Dutta:2021cip} will produce Cherenkov light via higher-energy electrons and photons in the final state.
Thus, Cherenkov light identification is a promising method for rejecting backgrounds in CCM and maximizing sensitivity to new physics.
The directional nature of Cherenkov light will also be helpful in requiring particles to originate from the tungsten target, which is not possible using only isotropic scintillation light. 

\begin{figure}[h!]
    \centering
     \begin{subfigure}[b]{0.45\textwidth}
         \centering
         \includegraphics[width=\textwidth]{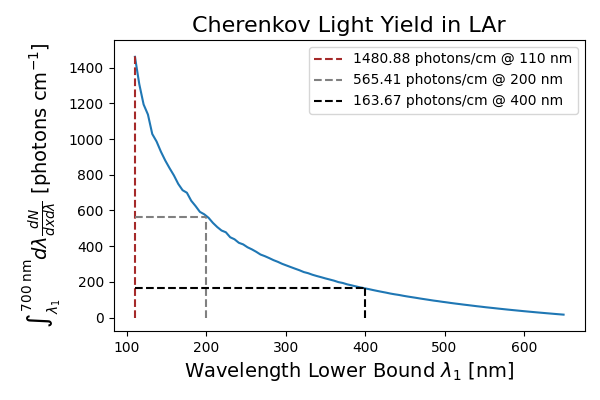}
         \caption{}
         \label{fig:CherenkovLightYield}
     \end{subfigure}
     \hfill
     \begin{subfigure}[b]{0.45\textwidth}
         \centering
         \includegraphics[width=\textwidth]{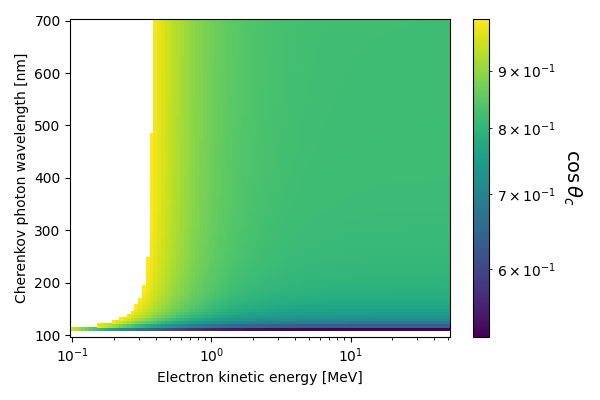}
         \caption{}
         \label{fig:CherenkovAngle}
     \end{subfigure}
     \hfill
        \caption{\Cref{fig:CherenkovLightYield} shows the integration in \cref{eq:Cherenkov_light_yield} as a function of $\lambda_1$ for $\lambda_2 = 700~{\rm nm}$ and $z=1$. \Cref{fig:CherenkovAngle} shows the Cherenkov cone angle $\cos \theta_C$ for an electron as a function of the photon wavelength and the electron kinetic energy.}
        \label{fig:CherenkovAnalytic}
\end{figure}

In order to determine the number of Cherenkov photons generated per unit length traveled by a charged particle, one must perform the integration \cref{eq:Cherenkov_light_yield}.
To do this, we use the parameterization of the liquid argon index of refraction given in Ref.~\cite{Grace:2015yta}.
We fix $\lambda_2 = 700~{\rm nm}$, as this is the upper bound on the detectable wavelength range of the CCM PMTs.
\Cref{fig:CherenkovLightYield} shows a numerical integration of \cref{eq:Cherenkov_light_yield} for an electron ($z=1$) as a function of $\lambda_1$.
One can see that sensitivity to lower wavelengths greatly increases the number of photons one can detect.
This is possible in CCM thanks to the use of TPB, which can covert the UV Cherenkov light into the visible range, greatly increasing the number of Cherenkov photons detected by each PMT.
The cone angle as a function of the electron kinetic energy and photon wavelength is shown in \cref{fig:CherenkovAngle}, which is $\cos \theta_C \approx 0.8$ over most of the parameter space.
This figure also indicates the threshold for Cherenkov emission; over most photon wavelengths, Cherenkov light is produced for electron kinetic energies above $T_{e^-} \sim 0.4~{\rm MeV}$.

A number of specialized studies have been performed investigating the detection of Cherenkov light in scintillation detector mediums, including water-based liquid scintillator~\cite{Kaptanoglu:2021prv,Caravaca:2020lfs}, slow-fluor scintillator~\cite{Dunger:2022gif}, and linear alkylbenzene~\cite{Gruszko:2018gzr}.
Two full-scale experiments have been able to perform a statistical measurement of Cherenkov radiation in a scintillating medium: Borexino, observing elastic scattering of sub-MeV solar $\nu_e$ in a large liquid scintillator detector~\cite{BOREXINO:2021efb}, and ICARUS, observing cosmic muons in a large liquid argon detector~\cite{Antonello:2004sx}.
However, no large-scale scintillation experiment has been able to observe Cherenkov light on an event-by-event basis.
This is because the Cherenkov light output is typically smaller than the scintillation light output by around two orders of magnitude, which can be understood by comparing the numbers in \cref{fig:CherenkovLightYield} to the $\mathcal{O}(10^4)$ scintillation photons per MeV produced in liquid argon~\cite{Jones:2013bca}.

Event-by-event detection of Cherenkov light might be possible in CCM for a few reasons.
First, liquid argon is transparent to UV light, which substantially increases the number of detectable Cherenkov photons as shown in \cref{fig:CherenkovLightYield}.
TPB-coated PMTs can detect these UV Cherenkov photons by shifting them into the visible range.
The wavelength-shifting efficiency of TPB actually increases by a factor of around two for $\lambda \sim 200~{\rm nm}$ compared to 128~nm argon scintillation light~\cite{Benson:2017vbw}, potentially boosting the response of CCM PMTs to Cherenkov light.
Second, the singlet lifetime of argon excimers is approximately 8~ns~\cite{DEAP:2020hms}, which is much slower than the Cherenkov light emission timescale ($< 1~{\rm ns}$)~\cite{Bae:2022dti}.
The 2~ns sampling time of the CCM digitizers may allow for the separation in time of prompt Cherenkov photons from the first scintillation photons.
Finally, 20\% of the PMTs in CCM are not coated in TPB.
This means they are only sensitive to scintillation photons after they have been emitted from TPB somewhere else in the detector, delaying their arrival time in these PMTs.
In contrast, Cherenkov photons in the visible regime will indeed produce signals on the uncoated PMTs.
Though the larger UV component of the Cherenkov spectrum will not be accessible, even just a few photoelectrons detected by uncoated PMTs in the prompt time region would be a strong indication of Cherenkov light.

\subsection{Simulation-Based Sensitivity Estimation} \label{sec:CCM_cherenkov_simulation}

In order to assess the sensitivity of CCM to Cherenkov light, we use the \texttt{Geant4} simulation of CCM200 described in Ref.~\cite{EdThesis}.
We inject electrons at the center of the detector, traveling along the beam direction with kinetic energies of $\{1,2,3,4,5\}~{\rm MeV}$.
In addition to the usual scintillation light, Cherenkov light is emitted along the electron path within the 100~nm to 700~nm wavelength range.
Both scintillation and Cherenkov photons are subject to wavelength-dependent effects within the detector, including absorption by impurities~\cite{WArP:2008dyo,WArP:2008rgv} and Rayleigh scattering~\cite{Grace:2015yta}.
The TPB response in the simulation has not been carefully evaluated in the $200 \leq \lambda~[{\rm nm}] \leq 400$ regime, so we have turned the TPB off for the purposes of this study.
Instead, we apply a time offset for coated PMTs reflecting the TPB emission, which is sampled from an exponential with a time constant of 1.7~ns~\cite{FLOURNOY1994349}.
For both coated and uncoated PMTs, we apply an additional time offset related to the rise time of the PMT $\tau_{\rm PMT}$, which is about 2.5~ns in CCM.
This is sampled from a Gaussian with a mean and width of $\tau_{\rm PMT}/2$.
For each photon that hits a PMT surface, we apply a weight reflecting the detection probability $P_{\rm det}$, which is calculated as
\begin{equation}
\begin{split}
&P_{\rm det}^{\rm uncoated} = P_{\rm PMT} \\
&P_{\rm det}^{\rm coated} = P_{\rm PMT} P_{\rm TPB},
\end{split}
\end{equation}
where $P_{\rm PMT}$ reflects the PMT quantum efficiency and $P_{\rm TPB}$ reflects the impact of TPB absorption and re-emission on the detection probability.
The former is given by $P_{\rm PMT} = 0.15$ for coated PMTs and
\begin{equation}
P_{\rm PMT} = 
\begin{cases}
0.15 & 300 < \lambda~[{\rm nm}] < 650 \\
0 & {\rm otherwise}
\end{cases}
\end{equation}
for uncoated PMTs, where $0.15$ is the typical quantum efficiency of a cryogenic PMT~\cite{EdThesis,photonics2000photomultiplier}.
This reflects the approximation that TPB makes the coated PMTs sensitive to all photon wavelengths within 100~nm to 700~nm, while the uncoated PMTs are only sensitive to the visible range~\cite{photonics2000photomultiplier}.
The TPB transmission probability is approximated by
\begin{equation} \label{eq:tpb_efficiency}
\begin{split}
&P_{\rm TPB} = P_{\rm abs} P_{\rm geo} P_{\rm vis}, \\
&P_{\rm abs} = \begin{cases}
    0.4 & \lambda < 200~{\rm nm} \\
    0.66 & \lambda > 200~{\rm nm} \\
\end{cases}, \\
&P_{\rm geo} = 0.4, \\
&P_{\rm vis} = \begin{cases}
    1.0 & \lambda < 400~{\rm nm} \\
    0.8 & \lambda > 400~{\rm nm} \\
\end{cases}, \\
\end{split}
\end{equation}
where $P_{\rm abs}$ is a simple step function representation of the TPB absorption efficiency~\cite{Benson:2017vbw}, $P_{\rm geo}$ is a geometric efficiency reflecting the fact that TPB emits photons isotropically off of the hemispherical PMT surface, and $P_{\rm vis}$ reflects the tendency of TPB to absorb visible light to some degree, as evidenced by its cloudy nature.

\begin{figure}[h!]
    \centering
     \begin{subfigure}[b]{0.45\textwidth}
         \centering
         \includegraphics[width=\textwidth]{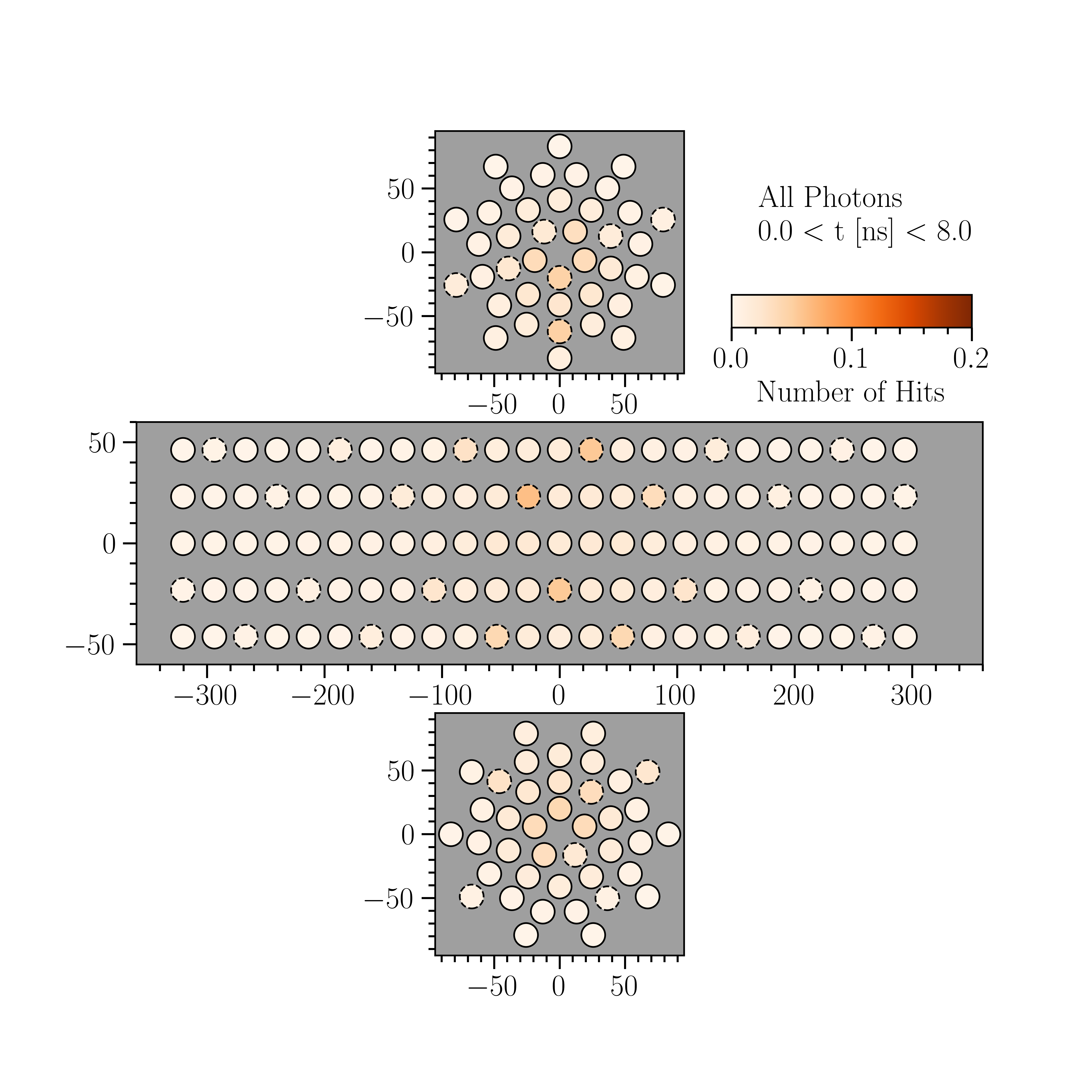}
         \caption{With Cherenkov photons; $T_{e^-} = 1~{\rm MeV}$}
         \label{fig:AllPhotons_E1}
     \end{subfigure}
     \hfill
     \begin{subfigure}[b]{0.45\textwidth}
         \centering
         \includegraphics[width=\textwidth]{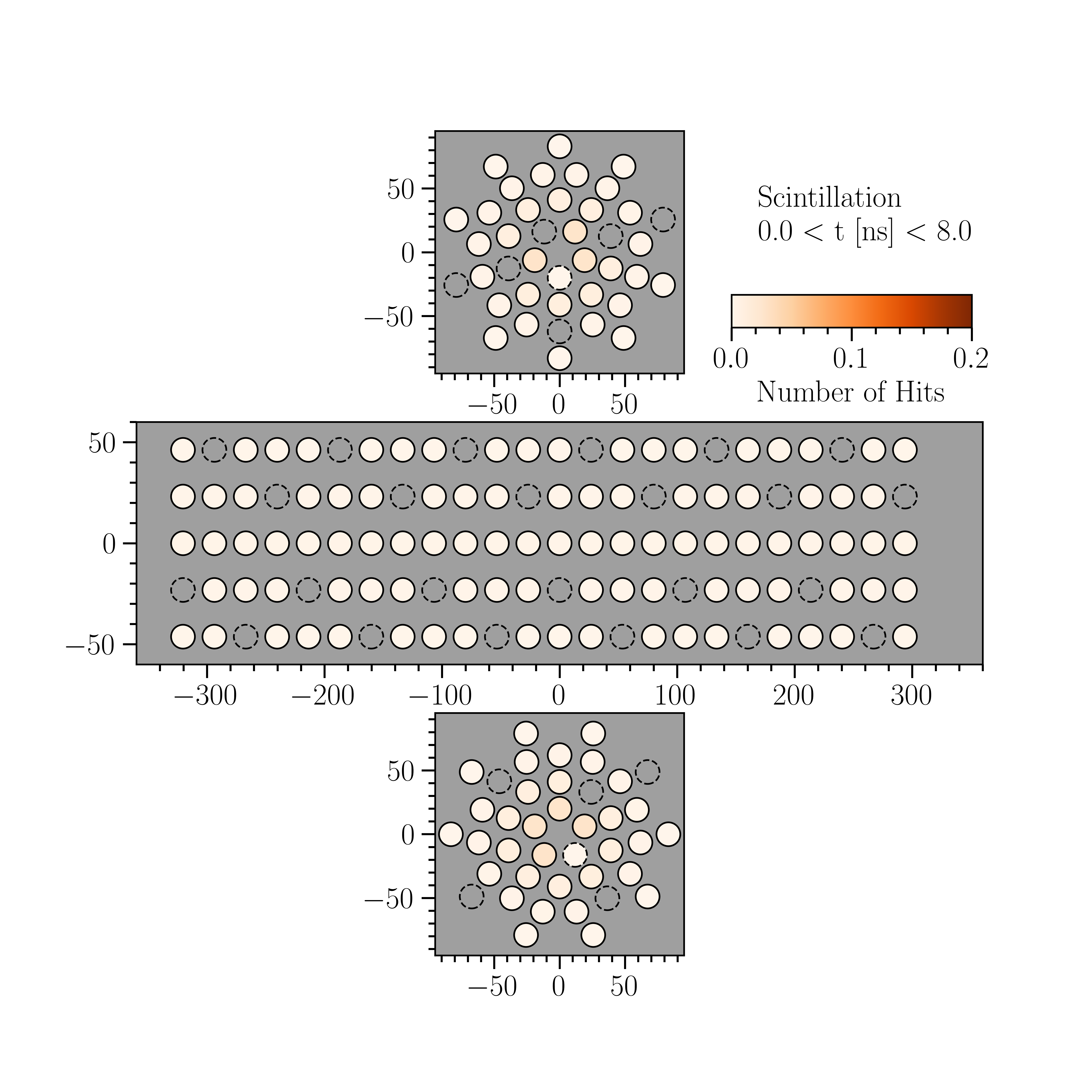}
         \caption{Without Cherenkov photons; $T_{e^-} = 1~{\rm MeV}$}
         \label{fig:ScintPhotons_E1}
     \end{subfigure}
     \hfill
     \begin{subfigure}[b]{0.45\textwidth}
         \centering
         \includegraphics[width=\textwidth]{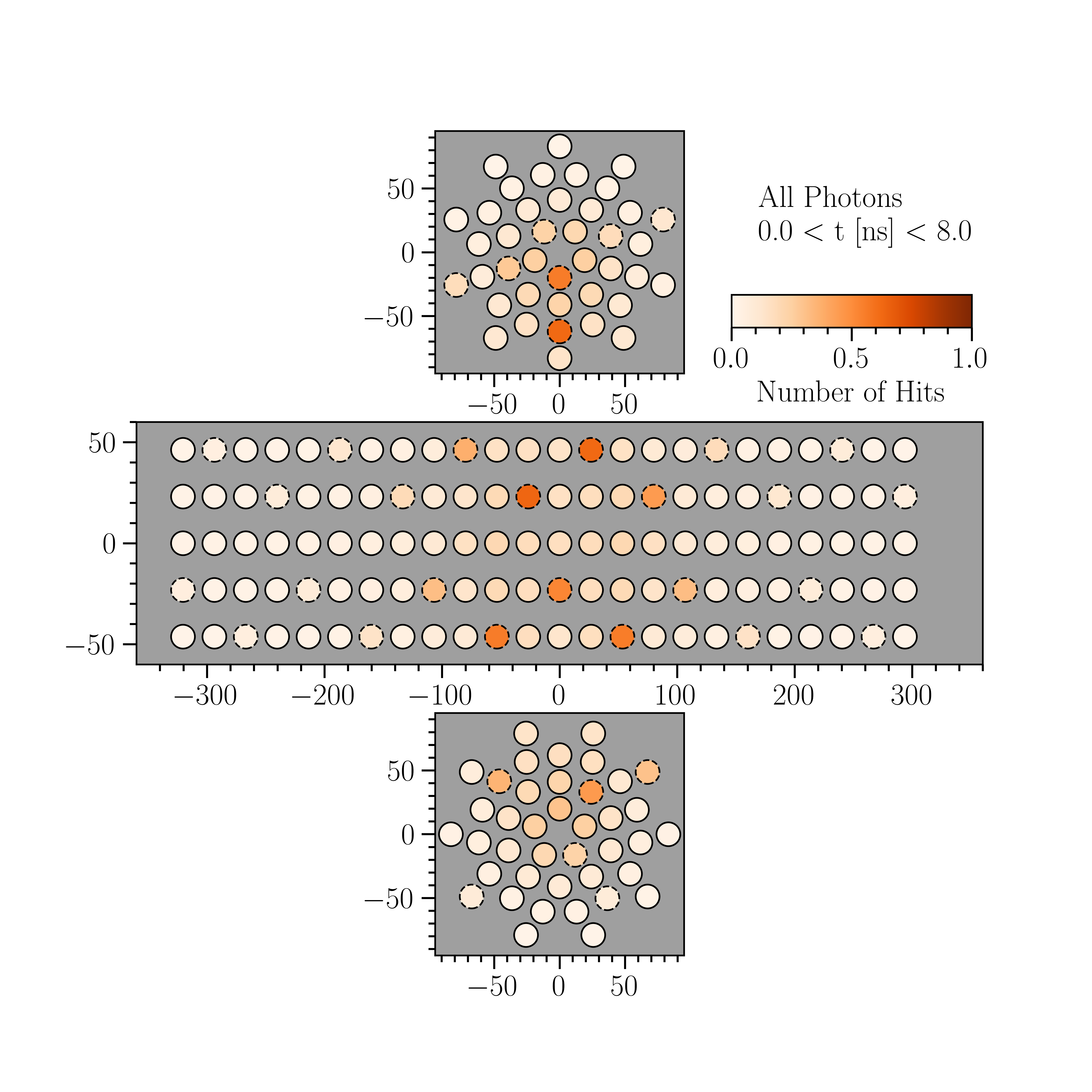}
         \caption{With Cherenkov photons; $T_{e^-} = 5~{\rm MeV}$}
         \label{fig:AllPhotons_E5}
     \end{subfigure}
     \hfill
     \begin{subfigure}[b]{0.45\textwidth}
         \centering
         \includegraphics[width=\textwidth]{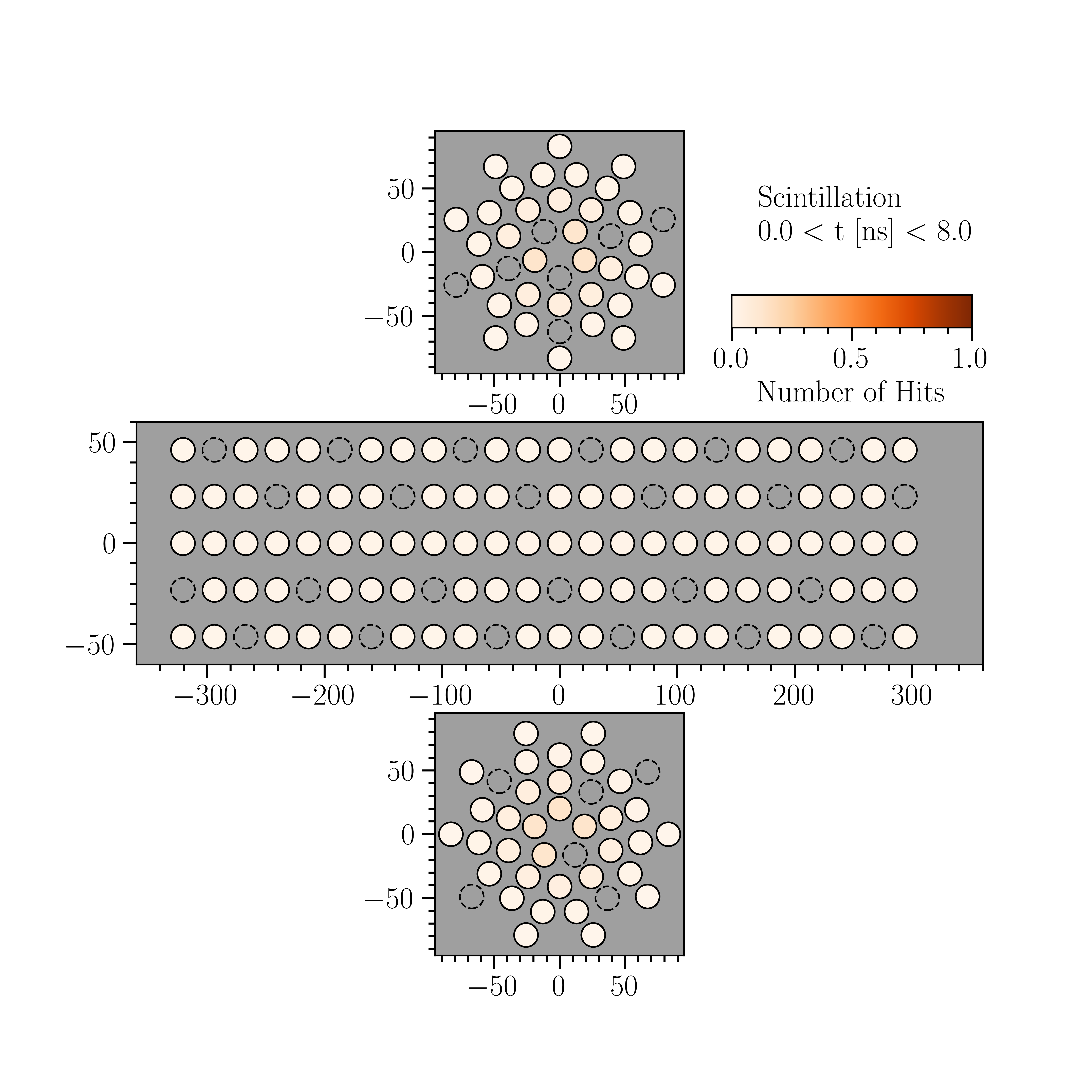}
         \caption{Without Cherenkov photons; $T_{e^-} = 5~{\rm MeV}$}
         \label{fig:ScintPhotons_E5}
     \end{subfigure}
     \hfill
        \caption{Templates of the average number of p.e. detected in each CCM PMT within the first 8~ns of an electron event. Different templates are shown for electron kinetic energies of $T_{e^-} = 1~{\rm MeV}$ and $T_{e^-} = 5~{\rm MeV}$, both with and without Cherenkov photons. Coated (uncoated) PMTs are indicated by the solid (dashed) circles. Grey PMTs indicate those which registered no hits within the first 8~ns across all simulations. The dimensions on each axis are in units of cm.}
        \label{fig:CherenkovAverageTemplates}
\end{figure}

\begin{figure}[h!]
    \centering
     \begin{subfigure}[b]{0.45\textwidth}
         \centering
         \includegraphics[width=\textwidth]{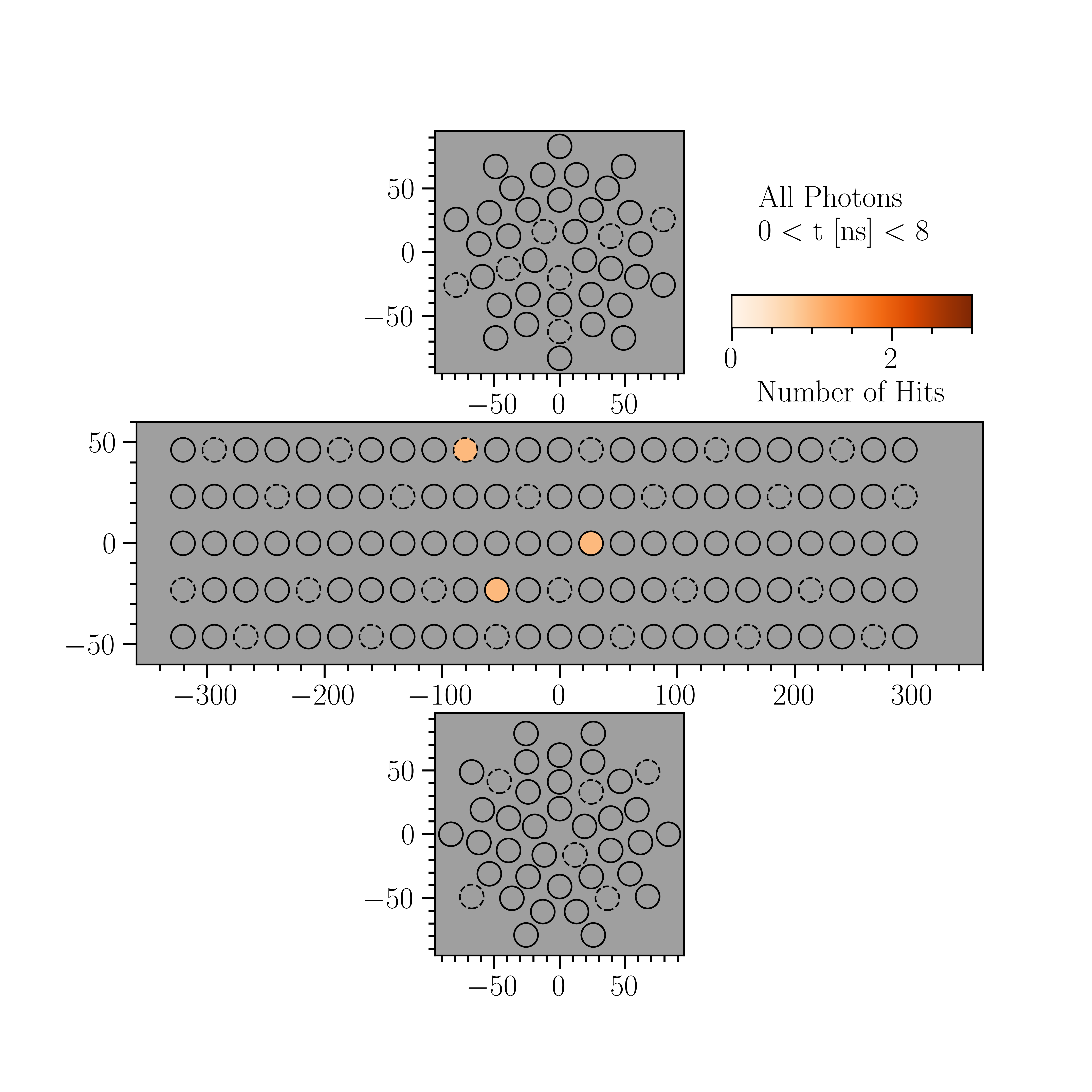}
         \caption{$T_{e^-} = 1~{\rm MeV}$}
         \label{fig:eventdisplay_E1}
     \end{subfigure}
     \hfill
     \begin{subfigure}[b]{0.45\textwidth}
         \centering
         \includegraphics[width=\textwidth]{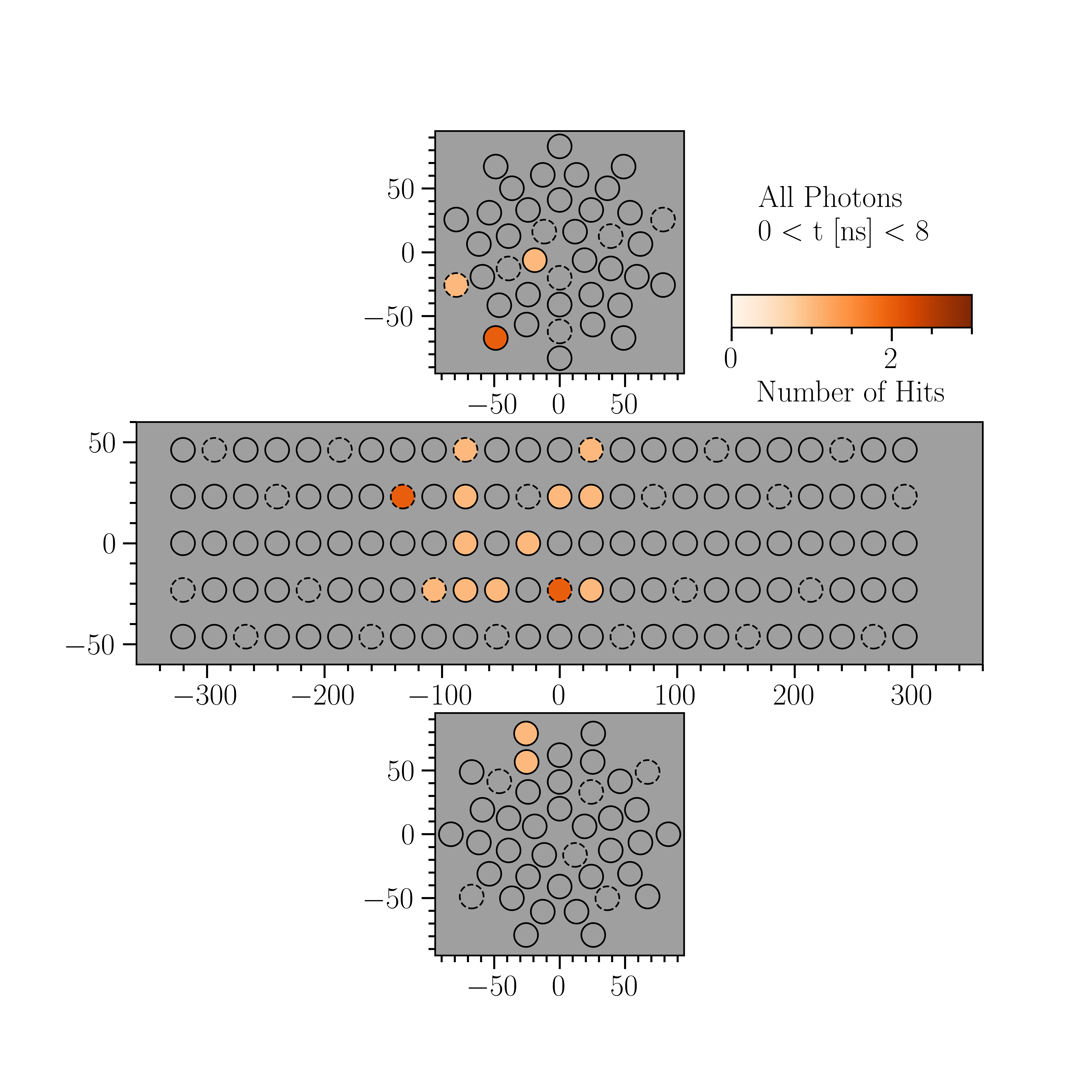}
         \caption{$T_{e^-} = 5~{\rm MeV}$}
         \label{fig:eventdisplay_E5}
     \end{subfigure}
     \hfill
        \caption{Example event displays showing the total number of p.e. detected in each CCM PMT within the first 8~ns of a single simulated electron event. Displays are shown for electron kinetic energies of $T_{e^-} = 1~{\rm MeV}$ and $T_{e^-} = 5~{\rm MeV}$. Coated (uncoated) PMTs are indicated by the solid (dashed) circles. Grey PMTs indicate those which registered no hits in the first 8~ns of this specific simulation.}
        \label{fig:CherenkovEventDisplays}
\end{figure}

Using the setup described above, we simulate $10^4$ electrons for each of the five tested kinetic energies to generate average templates of the expected number of registered hits in each PMT.
In \cref{fig:CherenkovAverageTemplates} we show templates for electrons with kinetic energies of $T_{e^-} = 1~{\rm MeV}$ and $T_{e^-} = 5~{\rm MeV}$.
Separate templates are shown including all photons and only scintillation photons (i.e., without Cherenkov light).
The positions of each PMT here can be thought of as an unfolding of the cylindrical CCM detector, where the center of the image ($(x,y) = (0,0)$ in the middle panel) corresponds to the beam direction.
Note that these templates only consider the first 8~ns after the electron starts traveling through the detector, as this is the region in which we expect Cherenkov light to stand out over scintillation light.
This is indeed the case; the PMTs in \cref{fig:AllPhotons_E1,fig:AllPhotons_E5} see much more light than those in \cref{fig:ScintPhotons_E1,fig:ScintPhotons_E5}.
This is especially true of the uncoated PMTs (dashed circles), which do not suffer the TPB detection inefficiency captured by \cref{eq:tpb_efficiency} nor the time delay from TPB re-emission.
The separation between the templates with and without Chernekov light is more obvious for $T_{e^-} = 5~{\rm MeV}$ than for $T_{e^-} = 1~{\rm MeV}$, indicating that Chernekov light will be a stronger discriminator in the higher energy region.
The directional nature of Cherenkov light is also clear here.
While scintillation photons are more-or-less isotropic, the event displays with Cherenkov photons included clearly show detected p.e. clustered around the beam direction.
\Cref{fig:CherenkovEventDisplays} shows example event displays of the number of p.e. detected in each CCM PMT in a single simulated electron event for both $T_{e^-} = 1~{\rm MeV}$ and $T_{e^-} = 5~{\rm MeV}$.

\begin{figure}[h!]
    \centering
     \begin{subfigure}[b]{0.45\textwidth}
         \centering
         \includegraphics[width=\textwidth]{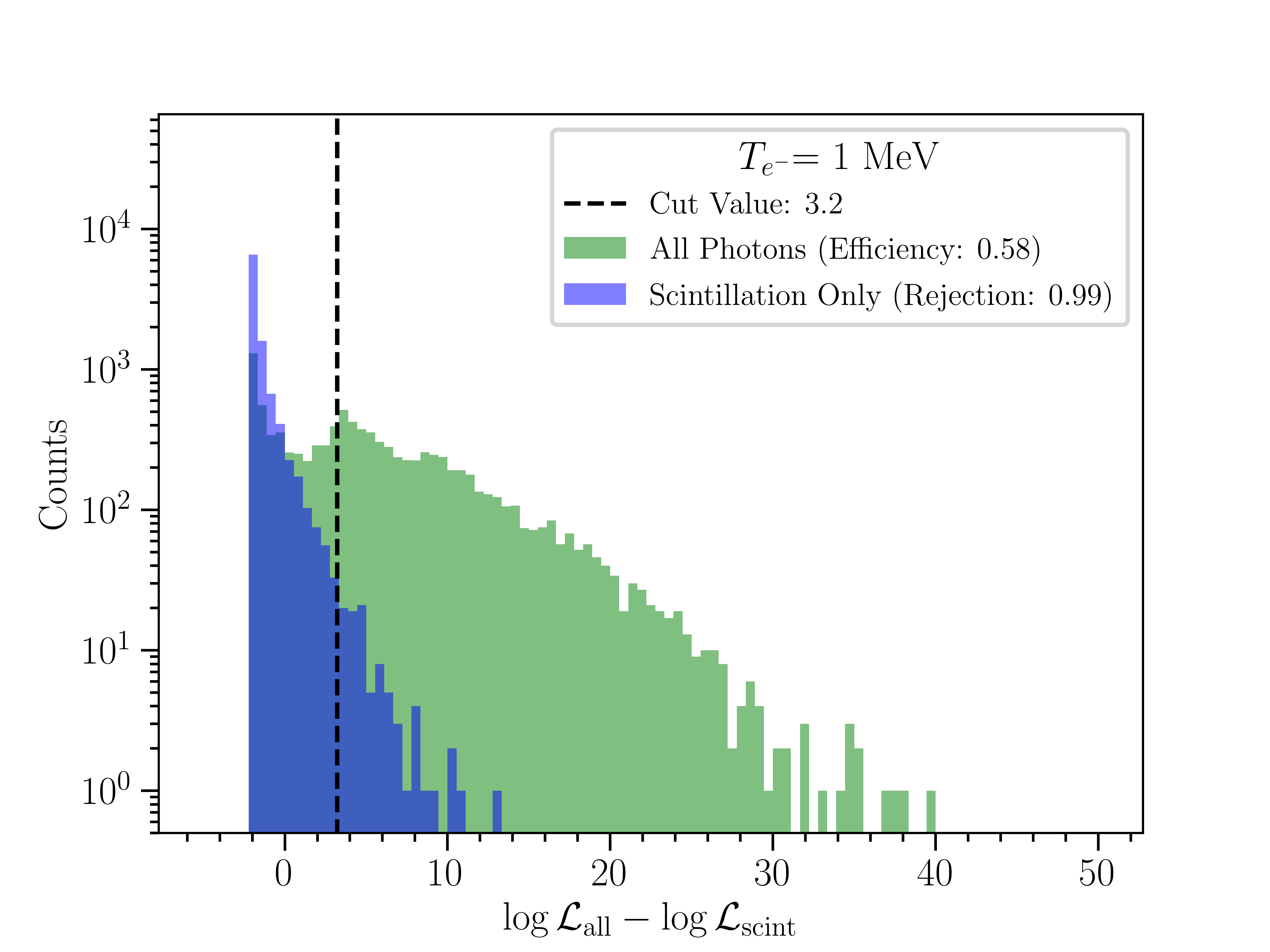}
         \caption{$T_{e^-} = 1~{\rm MeV}$}
         \label{fig:dllh_e1}
     \end{subfigure}
     \hfill
     \begin{subfigure}[b]{0.45\textwidth}
         \centering
         \includegraphics[width=\textwidth]{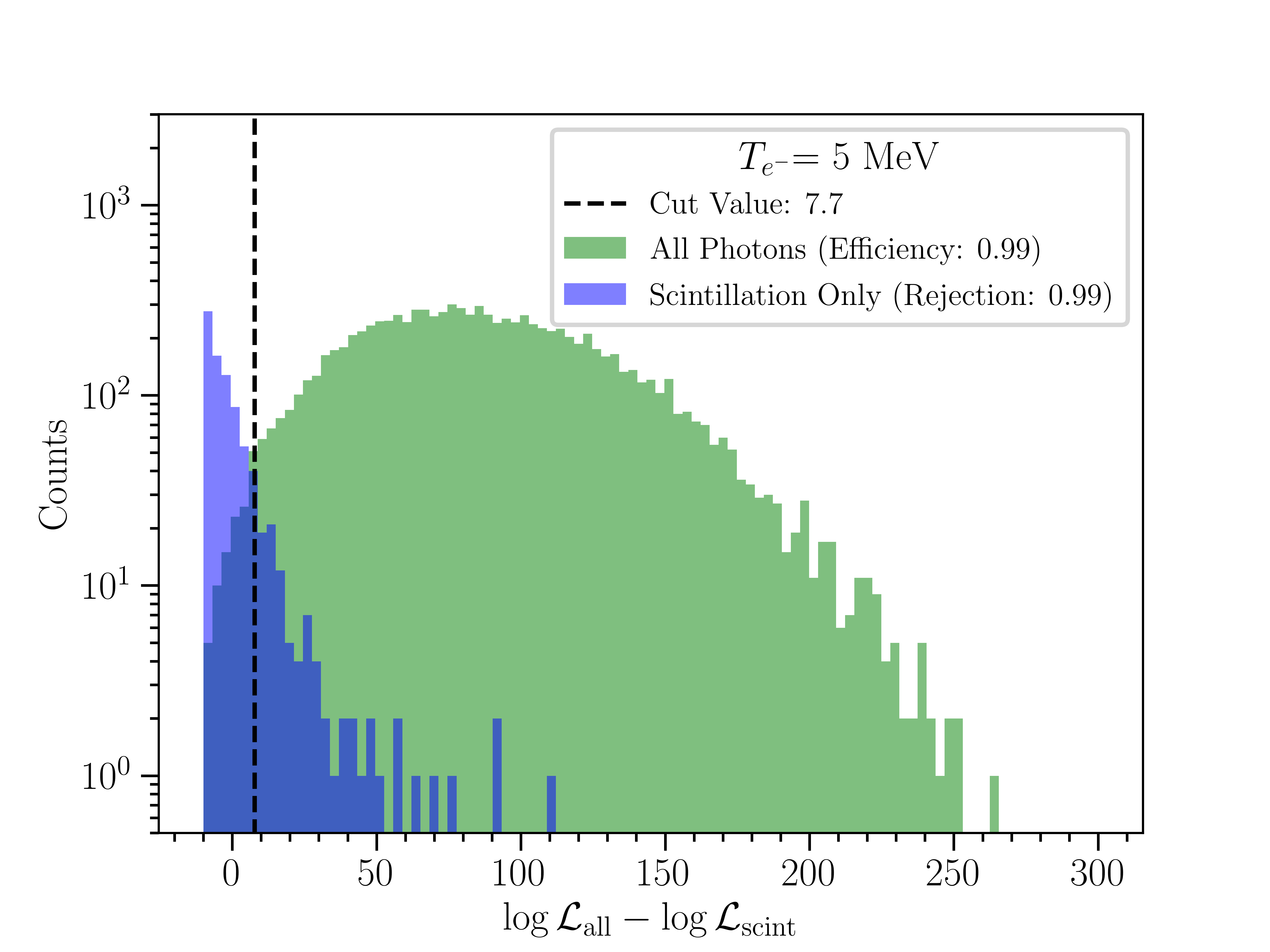}
         \caption{$T_{e^-} = 5~{\rm MeV}$}
         \label{fig:dllh_e5}
     \end{subfigure}
     \hfill
        \caption{Distributions of the test statistic in \cref{eq:cherenkov_test_statistic} over all $10^4$ simulations, considering either all photons or scintillation photons only. Distributions are shown for $T_{e^-} = 1~{\rm MeV}$ and $T_{e^-} = 5~{\rm MeV}$. The vertical line indicates the lower bound requirement which can reject 99\% of scintillation-only backgrounds.}
        \label{fig:dllh_distributions}
\end{figure}

\begin{figure}[h!]
    \centering
    \includegraphics[width=0.6\textwidth]{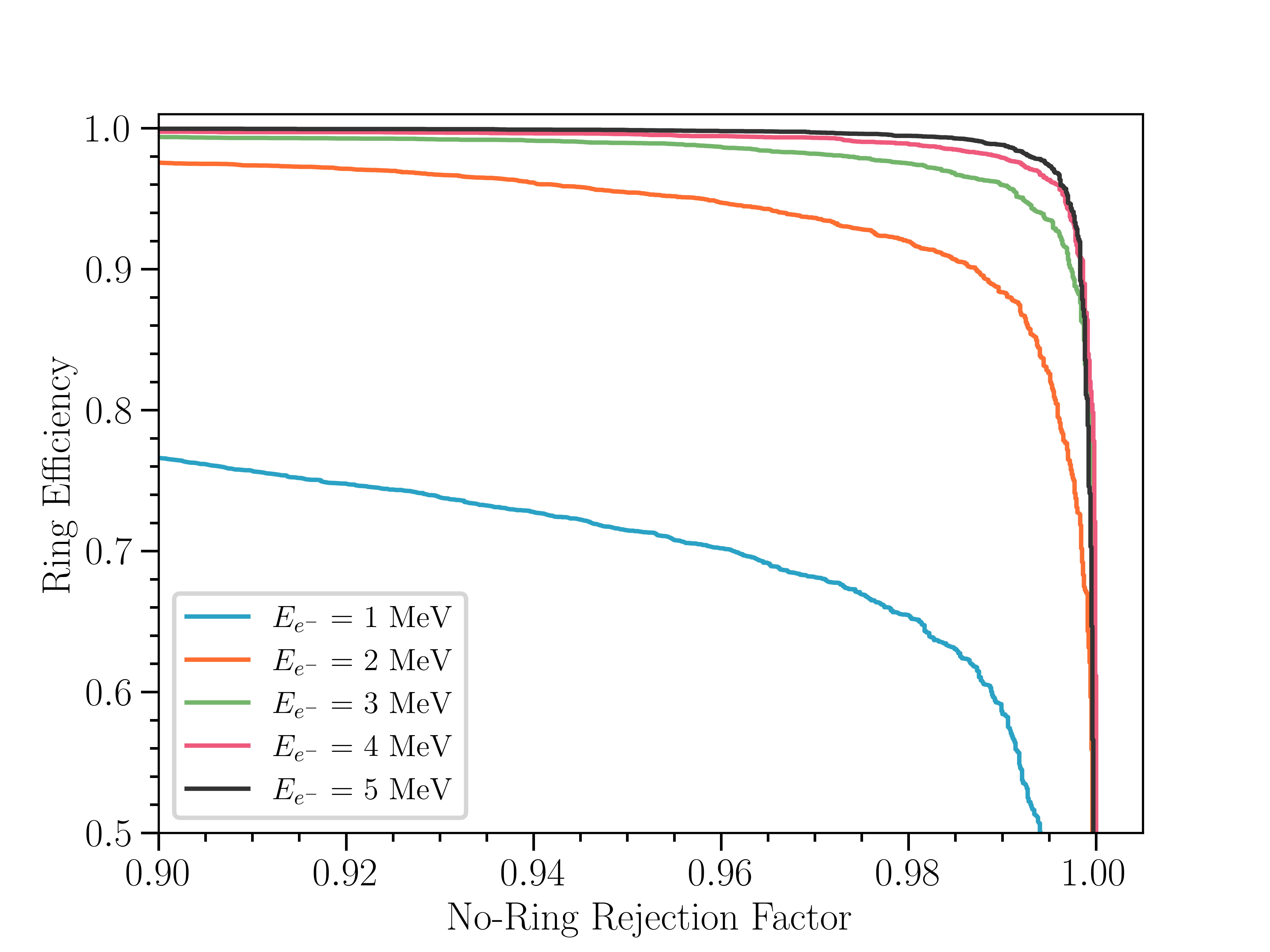}
    \caption{Curves of the efficiency to retain events with Cherenkov light (``Ring efficiency'') v.s. the fraction of events without Cherenkov light that can be rejected (``No-Ring Rejection Factor''), generated by considering successively larger lower bounds on $\Delta \log \mathcal{L}$. Different curves are shown for $T_{e^-} \in \{1,2,3,4,5\}~{\rm MeV}$.}
    \label{fig:cherenkov_ROC_curve}
\end{figure}

We use these simulations to build up a template-based Poisson likelihood reflecting the probability that a given event does or does not contain scintillation light.
We evaluate a separate Poisson likelihood in each PMT across five time bins of 2~ns width each, which reflects the temporal sampling rate in CCM.
The total likelihood is given by
\begin{equation}
\mathcal{L}_{\rm all/scint} = \prod_{\rm PMTs~i} \prod_{\rm time~bins~j} P_{\rm Poisson}(k_{ij}|\mu^{\rm all/scint}_{ij}),
\end{equation}
where $P_{\rm Poisson}$ is the Poisson probability, $k_{ij}$ is the observed number of p.e. in PMT $i$ and time bin $j$, and $\mu^{\rm all}_{ij}$ ($\mu^{\rm scint}_{ij}$) is the prediction for the templates with (without) Cherenkov light, as shown in \cref{fig:CherenkovAverageTemplates}.
In the case that no events are predicted in a given PMT $i$ and time bin $j$, we set $\mu_{ij} = 10^{-4}$, reflecting an uncertainty of one event over the $10^4$ simulations.
We can then use the detected p.e. in each simulation to build distributions of the log-likelihood-ratio test statistic
\begin{equation} \label{eq:cherenkov_test_statistic}
\Delta \log \mathcal{L} \equiv \log \mathcal{L}_{\rm all} - \log \mathcal{L}_{\rm scint}.
\end{equation}
In \cref{fig:dllh_distributions} we show distributions of this test statistic for $T_{e^-} = 1~{\rm MeV}$ and $T_{e^-} = 5~{\rm MeV}$, both with and without Cherenkov light.
Given the overall sign of \cref{eq:cherenkov_test_statistic}, $\Delta \log \mathcal{L}$ is larger when Cherenkov photons are included.
This can be seen in \cref{fig:dllh_distributions}, as the distributions with Cherenkov photons included sit at higher values of $\Delta \log \mathcal{L}$ compared to those without Cherenkov photons.
The vertical line in this plot indicates the lower bound on $\Delta \log \mathcal{L}$ that can reject 99\% of scintillation-only backgrounds, which can be thought of as, for example, beam-related neutrons depositing as many scintillation photons as an electron with the specified kinetic energy.
\Cref{fig:cherenkov_ROC_curve} shows curves of the efficiency to retain events with Cherenkov rings v.s. the rejection factor for events without Cherenkov rings, which are obtained by considering successively larger lower bounds on $\Delta \log \mathcal{L}$.
In the $T_{e^-} = 5~{\rm MeV}$ case, one can achieve a background rejection of $0.99$ while retaining a signal efficiency of $0.99$--thus, a Cherenkov likelihood cut has the potential to reduce backgrounds by around two orders of magnitude for free.

Of course, a few caveats must be mentioned regarding this study.
Most importantly, we have only constructed a likelihood for electrons of a few example kinetic energies originating at the center of the detector and traveling along the beam direction.
A full Cherenkov likelihood will need to be a function of the particle vertex $(x,y,z)$ and direction $(\theta,\phi)$ in addition to the particle energy.
In order to properly assess the background rejection power of a Cherenkov-based cut, one will need to simulate neutrons in CCM using a realistic flux from the Lujan target after moderation by the CCM shielding.
Finally, a more realistic TPB model within Geant4 itself will allow for a more robust prediction of the difference in coated and uncoated PMT response to Chernekov light.
That being said, the results from this simulation study indicate the power of a Cherenkov reconstruction algorithm in CCM and suggest that such a reconstruction is feasible for electrons with energies of a few MeV or greater.
Angular cuts based on the reconstructed direction of the Cherenkov cone will help further reduce backgrounds by correlating events with the beam direction.

\subsection{Identifying Cherenkov Light in Data} \label{sec:ccm_cherenkov_data}

\begin{figure}[h!]
    \centering
     \begin{subfigure}[b]{0.45\textwidth}
         \centering
         \includegraphics[width=\textwidth]{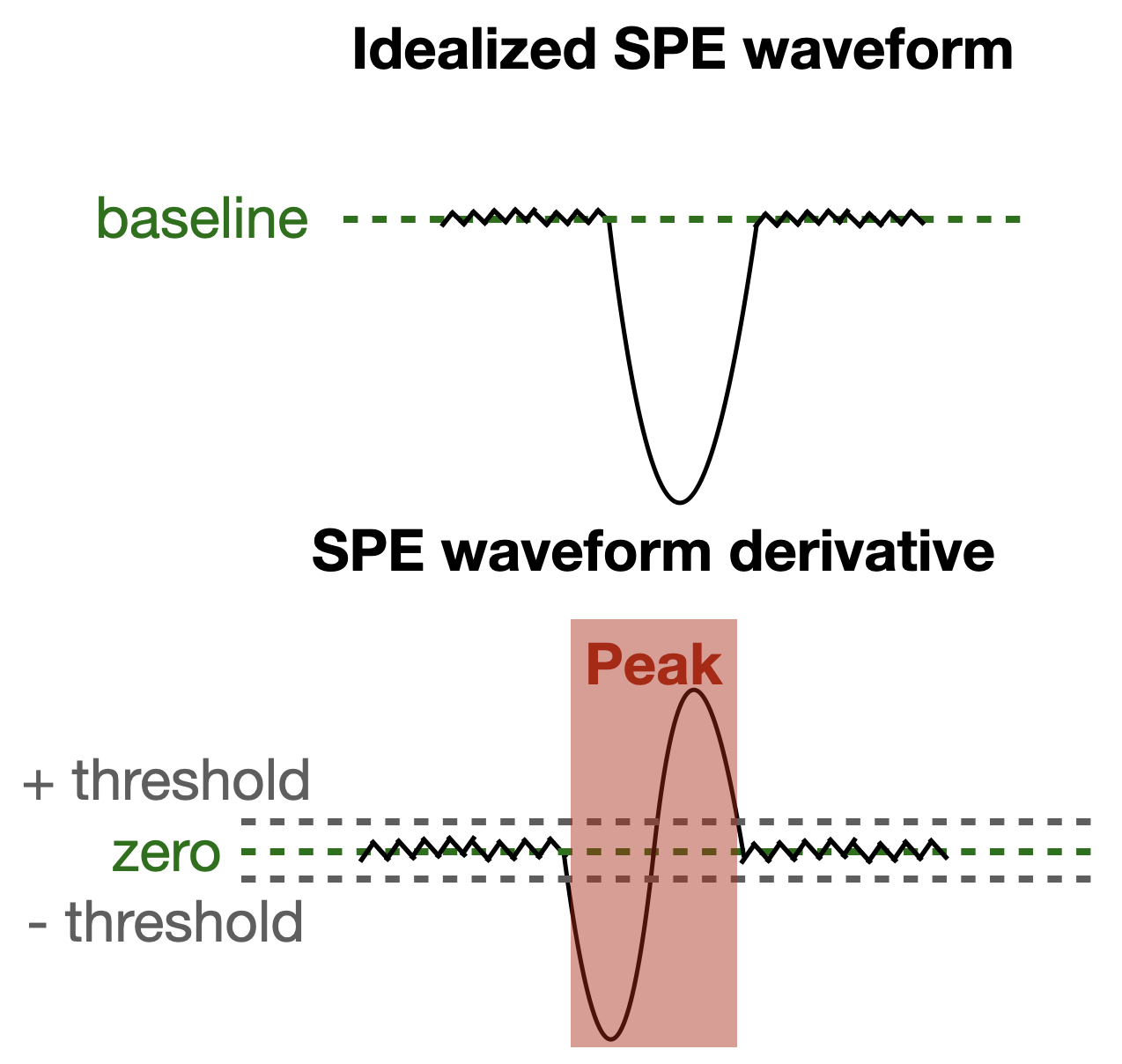}
         \caption{}
         \label{fig:CCM_DerivativeFilter}
     \end{subfigure}
     \hfill
     \begin{subfigure}[b]{0.45\textwidth}
         \centering
         \includegraphics[width=\textwidth]{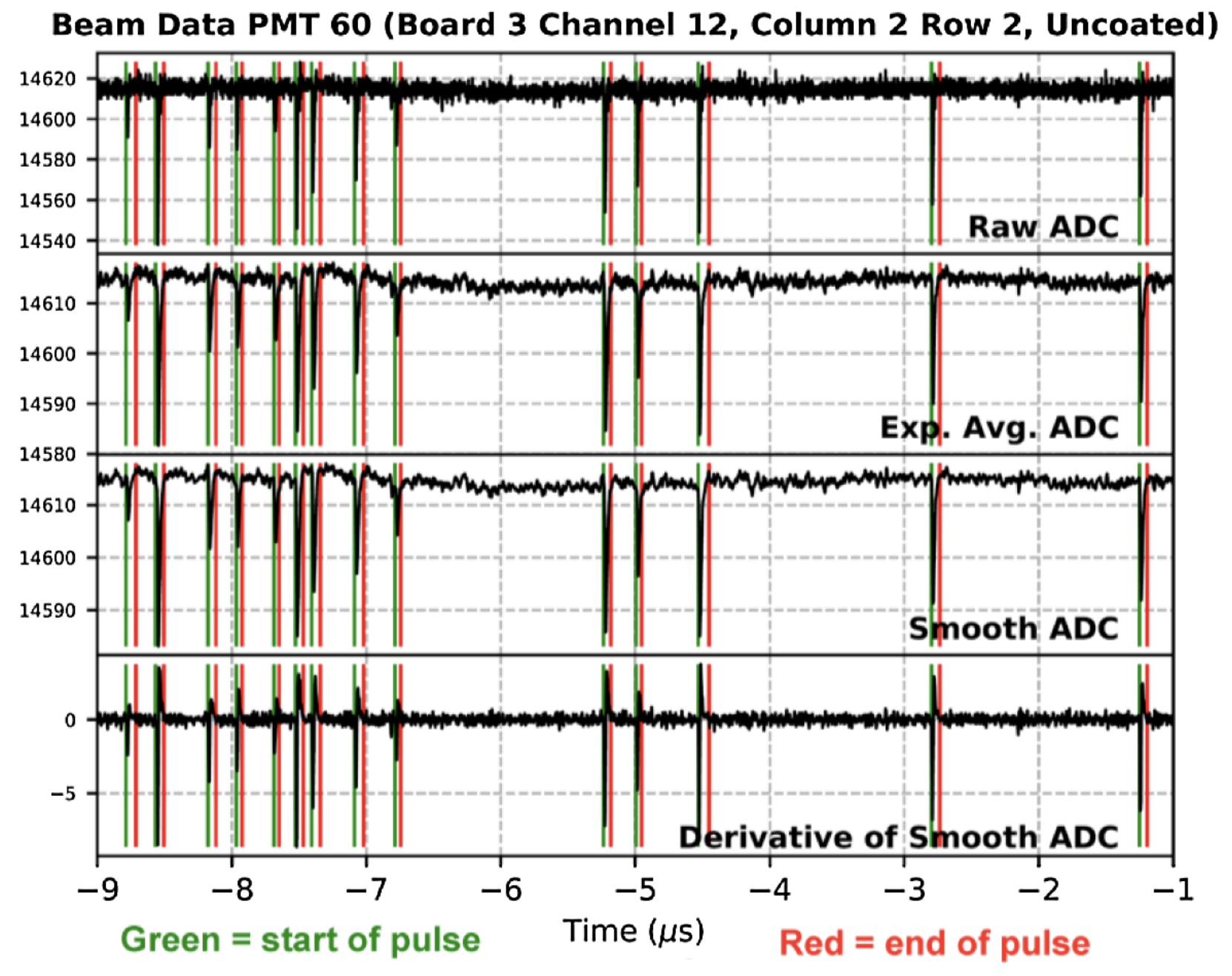}
         \caption{}
         \label{fig:CCM_OldPulseFinder}
     \end{subfigure}
     \hfill
        \caption{\Cref{fig:CCM_DerivativeFilter} shows a schematic depiction of the derivative-based pulse definition in the current CCM reconstruction. \Cref{fig:CCM_OldPulseFinder} shows an example of this pulse finder in a data event, from Ref.~\cite{CCM:2021leg}.}
        \label{fig:CCM_old_pulses}
\end{figure}

Having established the feasibility of a dedicated Cherenkov light reconstruction, we now discuss the path toward realizing this in CCM data.
Cherenkov reconstruction requires a more detailed treatment of the timing information contained in PMT signals.
The existing CCM reconstruction uses exponential smoothing, averaging, and derivative filters to determine regions of each PMT waveform with potential activity.
A pulse is defined as a region in which the derivative goes below a given negative threshold, then becomes positive, then relaxes below the positive threshold, as shown in \cref{fig:CCM_DerivativeFilter}.
\Cref{fig:CCM_OldPulseFinder} shows the identified pulses in an example CCM120 data event using this technique.
Each pulse is then approximated by a triangle with a length and height given by the pulse length and absolute value of the integral of the waveform derivative over the pulse, respectively.
Pulses were further required to be at least 20~ns in length to reduce the impact of noise.
While this method was sufficient for the scintillation-based CCM120 analyses, which were mainly concerned with the many-p.e. scintillation pulse summed over all PMTs, identification of Cherenkov light will require a new pulse-finding algorithm that treats single p.e. pulses more carefully.
It must also be able to retain pulses shorter than 20~ns to be sensitive to the Cherenkov-dominated early time window studied in \cref{sec:CCM_cherenkov_simulation}.

\begin{figure}[h!]
    \centering
     \begin{subfigure}[b]{0.45\textwidth}
         \centering
         \includegraphics[width=\textwidth]{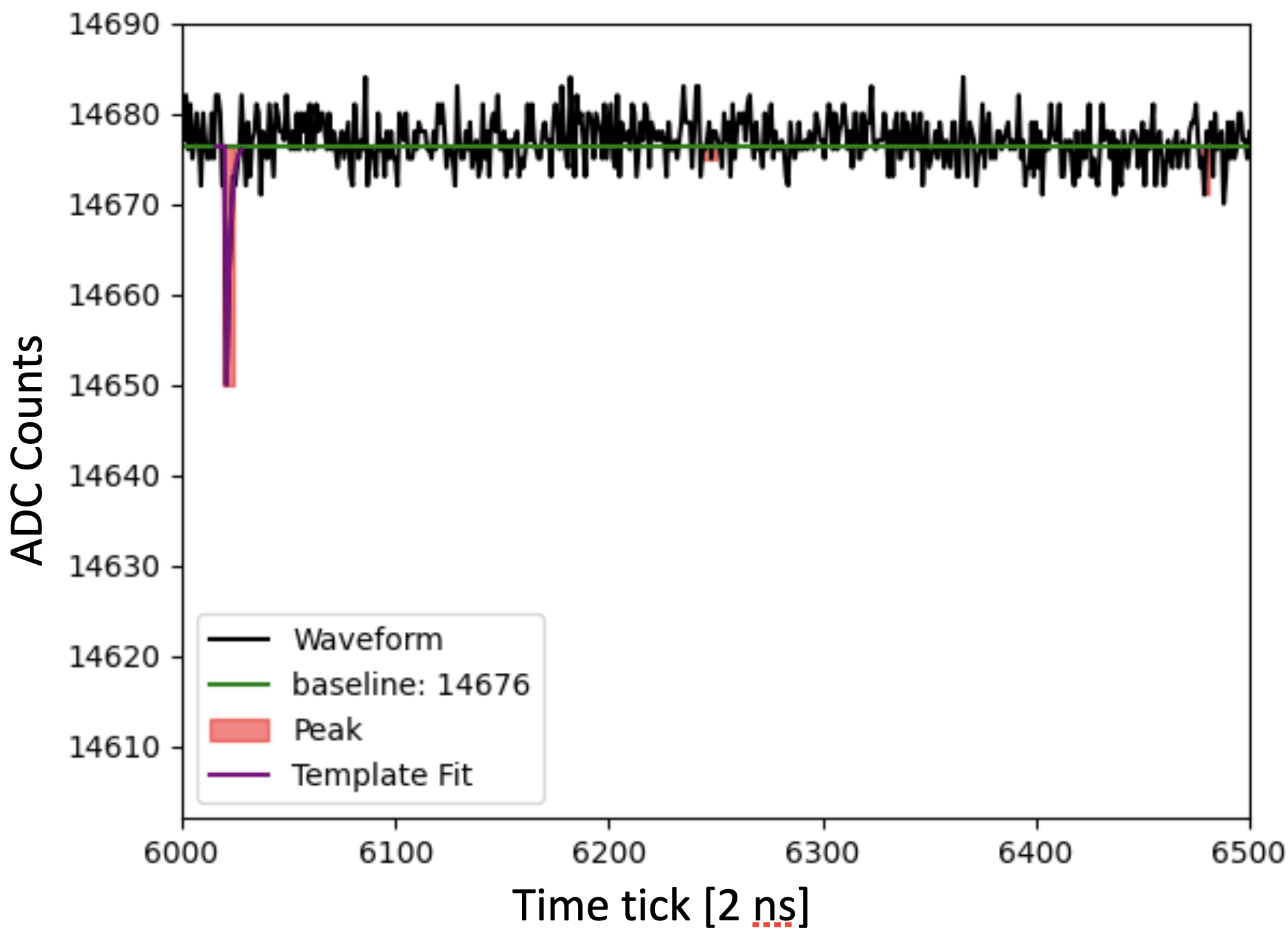}
         \label{fig:channel23_event1}
     \end{subfigure}
     \hfill
     \begin{subfigure}[b]{0.45\textwidth}
         \centering
         \includegraphics[width=\textwidth]{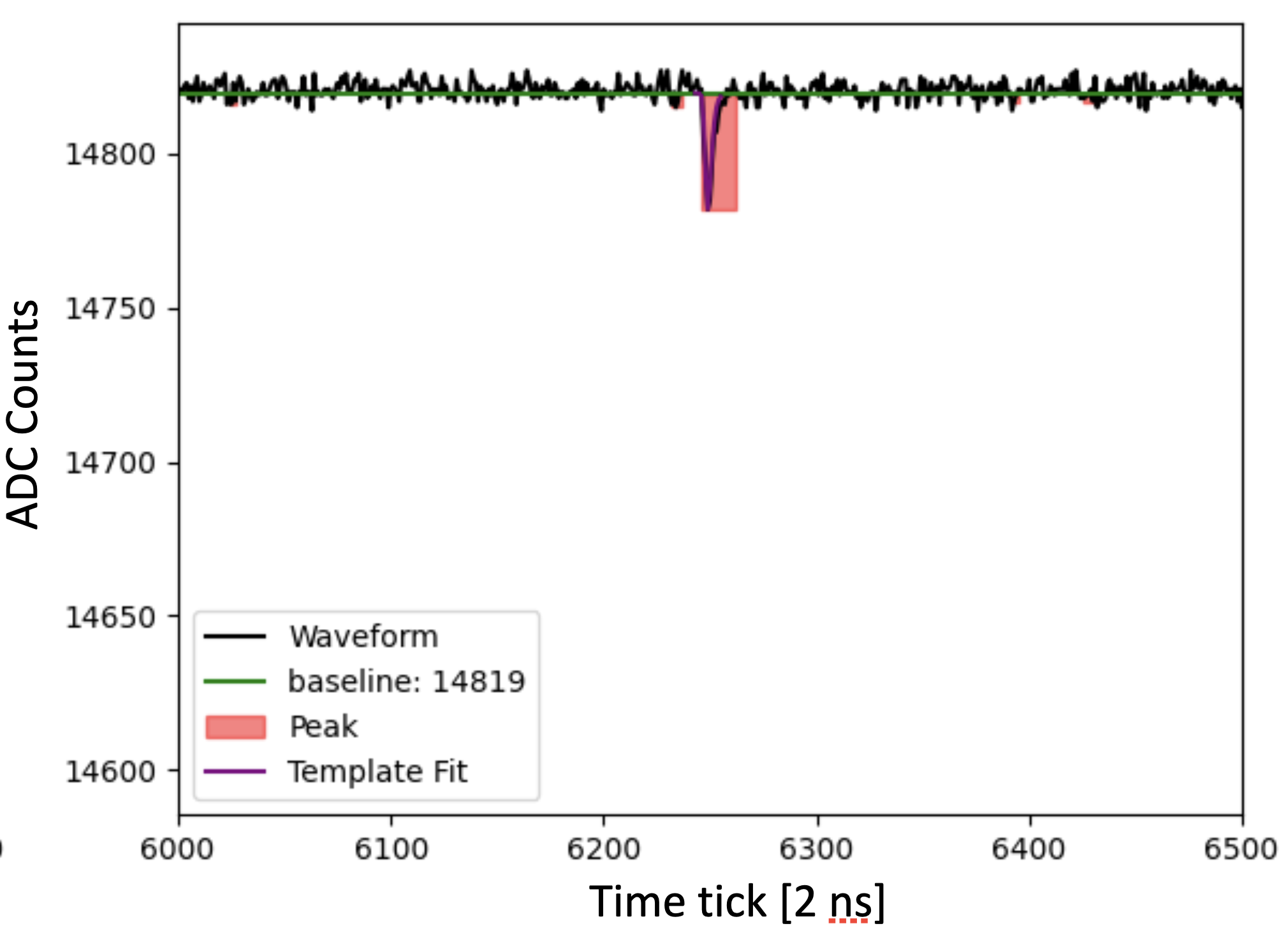}
         \label{fig:channel35_event1}
     \end{subfigure}
     \hfill
        \caption{Two example waveforms from CCM200 beam data. The regions identified by the derivative filter are indicated in red and the result of the fit to \cref{eq:spe_template} is indicated by the purple curves.}
        \label{fig:example_spe_template}
\end{figure}

To address these issues, we have begun revamping the pulse-finding algorithm in preparation for CCM200.
This includes the use of a more realistic single p.e. template to replace the triangle approximation.
We use a parameterization of the single p.e. waveform based on that used by the IceCube collaboration~\cite{IceCube:2020nwx},
\begin{equation} \label{eq:spe_template}
w(t) = \frac{c}{[e^{-(t-t_0)/b_1} + e^{(t-t_0)/b_2}]^8},
\end{equation}
where $c$, $t_0$, $b_1$ and $b_2$ are free parameters to be fit using data.
To determine the best-fit single p.e. template for each PMT, we begin by isolating candidate pulses using the derivative filter in \cref{fig:CCM_DerivativeFilter}.
We restrict to candidate pulses with an amplitude between 20 and 40 ADC counts, which corresponds roughly to the single p.e. amplitude range in CCM.
We fix two of the template parameters using the observed peak amplitude $w_{\rm max}$ and peak time $t_{\rm max}$ for each candidate pulse,
\begin{equation}
\begin{split}
&w_{\rm max} = \frac{c}{\Big[\Big(\frac{b_1}{b_2}\Big)^{\frac{b_2}{b_1+b_2}} + \Big(\frac{b_2}{b_1}\Big)^{\frac{b_1}{b_1+b_2}}\Big]^8}, \\
&t_{\rm max} = t_0 + \frac{b_1 b_2}{b_1 + b_+2} \ln(b_2/b_1). \\
\end{split}
\end{equation}
\Cref{fig:example_spe_template} shows the result of this template fit for candidate pulses on two different PMTs in CCM200 beam data.
The next step is to determine a robust single p.e. template for each PMT in CCM200.
These templates can then be used to unfold the observed waveform on every PMT for a given event into a series of single p.e. pulses at different times and with different amplitudes.
Such a reconstruction will help pick out an excess of single p.e. pulses in the early time window, which would be an indication of Cherenkov light.

\begin{figure}[h!]
    \centering
     \begin{subfigure}[b]{0.45\textwidth}
         \centering
         \includegraphics[width=\textwidth]{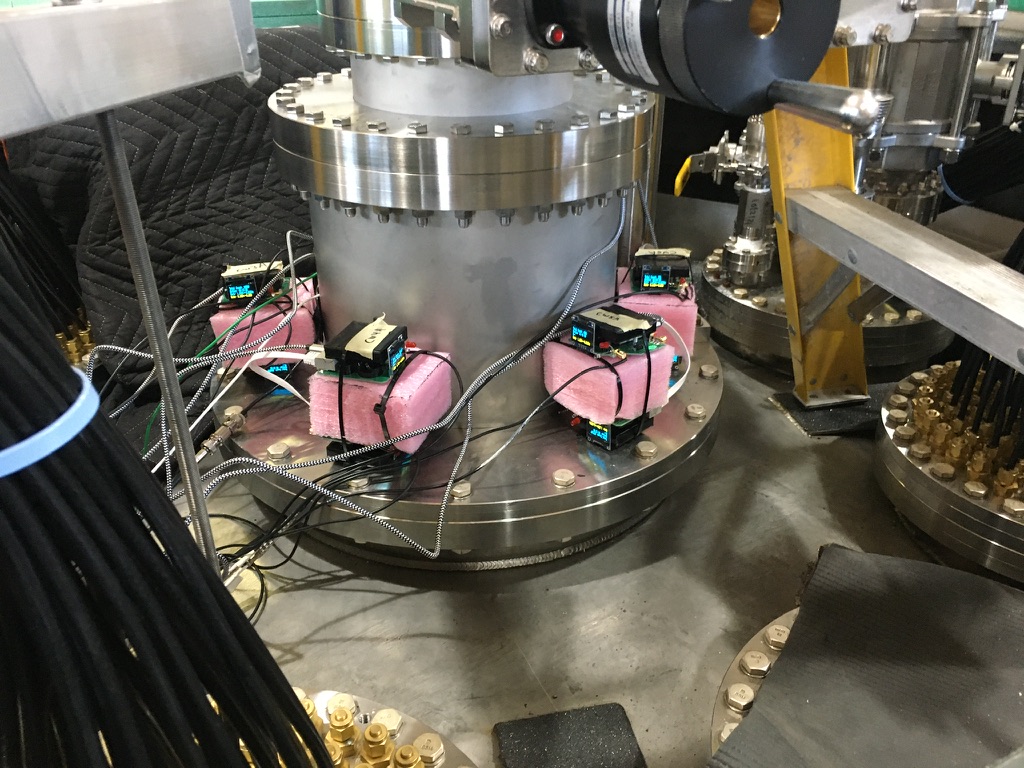}
         \caption{}
         \label{fig:cosmic_watches_on_CCM}
     \end{subfigure}
     \hfill
     \begin{subfigure}[b]{0.45\textwidth}
         \centering
         \includegraphics[width=\textwidth]{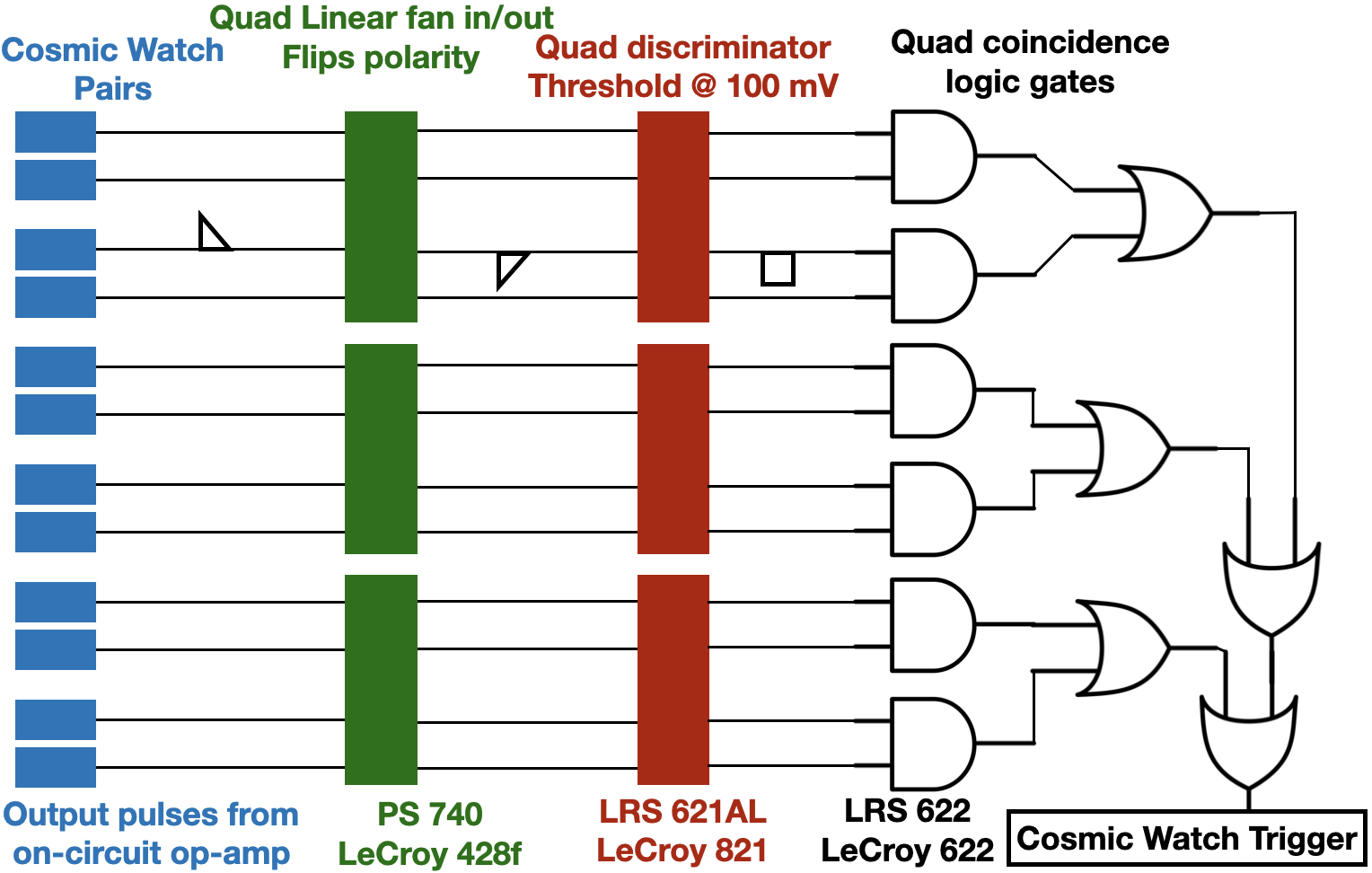}
         \label{fig:CosmicTriggerSchematic}
     \end{subfigure}
     \hfill
        \caption{\Cref{fig:cosmic_watches_on_CCM} shows an image of the six CosmicWatch pairs on top of the CCM detector. \Cref{fig:CosmicTriggerSchematic} shows a schematic diagram of the cosmic muon trigger in CCM200.}
        \label{fig:cosmicwatch_ccm}
\end{figure}

In order to inform the development of our Cherenkov reconstruction, we have isolated a new data sample in CCM200: cosmic muons.
These are useful for two reasons: (1) downward-going cosmic muons will produce Cherenkov light along a known direction, and (2) isotropic emission of Michel electrons from muon decay-at-rest will help test the angular resolution of the Cherenkov reconstruction.
Michel electrons can also serve as an important calibration point for the scintillation-based energy reconstruction, as they are emitted with a known energy distribution peaking at $m_\mu/2$~\cite{MicroBooNE:2021nss}.
To tag cosmic muons, we have assembled six pairs of CosmicWatch detectors~\cite{cosmicwatch} on top of CCM.
Each detector consists of a piece of plastic scintillator and a silicon photomultiplier, which detects photons produced in the scintillator.
Two detectors can be stacked on top of one another and set to trigger on the coincident observation of light between the pair, indicative of a crossing cosmic muon.
\Cref{fig:cosmic_watches_on_CCM} shows an image of the six CosmicWatch pairs on top of CCM.
The detectors in each pair are separated by approximately 6~in. of foam to ensure that crossing cosmic muons are sufficiently downward-going to pass through the bottom of the detector.
We have set up a dedicated cosmic muon trigger in CCM200 that saves an event whenever any of the six CosmicWatch pairs on top of CCM register a coincident signal.
A diagram of this trigger is shown in \cref{fig:CosmicTriggerSchematic}.

\begin{figure}[h!]
    \centering
     \begin{subfigure}[b]{0.45\textwidth}
         \centering
         \includegraphics[width=\textwidth]{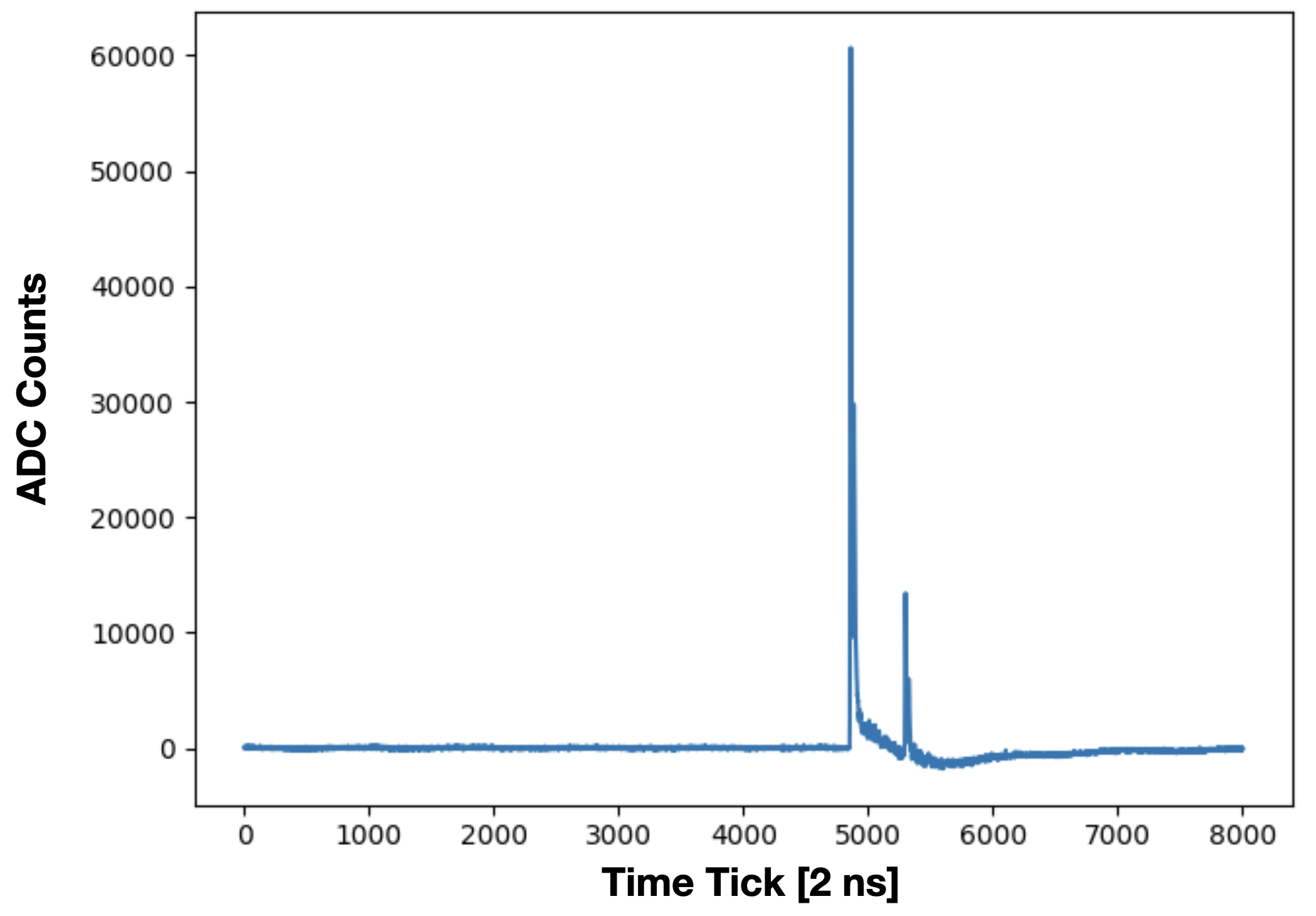}
         \caption{}
         \label{fig:example_cosmic}
     \end{subfigure}
     \hfill
     \begin{subfigure}[b]{0.45\textwidth}
         \centering
         \includegraphics[width=\textwidth]{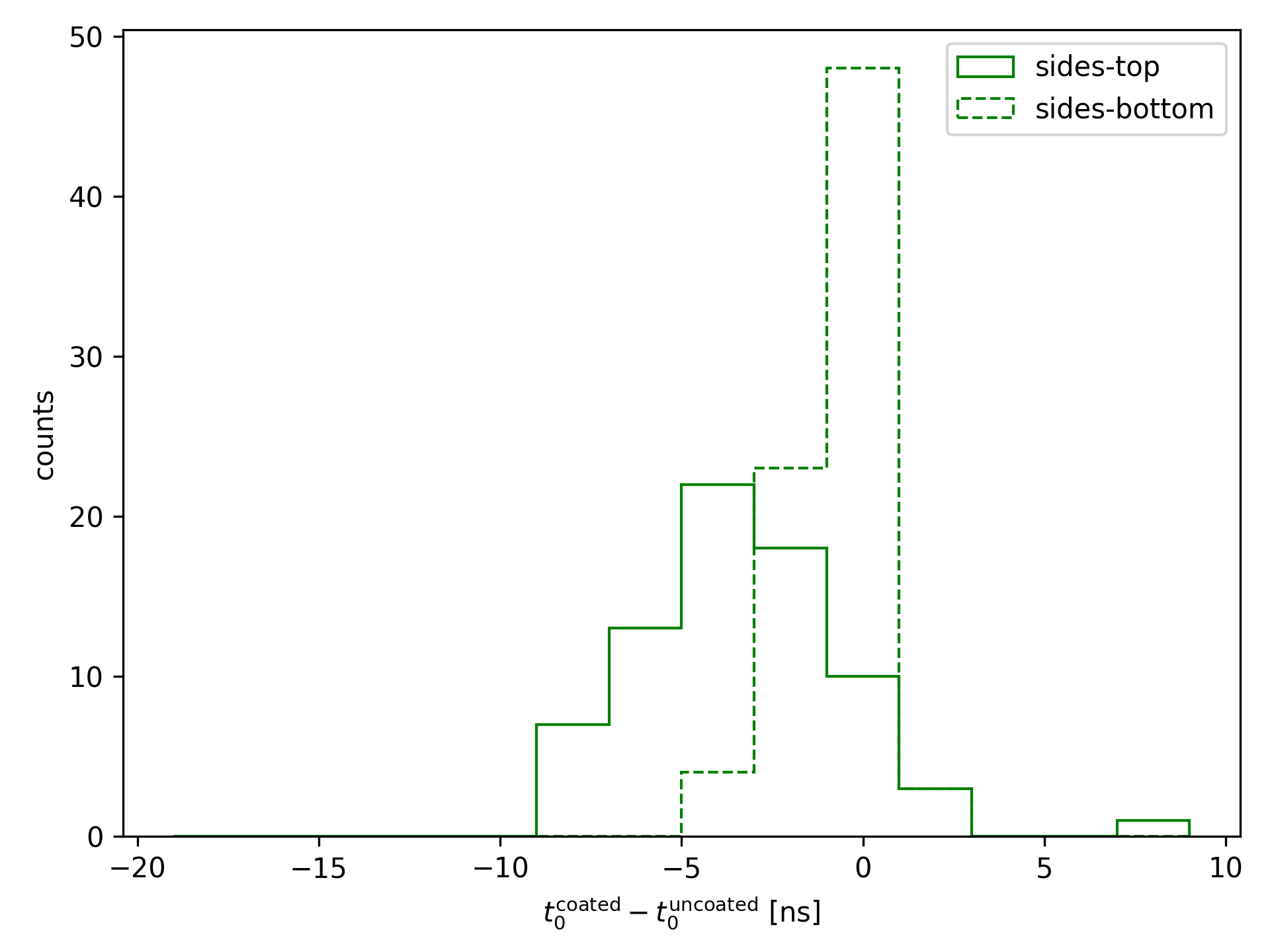}
         \label{fig:CFD_t0_dists}
     \end{subfigure}
     \hfill
        \caption{\Cref{fig:example_cosmic} shows the summed waveform across all PMTs in CCM for a single example cosmic muon trigger. The delayed signal from a Michel electron can also be seen. \Cref{fig:CFD_t0_dists} shows the difference in rise times between coated and uncoated PMT signals in the top and bottom halves of the barrel of the detector (labeled ``sides-top'' and ``sides-bottom'', respectively), as described in the text.}
        \label{fig:cosmic_data}
\end{figure}

The cosmic muon trigger became operational in Summer 2022 and can run concurrently with the beam, strobe, and LED triggers.
The total trigger rate is a bit under 1~Hz, meaning that $\mathcal{O}(10^6)$ cosmic muons have already been collected over the first few months of CCM200 data.
\Cref{fig:example_cosmic} shows the summed waveform across all PMTs in CCM200 for a single cosmic muon trigger.
One can also see a delayed signal from what is likely a Michel electron.
Though we are still in the preliminary phase of analyzing data from the cosmic muon trigger, we touch briefly on a promising initial study of this dataset.
This analysis examines the summed waveform in coated and uncoated PMTs separately in the top half and bottom half of the barrel of the detector.
We look specifically at the initial rise time $t_0$ of each summed waveform, defined as the time tick at which the waveform passes 0.5\% of its maximum value.
In the scintillation only case, the coated PMTs should see a signal earlier than the uncoated PMTs ($t_0^{\rm coated} < t_0^{\rm uncoated}$), as only photons reflected from elsewhere in the detector will register a p.e. on the uncoated PMTs.
This difference should diminish when Cherenkov light is involved, as both coated and uncoated PMTs will observe Cherenkov light to some level.
We set the threshold to 0.5\% as this reflects roughly the total number of Cherenkov photons generated by a crossing cosmic muon compared to scintillation photons. 
\Cref{fig:CFD_t0_dists} shows the distribution of $t_0^{\rm coated} - t_0^{\rm uncoated}$ for PMTs in the top and bottom of the barrel over $\sim 75$ cosmic muon events.
Given the angle of the Chernekov cone emitted by a cosmic muon traveling downward along the central axis of the detector, only PMTs in the bottom half of the barrel will observe Cherenkov light.
This is consistent with the distributions in \cref{fig:CFD_t0_dists}--the PMTs in the upper half of the barrel peak at $t_0^{\rm coated} - t_0^{\rm uncoated} \sim -5~{\rm ns}$, while PMTs in the lower half of the barrel peak around zero.
Thus, the delay of the signal in uncoated PMTs appears to disappear for PMTs sensitive to Cherenkov light.
This is far from a definitive detection of Cherenkov light in the CCM detector, as one would need to carefully consider geometric effects in the CCM detector to predict $t_0^{\rm coated} - t_0^{\rm uncoated}$ distributions with and without Cherenkov light.
However, it is a promising initial indication that Cherenkov reconstruction might be possible in CCM by leveraging differences in the coated and uncoated PMT signals.

\section{Neutrissimos in CCM} \label{sec:ccm_neutrissimos}

We close our CCM discussion with an investigation of the sensitivity of CCM200 to the neutrissimo model of \cref{ch:neutrissimos}.
The idea of looking for dipole-coupled HNLs at CE$\nu$NS experiments has already been proposed in the literature~\cite{Bolton:2021pey}.
The low energy threshold of these experiments enables a powerful signal definition: the coincidence of a nuclear recoil from upscattering ($\nu A \to \mathcal{N}A$) followed by a higher energy photon from neutrissimo decay ($\mathcal{N} \to \nu \gamma$).
In particular, the upcoming NUCLEUS experiment is projected to have strong sensitivity to $d_{e \mathcal{N}}$ for $m_{\mathcal{N}} \lesssim 10~{\rm MeV}$~\cite{Bolton:2021pey}.
In this section, we investigate whether CCM200 might also have sensitivity to neutrissimos in the $m_{\mathcal{N}} = \mathcal{O}(10~{\rm MeV})$ regime.
The important advantage of CCM200 is that it is already running; projected sensitivities derived here will become limits within a few years.

\begin{figure}[h!]
    \centering
    \includegraphics[width=0.6\textwidth]{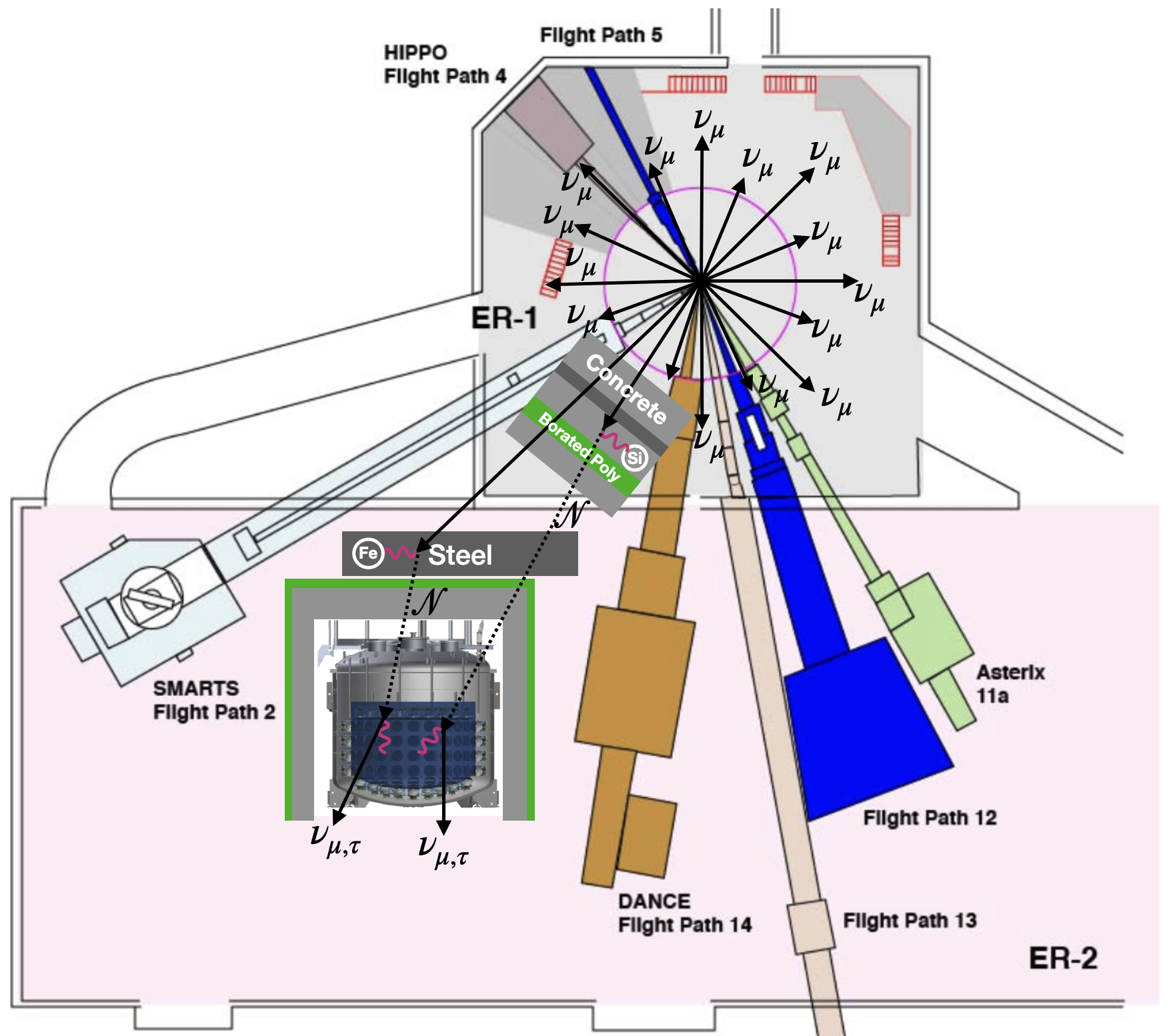}
    \caption{Schematic depiction of prompt $\nu_\mu$ from $\pi^+$ decay in the Lujan target upscattering to neutrissimos within shielding along the path to CCM200 and decaying to photons in the detector. The pink circle represents the TMRS shown in \cref{fig:lujan_target}.}
    \label{fig:ccm_neutrissimo_diagram}
\end{figure}

The beam-dump source at Lujan is an ideal location to look for neutrissimos.
The prompt $\nu_\mu$ flux at CCM's location is $4.74 \times 10^5~\nu/{\rm cm}^2/s$~\cite{CCM:2021leg}, corresponding to roughly $3.76 \times 10^{-2}~\nu/{\rm POT}$.
These neutrinos can undergo Primakoff upscattering in the shielding around the target and along the flight path to the detector, creating a flux of neutrissimos.
These neutrissimos can decay within the CCM detector to photons with $E_\gamma \sim 15~{\rm MeV}$, which is relatively high energy compared to the background energy distribution~\cite{CCM:2021lhc}.
A schematic depiction of this process is shown in \cref{fig:ccm_neutrissimo_diagram}.
In this section, we define the neutrissimo signal as a single high energy photon in the detector.
We do not evaluate our sensitivity to the coincidence channel discussed in Ref.~\cite{Bolton:2021pey}, as the energy threshold of CCM does not yet enable CE$\nu$NS detection.
Additionally, though we only consider neutrissimo production via Primakoff upscattering, neutrissimos can also be produced by the $\pi^0 \to \gamma (\gamma^* \to \nu \mathcal{N})$ and $\gamma \nu \to \mathcal{N}$ channels within the target.
Careful consideration of these additional detection and production channels may boost CCM's sensitivity to neutrissimos, especially once the energy threshold is sufficiently lowered to enable a CE$\nu$NS search.

As we rely on monoenergetic $\nu_\mu$ with $E_\nu \sim 30~{\rm MeV}$ to undergo upscattering, we are sensitive to neutrissimos with $m_{\mathcal{N}} \lesssim 30~{\rm MeV}$.
We simulate neutrissimo upscattering and decay using the updated \texttt{LeptonInjector}~\cite{LeptonInjector} simulation framework.
The public version of \texttt{LeptonInjector} includes a geometry file with a realistic description of the CCM200 detector and the surrounding Lujan facility, including the shielding shown in \cref{fig:ccm_neutrissimo_diagram} and the TMRS shown in \cref{fig:lujan_target} as well as the concrete floor.
Monoenergetic muon neutrinos are injected as an isotropic point source originating from the center of the lower tungsten target.
We sample an upscattering location along the flight path to CCM, considering only a cone surrounding the CCM detector to improve the simulation efficiency.
We then sample a neutrissimo decay location.
If the neutrissimo path crosses the CCM200 fiducial volume, its decay is required to occur within this volume to further improve the simulation efficiency.
We save the four-momentum and physical event weight (in units of POT$^{-1}$) for the final state photon.
By scaling to the expected CCM200 exposure of $2.25 \times 10^{22}$~POT, we can estimate the single photon event rate from neutrissimo decay in the total three-year CCM200 dataset.
This \texttt{LeptonInjector}-based CCM200 neutrissimo simulation was presented to the public during the 2023 CCM Workshop~\cite{CCMWorkshop}.

To assess sensitivity, we consider the background estimation from the ALP search~\cite{CCM:2021lhc}, as this analysis used a similar electromagnetic final state signal definition.
A sizable reduction in the background rate is expected for CCM200 compared to CCM120; the improved shielding already gives about an order-of-magnitude improvement~\cite{EdThesis}, and the Cherenkov reconstruction simulation study described in \cref{sec:CCM_cherenkov_simulation} suggests an additional two orders of magnitude in background reduction for a 5~MeV electron.
In reporting results, we show a projected sensitivity range based on a background reduction within $[10^{-4},10^{-2}]$.
The exact background reduction achieved by CCM200 will depend on the performance of the final Cherenkov reconstruction and argon filtration system as well as any improvements to the existing analysis cuts described in Ref.~\cite{CCM:2021lhc}.
We assume a neutrissimo detection efficiency of 1.0, which is consistent with the expected ALP detection efficiency after CCM200 upgrades~\cite{EdThesis}.
The photon energies are smeared according to an energy resolution of 15\%, such that the prediction in each reconstructed energy bin $\mu_i$ is given by
\begin{equation}
\mu_i = \sum_{j = 0}^{N_{\rm sim}} \int_{E_i^{\rm low}}^{E_i^{\rm high}} \frac{w_j}{\sqrt{2 \pi (0.15 E_j)^2}}  e^{\frac{-(E - E_j)^2}{2(0.15 E_j)^2}} dE,
\end{equation}
where $N_{\rm sim}$ is the number of simulated photons, $E_i^{\rm low}$ and $E_i^{\rm high}$ are the boundaries of reconstructed energy bin $i$, and $w_j$ and $E_j$ are the weight and energy of the $j^{\rm th}$ simulated photon from \texttt{LeptonInjector}.
This energy resolution is relatively consistent with the CCM120 energy resolution at $E_\gamma \sim 20~{\rm MeV}$~\cite{EdThesis} and a safe assumption once the updated Cherenkov reconstruction is in place.

\begin{figure}[h!]
    \centering
     \begin{subfigure}[b]{0.45\textwidth}
         \centering
         \includegraphics[width=\textwidth]{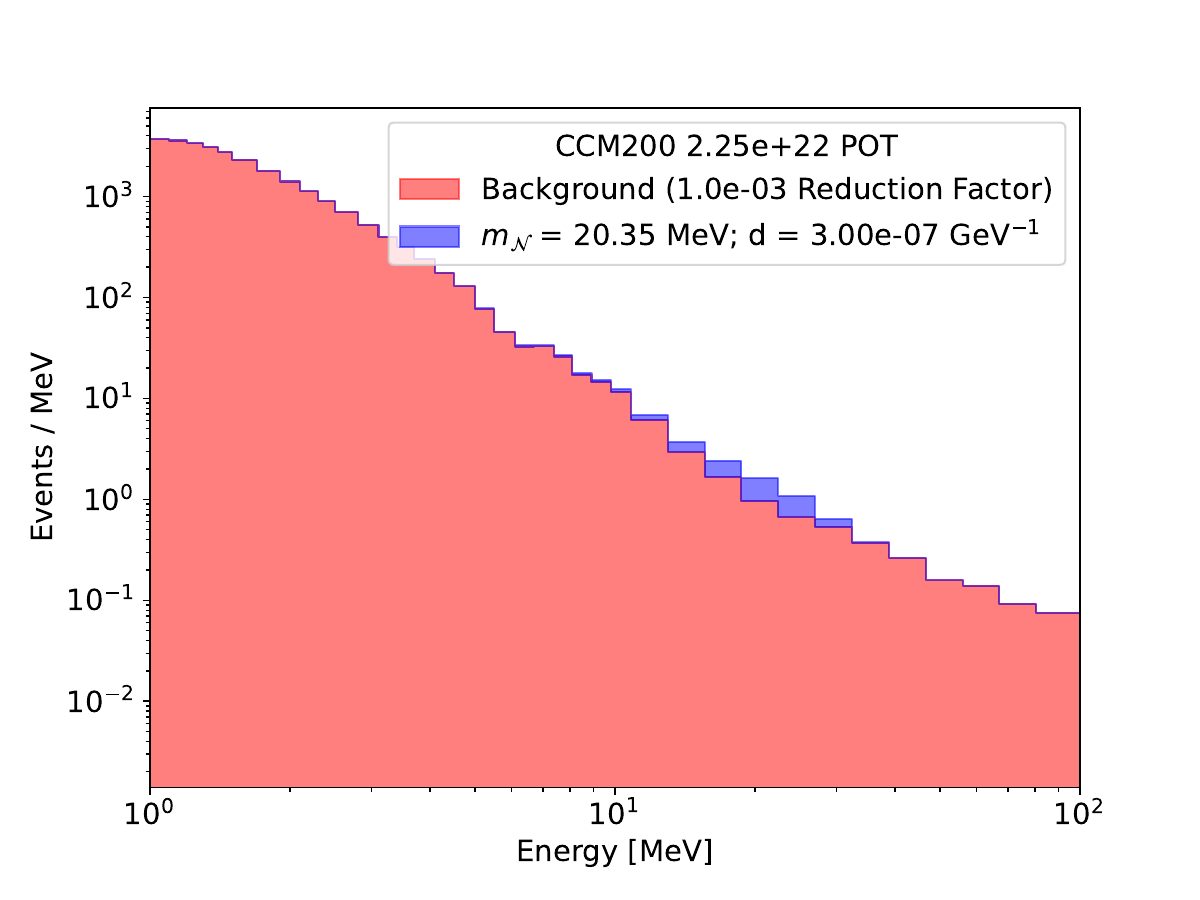}
         \caption{}
         \label{fig:CCM200_NeutrissimoRate}
     \end{subfigure}
     \hfill
     \begin{subfigure}[b]{0.45\textwidth}
         \centering
         \includegraphics[width=\textwidth]{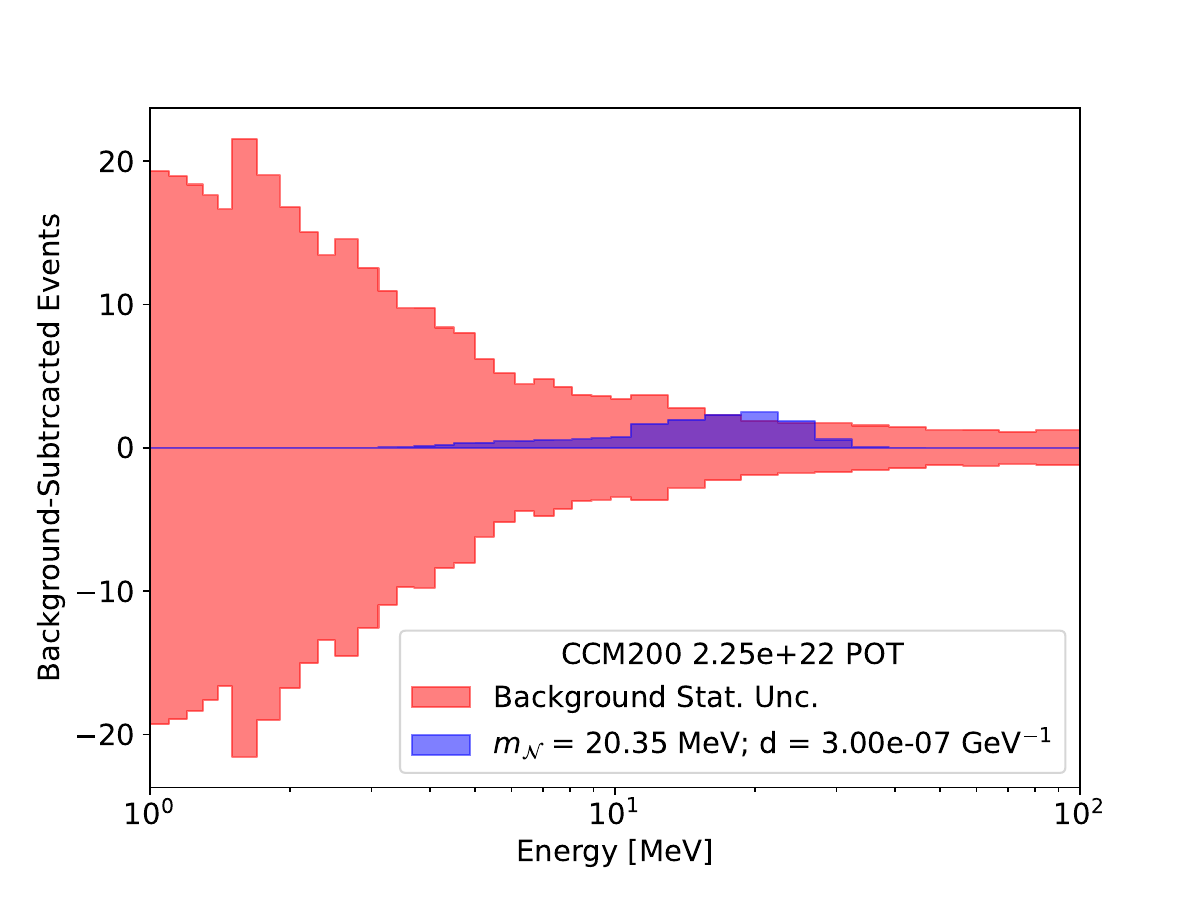}
         \label{fig:CCM200_NeutrissimoRate_BackgroundSubtraction}
     \end{subfigure}
     \hfill
        \caption{\Cref{fig:CCM200_NeutrissimoRate} shows the distribution of background and signal prediction in CCM200 for $m_{\mathcal{N}} = 20.35~{\rm MeV}$ and $d_{\mu {\mathcal{N}}} = 3 \times 10^{-7}~{\rm GeV}^{-1}$, considering a background reduction factor of $10^{-3}$ compared to CCM120. \Cref{fig:CCM200_NeutrissimoRate_BackgroundSubtraction} shows the background-subtracted plot, with a red band indicating the expected statistical uncertainty on the background.}
        \label{fig:CCM200_NeutrissimoPlots}
\end{figure}

\begin{figure}[h!]
    \centering
     \begin{subfigure}[b]{0.45\textwidth}
         \centering
         \includegraphics[width=\textwidth]{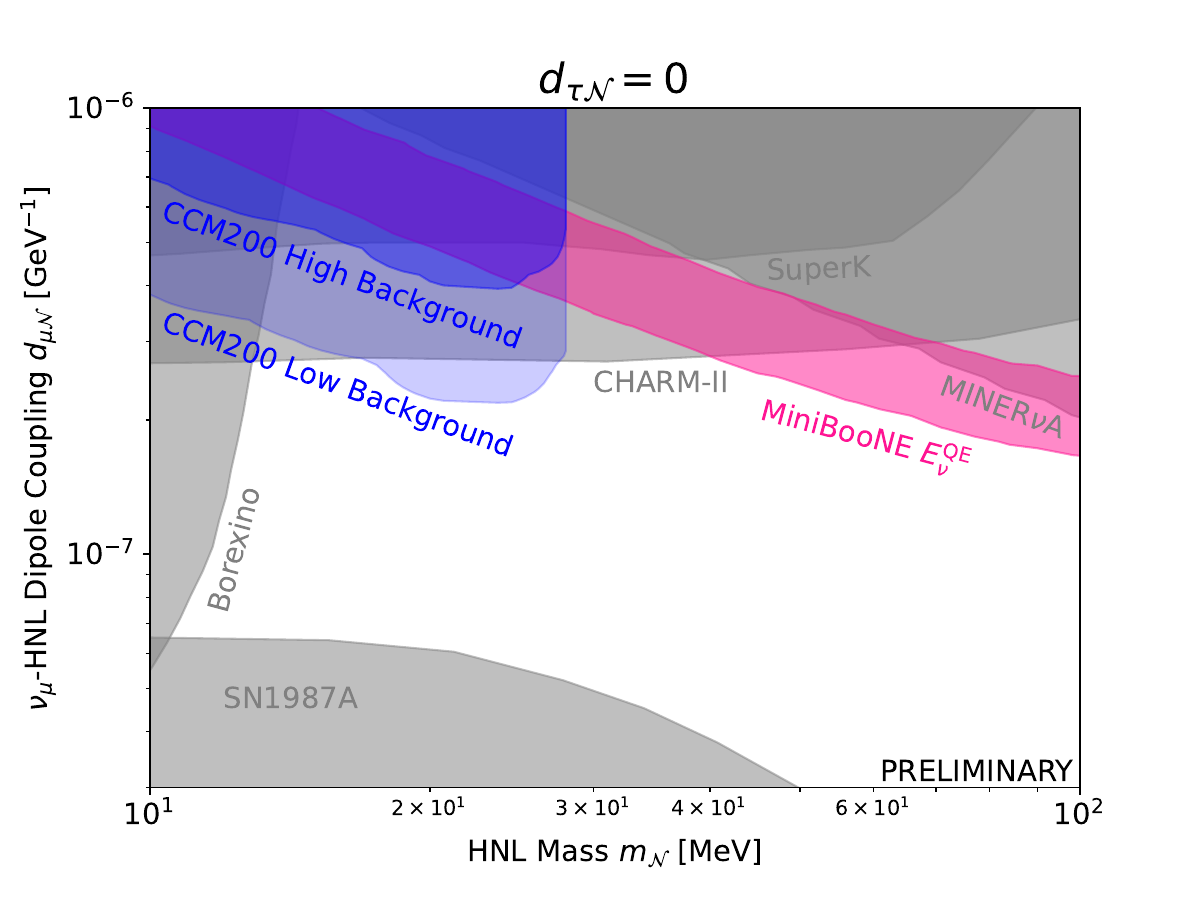}
         \caption{}
         \label{fig:CCMSensitivity}
     \end{subfigure}
     \hfill
     \begin{subfigure}[b]{0.45\textwidth}
         \centering
         \includegraphics[width=\textwidth]{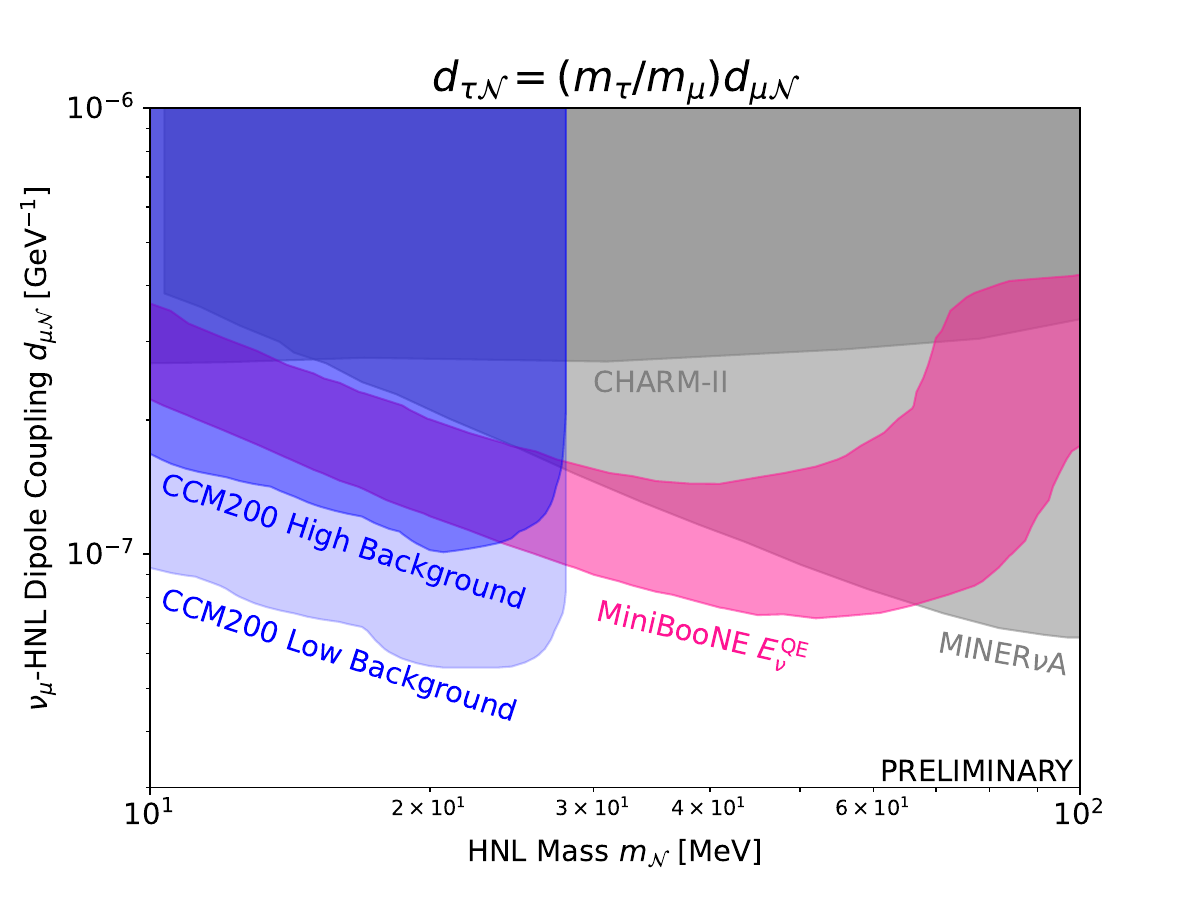}
         \caption{}
         \label{fig:CCMSensitivity_dtau}
     \end{subfigure}
     \hfill
        \caption{\Cref{fig:CCMSensitivity} shows the expected sensitivity of CCM200 to the neutrissimo model, where the blue band corresponds to a background reduction factor between $10^{-4}$ (``CCM200 Low Background'') and $10^{-2}$ (``CCM200 High Background''). The MiniBooNE $E_\nu^{\rm QE}$ allowed region (pink) and existing constraints (grey) come from Ref~\cite{Kamp:2022bpt}. \Cref{fig:CCMSensitivity_dtau} shows the same plot, but considering $\mathcal{N} \to \nu_\tau \gamma$ decays with $d_{\tau {\mathcal{N}}} = d_{\mu {\mathcal{N}}} m_\tau / m_\mu$.}
        \label{fig:CCMNeutrissimoSensitivity}
\end{figure}

We have run a series of simulations for neutrissimos with masses $m_{\mathcal{N}} [{\rm MeV}] \in [1,28]$ and dipole couplings $d_{\mu \mathcal{N}} [{\rm GeV}^{-1}] \in [10^{-7},10^{-5}]$.
\Cref{fig:CCM200_NeutrissimoPlots} shows the expected energy distribution of neutrissimo decay signal events and background events in CCM200 for $m_{\mathcal{N}} = 20.35~{\rm MeV}$ and $d_{\mu {\mathcal{N}}} = 3 \times 10^{-7}~{\rm GeV}^{-1}$, considering a background reduction factor of $10^{-3}$.
The single photons from neutrissimo decay peak at $E_\gamma \sim 20~{\rm MeV}$, as expected.
We calculate the CCM200 sensitivity across parameter space using the $\chi^2$ test statistic,
\begin{equation}
\chi^2(m_{\mathcal{N}},d_{\mu \mathcal{N}}) = \sum_i \bigg( \frac{\mu_i(m_{\mathcal{N}},d_{\mu \mathcal{N}})}{\sigma_i} \bigg)^2,
\end{equation}
where $\mu_i(m_{\mathcal{N}},d_{\mu \mathcal{N}})$ is the neutrissimo prediction in the $i^{\rm th}$ reconstructed energy in and $\sigma_i$ is the statistical $\oplus$ systematic error on the background, appropriately scaled from Table~6.1 of Ref.~\cite{EdThesis}.
We assume a $\chi^2$ distribution with two degrees of freedom to draw projected sensitivities at the 95\% confidence level (corresponding to a critical value of $\chi^2 = 6.18$).
\Cref{fig:CCMSensitivity} shows the preliminary projected CCM200 sensitivity to neutrissimo parameter space over the full three-year run.
If we achieve a background reduction factor closer to $10^{-4}$, CCM200 should be able to set world-leading constraints on the $d_{\mu \mathcal{N}}$ dipole coupling for $m_{\mathcal{N}} \sim 20~{\rm MeV}$.
As discussed in \cref{sec:neutrissimo_paper}, flavor-conserving UV completions of the dipole operator predict larger branching ratios for the decay $\mathcal{N} \to \nu_\tau \gamma$.
As the neutrissimo single photon rate in CCM200 is primarily limited by the neutrissimo decay probability within the detector, CCM200 will be more sensitive to a scenario with such flavor-dependent dipole couplings.
In \cref{fig:CCMSensitivity_dtau} we show a preliminary estimate of CCM200's sensitivity to the $d_{\tau {\mathcal{N}}} = d_{\mu {\mathcal{N}}} m_\tau / m_\mu$ discussed in \cref{sec:neutrissimo_paper}.
The full CCM200 dataset will exclude a large region of the available parameter space in this flavor-dependent dipole model.

A few points are worth noting regarding these preliminary sensitivities.
First, while the \texttt{LeptonInjector} simulation of CCM200 provides a relatively precise initial estimate of the neutrissimo decay rate in the detector, the geometric description of the Lujan facility will need to be verified before these sensitivities are ready for publication.
This is especially true considering the strong dependence of the neutrissimo event rate on the exact shielding configuration between the tungsten target and the detector.
That being said, any updates to the \texttt{LeptonInjector} CCM200 geometry made during this verification process are likely to be small and thus will not drastically shift the sensitivities in \cref{fig:CCMNeutrissimoSensitivity}.
The sensitivity might also improve after considering the additional neutrissimo production and detection mechanisms discussed earlier: $\pi^0 \to \gamma \nu \mathcal{N}$ and $\gamma \nu \to \mathcal{N}$ within the target, and nuclear recoil from Primakoff upscattering within the detector.
These are all possible to simulate within \texttt{LeptonInjector}.
The $\pi^0$ and $\gamma$ flux were calculated for the ALP analysis~\cite{CCM:2021lhc}, and nuclear recoil kinematics are already simulated within \texttt{LeptonInjector}.
These channels will be carefully evaluated as the CCM200 neutrissimo analysis is prepared for publication.
Further improvements to CCM200's sensitivity can be made for modified shielding configurations that introduce some amount of high-$Z$ material, such as lead ($Z = 82$), in order to benefit from the $Z^2$ coherent enhancement of the Primakoff upscattering cross section~\cite{Kamp:2022bpt}.
Potential improved shielding configurations can and will be studied within \texttt{LeptonInjector}.

\begin{figure}[h!]
    \centering
    \includegraphics[width=0.6\textwidth]{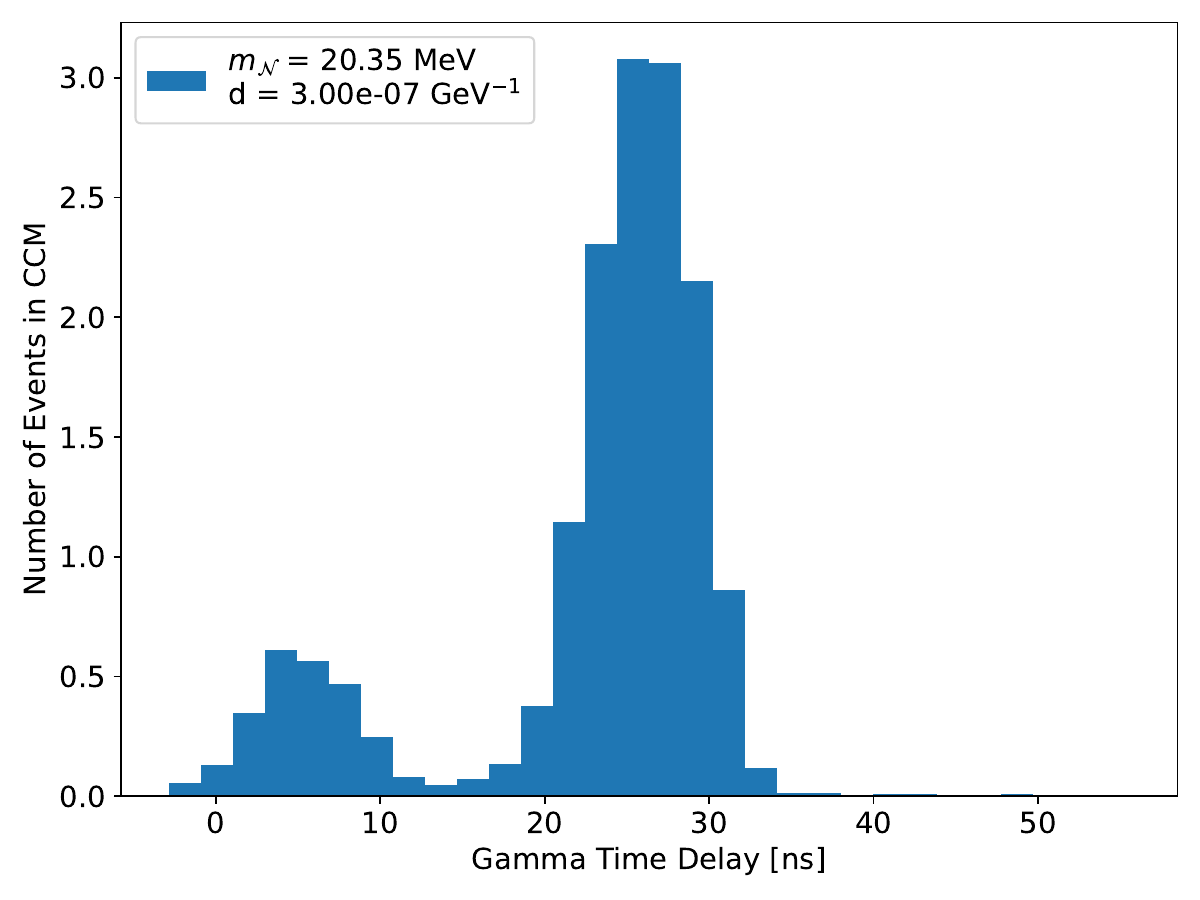}
    \caption{The time delay of neutrissimo single photon decays with the CCM detector for the indicated mass and coupling, as calculated using \texttt{LeptonInjector}.}
    \label{fig:CCM200_NeutrissimoTimeDelay}
\end{figure}

One must also consider the timing distribution of the photons from neutrissimo decay within the detector.
CCM relies on strict timing cuts to remove neutron backgrounds; thus, it is important to ensure that the added travel time of non-relativistic neutrissimos does not push these photons outside of the CCM ROI.
This is easily studied within \texttt{LeptonInjector}, as the entire path of the neutrino and neutrissimo is computed for each simulated event.
\Cref{fig:CCM200_NeutrissimoTimeDelay} shows the added time delay from the neutrissimo flight path for $m_{\mathcal{N}} = 20.35~{\rm MeV}$ (note that the time delay does not depend on the dipole coupling to first order).
This is the most important neutrissimo mass to check, as it corresponds to the strongest projected constraints in \cref{fig:CCMNeutrissimoSensitivity}.
Two distinct populations can be seen at 5~ns and 26~ns, corresponding to upscattering in the shielding surrounding CCM and the TMRS, respectively.
This suggests characteristic time delays $\Delta t \lesssim 30~{\rm ns}$, which is comfortably within the 150~ns ROI of the ALP analysis~\cite{CCM:2021lhc}.
Thus, the CCM timing cuts should not be an issue for the neutrissimo model.

The analysis presented here can in principle be used to set constraints using CCM120 data.
However, the signal efficiency of neutrissimo events must be more carefully evaluated for this dataset.
The CCM120 ALP analysis used a variety of cuts on the event length, reconstructed radius, and PMT charge uniformity to remove backgrounds.
While the improvements made for CCM200 are likely to push the ALP (and thus neutrissimo) selection efficiency to greater than $0.95$, the CCM120 ALP selection efficiency fell roughly within $[0.2,0.4]$ depending on the ALP energy~\cite{CCM:2021lhc}.
Further, the CCM120 analysis used only the first 38~ns of each event to calculate the reconstructed energy~\cite{CCM:2021lhc}, which will have a nontrivial impact on the assumed 15\% symmetric energy resolution.
Finally, the shielding configuration was updated between CCM120 and CCM200, and the CCM120 version is not currently modeled in \texttt{LeptonInjector}.
We will need to study the impact of all of these effects on single photon events before calculating CCM120 limits on the neutrissimo model.

The analysis presented in this section highlights the strong capability of CCM in constraining BSM physics.
The high-intensity beam dump at the Lujan center can produce a significant number of potential BSM particles, while the extremely low duty factor helps isolate these particles from SM backgrounds.
The large size, good energy resolution, and fast timing response of the CCM detector are well-suited for measuring any potential interactions of these BSM particles.
The neutrissimo model discussed here is just one of many BSM scenarios on which CCM200 will place world-leading constraints~\cite{CCM:2021leg,CCM:2021lhc,CCM:2021yzc,Dutta:2021cip}.
Additionally, the neutrissimo model specifically emphasizes the significant impact that Chernekov light reconstruction will have on CCM200 sensitivity to BSM interactions with electromagnetic final states (which also occur in the ALP model).

\chapter{Conclusions and Future Prospects}

This thesis presents an array of experimental and phenomenological investigations into the nature of the MiniBooNE low energy excess (LEE).
The MiniBooNE LEE, described in detail in \cref{ch:miniboone}, refers to the $4.8\sigma$ excess of electron-like neutrino interactions observed by the MiniBooNE detector along the Booster Neutrino Beamline (BNB)~\cite{MiniBooNE:2020pnu}.
This is one of the most significant and longest-standing anomalies with respect to the Standard Model (SM) of particle physics.
The excess has remained consistent over the full 17-year dataset, corresponding to $18.75 \times 10^{20}$ ($11.27 \times 10^{20}$) protons-on-target (POT) in neutrino (antineutrino) mode.
Due to the nature of a Cherenkov detector, MiniBooNE could not distinguish between electrons, photons, or collimated $e^+e^-$ pairs--thus, any of these can contribute to the MiniBooNE LEE.
Though SM explanations of the excess have been investigated thoroughly, both internally within MiniBooNE~\cite{MiniBooNE:2020pnu} and by the external community~\cite{Brdar:2021ysi,MicroBooNE:2021zai}, no promising SM candidates have emerged.
Thus, the MiniBooNE LEE may be an indication of beyond-the-SM (BSM) physics.
The most common BSM interpretation of the MiniBooNE anomaly is the $3+1$ model, which predicts $\nu_\mu \to \nu_e$ and $\overline{\nu}_\mu \to \overline{\nu}_e$ oscillations at short baseline through an eV-scale sterile neutrino.
This is an attractive solution as it would also explain the anomalous excess of $\overline{\nu}_e$ interactions observed by the LSND experiment, which was the original physics motivation behind the MiniBooNE experiment.

The MicroBooNE experiment was conceived to follow up on the MiniBooNE LEE.
MicroBooNE uses a liquid argon time projection chamber (LArTPC) detector~\cite{MicroBooNE:2016pwy}, described in detail in \cref{ch:microboone_detector}, to observe the interactions of neutrinos along the BNB.
The detailed images of neutrino interactions produced by the LArTPC enable the separation of events with electrons and photons in the final state, allowing a suite of analyses looking for different explanations of the MiniBooNE excess.
My research within MicroBooNE focused on the ``two-body CCQE analysis'': a search for $\nu_e$ charged-current quasi-elastic (CCQE) interactions consistent with two-body scattering kinematics~\cite{MicroBooNE:2021pvo}.
This analysis is described in detail in \cref{ch:microboone_selection}.
It uses a variety of techniques~\cite{MicroBooNE:2020sar,MicroBooNE:2021nss}, including several deep-learning-based methods~\cite{MicroBooNE:2020yze,MicroBooNE:2020hho}, to reconstruct and isolate signal events from backgrounds.
The electromagnetic shower reconstruction, which drives the reconstruction of the original neutrino energy, is validated using two data-driven standard candles: the $\pi^0$ invariant mass peak and the Michel electron energy spectrum cutoff~\cite{MicroBooNE:2021nss}.
This study is presented in \cref{sec:shower_publication}.
The most powerful technique used in the signal selection leverages an ensemble of boosted decision trees (BDTs) trained on a collection of variables describing the topology and kinematics of candidate $\nu_e$ CCQE interactions.
This BDT ensemble is specifically tailored to select events consistent with a clean two-body interaction, which helps reduce the impact of complicated nuclear effects on our signal prediction.

The two-body CCQE analysis observed 25 events passing all signal selection requirements in the first $6.67 \times 10^{20}$~POT of BNB data~\cite{MicroBooNE:2021pvo}.
The statistical results from this dataset are presented in \cref{ch:microboone_results}.
The two-body CCQE analysis does not observe significant excess of $\nu_e$ CCQE events at the lowest energies consistent with the MiniBooNE LEE.
Using a two-hypothesis test and the CL$_s$ test statistic, we are able to rule out the median expectation from the MiniBooNE excess at the $2.4\sigma$ confidence level.
A signal strength scaling test performed using the Feldman-Cousins procedure sets an upper bound of $0.38$ on the $\nu_e$ fractional contribution to the MiniBooNE LEE at the $2\sigma$ confidence level.
The results obtained by our analysis are consistent with the other two MicroBooNE $\nu_e$ analysis--no analysis observes a significant low-energy excess of charged-current $\nu_e$ interactions like that predicted by the MiniBooNE LEE~\cite{MicroBooNE:2021tya}.

Two external follow-studies regarding the MicroBooNE results are presented.
The first of these is a combined fit of MicroBooNE and MiniBooNE data to the $3+1$ model~\cite{MiniBooNE:2022emn}, presented in \cref{sec:MBuB_sterile_paper}.
This study shows that despite the strong constraints from MicroBooNE on $\nu_e$ interpretations of the MiniBooNE LEE, closed contours in $3+1$ parameter space still exist at the $3\sigma$ confidence level in the combined MiniBooNE-MicroBooNE fit.
Thus, the MicroBooNE results do not definitively rule out the $3+1$ interpretation of the MiniBooNE excess.
The second study, presented in \cref{sec:miniboone_antinu}, looked into the implications of a $\overline{\nu}_e$ explanation of the MiniBooNE excess~\cite{Kamp:2023mjn}.
Due to the isoscalar and non-isoscalar nature of the carbon and argon nucleus, respectively, the $\overline{\nu}_e$ cross section tends to be much more suppressed at low energy in argon compared to carbon, while the same is not true for $\nu_e$ interactions.
Thus, MicroBooNE will be less sensitive to the MiniBooNE LEE if it comes from $\overline{\nu}_e$ interactions.
Ref.~\cite{Kamp:2023mjn} shows that the MicroBooNE data are consistent at the $2\sigma$ confidence level with a scenario in which the MiniBooNE excess is comprised entirely of $\overline{\nu}_e$ events.

In \cref{ch:neutrissimos} we discuss a mixed model comprising an eV-scale sterile neutrino and an MeV-scale heavy neutral lepton with a dipole coupling to active neutrinos, referred to as a neutrissimo $\mathcal{N}$~\cite{Vergani:2021tgc,Kamp:2022bpt}.
It is shown that this model can significantly relax the internal tension in global $3+1$ by explaining the bulk of the MiniBooNE excess through neutrissimo decays to single photons.
As discussed in \cref{sec:neutrissimos_in_MB}, the neutrissimo with $m_\mathcal{N} \sim 500~{\rm MeV}$ can simultaneously provide a good explanation to the energy and angular distributions, which is not possible within the $3+1$ model alone.
In Ref.~\cite{Kamp:2022bpt}, we have calculated world-leading constraints on the neutrissimo model using the high $dE/dX$ sideband of the MINER$\nu$A elastic scattering analyses.
This study is presented in \cref{sec:neutrissimo_paper}.
While the MINER$\nu$A constraints are indeed very strong in the $m_\mathcal{N} = \mathcal{O}(100~{\rm MeV})$ region, they do not rule out the MiniBooNE-preferred region in neutrissimo parameter space at the 95\% confidence level.
This is because the kinematic cuts in the MINER$\nu$A analysis remove a majority of neutrissimo events for $m_{\mathcal{N}} \sim 500~{\rm MeV}$.
A dedicated MINER$\nu$A analysis would likely be able to make a definitive statement on the neutrissimo explanation of the MiniBooNE LEE.

Finally, we close with the Coherent CAPTAIN-Mills (CCM) experiment in \cref{ch:ccm}.
CCM uses a liquid argon detector at the Lujan proton beam dump facility of the Los Alamos Neutron Science Center (LANSCE) to look for the interactions of potential beyond-the-Standard-Model (BSM) particles.
The CCM detector uses an array of photo-multiplier tubes (PMTs) to observe the scintillation light produced in these interactions.
CCM leverages the low duty factor of the Lujan beam dump to make strict timing cuts on signals coming from the beam, isolating prompt signal events corresponding to BSM particles from slower beam-related neutron backgrounds.
This strategy has allowed CCM to place strong limits on a wide variety of BSM scenarios, including light vector-portal dark matter~\cite{CCM:2021leg}, leptophobic dark matter~\cite{CCM:2021yzc}, and axion-like-particles~\cite{CCM:2021lhc}, using only data from a six-week 120-PMT engineering run.
The full 200-PMT detector, ``CCM200'', started taking data in Summer 2022 for its nominal three-year run at the Lujan facility.
CCM200 will make a number of improvements upon the 120-PMT engineering run, including upgraded shielding and the deployment of an argon filtration system, which will help set world-leading limits on these models.

One of the most promising improvements is the development of a Cherenkov light reconstruction algorithm, which can help separate BSM events with electromagnetic final states from SM backgrounds.
\Cref{sec:CCM_cherenkov_simulation} presents a simulation-based study that indicates the feasibility of Cherenkov reconstruction in CCM thanks to the combination of PMTs coated and uncoated with wavelength-shifting tetraphenyl butadiene.
The study suggests that uncoated PMTs will be especially sensitive to Cherenkov light in the visible regime for the first $\sim 10~{\rm ns}$ of a given electromagnetic event.
\Cref{sec:ccm_cherenkov_data} discusses some of the initial steps toward developing the Cherenkov reconstruction in data, including the deployment of a dedicated cosmic muon trigger and improvements to the CCM pulse finding algorithm.
We discuss a potential use-case for the Cherenkov reconstruction in \cref{sec:ccm_neutrissimos}, which presents a calculation of CCM200's sensitivity to the neutrissimo model discussed above.
Preliminary estimates suggest that CCM200 will be sensitive to new regions of neutrissimo parameter space for $m_{\mathcal{N}} \sim 20~{\rm MeV}$ over its full three-year dataset.

Neutrinos have never failed to surprise us.
Since the first detection of neutrinos in 1956, experimental anomalies like the solar neutrino problem and the atmospheric neutrino anomaly have provided a guiding light toward a more complete understanding of the ``little neutral one''.
The experimentally-established fact that neutrinos have nonzero mass is not predicted by the Standard Model, and the comparatively tiny neutrino mass with respect to other particles is a strong motivation for BSM physics.
Anomalies in short baseline neutrino experiments could very well be hinting that the neutrino has secrets yet to reveal.
This thesis takes a few concrete steps along the path lit by the MiniBooNE anomaly.
Only time--and clever experimental analyses--will tell where this path leads.
\appendix
\chapter{Publication: \textit{Convolutional neural networks for shower energy
prediction in liquid argon time projection chambers}}

My first research project as a graduate student involved the development of a convolutional neural network for electromagnetic shower energy reconstruction in MicroBooNE.
While the MicroBooNE network never saw the light of day, the network was adapted for use in general LArTPC experiments by an MIT undergraduate student, Kiara Carloni.
Kiara led the study discussed in the \textit{JINST} publication below~\cite{Carloni:2021zbc}.
I served mainly as a mentor for this project, as it was extended from my early MicroBooNE work. 

\includepdf[pages=-]{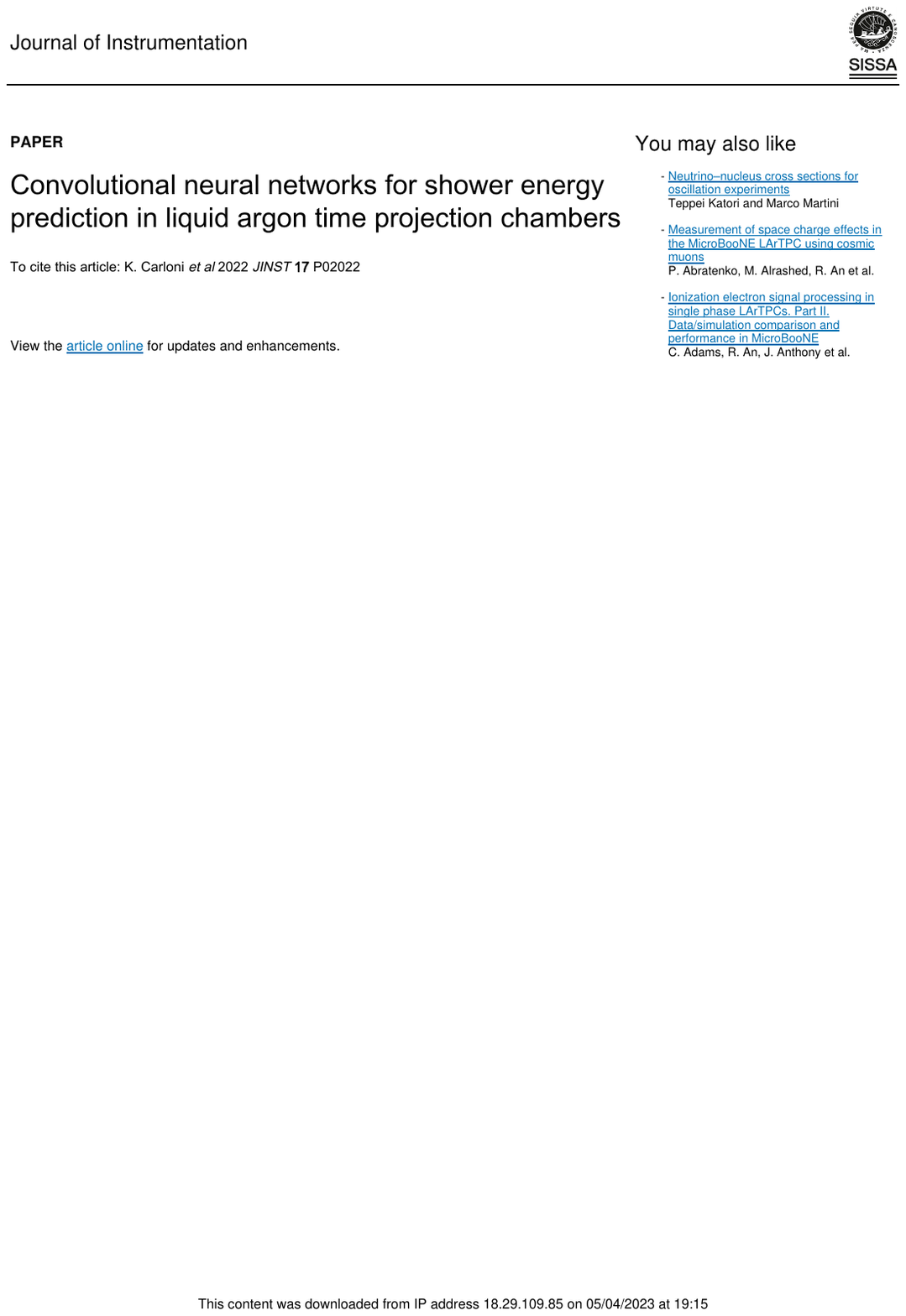}
\clearpage
\newpage

\chapter{Signal Event Displays from the Two-Body CCQE Analysis}

\begin{figure*}
    \centering
    \includegraphics[width=0.92\linewidth]{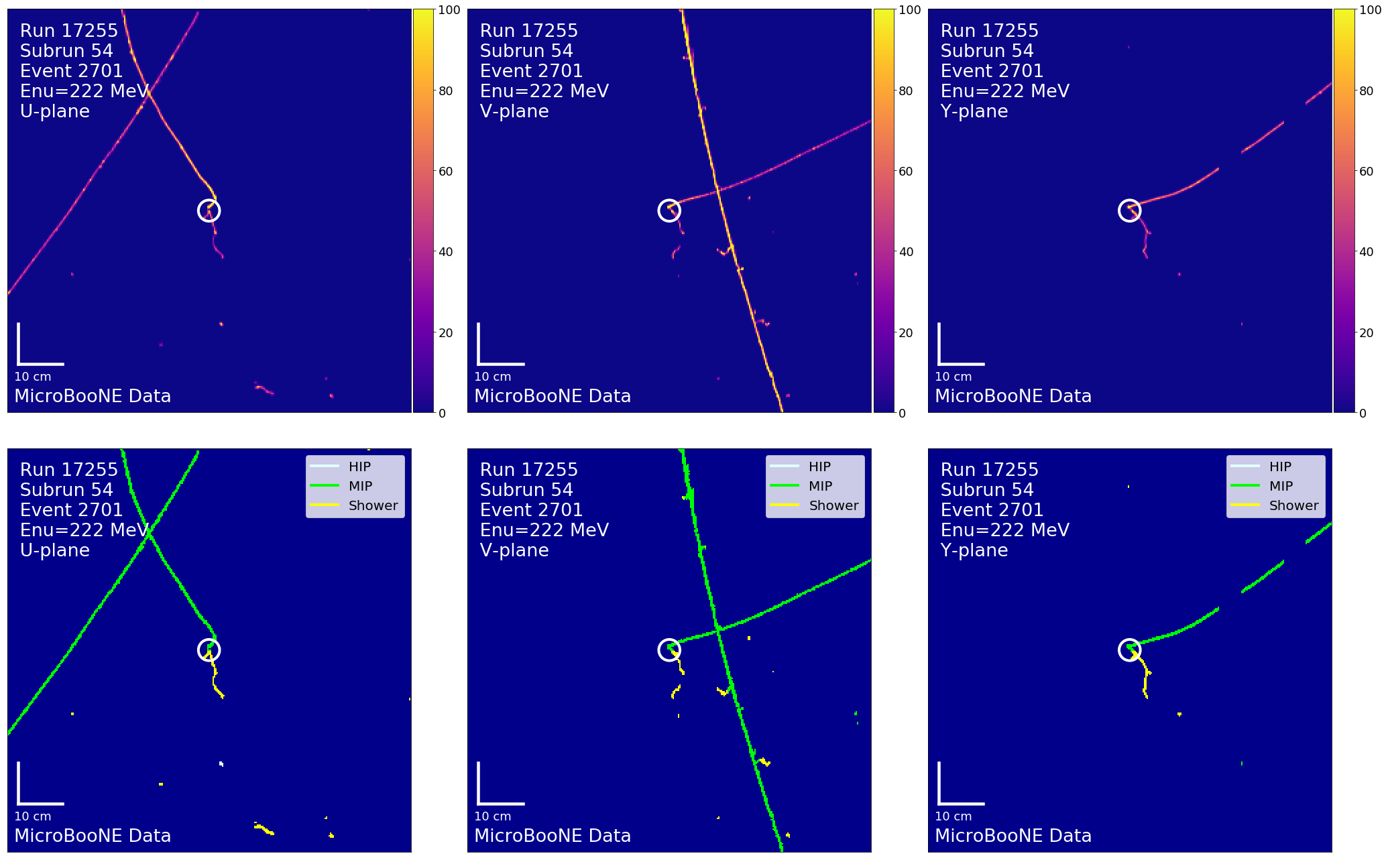}
    \caption{Top: pixel intensity; Bottom: \texttt{SparseSSNet} labels; Left to right: U, V, Y, planes. The white circle indicates the reconstructed vertex.}
    \label{fig:evd2}
\end{figure*}

\begin{figure*}
    \centering
    \includegraphics[width=0.92\linewidth]{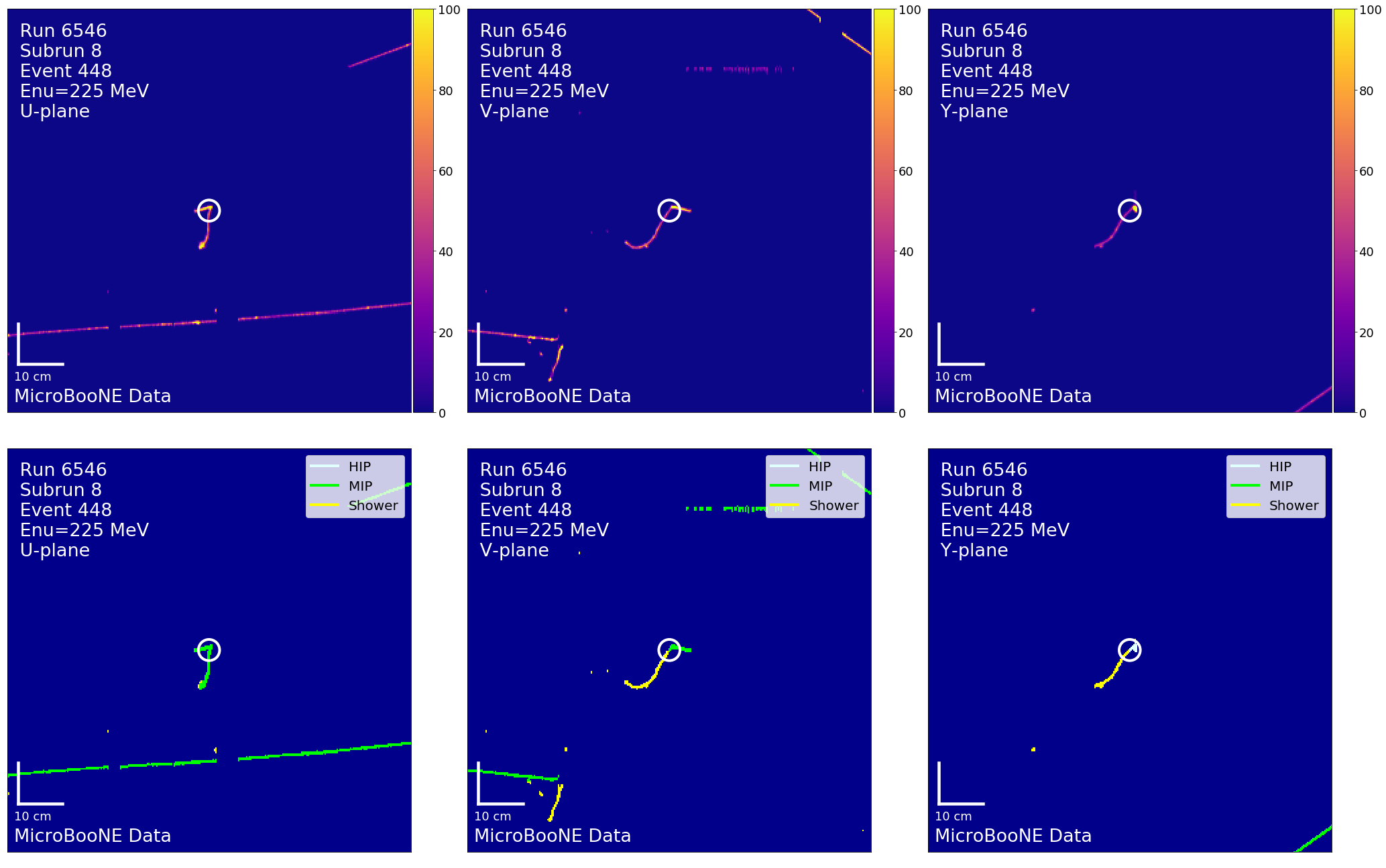}
    \caption{Top: pixel intensity; Bottom: \texttt{SparseSSNet} labels; Left to right: U, V, Y, planes. The white circle indicates the reconstructed vertex.}
    \label{fig:evd3}
\end{figure*}

\begin{figure*}
    \centering
    \includegraphics[width=0.92\linewidth]{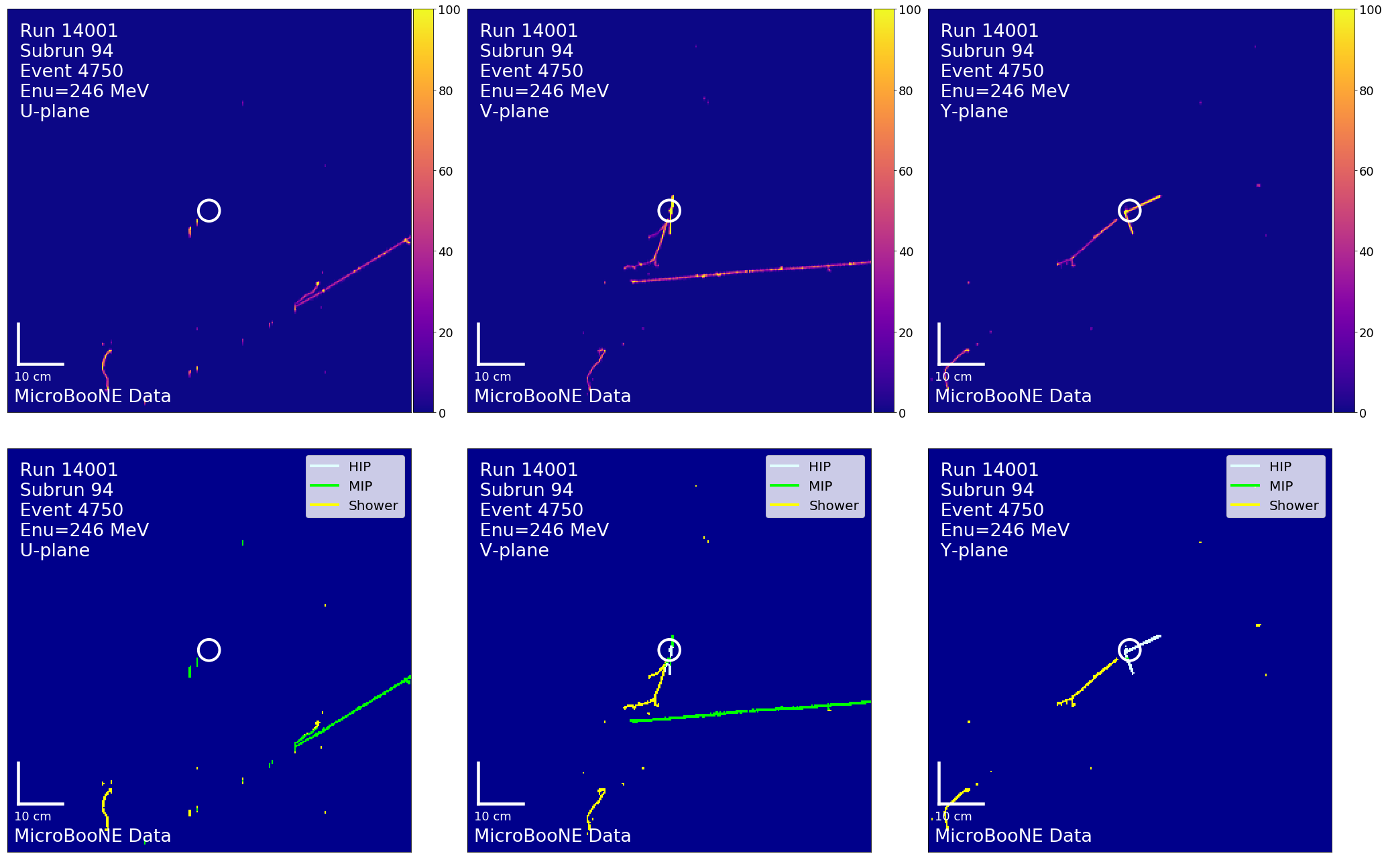}
    \caption{Top: pixel intensity; Bottom: \texttt{SparseSSNet} labels; Left to right: U, V, Y, planes. The white circle indicates the reconstructed vertex.}
    \label{fig:evd4}
\end{figure*}

\begin{figure*}
    \centering
    \includegraphics[width=0.92\linewidth]{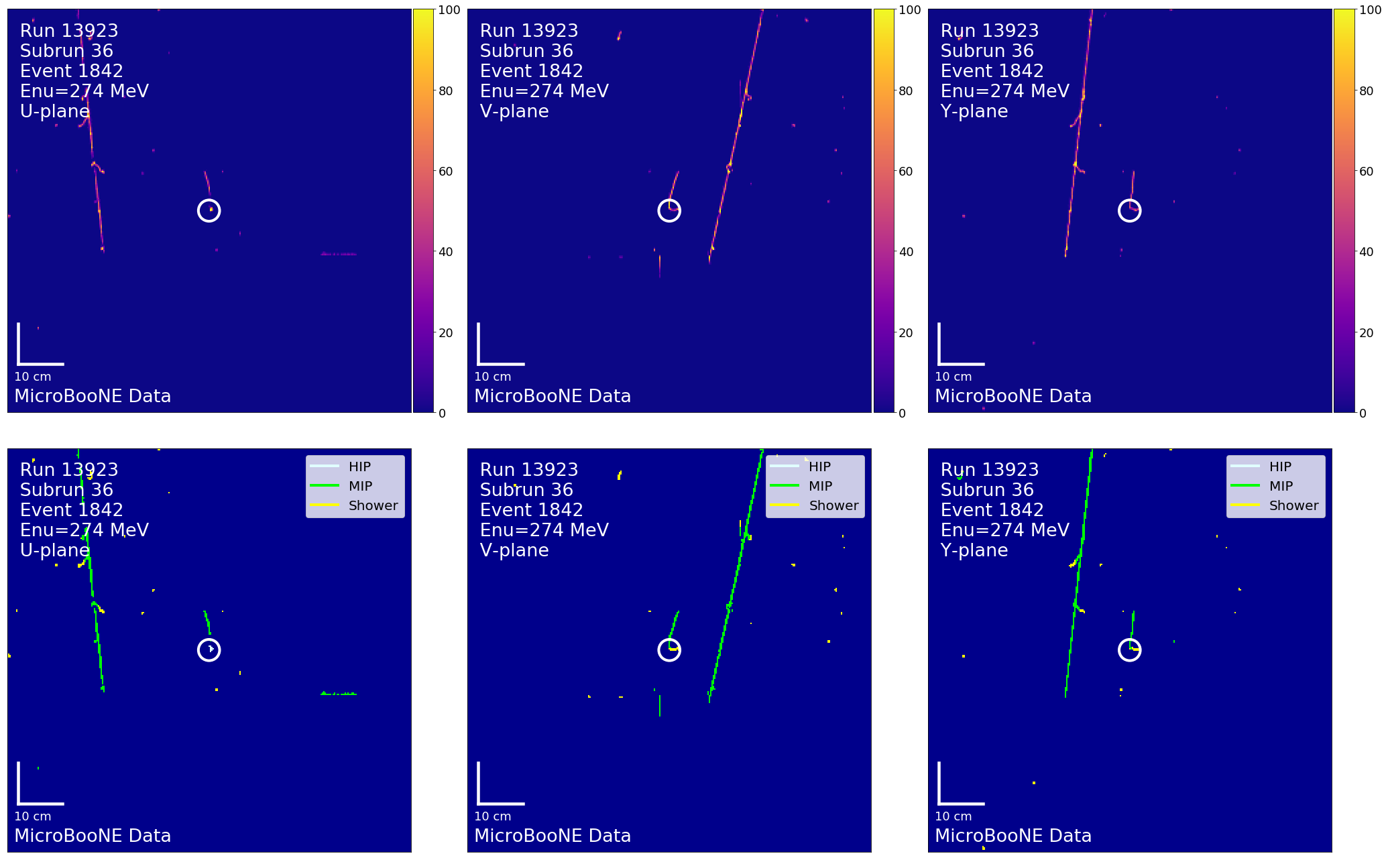}
    \caption{Top: pixel intensity; Bottom: \texttt{SparseSSNet} labels; Left to right: U, V, Y, planes. The white circle indicates the reconstructed vertex.}
    \label{fig:evd5}
\end{figure*}

\begin{figure*}
    \centering
    \includegraphics[width=0.92\linewidth]{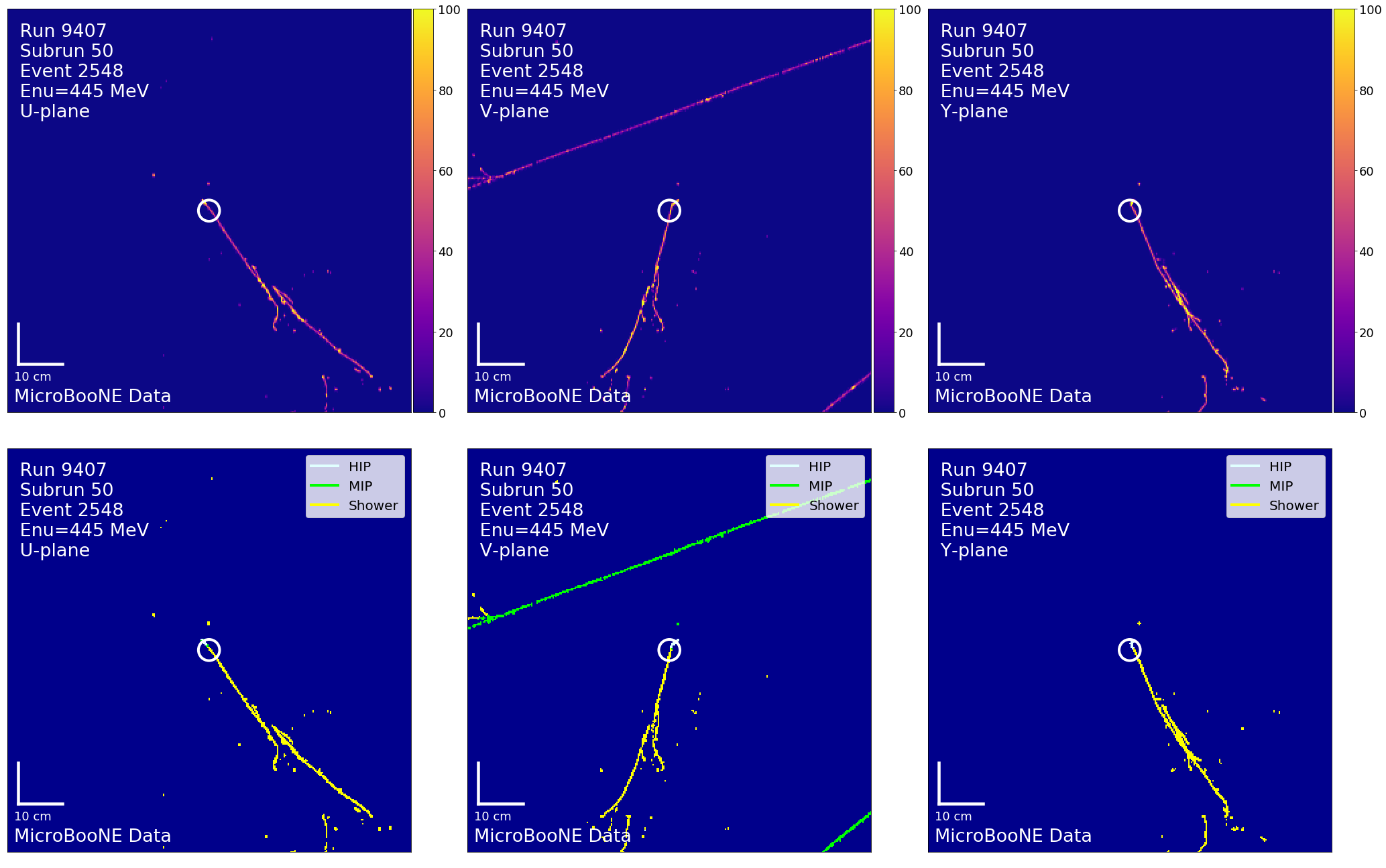}
    \caption{Top: pixel intensity; Bottom: \texttt{SparseSSNet} labels; Left to right: U, V, Y, planes. The white circle indicates the reconstructed vertex.}
    \label{fig:evd6}
\end{figure*}

\begin{figure*}
    \centering
    \includegraphics[width=0.92\linewidth]{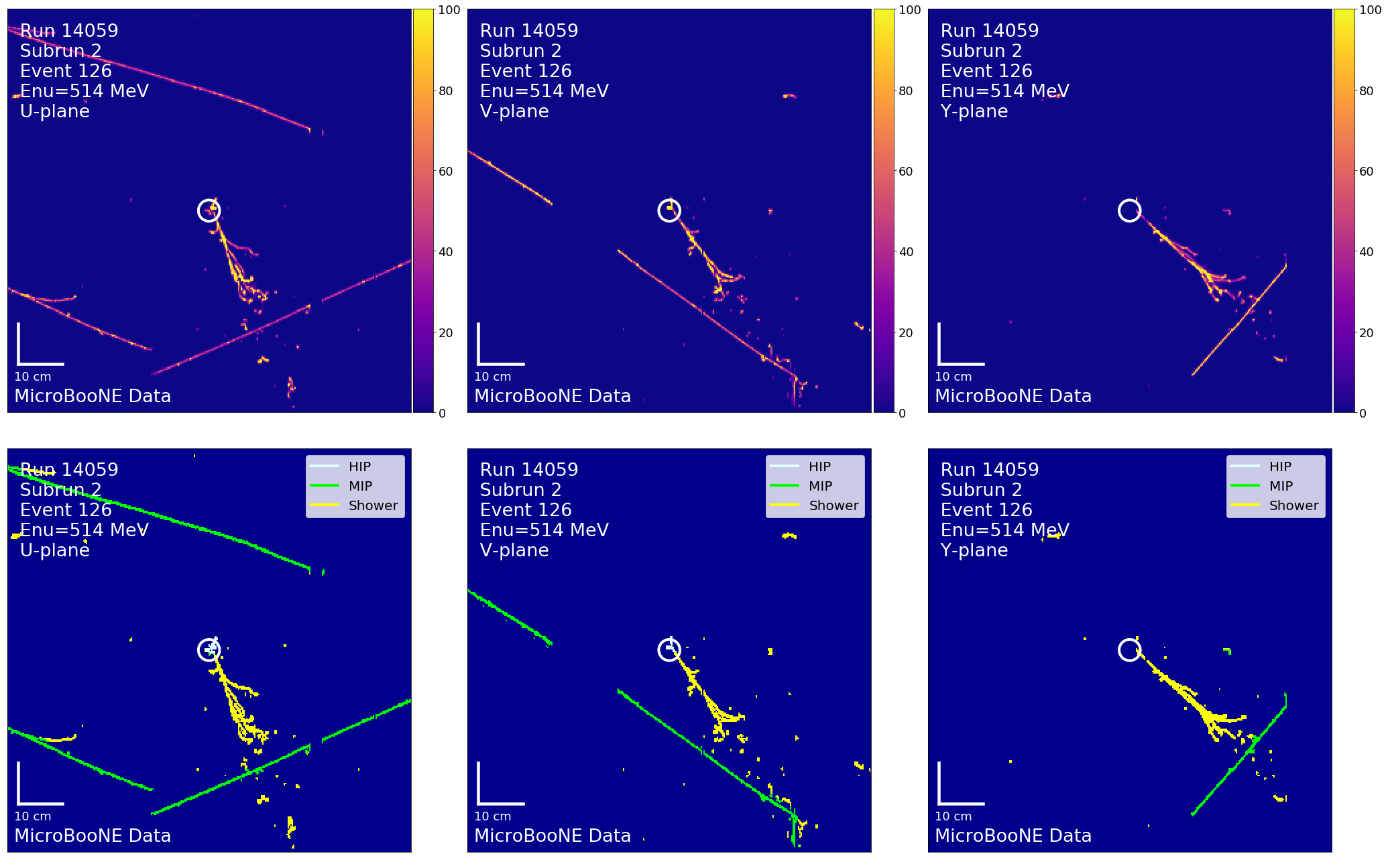}
    \caption{Top: pixel intensity; Bottom: \texttt{SparseSSNet} labels; Left to right: U, V, Y, planes. The white circle indicates the reconstructed vertex.}
    \label{fig:evd7}
\end{figure*}

\begin{figure*}
    \centering
    \includegraphics[width=0.92\linewidth]{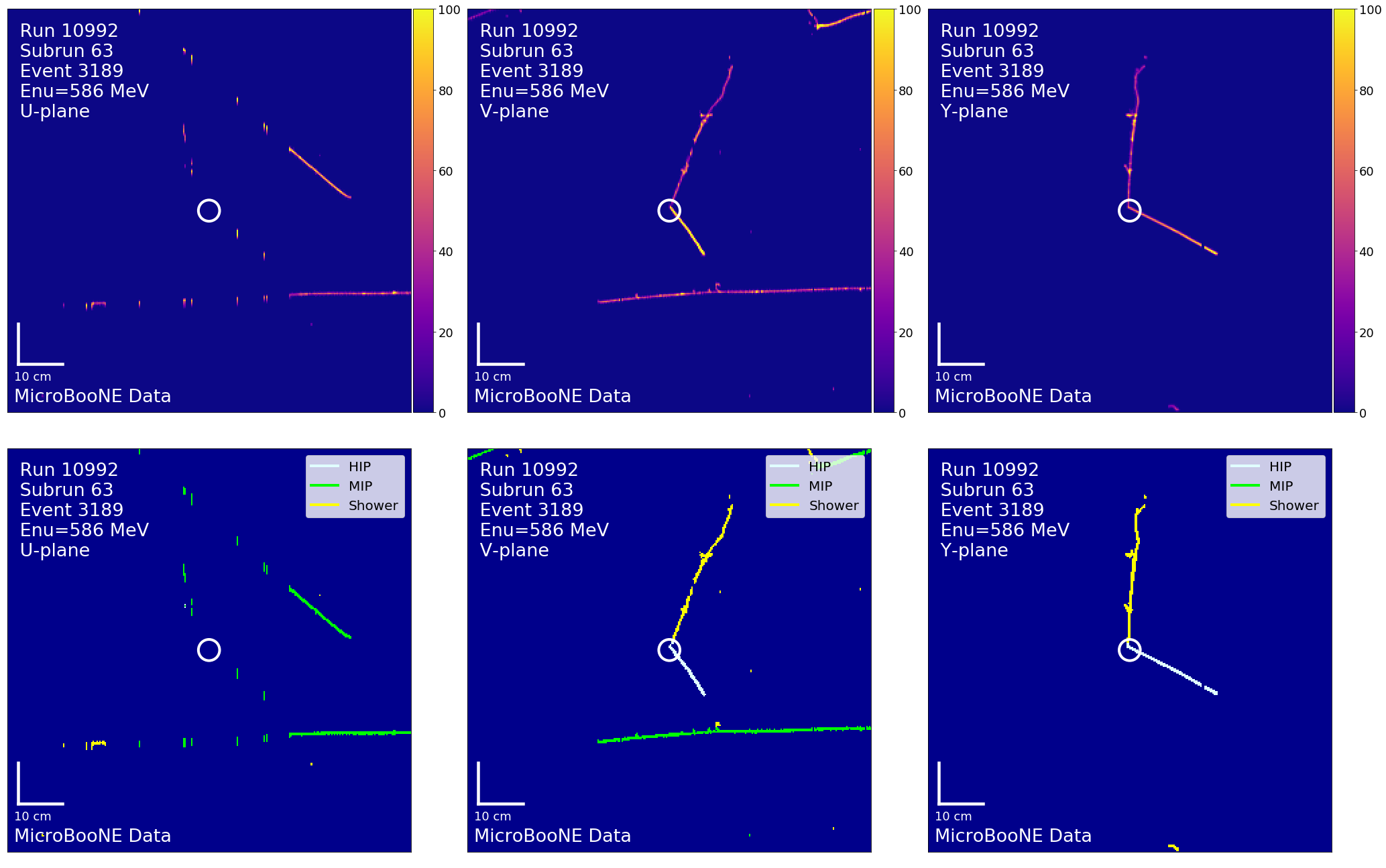}
    \caption{Top: pixel intensity; Bottom: \texttt{SparseSSNet} labels; Left to right: U, V, Y, planes. The white circle indicates the reconstructed vertex.}
    \label{fig:evd8}
\end{figure*}

\begin{figure*}
    \centering
    \includegraphics[width=0.92\linewidth]{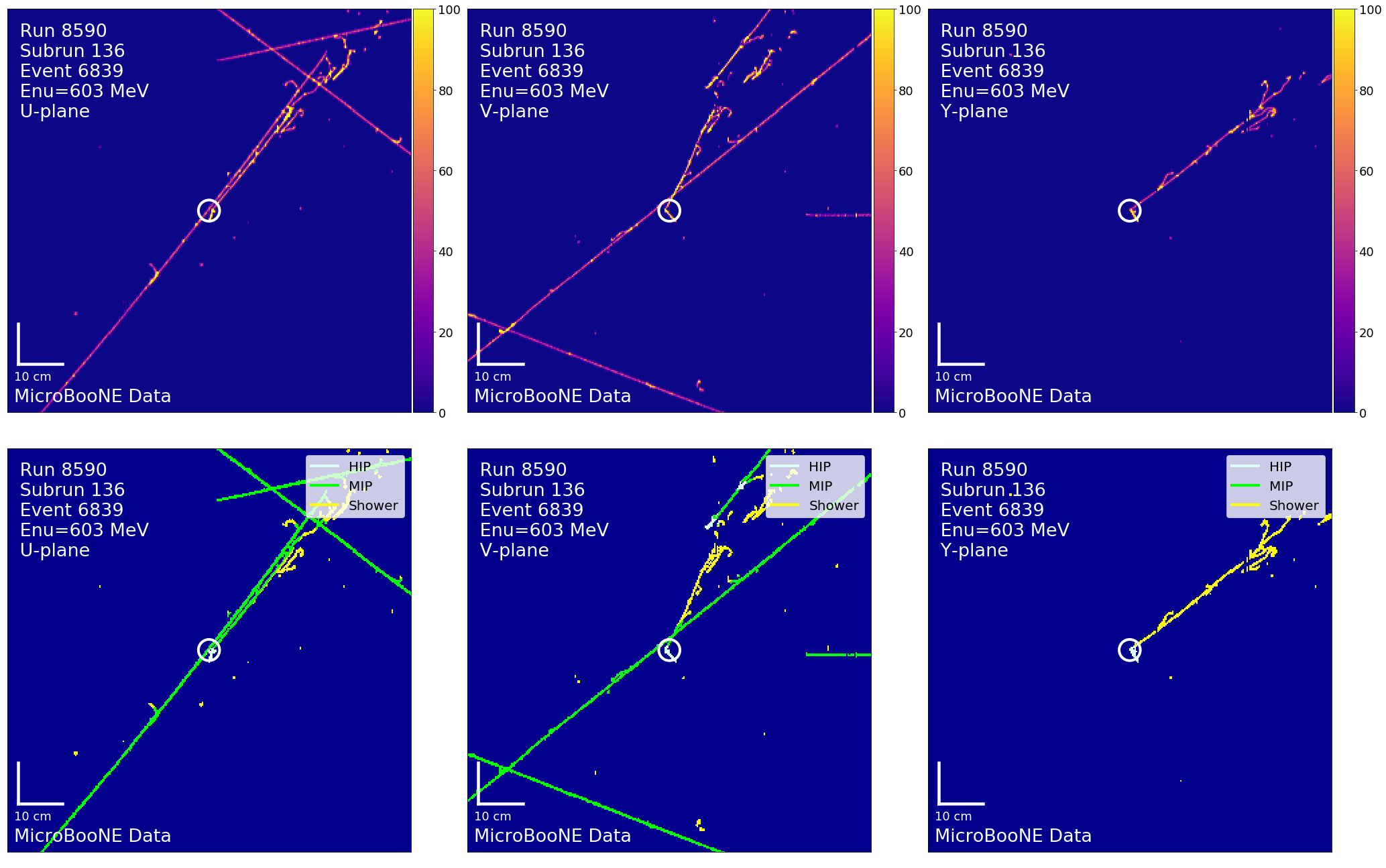}
    \caption{Top: pixel intensity; Bottom: \texttt{SparseSSNet} labels; Left to right: U, V, Y, planes. The white circle indicates the reconstructed vertex.}
    \label{fig:evd9}
\end{figure*}

\begin{figure*}
    \centering
    \includegraphics[width=0.92\linewidth]{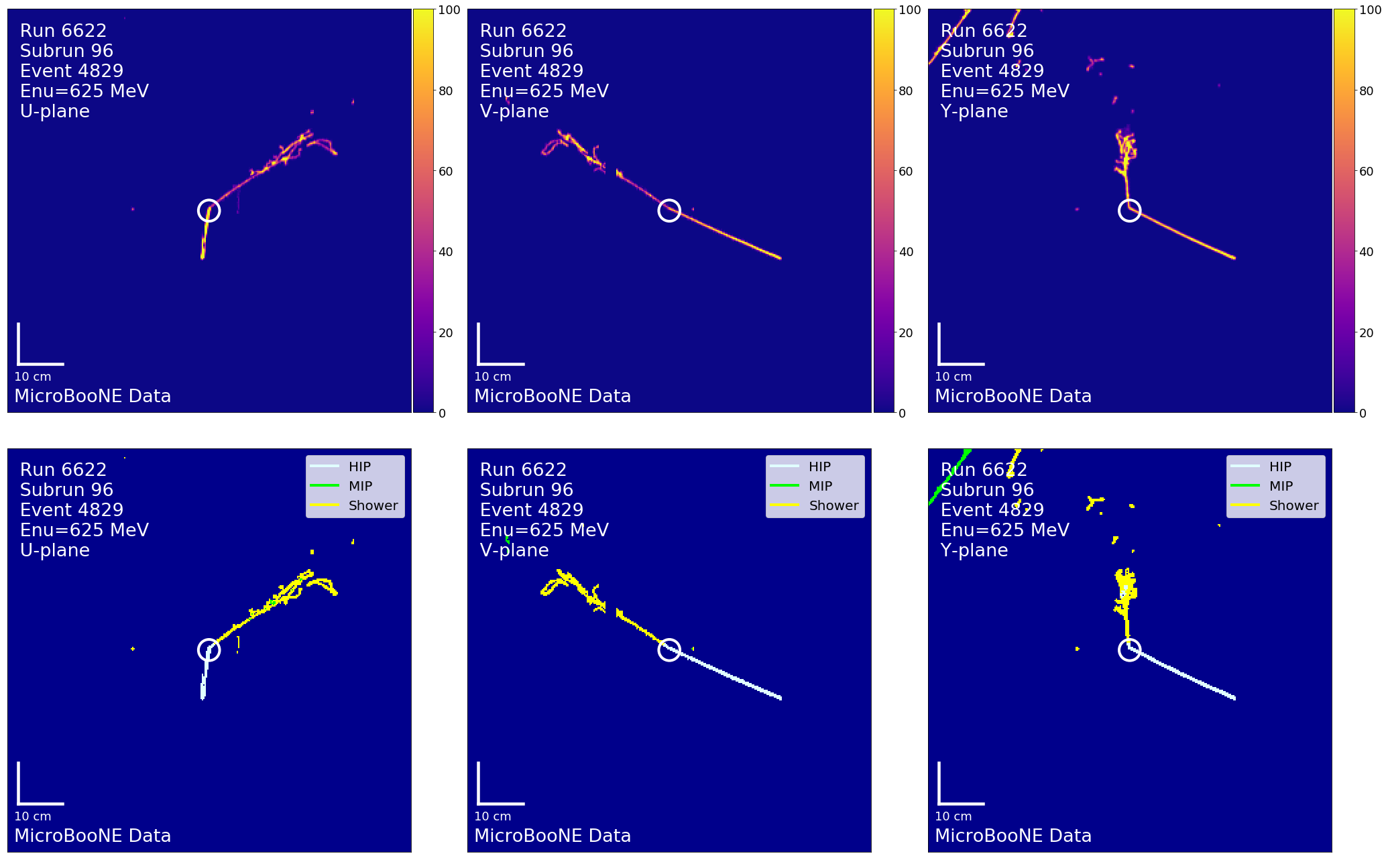}
    \caption{Top: pixel intensity; Bottom: \texttt{SparseSSNet} labels; Left to right: U, V, Y, planes. The white circle indicates the reconstructed vertex.}
    \label{fig:evd10}
\end{figure*}

\begin{figure*}
    \centering
    \includegraphics[width=0.92\linewidth]{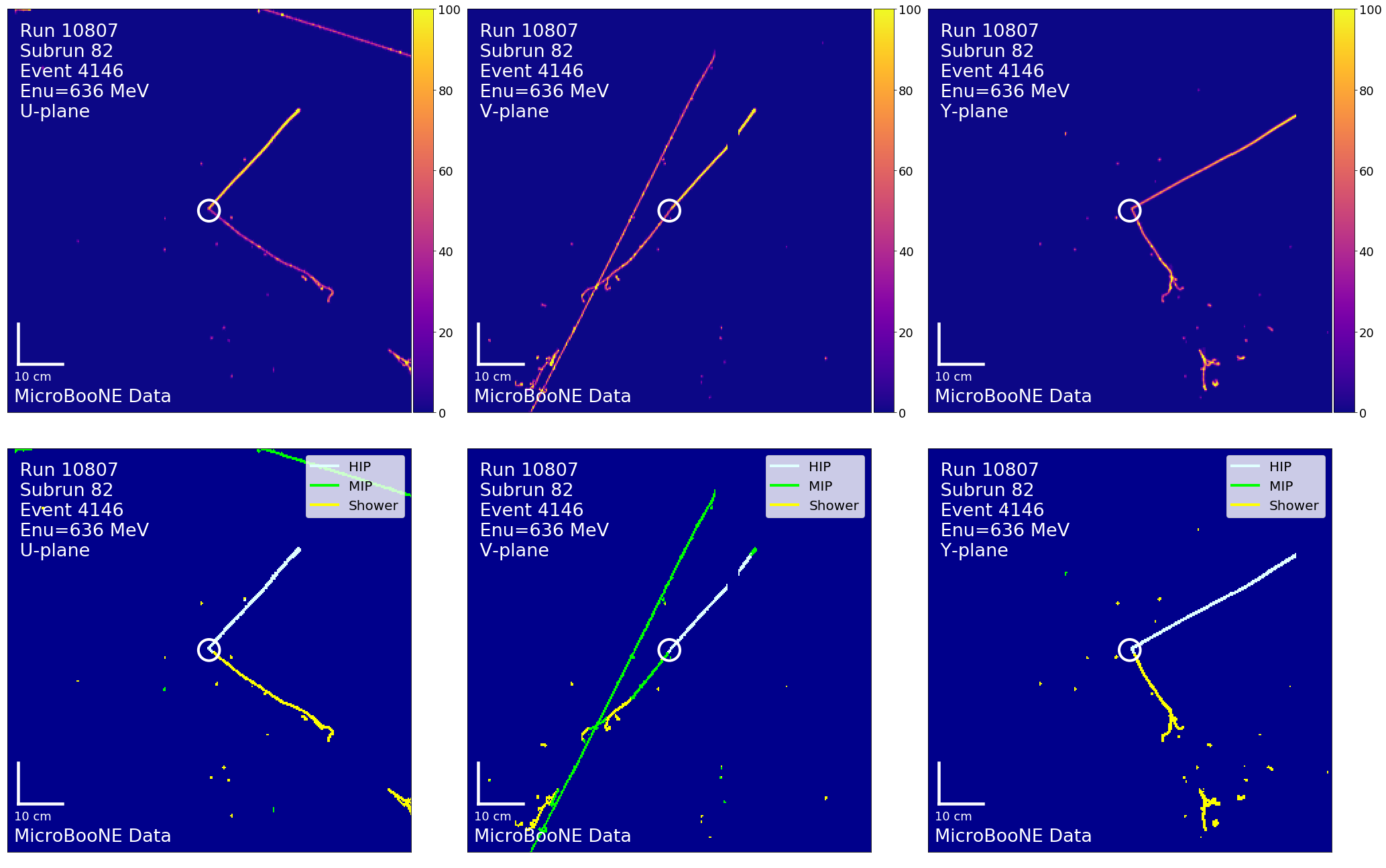}
    \caption{Top: pixel intensity; Bottom: \texttt{SparseSSNet} labels; Left to right: U, V, Y, planes. The white circle indicates the reconstructed vertex.}
    \label{fig:evd11}
\end{figure*}

\begin{figure*}
    \centering
    \includegraphics[width=0.92\linewidth]{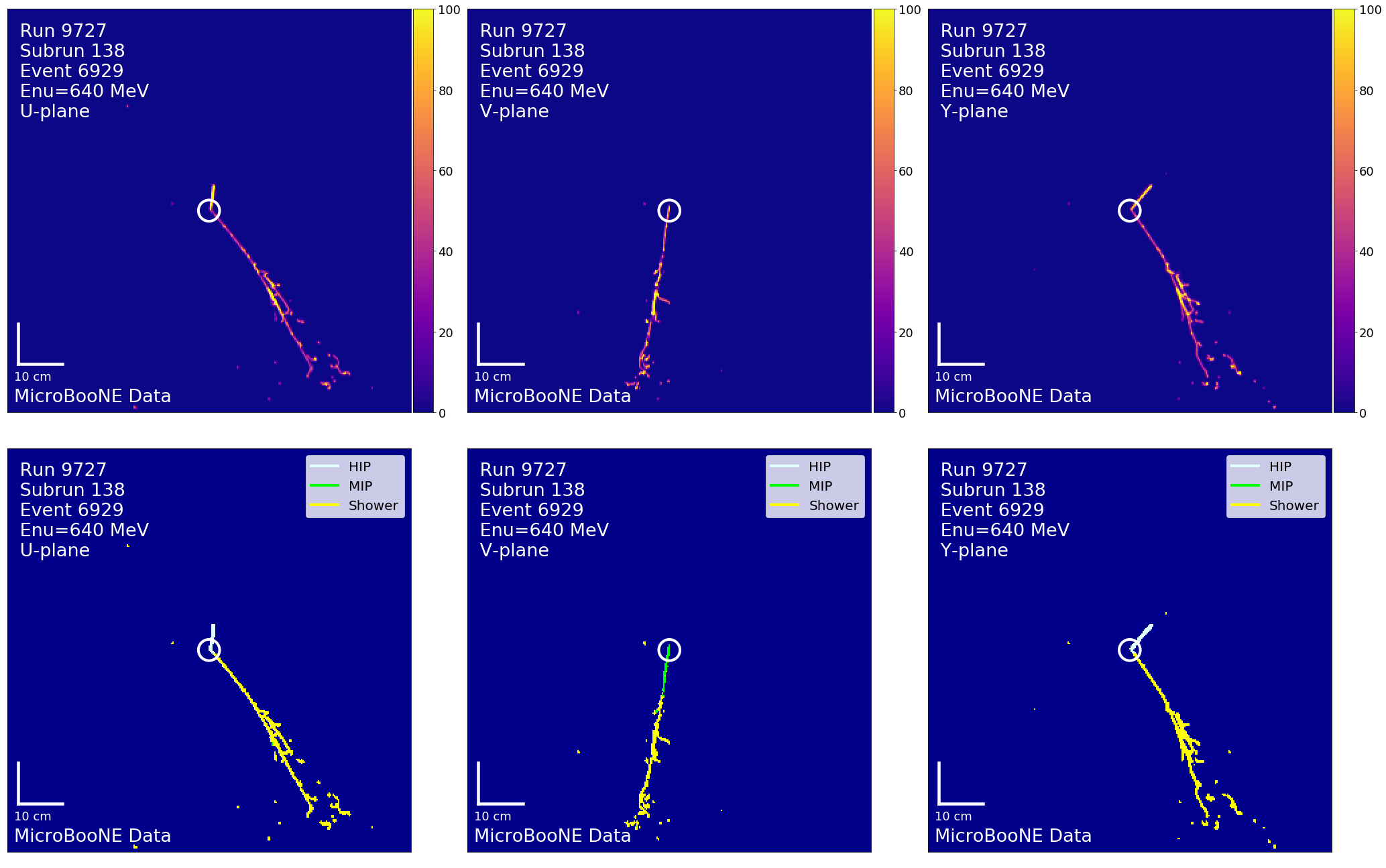}
    \caption{Top: pixel intensity; Bottom: \texttt{SparseSSNet} labels; Left to right: U, V, Y, planes. The white circle indicates the reconstructed vertex.}
    \label{fig:evd12}
\end{figure*}

\begin{figure*}
    \centering
    \includegraphics[width=0.92\linewidth]{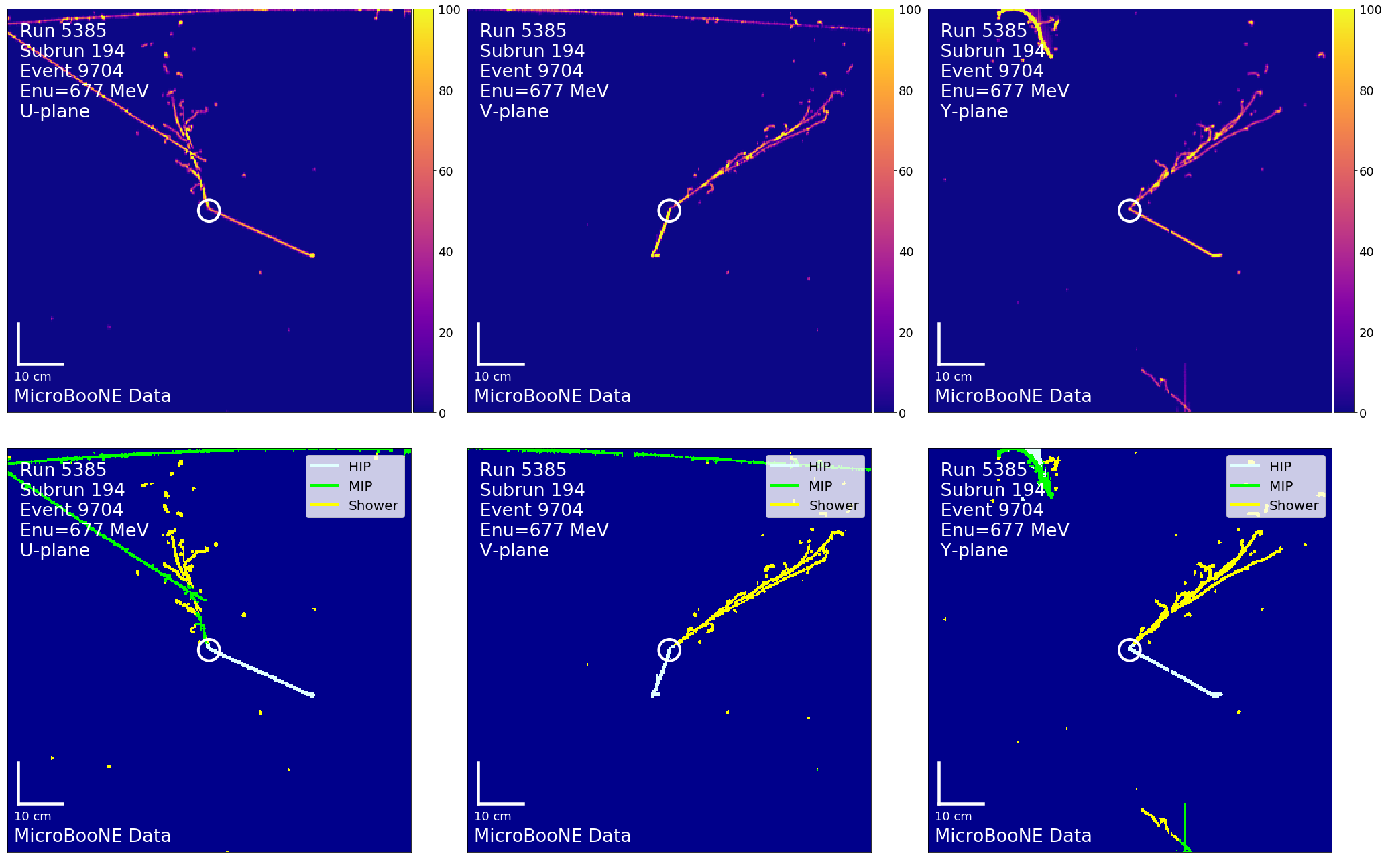}
    \caption{Top: pixel intensity; Bottom: \texttt{SparseSSNet} labels; Left to right: U, V, Y, planes. The white circle indicates the reconstructed vertex.}
    \label{fig:evd13}
\end{figure*}

\begin{figure*}
    \centering
    \includegraphics[width=0.92\linewidth]{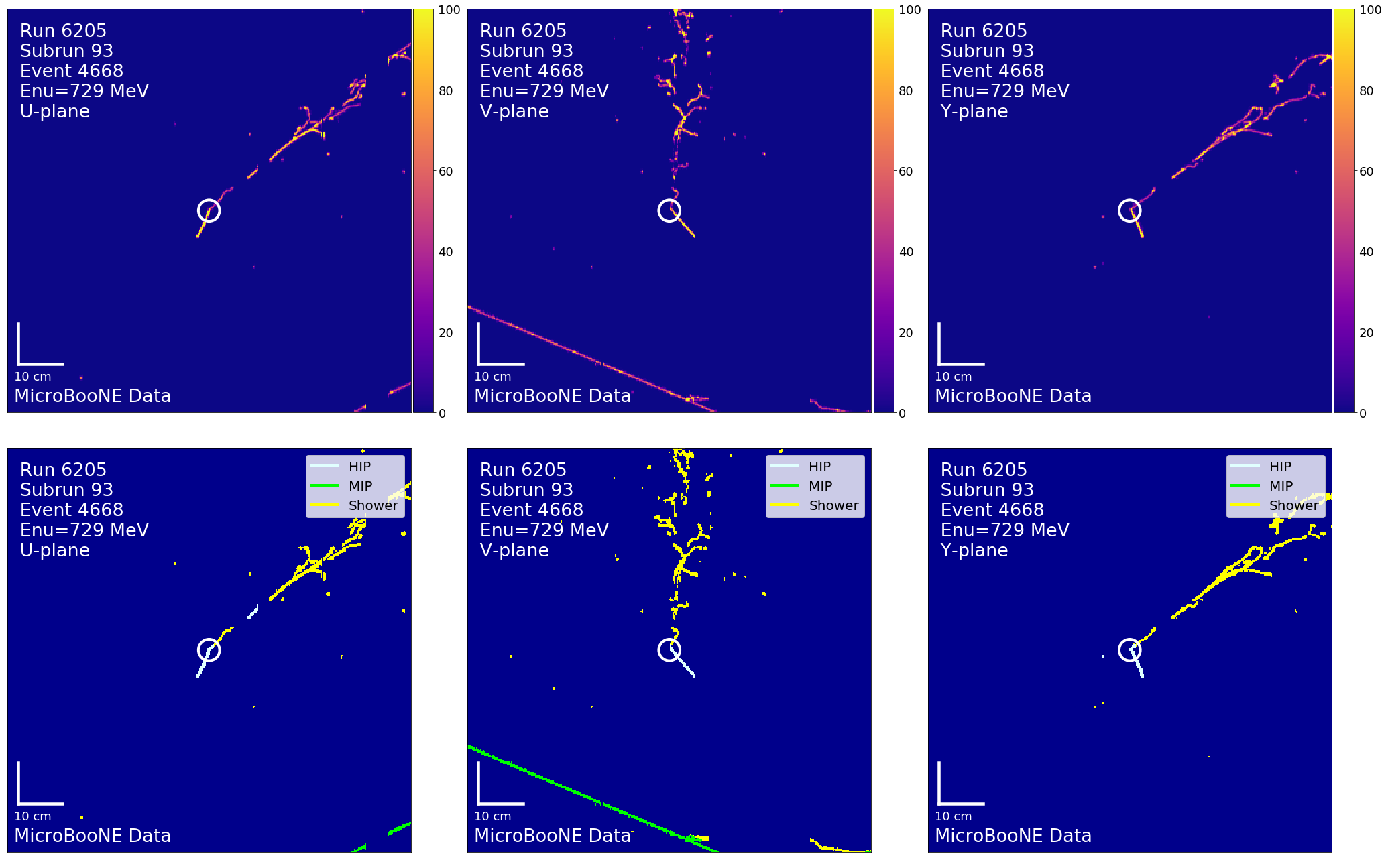}
    \caption{Top: pixel intensity; Bottom: \texttt{SparseSSNet} labels; Left to right: U, V, Y, planes. The white circle indicates the reconstructed vertex.}
    \label{fig:evd14}
\end{figure*}

\begin{figure*}
    \centering
    \includegraphics[width=0.92\linewidth]{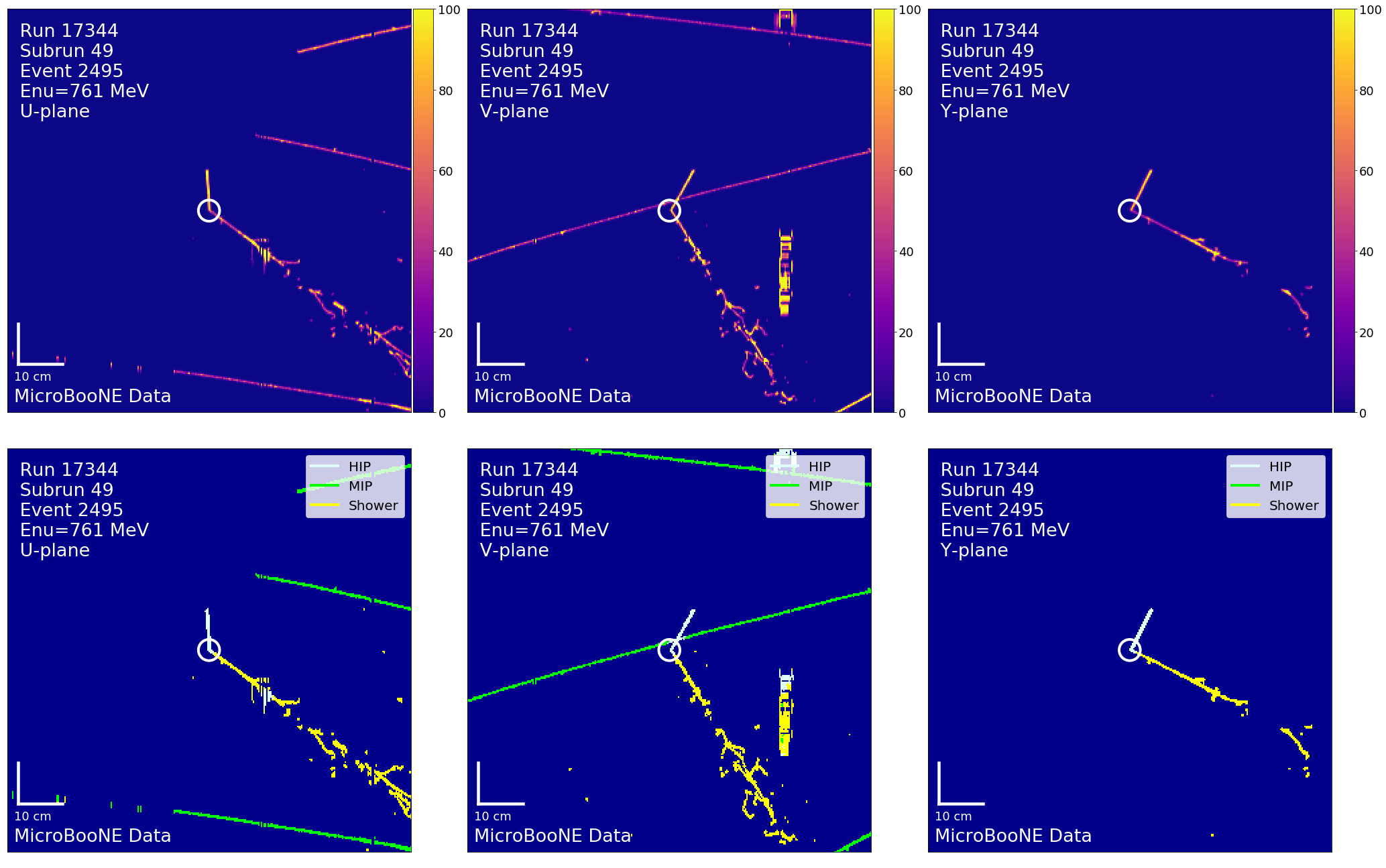}
    \caption{Top: pixel intensity; Bottom: \texttt{SparseSSNet} labels; Left to right: U, V, Y, planes. The white circle indicates the reconstructed vertex.}
    \label{fig:evd15}
\end{figure*}

\begin{figure*}
    \centering
    \includegraphics[width=0.92\linewidth]{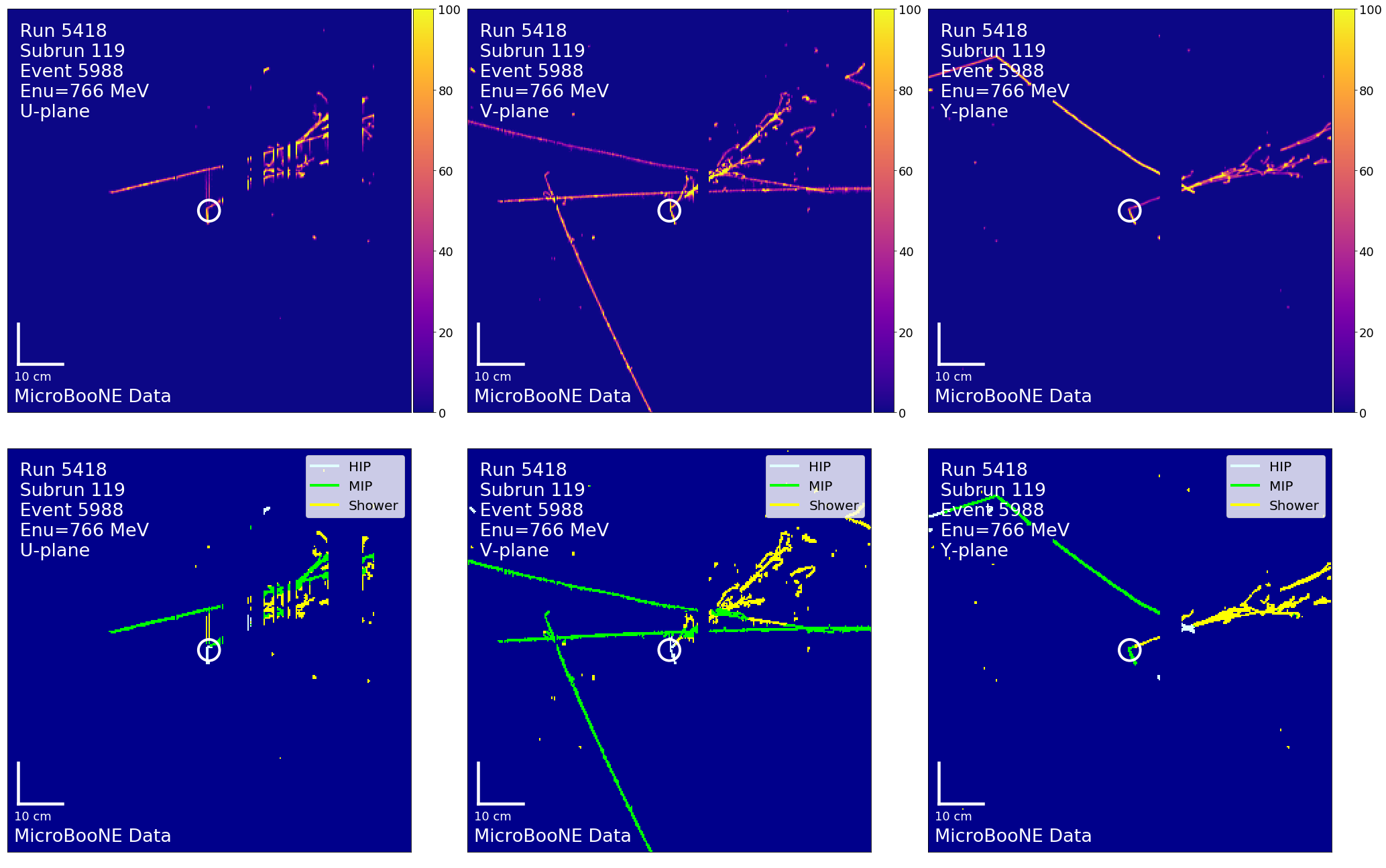}
    \caption{Top: pixel intensity; Bottom: \texttt{SparseSSNet} labels; Left to right: U, V, Y, planes. The white circle indicates the reconstructed vertex.}
    \label{fig:evd16}
\end{figure*}

\begin{figure*}
    \centering
    \includegraphics[width=0.92\linewidth]{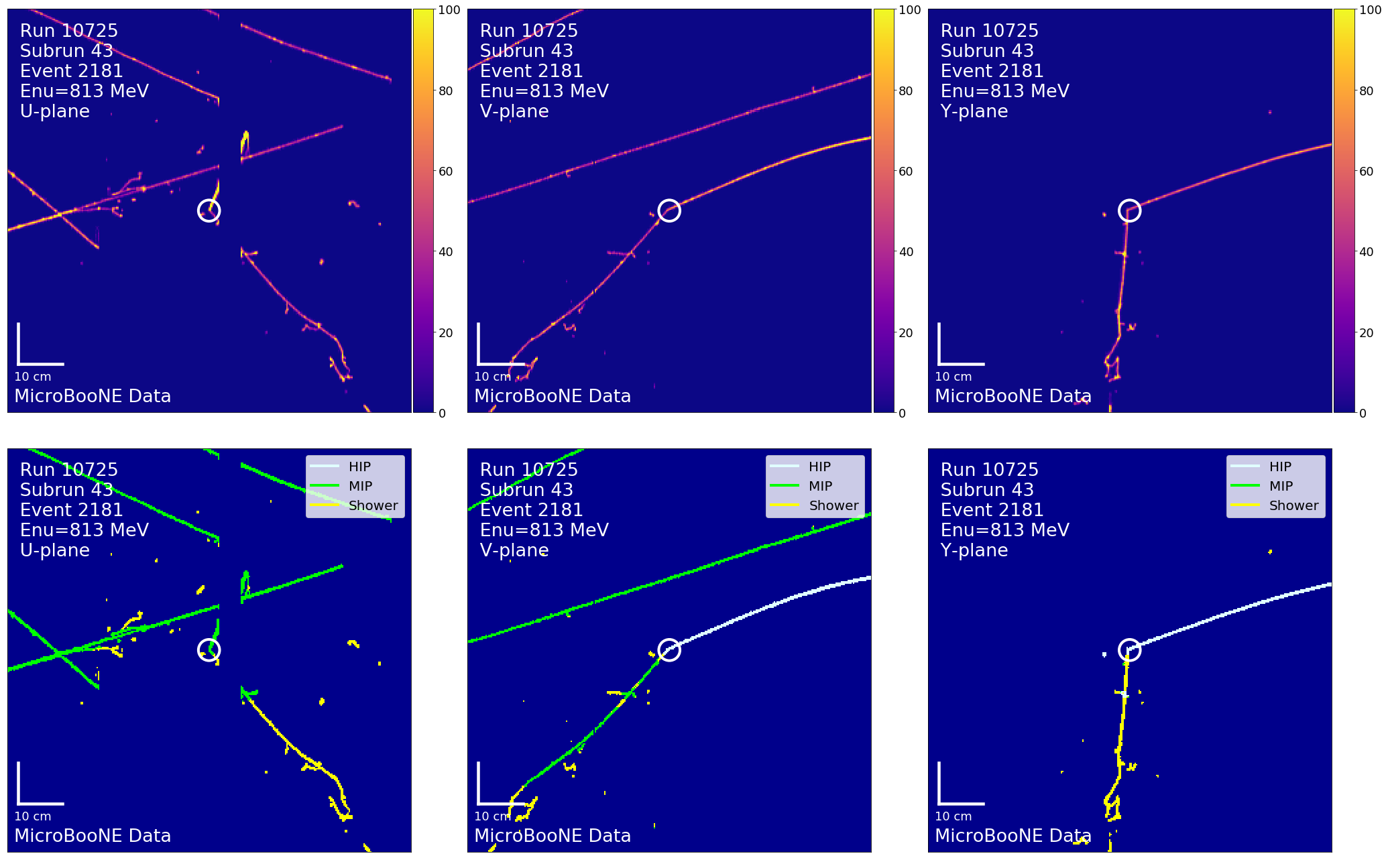}
    \caption{Top: pixel intensity; Bottom: \texttt{SparseSSNet} labels; Left to right: U, V, Y, planes. The white circle indicates the reconstructed vertex.}
    \label{fig:evd17}
\end{figure*}

\begin{figure*}
    \centering
    \includegraphics[width=0.92\linewidth]{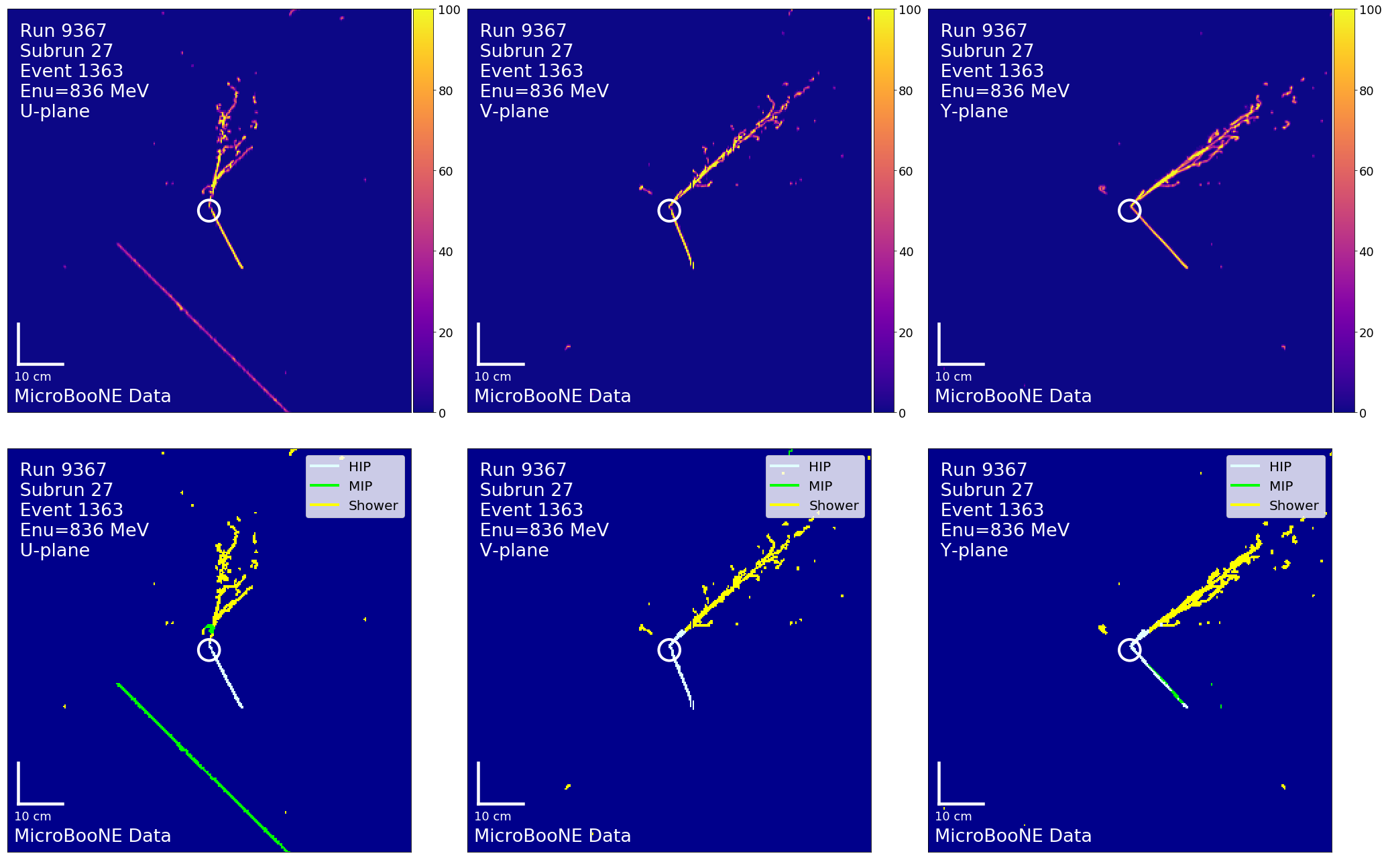}
    \caption{Top: pixel intensity; Bottom: \texttt{SparseSSNet} labels; Left to right: U, V, Y, planes. The white circle indicates the reconstructed vertex.}
    \label{fig:evd18}
\end{figure*}

\begin{figure*}
    \centering
    \includegraphics[width=0.92\linewidth]{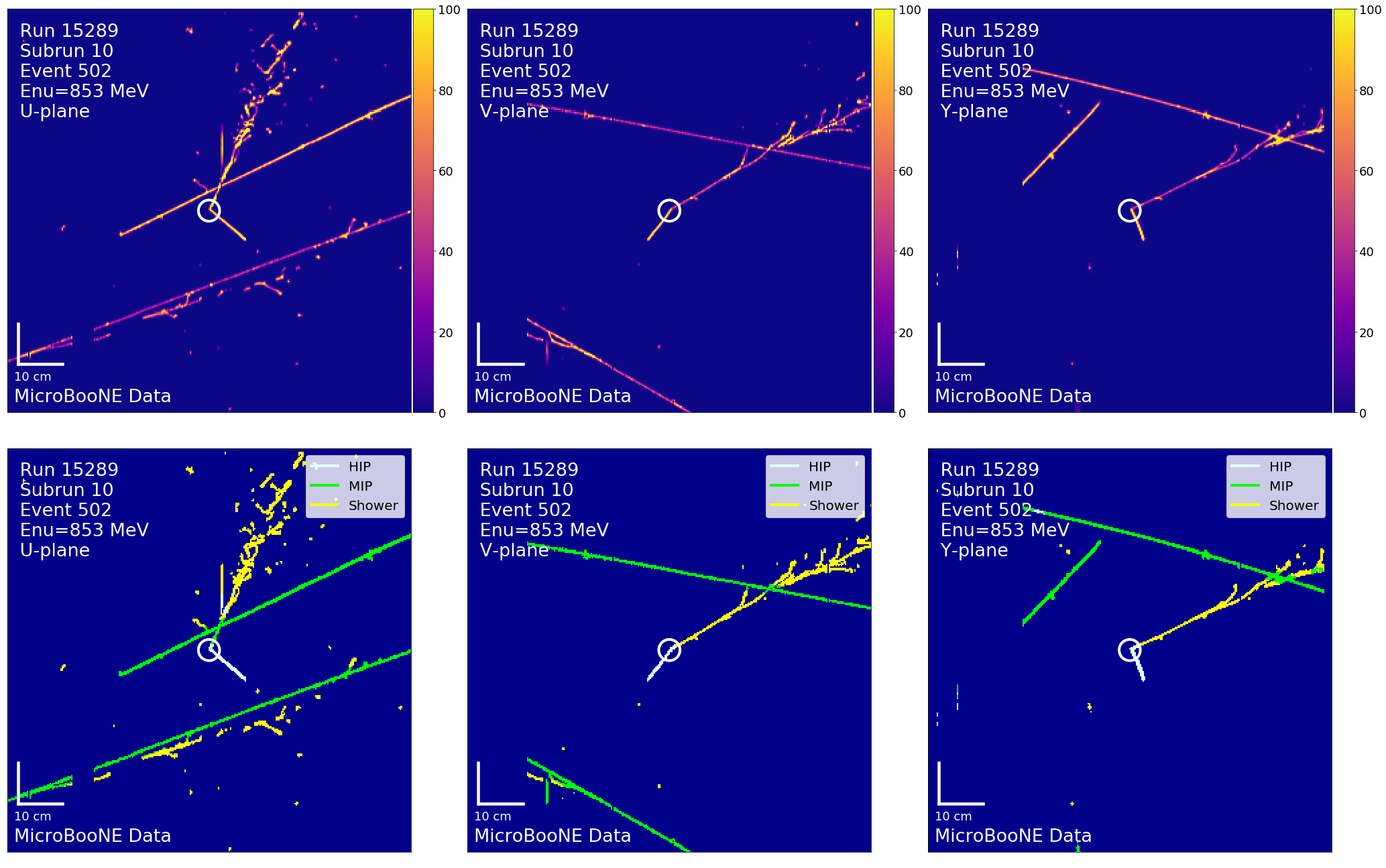}
    \caption{Top: pixel intensity; Bottom: \texttt{SparseSSNet} labels; Left to right: U, V, Y, planes. The white circle indicates the reconstructed vertex.}
    \label{fig:evd19}
\end{figure*}

\begin{figure*}
    \centering
    \includegraphics[width=0.92\linewidth]{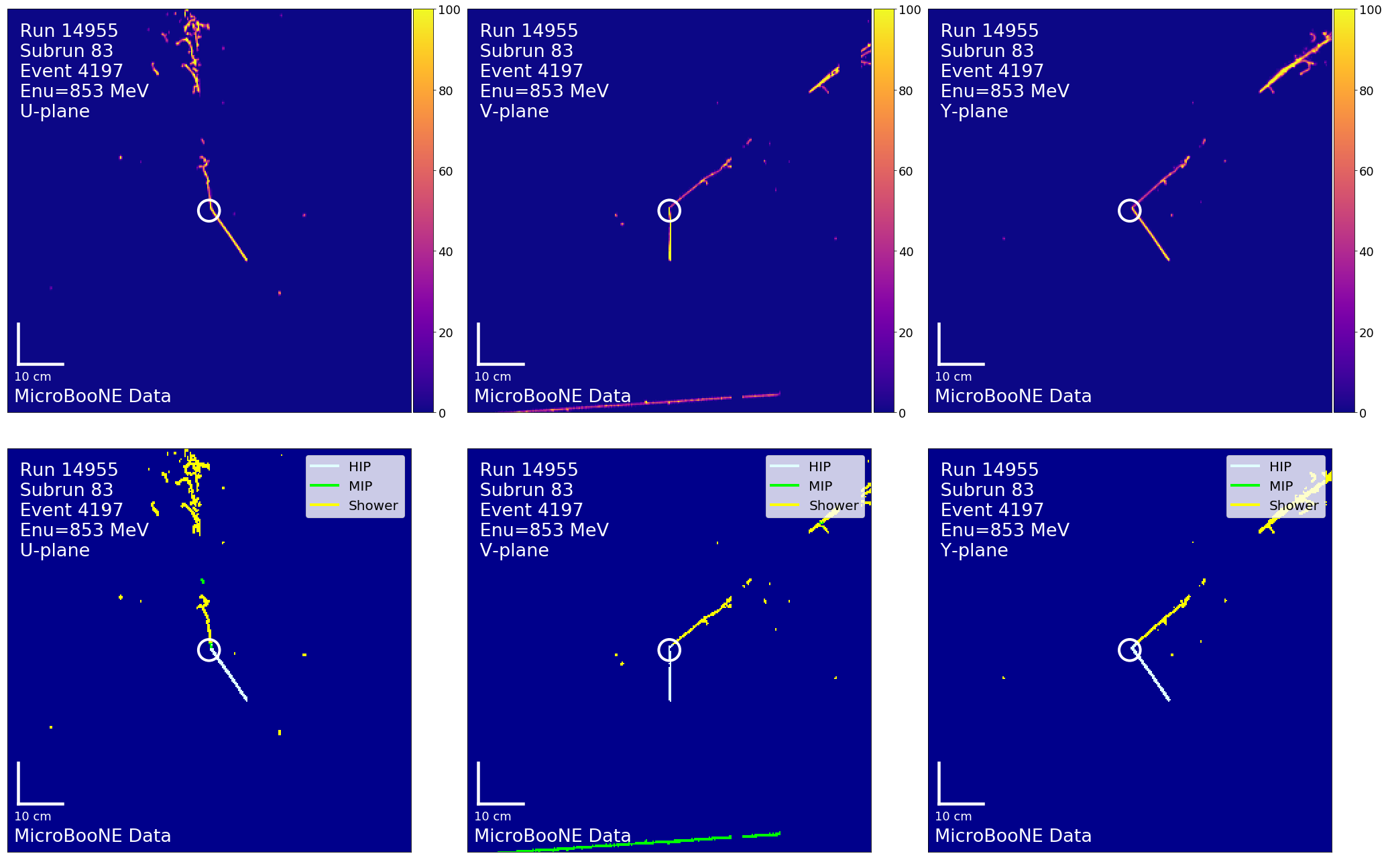}
    \caption{Top: pixel intensity; Bottom: \texttt{SparseSSNet} labels; Left to right: U, V, Y, planes. The white circle indicates the reconstructed vertex.}
    \label{fig:evd20}
\end{figure*}

\begin{figure*}
    \centering
    \includegraphics[width=0.92\linewidth]{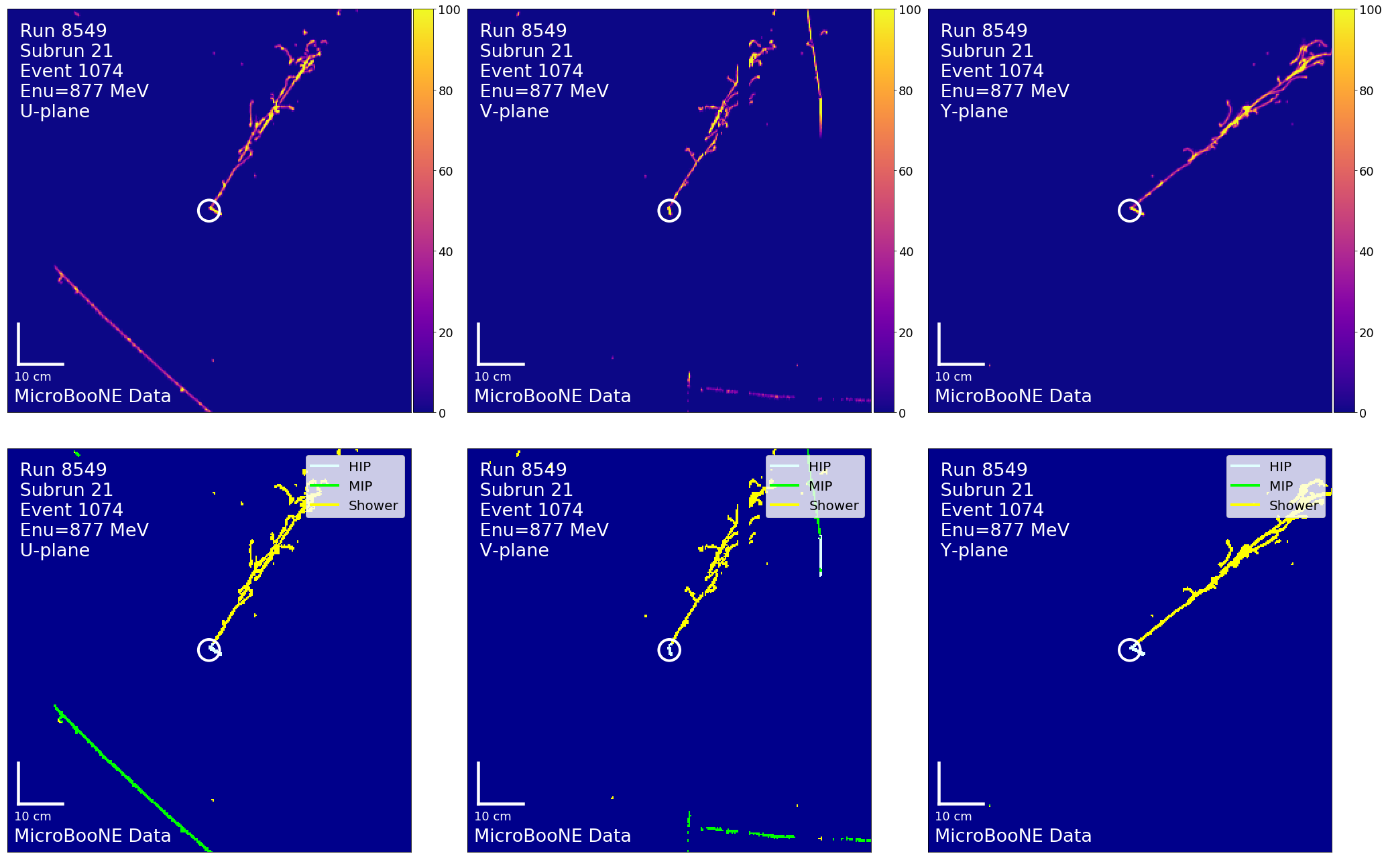}
    \caption{Top: pixel intensity; Bottom: \texttt{SparseSSNet} labels; Left to right: U, V, Y, planes. The white circle indicates the reconstructed vertex.}
    \label{fig:evd21}
\end{figure*}

\begin{figure*}
    \centering
    \includegraphics[width=0.92\linewidth]{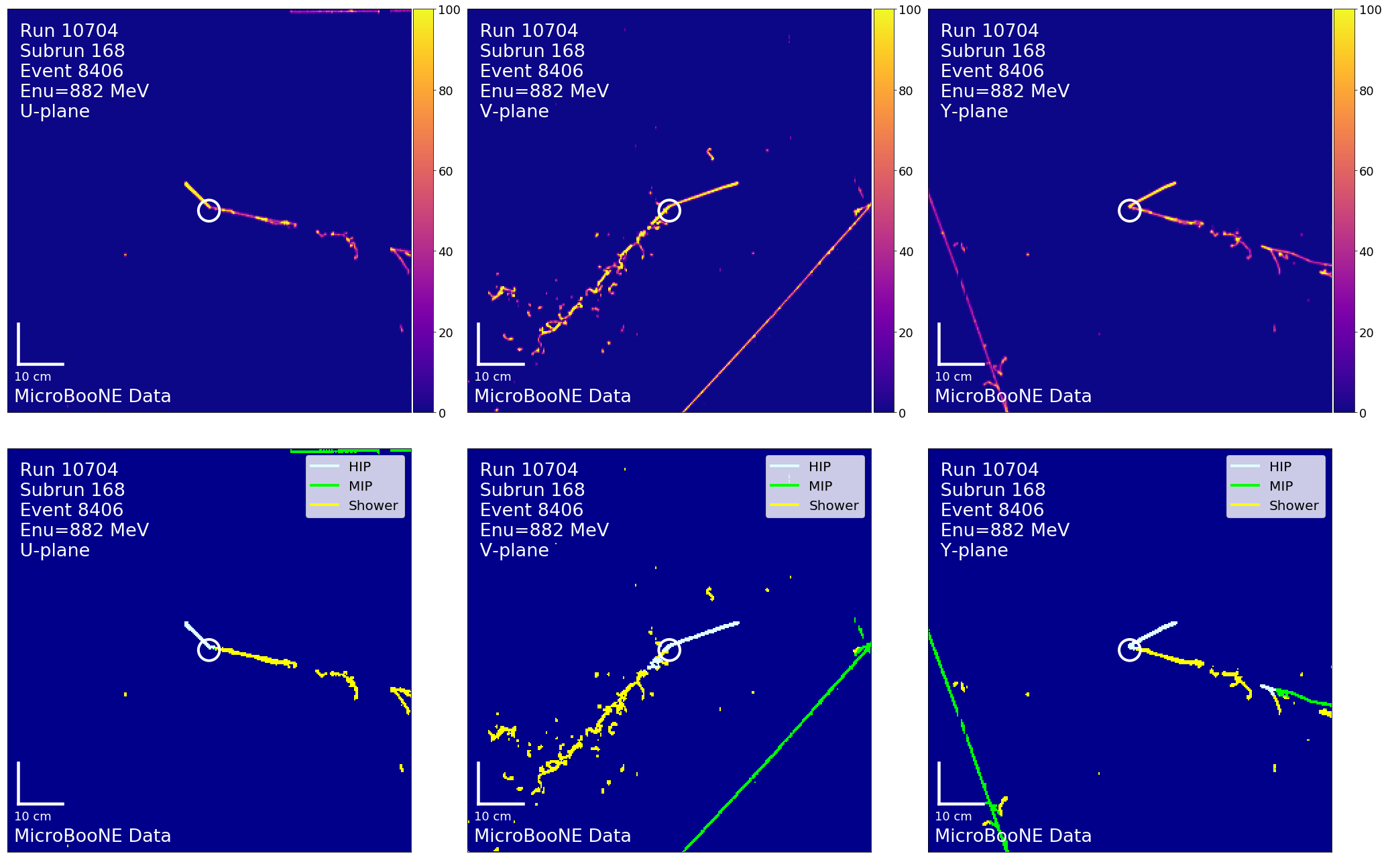}
    \caption{Top: pixel intensity; Bottom: \texttt{SparseSSNet} labels; Left to right: U, V, Y, planes. The white circle indicates the reconstructed vertex.}
    \label{fig:evd22}
\end{figure*}

\begin{figure*}
    \centering
    \includegraphics[width=0.92\linewidth]{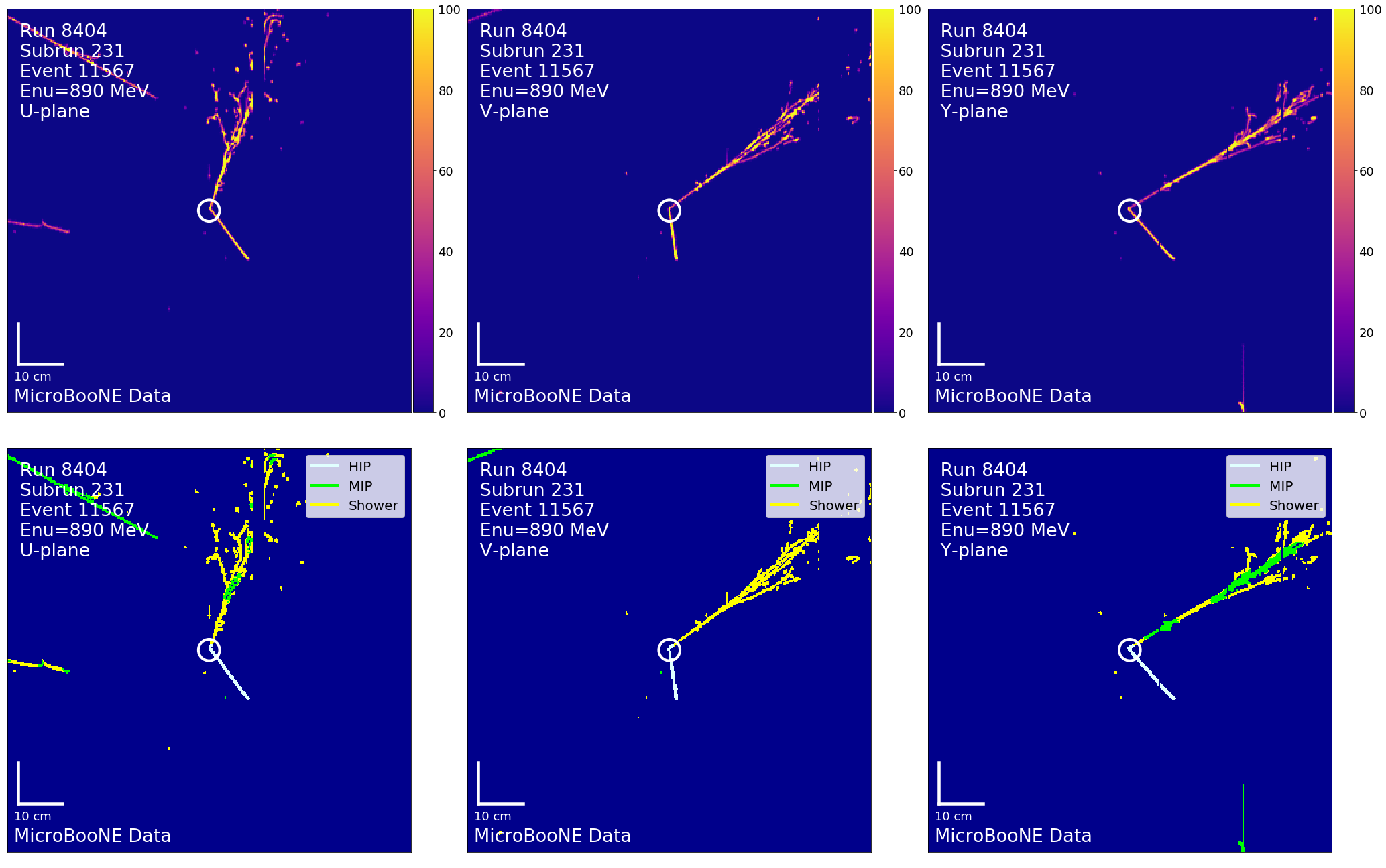}
    \caption{Top: pixel intensity; Bottom: \texttt{SparseSSNet} labels; Left to right: U, V, Y, planes. The white circle indicates the reconstructed vertex.}
    \label{fig:evd23}
\end{figure*}

\begin{figure*}
    \centering
    \includegraphics[width=0.92\linewidth]{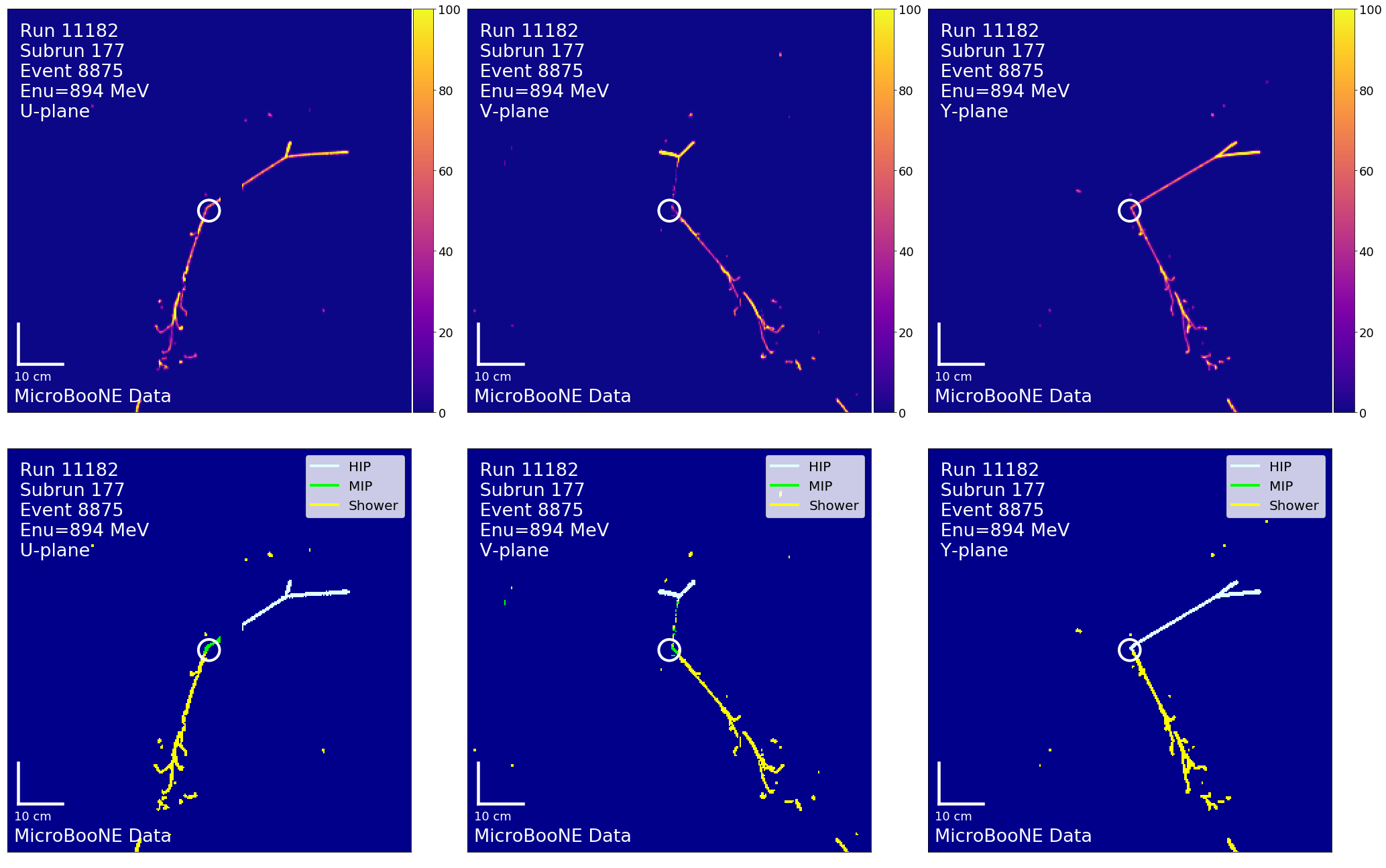}
    \caption{Top: pixel intensity; Bottom: \texttt{SparseSSNet} labels; Left to right: U, V, Y, planes. The white circle indicates the reconstructed vertex.}
    \label{fig:evd24}
\end{figure*}

\begin{figure*}
    \centering
    \includegraphics[width=0.92\linewidth]{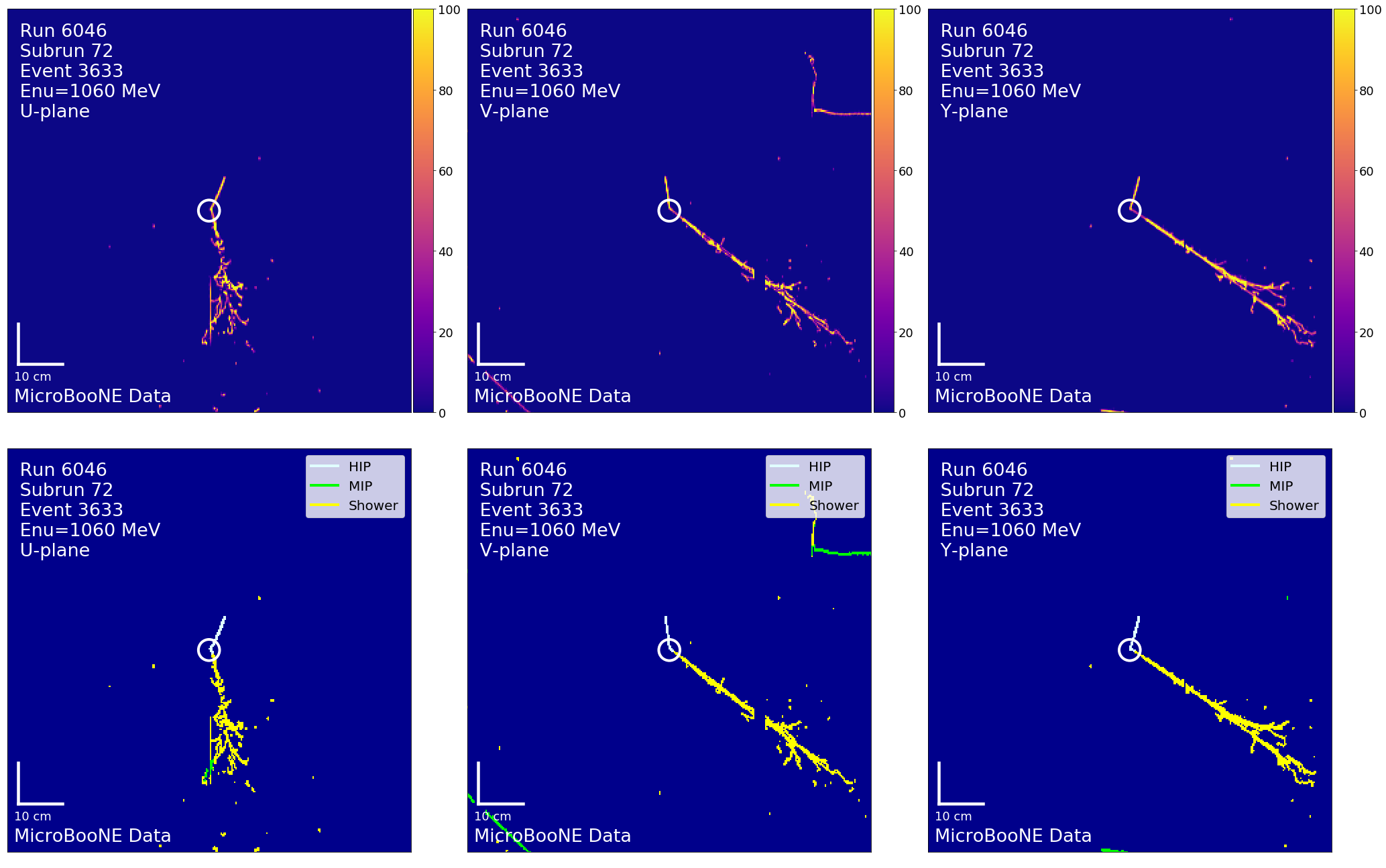}
    \caption{Top: pixel intensity; Bottom: \texttt{SparseSSNet} labels; Left to right: U, V, Y, planes. The white circle indicates the reconstructed vertex.}
    \label{fig:evd25}
\end{figure*}

\Cref{fig:evd1} in \cref{ch:microboone_results} shows the event display for one of the 25 $\nu_e$-candidate selected events in the $1e1p$ dataset.
We provide the remaining 24 event displays in \cref{fig:evd2} to \cref{fig:evd25}, in ascending order in energy.
The top row of each display shows the pixel intensity image, while the bottom row shows the \texttt{SparseSSNet} pixel labels.
From left to right, the columns correspond to the U,V, and Y plane images. 

\begin{singlespace}
\bibliography{main}

\begin{thebibliography}{100}

\bibitem{Bahcall:2004qv}
John~N. Bahcall.
\newblock {Solar models and solar neutrinos: Current status}.
\newblock {\em Phys. Scripta T}, 121:46--50, 2005.

\bibitem{Super-Kamiokande:1998kpq}
Y.~Fukuda et~al.
\newblock {Evidence for oscillation of atmospheric neutrinos}.
\newblock {\em Phys. Rev. Lett.}, 81:1562--1567, 1998.

\bibitem{SNO:2002tuh}
Q.~R. Ahmad et~al.
\newblock {Direct evidence for neutrino flavor transformation from neutral
  current interactions in the Sudbury Neutrino Observatory}.
\newblock {\em Phys. Rev. Lett.}, 89:011301, 2002.

\bibitem{Formaggio:2012cpf}
J.~A. Formaggio and G.~P. Zeller.
\newblock {From eV to EeV: Neutrino Cross Sections Across Energy Scales}.
\newblock {\em Rev. Mod. Phys.}, 84:1307--1341, 2012.

\bibitem{LSND:2001aii}
A.~Aguilar-Arevalo et~al.
\newblock {Evidence for neutrino oscillations from the observation of
  $\bar{\nu}_e$ appearance in a $\bar{\nu}_\mu$ beam}.
\newblock {\em Phys. Rev. D}, 64:112007, 2001.

\bibitem{Abazajian:2012ys}
K.~N. Abazajian et~al.
\newblock {Light Sterile Neutrinos: A White Paper}.
\newblock 4 2012.

\bibitem{Giunti:2022btk}
C.~Giunti, Y.~F. Li, C.~A. Ternes, O.~Tyagi, and Z.~Xin.
\newblock {Gallium Anomaly: critical view from the global picture of
  \ensuremath{\nu}$_{e}$ and $ {\overline{\nu}}_e $ disappearance}.
\newblock {\em JHEP}, 10:164, 2022.

\bibitem{Mention:2011rk}
G.~Mention, M.~Fechner, Th. Lasserre, Th.~A. Mueller, D.~Lhuillier, M.~Cribier,
  and A.~Letourneau.
\newblock {The Reactor Antineutrino Anomaly}.
\newblock {\em Phys. Rev. D}, 83:073006, 2011.

\bibitem{Barinov:2021asz}
V.~V. Barinov et~al.
\newblock {Results from the Baksan Experiment on Sterile Transitions (BEST)}.
\newblock {\em Phys. Rev. Lett.}, 128(23):232501, 2022.

\bibitem{MiniBooNE:2020pnu}
A.~A. Aguilar-Arevalo et~al.
\newblock {Updated MiniBooNE neutrino oscillation results with increased data
  and new background studies}.
\newblock {\em Phys. Rev. D}, 103(5):052002, 2021.

\bibitem{Dentler:2018sju}
Mona Dentler, \'Alvaro Hern\'andez-Cabezudo, Joachim Kopp, Pedro A.~N. Machado,
  Michele Maltoni, Ivan Martinez-Soler, and Thomas Schwetz.
\newblock {Updated Global Analysis of Neutrino Oscillations in the Presence of
  eV-Scale Sterile Neutrinos}.
\newblock {\em JHEP}, 08:010, 2018.

\bibitem{Hardin:2022muu}
J.~M. Hardin, I.~Martinez-Soler, A.~Diaz, M.~Jin, N.~W. Kamp, C.~A.
  Arg\"uelles, J.~M. Conrad, and M.~H. Shaevitz.
\newblock {New Clues About Light Sterile Neutrinos: Preference for Models with
  Damping Effects in Global Fits}.
\newblock 11 2022.

\bibitem{MiniBooNEDM:2018cxm}
A.~A. Aguilar-Arevalo et~al.
\newblock {Dark Matter Search in Nucleon, Pion, and Electron Channels from a
  Proton Beam Dump with MiniBooNE}.
\newblock {\em Phys. Rev. D}, 98(11):112004, 2018.

\bibitem{MiniBooNE:2008hfu}
A.~A. Aguilar-Arevalo et~al.
\newblock {The Neutrino Flux prediction at MiniBooNE}.
\newblock {\em Phys. Rev. D}, 79:072002, 2009.

\bibitem{MiniBooNE:2008paa}
A.~A. Aguilar-Arevalo et~al.
\newblock {The MiniBooNE Detector}.
\newblock {\em Nucl. Instrum. Meth. A}, 599:28--46, 2009.

\bibitem{MicroBooNE:2016pwy}
R.~Acciarri et~al.
\newblock {Design and Construction of the MicroBooNE Detector}.
\newblock {\em JINST}, 12(02):P02017, 2017.

\bibitem{Jones:2015bya}
Benjamin J.~P. Jones.
\newblock {\em {Sterile Neutrinos in Cold Climates}}.
\newblock PhD thesis, MIT, 2015.

\bibitem{MicroBooNE:2017qiu}
R.~Acciarri et~al.
\newblock {Noise Characterization and Filtering in the MicroBooNE Liquid Argon
  TPC}.
\newblock {\em JINST}, 12(08):P08003, 2017.

\bibitem{MicroBooNE:2018swd}
C.~Adams et~al.
\newblock {Ionization electron signal processing in single phase LArTPCs. Part
  I. Algorithm Description and quantitative evaluation with MicroBooNE
  simulation}.
\newblock {\em JINST}, 13(07):P07006, 2018.

\bibitem{osti_1573217}
None None.
\newblock Microboone low-energy excess signal prediction from unfolding
  miniboone monte-carlo and data.
\newblock 7 2018.

\bibitem{MicroBooNE:2020yze}
P.~Abratenko et~al.
\newblock {Semantic segmentation with a sparse convolutional neural network for
  event reconstruction in MicroBooNE}.
\newblock {\em Phys. Rev. D}, 103(5):052012, 2021.

\bibitem{MicroBooNE:2020hho}
P.~Abratenko et~al.
\newblock {Convolutional neural network for multiple particle identification in
  the MicroBooNE liquid argon time projection chamber}.
\newblock {\em Phys. Rev. D}, 103(9):092003, 2021.

\bibitem{Carloni:2021zbc}
Kiara Carloni, Nicholas~W. Kamp, Austin Schneider, and Janet~M. Conrad.
\newblock {Convolutional neural networks for shower energy prediction in liquid
  argon time projection chambers}.
\newblock {\em JINST}, 17(02):P02022, 2022.

\bibitem{MicroBooNE:2020sar}
P.~Abratenko et~al.
\newblock {Vertex-finding and reconstruction of contained two-track neutrino
  events in the MicroBooNE detector}.
\newblock {\em JINST}, 16(02):P02017, 2021.

\bibitem{Feldman:1997qc}
Gary~J. Feldman and Robert~D. Cousins.
\newblock {A Unified approach to the classical statistical analysis of small
  signals}.
\newblock {\em Phys. Rev. D}, 57:3873--3889, 1998.

\bibitem{Wilks:1938dza}
S.~S. Wilks.
\newblock {The Large-Sample Distribution of the Likelihood Ratio for Testing
  Composite Hypotheses}.
\newblock {\em Annals Math. Statist.}, 9(1):60--62, 1938.

\bibitem{Vergani:2021tgc}
Stefano Vergani, Nicholas~W. Kamp, Alejandro Diaz, Carlos~A. Arg\"uelles,
  Janet~M. Conrad, Michael~H. Shaevitz, and Melissa~A. Uchida.
\newblock {Explaining the MiniBooNE excess through a mixed model of neutrino
  oscillation and decay}.
\newblock {\em Phys. Rev. D}, 104(9):095005, 2021.

\bibitem{ZAVORKA2018189}
Lukas Zavorka, Michael~J. Mocko, and Paul~E. Koehler.
\newblock Physics design of the next-generation spallation neutron
  target-moderator-reflector-shield assembly at lansce.
\newblock {\em Nuclear Instruments and Methods in Physics Research Section A:
  Accelerators, Spectrometers, Detectors and Associated Equipment},
  901:189--197, 2018.

\bibitem{Baxter:2019mcx}
D.~Baxter et~al.
\newblock {Coherent Elastic Neutrino-Nucleus Scattering at the European
  Spallation Source}.
\newblock {\em JHEP}, 02:123, 2020.

\bibitem{CCM:2021leg}
A.~A. Aguilar-Arevalo et~al.
\newblock {First dark matter search results from Coherent CAPTAIN-Mills}.
\newblock {\em Phys. Rev. D}, 106(1):012001, 2022.

\bibitem{Kamp:2022bpt}
Nicholas~W. Kamp, Matheus Hostert, Austin Schneider, Stefano Vergani, Carlos~A.
  Arg\"uelles, Janet~M. Conrad, Michael~H. Shaevitz, and Melissa~A. Uchida.
\newblock {Dipole-coupled heavy-neutral-lepton explanations of the MiniBooNE
  excess including constraints from MINERvA data}.
\newblock {\em Phys. Rev. D}, 107(5):055009, 2023.

\bibitem{EarthCrust}
F.~W. Clarke and H.~S. Washington.
\newblock The composition of the earth's crust.
\newblock Technical report, 1924.

\bibitem{Pauli:1930pc}
W.~Pauli.
\newblock {Dear radioactive ladies and gentlemen}.
\newblock {\em Phys. Today}, 31N9:27, 1978.

\bibitem{Fermi:1934hr}
E.~Fermi.
\newblock {An attempt of a theory of beta radiation. 1.}
\newblock {\em Z. Phys.}, 88:161--177, 1934.

\bibitem{Reines:1953pu}
F.~Reines and C.~L. Cowan.
\newblock {Detection of the free neutrino}.
\newblock {\em Phys. Rev.}, 92:830--831, 1953.

\bibitem{Cowan:1956rrn}
C.~L. Cowan, F.~Reines, F.~B. Harrison, H.~W. Kruse, and A.~D. McGuire.
\newblock {Detection of the free neutrino: A Confirmation}.
\newblock {\em Science}, 124:103--104, 1956.

\bibitem{Pontecorvo:1957qd}
B.~Pontecorvo.
\newblock {Inverse beta processes and nonconservation of lepton charge}.
\newblock {\em Zh. Eksp. Teor. Fiz.}, 34:247, 1957.

\bibitem{Danby:1962nd}
G.~Danby, J.~M. Gaillard, Konstantin~A. Goulianos, L.~M. Lederman, Nari~B.
  Mistry, M.~Schwartz, and J.~Steinberger.
\newblock {Observation of High-Energy Neutrino Reactions and the Existence of
  Two Kinds of Neutrinos}.
\newblock {\em Phys. Rev. Lett.}, 9:36--44, 1962.

\bibitem{Maki:1962mu}
Ziro Maki, Masami Nakagawa, and Shoichi Sakata.
\newblock {Remarks on the unified model of elementary particles}.
\newblock {\em Prog. Theor. Phys.}, 28:870--880, 1962.

\bibitem{Pontecorvo:1967fh}
B.~Pontecorvo.
\newblock {Neutrino Experiments and the Problem of Conservation of Leptonic
  Charge}.
\newblock {\em Zh. Eksp. Teor. Fiz.}, 53:1717--1725, 1967.

\bibitem{Bilenky:2016pep}
S.~Bilenky.
\newblock {Neutrino oscillations: From a historical perspective to the present
  status}.
\newblock {\em Nucl. Phys. B}, 908:2--13, 2016.

\bibitem{Davis:1968cp}
Raymond Davis, Jr., Don~S. Harmer, and Kenneth~C. Hoffman.
\newblock {Search for neutrinos from the sun}.
\newblock {\em Phys. Rev. Lett.}, 20:1205--1209, 1968.

\bibitem{Bahcall:1968hc}
John~N. Bahcall, Neta~A. Bahcall, and G.~Shaviv.
\newblock {Present status of the theoretical predictions for the Cl-36 solar
  neutrino experiment}.
\newblock {\em Phys. Rev. Lett.}, 20:1209--1212, 1968.

\bibitem{Casper:1990ac}
D.~Casper et~al.
\newblock {Measurement of atmospheric neutrino composition with IMB-3}.
\newblock {\em Phys. Rev. Lett.}, 66:2561--2564, 1991.

\bibitem{Kamiokande-II:1992hns}
K.~S. Hirata et~al.
\newblock {Observation of a small atmospheric muon-neutrino / electron-neutrino
  ratio in Kamiokande}.
\newblock {\em Phys. Lett. B}, 280:146--152, 1992.

\bibitem{Learned:2000qq}
John~G. Learned.
\newblock {The Atmospheric neutrino anomaly: Muon neutrino disappearance}.
\newblock pages 89--130, 7 2000.

\bibitem{Super-Kamiokande:2002weg}
Y.~Fukuda et~al.
\newblock {The Super-Kamiokande detector}.
\newblock {\em Nucl. Instrum. Meth. A}, 501:418--462, 2003.

\bibitem{SNO:1999crp}
J.~Boger et~al.
\newblock {The Sudbury neutrino observatory}.
\newblock {\em Nucl. Instrum. Meth. A}, 449:172--207, 2000.

\bibitem{Mikheyev:1985zog}
S.~P. Mikheyev and A.~Yu. Smirnov.
\newblock {Resonance Amplification of Oscillations in Matter and Spectroscopy
  of Solar Neutrinos}.
\newblock {\em Sov. J. Nucl. Phys.}, 42:913--917, 1985.

\bibitem{Wolfenstein:1977ue}
L.~Wolfenstein.
\newblock {Neutrino Oscillations in Matter}.
\newblock {\em Phys. Rev. D}, 17:2369--2374, 1978.

\bibitem{KamLAND:2004mhv}
T.~Araki et~al.
\newblock {Measurement of neutrino oscillation with KamLAND: Evidence of
  spectral distortion}.
\newblock {\em Phys. Rev. Lett.}, 94:081801, 2005.

\bibitem{McDonald:2016ixn}
Arthur~B. McDonald.
\newblock {Nobel Lecture: The Sudbury Neutrino Observatory: Observation of
  flavor change for solar neutrinos}.
\newblock {\em Rev. Mod. Phys.}, 88(3):030502, 2016.

\bibitem{Kajita:2016cak}
Takaaki Kajita.
\newblock {Nobel Lecture: Discovery of atmospheric neutrino oscillations}.
\newblock {\em Rev. Mod. Phys.}, 88(3):030501, 2016.

\bibitem{deSalas:2020pgw}
P.~F. de~Salas, D.~V. Forero, S.~Gariazzo, P.~Mart\'\i{}nez-Mirav\'e, O.~Mena,
  C.~A. Ternes, M.~T\'ortola, and J.~W.~F. Valle.
\newblock {2020 global reassessment of the neutrino oscillation picture}.
\newblock {\em JHEP}, 02:071, 2021.

\bibitem{Esteban:2020cvm}
Ivan Esteban, M.~C. Gonzalez-Garcia, Michele Maltoni, Thomas Schwetz, and
  Albert Zhou.
\newblock {The fate of hints: updated global analysis of three-flavor neutrino
  oscillations}.
\newblock {\em JHEP}, 09:178, 2020.

\bibitem{Capozzi:2021fjo}
Francesco Capozzi, Eleonora Di~Valentino, Eligio Lisi, Antonio Marrone,
  Alessandro Melchiorri, and Antonio Palazzo.
\newblock {Unfinished fabric of the three neutrino paradigm}.
\newblock {\em Phys. Rev. D}, 104(8):083031, 2021.

\bibitem{T2K:2019bcf}
K.~Abe et~al.
\newblock {Constraint on the matter\textendash{}antimatter symmetry-violating
  phase in neutrino oscillations}.
\newblock {\em Nature}, 580(7803):339--344, 2020.
\newblock [Erratum: Nature 583, E16 (2020)].

\bibitem{NOvA:2019cyt}
M.~A. Acero et~al.
\newblock {First Measurement of Neutrino Oscillation Parameters using Neutrinos
  and Antineutrinos by NOvA}.
\newblock {\em Phys. Rev. Lett.}, 123(15):151803, 2019.

\bibitem{Hyper-Kamiokande:2016srs}
K.~Abe et~al.
\newblock {Physics potentials with the second Hyper-Kamiokande detector in
  Korea}.
\newblock {\em PTEP}, 2018(6):063C01, 2018.

\bibitem{DUNE:2015lol}
R.~Acciarri et~al.
\newblock {Long-Baseline Neutrino Facility (LBNF) and Deep Underground Neutrino
  Experiment (DUNE)}: {Conceptual Design Report, Volume 2: The Physics Program
  for DUNE at LBNF}.
\newblock 12 2015.

\bibitem{JUNO:2015zny}
Fengpeng An et~al.
\newblock {Neutrino Physics with JUNO}.
\newblock {\em J. Phys. G}, 43(3):030401, 2016.

\bibitem{Mohapatra:1998rq}
R.~N. Mohapatra and P.~B. Pal.
\newblock {\em {Massive neutrinos in physics and astrophysics. Second
  edition}}, volume~60.
\newblock 1998.

\bibitem{LlewellynSmith:1971uhs}
C.~H. Llewellyn~Smith.
\newblock {Neutrino Reactions at Accelerator Energies}.
\newblock {\em Phys. Rept.}, 3:261--379, 1972.

\bibitem{Schwartz:2014sze}
Matthew~D. Schwartz.
\newblock {\em {Quantum Field Theory and the Standard Model}}.
\newblock Cambridge University Press, 3 2014.

\bibitem{Yanagida:1979as}
Tsutomu Yanagida.
\newblock {Horizontal gauge symmetry and masses of neutrinos}.
\newblock {\em Conf. Proc. C}, 7902131:95--99, 1979.

\bibitem{KATRIN:2021uub}
M.~Aker et~al.
\newblock {Direct neutrino-mass measurement with sub-electronvolt sensitivity}.
\newblock {\em Nature Phys.}, 18(2):160--166, 2022.

\bibitem{KamLAND-Zen:2022tow}
S.~Abe et~al.
\newblock {Search for the Majorana Nature of Neutrinos in the Inverted Mass
  Ordering Region with KamLAND-Zen}.
\newblock {\em Phys. Rev. Lett.}, 130(5):051801, 2023.

\bibitem{CUORE:2017tlq}
C.~Alduino et~al.
\newblock {First Results from CUORE: A Search for Lepton Number Violation via
  $0\nu\beta\beta$ Decay of $^{130}$Te}.
\newblock {\em Phys. Rev. Lett.}, 120(13):132501, 2018.

\bibitem{GERDA:2018pmc}
M.~Agostini et~al.
\newblock {Improved Limit on Neutrinoless Double-$\beta$ Decay of $^{76}$Ge
  from GERDA Phase II}.
\newblock {\em Phys. Rev. Lett.}, 120(13):132503, 2018.

\bibitem{Kayser:1981ye}
Boris Kayser.
\newblock {On the Quantum Mechanics of Neutrino Oscillation}.
\newblock {\em Phys. Rev. D}, 24:110, 1981.

\bibitem{ParticleDataGroup:2020ssz}
P.~A. Zyla et~al.
\newblock {Review of Particle Physics}.
\newblock {\em PTEP}, 2020(8):083C01, 2020.

\bibitem{Abrams:1989yk}
G.~S. Abrams et~al.
\newblock {Measurements of $Z$ Boson Resonance Parameters in $e^{+} e^{-}$
  Annihilation}.
\newblock {\em Phys. Rev. Lett.}, 63:2173, 1989.

\bibitem{JSNS2:2013jdh}
M.~Harada et~al.
\newblock {Proposal: A Search for Sterile Neutrino at J-PARC Materials and Life
  Science Experimental Facility}.
\newblock 10 2013.

\bibitem{Maruyama:2022juu}
Takasumi Maruyama.
\newblock {The status of JSNS$^2$ and JSNS$^2$-II}.
\newblock {\em PoS}, NuFact2021:159, 2022.

\bibitem{JSNS2:2021hyk}
S.~Ajimura et~al.
\newblock {The JSNS2 detector}.
\newblock {\em Nucl. Instrum. Meth. A}, 1014:165742, 2021.

\bibitem{MiniBooNE:2007uho}
A.~A. Aguilar-Arevalo et~al.
\newblock {A Search for Electron Neutrino Appearance at the $\Delta m^2 \sim 1
  eV^2$ Scale}.
\newblock {\em Phys. Rev. Lett.}, 98:231801, 2007.

\bibitem{Patterson:2009ki}
R.~B. Patterson, E.~M. Laird, Y.~Liu, P.~D. Meyers, I.~Stancu, and H.~A.
  Tanaka.
\newblock {The Extended-track reconstruction for MiniBooNE}.
\newblock {\em Nucl. Instrum. Meth. A}, 608:206--224, 2009.

\bibitem{MiniBooNE:2008yuf}
A.~A. Aguilar-Arevalo et~al.
\newblock {Unexplained Excess of Electron-Like Events From a 1-GeV Neutrino
  Beam}.
\newblock {\em Phys. Rev. Lett.}, 102:101802, 2009.

\bibitem{MiniBooNE:2018esg}
A.~A. Aguilar-Arevalo et~al.
\newblock {Significant Excess of ElectronLike Events in the MiniBooNE
  Short-Baseline Neutrino Experiment}.
\newblock {\em Phys. Rev. Lett.}, 121(22):221801, 2018.

\bibitem{MicroBooNE:2021zai}
P.~Abratenko et~al.
\newblock {Search for Neutrino-Induced Neutral-Current \ensuremath{\Delta}
  Radiative Decay in MicroBooNE and a First Test of the MiniBooNE Low Energy
  Excess under a Single-Photon Hypothesis}.
\newblock {\em Phys. Rev. Lett.}, 128:111801, 2022.

\bibitem{Brdar:2021ysi}
Vedran Brdar and Joachim Kopp.
\newblock {Can standard model and experimental uncertainties resolve the
  MiniBooNE anomaly?}
\newblock {\em Phys. Rev. D}, 105(11):115024, 2022.

\bibitem{Kelly:2022uaa}
Kevin~J. Kelly and Joachim Kopp.
\newblock {More Ingredients for an Altarelli Cocktail at MiniBooNE}.
\newblock 10 2022.

\bibitem{Karagiorgi:2022fgf}
G.~Karagiorgi, B.~R. Littlejohn, P.~Machado, and Alexandre Sousa.
\newblock {Snowmass Neutrino Frontier: NF02 Topical Group Report on
  Understanding Experimental Neutrino Anomalies}.
\newblock In {\em {2022 Snowmass Summer Study}}, 9 2022.

\bibitem{Huber:2011wv}
Patrick Huber.
\newblock {On the determination of anti-neutrino spectra from nuclear
  reactors}.
\newblock {\em Phys. Rev. C}, 84:024617, 2011.
\newblock [Erratum: Phys.Rev.C 85, 029901 (2012)].

\bibitem{Mueller:2011nm}
Th.~A. Mueller et~al.
\newblock {Improved Predictions of Reactor Antineutrino Spectra}.
\newblock {\em Phys. Rev. C}, 83:054615, 2011.

\bibitem{Estienne:2019ujo}
M.~Estienne et~al.
\newblock {Updated Summation Model: An Improved Agreement with the Daya Bay
  Antineutrino Fluxes}.
\newblock {\em Phys. Rev. Lett.}, 123(2):022502, 2019.

\bibitem{Hayen:2019eop}
L.~Hayen, J.~Kostensalo, N.~Severijns, and J.~Suhonen.
\newblock {First-forbidden transitions in the reactor anomaly}.
\newblock {\em Phys. Rev. C}, 100(5):054323, 2019.

\bibitem{Kopeikin:2021ugh}
V.~Kopeikin, M.~Skorokhvatov, and O.~Titov.
\newblock {Reevaluating reactor antineutrino spectra with new measurements of
  the ratio between U235 and Pu239 \ensuremath{\beta} spectra}.
\newblock {\em Phys. Rev. D}, 104(7):L071301, 2021.

\bibitem{Giunti:2021kab}
C.~Giunti, Y.~F. Li, C.~A. Ternes, and Z.~Xin.
\newblock {Reactor antineutrino anomaly in light of recent flux model
  refinements}.
\newblock {\em Phys. Lett. B}, 829:137054, 2022.

\bibitem{SAGE:2009eeu}
J.~N. Abdurashitov et~al.
\newblock {Measurement of the solar neutrino capture rate with gallium metal.
  III: Results for the 2002--2007 data-taking period}.
\newblock {\em Phys. Rev. C}, 80:015807, 2009.

\bibitem{Kaether:2010ag}
F.~Kaether, W.~Hampel, G.~Heusser, J.~Kiko, and T.~Kirsten.
\newblock {Reanalysis of the GALLEX solar neutrino flux and source
  experiments}.
\newblock {\em Phys. Lett. B}, 685:47--54, 2010.

\bibitem{Giunti:2006bj}
Carlo Giunti and Marco Laveder.
\newblock {Short-Baseline Active-Sterile Neutrino Oscillations?}
\newblock {\em Mod. Phys. Lett. A}, 22:2499--2509, 2007.

\bibitem{Diaz:2019fwt}
A.~Diaz, C.~A. Arg\"uelles, G.~H. Collin, J.~M. Conrad, and M.~H. Shaevitz.
\newblock {Where Are We With Light Sterile Neutrinos?}
\newblock {\em Phys. Rept.}, 884:1--59, 2020.

\bibitem{NEOS:2016wee}
Y.~J. Ko et~al.
\newblock {Sterile Neutrino Search at the NEOS Experiment}.
\newblock {\em Phys. Rev. Lett.}, 118(12):121802, 2017.

\bibitem{RENO:2020uip}
J.~H. Choi et~al.
\newblock {Search for Sub-eV Sterile Neutrinos at RENO}.
\newblock {\em Phys. Rev. Lett.}, 125(19):191801, 2020.

\bibitem{DayaBay:2016qvc}
Feng~Peng An et~al.
\newblock {Improved Search for a Light Sterile Neutrino with the Full
  Configuration of the Daya Bay Experiment}.
\newblock {\em Phys. Rev. Lett.}, 117(15):151802, 2016.

\bibitem{STEREO:2022nzk}
H.~Almaz\'an et~al.
\newblock {STEREO neutrino spectrum of $^{235}$U fission rejects sterile
  neutrino hypothesis}.
\newblock {\em Nature}, 613(7943):257--261, 2023.

\bibitem{DANSS:2021raa}
I.~Alekseev et~al.
\newblock {Optimized scintillation strip design for the DANSS upgrade}.
\newblock {\em JINST}, 17(04):P04009, 2022.

\bibitem{PROSPECT:2020sxr}
M.~Andriamirado et~al.
\newblock {Improved short-baseline neutrino oscillation search and energy
  spectrum measurement with the PROSPECT experiment at HFIR}.
\newblock {\em Phys. Rev. D}, 103(3):032001, 2021.

\bibitem{RENO:2020hva}
Z.~Atif et~al.
\newblock {Search for sterile neutrino oscillations using RENO and NEOS data}.
\newblock {\em Phys. Rev. D}, 105(11):L111101, 2022.

\bibitem{KATRIN:2022ith}
M.~Aker et~al.
\newblock {Improved eV-scale sterile-neutrino constraints from the second
  KATRIN measurement campaign}.
\newblock {\em Phys. Rev. D}, 105(7):072004, 2022.

\bibitem{MiniBooNE:2022emn}
A.~A. Aguilar-Arevalo et~al.
\newblock {MiniBooNE and MicroBooNE Combined Fit to a 3+1 Sterile Neutrino
  Scenario}.
\newblock {\em Phys. Rev. Lett.}, 129(20):201801, 2022.

\bibitem{MINOS:2017cae}
P.~Adamson et~al.
\newblock {Search for sterile neutrinos in MINOS and MINOS+ using a
  two-detector fit}.
\newblock {\em Phys. Rev. Lett.}, 122(9):091803, 2019.

\bibitem{MINOS:2020iqj}
P.~Adamson et~al.
\newblock {Improved Constraints on Sterile Neutrino Mixing from Disappearance
  Searches in the MINOS, MINOS+, Daya Bay, and Bugey-3 Experiments}.
\newblock {\em Phys. Rev. Lett.}, 125(7):071801, 2020.

\bibitem{Stockdale:1984cg}
I.~E. Stockdale et~al.
\newblock {Limits on Muon Neutrino Oscillations in the Mass Range 55-eV**2
  \ensuremath{<} Delta m**2 \ensuremath{<} 800-eV**2}.
\newblock {\em Phys. Rev. Lett.}, 52:1384, 1984.

\bibitem{IceCube:2020phf}
M.~G. Aartsen et~al.
\newblock {eV-Scale Sterile Neutrino Search Using Eight Years of Atmospheric
  Muon Neutrino Data from the IceCube Neutrino Observatory}.
\newblock {\em Phys. Rev. Lett.}, 125(14):141801, 2020.

\bibitem{Conrad:2013mka}
Janet~M. Conrad, William~C. Louis, and Michael~H. Shaevitz.
\newblock {The LSND and MiniBooNE Oscillation Searches at High $\Delta m^2$}.
\newblock {\em Ann. Rev. Nucl. Part. Sci.}, 63:45--67, 2013.

\bibitem{HARP:2007dqt}
M.~G. Catanesi et~al.
\newblock {Measurement of the production cross-section of positive pions in the
  collision of 8.9-GeV/c protons on beryllium}.
\newblock {\em Eur. Phys. J. C}, 52:29--53, 2007.

\bibitem{E910:2007puw}
I.~Chemakin et~al.
\newblock {Pion production by protons on a thin beryllium target at 6.4-Ge,
  12.3-GeV/c, and 17.5-GeV/c incident proton momenta}.
\newblock {\em Phys. Rev. C}, 77:015209, 2008.
\newblock [Erratum: Phys.Rev.C 77, 049903 (2008)].

\bibitem{Wang:1970bn}
C.~L. Wang.
\newblock {Pion, kaon, and anti-proton production between 10 and 70 bev}.
\newblock {\em Phys. Rev. Lett.}, 25:1068--1072, 1970.

\bibitem{MiniBooNE:2010bsu}
A.~A. Aguilar-Arevalo et~al.
\newblock {First Measurement of the Muon Neutrino Charged Current Quasielastic
  Double Differential Cross Section}.
\newblock {\em Phys. Rev. D}, 81:092005, 2010.

\bibitem{MiniBooNE:2007iti}
A.~A. Aguilar-Arevalo et~al.
\newblock {Measurement of muon neutrino quasi-elastic scattering on carbon}.
\newblock {\em Phys. Rev. Lett.}, 100:032301, 2008.

\bibitem{MiniBooNE:2009dxl}
Alexis~A. Aguilar-Arevalo et~al.
\newblock {Measurement of $\nu_\mu$ and $\bar{\nu}_\mu$ induced neutral current
  single $\pi^0$ production cross sections on mineral oil at $E_\nu \sim {\cal
  O}(1 {\rm GeV})$}.
\newblock {\em Phys. Rev. D}, 81:013005, 2010.

\bibitem{MiniBooNE:2008mmr}
A.~A. Aguilar-Arevalo et~al.
\newblock {First Observation of Coherent $\pi^0$ Production in Neutrino Nucleus
  Interactions with $E_{\nu}<$ 2 GeV}.
\newblock {\em Phys. Lett. B}, 664:41--46, 2008.

\bibitem{MiniBooNE:2013qnd}
A.~A. Aguilar-Arevalo et~al.
\newblock {First measurement of the muon antineutrino double-differential
  charged-current quasielastic cross section}.
\newblock {\em Phys. Rev. D}, 88(3):032001, 2013.

\bibitem{MicroBooNE:2021tya}
P.~Abratenko et~al.
\newblock {Search for an Excess of Electron Neutrino Interactions in MicroBooNE
  Using Multiple Final-State Topologies}.
\newblock {\em Phys. Rev. Lett.}, 128(24):241801, 2022.

\bibitem{Bai:2015ztj}
Yang Bai, Ran Lu, Sida Lu, Jordi Salvado, and Ben~A. Stefanek.
\newblock {Three Twin Neutrinos: Evidence from LSND and MiniBooNE}.
\newblock {\em Phys. Rev. D}, 93(7):073004, 2016.

\bibitem{deGouvea:2019qre}
Andr\'e de~Gouv\^ea, O.~L.~G. Peres, Suprabh Prakash, and G.~V. Stenico.
\newblock {On The Decaying-Sterile Neutrino Solution to the Electron
  (Anti)Neutrino Appearance Anomalies}.
\newblock {\em JHEP}, 07:141, 2020.

\bibitem{Dentler:2019dhz}
Mona Dentler, Ivan Esteban, Joachim Kopp, and Pedro Machado.
\newblock {Decaying Sterile Neutrinos and the Short Baseline Oscillation
  Anomalies}.
\newblock {\em Phys. Rev. D}, 101(11):115013, 2020.

\bibitem{Pas:2005rb}
Heinrich Pas, Sandip Pakvasa, and Thomas~J. Weiler.
\newblock {Sterile-active neutrino oscillations and shortcuts in the extra
  dimension}.
\newblock {\em Phys. Rev. D}, 72:095017, 2005.

\bibitem{Carena:2017qhd}
Marcela Carena, Ying-Ying Li, Camila~S. Machado, Pedro A.~N. Machado, and
  Carlos E.~M. Wagner.
\newblock {Neutrinos in Large Extra Dimensions and Short-Baseline $\nu_e$
  Appearance}.
\newblock {\em Phys. Rev. D}, 96(9):095014, 2017.

\bibitem{Doring:2018ncz}
Dominik D\"oring and Heinrich P\"as.
\newblock {Sterile Neutrino Shortcuts in Asymmetrically Warped Extra
  Dimensions}.
\newblock {\em Eur. Phys. J. C}, 79(7):604, 2019.

\bibitem{Gninenko:2009ks}
S.~N. Gninenko.
\newblock {The MiniBooNE anomaly and heavy neutrino decay}.
\newblock {\em Phys. Rev. Lett.}, 103:241802, 2009.

\bibitem{Gninenko:2010pr}
Sergei~N. Gninenko.
\newblock {A resolution of puzzles from the LSND, KARMEN, and MiniBooNE
  experiments}.
\newblock {\em Phys. Rev. D}, 83:015015, 2011.

\bibitem{Dib:2011jh}
Claudio Dib, Juan~Carlos Helo, Sergey Kovalenko, and Ivan Schmidt.
\newblock {Sterile neutrino decay explanation of LSND and MiniBooNE anomalies}.
\newblock {\em Phys. Rev. D}, 84:071301, 2011.

\bibitem{Gninenko:2012rw}
S.~N. Gninenko.
\newblock {New limits on radiative sterile neutrino decays from a search for
  single photons in neutrino interactions}.
\newblock {\em Phys. Lett. B}, 710:86--90, 2012.

\bibitem{Masip:2012ke}
Manuel Masip, Pere Masjuan, and Davide Meloni.
\newblock {Heavy neutrino decays at MiniBooNE}.
\newblock {\em JHEP}, 01:106, 2013.

\bibitem{Radionov:2013mca}
Alexander Radionov.
\newblock {Constraints on electromagnetic properties of sterile neutrinos from
  MiniBooNE results}.
\newblock {\em Phys. Rev. D}, 88(1):015016, 2013.

\bibitem{Ballett:2016opr}
Peter Ballett, Silvia Pascoli, and Mark Ross-Lonergan.
\newblock {MeV-scale sterile neutrino decays at the Fermilab Short-Baseline
  Neutrino program}.
\newblock {\em JHEP}, 04:102, 2017.

\bibitem{Magill:2018jla}
Gabriel Magill, Ryan Plestid, Maxim Pospelov, and Yu-Dai Tsai.
\newblock {Dipole Portal to Heavy Neutral Leptons}.
\newblock {\em Phys. Rev. D}, 98(11):115015, 2018.

\bibitem{Balantekin:2018ukw}
A.~Baha Balantekin, Andr\'e de~Gouv\^ea, and Boris Kayser.
\newblock {Addressing the Majorana vs. Dirac Question with Neutrino Decays}.
\newblock {\em Phys. Lett. B}, 789:488--495, 2019.

\bibitem{Balaji:2019fxd}
Shyam Balaji, Maura Ramirez-Quezada, and Ye-Ling Zhou.
\newblock {CP violation and circular polarisation in neutrino radiative decay}.
\newblock {\em JHEP}, 04:178, 2020.

\bibitem{Balaji:2020oig}
Shyam Balaji, Maura Ramirez-Quezada, and Ye-Ling Zhou.
\newblock {CP violation in neutral lepton transition dipole moment}.
\newblock {\em JHEP}, 12:090, 2020.

\bibitem{Fischer:2019fbw}
Oliver Fischer, \'Alvaro Hern\'andez-Cabezudo, and Thomas Schwetz.
\newblock {Explaining the MiniBooNE excess by a decaying sterile neutrino with
  mass in the 250 MeV range}.
\newblock {\em Phys. Rev. D}, 101(7):075045, 2020.

\bibitem{Alvarez-Ruso:2021dna}
Luis Alvarez-Ruso and Eduardo Saul-Sala.
\newblock {Neutrino interactions with matter and the MiniBooNE anomaly}.
\newblock {\em Eur. Phys. J. ST}, 230(24):4373--4389, 2021.

\bibitem{Bertuzzo:2018ftf}
Enrico Bertuzzo, Sudip Jana, Pedro A.~N. Machado, and Renata
  Zukanovich~Funchal.
\newblock {Neutrino Masses and Mixings Dynamically Generated by a Light Dark
  Sector}.
\newblock {\em Phys. Lett. B}, 791:210--214, 2019.

\bibitem{Ballett:2018ynz}
Peter Ballett, Silvia Pascoli, and Mark Ross-Lonergan.
\newblock {U(1)' mediated decays of heavy sterile neutrinos in MiniBooNE}.
\newblock {\em Phys. Rev. D}, 99:071701, 2019.

\bibitem{Bertuzzo:2018itn}
Enrico Bertuzzo, Sudip Jana, Pedro A.~N. Machado, and Renata
  Zukanovich~Funchal.
\newblock {Dark Neutrino Portal to Explain MiniBooNE excess}.
\newblock {\em Phys. Rev. Lett.}, 121(24):241801, 2018.

\bibitem{Ballett:2019pyw}
Peter Ballett, Matheus Hostert, and Silvia Pascoli.
\newblock {Dark Neutrinos and a Three Portal Connection to the Standard Model}.
\newblock {\em Phys. Rev. D}, 101(11):115025, 2020.

\bibitem{Abdullahi:2020nyr}
Asli Abdullahi, Matheus Hostert, and Silvia Pascoli.
\newblock {A dark seesaw solution to low energy anomalies: MiniBooNE, the muon
  (g - 2), and BaBar}.
\newblock {\em Phys. Lett. B}, 820:136531, 2021.

\bibitem{Datta:2020auq}
Alakabha Datta, Saeed Kamali, and Danny Marfatia.
\newblock {Dark sector origin of the KOTO and MiniBooNE anomalies}.
\newblock {\em Phys. Lett. B}, 807:135579, 2020.

\bibitem{Dutta:2020scq}
Bhaskar Dutta, Sumit Ghosh, and Tianjun Li.
\newblock {Explaining $(g-2)_{\mu,e}$, the KOTO anomaly and the MiniBooNE
  excess in an extended Higgs model with sterile neutrinos}.
\newblock {\em Phys. Rev. D}, 102(5):055017, 2020.

\bibitem{Abdallah:2020biq}
Waleed Abdallah, Raj Gandhi, and Samiran Roy.
\newblock {Understanding the MiniBooNE and the muon and electron $g − 2$
  anomalies with a light $Z′$ and a second Higgs doublet}.
\newblock {\em JHEP}, 12:188, 2020.

\bibitem{Abdallah:2020vgg}
Waleed Abdallah, Raj Gandhi, and Samiran Roy.
\newblock {Two-Higgs doublet solution to the LSND, MiniBooNE and muon g-2
  anomalies}.
\newblock {\em Phys. Rev. D}, 104(5):055028, 2021.

\bibitem{Dutta:2021cip}
Bhaskar Dutta, Doojin Kim, Adrian Thompson, Remington~T. Thornton, and
  Richard~G. Van~de Water.
\newblock {Solutions to the MiniBooNE Anomaly from New Physics in Charged Meson
  Decays}.
\newblock {\em Phys. Rev. Lett.}, 129(11):111803, 2022.

\bibitem{Arguelles:2018mtc}
Carlos~A. Arg\"uelles, Matheus Hostert, and Yu-Dai Tsai.
\newblock {Testing New Physics Explanations of the MiniBooNE Anomaly at
  Neutrino Scattering Experiments}.
\newblock {\em Phys. Rev. Lett.}, 123(26):261801, 2019.

\bibitem{Jordan:2018qiy}
Johnathon~R. Jordan, Yonatan Kahn, Gordan Krnjaic, Matthew Moschella, and
  Joshua Spitz.
\newblock {Severe Constraints on New Physics Explanations of the MiniBooNE
  Excess}.
\newblock {\em Phys. Rev. Lett.}, 122(8):081801, 2019.

\bibitem{MicroBooNE:2021pvo}
P.~Abratenko et~al.
\newblock {Search for an anomalous excess of charged-current quasielastic
  \ensuremath{\nu}e interactions with the MicroBooNE experiment using
  Deep-Learning-based reconstruction}.
\newblock {\em Phys. Rev. D}, 105(11):112003, 2022.

\bibitem{Willis:1974gi}
W.~J. Willis and V.~Radeka.
\newblock {Liquid Argon Ionization Chambers as Total Absorption Detectors}.
\newblock {\em Nucl. Instrum. Meth.}, 120:221--236, 1974.

\bibitem{Rubbia:1977zz}
C.~Rubbia.
\newblock {The Liquid Argon Time Projection Chamber: A New Concept for Neutrino
  Detectors}.
\newblock 5 1977.

\bibitem{Nygren:1974nfi}
D.~R. Nygren.
\newblock {The Time Projection Chamber: A New 4 pi Detector for Charged
  Particles}.
\newblock {\em eConf}, C740805:58, 1974.

\bibitem{Charpak:1970az}
G.~Charpak, D.~Rahm, and H.~Steiner.
\newblock {Some developments in the operation of multiwire proportional
  chambers}.
\newblock {\em Nucl. Instrum. Meth.}, 80:13--34, 1970.

\bibitem{ICARUS:2004wqc}
S.~Amerio et~al.
\newblock {Design, construction and tests of the ICARUS T600 detector}.
\newblock {\em Nucl. Instrum. Meth. A}, 527:329--410, 2004.

\bibitem{Rubbia:2011ft}
C.~Rubbia et~al.
\newblock {Underground operation of the ICARUS T600 LAr-TPC: first results}.
\newblock {\em JINST}, 6:P07011, 2011.

\bibitem{ICARUS:2013cwr}
M.~Antonello et~al.
\newblock {Search for anomalies in the ${\nu}_e$ appearance from a
  ${\nu}_{\mu}$ beam}.
\newblock {\em Eur. Phys. J. C}, 73:2599, 2013.

\bibitem{ArgoNeuT:2014rlj}
R.~Acciarri et~al.
\newblock {Measurements of Inclusive Muon Neutrino and Antineutrino Charged
  Current Differential Cross Sections on Argon in the NuMI Antineutrino Beam}.
\newblock {\em Phys. Rev. D}, 89(11):112003, 2014.

\bibitem{Baibussinov:2009gs}
B.~Baibussinov et~al.
\newblock {Free electron lifetime achievements in Liquid Argon Imaging TPC}.
\newblock {\em JINST}, 5:P03005, 2010.

\bibitem{Jones:2013bca}
B.~J.~P. Jones, C.~S. Chiu, J.~M. Conrad, C.~M. Ignarra, T.~Katori, and
  M.~Toups.
\newblock {A Measurement of the Absorption of Liquid Argon Scintillation Light
  by Dissolved Nitrogen at the Part-Per-Million Level}.
\newblock {\em JINST}, 8:P07011, 2013.
\newblock [Erratum: JINST 8, E09001 (2013)].

\bibitem{Suzuki:1979km}
M.~Suzuki and S.~Kubota.
\newblock {MECHANISM OF PROPORTIONAL SCINTILLATION IN ARGON, KRYPTON AND
  XENON}.
\newblock {\em Nucl. Instrum. Meth.}, 164:197--199, 1979.

\bibitem{DEAP:2020hms}
P.~Adhikari et~al.
\newblock {The liquid-argon scintillation pulseshape in DEAP-3600}.
\newblock {\em Eur. Phys. J. C}, 80(4):303, 2020.

\bibitem{Kubota_1978}
S~Kubota, M~Hishida, and J~Raun.
\newblock Evidence for a triplet state of the self-trapped exciton states in
  liquid argon, krypton and xenon.
\newblock {\em Journal of Physics C: Solid State Physics}, 11(12):2645, jun
  1978.

\bibitem{Einstein:1905cc}
Albert Einstein.
\newblock {Concerning an heuristic point of view toward the emission and
  transformation of light}.
\newblock {\em Annalen Phys.}, 17:132--148, 1905.

\bibitem{Briese:2013wua}
T.~Briese et~al.
\newblock {Testing of Cryogenic Photomultiplier Tubes for the MicroBooNE
  Experiment}.
\newblock {\em JINST}, 8:T07005, 2013.

\bibitem{WArP:2008dyo}
R.~Acciarri et~al.
\newblock {Oxygen contamination in liquid Argon: Combined effects on ionization
  electron charge and scintillation light}.
\newblock {\em JINST}, 5:P05003, 2010.

\bibitem{WArP:2008rgv}
R.~Acciarri et~al.
\newblock {Effects of Nitrogen contamination in liquid Argon}.
\newblock {\em JINST}, 5:P06003, 2010.

\bibitem{Asaadi:2018ixs}
J.~Asaadi, B.~J.~P. Jones, A.~Tripathi, I.~Parmaksiz, H.~Sullivan, and Z.~G.~R.
  Williams.
\newblock {Emanation and bulk fluorescence in liquid argon from tetraphenyl
  butadiene wavelength shifting coatings}.
\newblock {\em JINST}, 14(02):P02021, 2019.

\bibitem{Grace:2015yta}
Emily Grace and James~A. Nikkel.
\newblock {Index of refraction, Rayleigh scattering length, and Sellmeier
  coefficients in solid and liquid argon and xenon}.
\newblock {\em Nucl. Instrum. Meth. A}, 867:204--208, 2017.

\bibitem{MicroBooNE:2018vro}
C.~Adams et~al.
\newblock {Ionization electron signal processing in single phase LArTPCs. Part
  II. Data/simulation comparison and performance in MicroBooNE}.
\newblock {\em JINST}, 13(07):P07007, 2018.

\bibitem{10.7551/mitpress/2946.001.0001}
Norbert Wiener.
\newblock {\em {Extrapolation, Interpolation, and Smoothing of Stationary Time
  Series: With Engineering Applications}}.
\newblock The MIT Press, 08 1949.

\bibitem{MicroBooNE:2021nxr}
P.~Abratenko et~al.
\newblock {Search for an anomalous excess of inclusive charged-current $\nu_e$
  interactions in the MicroBooNE experiment using Wire-Cell reconstruction}.
\newblock {\em Phys. Rev. D}, 105(11):112005, 2022.

\bibitem{TheMicroBooNECollaboration:2021cjf}
P.~Abratenko et~al.
\newblock {Search for an anomalous excess of charged-current \ensuremath{\nu}e
  interactions without pions in the final state with the MicroBooNE
  experiment}.
\newblock {\em Phys. Rev. D}, 105(11):112004, 2022.

\bibitem{MicroBooNE:2021nss}
P.~Abratenko et~al.
\newblock {Electromagnetic shower reconstruction and energy validation with
  Michel electrons and \ensuremath{\pi}$^{0}$ samples for the
  deep-learning-based analyses in MicroBooNE}.
\newblock {\em JINST}, 16(12):T12017, 2021.

\bibitem{Andreopoulos:2009rq}
C.~Andreopoulos et~al.
\newblock {The GENIE Neutrino Monte Carlo Generator}.
\newblock {\em Nucl. Instrum. Meth. A}, 614:87--104, 2010.

\bibitem{Andreopoulos:2015wxa}
Costas Andreopoulos, Christopher Barry, Steve Dytman, Hugh Gallagher, Tomasz
  Golan, Robert Hatcher, Gabriel Perdue, and Julia Yarba.
\newblock {The GENIE Neutrino Monte Carlo Generator: Physics and User Manual}.
\newblock 10 2015.

\bibitem{GENIE:2021npt}
Luis Alvarez-Ruso et~al.
\newblock {Recent highlights from GENIE v3}.
\newblock {\em Eur. Phys. J. ST}, 230(24):4449--4467, 2021.

\bibitem{GENIE:2021zuu}
J\'ulia Tena-Vidal et~al.
\newblock {Neutrino-nucleon cross-section model tuning in GENIE v3}.
\newblock {\em Phys. Rev. D}, 104(7):072009, 2021.

\bibitem{Nieves:2011pp}
J.~Nieves, I.~Ruiz~Simo, and M.~J. Vicente~Vacas.
\newblock {Inclusive Charged--Current Neutrino--Nucleus Reactions}.
\newblock {\em Phys. Rev. C}, 83:045501, 2011.

\bibitem{Bodek:1980ar}
A.~Bodek and J.~L. Ritchie.
\newblock {Fermi Motion Effects in Deep Inelastic Lepton Scattering from
  Nuclear Targets}.
\newblock {\em Phys. Rev. D}, 23:1070, 1981.

\bibitem{Pandharipande:1992zz}
V.~R. Pandharipande and Steven~C. Pieper.
\newblock {Nuclear transparency to intermediate-energy nucleons from (e, e'p)
  reactions}.
\newblock {\em Phys. Rev. C}, 45:791--798, 1992.

\bibitem{MicroBooNE:2021ccs}
P.~Abratenko et~al.
\newblock {New $CC0\pi$ GENIE model tune for MicroBooNE}.
\newblock {\em Phys. Rev. D}, 105(7):072001, 2022.

\bibitem{T2K:2016jor}
Ko~Abe et~al.
\newblock {Measurement of double-differential muon neutrino charged-current
  interactions on C$_8$H$_8$ without pions in the final state using the T2K
  off-axis beam}.
\newblock {\em Phys. Rev. D}, 93(11):112012, 2016.

\bibitem{Ankowski:2020qbe}
Artur~M. Ankowski and Alexander Friedland.
\newblock {Assessing the accuracy of the GENIE event generator with
  electron-scattering data}.
\newblock {\em Phys. Rev. D}, 102(5):053001, 2020.

\bibitem{electronsforneutrinos:2020tbf}
A.~Papadopoulou et~al.
\newblock {Inclusive Electron Scattering And The GENIE Neutrino Event
  Generator}.
\newblock {\em Phys. Rev. D}, 103:113003, 2021.

\bibitem{CLAS:2021neh}
M.~Khachatryan et~al.
\newblock {Electron-beam energy reconstruction for neutrino oscillation
  measurements}.
\newblock {\em Nature}, 599(7886):565--570, 2021.

\bibitem{GEANT4:2002zbu}
S.~Agostinelli et~al.
\newblock {GEANT4--a simulation toolkit}.
\newblock {\em Nucl. Instrum. Meth. A}, 506:250--303, 2003.

\bibitem{Snider_2017}
E.L. Snider and G.~Petrillo.
\newblock Larsoft: toolkit for simulation, reconstruction and analysis of
  liquid argon tpc neutrino detectors.
\newblock {\em Journal of Physics: Conference Series}, 898(4):042057, oct 2017.

\bibitem{DAgostini:1994fjx}
G.~D'Agostini.
\newblock {A Multidimensional unfolding method based on Bayes' theorem}.
\newblock {\em Nucl. Instrum. Meth. A}, 362:487--498, 1995.

\bibitem{Arguelles:2021meu}
C.~A. Arg\"uelles, I.~Esteban, M.~Hostert, Kevin~J. Kelly, J.~Kopp, P.~A.~N.
  Machado, I.~Martinez-Soler, and Y.~F. Perez-Gonzalez.
\newblock {MicroBooNE and the \ensuremath{\nu}e Interpretation of the MiniBooNE
  Low-Energy Excess}.
\newblock {\em Phys. Rev. Lett.}, 128(24):241802, 2022.

\bibitem{Benhar:2015ula}
Omar Benhar, Alessandro Lovato, and Noemi Rocco.
\newblock {Contribution of two-particle\textendash{}two-hole final states to
  the nuclear response}.
\newblock {\em Phys. Rev. C}, 92(2):024602, 2015.

\bibitem{Thomson:2013zua}
Mark Thomson.
\newblock {\em {Modern particle physics}}.
\newblock Cambridge University Press, New York, 2013.

\bibitem{Ankowski:2005wi}
Artur~M. Ankowski and Jan~T. Sobczyk.
\newblock {Argon spectral function and neutrino interactions}.
\newblock {\em Phys. Rev. C}, 74:054316, 2006.

\bibitem{MicroBooNE:2020vry}
P.~Abratenko et~al.
\newblock {Neutrino event selection in the MicroBooNE liquid argon time
  projection chamber using Wire-Cell 3D imaging, clustering, and charge-light
  matching}.
\newblock {\em JINST}, 16(06):P06043, 2021.

\bibitem{MicroBooNE:2021zul}
P.~Abratenko et~al.
\newblock {Cosmic Ray Background Rejection with Wire-Cell LArTPC Event
  Reconstruction in the MicroBooNE Detector}.
\newblock {\em Phys. Rev. Applied}, 15(6):064071, 2021.

\bibitem{ronneberger2015u}
Olaf Ronneberger, Philipp Fischer, and Thomas Brox.
\newblock U-net: Convolutional networks for biomedical image segmentation.
\newblock In {\em Medical Image Computing and Computer-Assisted
  Intervention--MICCAI 2015: 18th International Conference, Munich, Germany,
  October 5-9, 2015, Proceedings, Part III 18}, pages 234--241. Springer, 2015.

\bibitem{he2016deep}
Kaiming He, Xiangyu Zhang, Shaoqing Ren, and Jian Sun.
\newblock Deep residual learning for image recognition.
\newblock In {\em Proceedings of the IEEE conference on computer vision and
  pattern recognition}, pages 770--778, 2016.

\bibitem{graham20183d}
Benjamin Graham, Martin Engelcke, and Laurens Van Der~Maaten.
\newblock 3d semantic segmentation with submanifold sparse convolutional
  networks.
\newblock In {\em Proceedings of the IEEE conference on computer vision and
  pattern recognition}, pages 9224--9232, 2018.

\bibitem{MicroBooNE:2018kka}
C.~Adams et~al.
\newblock {Deep neural network for pixel-level electromagnetic particle
  identification in the MicroBooNE liquid argon time projection chamber}.
\newblock {\em Phys. Rev. D}, 99(9):092001, 2019.

\bibitem{szegedy2015going}
Christian Szegedy, Wei Liu, Yangqing Jia, Pierre Sermanet, Scott Reed, Dragomir
  Anguelov, Dumitru Erhan, Vincent Vanhoucke, and Andrew Rabinovich.
\newblock Going deeper with convolutions.
\newblock In {\em Proceedings of the IEEE conference on computer vision and
  pattern recognition}, pages 1--9, 2015.

\bibitem{chen2016xgboost}
Tianqi Chen and Carlos Guestrin.
\newblock Xgboost: A scalable tree boosting system.
\newblock In {\em Proceedings of the 22nd acm sigkdd international conference
  on knowledge discovery and data mining}, pages 785--794, 2016.

\bibitem{10.1145/2009916.2009932}
Yasser Ganjisaffar, Rich Caruana, and Cristina~Videira Lopes.
\newblock Bagging gradient-boosted trees for high precision, low variance
  ranking models.
\newblock In {\em Proceedings of the 34th International ACM SIGIR Conference on
  Research and Development in Information Retrieval}, SIGIR '11, page 85–94,
  New York, NY, USA, 2011. Association for Computing Machinery.

\bibitem{Kamp:2023mjn}
Nicholas~W. Kamp, Matheus Hostert, Carlos~A. Arg\"uelles, Janet~M. Conrad, and
  Michael~H. Shaevitz.
\newblock {Implications of MicroBooNE\textquoteright{}s low sensitivity to
  electron antineutrino interactions in the search for the MiniBooNE excess}.
\newblock {\em Phys. Rev. D}, 107(9):092002, 2023.

\bibitem{moyalapprox}
J.~E. Moyal.
\newblock Stochastic processes and statistical physics.
\newblock {\em Journal of the Royal Statistical Society: Series B
  (Methodological)}, 11(2):150--210, 1949.

\bibitem{cholesky}
Nicholas Higham.
\newblock Cholesky factorization.
\newblock {\em Wiley Interdisciplinary Reviews: Computational Statistics},
  1:251 -- 254, 09 2009.

\bibitem{MicroBooNE:2021roa}
P.~Abratenko et~al.
\newblock {Novel approach for evaluating detector-related uncertainties in a
  LArTPC using MicroBooNE data}.
\newblock {\em Eur. Phys. J. C}, 82(5):454, 2022.

\bibitem{MicroBooNE:2019nio}
P.~Abratenko et~al.
\newblock {First Measurement of Inclusive Muon Neutrino Charged Current
  Differential Cross Sections on Argon at $E_\nu\sim$0.8 GeV with the
  MicroBooNE Detector}.
\newblock {\em Phys. Rev. Lett.}, 123(13):131801, 2019.

\bibitem{Day:2012gb}
Melanie Day and Kevin~S. McFarland.
\newblock {Differences in Quasi-Elastic Cross-Sections of Muon and Electron
  Neutrinos}.
\newblock {\em Phys. Rev. D}, 86:053003, 2012.

\bibitem{Calcutt:2021zck}
J.~Calcutt, C.~Thorpe, K.~Mahn, and Laura Fields.
\newblock {Geant4Reweight: a framework for evaluating and propagating hadronic
  interaction uncertainties in Geant4}.
\newblock {\em JINST}, 16(08):P08042, 2021.

\bibitem{10.1214/aoms/1177704472}
Emanuel Parzen.
\newblock {On Estimation of a Probability Density Function and Mode}.
\newblock {\em The Annals of Mathematical Statistics}, 33(3):1065 -- 1076,
  1962.

\bibitem{10.1214/aoms/1177728190}
Murray Rosenblatt.
\newblock {Remarks on Some Nonparametric Estimates of a Density Function}.
\newblock {\em The Annals of Mathematical Statistics}, 27(3):832 -- 837, 1956.

\bibitem{Ji:2019yca}
Xiangpan Ji, Wenqiang Gu, Xin Qian, Hanyu Wei, and Chao Zhang.
\newblock {Combined Neyman\textendash{}Pearson chi-square: An improved
  approximation to the Poisson-likelihood chi-square}.
\newblock {\em Nucl. Instrum. Meth. A}, 961:163677, 2020.

\bibitem{Junk:1999kv}
Thomas Junk.
\newblock {Confidence level computation for combining searches with small
  statistics}.
\newblock {\em Nucl. Instrum. Meth. A}, 434:435--443, 1999.

\bibitem{Read:451614}
A~L Read.
\newblock {Modified frequentist analysis of search results (the $CL_{s}$
  method)}.
\newblock 2000.

\bibitem{Denton:2021czb}
Peter~B. Denton.
\newblock {Sterile Neutrino Search with MicroBooNE\textquoteright{}s Electron
  Neutrino Disappearance Data}.
\newblock {\em Phys. Rev. Lett.}, 129(6):061801, 2022.

\bibitem{MicroBooNE:2022sdp}
P.~Abratenko et~al.
\newblock {First Constraints on Light Sterile Neutrino Oscillations from
  Combined Appearance and Disappearance Searches with the MicroBooNE Detector}.
\newblock {\em Phys. Rev. Lett.}, 130(1):011801, 2023.

\bibitem{Loinaz:2004qc}
Will Loinaz, Naotoshi Okamura, Saifuddin Rayyan, Tatsu Takeuchi, and L.~C.~R.
  Wijewardhana.
\newblock {The NuTeV anomaly, lepton universality, and nonuniversal neutrino
  gauge couplings}.
\newblock {\em Phys. Rev. D}, 70:113004, 2004.

\bibitem{NuSOnG:2008weg}
T.~Adams et~al.
\newblock {Terascale Physics Opportunities at a High Statistics, High Energy
  Neutrino Scattering Experiment: NuSOnG}.
\newblock {\em Int. J. Mod. Phys. A}, 24:671--717, 2009.

\bibitem{MINERvA:2015nqi}
J.~Park et~al.
\newblock {Measurement of Neutrino Flux from Neutrino-Electron Elastic
  Scattering}.
\newblock {\em Phys. Rev. D}, 93(11):112007, 2016.

\bibitem{Valencia:2019mkf}
E.~Valencia et~al.
\newblock {Constraint of the MINER$\nu$A medium energy neutrino flux using
  neutrino-electron elastic scattering}.
\newblock {\em Phys. Rev. D}, 100(9):092001, 2019.

\bibitem{MINERvA:2022vmb}
L.~Zazueta et~al.
\newblock {Improved constraint on the MINER\ensuremath{\nu}A medium energy
  neutrino flux using
  \ensuremath{\nu}\textasciimacron{}e-\textrightarrow{}\ensuremath{\nu}\textasciimacron{}e-
  data}.
\newblock {\em Phys. Rev. D}, 107(1):012001, 2023.

\bibitem{Shrock:1982sc}
Robert~E. Shrock.
\newblock {Electromagnetic Properties and Decays of Dirac and Majorana
  Neutrinos in a General Class of Gauge Theories}.
\newblock {\em Nucl. Phys. B}, 206:359--379, 1982.

\bibitem{Pal:1981rm}
Palash~B. Pal and Lincoln Wolfenstein.
\newblock {Radiative Decays of Massive Neutrinos}.
\newblock {\em Phys. Rev. D}, 25:766, 1982.

\bibitem{Babu:2020ivd}
K.~S. Babu, Sudip Jana, and Manfred Lindner.
\newblock {Large Neutrino Magnetic Moments in the Light of Recent Experiments}.
\newblock {\em JHEP}, 10:040, 2020.

\bibitem{Georgi:1990za}
Howard Georgi and Michael~E. Luke.
\newblock {Neutrino Moments, Masses and Custodial SU(2) Symmetry}.
\newblock {\em Nucl. Phys. B}, 347:1--11, 1990.

\bibitem{Lindner:2017uvt}
Manfred Lindner, Branimir Radov\v{c}i\'c, and Johannes Welter.
\newblock {Revisiting Large Neutrino Magnetic Moments}.
\newblock {\em JHEP}, 07:139, 2017.

\bibitem{Giunti:2008ve}
Carlo Giunti and Alexander Studenikin.
\newblock {Neutrino electromagnetic properties}.
\newblock {\em Phys. Atom. Nucl.}, 72:2089--2125, 2009.

\bibitem{deGouvea:2006hfo}
Andre de~Gouvea and James Jenkins.
\newblock {What can we learn from neutrino electron scattering?}
\newblock {\em Phys. Rev. D}, 74:033004, 2006.

\bibitem{Balantekin:2013sda}
A.~B. Balantekin and N.~Vassh.
\newblock {Magnetic moments of active and sterile neutrinos}.
\newblock {\em Phys. Rev. D}, 89(7):073013, 2014.

\bibitem{Vogel:1989iv}
P.~Vogel and J.~Engel.
\newblock {Neutrino Electromagnetic Form-Factors}.
\newblock {\em Phys. Rev. D}, 39:3378, 1989.

\bibitem{Kayser:1982br}
Boris Kayser.
\newblock {Majorana Neutrinos and their Electromagnetic Properties}.
\newblock {\em Phys. Rev. D}, 26:1662, 1982.

\bibitem{Babu:2021jnu}
K.~S. Babu, Sudip Jana, Manfred Lindner, and Vishnu~P. K.
\newblock {Muon g - 2 anomaly and neutrino magnetic moments}.
\newblock {\em JHEP}, 10:240, 2021.

\bibitem{Gninenko:1998nn}
S.~N. Gninenko and N.~V. Krasnikov.
\newblock {Limits on the magnetic moment of sterile neutrino and two photon
  neutrino decay}.
\newblock {\em Phys. Lett. B}, 450:165--172, 1999.

\bibitem{Coloma:2017ppo}
Pilar Coloma, Pedro A.~N. Machado, Ivan Martinez-Soler, and Ian~M. Shoemaker.
\newblock {Double-Cascade Events from New Physics in Icecube}.
\newblock {\em Phys. Rev. Lett.}, 119(20):201804, 2017.

\bibitem{Plestid:2020vqf}
Ryan Plestid.
\newblock {Luminous solar neutrinos I: Dipole portals}.
\newblock {\em Phys. Rev. D}, 104:075027, 2021.

\bibitem{Schwetz:2020xra}
Thomas Schwetz, Albert Zhou, and Jing-Yu Zhu.
\newblock {Constraining active-sterile neutrino transition magnetic moments at
  DUNE near and far detectors}.
\newblock {\em JHEP}, 21:200, 2020.

\bibitem{Atkinson:2021rnp}
Mack Atkinson, Pilar Coloma, Ivan Martinez-Soler, Noemi Rocco, and Ian~M.
  Shoemaker.
\newblock {Heavy Neutrino Searches through Double-Bang Events at
  Super-Kamiokande, DUNE, and Hyper-Kamiokande}.
\newblock {\em JHEP}, 04:174, 2022.

\bibitem{Bolton:2021pey}
Patrick~D. Bolton, Frank~F. Deppisch, K\r{a}re Fridell, Julia Harz, Chandan
  Hati, and Suchita Kulkarni.
\newblock {Probing active-sterile neutrino transition magnetic moments with
  photon emission from CE\ensuremath{\nu}NS}.
\newblock {\em Phys. Rev. D}, 106(3):035036, 2022.

\bibitem{Gustafson:2022rsz}
R.~Andrew Gustafson, Ryan Plestid, and Ian~M. Shoemaker.
\newblock {Neutrino Portals, Terrestrial Upscattering, and Atmospheric
  Neutrinos}.
\newblock 5 2022.

\bibitem{Ovchynnikov:2022rqj}
Maksym Ovchynnikov, Thomas Schwetz, and Jing-Yu Zhu.
\newblock {Dipole portal and neutrinophilic scalars at DUNE revisited: The
  importance of the high-energy neutrino tail}.
\newblock {\em Phys. Rev. D}, 107(5):055029, 2023.

\bibitem{Zhang:2023nxy}
Yu~Zhang and Wei Liu.
\newblock {Probing active-sterile neutrino transition magnetic moments at LEP
  and CEPC}.
\newblock 1 2023.

\bibitem{Brdar:2023tmi}
Vedran Brdar, Andr\'e de~Gouv\^ea, Ying-Ying Li, and Pedro A.~N. Machado.
\newblock {Neutrino magnetic moment portal and supernovae: New constraints and
  multimessenger opportunities}.
\newblock {\em Phys. Rev. D}, 107(7):073005, 2023.

\bibitem{Mohapatra:1986aw}
R.~N. Mohapatra.
\newblock {Mechanism for Understanding Small Neutrino Mass in Superstring
  Theories}.
\newblock {\em Phys. Rev. Lett.}, 56:561--563, 1986.

\bibitem{Mohapatra:1986bd}
R.~N. Mohapatra and J.~W.~F. Valle.
\newblock {Neutrino Mass and Baryon Number Nonconservation in Superstring
  Models}.
\newblock {\em Phys. Rev. D}, 34:1642, 1986.

\bibitem{Arguelles:2021dqn}
Carlos~A. Arg\"uelles, Nicol\`o Foppiani, and Matheus Hostert.
\newblock {Heavy neutral leptons below the kaon mass at hodoscopic neutrino
  detectors}.
\newblock {\em Phys. Rev. D}, 105(9):095006, 2022.

\bibitem{T2K:2019jwa}
K.~Abe et~al.
\newblock {Search for heavy neutrinos with the T2K near detector ND280}.
\newblock {\em Phys. Rev. D}, 100(5):052006, 2019.

\bibitem{E949:2014gsn}
A.~V. Artamonov et~al.
\newblock {Search for heavy neutrinos in $K^+\to\mu^+\nu_H$ decays}.
\newblock {\em Phys. Rev. D}, 91(5):052001, 2015.
\newblock [Erratum: Phys.Rev.D 91, 059903 (2015)].

\bibitem{Bernardi:1987ek}
G.~Bernardi et~al.
\newblock {FURTHER LIMITS ON HEAVY NEUTRINO COUPLINGS}.
\newblock {\em Phys. Lett. B}, 203:332--334, 1988.

\bibitem{Voloshin:1987qy}
M.~B. Voloshin.
\newblock {On Compatibility of Small Mass with Large Magnetic Moment of
  Neutrino}.
\newblock {\em Sov. J. Nucl. Phys.}, 48:512, 1988.

\bibitem{Babu:1989wn}
K.~S. Babu and R.~N. Mohapatra.
\newblock {Model for Large Transition Magnetic Moment of the $\nu_e$}.
\newblock {\em Phys. Rev. Lett.}, 63:228, 1989.

\bibitem{Leurer:1989hx}
Miriam Leurer and Neil Marcus.
\newblock {A Model for a Large Neutrino Magnetic Transition Moment and
  Naturally Small Mass}.
\newblock {\em Phys. Lett. B}, 237:81--87, 1990.

\bibitem{Babu:1989px}
K.~S. Babu and R.~N. Mohapatra.
\newblock {Supersymmetry and Large Transition Magnetic Moment of the Neutrino}.
\newblock {\em Phys. Rev. Lett.}, 64:1705, 1990.

\bibitem{Barbieri:1988fh}
Riccardo Barbieri and Rabindra~N. Mohapatra.
\newblock {A Neutrino With a Large Magnetic Moment and a Naturally Small Mass}.
\newblock {\em Phys. Lett. B}, 218:225--229, 1989.

\bibitem{Gariazzo:2017fdh}
S.~Gariazzo, C.~Giunti, M.~Laveder, and Y.~F. Li.
\newblock {Updated Global 3+1 Analysis of Short-BaseLine Neutrino
  Oscillations}.
\newblock {\em JHEP}, 06:135, 2017.

\bibitem{Maltoni:2003cu}
M.~Maltoni and T.~Schwetz.
\newblock {Testing the statistical compatibility of independent data sets}.
\newblock {\em Phys. Rev. D}, 68:033020, 2003.

\bibitem{McKeen:2010rx}
David McKeen and Maxim Pospelov.
\newblock {Muon Capture Constraints on Sterile Neutrino Properties}.
\newblock {\em Phys. Rev. D}, 82:113018, 2010.

\bibitem{Shoemaker:2018vii}
Ian~M. Shoemaker and Jason Wyenberg.
\newblock {Direct Detection Experiments at the Neutrino Dipole Portal
  Frontier}.
\newblock {\em Phys. Rev. D}, 99(7):075010, 2019.

\bibitem{Fricke:1995zz}
G.~Fricke, C.~Bernhardt, K.~Heilig, L.~A. Schaller, L.~Schellenberg, E.~B.
  Shera, and C.~W. de~Jager.
\newblock {Nuclear Ground State Charge Radii from Electromagnetic
  Interactions}.
\newblock {\em Atom. Data Nucl. Data Tabl.}, 60:177--285, 1995.

\bibitem{DeVries:1987atn}
H.~De~Vries, C.~W. De~Jager, and C.~De~Vries.
\newblock {Nuclear charge and magnetization density distribution parameters
  from elastic electron scattering}.
\newblock {\em Atom. Data Nucl. Data Tabl.}, 36:495--536, 1987.

\bibitem{DeJager:1974liz}
C.~W. De~Jager, H.~De~Vries, and C.~De~Vries.
\newblock {Nuclear charge and magnetization density distribution parameters
  from elastic electron scattering}.
\newblock {\em Atom. Data Nucl. Data Tabl.}, 14:479--508, 1974.
\newblock [Erratum: Atom.Data Nucl.Data Tabl. 16, 580--580 (1975)].

\bibitem{VT_NDT}
\url{http://discovery.phys.virginia.edu/research/groups/ncd/index.html}.

\bibitem{IceCube:2020tcq}
R.~Abbasi et~al.
\newblock {LeptonInjector and LeptonWeighter: A neutrino event generator and
  weighter for neutrino observatories}.
\newblock {\em Comput. Phys. Commun.}, 266:108018, 2021.

\bibitem{CCMWorkshop}
{CCM} workshop 2023.
\newblock \url{https://ccm.mit.edu/ccm-workshop-2023}.

\bibitem{LeptonInjector}
\texttt{LeptonInjector}.
\newblock \url{https://github.com/Harvard-Neutrino/LeptonInjector}.

\bibitem{10.5555/1593511}
Guido Van~Rossum and Fred~L. Drake.
\newblock {\em Python 3 Reference Manual}.
\newblock CreateSpace, Scotts Valley, CA, 2009.

\bibitem{Abdullahi:2022cdw}
Asli~M. Abdullahi, Jaime Hoefken~Zink, Matheus Hostert, Daniele Massaro, and
  Silvia Pascoli.
\newblock {DarkNews: a Python-based event generator for heavy neutral lepton
  production in neutrino-nucleus scattering}.
\newblock 7 2022.

\bibitem{MiniBooNE_eff}
Z.~Pavlovic, R.G. Van~de Water, and S.~Zeller.
\newblock Miniboone gamma-ray and electron efficiencies.
\newblock Technical report, 2012.

\bibitem{MBres}
Private communication with MiniBooNE Collaboration.

\bibitem{Shaevitz:2008zza}
Michael~H. Shaevitz.
\newblock {MiniBooNE oscillation results and implications}.
\newblock {\em J. Phys. Conf. Ser.}, 120:052003, 2008.

\bibitem{hepdata.114365}
{MiniBooNE Collaboration}.
\newblock {Updated MiniBooNE Neutrino Oscillation Results with Increased Data
  and New Background Studies}.
\newblock {HEPData (collection)}, 2021.
\newblock \url{https://doi.org/10.17182/hepdata.114365}.

\bibitem{COHERENT:2017ipa}
D.~Akimov et~al.
\newblock {Observation of Coherent Elastic Neutrino-Nucleus Scattering}.
\newblock {\em Science}, 357(6356):1123--1126, 2017.

\bibitem{CCM:2021yzc}
A.~A. Aguilar-Arevalo et~al.
\newblock {First Leptophobic Dark Matter Search from the
  Coherent\textendash{}CAPTAIN-Mills Liquid Argon Detector}.
\newblock {\em Phys. Rev. Lett.}, 129(2):021801, 2022.

\bibitem{CCM:2021lhc}
A.~A. Aguilar-Arevalo et~al.
\newblock {Axion-Like Particles at Coherent CAPTAIN-Mills}.
\newblock 12 2021.

\bibitem{NELSON2012172}
A.T. Nelson, J.A. O'Toole, R.A. Valicenti, and S.A. Maloy.
\newblock Fabrication of a tantalum-clad tungsten target for lansce.
\newblock {\em Journal of Nuclear Materials}, 431(1):172--184, 2012.
\newblock Special Issue of the Tenth International Workshop on Spallation
  Materials Technology, (IWSMT-10).

\bibitem{LISOWSKI2006910}
Paul~W. Lisowski and Kurt~F. Schoenberg.
\newblock The los alamos neutron science center.
\newblock {\em Nuclear Instruments and Methods in Physics Research Section A:
  Accelerators, Spectrometers, Detectors and Associated Equipment},
  562(2):910--914, 2006.
\newblock Proceedings of the 7th International Conference on Accelerator
  Applications.

\bibitem{CAPTAIN:2013irr}
H.~Berns et~al.
\newblock {The CAPTAIN Detector and Physics Program}.
\newblock In {\em {Snowmass 2013}: {Snowmass on the Mississippi}}, 9 2013.

\bibitem{EdThesis}
Edward Dunton.
\newblock {\em A Search for Axion-like Particles at the Coherent CAPTAIN Mills
  Experiment}.
\newblock PhD thesis, Columbia University, 2022.

\bibitem{Simon:2002cw}
E.~Simon et~al.
\newblock {SICANE: A Detector array for the measurement of nuclear recoil
  quenching factors using a monoenergetic neutron beam}.
\newblock {\em Nucl. Instrum. Meth. A}, 507:643--656, 2003.

\bibitem{COHERENT:2020iec}
D.~Akimov et~al.
\newblock {First Measurement of Coherent Elastic Neutrino-Nucleus Scattering on
  Argon}.
\newblock {\em Phys. Rev. Lett.}, 126(1):012002, 2021.

\bibitem{CAPTAIN:2020pup}
C.~E. Taylor et~al.
\newblock {The Mini-CAPTAIN liquid argon time projection chamber}.
\newblock {\em Nucl. Instrum. Meth. A}, 1001:165131, 2021.

\bibitem{Grupen:2008zz}
Claus Grupen and Boris Schwartz.
\newblock {\em {Particle detectors}}.
\newblock Cambridge Univ. Pr., Cambridge, UK, 2008.

\bibitem{Kaptanoglu:2021prv}
T.~Kaptanoglu, E.~J. Callaghan, M.~Yeh, and G.~D.~Orebi Gann.
\newblock {Cherenkov and scintillation separation in water-based liquid
  scintillator using an LAPPD$^{TM}$}.
\newblock {\em Eur. Phys. J. C}, 82(2):169, 2022.

\bibitem{Caravaca:2020lfs}
J.~Caravaca, B.~J. Land, M.~Yeh, and G.~D. Orebi~Gann.
\newblock {Characterization of water-based liquid scintillator for Cherenkov
  and scintillation separation}.
\newblock {\em Eur. Phys. J. C}, 80(9):867, 2020.

\bibitem{Dunger:2022gif}
Jack Dunger, Edward~J. Leming, and Steven~D. Biller.
\newblock {Slow-fluor scintillator for low energy solar neutrinos and
  neutrinoless double beta decay}.
\newblock {\em Phys. Rev. D}, 105(9):092006, 2022.

\bibitem{Gruszko:2018gzr}
Julieta Gruszko, Brian Naranjo, Byron Daniel, Andrey Elagin, Diana Gooding,
  Chris Grant, Jonathan Ouellet, and Lindley Winslow.
\newblock {Detecting Cherenkov light from 1\textendash{}2 MeV electrons in
  linear alkylbenzene}.
\newblock {\em JINST}, 14(02):P02005, 2019.

\bibitem{BOREXINO:2021efb}
M.~Agostini et~al.
\newblock {First Directional Measurement of Sub-MeV Solar Neutrinos with
  Borexino}.
\newblock {\em Phys. Rev. Lett.}, 128(9):091803, 2022.

\bibitem{Antonello:2004sx}
M.~Antonello et~al.
\newblock {Detection of Cherenkov light emission in liquid argon}.
\newblock {\em Nucl. Instrum. Meth. A}, 516:348--363, 2004.

\bibitem{Benson:2017vbw}
Christopher Benson, Gabriel Orebi~Gann, and Victor Gehman.
\newblock {Measurements of the intrinsic quantum efficiency and absorption
  length of tetraphenyl butadiene thin films in the vacuum ultraviolet regime}.
\newblock {\em Eur. Phys. J. C}, 78(4):329, 2018.

\bibitem{Bae:2022dti}
Junghyun Bae and Stylianos Chatzidakis.
\newblock {Fieldable muon spectrometer using multi-layer pressurized gas
  Cherenkov radiators and its applications}.
\newblock {\em Sci. Rep.}, 12(1):2559, 2022.

\bibitem{FLOURNOY1994349}
J.M. Flournoy, I.B. Berlman, B.~Rickborn, and R.~Harrison.
\newblock Substituted tetraphenylbutadienes as fast scintillator solutes.
\newblock {\em Nuclear Instruments and Methods in Physics Research Section A:
  Accelerators, Spectrometers, Detectors and Associated Equipment},
  351(2):349--358, 1994.

\bibitem{photonics2000photomultiplier}
Hamamatsu Photonics.
\newblock {\em Photomultiplier tubes}.
\newblock Hamamatsu, 2000.

\bibitem{IceCube:2020nwx}
M.~G. Aartsen et~al.
\newblock {In-situ calibration of the single-photoelectron charge response of
  the IceCube photomultiplier tubes}.
\newblock {\em JINST}, 15(06):P06032, 2020.

\bibitem{cosmicwatch}
Cosmicwatch.
\newblock \url{http://www.cosmicwatch.lns.mit.edu/}.

\end{thebibliography}
\bibliographystyle{unsrt}
\end{singlespace}

\end{document}